\documentclass[review]{elsarticle}
\usepackage[colorlinks,citecolor=blue,linktoc=all,linkcolor=cyan]{hyperref}
\usepackage{graphicx}

\usepackage[T1]{fontenc}
\usepackage{dsfont}               % use mathds instead of mathbb for outline fonts
\usepackage{mathrsfs}             % provides mathscr without overwriting mathcal
\usepackage{slashed}              % For Dirac slash notation.
\usepackage{amsmath}
\usepackage{amssymb}
\usepackage{amsbsy}
\usepackage{amsfonts}

\usepackage{graphicx,subfigure,xspace}
\usepackage{feynmp-auto}
\usepackage{multicol,multirow,setspace}
\usepackage[usenames,dvipsnames]{color,xcolor}
\usepackage[utf8]{inputenc}
\usepackage{float}
\usepackage{orcidlink}
\usepackage{graphicx,color,rotating}% Include figure files
\usepackage{dcolumn}% Align table columns on decimal point
\usepackage{bm}
\usepackage{units,comment}
\usepackage[colorinlistoftodos]{todonotes}
\usepackage{subcaption,capt-of} 
\usepackage{wrapfig}
\usepackage[export]{adjustbox}
\usepackage[permil]{overpic}
\usepackage{titlesec}
\usepackage{ulem}
\usepackage[rightcaption]{sidecap}

        \def\longlonglongrightarrow{
        \relbar\joinrel\relbar\joinrel\relbar\joinrel\relbar\joinrel\relbar\joinrel\relbar\joinrel\rightarrow}

\numberwithin{equation}{section}
\numberwithin{table}{section}
\numberwithin{figure}{section}

\journal{Progress in Particle and Nuclear Physics}

\usepackage{titlesec}
\usepackage{sectsty}
\titleformat{\section}{\normalfont\Large\bfseries}{\thesection}{1em}{}
\titleformat{\subsection}{\normalfont\large\bfseries}{\thesubsection}{1em}{}
\titleformat{\subsubsection}{\normalfont\normalsize\bfseries}{\thesubsubsection}{1em}{}
\biboptions{sort&compress}
\begin{document}

\def\Ast{\hbox{$\ast$}}
\def\twoast{\hbox{\Ast\Ast}}
\def\fourast{\hbox{\Ast\Ast\Ast\Ast}}
\def\threeast{\hbox{\Ast\Ast\Ast}}
\newcommand{\nonett}[9]
{
\setlength{\unitlength}{0.8mm}
\begin{picture}(150.00,90.00)
\put(10.00,45.00){\vector(1,0){70.00}}
\put(45.00,10.00){\vector(0,1){70.00}}
\put(110.00,45.00){\vector(1,0){30.00}}
\put(125.00,30.00){\vector(0,1){30.00}}
\put(82.50,42.50){\makebox(5.00,5.00){$I_3$}}
\put(142.50,42.50){\makebox(5.00,5.00){$I_3$}}
\put(127.50,60.00){\makebox(5.00,5.00){\quad\ S\quad\ Singlet }}
\put(35.50,78.00){\makebox(5.00,5.00){\quad\ S }}
\put(45.50,78.00){\makebox(5.00,5.00){\quad\ \quad\ Octet }}
\put(45.00,45.00){\circle*{2.00}}
\put(70.00,45.00){\circle*{2.00}}
\put(20.00,45.00){\circle*{2.00}}
\put(32.50,70.00){\rd\circle*{2.00}}
\put(57.50,70.00){\rd\circle*{2.00}}
\put(32.50,20.00){\circle*{2.00}}
\put(57.50,20.00){\circle*{2.00}}
\put(125.00,45.00){\circle*{2.00}}
\put(45.00,45.00){\circle{0.00}}
\put(45.00,45.00){\circle{5.00}}
\put(32.50,70.00){\line(1,0){25.00}}
\put(57.50,70.00){\line(1,-2){12.50}}
\put(70.00,45.00){\line(-1,-2){12.50}}
\put(57.50,20.00){\line(-1,0){25.00}}
\put(32.50,20.00){\line(-1,2){12.50}}
\put(20.00,45.00){\line(1,2){12.50}}
\put(60.00,70.00){\makebox(15.00,5.00)[l]{#2}}
\put(15.00,70.00){\makebox(15.00,5.00)[r]{#1}}
\put(15.00,15.00){\makebox(15.00,5.00)[r]{#3}}
\put(6.75,37.50){\makebox(13.25,5.00)[r]{#5}}
\put(60.00,15.00){\makebox(15.00,5.00)[l]{#4}}
\put(70.00,37.50){\makebox(13.75,5.00)[l]{#6}}
\put(47.50,37.50){\makebox(12.50,5.00)[l]{#7}}
\put(47.50,47.50){\makebox(12.50,5.00)[l]{#8}}
\put(127.50,47.50){\makebox(12.50,5.00)[l]{#9}}
\end{picture}
} 
\newcommand{\oo}{$^o$}
\newcommand {\be} {\begin{equation}}
\newcommand {\ee} {\end{equation}}
\newcommand {\ba} {\begin{eqnarray}}
\newcommand {\ea} {\end{eqnarray}}
\newcommand {\bq} {\begin{eqnarray}}
\newcommand {\eq} {\end{eqnarray}}
\newcommand {\bc} {\begin{center}}
\newcommand {\ec} {\end{center}}
\newcommand {\bea} {\begin{eqnarray}}
\newcommand {\eea} {\end{eqnarray}}
\newcommand{\no}{\nonumber \\}
\newcommand{\nnb}{\nonumber}
\newcommand{\nn}{\nonumber}
\newcommand{\ra}{\rightarrow}
\newcommand {\er} {$\pm$}
\newcommand{\MeV}{\,\text{MeV}}
\newcommand{\GeV}{\,\text{GeV}}
\definecolor{darkred}{rgb}{0.7,0,0}
\definecolor{red2}{rgb}{0.9,0.0,0.6}
\definecolor{darkgreen}{rgb}{0,0.4,0}
\definecolor{lightgreen}{rgb}{0.55,0.71,0.0}
\definecolor{darkblue}{rgb}{0,0,1.0}
\definecolor{bl2}{rgb}{0,0.7,1.}
\definecolor{darkbrown}{rgb}{0.5,0,0}
\definecolor{br2}{rgb}{0.8,0.6,0.5}
\definecolor{darkcyan}{cmyk}{1,0.3,0.3,0.3}
\definecolor{midgreen}{rgb}{0,0.6,0}
\definecolor{grey}{rgb}{0.5,0.5,0.5}
\newcommand {\rd}{\color{red}}
\newcommand {\rdl}{\color{pink}}
\newcommand {\bl}{\color{blue}}
\newcommand {\bll}{\color{bl2}}
\newcommand {\gr}{\color{midgreen}}
\newcommand {\grl}{\color{lightgreen}}
\newcommand {\br}{\color{darkbrown}}
\newcommand {\cy}{\color{cyan}}
\newcommand {\mg}{\color{magenta}}
\newcommand {\bk}{\color{black}}
\def\p{\partial}
\newcommand{\conjg}[1]{\ensuremath{\hspace{1pt}\overline{\hspace{-1pt}#1\hspace{-1pt}}}\hspace{1pt}}
\newcommand{\vect}[1]{{\mbox{\boldmath $#1$}}}
%\numberwithin{equation}{subsection}
\newcommand{\ph}{\phantom}

\newcommand{\GE}[1]{{\color{magenta}[GE: #1]}}
\newcommand{\EK}[1]{{\color{red}[EK: #1]}}

        \def\mA{\ensuremath{\mathcal{A}}}
        \def\mC{\ensuremath{\mathcal{C}}}
        \def\mD{\ensuremath{\mathcal{D}}}
        \def\mF{\ensuremath{\mathcal{F}}}
        \def\mL{\ensuremath{\mathcal{L}}}
        \def\mM{\ensuremath{\mathcal{M}}}
        \def\mN{\ensuremath{\mathcal{N}}}
        \def\mO{\ensuremath{\mathcal{O}}}
        \def\mP{\ensuremath{\mathcal{P}}}
        \def\mS{\ensuremath{\mathcal{S}}}
        \def\mT{\ensuremath{\mathcal{T}}}

\newcommand{\EPJA}{With kind permission of The European Physical Journal (EPJ). }
\newcommand{\APS}[2]{Reprinted figure with permission from #1. Copyright #2 by the American Physical Society. }
\newcommand{\FewBody}[2]{Reprinted by permission from Few-Body Systems: #1, Copyright #2. }
\newcommand{\NuclPhysA}[2]{Reprinted by permission from Nuclear Physics A: #1, Copyright #2. }

\setlength{\baselineskip}{12pt}
\begin{frontmatter}

\title{The impact of \texorpdfstring{$\gamma N$}~ and \texorpdfstring{$\gamma^* N$}\ \ interactions on our
understanding of nucleon excitations }

\author[JLab]{Volker Burkert} \ead{Bukert@jlab.org}
\author[Graz]{Gernot Eichmann} \ead{gernot.eichmann@uni-graz.at}
\author[JLab,Bonn]{Eberhard~Klempt}
		\ead{Klempt@hiskp.uni-bonn.de}

\address[JLab]{Thomas Jefferson National Accelerator Facility, Newport News, VA 23606, USA}

\address[Graz]{Institute of Physics, University of Graz, NAWI Graz, Universit\"atsplatz 5, 8010 Graz, Austria}

\address[Bonn]{Helmholtz-Institut f\"ur Strahlen- und Kernphysik 
der Rheinischen Friedrich-Wilhelms Universit\"at, Nussallee 14 - 16, 53115 Bonn, Germany}

\date{\today}% It is always \today, today,

\begin{abstract}

We review recent progress in our understanding of the nucleon excitation spectrum. Thanks to dedicated efforts at facilities such as ELSA, MAMI and Jefferson Lab, several new nucleon resonances have been discovered, and evidence for previously elusive states has been significantly improved. Numerous decay channels have  been observed for the first time, and resonance properties are being extracted from these data by several groups through coupled-channel analyses of varying complexity. Electroproduction experiments have provided further insights into the internal structure of light baryon resonances -- for example, the long-debated Roper resonance $N(1440)$ is observed as a three-quark state with a significant meson-cloud component. While the non-relativistic quark model remains a valuable tool for organizing the spectrum of nucleon and $\Delta$ resonances, a variety of theoretical frameworks have emerged to offer deeper understanding, including phenomenological quark models, holographic QCD, functional methods, effective field theories, and lattice QCD. We examine the interplay between these approaches, highlight their respective strengths and explore how they complement each other in shaping our knowledge of  light baryon resonances. We address several open questions in baryon spectroscopy, including the nature of the enigmatic $\Lambda(1405)$, ongoing searches  for exotic states such as hybrid baryons and pentaquarks, and the dichotomy between microscopic descriptions of baryons in terms of quarks and gluons versus effective hadronic descriptions based on meson-baryon dynamics.

\end{abstract}

\end{frontmatter}

	\thispagestyle{empty}
	\tableofcontents

\section{\label{intro}Introduction}
Quantum Chromodynamics (QCD), the theory describing the interaction between colored quarks and colored gluons \cite{Gross:2022hyw}, encompasses a vast range of physics. Over time, it explains how quarks and gluons formed a quark-gluon plasma (QGP) soon after the end
of the Universe’s inflationary phase and the subsequent formation of mesons and baryons approximately 10\,$\mu$s later. QCD is essential for understanding stellar evolution and the structure of protons and neutrons. Moreover, it plays a crucial role in various areas of modern physics, such as calculations of the ``background" for Higgs particle production at the LHC at 
multi-TeV energies and determining the hadronic contributions to the muon's anomalous magnetic moment~\cite{Aliberti:2025beg}.

\begin{figure}[bht]
\centering
\includegraphics[width=0.46\linewidth]{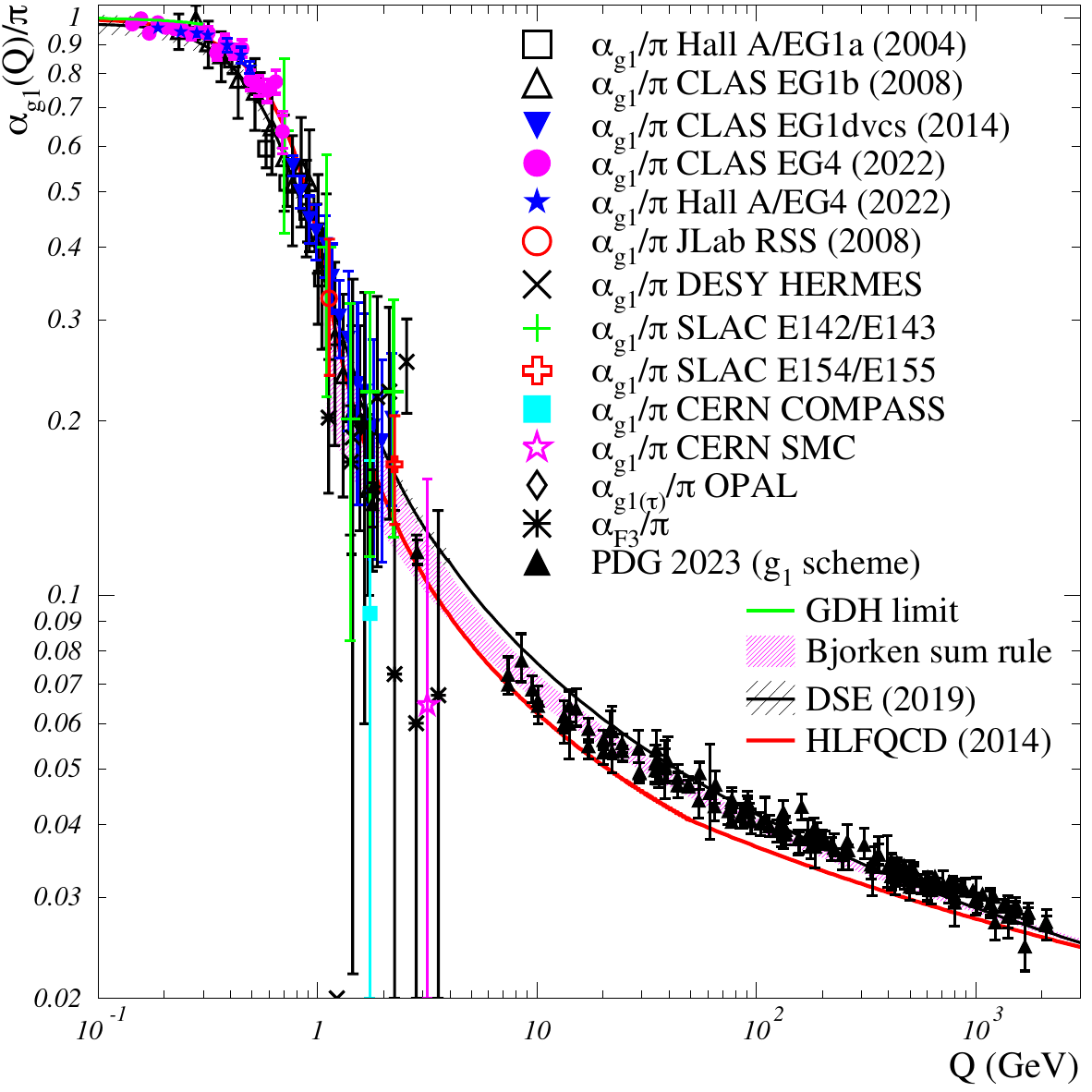} 
\caption{(color online) The strong interaction coupling constant, covering the non-perturbative to perturbative regions. References
to the data can be found in \protect \cite{Deur:2022msf}.}
\label{fig}
\end{figure}

Across this broad spectrum, the QCD coupling constant $\alpha_s$ varies significantly. At high momentum transfers, $\alpha_s$ is small,
allowing perturbative methods to be applied for computing physical quantities. However, at low momentum transfers, $\alpha_s$
increases, necessitating non-perturbative  
calculations. Despite these challenges, a comprehensive view of the
coupling constant is illuminating (see Fig.~\ref{fig}). Perturbative determinations are compiled by the Particle Data Group \cite{ParticleDataGroup:2024cfk}, while non-perturbative results 
with references to the relevant calculations can be found in Ref.~\cite{Deur:2022msf}.

Quarks confined within hadrons experience very low momentum transfers, making the hadron 
spectrum difficult to calculate and
understand. Gell-Mann \cite{Gell-Mann:1964ewy} and Zweig \cite{Zweig:1964jf} proposed to interpret mesons as bound states
of a quark and an antiquark, and baryons as bound states of three quarks. The meson spectrum already presents a challenge —
so why should we focus on baryons?

``Why $N^*$’s?” This was the question with which Nathan Isgur began his talk at $N^*$2000 \cite{Isgur:2000ad}, held at the 
Thomas Jefferson National Accelerator Facility in Newport News, VA, one year before his untimely passing. He provided three answers:
First, ``Nucleons are the building blocks of our world. As such, they must be central to any discussion of why the world
we experience has the characteristics it does."
Second, nucleons ``are the simplest system in which the quintessential non-Abelian character of QCD is manifest. After all,
there are $N_c$ quarks in a proton because there are $N_c$ colors.”
Third, ``baryons are sufficiently complex to reveal physics that remains hidden to us in the mesons." Gell-Mann and Zweig
did not fully develop their quark model based on mesons, whose simpler structure allowed for alternative interpretations.

\noindent
The combination of three quarks forms a baryon structure, which, within $SU(3)_f$ symmetry, gives the octet and decuplet
including the famous $\Omega^-$.

Isgur’s talk addressed topics still relevant today:
\begin{enumerate}
\item What are the relevant low-energy degrees of freedom?
\item Why is the non-relativistic quark model so successful?
\item What is the structure of excited states?
\item Is there an effective ``residual" quark-quark interaction in baryons on top of confinement?
\item Is the number of states reduced to those expected from quark-diquark models with a quasi-stable diquark?
\item Where are hybrid baryons?
\end{enumerate}

When Isgur gave his talk, new facilities had just come online, and most of our knowledge of nucleon excited states
still stemmed from research in the 1960s and 1970s. Additional questions have since arisen, some predating his talk,
others emerging afterward:
\begin{enumerate}
\item Do pentaquarks exist?
\item Is there evidence for three-body dynamics in baryons?
\item How do constituent quarks evolve from current quarks?
\item Is there chiral symmetry restoration in high-mass baryons?
\item What is the relationship between dynamically generated resonances and quark-model states?
\end{enumerate}

Several excellent reviews on nucleon excitations have been written. In 1982, Hey and Kelly \cite{Hey:1982aj} reviewed
symmetries and quark model calculations, comparing experimental results from $\pi N$ interactions with theoretical predictions. First results from the new electron accelerator facilities were discussed by Burkert and Lee~\cite{Burkert:2004sk}, where the formalism for electromagnetic meson production was laid out and theoretical models discussed.   
Early photoproduction experiments were summarized by Klempt and Richard \cite{Klempt:2009pi}, while Crede and
Roberts \cite{Crede:2013kia} discussed advancements in data analysis and new observables. Ireland, Pasyuk, and
Strakovsky \cite{Ireland:2019uwn} reviewed photon beam experiments and compiled an extensive set of experimental
results with comprehensive references. Eichmann delved into baryon theory, including quark-model wave functions and
advancements in solving the three-body Bethe-Salpeter equation \cite{Eichmann:2022zxn}. Thiel, Afzal, and
Wunderlich \cite{Thiel:2022xtb} provided an extensive review of baryon spectra, spin formalisms, and the state of
``complete experiments," summarizing major data sets and partial-wave analyses. Mai, Mei\ss ner,
and Urbach reviewed the current state of the art of our understanding of the spectrum of excited strongly interacting particles~\cite{Mai:2022eur}. Dynamical coupled channel approaches aiming at a reliable 
extraction of resonance spectra and their properties from data were discussed by Döring, Haidenbauer, 
Mai, and Sato~\cite{Doring:2025sgb}.
Aznauryan and Burkert \cite{Aznauryan:2011qj,Burkert:2022ioj} and Carman, Jo, and Mokeev \cite{Carman:2020qmb}
reviewed progress in understanding nucleon resonance structures through meson electroproduction data, exploring
the interplay between a core of three dressed quarks and the surrounding meson-baryon cloud. 
Ramalho and Pe\~na \cite{Ramalho:2023hqd} wrote a comprehensive review on recent experimental and 
theoretical achievements in our understanding of the electromagnetic structure of nucleons and nucleon excitations.
The authors of Ref.~\cite{Achenbach:2025kfx} focus on  nucleon resonance electroexcitation
and the emergence of hadron mass.

This review focuses on interpreting the results achieved thus far: How do theories of nucleon structure and
excitation align with recent findings? In Section~\ref{status}, we present key data on $N^*$ and $\Delta^*$ resonances 
and summarize the various analytical approaches to extract the properties of resonances. This is followed by a 
table of the resonances and a list of the decay modes that have been determined. 

In Section~\ref{Models}, we discuss the basics of baryon spectroscopy. A survey is provided of harmonic oscillator
wave functions which are then used to classify light-quark baryon resonances. 
The status of hyperon spectroscopy is briefly discussed, and  
a short list of references is included for baryon-decay calculations. 

In Section \ref{QCD}, approximations consistent with QCD are reviewed. At low energies,
QCD can be described by effective field theories in which hadrons are the elementary building blocks.
We compare the resonance spectrum to predictions of different quark models, starting from so-called {\it constituent} 
quarks with large constituent-quark masses. 
% The comparison reveals successes and failures of these models.
% The mass pattern of resonances provides support for instanton-induced interactions in the spectrum. 
The strong interaction constant saturates at low $Q^2$ (see Fig.\ref{fig}), a fact exploited in AdS/QCD,
which successfully predicts the full excitation spectrum of the nucleon, provided the AdS/QCD mass formula 
is given a new interpretation. 
The baryon spectrum can also be calculated using functional methods, which provide insights into
QCD's non-perturbative phenomena such as the generation of dynamical mass.
Significant progress has been achieved in recent years in calculating hadron masses in lattice QCD.
While the baryon ground-state masses have been calculated using quark masses close to or at their ``true" values,
studying the  excitation spectrum is still an ongoing endeavor. 
% For
% the full excitation spectrum, the quark masses lead to a pion mass of about 400,MeV.

Section~\ref{Structure} covers electroproduction data exploring the internal
structure of baryon resonances. 
The status of searches for baryonic hybrids and light-quark pentaquarks
is reviewed in Section~\ref{Out}. Specific questions, such as the $\Lambda(1405)$ and chiral symmetry
restoration are discussed.
The review concludes
with a summary in Section~\ref{Summary}.
Throughout this review, we offer personal perspectives on key issues.

\clearpage

\section{\label{status}Data and Analyses}

\subsection{\label{Data}Data}

\smallskip
\textbf{\quad\ Pion-induced reactions:}
For many years, most of our knowledge about the excited states of the nucleon was based on analyses of $\pi N$ elastic scattering conducted by three major groups: the Karlsruhe-Helsinki (KH) group, the Carnegie-Mellon-Berkeley (CMB) group, and the SAID group at George Washington University. Early data on $\pi N$ elastic scattering were compiled in \cite{Hohler:1977em}, with results reported in \cite{Hohler:1979yr,Koch:1980ay,Koch:1983ht,Hohler:1984ux,Koch:1985bp,Koch:1985bn,Hohler:1993lbk}.

The KH group constructed $\pi N$ partial wave amplitudes constrained by a simultaneous 
fixed momentum transfer analysis, thus satisfying $s-u$ crossing symmetry. Different types
of dispersion relations were imposed.

The CMB group emphasized the importance of performing a rigorous statistical analysis. They constructed a new dataset with predefined energy bins and solid angles \cite{Kelly:1979uf}. This method, termed ``amalgamation", accounted for normalization errors, momentum calibration errors, momentum resolution, electromagnetic corrections, threshold structures, and inconsistencies among datasets. The full covariance matrix of the amalgamated data was calculated, incorporating both statistical and systematic errors.

The SAID group has continuously updated the database. References to the data are available at 
Ref.~\cite{gwdac}.

Only a few elastic $\pi N$ scattering data have been reported since 1990. Notably, spin rotation parameter measurements conducted at ITEP, Moscow, made significant contributions 
\cite{Alekseev:2005zr,Alekseev:1996gs,ITEP-PNPI:2000gqp,Alekseev:2005zr,ITEP-PNPI:2008cmv}.

Inelastic pion-induced reactions have also contributed to our understanding. Pioneering work by Longacre and collaborators \cite{Herndon:1974xd,Longacre:1975qj,Longacre:1976ja,Longacre:1977ga}, as well as Novoseller \cite{Novoseller:1977xt,Novoseller:1977xs}, included data on inelastic channels and reported on $N^*$ and $\Delta^*$ couplings to $N\pi\pi$ final states.
Most data on inelastic pion scattering originate from bubble-chamber experiments, with analyses conducted by the Kent group. References to these data and their results are provided in \cite{Manley:1984zs,Manley:1992yb}. More recent counter experiments have focused on two-pion production in $\pi N$ scattering. At BNL, the reaction $\pi^-p \to \pi^0\pi^0p$ was measured in the beam momentum range 270–750~MeV using the Crystal Ball multiphoton spectrometer \cite{CrystalBall:2004qln}. The HADES Collaboration investigated $\pi^-p \to \pi^+\pi^-p$ at four pion beam momenta: 650, 685, 733, and 786~MeV~\cite{HADES:2020kce}.

Scattering of $\pi^+$ on protons is particularly suited for studying $\Delta^*$ resonances~\cite{Candlin:1984av}, as 
no $N^*$ resonances with charge $+2$ exist. Studies of the reaction $\pi^+p \to K^+\Sigma^+$ 
\cite{Edinburgh-Rutherford-Westfield:1983qsm} benefit from the ability to measure spin rotation 
parameters using a polarized target and by detecting $\Sigma^+$ polarization~\cite{Candlin:1988pn}.

\smallskip
\textbf{ Hyperon production:}
A comprehensive list of references to data on elastic and inelastic $\bar KN$ scattering in the resonance region can be found in Ref.~\cite{Matveev:2019igl}.

\smallskip
\textbf{Photoproduction of mesons:}
Over the past two decades, our understanding of nucleon excitations, including $N^*$ and $\Delta^*$ states, has advanced significantly due to efforts by a large scientific community studying photon-induced reactions on H$_2$ or D$_2$ targets.

In Bonn, the first 500~MeV electron synchrotron was constructed under the leadership of Wolfgang Paul between 1953 and 1958, with initial experiments on pion production from protons reported in \cite{Althoff:1963hfa}. A second synchrotron capable of accelerating electrons up to 2.5~GeV became operational in 1967 \cite{Althoff:1968twa}. Shortly thereafter, a butanol target with polarized protons, developed at CERN \cite{Mango:1969ww}, was installed in Bonn \cite{Althoff:1971zza}. Development of polarized electron sources in Bonn has a long tradition, with the first such source using photoionization of polarized lithium atoms \cite{Baum:1969es}. Later, spin-polarized electrons were produced using photoemission from Be-InGaAs/AlGaAs strained layer superlattice photocathodes, driven by circularly polarized laser light \cite{Hillert:2006yb}. The first acceleration of polarized electrons in the 2.5 GeV electron synchrotron occurred in the mid-1980's ~\cite{Brefeld:1984iv} to study the effect of the depolarizing resonances. 

The synchrotron initially delivered high-intensity electron bunches with an approximately 5\% duty cycle. To increase this value and make coincidence measurements feasible, the Electron Stretcher Ring ELSA, capable of operating at a maximum energy of 3.5~GeV, was constructed and became operational in 1987 \cite{Husmann:1988vg}. 
Several experiments relevant for baryon spectroscopy have been conducted there.

The PHOENICS experiment used a polarized frozen spin target to study pion and $\eta$ photoproduction off protons and deuterons. The detector was also used by the GDH collaboration, exploiting circularly polarized photons and longitudinally polarized protons to test the Gerasimov-Drell-Hearn sum rule \cite{Dutz:1996uc,Dutz:2004zz,Hoffmann-Rothe:1997jmq,Bock:1998rk,GDH:2003xhc,Naumann:2003vf,GDH:2005noz}. Data on the low-energy region were 
taken at Mainz (see below).

The ELAN experiment collected data on the four-momentum transfer dependence of the $N\to\Delta$ transition by measuring angular distributions of the pion production cross section. Preliminary results are reported in~\cite{Gothe:2001yjy}.

The first data at ELSA were taken with the SAPHIR detector~\cite{Schwille:1994vg}. This experiment was the first counter experiment on photoproduction in the resonance region, covering the full angular range. Results were reported in Refs.~\cite{Bockhorst:1994jf,Mirazita:1997yp,Bennhold:1997mg,SAPHIR:1998noz,SAPHIR:1998fev,%
Muccifora:1998ct,SAPHIR:1999wfu,Barth:2001cb,Barth:2003kv,Barth:2003bq,%
SAPHIR:2003lnh,Glander:2003jw,Wu:2005wf,Lawall:2005np,Wieland:2010run}.

In 1998, a current coil of the SAPHIR dipole magnet broke down, and a new detector for the CBELSA/TAPS experiment was ready for installation. As a result, the SAPHIR experiment was dismantled and replaced by the CBELSA/TAPS experiment, while ELAN continued to take data.

The Crystal Barrel detector~\cite{CrystalBarrel:1992qav} had previously been used at the CERN Low Energy Antiproton Ring. The Two-Arm-Photon Spectrometer~\cite{Novotny:1991ht} was dismantled at Mainz and installed in Bonn to provide better granularity for the forward hole of the Crystal Barrel detector. Data were collected on meson photoproduction off protons or neutrons, both without polarization and with various combinations of photon and target polarizations \cite{CB-ELSA:2003rxy,CB-ELSA:2004sqg,CBELSATAPS:2005iwc,CB-ELSA:2007htf,CB-ELSA:2007wep,%
CBELSATAPS:2007oqn,CBELSA:2007vce,Thoma:2007bm,Sarantsev:2007aa,CB-ELSA:2007xbv,CBELSA:2008wwq,% CBELSATAPS:2008pjl,CBELSATAPS:2008mpu,CBELSA:2008epm,CBELSATAPS:2008jry,% CBELSA:2008nzn,CBELSATAPS:2009ntt,CBELSA:2009irj,CBELSATAPS:2010mtq,CBELSATAPS:2010fcv,% CBELSATAPS:2010zru,CBELSATAPS:2010sqr,CBELSATAPS:2011nwh,Jaegle:2011sw,CBELSATAPS:2011gly,% CBELSATAPS:2012few,CBELSATAPS:2013waf,CBELSATAPS:2013btn,CBELSATAPS:2014wvh,CBELSATAPS:2014ibm,% CBELSATAPS:2014itx,Hartmann:2014mya,CBELSATAPS:2015taz,CBELSATAPS:2015wwn,Eberhardt:2015lwa,% CBELSATAPS:2015tyg,CBELSATAPS:2015rtp,CBELSATAPS:2015kka,CBELSATAPS:2015ftl,CBELSATAPS:2016lkm,% CBELSATAPS:2016qdi,Friedrich:2016cms,CBELSATAPS:2017lap,CBELSATAPS:2018sck,CBELSATAPS:2019hhr,% 
CBELSATAPS:2019ylw,CBELSATAPS:2020cwk,CBELSATAPS:2021osa,CBELSATAPS:2022uad}.

The first accelerator built in Mainz (1964–1966) was a 300~MeV linear electron accelerator \cite{Ehrenberg:1972tg} for nuclear structure experiments. In the summer of 1976, a decision was made to construct a multilevel racetrack Microtron (MAMI). A fit stage, which accelerated electrons from a 2.1~MeV (later 3.46~MeV) linear accelerator to 14~MeV, was successfully tested in 1979. The second stage, with an energy of 175~MeV, went into operation \cite{Herminghaus:1983nv}. The third stage, completed in 1990, accelerated electrons up to 855~MeV and delivered an intense beam \cite{Euteneuer:1992qe}. The final stage, MAMI-C, was completed in 1998 and reached an energy of 1508~MeV \cite{Kaiser:2008zza}, covering the mass range up to the fourth resonance region. It provides a beam current of 100\,$\mu$A within a narrow energy window. Polarized electrons, typically 80\% polarized, achieve a beam current of up to 20\,$\mu$A. Tagged photons hit a target, and the final state of the interaction is detected in a system composed of the Crystal Ball (CB) \cite{Chan:1978ck} and TAPS \cite{Novotny:1991ht} photon spectrometers. Results were reported in Refs. 
\cite{CrystalBallatMAMI:2008cye,CrystalBallatMAMI:2009lze,Starostin:2009zz,CrystalBallatMAMI:2009iym,Schumann:2010js,CrystalBallatMAMI:2010slt,% CrystalBallatMAMI:2010cug,Berghauser:2011zz,Starostin:2011kx,Pheron:2012aj,Kashevarov:2012wy,MAMI:2012bag,%
Zehr:2012tj,A2:2012lnr,Thiel:2013cea,Robinson:2013sxa,CrystalBall:2013oow,Oberle:2013kvb,Maghrbi:2013xqd,%
A2:2013cqk,AguarBartolome:2013mga,CrystalBallatMAMI:2013iig,A2:2013wad,Sikora:2013vfa,A2:2013tbo,% A2:2013wkp,A2:2014snn,A2:2014fca,A2atMAMI:2014zdf,Strakovsky:2014wja,A2:2014pie,A2:2014iky,A2atMAMI:2014rwc,%
A2:2014bam,A2:2015mqi,A2:2015ocx,A2:2015mhs,% Kaser:2015bum,MAINZ-A2:2015yzu,A2:2015pgk,A2:2016tkj,%
MAINZ-A2:2016iua,A2:2016vzp,Adlarson:2016hpp,A2:2016bij,A2:2016nio,A2:2016sjm,A2:2017gwp,A2:2017auj,% Dieterle:2017myg,Adlarson:2017wlz,A2:2018vbv,A2:2018pjo,A2:2018jcd,Bashkanov:2018ftd,Kaser:2018bba,%
A2:2018doh,A2:2019fvn,A2:2019yud,A2:2019bqm,A2:2019arr,Dieterle:2020vug,A2CollaborationatMAMI:2021vfy,%
A2:2022ycp,A2:2022kkx,A2:2022ipx,Mornacchi:2023oir}.

The Thomas Jefferson National Accelerator Facility Laboratory (JLab) in Newport News, Virginia, delivered its first beam in 1994. A central topic in its wide program is pursued with the CLAS detector \cite{CLAS:2003umf}, measuring a variety of meson final states using unpolarized, linearly polarized, and circularly polarized photon beams with energies up to 5.6~GeV. Results are documented in 
\cite{CLAS:2023ddn,CLAS:2021jhm,CLAS:2021hex,CLAS:2021osv,CLAS:2020ngl,CLAS:2020rdz,CLAS:2020spy,%
Mokeev:2020hhu,CLAS:2018mmb,CLAS:2018kvn,CLAS:2018azo,CLAS:2018avi,CLAS:2018drk,CLAS:2018gxz,%
CLAS:2020rdz,CLAS:2018xbd,Anisovich:2018yoo,CLAS:2017kyf,CLAS:2017vxx,CLAS:2017jrx,CLAS:2017sgi,%
Anisovich:2017pox,Anisovich:2015gia,Anisovich:2017ygb,Anisovich:2017xqg,CLAS:2017gsu,CLAS:2017dco,%
CLAS:2017kua,CLAS:2017yjv,CLAS:2017rxe,CLAS:2016zjy,CLAS:2016wrl,CLAS:2015pjm,CLAS:2015ykk,%
CLAS:2014tbc,Anisovich:2014yza,CLAS:2013pcs,CLAS:2013jlg,CLAS:2013rxx,CLAS:2013qgi,CLAS:2013rxx,%
CLAS:2013owj,Anisovich:2013jya,CLAS:2013rjt,Afanasev:2012fh,Price:2004hr,CLAS:2011aa,CLAS:2010gcv,%
CLAS:2010aen,Qian:2010rr,CLAS:2009rdi,CLAS:2009ngd,CLAS:2009tyz,CLAS:2009wde,CLAS:2009hpc,Chen:2009sda,%
CLAS:2008ycy,Dugger:2007bt,CLAS:2007xhu,Guo:2007dw,CLAS:2007kab,CLAS:2006pde,CLAS:2005koo,CLAS:2006czw,%
CLAS:2006rru,CLAS:2006anj,CLAS:2005rxe,CLAS:2005oqk,CLAS:2005lui,Burkert:2005ft,CLAS:2004gjf,%
CLAS:2003wfm,CLAS:2003zrd,CLAS:2003yuj,CLAS:2002cdi,CLAS:2002rxi,CLAS:2001zxv,CLAS:2000kid,Manak:2000ty,%
JeffersonLabE94014:1998czy,Frolov:1998pw}.

The GRAAL Collaboration scattered a beam of ultraviolet laser light off the electron beam
in the straight section of the European Synchrotron Radiation Facility, thus producing narrowly collimated $\gamma$-rays
with a maximum energy of 1.47~GeV. The energy distribution is flat, and the $\gamma$-rays carry
the polarization of the primary photons. GRAAL results can be found in 
Refs.~\cite{GRAAL:2000gnk,GRAAL:2000qng,GRAAL:2002mil,Assafiri:2003mv,Gurzadyan:2004rx,GRAAL:2005mor,%
Lleres:2007tx,GRAAL:2007gsc,Bartalini:2008zza,% GRAAL:2008jrm,Fantini:2008zz,DiSalvo:2009zz,%
GRAAL:2013tzy,LeviSandri:2014uhc}. The BGO calorimeter is a central part of the GRAAL detector. After the end of data taking
in Grenoble in 2008, the calorimeter
was transferred to Bonn and became the central component of
the BGO-OD experiment~\cite{Schmieden:2018yjy,BGO-OD:2019utx} at ELSA. Results of the BGO-OD experiment at Bonn can be found in 
Refs.~\cite{Jude:2020byj,Alef:2020yul,BGOOD:2021sog,BGOOD:2021oxp,Jude:2022atd,Figueiredo:2024zmh,Rosanowski:2024rww}.

At the 8~GeV synchrotron radiation facility Spring-8, situated in Harima Science Park, Hyōgo, 
an ultraviolet photon beam is backscattered off the circulating electron beam, producing 
collimated $\gamma$-rays up to 2.4~GeV \cite{Nakano:2000ku,Nakano:2001xp}.
Charged particles produced in $\gamma N$ reactions in the forward direction are momentum-analyzed 
in a dipole magnet, with particle identification provided by time-of-flight and Cherenkov counters. 
The facility started operating in 1999. Later, a new experiment was installed with an egg-shaped 
BGO calorimeter, forward drift chambers, and a time-of-flight wall~\cite{Muramatsu:2021bpl}.
Physics results obtained by the LEPS and LEPS2/BGOegg collaborations 
have been reported in 
Refs.~\cite{Matsumura:2003mw,LEPS:2003wug,LEPS:2003buk,CDF:2003epb,Nakano:2004cr,Ishikawa:2004id,%
LEPS:2005hax,LEPS:2005hji,Kohri:2006yx,Chang:2007fc,LEPS:2007nri,Sumihama:2007qa,Sumihama:2007gfo,Miwa:2007xk,%
Niiyama:2008rt,LEPS:2008ghm,LEPS:2008azb,Muramatsu:2009zp,LEPS:2009isz,LEPS:2009nuw,LEPS:2009pib,LEPS:2010ovn,%
Hwang:2012zza,Hwang:2013usa,Morino:2013raa,LEPS:2013dqu,LEPS:2016ljn,LEPS:2017jqw,LEPS:2017vas,LEPS:2017nqz,%
LEPS:2017pzl,LEPS:2018pbi,LEPS2:2019bek,LEPS2BGOegg:2020cth,LEPS2:2020ttk,LEPS2BGOegg:2022dop,LEPS2BGOegg:2023ssr}.

\smallskip
\textbf{Electroproduction:}
Meson photoproduction has become an essential tool in the search for new excited light-quark baryon states. As discussed in the previous section, many new excited states have been discovered thanks to high-precision photoproduction data in different final states~\cite{Anisovich:2017bsk}, and are now included in recent editions of the Review of Particle Physics (RPP)~\cite{ParticleDataGroup:2022pth}. The experimental exploration of resonance transition form factors spans over 60 years, with many
review articles describing this history. Here we refer to some recent ones~\cite{Aznauryan:2011qj,Stoler:1993yk,Burkert:2004sk,Aznauryan:2012ba}. A review of recent electroproduction experiments in hadron physics and their interpretation within modern approaches to strong interaction physics can be found in Ref. \cite{Proceedings:2020fyd}. The exploration of the internal structure
of excited states and the effective degrees of freedom contributing to $s$-channel resonance excitation requires the use of electron beams,
where the virtuality ($Q^2$) of the exchanged photon can be varied to penetrate through the peripheral meson cloud and probe the quark core and its spatial structure. Electroproduction can thus reveal whether a resonance is generated through short-distance photon interaction with the small quark core or through interaction with a more extended hadronic system. Electroproduction of final states with pseudoscalar mesons (e.g., $N\pi$, $p\eta$, $K\Lambda$, $p\pi^-\pi^+$) has been employed at Jefferson Laboratory mainly with the CEBAF Large Acceptance Spectrometer (CLAS) operating at an instantaneous luminosity of $10^{34}\,\mathrm{sec}^{-1}\mathrm{cm}^{-2}$. In Hall A and Hall C, pairs of individual well-shielded focusing magnetic spectrometers are employed with more specialized aims and limited acceptance, but operating at much higher luminosity. This experimental program has led to new insights into the scale dependence of effective degrees of freedom, e.g., meson-baryon, constituent quarks, and dressed quark contributions.
Data on meson electroproduction can be found in Refs.~\cite{CLAS:2023mfc,Mokeev:2023zhq,CLAS:2022iqy,CLAS:2022kta,CLAS:2021cvy,%
Mokeev:2020hhu,CLAS:2021jhm,CLAS:2021ovm,CLAS:2019cpp,CLAS:2019uzc,CLAS:2018fon,Aznauryan:2018okk,Aznauryan:2017nkz,%
Aznauryan:2016wwm,CLAS:2017rgp,CLAS:2017fja,CLAS:2017jjr,CLAS:2016tqs,CLAS:2016cks,CLAS:2016dlz,CLAS:2016qxj,%
CLAS:2016ikd,Kim:2015pkf,Skorodumina:2015ccu,Mokeev:2015lda,CLAS:2014fml,CLAS:2014jpc,%
CLAS:2014udv,Aznauryan:2015zta,%
Aznauryan:2014xea,CLAS:2012qga,Aznauryan:2012ba,CLAS:2012wxw,CLAS:2012cna,Aznauryan:2012ec,Aznauryan:2011qj,%
CLAS:2009ces,CLAS:2009sbn,CLAS:2008roe,CLAS:2008wls,CLAS:2008agj,CLAS:2007jvi,CLAS:2008ihz,CLAS:2008rpm,%
CLAS:2008cms,CLAS:2007bvs,CLAS:2006sjw,CLAS:2006ogr,CLAS:2006ezq,CLAS:2005oqk,Aznauryan:2005tp,CLAS:2005vxa,%
CLAS:2005nkx,CLAS:2004cri,Aznauryan:2004jd,CLAS:2004ncx,Burkert:2004sk,CLAS:2002xbv,CLAS:2002rxi,CLAS:2003hro,%
CLAS:2002zlc,CLAS:2001cbm,CLAS:2000mbw,Breuker:1977vy,Breuker:1982nw,Beck:1974wd,%
Siddle:1971ug,Alder:1975tv,Brasse:1975bg,Alder:1975na,Alder:1975xt,Brasse:1976bf,Brasse:1977as,Haidan:1979yqa}.

\subsection{Analyses}
\label{PWA-DCC}
The three aforementioned groups (KH, CMB and SAID) determined 
the complex scattering amplitudes decomposed into partial wave contributions
from the $\pi N$ elastic scattering data. Since there are two processes
involved, spin-flip and  non-flip scattering, two complex amplitudes
need to be determined. Measurements of the angular distributions do not provide
sufficient information, and data with target polarization exist only over a limited
range. Furthermore, for any energy, there is an undetermined overall phase. Hence
further input is required to determine the amplitudes.

The main KH results can be found in
\cite{Hohler:1979yr}, with an update of the results presented in
\cite{Koch:1980ae}. A comprehensive description of methods and
results can be found in the handbook \cite{Hohler:1984ux} published
in the Landold-B\"ornstein series. H\"ohler presented updated
resonance masses and widths using the so-called speed-plot method
\cite{Hohler:1992ru}.
The KH partial-wave analysis was focused on the intention
to overcome the ambiguities due to incomplete experimental
information by exploiting dispersion relations for fixed squared momentum transfer
at small $t$, for fixed lab angles, and backward
dispersion relations using $\bar pp\to\pi^+\pi^-$ data and the
$\pi\pi$ phase shifts. 
Unitarity was imposed by adding an appropriate penalty term to the
overall $\chi^2$ of the fit.

In the partial-wave analysis at CMB, real and imaginary parts of the
invariant amplitudes were both parameterized by a sum of a
term describing the Regge behavior at large
energies and a series of ``Born
terms", representing $t$-channel Pomeron, $\rho$, and $f_2$-exchange
and $u$-channel $N$ and $\Delta$ exchange. 
In the imaginary part, a diffractive Pomeron term was included and a
suitable form representing two-pion exchange. Unitarity was enforced
in the fits \cite{Cutkosky:1979fy}. Ambiguities were resolved
by forcing the amplitudes to be analytic in
the Mandelstam plane along a few carefully chosen hyperbolas. In
addition, some forward dispersion relations were used as part of the
input.
The resulting partial-wave amplitudes were described by
coupled-channel Breit-Wigner amplitudes, whose pole positions and residues
are presented and discussed in \cite{Cutkosky:1979zv}. The results
were updated at Baryon 1980 \cite{Cutkosky:1980rh}.

A long tradition of partial-wave analyses was initiated in 1969, when
Arndt moved to the Virginia Polytechnic Institute (VPI) to
collaborate with Roper \cite{Arndt:1970wr}. 
Already at a very early stage, the group enabled external users to
access the full data base and solution files through a Scattering
Analysis Interactive Dial-in (SAID) computer system. In addition to
their own solutions, SAID also encodes the Karlsruhe-Helsinki
solution, the Carnegie-Mellon-Berkeley solution, and production
partial waves. The system is an extremely useful tool to discuss
experimental results and to plan new experiments. 

The SAID analysis started with an energy-dependent fit to the data based
on a relativistically invariant T-matrix, which was parameterized in the form
of a $4\times 4$ real and symmetric K-matrix for each partial wave.
The four channels considered were $N\pi$, $N\eta$, $\Delta\pi$, and
$N\rho$. The K-matrices contained one pole and a low-order polynomial.
The amplitudes thus obtained were used as starting values for the
energy-independent partial-wave analysis. In a second step, data
within a prefixed energy bin were fitted with  an S-matrix of the
form\\[-4ex]
\begin{eqnarray}
\mathrm{S_e = (1 + 2iT_e)} = \cos (\rho) \,e^{2 i \delta},\label{2}\\[-4ex]
\nonumber\end{eqnarray}
where $\rm T_e$ is the T-matrix of the energy-dependent analysis 
calculated at the given energy. The phase parameters
$\delta$ and $\rho$ were expanded as linear functions around the
given energy,  with a slope (energy derivative) fixed by the
energy-dependent solution. These single-energy results were then
compared to the global fit in order to search for structure which
could be missing. This procedure was iterated until a stable best-fit
to the data was found. The solutions were constrained by 
dispersion relations.
Milestones of their progress are documented in
\cite{Arndt:1985vj,Arndt:1990bp,Arndt:1995bj,Arndt:2003if,Arndt:2006bf,Workman:2008iv,Workman:2011vb,%
Workman:2012jf,Workman:2012hx,Briscoe:2012ni,Workman:2016ysf,Briscoe:2020qat,Briscoe:2021siu,%
Strakovsky:2022tvu,Briscoe:2023gmb}.

A large number of groups, including SAID, have analyzed photo-induced reactions. Often, the intention of the work
is a test of dynamical assumptions. Resonances are introduced with known properties, and just
photocouplings are determined. In some
cases, the agreement between different data is not really satisfactory. In the following we only discuss analyses
that have been included by the Particle Data Group. An excellent summary of the methods
applied by the different groups can be found in Ref.~\cite{Thiel:2022xtb}.

The MAID group, led by Drechsel and Tiator at Mainz, developed isobar models for analyses
of photo- and electroproduction of pions~\cite{A2:2015mhs,Drechsel:1998hk,Kamalov:2000en,Tiator:2008kd,Tiator:2010rp,Stajner:2017fmh}, $\eta$~\cite{A2atMAMI:2014rwc,Drechsel:2007if,Chiang:2001as} 
and $\eta'$~\cite{Tiator:2018heh}, two-pions~\cite{Fix:2005if}, $\pi\eta$~\cite{Fix:2010nv}, and $K\Lambda$~\cite{Mart:1999ed}
production.  MAID provides online programs~\cite{MAIDkph} for real-time
calculations of observables, amplitudes and partial waves (multipoles). 

The Kent State University group, led by Manley, contributed very significantly to the study of $N^*$ and $\Delta^*$
resonances and to the search for new states 
\cite{Manley:1984zs,Manley:1984ih,Manley:1984jz,Manley:1992yb,Shrestha:2012va,Hunt:2018tvt,Hunt:2018mrt,Hunt:2018wqz}.
In their latest publications, Hunt and Manley report a study of $\gamma p\to \eta p$,  $\gamma n\to \eta n$ \cite{Hunt:2018tvt},
and $\gamma p\to K^+\Lambda$ \cite{Hunt:2018mrt}, and compare their partial-wave amplitudes 
with those obtained from other groups. The spread of results looks alarming. However, often the spread in masses, widths,
and branching ratios is considerably smaller than one might suppose looking at the moments alone. In Ref.~\cite{Hunt:2018wqz},
Hunt and Manley present a fit to data on $\pi N\to \pi N$, $\to \pi\pi N$, $\to \eta N$,
to $K\Lambda$, and to $\gamma N\to \pi N$, $\to \eta N$, and $\to K\Lambda$. A new $N(2200)7/2^+$ resonance with significant
couplings to $\pi N$ and $K\Lambda$ is suggested. Masses, widths and decay properties of further 21 known $N^*$'s and twelve
known $\Delta^*$'s are reported.

Mosel at Gie\ss en initiated coupled-channel analyses of pion and photoinduced reactions off nucleons in the late
70ies of the last century. The first fits included resonances with spin 1/2 and 3/2 only \cite{Penner:2001fu,Penner:2002ma,Penner:2002md}. 
In \cite{Shklyar:2004dy}, spin 5/2 was introduced. The further fits included eleven $N^*$ and nine $\Delta^*$ resonances with spins $J\leq 5/2$
and reported masses, width and branching ratios for decays into $\pi N$ \cite{Shklyar:2004ba},  
$K\Lambda$ \cite{Shklyar:2005xg}, $\eta N$ \cite{Shklyar:2006xw,Shklyar:2012js}, and $K\Sigma$ \cite{Cao:2013psa}. 
No errors were evaluated. In Ref.~\cite{Shklyar:2014kra}, the emphasis of the analysis is to study the isobars in the $N(1440)\to N\pi\pi$ decays. 

The Bonn-Gatchina (BnGa) group was initiated in 2002 by one of the authors (E.K.). The group 
performed coupled-channel analyses of a very large data set on pion and photo-induced 
reactions and discovered several new resonances or improved the evidence for their existence. 
The method is described in \cite{Anisovich:2004zz,Anisovich:2006bc,Burkert:2022bqo} and 
results are presented in Refs.~%
\cite{Anisovich:2018yoo,Anisovich:2017pox,Anisovich:2015gia,Anisovich:2017ygb,Anisovich:2017xqg,Anisovich:2014yza,Anisovich:2013jya,Sarantsev:2005tg,Anisovich:2005tf,Klempt:2006sa,Nikonov:2007br,%
Anisovich:2007bq,Anisovich:2008wd,Anisovich:2010mks,%
Anisovich:2010an,Anisovich:2011ye,Anisovich:2011sv,Anisovich:2011su,Anisovich:2011fc,Anisovich:2012ct,Anisovich:2013sva,%
Anisovich:2013tij,Anisovich:2013vpa,Anisovich:2015tla,%
Denisenko:2016ugz,Anisovich:2016vzt,Anisovich:2017afs}. 
The distinctive
features of the BnGa analysis are the use of event-based likelihood fits to three-body final states and the rapid convergence
of the fitting procedure. This aim is achieved by implementing derivatives of fit quantities and a multilevel
structure for fitting parameters. The speed of the fit allows
the group to systematically search for new resonances in all partial waves. Systematic errors are evaluated by adding further resonances
at high masses and by changing the weights with which data sets are included in the fits. The group
found that including or excluding high-mass resonances provides the largest contribution to the final uncertainties.
The data base and fit curves can be found on the web page~\cite{pwahiskp}.

The ANL-Osaka group deveolped a dynamical coupled-channel model for the combined
analysis of pion- and photon-induced reactions. The progress is reported in a series of publications
\cite{Sato:1996gk,Sato:2000jf,Julia-Diaz:2006ios,Julia-Diaz:2007qtz,Matsuyama:2006rp,Suzuki:2008rp,Kamano:2008gr,%
Julia-Diaz:2009dnz,Suzuki:2009nj,Kamano:2009im,Sandorfi:2010uv,Suzuki:2010yn,Kamano:2010ud,Kamano:2013iva,%
Nakamura:2015rta,Kamano:2016bgm,Kamano:2019gtm}. In the latest works, a very large data set on pion and photo-induced 
reactions is used to determine properties of ten $N^*$ and nine $\Delta^*$ resonances. The approach concentrates on
a rigorous definition of the scattering and production amplitudes, but at the cost of a large amount of computing time
which limits the feasability of systematic studies. 
 
The Jülich-Bonn (JüBo) group 
developed a dynamical coupled-channel model for the combined analysis of a very large
data set on pion- and photo-induced reactions, referred to as the Jülich-Bonn-Washington model~\cite{Mai:2021vsw,Mai:2021aui,Mai:2023cbp}. 
The analytic properties of a resonance were discussed \cite{Doring:2009yv} and the dependence of the pole properties on the background amplitudes \cite{Doring:2009bi}. 
The phase and pole structure of $N(1535)1/2^-$ by Döring
and Nakayama~\cite{Doring:2009uc}.
In a study of data on $I_s$, $I_c$ for the reaction $\gamma p\to\pi^0\eta p$~\cite{CBELSATAPS:2015tyg}, prominent contributions due to 
$\Delta(1700)3/2^-$ formation was found, 
interpreted as a dynamically generated resonance in Ref.~\cite{Doring:2005bx}. 

In a series of publications, the JüBo group included an increasing data base 
in a coupled-channel fit \cite{Doring:2010ap,Ronchen:2012eg,Ronchen:2014cna,Ronchen:2015vfa,Ronchen:2018ury,Ronchen:2022hqk}
resulting in a full spectrum of $N^*$ and $\Delta^*$ resonances with their decay properties. 
Figures representing the fit results including a display of the data can be found in Ref.~\cite{collaborations.fz-juelich}.
The dynamical coupled-channel approaches applied by the ANL-Osaka and Jülich-Bonn groups are reviewed in Ref.~\cite{Doring:2025sgb}.

A simple method to extract pole parameters from the data, called the Laurent-Pietarinen (L + P) method, has been
developed by \v{S}varc and collaborators. Data, mostly single channels, were fitted with a function
containing poles, branch points and cuts \cite{Svarc:2012pt,Svarc:2013laa,Svarc:2014zja,Svarc:2014sqa,%
Svarc:2014aga,Svarc:2015usk,Anisovich:2017bsk,Anisovich:2017ygb,%Svarc:2017yuo,Svarc:2017rii,%
Svarc:2018eol,Svarc:2018aay,Svarc:2020waq,Svarc:2020cic,Svarc:2021zkb,Svarc:2021gcs,Svarc:2022buh}.
The results are pole positions, but no branching ratios or transition residues have been derived.

In photoproduction, the nucleon and the photon spin can be parallel or opposite,
the nucleon spin can flip or not. Thus there are four independent  amplitudes
describing the process. In single pseudoscalar meson photoproduction, there
are sixteen linearly independent measurable quantities; the four complex
amplitudes require at least eight independent measurements. In energy-dependent 
analyses, these four amplitudes are constrained by a model. In the recent years, 
progress has been made in energy-independent reconstructions of amplitudes
in truncated partial-wave analyses in which the number of required measurements
is reduced to five~\cite{Wunderlich:2013iga,Workman:2016irf}. The method has been applied
to single-pion photoproduction~\cite{Fix:2022shn}, to $\gamma p\to \Lambda K^+$ 
\cite{Svarc:2021gcs}, and to $\gamma p\to p\eta$~\cite{Svarc:2020cic,Kroenert:2023ovd}.
In the future, the constructed amplitudes could be used to extract the pole position of resonances
in the L + P formalism. 

\begin{figure}[t!]
\centering
\begin{tabular}{cc}
\hspace{-5mm}   \includegraphics[width=0.5\linewidth,height=0.31\linewidth]{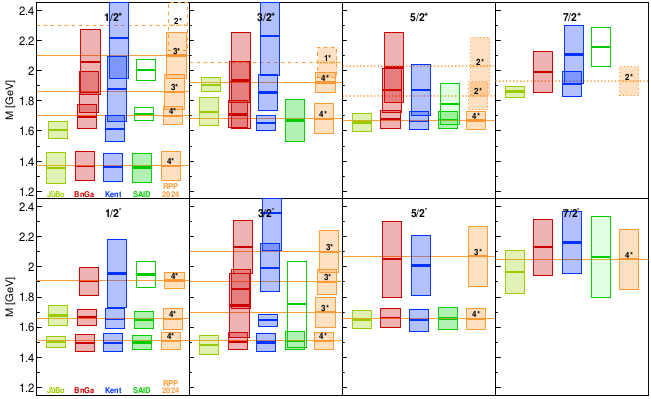}&
\hspace{-3mm}   \includegraphics[width=0.5\linewidth,height=0.31\linewidth]{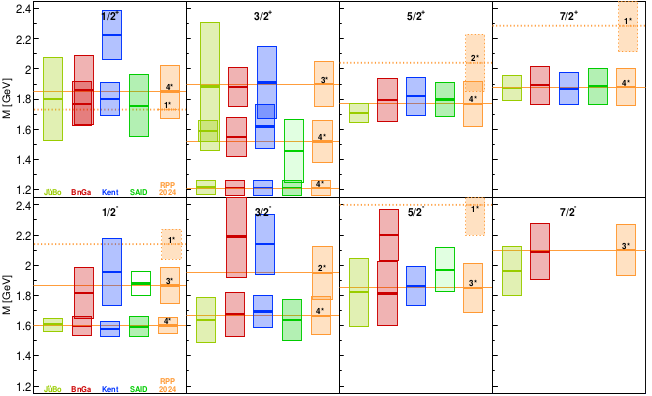}
\vspace{-2mm}
\end{tabular}
    \caption{Comparison of PWA results on pole positions ($M$ and $\Gamma$) of $N^*$ (left) and $\Delta^*$ (right)
     resonances, including
    JüBo~\cite{Ronchen:2022hqk}, BnGa~\cite{CLAS:2024iir,Sarantsev:2025lik}, Kent~\cite{Hunt:2018wqz},
    SAID~\cite{Arndt:2006bf}, and the RPP 2024~\cite{ParticleDataGroup:2024cfk}. Graphics credit: Ulrike Thoma.}
    \label{fig:Thoma}
\vspace{-2mm}
\end{figure}

The RPP lists the masses and width of $N^*$ and $\Delta^*$ resonances resulting from the different PWAs.  
 Figure~\ref{fig:Thoma} shows a selection of the PWA results on the pole position (with $M$ as central 
 value and $\Gamma$ as range) in comparison with the PDG result. 
Based on distribution of the results, the RPP quotes a central value as best guess for an observable  and a range. 
The range is selected to cover at least most of the measurements.
The consistency of the data with the range varies significantly. We have calculated a mean $\chi^2$ 
for the different PWA for the pole positions of the $N^*$ and $\Delta^*$ resonances 
with four stars. 
The $\chi^2$ contribution of a resonance is calculated from the difference between the PWA results
and the range given in the RPP. The results are shown in Table~\ref{PWA-compare}.
Since the range is chosen to cover most of the reliable results, we expect
a $\chi^2$ per data point below 1 or close to 1. This is indeed the case for most analyses. However, for
some analyses, a significantly larger $\chi^2$ is obtained. Of course, the ``true" value cannot
be determined by a majority decision, but the deviations from the RPP mean values in different approaches
are spread over different resonances. The difficulty is the unknown influence of a significantly ``wrong"
result on the extraction of other measurable quantities. 

{\bf Hyperon resonances:}
Most hyperon resonances were discovered in the 70s of the last century in single-channel analyses.
The results obtained so far were summarized by Gopal at Baryon 1980, the 4th International Conference on 
Baryon Resonances, held in Toronto, Canada, in July 1980 \cite{Gopal:1980ur}. Then, the field had a standstill.
Manley and his group at KSU collected a large set of data on $K^-p$ interactions and performed a coupled-channel analysis~\cite{Zhang:2013cua,Zhang:2013sva}. The KSU partial-wave
amplitudes were also fitted by the JPAC group in a coupled-channel fit \cite{Fernandez-Ramirez:2015tfa}. The
Osaka-ANL group analyzed the same data in a dynamical coupled-channel approach \cite{Kamano:2014zba,Kamano:2015hxa}. 
The BnGa group added further published data and performed a coupled-channel analysis \cite{Matveev:2019igl,Sarantsev:2019xxm}.
\bigskip

\begin{figure}[!t]
  \centering
  \includegraphics[width=0.98\textwidth]{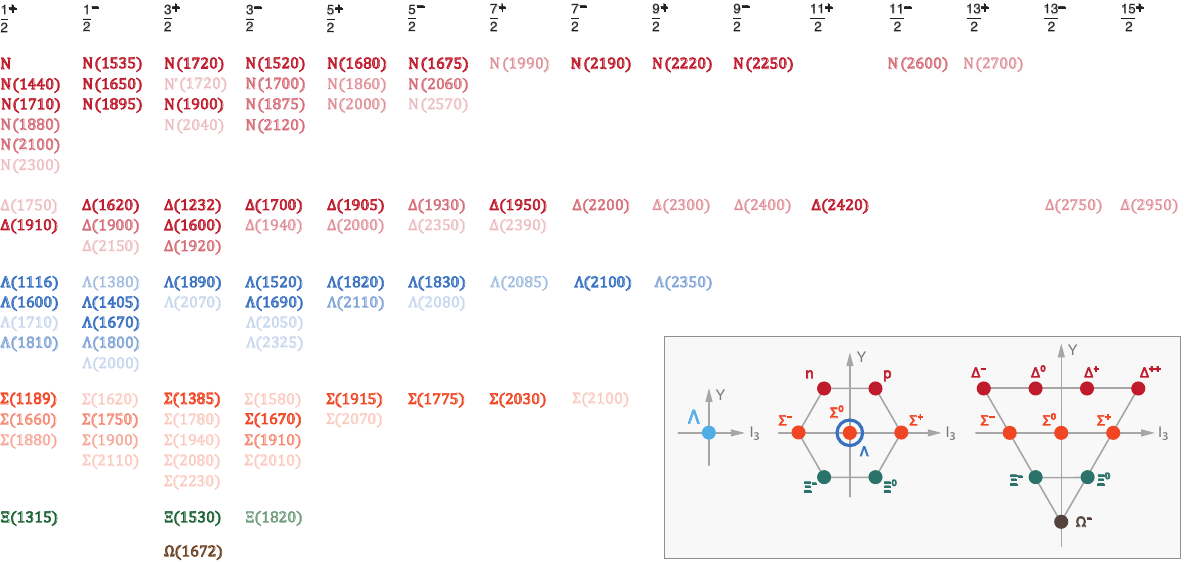}
\caption{(color online) Light and strange baryon spectrum from the PDG~\cite{ParticleDataGroup:2024cfk}.
         The columns correspond to different $J^P$ and the colors to isospin and hypercharge. The different font weights represent four-, three-, two- and one-star resonances.\vspace{-0mm}}
\label{fig:spectrum-1}
\end{figure}
  
\begin{figure}[pb]
\begin{minipage}{.55\linewidth}
    \captionof{table}{$\chi^2$ per data point for the consistency of the results on the pole positions
    of 4* resonances from different PWAs with the RPP mean values within the estimated ranges. 
    BnGa = Bonn-Gatchina, JüBo = Jülich-Bonn, Gi = Gie\ss en, CM = Carnegie-Mellon, PiA = Pittsburgh-Argonne, KH = Karlsruhe-Helsinki. In most analyses, $N_{\rm data}=28$ or 30.
    \vspace{-1mm}
    }
    \label{PWA-compare}
\renewcommand{\arraystretch}{1.4}
    \centering
    \footnotesize
    \begin{tabular}{lccclccclcc}
    \hline\hline
BnGa &\cite{Sarantsev:2025lik}: &0.66&JüBo& \cite{Ronchen:2022hqk}:&3.96& BnGa&\cite{CBELSATAPS:2015kka}: &0.55 \\[-1ex]
L+P  &\cite{Svarc:2014zja}:      &0.47& CM    & \cite{Cutkosky:1980rh}:&0.78& Kent&\cite{Hunt:2018wqz}:       &1.28\\[-1ex]
JüBo &\cite{Ronchen:2015vfa}:    &4.06& Gi    & \cite{Shklyar:2012js}: &8.11& BnGa& \cite{Anisovich:2011fc}:  &0.77      \\[-1ex]
SAID &\cite{Arndt:2006bf}:       &1.58& PiA   & \cite{Vrana:1999nt}:   &4.36& KH  &\cite{Hohler:1993lbk}:     &1.18\\
\hline\hline\vspace{8mm}
    \end{tabular}
    \captionof{table}{\label{highmass-piN} Right: $N^*$'s and $\Delta^*$'s known from $\pi N$ 
    elastic scattering only. Masses and widths are given by the pole positions; the index $^a$ 
    indicates Breit-Wigner parameters.
}
\end{minipage}
\hspace{5mm}\begin{minipage}{.38\linewidth}
\renewcommand{\arraystretch}{1.4}
\begin{tabular}{ l @{\quad} r @{\quad} c @{\quad} l @{\quad} l}
\hline\hline
& $J^P$ & & $M$ [MeV] & $\Gamma$ [MeV] \\ \hline\hline
$N(2600)$ & ${{11}/2}^-$  &\threeast  &2650\er100$^a$&650\er150$^a$\\
$N(2700)$ & ${{13}/2}^+$  &\twoast    &2612\er 45$^a$&350\er50$^a$\\ \hline
$\Delta(2300)$ & ${9/2}^+$& \twoast   & 2370\er80  &420\er160\\
$\Delta(2350)$ & ${5/2}^-$& \Ast      & 2400\er125 &400\er150\\
$\Delta(2390)$ & ${7/2}^+$& \Ast      & 2230\er100 &300\er140\\
$\Delta(2400)$ & ${9/2}^-$&  \twoast  & 2360\er100 &300\er150\\
$\Delta(2420)$ & ${{11}/2}^+$&\fourast& 2400\er100 &450\er100\\
$\Delta(2750)$ & ${{13}/2}^-$& \twoast& 2794\er80  &350\er100$^a$\\
$\Delta(2950)$ & ${{15}/2}^+$& \twoast& 2990\er100 &330\er100$^a$\\
\hline\hline\vspace{-6mm}
\end{tabular}
\end{minipage}
\end{figure}

\subsection{Present status}

The efforts at the various laboratories discussed above have significantly increased the number of known 
excitations of the nucleon and their decay modes. 
A survey of the known light-quark baryons including hyperons is shown in Fig.~\ref{fig:spectrum-1}. 
Table~\ref{tab:statusN} compares the current status with the status as of
20 years ago \cite{ParticleDataGroup:2004fcd}. In the mass region from 1850~MeV to 2150~MeV, 
we now recognize 11 $N^*$ resonances, compared to four known in 2004.
These 11 resonances have a total overall star rating of 30 stars; in 2004, they had 7 overall stars. The star rating was 
attributed to their $N\pi$ coupling, with only 11 additional stars given for non-$N\pi$ decay modes, all with one-star evidence. Currently, 
for resonances in the mass range from 1850~MeV to 2150~MeV, 107 stars are attributed to non-$N\pi$ decay modes.
Our knowledge of high-mass resonances comes primarily from elastic scattering of $\pi N$. Table~\ref{highmass-piN} lists  their PDG masses, 
widths and uncertainties given in the RPP. For 1* and 2* resonances, the uncertainties are from the 
measurements; for multiple measurements, the uncertainties span the range of reported data.

\begin{table*}[ph]
\caption{$N^\ast$ and $\Delta^*$ resonances. 
Pole positions in terms of the mass $M$ and width $\Gamma$ (in MeV) are taken
from the BnGa analysis~\cite{CLAS:2024iir,Sarantsev:2025lik}. Their decay channels as of 2004 \cite{ParticleDataGroup:2004fcd} are shown in black and those from 2024 in red and 
in brown. Recent findings are highlighted in blue. Blue stars are suggestions by us.
A candidate reported as observed in electroproduction is shown in green~\cite{Mokeev:2020hhu}.
Resonances in {\cy cyan} are not seen in Ref.~\cite{CLAS:2024iir,Sarantsev:2025lik}. 
Their pole positions are from Ref.~\cite{ParticleDataGroup:2024cfk}. We use the symbol 
$\sigma$ for the $f_0(500)$ throughout this paper.}
\label{tab:statusN}
\includegraphics[width=\linewidth]{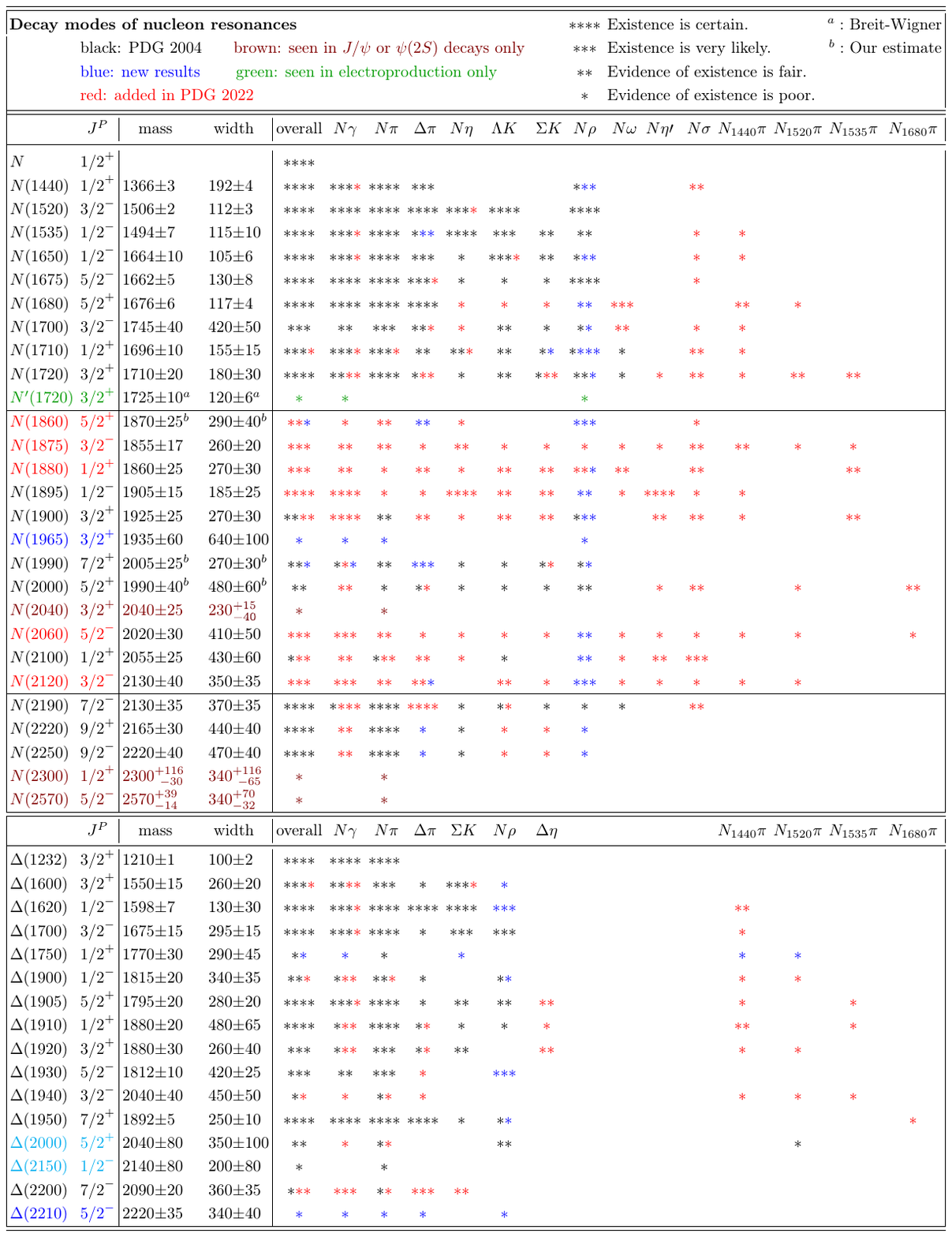}
\end{table*}
\clearpage

\begin{figure}[!t]
\centering
\hspace{-15mm}{\includegraphics[width=0.50\linewidth]{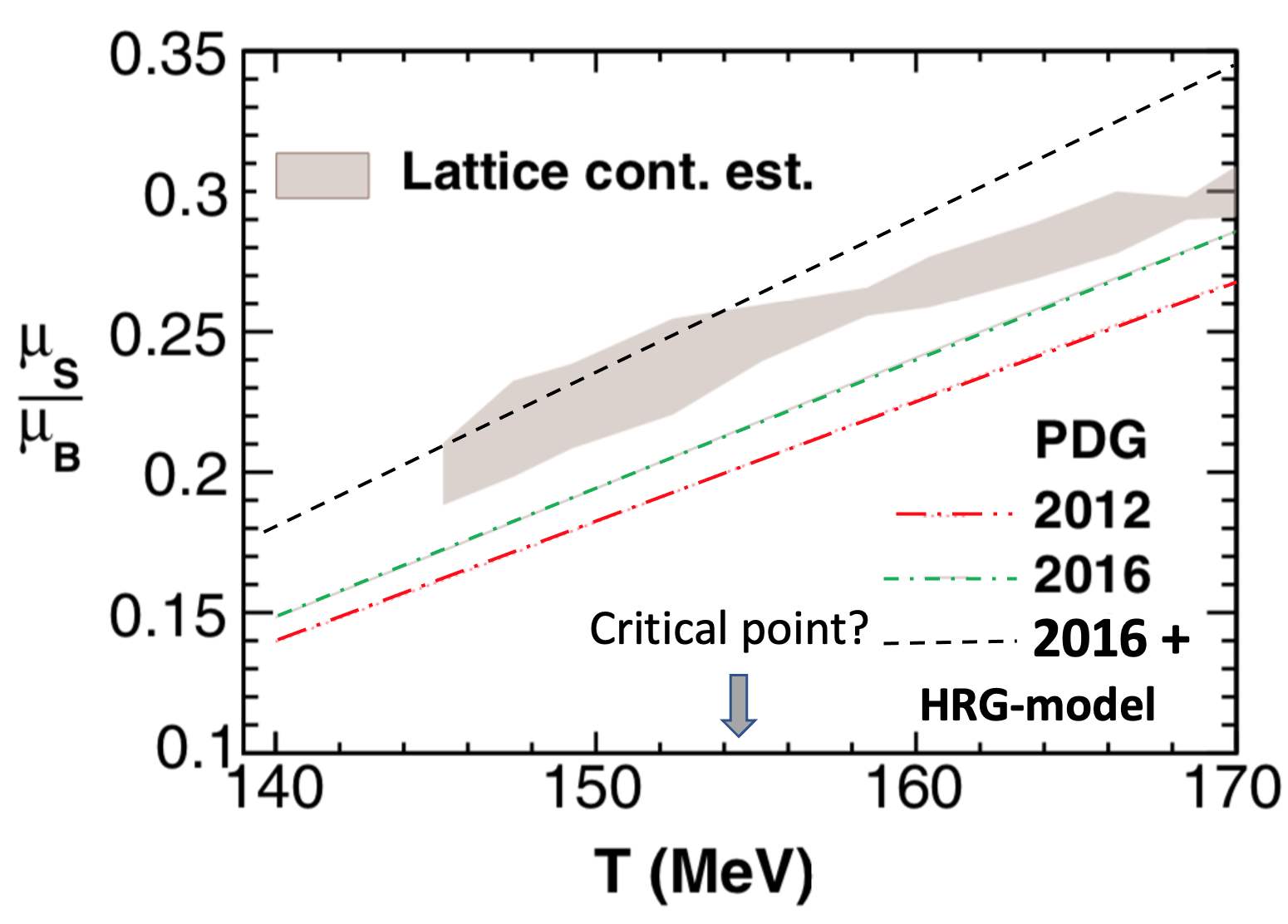}
\hspace{+10mm}
\includegraphics[width=0.257\linewidth]{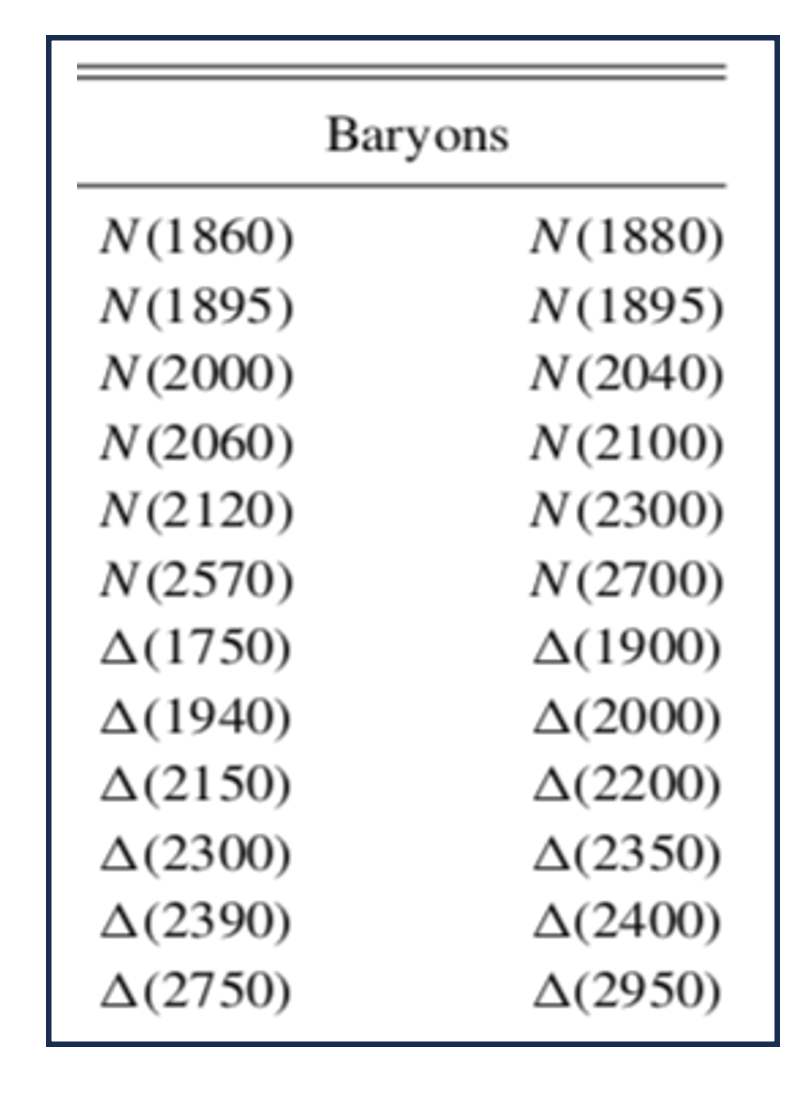}}
\caption{(Color online) Modeling the transition of the universe from the quark-gluon phase to the hadron phase during the first microseconds after the Big Bang. Left: Ratio of strangeness to baryon chemical potential near the critical point in the transition. The shaded area shows the projections from %hot 
Lattice QCD. The dashed lines represent the hadron resonance gas model projections with all known hadrons as cataloged by the PDG in the RPP editions in 2012 (red dashed-dotted) and 2016 (green dashed-dotted), including $^{***}$ and $^{****}$ states. Right: The $N^*$ and $\Delta^*$ resonances are the additional 2016 states with only $^{*}$ and $^{**}$ states, which are unconfirmed. If a number of those could be confirmed experimentally they could have a significant impact on a more complete understanding of this fundamental hadron phase transition, as the black dashed line indicates. The graphics is adapted from ~\cite{Chatterjee:2017yhp}.  }
\label{universe}
\end{figure}

The enlarged number of resonances is not only important in understanding the dynamics of the nucleon's excitation spectrum. The number of resonances also plays an important role in the development of the early universe at the time when the transition occurred from a quark-gluon phase to the hadron phase. 
In high-energy heavy ions collisions at RHIC, at CERN, and other places, measurements have been performed that try to reproduce aspects of this phase transition. As we know today, this transition is strongly influenced by the hadron excitation spectrum, both from mesons and baryons as well as from the quark content (light or heavy quarks). Figure~\Ref{universe} shows the dependence of the strangeness chemical potential over the baryon chemical potential in a temperature range around the critical point. The various areas and lines show the sensitivity of this ratio to the baryon resonance content. 
One can see that the spectrum of resonances as known in 2012 is not sufficiently rich to reproduce the freeze-out after
which the number of baryons is frozen: The newly found resonances were involved in shaping the world.

\subsection{Electroproduction of pseudoscalar mesons}
\label{formalism}

The analysis tools for pion and photoproduction of mesons off nucleons are described in detail in
recent review articles, in particular in~\cite{Thiel:2022xtb}. The formalism used to analyze the electroproduction
data is accessible only in the original articles. Therefore, the methods used are
described in some detail in this review.

\subsubsection{Formalism}

The simplest process used in the study of resonance transition amplitudes is single pion production off proton targets, e.g. $ep \to e\pi^0 p^\prime$ or $ep \to e \pi^+ n$, and $ep\to ep^\prime \eta$. These processes are most suitable for the study of resonances in the mass range up to about 1.7~GeV. At higher masses, two-pion channels begin to dominate their hadronic decays. While many of these higher-mass states have been studied in photoproduction, there is very little information about the $Q^2$ dependence of their transition amplitudes, and we do not discuss them in this article. 
 Resonances in the mass range up to about 1.7~GeV and the $Q^2$ dependence of their transition amplitudes are
the topics we will elucidate in the following sections, together with their relevance for (approximations to) 
QCD. First, we briefly discuss the formalism needed for a quantitative analysis of data on electroproduction
of single pseudoscalar mesons.
%\clearpage

The unpolarized differential cross section for single pseudoscalar meson production 
can be written in the one-photon exchange approximation as %\\[-4ex]
\begin{equation}
 \frac {d^5\sigma} {dE_f d\Omega_e d\Omega_{\pi}} = \Gamma {\frac {d^2\sigma}  {d\Omega_{\pi}}}\,,  
\end{equation}    
where $\Gamma$ is the virtual photon flux, %\\[-4ex]
\begin{equation}
\Gamma = \frac{d^3\sigma}{dE_fd\Omega_e}%\nonumber \\
 = \frac{\alpha_{em}}{2\pi^2Q^2} \frac{(W^2-M^2)E_f}{2ME_i} \frac{1}{1 - \epsilon }\,, 
\end{equation}
$M$ is the proton mass, $W$ the mass of the hadronic final state, $\epsilon$ 
is the photon polarization parameter, $Q^2$ the photon virtuality, and $E_i$ and $E_f$ represent the initial and the final electron energies, respectively. Moreover,%\\[-7ex]
\begin{align} 
\epsilon &= \left[ 1 + 2\left(1 + \frac{\nu^2}{Q^2}\right) \tan^2 \frac{\theta_e}{2}\right]^{-1} \,,\\
\frac{d\sigma}{d\Omega_{\pi}} &=\sigma_T + \epsilon \,\sigma_L + \epsilon\,\sigma_{TT}\cos2\phi_{\pi} + \sqrt{2\epsilon(1+\epsilon)}\,\sigma_{LT} \cos{\phi}_{\pi}\,. 
\label{eq:response}
\end{align}

\begin{figure}[b!]
\centering
\includegraphics[width=0.5\linewidth]{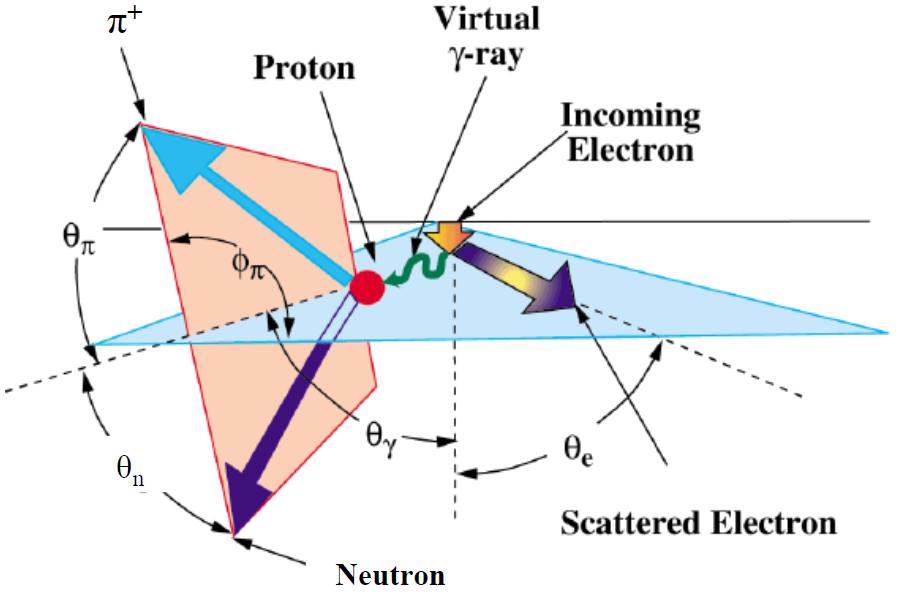}
\caption{(color online) The kinematics of single $\pi^+$ electroproduction off protons in the laboratory system.}
\label{kine}
\end{figure}

The kinematics for single $\pi^+$ production is shown in Fig.~\ref{kine}.
\noindent The observables of the process $\gamma^\ast p \to \pi N'$ are often expressed  in terms of six helicity amplitudes~\cite{Aznauryan:2011qj,Walker:1968xu,Berends:1967vi} \vspace{-2mm}
 \begin{eqnarray}
{H_i = \left<\lambda_\pi;\lambda_N|T|\lambda_{\gamma^\star} ; \lambda_p\right> , }\vspace{-2mm}
\end{eqnarray} 
where $\lambda$ denotes the helicity of the respective particle, $\lambda_\pi = 0$, $\lambda_N = \pm \frac{1}{2}$,  $\lambda_{\gamma^\star} = \pm 1, 0$, and $\lambda_p = \pm \frac{1}{2}$, and $H_i$ are complex functions of $Q^2$, $W$, and $\theta^*_\pi$.  %\vspace{-2mm}

The response functions in Eq.~(\ref{eq:response}) are given by
\begin{equation}
\begin{array}{rl}
\sigma_T &\!\!\!= \displaystyle\frac{|\vect{p}_\pi| W}{2KM} \,(|H_1|^2 + |H_2|^2 + |H_3|^2 + |H_4|^2)\,, \\[4mm]
\sigma_{TT} &\!\!\!= \displaystyle\frac{|\vect{p}_\pi|  W}{2KM} \,\text{Re}\, (H_2H_3^* - H_1H_4^*)\,, 
\end{array}\qquad
\begin{array}{rl}
\sigma_L &\!\!\!= \displaystyle \frac{|\vect{p}_\pi|  W}{2KM} \,(|H_5|^2 + |H_6|^2)\,,\\[4mm]
\sigma_{LT} &\!\!\!= \displaystyle \frac{|\vect{p}_\pi|  W}{2KM} \,\text{Re}\, [H_5^*(H_1-H_4) + H_6^* (H_2 + H_3)]\,,
\end{array}
\end{equation}
where $|\vect{p}_\pi|$ is the pion three-momentum in the hadronic center-of-mass system, and $K$ is the equivalent real photon lab energy
needed to generate a state with mass $W$: 
\begin{equation} 
K = \frac{W^2 - M^2}{2M}\,.
\end{equation}
The helicity amplitudes $H_i$, $i=1\dots 6$, can be expanded into Legendre polynomials, 
\begin{equation}
\begin{array}{rl}
H_1&\!\!\!=\displaystyle\frac{1}{\sqrt{2}}\sin\theta\cos{\frac{\theta}{2}} \sum_{l=1}^\infty (B_{l+}-B_{(l+1)-})(P^{\prime\prime}_l-P^{\prime\prime}_{l+1})\,, \\[4mm] 
H_2&\!\!\!=\displaystyle\sqrt{2}\cos{\frac{\theta}{2}} \sum_{l=1}^\infty (A_{l+}-A_{(l+1)-})(P^{\prime}_l-P^{\prime}_{l+1})\,, \\[4mm]
H_3&\!\!\!=\displaystyle\frac{1}{\sqrt{2}}\sin\theta\sin{\frac{\theta}{2}} \sum_{l=1}^\infty (B_{l+}+B_{(l+1)-})(P^{\prime\prime}_l+P^{\prime\prime}_{l+1})\,,
\end{array}\qquad
\begin{array}{rl}
H_4&\!\!\!=\displaystyle\sqrt{2}\sin{\frac{\theta}{2}} \sum_{l=1}^\infty (A_{l+}+A_{(l+1)-})(P^{\prime}_l+P^{\prime}_{l+1})\,,\\[4mm]
H_5&\!\!\!=\displaystyle\sqrt{2}\cos{\frac{\theta}{2}} \sum_{l=1}^\infty (C_{l+}-C_{(l+1)-})(P^{\prime}_l-P^{\prime}_{l+1})\,,\\[4mm]
H_6&\!\!\!=\displaystyle\sqrt{2}\sin{\frac{\theta}{2}} \sum_{l=1}^\infty (C_{l+}+C_{(l+1)-})(P^{\prime}_l+P^{\prime}_{l+1})\,,
\end{array} 
\end{equation}
where the $A_{l+}$ and $B_{l+}$ etc., are the transverse partial wave helicity elements for $\lambda_{\gamma p} =  \frac{1}{2}$  and 
$\lambda_{\gamma p} = \frac{3}{2}$, and $C_{\pm}$  the longitudinal partial wave helicity elements. 
In the subscripts, $l+$ and $(l+1)-$ define the $\pi$ orbital angular momenta, and the sign
$\pm$ is related to the total angular momentum $J = l_{\pi} \pm \frac{1}{2}$.

In the analysis of data on the $N\gamma^\ast \to \Delta(1232)$ transition, linear combinations of partial wave helicity 
elements are expressed in terms of electromagnetic multipoles, 
\begin{equation}
\begin{array}{rl}
M_{l+} &\!\!\!=\displaystyle \frac{1}{2(l+1)} \,[2A_{l+} - (l+2)B_{l+}]\,, \\[4mm]
M_{l+1,-} &\!\!\!=\displaystyle \frac{1}{2(l+1)} \,(2A_{l+1,-} - lB_{l+1,-1})\,, \\[4mm]
S_{l+} &\!\!\!=\displaystyle \frac{1}{ l+1} \sqrt{\frac{{\vect{Q^*}^2}}{Q^2}}\,C_{l+}\,,
\end{array}\qquad
\begin{array}{rl}
E_{l+} &\!\!\!=\displaystyle\frac{1}{2(l+1)} \,(2A_{l+} + lB_{l+})\,, \\[4mm]
E_{l+1,-} &\!\!\!=\displaystyle\frac{1}{2(l+1)} \,[-2A_{l+1,-} + (l+2)B_{l+1,-}] \,, \\[4mm]
S_{l+1,-} &\!\!\!=\displaystyle \frac{1}{ l+1} \sqrt{\frac{{\vect{Q^*}^2}}{Q^2}}\,C_{l+1,-}\,,
\end{array}
\end{equation}
where $\vect{Q}^*$ is the photon 3-momentum in the hadronic rest frame. 
Electromagnetic multipoles are often used to describe the transition from the nucleon ground state to the $\Delta(1232)$, which is predominantly %described as 
a magnetic dipole transition $M_{1+}$.      
The electromagnetic multipoles as well as the partial-wave helicity elements are complex
quantities and contain both non-resonant and resonant contributions. 
In order to compare the results to model predictions and lattice QCD, an additional analysis must 
be performed to separate the resonant parts $\hat{A}_{\pm}$,  $\hat{B}_{\pm}$, etc., 
from the non-resonant parts of the amplitudes. In a final step, the
known hadronic properties of a given resonance can be used to determine
photocoupling helicity amplitudes that characterize the electromagnetic vertex,
\begin{equation}
\begin{array}{rl}
\hat{A}_{l\pm} &\!\!\!=\displaystyle \mp F\,C^I_{\pi N} \,A_{1/2}\,, \\[4mm]
\hat{S}_{l\pm} &\!\!\!=\displaystyle - F \,\frac{2\sqrt{2}}{2J+1} \,C^I_{\pi N} \,S_{1/2}\,,   
\end{array}\qquad
\begin{array}{rl}
\hat{B}_{l\pm} &\!\!\!=\displaystyle \pm F \,\sqrt{\frac {16} {(2j -1)(2j + 3)} } \,C^I_{\pi N} \,A_{3/2}\,, \\[4mm]
F &\!\!\!=\displaystyle \sqrt{\frac{1}{(2j + 1)}\pi} \,\frac {K}{p_\pi} \,\frac{\Gamma_\pi}{\Gamma^2} \,,
\end{array}
\end{equation}
where the $C^I_{\pi N}$  are isospin coefficients.   
The subscripts $1/2$ and $3/2$ represent the total helicity $\pm 1/2$ and $\pm 3/2$ of the photon-proton system $\gamma^* p$ in the initial state.  
The total transverse absorption cross section for the transition into a specific resonance is given by \begin{equation}
\sigma_T = \frac{2M}{W_R\Gamma} (A^2_{1/2} + A^2_{3/2})\,, 
\end{equation} 
and similarly for the longitudinal (scalar) total cross section $\sigma_L$. The quantities $A_{1/2},~A_{3/2},~S_{1/2}$ are the transverse and scalar transition amplitudes for resonances excited through photon-proton interactions. Resonances with $J = {1/2}$ have only $A_{1/2}$ and $S_{1/2}$ contributions, while resonances with $ J \ge 3/2 $ can also be excited through $A_{3/2}$ contributions. 
There are predictions for the $Q^2$ dependence of the $A_{1/2}$ and $A_{3/2}$ amplitudes in asymptotic QCD: \begin{equation} 
A_{1/2} \propto \frac{a}{Q^3}, \qquad A_{3/2} \propto \frac{b}{Q^5}, \qquad Q^2 \to \infty\,.
\label{pQCD}
\end{equation} 
Experiments in the region of the $\Delta(1232)\frac{3}{2}^+$ resonance often determine the electric and scalar quadrupole ratios $R_{EM}$ and $R_{SM}$, \begin{equation}
R_{EM} = \frac{\text{Im}\,E_{1+}}{\text{Im}\,M_{1+}}, \qquad
R_{SM} = \frac{\text{Im}\,S_{1+}}{\text{Im}\,M_{1+}} \,, \label{QR}
\end{equation}
where $E_{1+}$, $S_{1+}$, and $M_{1+}$ are the electromagnetic transition multipoles at the mass of the $\Delta(1232)\nicefrac{3}{2}^+$ resonance. 

\subsubsection{\bf Resonance analysis tools}

The analysis of electroproduction data at fixed $Q^2$
makes use of the known properties of the resonances: masses and widths (i.e., pole positions), and couplings to the 
pseudoscalar mesons in the final states $N^{res} \to \pi N$, $N \eta$, $K\Lambda$. 

A model-independent determination of the amplitudes contributing to the electro-excitation of resonances in single pseudoscalar pion production $ep \to e^\prime N \pi$ (see kinematics  
of single pion production in Fig.~\ref{kine}) requires 
a large number of independent measurements at each value of the electron kinematics $W$, $Q^2$, the hadronic cms angle $\cos{\theta^\pi}$, and the azimuthal angle $\phi^\pi$ describing 
the angle between the electron scattering plane and the hadronic decay plane. Such a measurement
requires full exclusivity of the final state and employing both polarized electron beams and the measurements of the nucleon recoil polarization. 

\begin{figure}[!t]
\vspace{-3mm}
\centering
\includegraphics[width=0.8\linewidth]{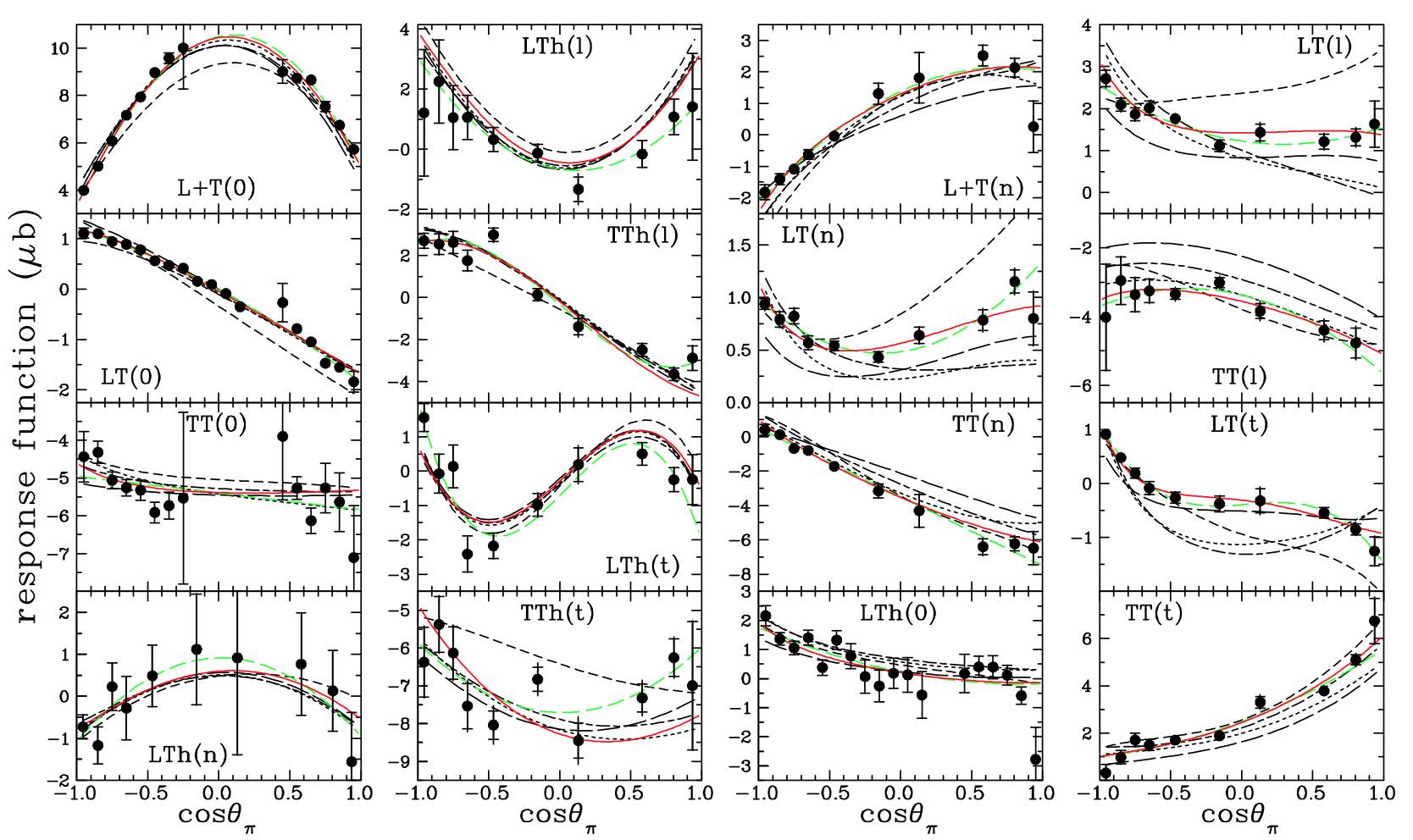}
\caption{JLab/Hall A data for $\vec{e}p  \to e\vec{p}\pi^0$ response functions at W=1.232 GeV and $Q^2=1.0$~\cite{Kelly:2005jy}. Notations refer to transverse (t), normal (n) and longitudinal (l) components of the proton recoil polarization. The curves correspond to results obtained using SAID (short dashed), MAID (dashed-dotted), and the dynamical models DMT~\cite{Kamalov:2000en} (dotted), and SL~\cite{Sato:2000jf}   (long-dashed/green). The other curves correspond to Legendre and multipole fits performed by the authors.}
\label{Hall-A}
\end{figure}
Exclusive measurements would in general require full $4\pi$ coverage for the hadronic final state. The only measurement that could claim to be complete was carried out at Jefferson Lab in Hall A~\cite{Kelly:2005jy} employing a limited kinematics centered at resonance  for $\vec{e} p \to e^{\prime} \vec{p} \pi^0$ at $W = 1.232$~GeV, and $Q^2 \approx 1$~GeV$^2$. Figure~\ref{Hall-A} shows the 16 response functions extracted from this measurement. The results of this measurement in terms of the magnetic $N\Delta$ transition form factor and the quadrupole ratios are included in 
Fig.~\ref{Delta-GM-REM-RSM} in Section~\ref{Structure} among other data. They coincide very well with results of other experiments~\cite{CLAS:2009ces,CLAS:2006ezq,CLAS:2001cbm,Frolov:1998pw} using different analysis techniques that may be also applied to broader kinematic conditions, especially higher-mass resonances. Details of the latter are discussed in ~\cite{Aznauryan:2011qj,Tiator:2011pw}.
A similar analysis has recently been carried out within the J\"ulich-Bonn-Washington model for pion electroproduction multipoles~\cite{Mai:2021vsw}, and for combined pion and eta electroproduction~\cite{Mai:2021aui}. Their results for pion production at the $\Delta(1232)$ compare generally well with the ones used in the analysis~\cite{Kelly:2005jy} but fitted to parameterized amplitudes, while the original analysis was performed at each center-of-mass polar angle separately.  

We can also make use of theory constraints, as briefly described here:     
   \begin{itemize}
    \item {\sl Dispersion Relations (DR)} have been employed in two ways: One is based on fixed-$t$ dispersion relations 
    for the invariant amplitudes and was successfully used thr\-oughout the nucleon resonance region. Another way is 
   based on DR for the multipole amplitudes of the $\Delta(1232)$ resonance and allows one to get functional forms
    of these amplitudes with one free parameter for each of them. It was employed for the analysis of 
    the more recent data.   
    \item {\sl The Unitary Isobar Model (UIM) } was developed in ~\cite{Drechsel:1998hk} from the effective Lagrangian 
    approach for pion photoproduction~\cite{Peccei:1969sb}. Background contributions from $t$-channel $\rho$ and $\omega$ 
    exchanges are introduced and the overall amplitude is unitarized in a K-matrix approximation. 
    \item {\sl Dynamical Models} have been developed, such as e.g. SAID from pion photoproduction data~\cite{Arndt:2002xv}. The essential feature of the Sato-Lee model  developed in~\cite{Sato:1996gk} is the consistent description of $\pi N$ scattering and pion 
     electroproduction off nucleons. It 
    was utilized in the study of $\Delta(1232)$ excitations in the $ep\to ep\pi^0$ channel~\cite{Sato:2000jf}. The Dubna-Mainz-Taipei model~\cite{Kamalov:1999hs} builds in unitarity by directly including the $\pi N$ final state in the T-matrix of photo- and electroproduction. 
\end{itemize}  
For some recent reviews of resonance electroproduction and analysis we refer to~\cite{Ramalho:2023hqd,Aznauryan:2011qj}. 
The results presented here have been obtained by using both the dispersion relation {\sl DR} approach as well as the {\sl UIM} approach. 
The difference of the two results is the most important contribution to the systematic uncertainty in an evaluation
of the model dependence.

\begin{figure*}[t]
\centering
\resizebox{1.\columnwidth}{!}{\includegraphics{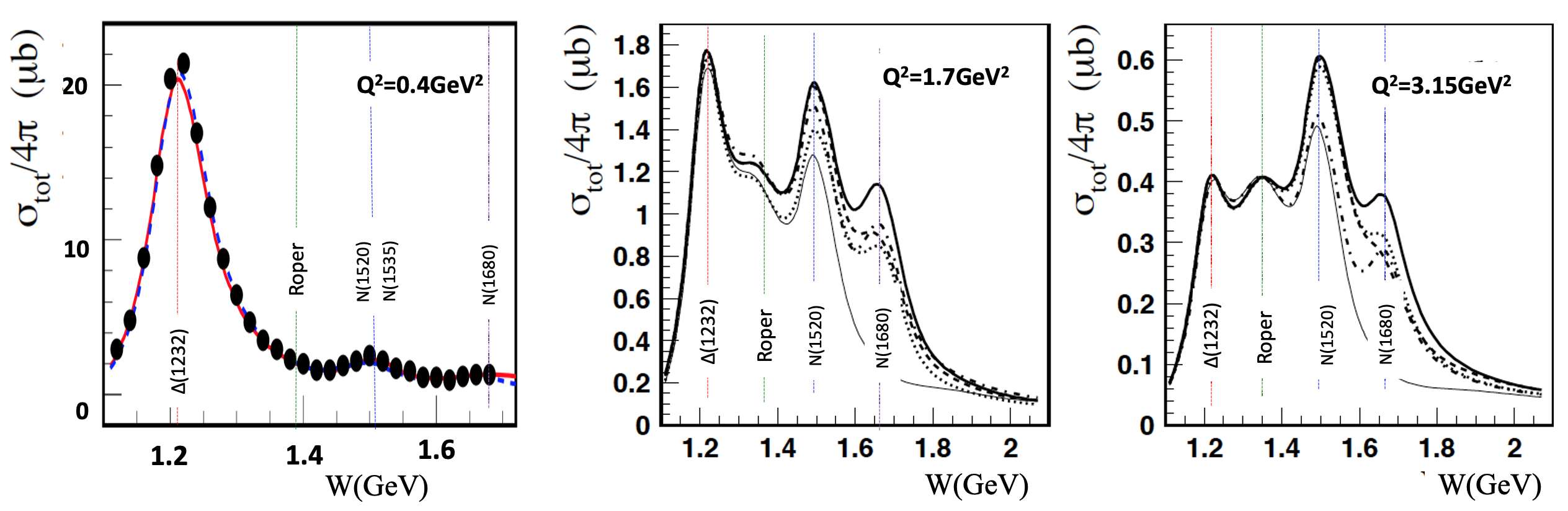}}
\caption{\small Samples of total virtual photon absorption cross sections for different $Q^2$ values. The different dependencies on $Q^2$ of the different resonance regions are quite evident. At low $Q^2$ the $\Delta(1232)\nicefrac32^+$ is the absolute dominant contribution, while the Roper $N(1440)\nicefrac12^+$ is hardly visible. At high $Q^2$ the higher mass states, even the Roper resonance, become dominant over the $\Delta(1232)$.
The bold solid line indicates the total contribution of all resonances. The dashed, dotted, dashed-dotted and thin solid curves indicate the total contributions when specific resonance contributions are excluded.}
\label{total-crs}
\end{figure*}

In Fig.~\ref{total-crs} we show the total photo-absorption cross section and various values of the virtuality $Q^2$ of the photon. There is obviously a very large dependence of the excitation strength in the various resonance regions on the photon virtuality parameter $Q^2$. 
All the states in this mass range have been assigned to states in multiplets of the [SU(6)$\otimes $O(3)] representation of the symmetric quark model. However, several of these
states have also been discussed as meson-baryon resonances, or composite states. In particular, several of the low mass resonances $N(1440)\nicefrac12^+$, $N(1535)\nicefrac12^-$, $N(1520)\nicefrac32^-$, and $N(1710)\nicefrac12^+$ have been assigned as ''tend to be composite" states~\cite{Wang:2024jns}. 

How may meson electroproduction help reveal the nature of the excited states? 
In the following study we use the aforementioned $Q^2$ dependence of the excitation strength as a means of revealing the different contributions to the resonance strength and to identify the quark core and meson-baryon contributions.   

\clearpage

\section{\label{Models} Baryon Spectrum}
\subsection{Historical remarks} %Origins of the quark model}
It took a long path before all the resonances listed in Tables~\ref{tab:statusN} and~\ref{highmass-piN} were identified.  
The RPP 2023 lists 116 light-quark baryons. In the first RPP issue in 1957 \cite{Gell-Mann:1957uuj}, no baryon resonance was included,  
even though the $\Delta^{++}(1232)$ isobar had been observed more than five years earlier. Enrico Fermi had used a $\pi^+$ meson beam  
scattering off protons in a hydrogen target~\cite{Anderson:1952nw}. The cross section showed a sharp rise from the $\pi$ production  
threshold up to about 1200~MeV. The pion beam's energy was insufficient to reveal the full resonance profile, but there was a strong  
indication of the first baryon resonance. In 1961, the PDG recognized the existence of short-lived resonances, reporting the properties  
of the isospin $I=\nicefrac32$ resonance $\Delta(1232)$ and early evidence for the isospin $I=1$ 
resonance $\Sigma(1385)$~\cite{barkas1961data}. The discovery  
of a double-strange $\Xi(1530)$ \cite{Pjerrou:1962oxa} as an isospin $I=\nicefrac12$ resonance helped predict an isospin-0 resonance  
$\Omega^-$ at the correct mass \cite{Gell-Mann:1962yej}, paving the way for quarks to be proposed as the fundamental constituents of  
mesons and baryons~\cite{Gell-Mann:1964ewy,Zweig:1964jf}.  

In the quark model, baryons consist of three constituent quarks, each with a spin  $s=\nicefrac12$. In $N^*$ and $\Delta^*$ resonances, all three  
quarks possess the isospin $I=\nicefrac12$, with the up ($u$) quark having an isospin component $I_3=+\nicefrac12$ and 
charge $Q=+\nicefrac{2}{3}$ (in units of the elementary charge $e$), while the down ($d$) quark has $I_3=-\nicefrac12$ and $Q=-\nicefrac{1}{3}$.
The total quark spin can be $S=\nicefrac12$ or $S=\nicefrac32$, and the 
total isospin $I=\nicefrac12$ or $I=\nicefrac32$. Baryons with isospin $I=\nicefrac12$ are  
referred to as nucleons, while $\Delta$ baryons have $I=\nicefrac32$. The $\Delta^{++}(1232)$ corresponds to  
$|u{\uparrow}u{\uparrow}u{\uparrow}\rangle$, with the three quark spins aligned. This alignment should be forbidden in a ground state  
with all quarks in an $S$ wave, as the overall wave function would be symmetric, contradicting the Pauli principle. 
Gell-Mann warned:  
``Such particles [quarks] presumably are not real, but we may use them in our field theory anyway \ldots"~\cite{Gell-Mann:1964hhf}.  
It took the introduction of para-Fermi statistics~\cite{Greenberg:1964pe}, later coined ``color" by Gell-Mann, to make the overall wave function  
anti-symmetric. Thus, the $\Delta^{++}(1232)$ resonance may be seen as a harbinger of the development of QCD.  

In strange baryons, an up or down quark is replaced by a strange ($s$) quark with 
isospin $I=0$ and charge $Q=-\nicefrac{1}{3}$. In cascade baryons, two  strange quarks are combined with an up or down quark. 
The $\Xi^0$ and $\Xi^-$ baryons decay  mostly in a {\it cascade} into $\Lambda\pi \to N\pi\pi$. The strangeness 
quantum number $S=-1$ is assigned to $\Sigma$, $S=-2$ to $\Xi$ baryons. Later, we will shortly discuss properties
of baryons with charm and with bottomness, with a charm or bottom quark, $c$ with $Q=\nicefrac{2}{3}$ and $b$
with $Q=-\nicefrac{1}{3}$. Top ($t$) quarks are presumably too short-lived to form baryons or mesons, although recent experimental results have sparked a new debate on this matter~\cite{CMS:2025kzt}.

The isospin, and thus $SU(2)$ symmetry, was introduced 1932 by Heisenberg \cite{Heisenberg:1932aaa}. Ne'eman and 
Gell-Mann~\cite{Neeman:1961jhl,Gell-Mann:1961omu} organized mesons and baryons in a new symmetry group, 
$SU(3)$, today called flavor $SU(3)_f$, see the inset in Fig.~\ref{fig:spectrum-1}. Based on this scheme,
Gell-Mann and Okubo \cite{Gell-Mann:1961omu,Okubo:1961jc} derived a mass formula, which for baryons can be cast into the form 
\begin{eqnarray}
\frac{1}{2}\,(M_N+M_\Xi)=\frac{1}{4}\,(3M_\Lambda+M_\Sigma)
\end{eqnarray}
and holds to sub-percent accuracy. For the ground-state  decuplet baryons, the equal-spacing rule 
\begin{eqnarray}
M_\Sigma^*-M_\Delta=M_\Xi^*-M_\Sigma^*=M_\Omega-M_\Xi^*\approx 147\,{\rm MeV}
\end{eqnarray}
can be thought of as the difference in the strange and light $(u,d)$ quark masses.

The seemingly simple understanding of the magnetic moments of protons and neutrons was another success of the
quark model. The spin and flavor wave functions for protons and neutrons are given in Eq.~\ref{spin-1/2} and
Table~\ref{tab:baryons-flavor-2} below; taking them into account using quark magnetic moments $e_q/(2m_q)$ one finds
\begin{eqnarray}
\mu_p=\frac{4}{3}\,\mu_u-\frac{1}{3}\,\mu_d = \frac{e}{2m_q}\,, \qquad
\mu_n=\frac{2}{3}\,\mu_d-\frac{2}{3}\,\mu_u =-\frac{2}{3}\frac{e}{2m_q}\,, \qquad 
\frac{\mu_n}{\mu_p}=-\frac23 \,.
\end{eqnarray}
The relation $\mu_p=e/(2m_q)$ immediately explains why the $g$ factor of the proton
is not 2 (or slightly larger) but 5.586 (known with many more digits!). The relevant mass
is not the proton mass but the quark mass. Scaling the masses with the $g$ factor, one can estimate the quark masses to  
$m_{u,d} \approx 340\,{\rm MeV}$.
The ratio $\mu_n/\mu_p$ can be compared to the experimental value  $-0.685$. Given this agreement, it came
as a surprise when the EMC Collaboration reported that quarks contribute only little to the
proton's spin~\cite{EuropeanMuon:1987isl}. For many years, this question remained a topical issue.
The spin of quarks and gluons and intrinsic orbital angular momenta all contribute significantly to the 
proton's spin. New experiments are planned at several laboratories to improve the accuracy of these 
determinations, see e.g.~\cite{Ji:2020ena} for a review.

Back to history: At the Les Houches Summer School in 1965, Dalitz reported on {\it Quark models for the ‘elementary particles’}  
and sketched the pattern of negative- and positive-parity excitations~\cite{Dalitz:1965fb}. More details were  
presented one year later at the Oxford International Conference on Elementary Particles~\cite{Dalitz:1966fd}.  
For the first time, an interpretation was given within the quark model for the (so far few) nucleon and hyperon resonances.  
In 1972, color charge was identified as the source of the strong interaction~\cite{Fritzsch:1972jv}. The $SU(3)$ structure of color, together with  
the requirement that the interaction be invariant under local gauge transformations, defined Quantum Chromodynamics (QCD) as the theory  
of strong interactions~\cite{Fritzsch:1973pi}, with ``asymptotic freedom" and ``infrared slavery"~\cite{Gross:1973id,Politzer:1973fx}.  
The discovery of $J/\psi$ at Brookhaven~\cite{E598:1974sol} and SLAC~\cite{SLAC-SP-017:1974ind} in 1974, the observation of the $\psi'$  
at SLAC~\cite{Abrams:1974yy}, and the interpretation of these narrow resonances using the Cornell potential~\cite{Eichten:1974af},  
\begin{eqnarray}  
   V(r) = -\frac{4}{3}\frac{\alpha_s}{r} + br  \,,
   \label{Eichten}  
\end{eqnarray}  
finally established the quark model, where the first term in Eq.~\eqref{Eichten} is attributed to one-gluon exchange and  
the second term enforces confinement.  
The combination of the confinement potential and some kind of effective interactions between quarks forms the basis of the quark models   
discussed in more detail in Sec.~\ref{pheno} below. Here, we briefly introduce elements of the non-relativistic quark model needed for an interpretation  
of the mass spectra of light-quark baryon resonances.

\subsection{Baryon wave functions}
\label{Construction}

Baryon resonances are characterized by their mass, their spin-parity $J^P$, and their flavor, i.e., their quark content. In the $SU(6)$-symmetric quark model,
the wave function of baryons contains four parts: 
$|\Psi\rangle = |\psi_{\rm space}\rangle \otimes |\chi_{\rm spin}\rangle \otimes |\tau_{\rm flavor}\rangle \otimes |\kappa_{\rm color}\rangle$.
For $N^*$ and $\Delta^*$ resonances, flavor
is reduced to isospin. 
The overall wave function must be antisymmetric with respect to the exchange
of any quark pair. The color wave function $|\kappa_{\rm color}\rangle = \frac{1}{\sqrt6}\,|rgb - rbg + brg - bgr + gbr -grb\rangle$,
where $r,g,b$ stand for the three colors,
is fully antisymmetric, so the remaining space-spin-flavor
wave function needs to be symmetric. \\[-4ex]

\subsubsection{Spin-flavor $SU(6)$}\label{sec:su(6)}
The flavor wave functions $|\tau_{\rm flavor}\rangle$ of baryons with light and strange quarks can be
classified in $SU(3)_f$ flavor representations according to the decomposition
$\bm{3} \otimes \bm{3} \otimes \bm{3} = \bm{10}_\mathcal{S} \oplus \bm{8}_{\mM_\mS} \oplus \bm{8}_{\mM_\mA} \oplus \bm{1}_\mathcal{A}$,
\textit{i.e.}, a decuplet which is symmetric with respect to the exchange of
any two quark flavors, a singlet which is antisymmetric and two octets of
mixed symmetry. 
This is straightforward to derive using the properties of the permutation group S$_3$, whose representations form symmetric singlets $\mS$, mixed symmetric doublets $\mD = ( \mM_\mA, \mM_\mS)^T$, and antisymmetric singlets $\mA$.
The upper component $\mM_\mA$ of a doublet $\mD$ is antisymmetric in the first two entries and the lower component $\mM_\mS$ is symmetric.
Applied to the quark flavors $u$, $d$, $s$,
the 27 possible combinations can then be arranged into the flavor wave functions in Table~\ref{tab:baryons-flavor-2}.
The eight doublets in the upper block of the table are the flavor octet wave functions,
the ten singlets in the second block the flavor decuplet wave functions, and the antisinglet in the third block is the flavor singlet wave function. Thus, the correspondence is $\mathbf{8} \; \leftrightarrow \; \mD_f$, $\mathbf{10} \; \leftrightarrow \; \mS_f$ and $\mathbf{1} \; \leftrightarrow \; \mA_f$,
with a subscript $f$ for flavor. 
This yields six `families' of baryons characterized by the same isospin $I$ and strangeness $S$:
nucleons (proton and neutron) in the octet, $\Delta$ baryons in the decuplet, $\Lambda$ baryons in the octet and singlet,
$\Sigma$ and $\Xi$ baryons both in the octet and decuplet, and $\Omega$ baryons in the decuplet.

             \begin{table*}[ht]

               \caption{$SU(3)_f$ flavor wave functions for octet, decuplet and singlet baryons. $I$ and $I_3$ denote the isospin and isospin-3 component and S the strangeness.}
               \label{tab:baryons-flavor-2}
                \centering
                \footnotesize
                \setlength\tabcolsep{3mm}
                \begin{tabular}{   |  c |  l | c | r | r | c | }  \hline \rule{-1mm}{0.35cm}

                                              Flavor content  & Baryon & $I$  & $I_3$  & S & Flavor wave function     \\[1mm] \hline\hline  \rule{-1mm}{0.6cm}

                                          $uud$  &  $p$ & $\nicefrac{1}{2}$  & $\nicefrac{1}{2}$  & $0$ &  $\frac{1}{\sqrt{2}}\left[ \begin{array}{c} udu-duu \\ -\frac{1}{\sqrt{3}}\,(udu+duu-2uud) \end{array}\right]$  \\[4mm]

                                          $udd$  &  $n$ & $\nicefrac{1}{2}$  & $-\nicefrac{1}{2}$  & $0$ &  $\frac{1}{\sqrt{2}}\left[ \begin{array}{c} udd-dud \\  \frac{1}{\sqrt{3}}\,(dud+udd-2ddu) \end{array}\right]$  \\[4mm] \hline  \rule{-1mm}{0.6cm}

                                          $uus$  &  $\Sigma^{+}$ & $1$  & $1$  & $-1$ &  $\frac{1}{\sqrt{2}}\left[ \begin{array}{c} usu-suu \\ -\frac{1}{\sqrt{3}}\,(usu+suu-2uus) \end{array}\right]$  \\[4mm]
                                          $uds$  &  $\Sigma^{0}$ & $1$  & $0$  & $-1$ & $\frac{1}{2}\left[ \begin{array}{c} sud-usd+sdu-dsu \\ \frac{1}{\sqrt{3}}\,(sud+usd+sdu+dsu-2uds-2dus) \end{array}\right]$  \\[4mm]
                                          $dds$  &  $\Sigma^{-}$ & $1$  & $-1$  & $-1$ & $\frac{1}{\sqrt{2}}\left[ \begin{array}{c} dsd-sdd \\ -\frac{1}{\sqrt{3}}\,(dsd+sdd-2dds) \end{array}\right]$  \\[4mm] \hline  \rule{-1mm}{0.6cm}

                                          $uds$  &  $\Lambda^0$ & $0$  & $0$  & $-1$ & $\frac{1}{2}\left[ \begin{array}{c} \frac{1}{\sqrt{3}}\,(2uds-2dus+usd-dsu+sdu-sud) \\ usd-dsu+sud-sdu \end{array}\right]$  \\[4mm] \hline  \rule{-1mm}{0.6cm}

                                          $uss$  &  $\Xi^{0}$ & $\nicefrac{1}{2}$  & $\nicefrac{1}{2}$  & $-2$ & $\frac{1}{\sqrt{2}}\left[ \begin{array}{c} uss-sus \\  \frac{1}{\sqrt{3}}\,(sus+uss-2ssu) \end{array}\right]$  \\[4mm]
                                          $dss$  &  $\Xi^{-}$ & $\nicefrac{1}{2}$  & $-\nicefrac{1}{2}$  & $-2$ & $\frac{1}{\sqrt{2}}\left[ \begin{array}{c} dss-sds \\  \frac{1}{\sqrt{3}}\,(sds+dss-2ssd) \end{array}\right]$ \\[4mm] \hline\hline  \rule{-1mm}{0.4cm}

                                          $uuu$  &  $\Delta^{++}$ & $\nicefrac{3}{2}$  & $\nicefrac{3}{2}$  & $0$ &  $uuu$  \\
                                          $uud$  &  $\Delta^{+}$ & $\nicefrac{3}{2}$  & $\nicefrac{1}{2}$  & $0$ &  $\frac{1}{\sqrt{3}}\,(uud+udu+duu)$  \\
                                          $udd$  &  $\Delta^{0}$ & $\nicefrac{3}{2}$  & $-\nicefrac{1}{2}$  & $0$ &  $\frac{1}{\sqrt{3}}\,(udd+dud+ddu)$  \\
                                          $ddd$  &  $\Delta^{-}$ & $\nicefrac{3}{2}$  & $\nicefrac{3}{2}$  & $0$ &  $ddd$  \\[2mm] \hline  \rule{-1mm}{0.4cm}

                                          $uus$  &  $\Sigma^{+}$ & $1$  & $1$  & $-1$ &  $\frac{1}{\sqrt{3}}\,(uus+usu+suu)$  \\
                                          $uds$  &  $\Sigma^{0}$ & $1$  & $0$  & $-1$ & $\frac{1}{\sqrt{6}}\,(uds+sud+dsu+dus+usd+sdu)$  \\
                                          $dds$  &  $\Sigma^{-}$ & $1$  & $-1$  & $-1$ & $\frac{1}{\sqrt{3}}\,(dds+dsd+sdd)$  \\[2mm] \hline  \rule{-1mm}{0.4cm}

                                          $uss$  &  $\Xi^{0}$ & $\nicefrac{1}{2}$  & $\nicefrac{1}{2}$  & $-2$ & $\frac{1}{\sqrt{3}}\,(uss+sus+ssu)$  \\
                                          $dss$  &  $\Xi^{-}$ & $\nicefrac{1}{2}$  & $-\nicefrac{1}{2}$  & $-2$ & $\frac{1}{\sqrt{3}}\,(dss+sds+ssd)$  \\[2mm] \hline  \rule{-1mm}{0.4cm}

                                          $sss$  &  $\Omega^{-}$ & $0$  & $0$  & $-3$ & $sss$\\[2mm] \hline\hline  \rule{-1mm}{0.4cm}

                                          $uds$  &  $\Lambda^0$ & $0$  & $0$  & $-1$ &  $\frac{1}{\sqrt{6}}\,(uds+sud+dsu-dus-usd-sdu)$  \\[2mm] \hline

                \end{tabular}
                \renewcommand{\arraystretch}{1.0}
        \end{table*}

In the non-relativistic quark model one goes a step further by combining the $SU(2)_s$ spin
and $SU(3)_f$ flavor wave functions into $SU(2)_s\otimes SU(3)_f \simeq SU(6)$ representations.
The spins $\bm{s}_i$ of the three quarks combine to form the total quark spin $\bm{S}$ with $S=\nicefrac{1}{2}$ or $\nicefrac{3}{2}$.
The $SU(2)_s$ spin wave functions, with index $s$ for spin, can be read off from the entries with $uuu$, $uud$, $udd$, $ddd$ in Table~\ref{tab:baryons-flavor-2}
with the replacements $u \to$ spin up ($\uparrow$) and $d \to$ spin down ($\downarrow$).
This yields two doublets for total quark spin $S=\frac{1}{2}$, 
\begin{align}\label{spin-1/2}
    \begin{split}
        \mD_s\, =\,  & \frac{1}{\sqrt{2}}\left[ \left|\,\uparrow \downarrow \uparrow - \downarrow \uparrow \uparrow \,\right\rangle\atop -\frac{1}{\sqrt{3}}\left|\, \uparrow \downarrow \uparrow + \downarrow\uparrow \uparrow - 2\uparrow\uparrow \downarrow\,\right\rangle \right], \quad \frac{1}{\sqrt{2}}\left[ \left|\,\uparrow \downarrow\downarrow - \downarrow \uparrow \downarrow\,\right\rangle \atop \frac{1}{\sqrt{3}}\left| \,\uparrow \downarrow\downarrow + \downarrow \uparrow \downarrow - 2\downarrow\downarrow\uparrow\,\right\rangle \right],
    \end{split}
   \end{align} 
and four singlets for $S=\frac{3}{2}$, which together form the spin wave functions $|\chi_\text{spin}\rangle$: % in Eq.~\eqref{total-wf-generic}:
\begin{align}\label{spin-3/2}
    \begin{split}
       \mS_s \,=\; &|\uparrow\uparrow\uparrow\,\rangle\,, \quad
       \frac{1}{\sqrt{3}}\,|\uparrow\uparrow\downarrow + \uparrow\downarrow\uparrow + \downarrow \uparrow\uparrow\,\rangle\,, \quad
       \frac{1}{\sqrt{3}}\,|\uparrow\downarrow\downarrow + \downarrow \uparrow\downarrow + \downarrow\downarrow \uparrow\,\rangle\,, \quad
       |\downarrow\downarrow\downarrow\,\rangle\,.
    \end{split}
   \end{align}

\begin{figure*}[!t]
      \centering
      \includegraphics[width=1\textwidth]{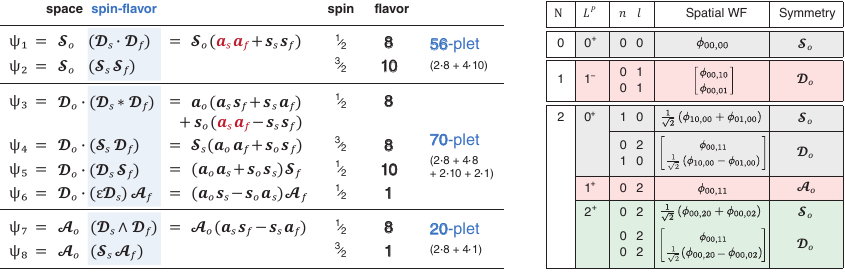}
      \caption{\textit{Left:} Space-spin-flavor wave functions in the SU(6)-symmetric quark model. Marked in red are components in the wave functions that are mixed-antisymmetric both in spin and  flavor. \textit{Right:} Oscillator states for $\text{N} \leq 2$ with definite S$_3$ symmetry. 
      ${\rm N}$ denotes the number of
    oscillator excitations, $L$ the total orbital angular momentum, $P$ the parity,
    and $n$ and $l$ are defined below Eq.~\eqref{oscillator-wfs}.
    The spatial wave functions form symmetric singlets ($\mS_o$), mixed (anti-)symmetric doublets ($\mD_o$) and antisymmetric singlets ($\mA_o$), with subscript~$o$ for `orbital'.}
      \vspace{-1mm}
      \label{fig:flavor-wfs}
    \end{figure*}

The combination of spin ($\mD_s$, $\mS_s$) and flavor wave functions ($\mD_f$, $\mS_f$, $\mA_f$) produces again singlets, doublets and antisinglets in spin-flavor space. 
For example, the only singlets that can be formed from ($\mD_s$, $\mS_s$) $\times$ ($\mD_f$, $\mS_f$, $\mA_f$)
are $\mD_s\cdot\mD_f$ and $\mS_s \,\mS_f$. Because there are two spin doublets $\mD_s$ in Eq.~\eqref{spin-1/2}  with $S=\tfrac{1}{2}$  
and eight flavor doublets in Table~\ref{tab:baryons-flavor-2}, which form the baryon octet, the combination $\mD_s\cdot\mD_f$ gives 16 possibilities.
Likewise, there are four spin singlets $\mS_s$ in Eq.~\eqref{spin-3/2} with $S=\tfrac{3}{2}$ and ten flavor singlets $\mS_f$ in Table~\ref{tab:baryons-flavor-2}
forming the decuplet, which results in 40 possibilities. Taken together, this returns the  $\mathbf{56}$-plet in $SU(6)$: $\bm{56} = {^2}\bm{8} \oplus {^4}\bm{10}$,
where the regular-sized symbols refer to the $SU(3)_f$ representation and the superscripts to the $SU(2)_s$ representation.
For the remaining multiplets one must find all possible spin-flavor doublets and antisinglets.
The possible combinations are collected in the left of Fig.~\ref{fig:flavor-wfs}, where we defined
\begin{eqnarray} \renewcommand{\arraystretch}{0.7}
    \mD = \left[  \begin{array}{c} \mM_\mA \\  \mM_\mS \end{array}\right] = \left[ \begin{array}{c} a \\  s  \end{array} \right], \quad 
    \mD' = \left[ \begin{array}{c} a' \\ s' \end{array} \right], \quad
    \begin{array}{rl}
     \mD \cdot \mD' &\!\!\!= aa' + ss'\,, \\
    \mD \wedge \mD' &\!\!\!= as' - sa'\,, 
    \end{array}\quad 
       \mD \ast \mD' = \left[ \begin{array}{c} as' + sa' \\ aa' - ss' \end{array}\right],
        \quad
       \varepsilon = \left( \begin{array}{r @{\;\;\;}r} 0 & 1 \\ -1 & 0 \end{array}\right).
\end{eqnarray} 
For example, $\mD_s\ast\mD_f$ and ($\varepsilon\mD_s)\mA_f$ are doublets and $\mD_s\wedge\mD_f$ is an antisinglet.
In total, this gives  the $SU(6)$ spin-flavor decomposition
$\bm{6} \otimes  \bm{6} \otimes  \bm{6}
  =
  \bm{56}_\mathcal{S} \oplus  \bm{70}_\mathcal{M} \oplus  \bm{70}_\mathcal{M} \oplus  \bm{20}_\mathcal{A}$\,,
where the $SU(6)$ representations are decomposed as 
\begin{eqnarray}
  \bm{56} = {^2}\bm{8} \oplus {^4}\bm{10}\,, \qquad
  \bm{70} = {^2}\bm{8}  \oplus {^4}\bm{8} \oplus {^2}\bm{10} \oplus  {^2}\bm{1}\,, \qquad
  \bm{20} = {^2}\bm{8}  \oplus  {^4}\bm{1}\,.
\end{eqnarray}

Finally, the spin-flavor wave functions must be combined with spatial wave functions $|\psi_{\rm space}\rangle$ 
of the same symmetry to yield a totally symmetric spatial-spin-flavor state. 
This implies that the $\mathbf{56}$-plet must come with a symmetric orbital wave function ($\mS_o$), 
the $\mathbf{70}$-plet with a mixed-symmetric orbital part ($\mD_o$), and the $\mathbf{20}$-plet with an antisymmetric orbital part ($\mA_o$).
This is also shown in the left of Fig.~\ref{fig:flavor-wfs}:
e.g., the total wave function $\psi_1$ describes nucleons in the \textbf{56}-plet,
$\psi_2$ are $\Delta$'s in the \textbf{56}-plet, $\psi_3$ and $\psi_4$ are nucleons in the \textbf{70}-plet, %$\psi_5$ are $\Delta$'s in the 70-plet, 
and so on. %In Fig.~\ref{fig:flavor-wfs} the wave functions for $\mathbf{8}$ and $\mathbf{10}$ are written out in components. 
In particular, the wave function of the ground-state nucleon is $\psi_1$ and that of the ground-state $\Delta$ baryon is $\psi_2$.

It is  instructive to relate these wave functions with the underlying diquark structure. A diquark wave function satisfies similar constraints as 
the baryon wave function:
it must be antisymmetric under quark exchange due to the Pauli principle, and because the color part is antisymmetric, the remainder must be symmetric. Combining $u$ and $d$ quarks yields an antisymmetric isospin singlet $\frac{1}{\sqrt{2}}(ud-du)$ with $I=0$ and a symmetric triplet $\{ uu, \frac{1}{\sqrt{2}}(ud+du), dd \}$ with $I=1$. These must be combined with spin parts of the same symmetry. For ground states with $L=0$, one arrives at scalar diquarks ($J^P = 0^+$) with $S=0$, $I=0$ and axialvector diquarks ($J^P = 1^+$) with $S=1$, $I=1$.
These were nicknamed ``good'' and ``bad'' diquarks by Wilczek~\cite{Wilczek:2004im} and they play a significant role; e.g.,
in diquark models
axialvector diquarks are typically 200 \dots 300~MeV heavier than scalar diquarks~\cite{Anselmino:1992vg,Santopinto:2014opa,Barabanov:2020jvn}. Because of isospin, the ground-state nucleon ($I=0$) contains both scalar and axialvector diquarks whereas the $\Delta$ resonance ($I=1$) consists only of axialvector diquarks.
The diquark components enter in the spin and flavor wave functions of baryons (Fig.~\ref{fig:flavor-wfs}): the upper ($\mM_\mA$) components in the doublets are antisymmetric in the first two indices and can thus be related to the ``good'' diquarks, whereas the lower ($\mM_\mS$) components and the singlets $\mS$ are symmetric and associated with the ``bad'' diquarks. This will play a role in the quark models discussed below.

\subsubsection{ Spatial wave functions and the non-relativistic harmonic oscillator}   

To classify the baryon spectrum, we must also discuss the spatial or orbital wave functions $|\psi_{\rm space}\rangle$.  
The three-body problem is characterized by the time-dependent position of three particles. 
Instead of  the position vectors $\bm{x}_{i} \,
 (i=1,2,3)$ of the three particles, one usually defines the Jacobi coordinates 
\begin{eqnarray}\label{jacobi-coordinates}
  \bm\rho =  \frac{1}{\sqrt 2}\left(\bm{x}_1 - \bm{x}_2\right), \qquad
  \bm\lambda =   \frac{1}{\sqrt 6}\left(\bm{x}_1 + \bm{x}_2 - 2\bm{x}_3\right), \qquad
  \bm X =  \frac{1}{\sqrt 3}\left(\bm{x}_1 + \bm{x}_2 + \bm{x}_3\right)\,,
\end{eqnarray} 
which include the trivial center-of-mass (CM) motion and two oscillations represented by the variables $\bm\rho$ and $\bm\lambda$. The $\bm\rho$ oscillator describes the dynamics within a diquark and the $\bm\lambda$ oscillator the motion of the third quark relative to the diquark. Permutation symmetry ensures that all three quarks have identical functions.
Instead of Jacobian coordinates, 
some authors~\cite{Giannini:2001kb,Giannini:2002vp,Giannini:2003xx,DeSanctis:2005kt,Santopinto:2010zz,%
Santopinto:2012nq,Salehi:2013qga,Giannini:2015zia,Giannini:2016jta,Sattari:2019kkd,Tazimi:2021azh} 
use hypercentral coordinates given by the collective size variable $\sqrt{\rho^2+\lambda^2}$
and an angle $\arctan\frac\rho\lambda$.

In a series of papers, Isgur and Karl studied masses and mixing angles of low-mass baryons
in a model that contains a flavor-independent confinement potential and hyperfine interactions between quarks
due to one-gluon exchange~\cite{Isgur:1977ef,Isgur:1978xj,Isgur:1978xi,Isgur:1978wd,Isgur:1978xb,Isgur:1979be,Isgur:1979ee}. 
The confining potential is assumed to be harmonic, and the deviation due to the linearly rising confining potential
expected in QCD is treated perturbatively. In our following interpretation of the baryon resonance spectrum, we
will often use the symmetry properties of the nonrelativistic quark model with a harmonic oscillator as confining
potential since it predicts the number of states and, at least qualitatively, their masses. 
%This is why we will refer to this model in the following. 
The eigenfunctions of the Hamiltonian are then expanded
into eigenfunctions of the harmonic oscillator serving as a complete orthonormal set of basis functions.  

Employing a harmonic oscillator potential, the Hamiltonian turns into two uncoupled three-dimensional harmonic oscillators for $\bm{\rho}$ and $\bm{\lambda}$, 
\begin{eqnarray}\label{eq:harmosc}
H=\frac{\bm{p}_\rho^2}{2m}+\frac{\bm{p}_\lambda^2}{2m}+\frac{K}{2} (\bm{\rho}^2+\bm{\lambda}^2)~,
\end{eqnarray}
leading to the energy spectrum and eigenfunctions
\begin{eqnarray}\label{oscillator-wfs} %\label{eq:ho-levels}
\sqrt{\frac{K}{m}}\left(3+ l_\rho + 2 {n}_\rho+  l_\lambda +
2 {n}_\lambda\right)= \sqrt{\frac{K}{m}}\left(3+ {\rm N}\right) , \qquad 
\phi_{n_\rho l_\rho}(\bm{\rho}) \otimes \phi_{n_\lambda l_\lambda}(\bm{\lambda}) \equiv \phi_{n_\rho n_\lambda, l_\rho l_\lambda} \,.
\end{eqnarray}
With $n_\rho$, $n_\lambda$, $l_\rho$, $l_\lambda = 0, 1, 2, \dots$,
this allows for radial ($n_{\rho,\lambda} > 0$) and orbital ($l_{\rho,\lambda}> 0$) excitations.
Abbreviating $n = n_\rho + n_\lambda$ and $l = l_\rho + l_\lambda$, one finds the shell number $\text{N}=2n+l$. Thus, if the confining potential were harmonic,
and neglecting spin-spin and spin-orbit interactions and tensor forces, the energy levels would follow the sequence $M=M_0 + \text{N}\, \omega$. 
For baryons with one heavy quark $Q$, the $\lambda$ oscillator
has a reduced mass $\mu$
with $\mu^{-1}= (2 M^{-1}+ m^{-1})/3$ replacing $m$. Then the energy
levels are modified as
\begin{eqnarray}\label{eq:ho-levels1}
\sqrt{\frac{K}{m}}\left(\frac32 + l_\rho + 2 {n}_\rho\right)
+ \sqrt{\frac{K}{\mu}}\left(\frac32 + l_\lambda + 2 n_\lambda\right)\,.
\end{eqnarray}
Hence the $\lambda $ excitations are lower than their $\rho$ analogs
for single-flavor baryons.

\begin{figure}[t]
    \centering
    \includegraphics[width=1\textwidth]{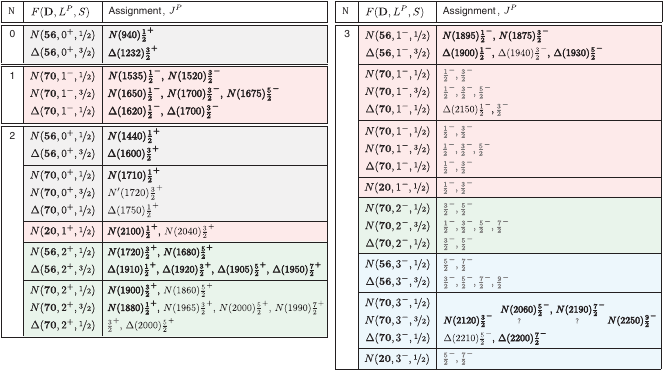}
    \caption{(color online) Light baryon spectrum in the harmonic oscillator model with band quantum number N,  spin $S$, orbital angular momentum $L$, total angular momentum $J$ and parity $P$.
               $F$ denotes the flavor multiplet ($N$ for octet, $\Delta$ for decuplet) and $\mathbf{D}$ the spin-flavor multiplet. 
               The assignment with experimental states is discussed in the text.
               Three- and four-star resonances are shown in bold font.
               the $J^P$ quantum numbers in gray do not have good experimental candidates.
               The background colors refer to the orbital angular momentum $L$.
               }
     \label{fig:missing-resonances}
\end{figure}

The wave functions within a given shell can be further rearranged to form combinations 
with definite S$_3$ symmetry, as shown in the right panel of Fig.~\ref{fig:flavor-wfs} %for states with $\text{N} \leq 2$ 
(see e.g. Ref.~\cite{Flamm:1982jv} for details). This yields the spatial wave functions $\mS_o$, $\mD_o$ and $\mA_o$ appearing
in the left panel  of Fig.~\ref{fig:flavor-wfs}.
The total orbital angular momentum $\bm{L}= \bm{l}_\rho +\bm{l}_\lambda$
can adopt values $L = |l_\rho - l_\lambda|, \dots , l_\rho+ l_\lambda$, and $\bm{L}$
combines with the quark spin $\bm{S}$ to the spin of the particle $\bm{J}=\bm{L}+\bm{S}$. 
The quantum numbers $J^P$ are then obtained from the parity $P=(-1)^l$.

%\subsubsection{ Total wave function}
Combining the spatial and spin-flavor wave functions, one can see from Fig.~\ref{fig:flavor-wfs} that there are six combinations leading to a totally symmetric wave function for octet and decuplet baryons, namely
$\psi_{1 \dots 5}$ and $\psi_7$.
Fig.~\ref{fig:missing-resonances} shows all resulting possible combinations  up to the third shell
($\text{N} \leq 3$).  
The $J^P$ quantum numbers then allow one to count the expected number of states.
The zeroth shell ($\text{N}=0$) contains the $N$ and $\Delta$ ground states.
The first shell ($\text{N}=1$) contains the orbital excitations with $l=1$ (that is, $l_\rho = 1$, $l_\lambda = 0$ or $l_\rho=0$, $l_\lambda=1$) and we expect five $N^*$'s and two $\Delta^*$'s. 
%For $\Delta^*$ resonances, states
%with $S=\nicefrac32$ are forbidden since spin and isospin would be symmetric but there is no symmetric orbital wave function.
In the second shell ($\text{N}=2$), we anticipate 13 $N^*$ resonances and eight $\Delta^*$ resonances.
Similarly, in the third shell ($\text{N}=3$) we would expect 30 $N^*$ resonances and 15 $\Delta^*$ resonances.

\begin{figure*}[!t]
%    \hspace{2mm}
    \includegraphics[width=0.48\textwidth]{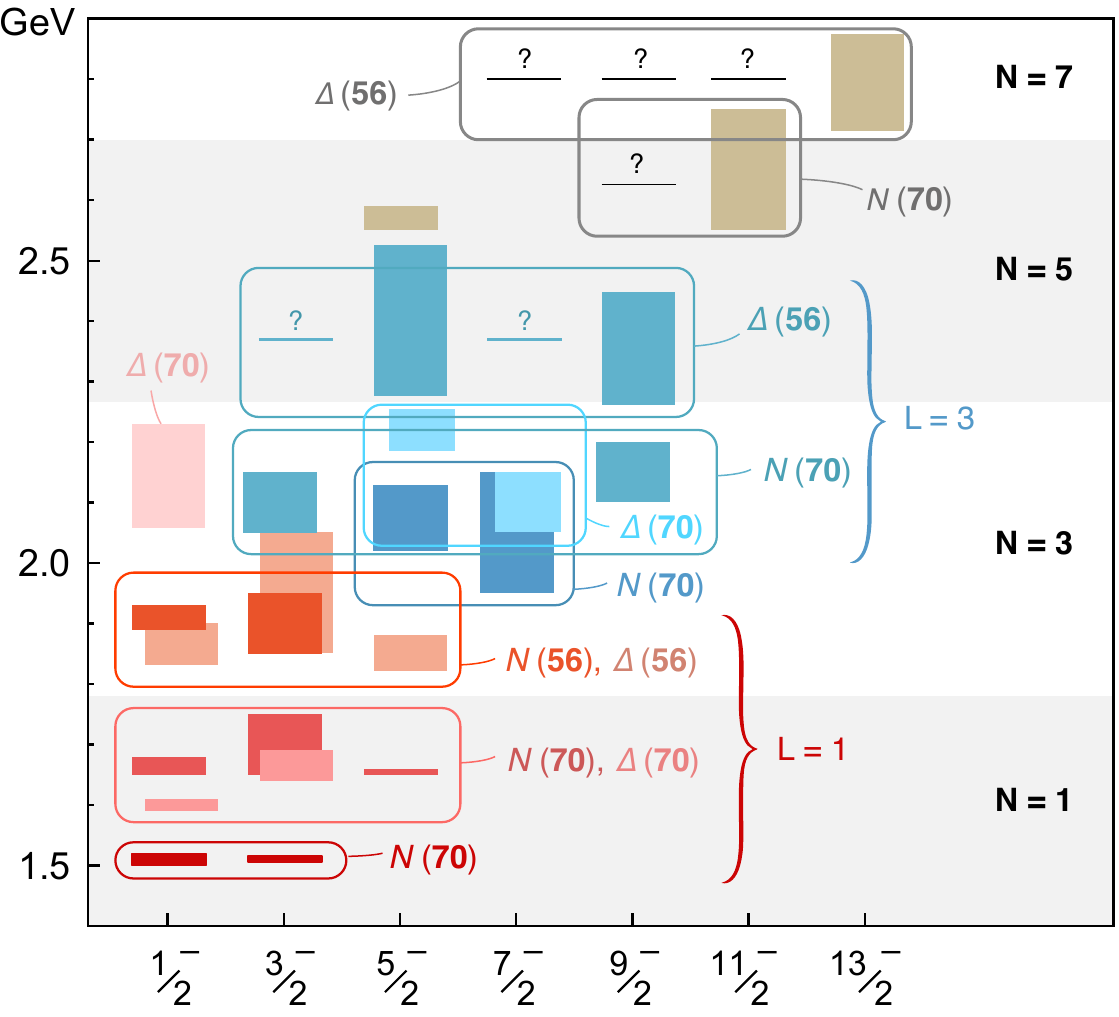}\hspace{3mm}
    \includegraphics[width=0.50\textwidth]{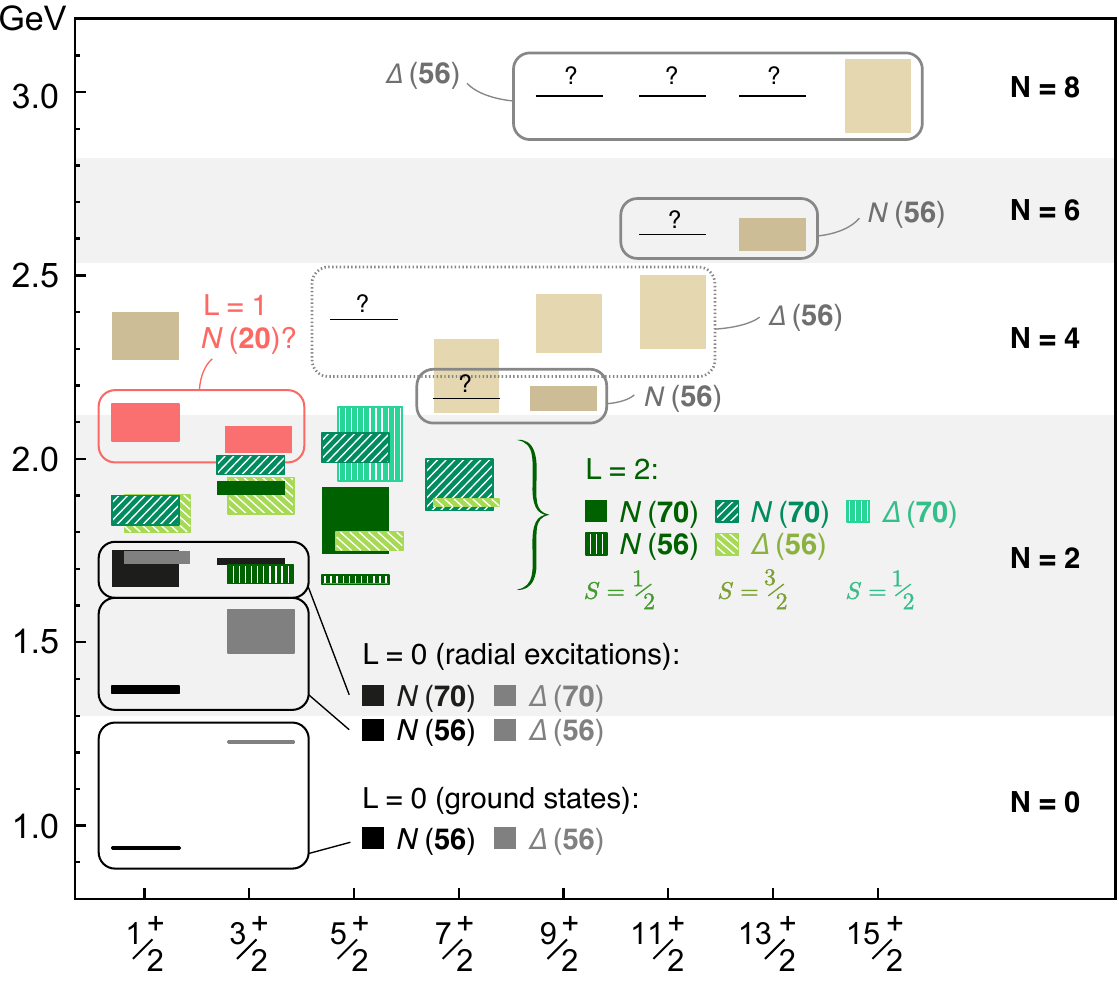}
    \vspace{-3mm}
     \caption{(color online) The experimental spectrum of negative-parity (left) and positive-parity (right) $N^*$ and $\Delta^*$ resonances
      as a function of the total angular momentum $J$. We also show the assignments to different $SU(6)$ multiplets according to Fig.~\ref{fig:missing-resonances}. 
     Thin lines with question marks are hypotheses.
     }
    \label{fig:neg-parity}
\end{figure*}

\subsection{Light baryon spectrum} \label{Global}

Based on the nonrelativistic quark model construction of the baryon wave functions  leading to Fig.~\ref{fig:missing-resonances}, 
let us now explore to what extent these features are visible in the observed baryon spectrum. 
Concrete quark models take a realistic confining potential and additional quark-quark interactions
into account; these will be discussed later in Sec.~\ref{pheno}.

\begin{figure}[b!]
 \centering
\includegraphics[width=0.45\linewidth]{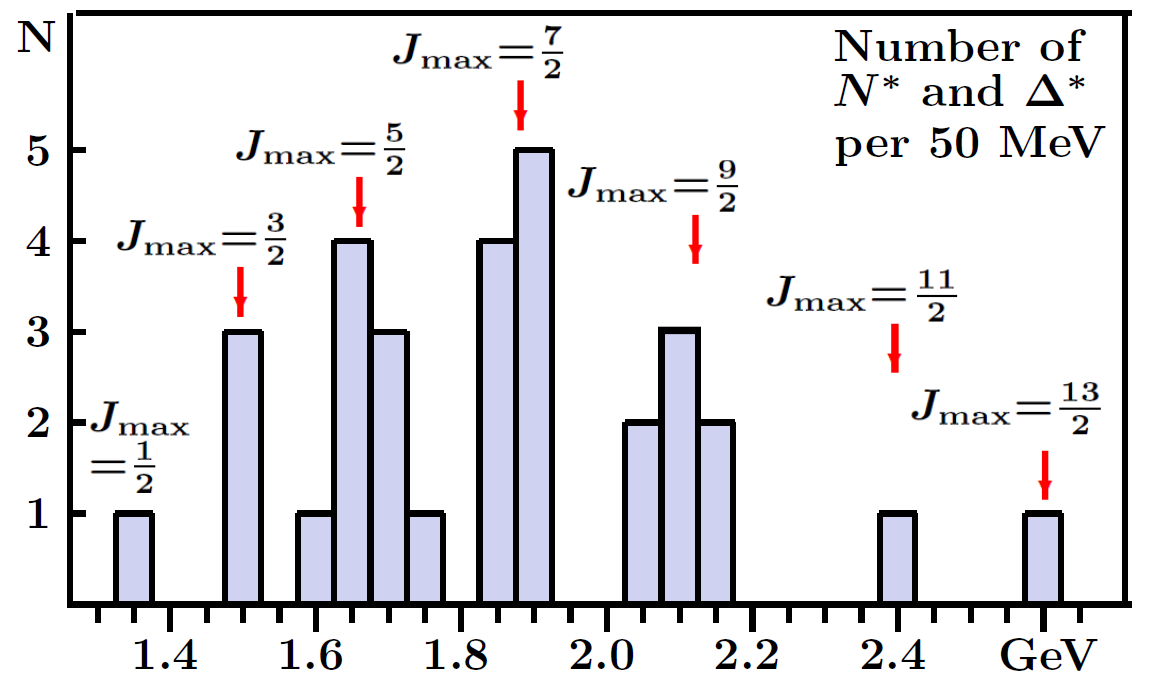} 
\caption{The number of 3* and 4* $N^*$ and $\Delta^*$ resonances in~\cite{ParticleDataGroup:2024cfk} above the ground states in 50~MeV bins.
The arrows mark the largest spin up to the given mass.}
\label{fig:shells}
\end{figure}

First of all, despite the fact that the confining potential is not harmonic and spin-spin and spin-orbit interactions cannot be neglected,
the observed spectrum still shows remnants of a shell structure.
The $N^*$ and $\Delta^*$ resonances span a broad mass range from 1 to 3~GeV, with
positive or negative parity and spins ranging from $J=\nicefrac12$ to $\nicefrac{15}{2}$. 
If the experimental spectrum is plotted over the total angular momentum $J$ 
as in Fig.~\ref{fig:neg-parity}, one can see that the baryon excitations roughly cluster in multiplets
whose entries have similar masses but differ in $J$, just like the 
quark model in  Fig.~\ref{fig:missing-resonances} predicts.
A spectrum that does not strongly depend  on $J$ also implies that the underlying spin-orbit forces are small. 
The shell structure can also be seen  in Fig.~\ref{fig:shells}, 
which shows the number of states observed with pole masses in 50~MeV bins. 
A notable feature is that the masses of the baryon excitations cluster at discrete values.
This suggests a shell structure resembling
similar patterns observed in atomic and nuclear physics, which point towards the compositeness of atoms and nuclei.

In each cluster there is a state with maximal spin.
These states are referred to as {\it stretched} states,  
where the total angular momentum is $J=L+S$.  
In the high-mass region, the stretched states dominate
the reaction, either on physical grounds or because the largest partial wave is the easiest one
to identify in angular distributions. 
The left panels in Fig.~\ref{fig:Regge-N} show the squared baryon mass with maximal spin $J$ in a given cluster, 
for positive-parity $N^*$ and $\Delta^*$ resonances, respectively, as a function of $J$.
One observes a Regge behavior, i.e.,  the squared masses 
of the stretched states depend linearly on $J$~\cite{Veneziano:1968yb}. 
Similar Regge trajectories can be drawn for
isoscalar mesons, including the
$\omega$, $f_2(1270)$, $\omega_3(1670)$, $f_4(1950)$, $\rho_5(2350)$, and $f_6(2510)$.
% The $\omega_5$ is not known and is replaced by its isospin partner.
Mesons and $\Delta^*$ baryons have very similar Regge slopes: 
writing $M^2 = M_0^2 + a J$, 
one finds
$a=1.14(2)$~GeV$^2$ for mesons and $a=0.98(2)$~GeV$^2$ for baryons.  
In mesons, the quark and antiquark are bound by the strong force between a colored quark and an antiquark carrying  
anti-color. 
%The centripetal forces in $\Delta^*$'s arise from the interaction between a colored quark and an antiquark.  % GE: ?
Two colored quarks in a baryon can form a diquark carrying the anti-color of the third quark. 
The similarity of the meson and baryon trajectories has led to the hypothesis that 
baryon resonances (more precisely, the {\it stretched} states) can be understood  
as quark-diquark oscillations.

In the following we  assign the $N^*$ and $\Delta^*$ resonances to $SU(6)$ multiplets with defined shell number
\text{N}, intrinsic orbital angular momentum $L$, total quark spin $S$ and parity $P$. 
To this end, we combine the masses of baryons in the same multiplet (which  differ only by $J$) 
to the center-of-gravity (c.o.g.) mass\footnote{The c.o.g. mass can be used for heavy quarkonia to
estimate the mass of the singlet meson from the three triplet states, provided spin-spin and tensor interactions
are small. It has been applied to light-quark mesons \cite{Klempt:2021nuf} even though
this is not really justified. The idea is to find a quantity that does not dependent on spin-orbit
interactions.} defined by 
\begin{eqnarray}
M_{\rm c.o.g.}=\frac{\sum_J J \cdot M_J}{\sum_J J},
\label{cog}
\end{eqnarray}
where we use the pole masses $M_J$ from Tables~\ref{highmass-piN} and~\ref{tab:statusN} and add the errors in quadrature.
The uncertainties are calculated by replacing $M_J$ with $(\delta M_J)^2$. 
The resulting assignment, to be discussed in detail below, is shown in Fig.~\ref{fig:missing-resonances}
and the c.o.g. masses are collected in the center of Fig.~\ref{fig:Regge-N}.

Before we proceed, we should issue a few words of caution. 
On the one hand, baryons with identical flavor and spin-parity can mix, 
but the mixing between these states is usually small. 
On the other hand, the three-quark spin $S$ and orbital angular momentum $L$ are frame-dependent and, in contrast to $J$, cannot be used to label the states. Relativistically, a baryon's (Bethe-Salpeter) wave function is a superposition of  tensors that transform under the Lorentz group and correspond to different $S$ and $L$,  e.g., in the baryon rest frame, where also new components arise that are nonrelativistically forbidden. 
However, as  will be discussed in Sec.~\ref{sec:fm}, the nonrelativistic wave functions in Fig.~\ref{fig:flavor-wfs} 
still appear as parts of the full tensor bases, 
and in the low-lying baryon spectrum they are typically the leading or at least important components.
Therefore, the assignment to definite N, $S$ and $L$ is still a useful bookkeeping device, and
in the remainder of this review we will sometimes refer to  specific baryon excitations using their $SU(6)$ harmonic-oscillator assignment
in Fig.~\ref{fig:missing-resonances}.

\subsubsection{ First excitation shell}

We start with the first excitation shell $\text{N}=1$, which appears in the negative-parity ($P=-1$) sector shown in the left panel in Fig.~\ref{fig:neg-parity}.
There is a low-mass cluster of two nearly mass degenerate $N^*$'s around 1.5~GeV,
the $N(1535)\nicefrac12^-$ and $N(1520)\nicefrac32^-$, where the former is the nucleon's parity partner.
In the $SU(6)$ assignment these correspond to
$(\mathbf{D}, L^P, S) = (\mathbf{70}, 1^-, \nicefrac{1}{2})$, i.e., 
 their wave functions have $SU(6)$ dimensionality $\mathbf{70}$ and thus  mixed symmetry, and
they carry total quark spin $S=\nicefrac{1}{2}$ and orbital angular momentum $L=1$ which are coupled to $J=\nicefrac{1}{2}$ and $\nicefrac{3}{2}$.
The next cluster around 1.65~GeV contains a $N^*$ triplet, the $N(1650)\nicefrac12^-$, $N(1700)\nicefrac32^-$, and $N(1675)\nicefrac52^-$
forming the $N(\mathbf{70}, 1^-, \nicefrac{3}{2})$, and a 
$\Delta^*$ pair, the $\Delta(1620)\nicefrac12^-$ and $\Delta(1700)\nicefrac32^-$ forming the $\Delta(\mathbf{70}, 1^-, \nicefrac{1}{2})$.
Note  that the first shell does not permit a $\Delta$ triplet.

\begin{figure}
\begin{overpic}[width=\textwidth]{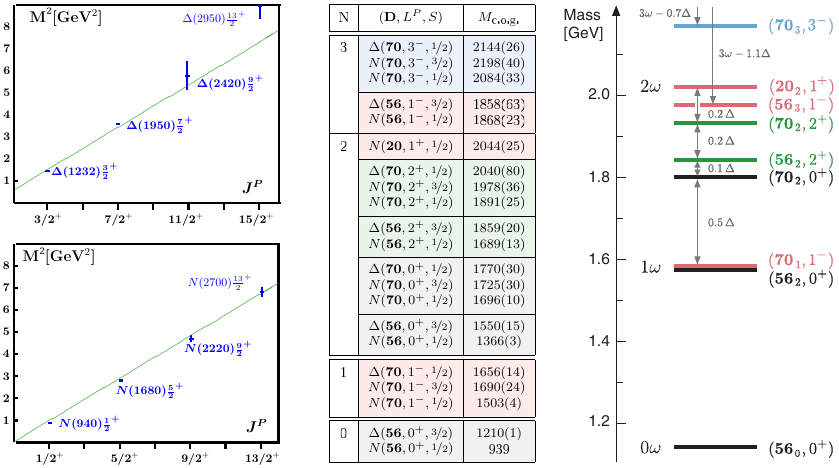}
\put(160,530){\bf a}%
\put(160,240){\bf b}%
\put(370,530){\bf c}%
\put(750,510){\bf d}%
\end{overpic}
\caption{Squared masses  of {\it stretched} $N^*$'s ({\bf a}) and $\Delta^*$'s ({\bf b}) as a function
of their spin $J$. {\bf c}: C.o.g. masses (in MeV) for the 
baryon excitations in Fig.~\ref{fig:missing-resonances} 
up to the second excitation shell.
{\bf d}: Mass splittings in the second and third excitation shell expected in the quark model,
see Refs.~\cite{Isgur:1977ef,Bowler:1980pfa,Bowler:1981xh}.}
\label{fig:Regge-N}
\end{figure}
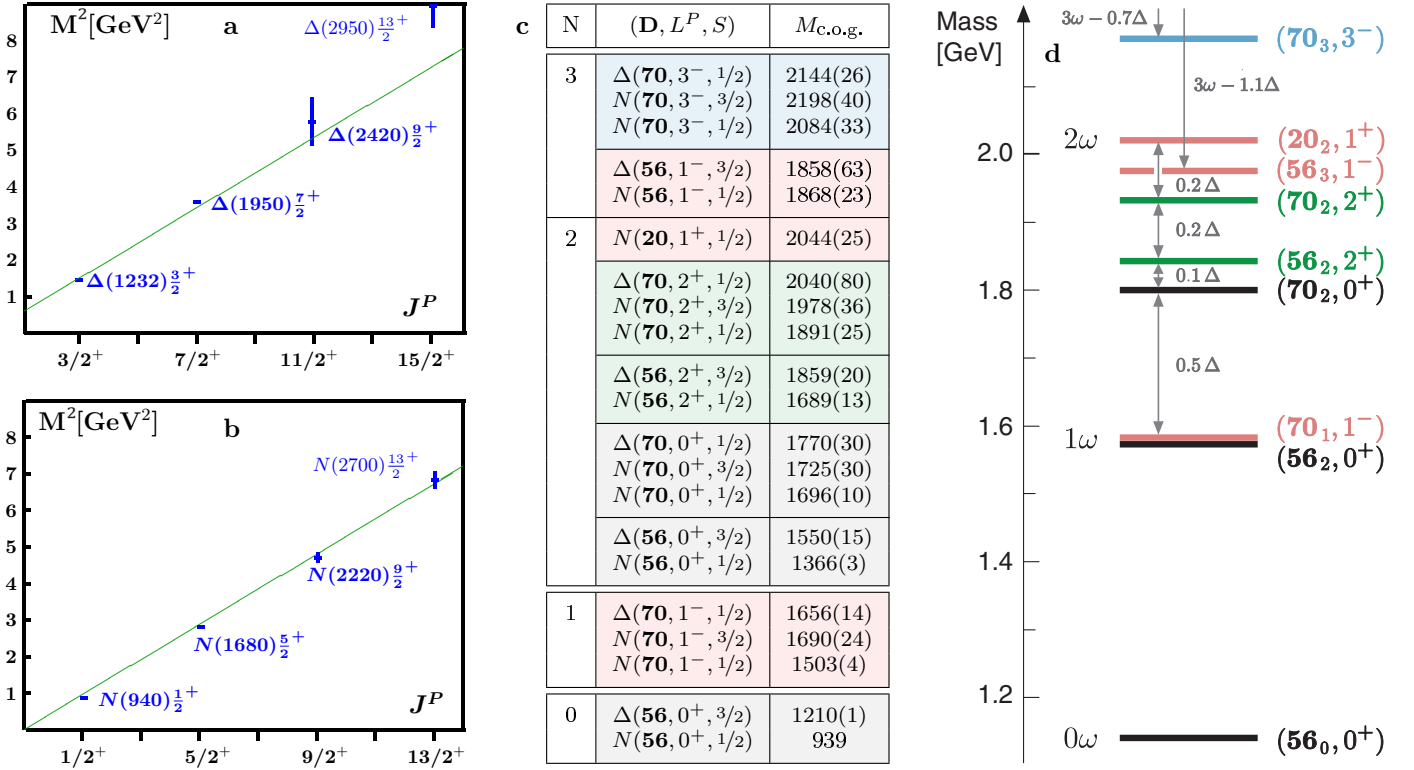

The c.o.g. masses of these states are collected in Fig.~\ref{fig:Regge-N}.
The small mass splittings within the two spin doublets and the spin triplet imply that  the $\bm{L}\cdot\bm{S}$ coupling is small.
The mass-square difference between the $N(\mathbf{70}, 1^-, \nicefrac{1}{2})$   
and the mean value of the five remaining states is 0.56(7)~GeV$^2$.
In principle, the states with identical quantum numbers can mix, but the mixing angle is small: Experimentally and in calculations, the mixing angle is determined
to be $-31,7^\circ$ for 
the two $\nicefrac12^-$ and +10$^\circ$ for the $\nicefrac32^-$ 
states~\cite{Isgur:1977ef}. Here, we caution the reader: the mixing angle
was derived assuming a small branching ratios of about 1\% for $N(1650)\nicefrac12^-\to N\eta$ from
unreliable data on $\pi^- p\to \eta p$; the later determination was reported in Ref.~\cite{Hunt:2018wqz}. 
Recent determinations using precise data on $\eta$ photoproduction find, however, 
a branching ratio of 28(11)\%~\cite{A2:2017gwp} and 33(4)\%~\cite{CBELSATAPS:2019ylw}.

\subsubsection{ Second excitation shell}

The states in the second shell $\text{N}=2$ have positive parity and are shown in the right panel of Fig.~\ref{fig:neg-parity}.
While for a pure harmonic-oscillator potential all states in 
a given shell would be mass-degenerate,
the binding forces do not grow
linearly with distance but can rather be assumed to be constant as a function of the
interquark separation.
This leads to a splitting of the mass eigenvalues sketched
in the right of Fig.~\ref{fig:Regge-N}, where the spin-spin, spin-orbit, and tensor forces would change the masses further. 

\smallskip
$(\mathbf{56},0^+)$:
The so-called radial excitations of the nucleon and  $\Delta(1232)$ belong to the \textbf{56}-plet,
so they have symmetric orbital and spin-flavor wave functions and carry orbital angular momentum $L=0$.
The low mass of  1366(3)~MeV for the Roper resonance $N(1440)\nicefrac{1}{2}^+$  
compared to the $N(1535)\nicefrac12^-$  has led to speculations concerning its nature~\cite{Burkert:2017djo}.
Its squared mass is 0.98(1)~GeV$^2$ above $M^2_N$.
The $\Delta(1600)\nicefrac{3}{2}^+$ with a pole mass of 1550(15)~MeV
is the analogue of the Roper and was nicknamed `Doper' by Nefkens 
(who also introduced the names `Loper' for $\Lambda(1600)\nicefrac12^+$, 
`Soper' for $\Sigma(1660)\nicefrac12^+$ and `Xoper' for the not yet identified spin-$\nicefrac12^+$ $\Xi$ resonance
in his talk at MENU2004~\cite{Nefkens:2005dh}). 
The $\Delta(1600)\nicefrac{3}{2}^+ - N(1440)\nicefrac{1}{2}^+$ mass-square difference of 0.54(4)~GeV$^2$ is compatible with the 
$\Delta(1232)-N(940)$ mass-square difference of  0.58~GeV$^2$.

\smallskip
$(\mathbf{70},0^+)$:
The remaining $\text{N} = 2$ mass region
is rather densely packed with baryon excitations.
%In the $(\mathbf{70},0^+)$ multiplets, t
The $N(1710)\nicefrac{1}{2}^+$ has a pole mass of 1696(10) MeV.
Given that its squared mass  splitting with the Roper resonance
of 1.01(4)~GeV$^2$ is of comparable size as  that of the Roper versus the nucleon, it could  be interpreted as
the second radial excitation of the nucleon in a \textbf{56}-plet. 
In quark models, however, the $N(1710)\nicefrac12^+$ is found 
%seen as a partner of the Roper with quark spin $S=\nicefrac12$  
in a \textbf{70}-plet, which  
has a spatial wave functions with mixed symmetry. % ($\mD_o$ in Fig.~\ref{fig:flavor-wfs}).
%In $M_{\mathcal A}$, 
In that case, both  $\rho$ and $\lambda$ oscillators carry one angular momentum excitation and %. In this case,
we expect  a sizable fraction of all decays to proceed via an intermediate excitation, i.e., 
via $N(1440)\nicefrac{1}{2}^+$ or $N\sigma$. In the RPP these two decay modes are not listed as {\it seen}, 
but in a recent analysis  the former branching ratio is reported with a 
branching ratio 22(6)\% and the latter  with 11(4)\%~\cite{CLAS:2024iir,Sarantsev:2025lik}.
Thus, in agreement with the quark model interpretation we consider the $N(1710)\nicefrac{1}{2}^+$ 
as the $N(\mathbf{70},0^+,\nicefrac{1}{2})$ state.
We then further expect a nucleon spin partner $N(\mathbf{70},0^+,\nicefrac{3}{2})$ with
$J=\nicefrac{3}{2}^+$ in this multiplet. This could be the $N'(1720)\nicefrac{3}{2}^+$  
suggested in a common photo- and electroproduction analysis of $p\pi^+\pi^-$ \cite{Mokeev:2015lda}, 
which is not included in the RPP 2024. 
The remaining decuplet state $\Delta(\mathbf{70},0^+,\nicefrac{1}{2})$ could  be the %can be identified with the 
1* resonance $\Delta(1750)\nicefrac{1}{2}^+$ with a mass of 1770(30) MeV.

\smallskip
$(\mathbf{56},2^+)$:
In the same mass region there are several further states:
a spin doublet $N(\mathbf{56},2^+,\nicefrac{1}{2})$ containing the $N(1720)\nicefrac{3}{2}^+$ and $N(1680)\nicefrac{5}{2}^+$  
with  c.o.g. mass 1689(13) MeV, 
and a spin quartet $\Delta(\mathbf{56},2^+,\nicefrac{3}{2})$ with the $\Delta(1910)\nicefrac12^+$, $\Delta(1920)\nicefrac32^+$, $\Delta(1905)\nicefrac52^+$ and $\Delta(1950)\nicefrac72^+$   
and c.o.g. mass 1859(20) MeV. We note that the $\Delta(1905)\nicefrac52^+$ has   
an unexpectedly low pole mass of 1795(25)~MeV.
The mass square from $\Delta(\mathbf{56},2^+,\nicefrac{3}{2})$ to $N(\mathbf{56},2^+,\nicefrac{1}{2})$ is 0.60(8)~GeV$^2$
and $2\times$ 1.00(8)~GeV$^2$ to the $\Delta(1232)\nicefrac32^+$.

\smallskip
$(\mathbf{70},2^+)$:
As can be seen in Fig.~\ref{fig:missing-resonances},
one expects eight states with masses around 1900~MeV in this multiplet: a spin doublet and a spin quartet 
of $N^*$'s, and a spin doublet of $\Delta^*$'s.
We organize the six observed $N^*$'s 
into the spin-doublet $N(\mathbf{70},2^+,\nicefrac{1}{2})$ consisting of the resonances $N(1900)\nicefrac32^+$ and $N(1860)\nicefrac52^+$ with 
c.o.g. mass 1891(25) MeV,
and a spin-quartet $N(\mathbf{70},2^+,\nicefrac{3}{2})$ formed by $N(1880)\nicefrac12^+$, $N(1965)\nicefrac32^+$, $N(2000)\nicefrac52^+$,  
$N(1990)\nicefrac72^+$ with c.o.g. mass 1978(36) MeV.
In the RPP 2024, only the $N(1880)\nicefrac12^+$ and $N(1900)\nicefrac32^+$ are established with a 3* and 4* rating.
The assignment of the $J^P=\nicefrac32^+$ and $\nicefrac52^+$ states to the doublet and quartet
is driven by the hypothesis that the spin-doublet has a lower mass than the quartet.
Only one positive-parity $\Delta^*$ has been identified at about this mass, the $\Delta(2000)\nicefrac{5}{2}^+$ with mass 2040(80) MeV.
 
 \smallskip
 $(\mathbf{20},1^+)$:
 This is the only multiplet in the second shell with $L=1$  despite having positive parity.
 There is a spin-doublet consisting of the $N(2100)\nicefrac12^+$ and $N(2040)\nicefrac32^+$, the latter state
being observed only in $J/\psi \to N\bar N\pi$. These two states might belong to the
second excitation shell and could be interpreted as members of the $N(\mathbf{20},1^+,\nicefrac{1}{2})$. 
However, in Sec.~\ref{hybrid} we will argue that the $N(2100)\nicefrac12^+$ fits better
in the $N(\mathbf{56},4^+,\nicefrac12)$ multiplet in the fourth shell. %$^2N(56,4^+)_4$ multiplet.

\subsubsection{ The third shell}

Concerning the negative-parity states in the third shell $\text{N}=3$, one can see in
Fig.~\ref{fig:missing-resonances} that the number of  states expected in the harmonic-oscillator 
quark model is very large. Only a handful of states are experimentally confirmed, which makes 
the assignments difficult (and questionable). In the non-relativistic quark model, the 
expected pattern for any potential is shown in the right of Fig.~\ref{fig:Regge-N}.

\smallskip
$(\mathbf{56},1^-)$:
There are five negative-parity states with masses that are difficult to reconcile in quark models. 
In~\cite{Capstick:1986ter,Loring:2001kx}, the lowest-mass 
$N^*$ doublet in the third shell is expected at about 1950~MeV and a $\Delta^*$ triplet at 
around 2100~MeV. Experimentally, we find a cluster of $N^*$'s forming a spin-doublet,
the $N(1895)\nicefrac12^-$ and $N(1875)\nicefrac32^-$ with c.o.g. mass 1868(23) MeV, and the spin triplet 
$\Delta(1900)\nicefrac12^-$, $\Delta(1940)\nicefrac32^-$, and
$\Delta(1930)\nicefrac52^-$ with c.o.g. mass 1858(63) MeV.
This would fit with the $N(\mathbf{56},1^-,\nicefrac{1}{2})$ and $\Delta(\mathbf{56},1^-,\nicefrac{3}{2})$ assignment.
To explain its low mass, one of these states, the $N(1895)\nicefrac12^-$, has been interpreted as a dynamically generated state
with a two-pole structure~\cite{Khemchandani:2013nma,Khemchandani:2020exc}.

\smallskip
$(\mathbf{70},3^-)$:
In addition to  these low-mass resonances, where a unique multiplet assignment seems possible, 
Table~\ref{tab:statusN} lists the four $N^*$ states  $N(2120)\nicefrac32^-, N(2060)\nicefrac52^-, N(2190)\nicefrac72^-, N(2250)\nicefrac92^-$, 
an isolated $\Delta(2150)\nicefrac12^-$, a spin doublet $\Delta(2210)\nicefrac52^-$ and $\Delta(2200)\nicefrac72^-$,
and the $\Delta(2400)\nicefrac92^-$. 
If the four $N^*$ states  form the quartet $N(\mathbf{70},3^-,\nicefrac{3}{2})$, then we should expect a spin doublet $N(\mathbf{70},3^-,\nicefrac{1}{2})$
in addition, like for the first shell where a spin triplet (with $S=\nicefrac32^-$) and a spin doublet is observed.
Therefore, two states with $J^P=\nicefrac52^-$ and $\nicefrac72^-$ seem to be missing as indicated in Fig.~\ref{fig:missing-resonances}.
On the other hand, the spin doublet of $\Delta(2210)\nicefrac52^-$ and $\Delta(2200)\nicefrac72^-$ 
can be assigned to $\Delta(\mathbf{70},3^-,\nicefrac{1}{2})$, which has a c.o.g. mass 2144(26) MeV.
This is very similar to the mean pole mass of the two states $N(2120)\nicefrac32^-$ and $N(2250)\nicefrac92^-$, 2198(40)~MeV,
while the two $N^*$' with $J^P=\nicefrac{5}{2}^-$ and $\nicefrac{7}{2}^-$ have a c.o.g. mass of 2084(33)~MeV.
Note that in the first excitation shell the masses of the $N^*$ triplet and the $\Delta^*$ doublet are also about mass degenerate.
%The two $N^*$' with $J^P=\nicefrac{5}{2}^-$ and $\nicefrac{7}{2}^-$ have a c.o.g. mass of 2084(33)~MeV; they likely contain 
%contributions from the spin doublet $^2N(70,3^-)_3$ and the spin quartet $^4N(70,3^-)_3$.
Finally, the 1* $\Delta(2150)\nicefrac12^-$ stands isolated and could be a member of the $\Delta(\mathbf{70},1^-,\nicefrac{1}{2})$.

\subsubsection{ Beyond the third shell}  \label{sec-beyond-3rd-shell}

\begin{figure*}[t]
\centering
\includegraphics[width=0.9\textwidth]{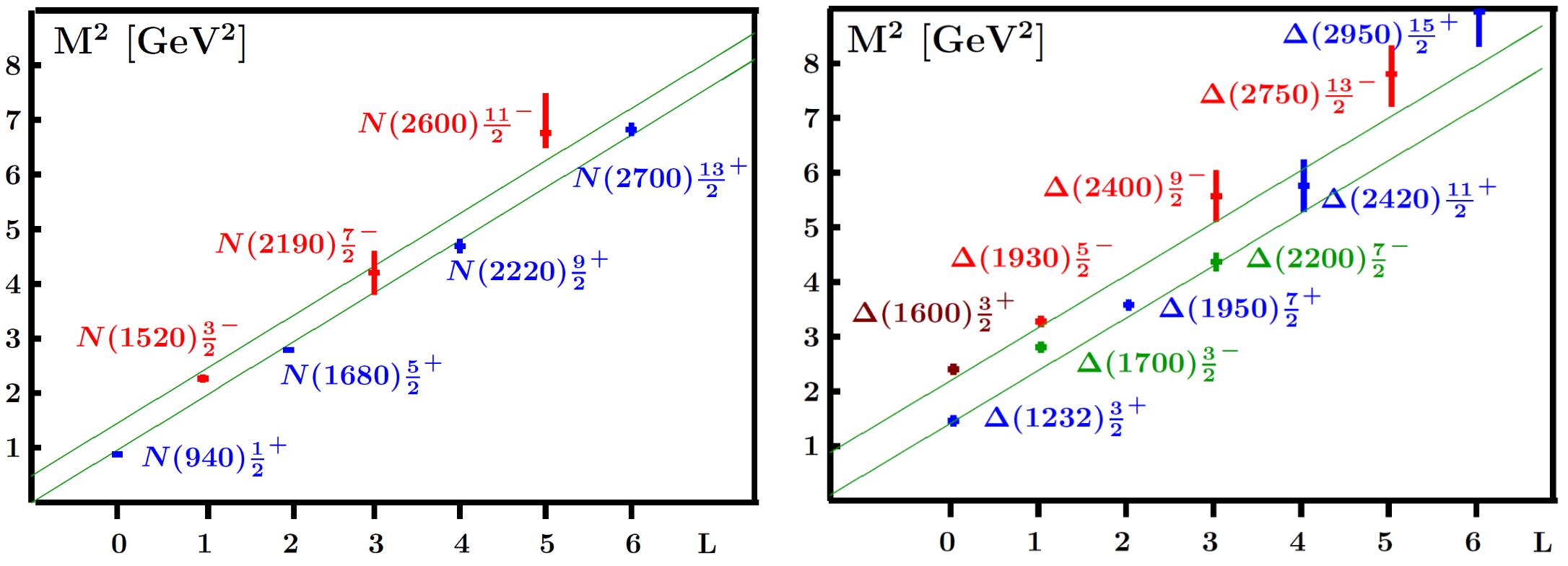}
\caption{\label{fig:Regge-two} 
Left: The squared mass of {\it stretched} $N^*$'s with total quark spin $S=\nicefrac12$ as a function
of the intrinsic orbital angular momentum $L$. The selected positive-parity $N^*$'s are in 56-plets,
negative-parity $N^*$'s in 70-plets. 
Right: The squared mass of {\it stretched} $\Delta^*$'s as a function
of $L$. The positive-parity $\Delta^*$'s (in blue) and the negative-parity $\Delta^*$'s 
(in red) have $S=\nicefrac32$ and belong to a 56-plet. $\Delta(1600)\nicefrac32^+$  and the negative-parity 
$\Delta^*$'s on the lower line (in green) belong to a 70-plet and have $S=1/2.$ 
}
\end{figure*}

Tables~\ref{highmass-piN} and ~\ref{tab:statusN} list five more $N^\ast$ and seven more $\Delta^\ast$ states 
in the high-mass region that we did not yet discuss. Some of these are well-established, with a three- or four-star rating, 
and carry large values of $J$. For their interpretation it is instructive to consider the Regge trajectories as 
functions of $L$ instead of $J$ as shown in Fig.~\ref{fig:Regge-two}. 

In the negative-parity sector, 
the $\Delta(2350)\nicefrac52^-$ and $\Delta(2400)\nicefrac92^-$  could be naturally
assigned to the $\Delta(\mathbf{56},3^-,\nicefrac32)$ multiplet in the third excitation shell (see Fig.~\ref{fig:missing-resonances}). 
However, the former 1*  resonance is observed at 2400(125)~MeV \cite{Cutkosky:1980rh}, while for the latter 2* resonance %$\Delta(2400)\nicefrac92^-$
two measurements have been reported,  2260(60)~MeV~\cite{Cutkosky:1980rh} and 2458(2)~MeV~\cite{Ronchen:2022hqk}.
The unweighted mean of these three values is 2375(100) MeV, which is considerably higher than the masses of the other states in the third shell.
Moreover, the masses of these two resonances fall onto a Regge trajectory with an offset compared to  main trajectory
given by the $\Delta(1600)\nicefrac32^+-\Delta(1232)\nicefrac32^+$ squared mass separation,
as shown in the right panel of Fig.~\ref{fig:Regge-two}. 
We thus prefer an interpretation of these states as members of % $\Delta(2350)\nicefrac52^-$ and $\Delta(2400)\nicefrac92^-$ as members of 
$\Delta(\mathbf{56}, 3^-, \nicefrac32)$ in the fifth excitation shell ($\text{N}=5$). 
For the same reasons one can assign the 3* resonance $N(2600)\nicefrac{11}{2}^-$ to the multiplet $N(\mathbf{70}, 5^-, \nicefrac12)$ in the fifth shell
and the $\Delta(2750)\nicefrac{13}{2}^-$ with mass 2794(80) MeV 
to $\Delta(\mathbf{56}, 5^-, \nicefrac32)$ in the seventh shell.

Concerning the positive-parity resonances, the $\Delta(2420)\nicefrac{11}{2}^+$, 
$\Delta(2300)\nicefrac{9}{2}^+$ and $\Delta(2390)\nicefrac{7}{2}^+$  have a c.o.g. mass of 2360(100) MeV and can be combined 
into a multiplet $\Delta(\mathbf{56}, 4^+, \nicefrac32)$ in the fourth shell, 
where the $J=\nicefrac52^+$ state is missing.
In the nucleon sector, the RPP lists the 4* resonance $N(2220)\nicefrac92^+$;
the $N(2200)\nicefrac72^+$  suggested by
Hunt and Manley~\cite{Hunt:2018wqz}  could serve as its spin partner to form a $N(\mathbf{56}, 4^+, \nicefrac12)$ in the fourth shell.
Finally, the $N(2700)\nicefrac{13}{2}^+$ and $\Delta(2950)\nicefrac{15}{2}^+$ are
the highest-mass states included in the RPP;
the Regge trajectories in Fig.~\ref{fig:Regge-two} suggest to place the former
in the sixth shell with $N(\mathbf{56},6^+,\nicefrac12)$ and the latter in the eight shell with $\Delta(\mathbf{56},6^+,\nicefrac32)$.

\begin{figure}
\begin{minipage}{.42\linewidth}
    \captionof{table}{Number of expected and (in parentheses) reported $N^*$ and $\Delta^*$ resonances with $J^P=\nicefrac12^\pm, \cdots , \nicefrac92^\pm$ in the first three excitation shells. The $N'(1720)\nicefrac32^+$ suggested in electroproduction~ \cite{Mokeev:2015lda} and $N(1965)\nicefrac32^+$ suggested in \cite{CLAS:2024iir,Sarantsev:2025lik} are not included in the counting.}
    \label{tab:nexp}
    \centering
    \footnotesize
    \renewcommand{\arraystretch}{1.4}
    \setlength\tabcolsep{1.5mm}
\begin{tabular}{cccccccc}
\hline\hline
$N$&Shell &$P$ & $J=\frac12$ & $J=\frac32$ & $J=\frac52$ & $J=\frac72$ & $J=\frac92$ 
\\\hline
&1$^{\rm st}$ &$-$&  2 (2) &  2 (2)  &  1 (1)  &       &\\
&2$^{\rm nd}$ &+&  4 (4) &  5 (3) &  3 (3) &  1 (1) & \\
&3$^{\rm rd}$ &$-$&  7 (1) &  9 (2) &  8 (1) &  5 (1)  &  1 (1) \\\hline
$\Delta$ & &&  $J=\frac12$ &  $J=\frac32$  &  $J=\frac52$ &  $J=\frac72$ &  $J=\frac92$ \\\hline
&1$^{\rm st}$ &$-$&  1 (1) &  1(1)  & & &\\
&2$^{\rm nd}$ &+& 2 (2) & 3 (2) & 2 (2)  & 1 (1)\\
&3$^{\rm rd}$ &$-$& 3 (2) & 5 (1) & 4 (2) & 2 (1) & 1 (0)\\
\hline\hline
\end{tabular}
\end{minipage} 
\hspace{2mm}\begin{minipage}{.30\linewidth}\vspace{3mm}
    \captionof{table}{Difference in squared masses of radial excitations. In some cases the S$U(6)$ multiplets
    are not identical on the left- and right-hand side; e.g.,
the $N(1440)\nicefrac12^+$ is in a 56-plet but the $N(1710)\nicefrac12^+$ in a 70-plet.}
    \centering
    \footnotesize
    \renewcommand{\arraystretch}{1.4}
    \setlength\tabcolsep{1.5mm}
\begin{tabular}{ccc}
\hline\hline
$X$ & $Y$ & $\delta M^2$ [GeV$^2$] \\ \hline\hline
$N$  & $N(1440)\frac12^+$ & $0.98 \pm 0.01$\\
$N(1440)\frac12^+$     & $N(1710)\frac12^+$    & 1.01\er 0.04\\
$\Delta(1232)\frac32^+$&$\Delta(1600)\frac32^+$& 0.94\er 0.05\\
$\Lambda$             &$\Lambda(1600)\frac12^+$& 1.16\er 0.03\\
$\Sigma$             &$\Sigma(1660)\frac12^+$& 1.26\er 0.06\\
$\Delta(1620)\frac12^-$&$\Delta(1900)\frac12^-$& 0.74\er 0.08\\
$\Delta(1700)\frac32^-$&$\Delta(1940)\frac32^-$& 1.36\er 0.17\\
\hline\hline\\[-2ex]
\end{tabular}
    \label{tab:radials}
\end{minipage} 
\hspace{2mm}\begin{minipage}{.24\linewidth}
    \captionof{table}{Difference in squared masses due to hyperfine interactions.}
    \centering
    \footnotesize
    \renewcommand{\arraystretch}{1.4}
    \setlength\tabcolsep{1.5mm}
\begin{tabular}{ccc}
\hline\hline
$X$ & $Y$ & $\delta M^2$ [GeV$^2$] \\ \hline\hline
$\Delta(1232)\frac32^+$  & $N$ & 0.58\\
$\Sigma(1385)\frac32^+$     & $\Sigma$    & 0.48\\
$\Xi(1530)\frac32^+$&$\Xi$& 0.61\\
$\rho(770)$             &$\pi$& 0.58\\
\hline\hline\\
   \end{tabular}
    \label{tab:hfs}
%     Also:\\[1ex]
% $M^2_{\Delta(\mathbf{56},2^+,\nicefrac32)} - M^2_{N(\mathbf{56},2^+,\nicefrac12)}$ = (0.60\er0.08)\,{\rm GeV}$^2$.\\
 \end{minipage} 
\end{figure}

\subsubsection{Hyperons}

The $\Lambda$, $\Sigma$, $\Xi$, and $\Omega$ resonances are related to the spectrum of nucleon and
$\Delta$ resonances by $SU(3)_f$, see Sec.~\ref{sec:su(6)}. 
A comparison of the nucleon and hyperon excitation spectra can thus offer valuable insight into the validity of $SU(3)_f$ symmetry and may help us in the spectroscopic interpretation
of resonances.

The simplest spectrum is that of the
$\Omega$ family. All $\Omega^*$'s belong to the decuplet, which has fewer resonances than octet baryons; 
they all have charge $-1$, and they are narrower:
the $\Omega(2012)$ with unknown $J^P$ (but most likely $\nicefrac12^-$)  has a width of less than 10~MeV, 
compared to 110~MeV for the $\Delta(1620)\nicefrac12^-$. However, experimentally very little is known: 
Only four excited states are known, and
two of them likely belong to higher bands. 

The $\Sigma^*$ and $\Xi^*$ spectra are the most complex ones, since they house both octet and decuplet states and thus the expected spectrum is very rich.
Experimentally, our knowledge is poor and often dates back several decades. Above the ground state,
ten $\Xi$ resonances were reported; for only two of them the spin and parity are known. 

\begin{figure}[ht]
    \centering
    \includegraphics[width=0.65\textwidth]{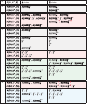}
    \caption{(color online) Hyperon spectrum in the $SU(6)$ multiplet arrangement analogous to Fig.~\ref{fig:missing-resonances}.
    In the first and second excitation band, all states expected from the quark model are shown, 
    while the third band lists only multiplets for which at least one $\Lambda$ or $\Sigma$ candidate exists. 
    The states with a dagger $\dagger$ are special: In the $\Lambda$ sector at about 2100~MeV, two states with $J^P=\nicefrac52^-$ and $\nicefrac72^-$
are expected with a mass separation of about 100~MeV but only one pair is known. In the $\Sigma$ sector,
one pair is expected at about 1750 to 1800~MeV and two pairs at about 2000 to 2050~MeV. However, only two pairs are 
found and the third pair is missing. The observed pairs of states are possibly each mixtures of these configurations~\cite{Klempt:2020bdu}. 
 }
     \label{Sum}
\end{figure}

Table~\ref{Sum} presents a quark-model interpretation of the $\Lambda$ and $\Sigma$ resonances
\cite{Klempt:2020bdu}
analogous to Fig.~\ref{fig:missing-resonances}. 
In empty slots, no resonance is expected. 
Most exciting are the $\Lambda^*$'s, the only family that contributes to the flavor singlet spectrum.
For each $N^*$ resonance, one $\Lambda^*$ octet state is expected. 
The $\Sigma^*$ resonances can belong to the $SU(3)_f$ octet or decuplet, so for each
$N^*$ or $\Delta^*$  a $\Sigma^*$ resonance should exist. 
New data are expected from CLAS12 at JLab~\cite{Burkert:2020akg,Ernst:2020hru}
PANDA at GSI/FAIR~\cite{PANDA:2020hmi}, and ELSA~\cite{Cluster:2025}.
For a recent review on hyperons we refer to Ref.~\cite{Crede:2024hur}.

\subsubsection{Baryon decays}

Quark model calculations do not only give masses and widths of baryon resonances but also
their wave functions 
(not only the harmonic oscillator basis wave functions). These can be used to calculate branching ratios,
photocouplings or the $Q^2$ dependence of transition form factors. Table~\ref{tab:NSdecay} gives a survey
of references to calculations. A comprehensive review of the branching ratio calculations
published before 2000 can be found in~\cite{Capstick:2000qj}.

\begin{figure}[t!]
\hspace{70mm}\begin{minipage}[r]{.32\linewidth}
  \captionof{figure}{In the $^3P_0$-model, a quark-antiquark pair is created with the quantum numbers of the vacuum.
    The sum of intrinsic orbital angular momentum $L=1$ (blue) and spins (red) vanishes. }
   \label{fig:excite}
\end{minipage}
\hspace{5mm}\begin{minipage}[r]{.32\linewidth}
\raisebox{5mm}{\includegraphics[width=0.5\linewidth]{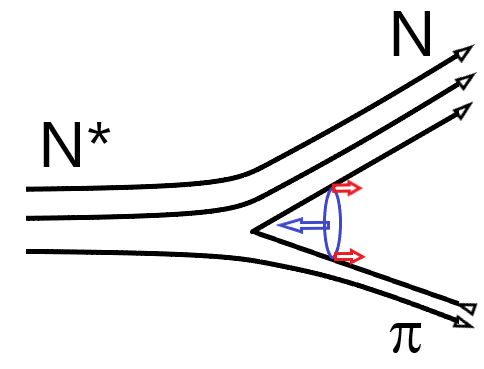}}
\end{minipage}
\vspace{-6mm}
\end{figure}

In baryon decays $N^*\to NM$, a new $q\bar q$ pair is created in a process as depicted
in Fig.~\ref{fig:excite}. The antiquark combines with a quark of the resonance
to form a meson. The newly created  $q\bar q$ pair does not form the meson $M$, which is forbidden
by the OZI rule \cite{Okubo:1963fa,Zweig:1964jf,Iizuka:1966fk}. Often, the new $q\bar q$ pair is
supposed to have the quantum numbers of the vacuum, $J^{PC}=0^{++}$. These $^3P_0$ models have 
met with considerable success.

Simple rules are at most approximately consistent with the experimental results. In hyperons, for example, the probability
to emit a kaon is only 1/3, while the $s$ quark ends up in the baryon with a probability of 2/3. 
 This makes the identification
of hyperons in $K^-p$ elastic scattering difficult. In Table~\ref{tab:BR} we compare
$N\bar K$ decays of $\Sigma$ hyperons with 4* rating with $N\pi$ decays of their non-strange partners.
The suppression of $\Sigma^*\to N\bar K$ decays relative to their \mbox{non-strange partners varies from 6 to 1.}
In Refs.~\cite{Fernandez-Ramirez:2015fbq,JPAC:2018zjz}, a novel attempt is made to
gain insight into the internal structure of baryons by modeling real and imaginary
parts of baryon resonances. For the fits, baryons are divided into several classes:
$N$, $\Delta$, $\Lambda$, and $\Sigma$ on the one hand, and according to their 
naturality and signature\footnote{The naturality is given by 
$\eta=+1$ if $P=(-1)^{J_p-1/2}$ and $\eta=-1$ if $P=-(-1)^{J_p-1/2}$;
the signature $\tau$ by $\tau = \nicefrac{P}{\eta}$.} on the other hand. 
The real parts of resonances in a given class show the expected linearity as 
functions of $J$, the imaginary parts a square root behavior. The square-root 
dependence is assigned to the contribution of the phase space
to the scattering amplitude, which is proportional to the momentum $q\sim\sqrt{s-s_t}$.  
The hope was that major deviations from the observed pattern would signal 
important components beyond compact $3q$ physics.
The authors of Ref.~\cite{Fernandez-Ramirez:2015fbq,JPAC:2018zjz} conclude that in the $\Lambda$ and $\Sigma$ sector a consistent picture emerges for the leading hyperon
Regge trajectories. The $\Lambda(1405)$ will be discussed below (see Section~\ref{sec:1405}),
where its two-pole nature is discussed. In \cite{Fernandez-Ramirez:2015fbq}, it
is found that the higher-mass pole of the $\Lambda(1405)$ fits into the general picture but the lower-mass pole
does not, so it is suggested that it might be a molecule. The $N(1680)\nicefrac52^+$, $N(1720)\nicefrac32^+$, 
$\Delta(1930)\nicefrac32^-$, and 
$\Delta(2400)\nicefrac92^-$ are suggested to have sizable physics beyond the compact $3q$ picture. 
However, a few open questions remain: The $N$ and $\Delta$ are well-separated while the octet and decuplet $\Sigma$ resonances
are not, and the decuplet $\Sigma(1385)\nicefrac32^+$ is shown jointly with the octet $\Sigma(1775)\nicefrac52^-$ 
(these quantum numbers are forbidden for a decuplet state) and with the decuplet $\Sigma(2030)\nicefrac72^+$. 
 The $N(1720)\nicefrac32^+$ is related to the $N(1990)\nicefrac72^+$; both belong to the second
 excitation shell, and both have $L=2$ as their leading 
orbital angular momentum. The $\Delta(1930)\nicefrac32^-$
 must have one unit of radial excitation, and the $\Delta(2400)\nicefrac92^-$ likely has the same. It seems questionable
 whether these states should be related to the $\Delta(1232)\nicefrac32^+$, $\Delta(1950)\nicefrac72^+$, and
 $\Delta(2420)\nicefrac{11}{2}^+$.

\begin{figure}[!t]
\begin{minipage}{.30\linewidth}
  \centering\vspace{-0mm}
  \captionof{table}{\label{tab:NSdecay}Calculations of branching ratios BR,
photocouplings $A$, and of the $Q^2$ dependence of transition form factors FF.\vspace{-2mm}}
\footnotesize
\renewcommand{\arraystretch}{1.4}
\begin{tabular}{cccc}
\hline\hline
Ref. & $A$& FF  & BR \\ \hline
\cite{Bijker:1994yr,Bijker:1995ii,Bijker:1996tr}                                               & yes & yes & yes\\
\cite{Giannini:2015zia,Merten:2002nz,Metsch:2003ix}                                            & yes & yes & no\\
\cite{Koniuk:1979vy,Koniuk:1979vw,Koniuk:1981ej,Capstick:1992th,Capstick:1993kb,%
Capstick:1998uh,Capstick:1998md,Goity:2007ft,Goity:2009wq,Jayalath:2011uc}                     & yes & no & yes\\
\cite{LeYaouanc:1973ldf,LeYaouanc:1974cvx,Stancu:1988gb,Stancu:1989iu,Stassart:1990zt,
Stancu:1993xz,Stassart:1995qf,Melde:2005hy,Faessler:2010zzc}                                   & no & no & yes\\
\cite{Theussl:2000sj,Wagenbrunn:2000es,Boffi:2001zb,Melde:2005hy,Melde:2007zz,Sengl:2007yq}  & no & yes & no \\
\hline\hline
\end{tabular}
\end{minipage}
\hspace{10mm}
\begin{minipage}[c]{.62\linewidth}\vspace{-11mm}
    \captionof{table}{Branching ratios of non-strange and strange baryons into $N\pi$ or $N\bar K$.\vspace{-2mm}}
    \label{tab:BR}
    \footnotesize
\renewcommand{\arraystretch}{1.4}
    \begin{tabular}{ccccccccc}
    \hline\hline
    $J^P$ & \multicolumn{2}{c}{$B^*\to N\pi$}&& \multicolumn{2}{c}{$B^*\to N\bar K$}&& \multicolumn{2}{c}{$B^*\to N\bar K$}\\
    \hline
$\frac32^-$&$N(1520)$ & (60\er5)\% && $\Sigma(1670)$ & (9\er3)\%  &&$\Lambda(1690)$ & (25\er5)\% \\
$\frac52^-$&$N(1675)$ & (40\er2)\% && $\Sigma(1775)$ & (40\er2)\% &&$\Lambda(1830)$ & (6\er2)\% \\
$\frac52^+$&$N(1680)$ & (65\er5)\% && $\Sigma(1915)$ & (10\er5)\% &&$\Lambda(1820)$ & (60\er5)\% \\
$\frac72^+$&$\Delta(1950)$ & (40\er5)\% && $\Sigma(2030)$ & (20\er3)\% && \\
    \hline\hline
    \end{tabular}
\end{minipage}
\vspace{-15mm}
\end{figure}  

The decays of octet and decuplet baryons into octet or decuplet baryon plus a pseudoscalar meson
are related by SU(3). Such analyses have been performed by Samios, Goldberg, and Meadows in 1974 \cite{Samios:1974tw}
and 30 years later by Guzey and Polyakov \cite{Guzey:2005vz}. Both analyses agree that the data are compatible
with SU(3) even though their assignments of baryons to multiplets are sometimes different and differ
from our assignment as well.

\begin{figure}[ht]
\begin{tabular}{ccc}
\includegraphics[height=0.2\textheight]{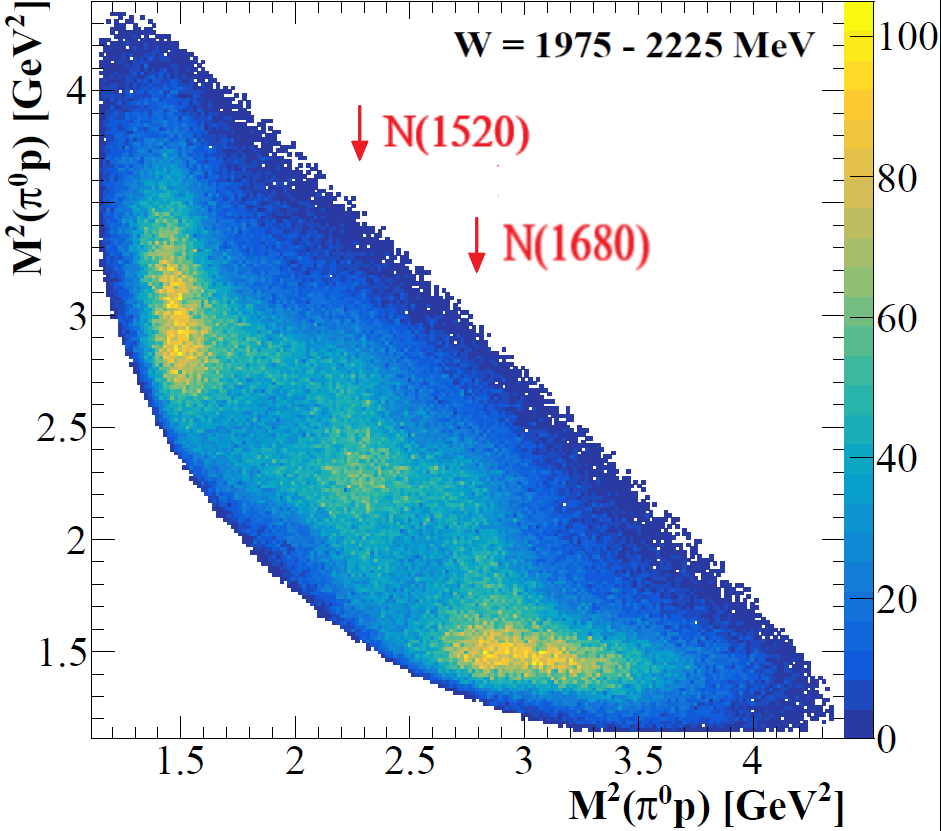} &
\hspace{3mm}\raisebox{10mm}{\includegraphics[height=0.16\textheight]{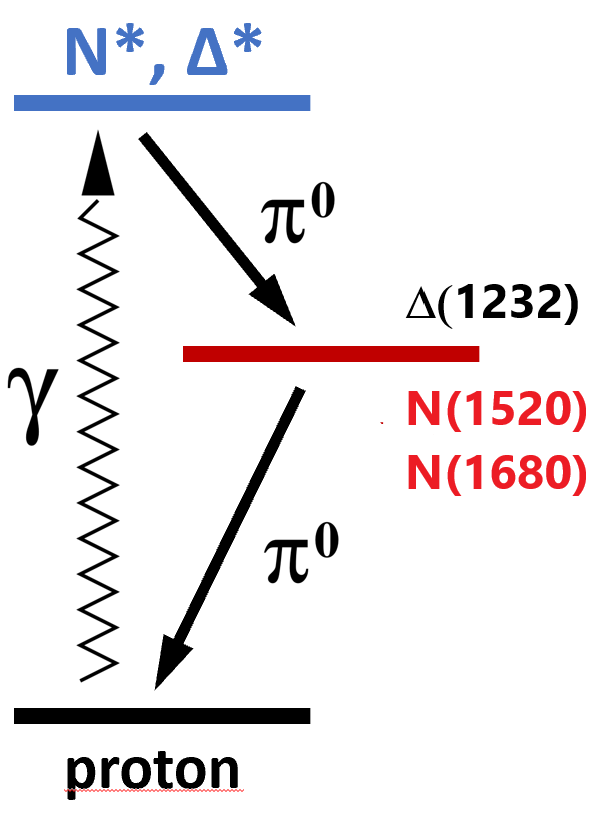}} &
\includegraphics[height=0.2\textheight]{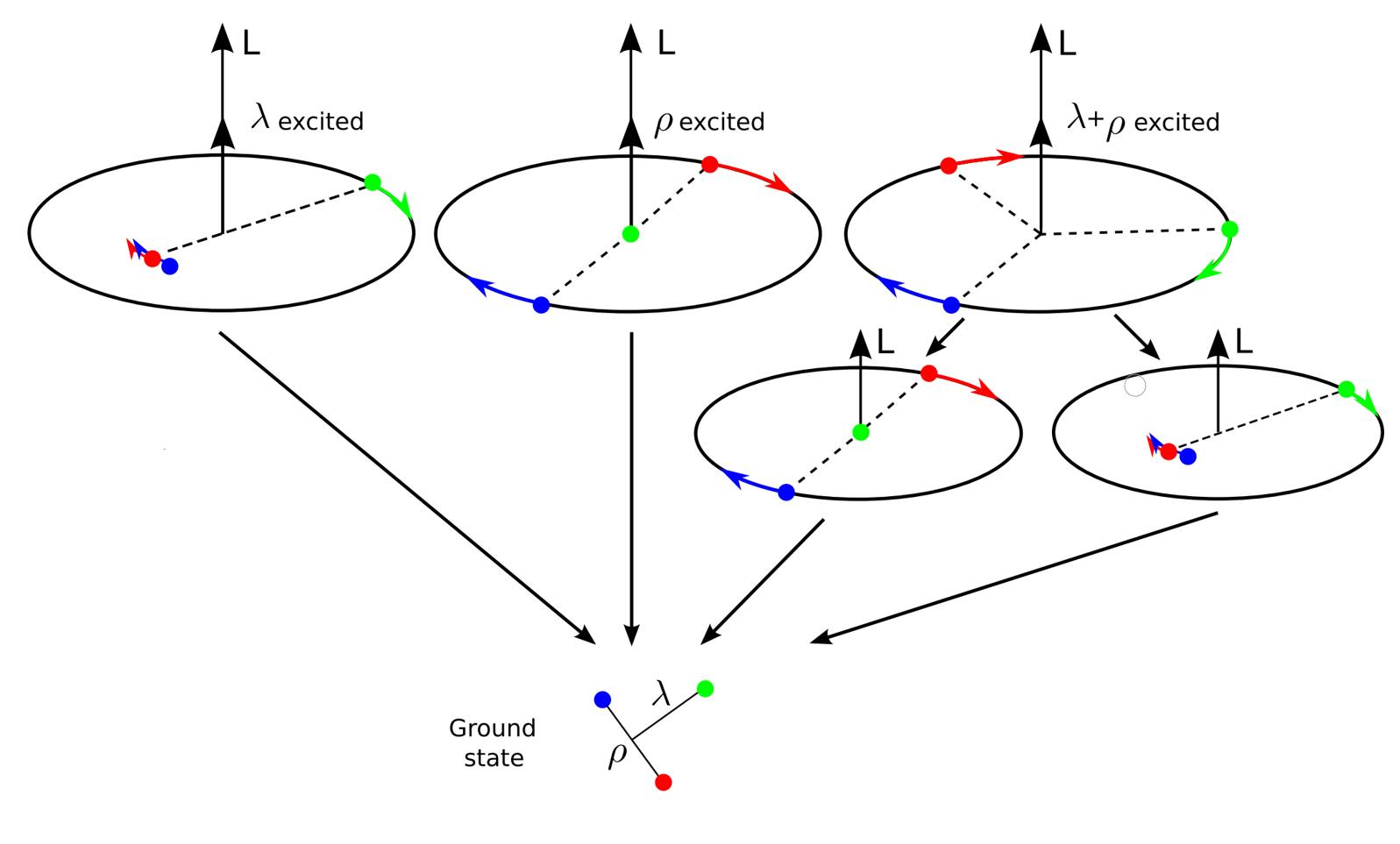}
\end{tabular}
\caption{\label{fig:dalitz}Left: $\gamma p \to p \pi^0\pi^0$ Dalitz plot for a selected $E_\gamma$ bin of 1975--2225~MeV (CBELSA/TAPS)~\cite{Sarantsev:2025lik}.  The $\Delta(1232)\nicefrac32^+$, $N(1520)\nicefrac32^+$, and $N(1680)\nicefrac52^+$
are easily identified as intermediate states. Center: Cascade decays of resonances via an immediate state. 
Right: Classical orbits of nucleon excitations with $L=2$ (upper row) and $L=1$ (middle row)
\cite{CBELSATAPS:2015taz}. The first two pictures in the upper row 
show excitations of the $\rho$ and $\lambda$ oscillators, while in the third picture both 
$\rho$ and $\lambda$ are excited. When both oscillators are excited, de-excitation leads to an excited intermediate state
(middle row).}
\vspace{-3mm}
\end{figure}

One can also use arguments based on cascade decays of baryon resonances. Figure~\ref{fig:dalitz}
shows a Dalitz plot for events due to the reaction $\gamma p\to p\pi^0\pi^0$ for the mass range $1900<\sqrt s < 2100$~MeV.
Clear bands are visible. They are understood as decays of baryons with masses of about 2~GeV
into intermediate resonances, e.g. a $N(1520)\nicefrac32^-$, plus a pion; the intermediate resonance
then decays into proton plus a second pion. 

In \cite{Thiel:2013cea} it was shown that cascade decays via intermediate  resonances with intrinsic
orbital angular momentum stem preferentially from baryons with a contribution in the wave function of 
the excited state in which both oscillators are excited simultaneously. Resonances like, e.g., the $N(1900)\nicefrac32^+$
have a mixed-symmetry spin-flavor wave function and both components
$\phi_{00,20}-\phi_{00,02}$ and $\phi_{00,11}$
contribute to the spatial wave function. When a pion is emitted from the latter part
from one oscillator, the other oscillator remains excited and de-excites in a second step. 
A coherent emission of the pion from both oscillators is, however,
possible because then the two pions can interact and form a $\sigma$ meson.  
In resonances like the $\Delta(1950)\nicefrac72^+$, the $\rho$ and the $\lambda$ oscillators are excited coherently to 
% $\nicefrac{1}{\sqrt2}$\{($\phi_{00,20}+\phi_{00,02}$)\}; 
$\phi_{00,20}+\phi_{00,02}$
which decays preferentially into two hadrons without orbital 
excitations. Both hypotheses
hold true only approximately. The two oscillators can decay coherently, e.g. into $\pi N(1520)\nicefrac32^-$, and this
intermediate state can re-scatter into $\rho N$. On the other hand, resonances like the $\Delta(1950)\nicefrac72^-$
can decay into $\pi N(1520)\nicefrac32^-$ in a $D$ wave.  

\subsubsection{Comments}\label{sec:spectrum-comments}

\textbf{\quad\ Number of states:} If we count all observations from the previous discussion, 
there is just one {\it missing resonance} in the first and second excitation shells.  
However, five of these resonances have a $1^*$ or $2^*$ assignment only and their 
existence is not beyond doubt, and two are not (yet) included in the RPP (see 
Table~\ref{tab:nexp}). It is certainly mandatory that the existence of these  
states is solidified. The assignment of the observed states to multiplets is 
often a pure guess. Sometimes, the multiplets show a rather irregular pattern: For 
instance, the masses of the $\Delta^*$ quartet with $J^P=\nicefrac12^+, \cdots,  
\nicefrac72^+$ are 1880\er20, 1880\er30, 1795\er20, 1892\er 5~MeV, respectively. 
There is no obvious reason for the low mass of $\Delta(1905)\nicefrac52^+$. 
The situation is even worse in the third excitation shell, since many resonances 
in the third shell are missing. Note  that only quark models give a 
prediction of the number of states.

\smallskip
\textbf{ Radial excitations:} The increase of the squared masses of radial excitations is approximately the
same as the Regge slope, see Table~\ref{tab:radials}. This is surprising since quark models predict
an approximate $2n+l$ dependence of the baryon masses. Indeed, quark models have met with difficulties
to explain the low $N(1440)\nicefrac12^+$ mass. Inspection of Table~\ref{tab:radials} shows that 
the small mass increase per unit of the radial excitation quantum number $n$  is not restricted
to the $N(1440)\nicefrac12^+$.

\smallskip
\textbf{ Hyperfine structure:}
In atomic physics, the splitting due to the spin-spin interaction is called hyperfine structure.
In hadron spectroscopy, hyperfine splittings appear  in many cases, most prominently in the mass difference between the nucleon and the $\Delta(1232)$ as shown in   Table~\ref{tab:hfs}. The states therein have the same intrinsic orbital angular momentum but different quark spin.
The states on the left have a symmetric wave function ($\psi_2$ in Fig.~\ref{fig:flavor-wfs}), while
those on the right contain a quark pair which is antisymmetric in spin and flavor, 
a {\it good diquark} with wave function $\psi_1$.
The hyperfine splittings are numerically very similar  throughout the spectrum, with values around $0.5 \dots 0.6$ GeV$^2$; also the squared mass difference between the corresponding states in the second oscillator shell is
$M^2_{\Delta(\mathbf{56},2^+,\nicefrac32)} - M^2_{N(\mathbf{56},2^+,\nicefrac12)}$ = 0.60(8)~{\rm GeV}$^2$.
This is also comparable to $M^2_\rho -M^2_\pi=$ 0.58(1)~{\rm GeV}$^2$, which is about half the Regge slope.
% The Regge slope (divided by 2), 
% the radial excitation energy, and the hyperfine splitting are numerically
% very similar in magnitude. 
% This may be seen as a hint
% that confinement and chiral symmetry breaking~\cite{tHooft:1986ooh} may be deeply connected. 
% Indeed, there is evidence that center vortices drive both dynamical chiral symmetry breaking and quark 
% confinement~\cite{Kamleh:2023gho}. 

The good agreement between the non-relativistic quark model predictions and the measured masses is, and
has always been, extremely astonishing. Even constituent quarks should move at relativistic velocities when
the nucleon is excited. Furthermore, the coupling of baryons to virtual baryon-meson intermediate states may
lead to large baryon self-energies and to mixing of baryons of the same quantum numbers~\cite{Morel:2002vk}. 
There is no reason at all to believe that the non-relativistic quark model may be useful in understanding the
spectrum, except one: it works surprisingly well. 
Possibly, the concept of rapidly moving constituent quarks with a fixed rest mass is misleading: 
when a gluon string evolves, a string of confined color field connecting quarks,
a large fraction of the excitation energy may be stored in the string, the effective
constituent quark masses increase, and the resulting motion could remain a classical one.

\smallskip
\textbf{Overlapping resonances:}
In the high-mass region above the second excitation shell, a large number of multiplets is predicted.
% In both the $N^*$ and the $\Delta^*$ sector, we observe exactly one resonance in each partial wave.
They all fall into a rather narrow mass window, 
where the resonances are often separated in mass by less than their widths. 
%Thus one can ask: 
What happens when multiple 
resonances with identical quantum numbers have nearby masses?
Well-known examples in the low-lying baryon spectrum are the $N(1535)\nicefrac12^-$
and $N(1650)\nicefrac12^-$, whose interference creates
a pattern that cannot be described by two Breit-Wigner amplitudes, and other methods like the $K$-matrix
approach are required. If resonances get closer, the observed pattern may change significantly.

In Ref.~\cite{Anisovich:1998au}, two resonances were considered with one decay channel only, in which case
one of the states becomes narrow and the other one broad. We know, however, no physical case corresponding
to this situation. The second possibility are two close-by states with strong couplings to two decay
modes. For example, the two mesons $K_{1A}$ and $K_{1B}$, the strange partners of $a_1(1260)$ and $b_1(1235)$,
have the decay modes $K^*\pi$ and $K\rho$ which are expected with similar rates. Due to mixing,
these two mesons repel each other, and the mixed state $K_{1}(1280)$ couples more strongly to $K\rho$
and $K_{1}(1400)$  very strongly to $K^*\pi$.  
On the other hand, these two scenarios -- decays into one dominant channel or mutual repulsion of pairs of states -- do not seem to be applicable in the high-lying baryon spectrum, where 
the number of negative-parity $N^*$ and $\Delta^*$ states expected in the third excitation shell
is rather large (see Table~\ref{fig:missing-resonances}). In the $N^*$ sector there is a low-mass doublet and in the $\Delta^*$ sector 
a low-mass triplet, and only these five low-mass resonances can be assigned to a unique multiplet. Above, there is exactly one observed state per partial wave from $J^P=\nicefrac12^-$
to $\nicefrac92^-$. In the quark model,
the four $\Delta\nicefrac32^-$ states above the $\Delta(1940)\nicefrac32^-$ in the third excitation shell have calculated masses within a 100~MeV wide interval and the seven 
$N\nicefrac32^-$ states above the $N(1875)\nicefrac32^-$ fall into a 200~MeV wide interval, and these resonances
are expected to have many decay modes.  

In principle, two further scenarios are possible. 
The multitude of states could form an unresolved cluster of states, although it seems difficult
to combine such a plethora of states in a common $K$ matrix without a massive distortion of
the final pole structure. Alternatively, a reduction in the number of observed states relative to quark model predictions could arise from a different mechanism, namely the coherent excitation of states into a common wave function $\phi_{00,30}\pm\phi_{00,03}$
with angular momentum $L=3$. 
This is certainly not the spatial wave function of
an energy eigenstate, because in the \textbf{70}-plets the energy eigenfunctions contain, e.g.,  
$\mS = \phi_{00,30} -\sqrt{3}\,\phi_{00,12}$,
$\mM_\mS = \sqrt{3}\,\phi_{00,03}     +\phi_{00,21}$,
$\mM_\mA = \phi_{00,30}     +\sqrt{3}\,\phi_{00,12}$
and
$\mA  = -\phi_{00,12}     +\sqrt{3}\,\phi_{00,30}$.
However, a coherent production of energy eigenstates could form a state like 
$\phi_{00,30}\pm\phi_{00,03}$,
which could then de-excite via its energy eigenfunctions.

\clearpage
\section{\label{QCD}Baryons and QCD}
In Sec.~\ref{Models} we discussed the experimentally known baryon spectrum.
As we have seen, the spectrum carries clear traces of the nonrelativistic quark model, 
which was historically also one of the pillars
for the development of QCD.
In  quark models, the light $u$ and $d$ quarks have masses of about $300 \dots 350$ MeV,
which are used to reproduce the masses of ground-state baryons  and their excitations. 
These massive quarks are called {\it constituent} quarks.
On the other hand, the  light current quarks in the QCD Lagrangian have masses of a few MeV,
which also appear in low-energy relations of QCD~\cite{Weinberg:1978kz}
such as the Gell-Mann-Oakes-Renner relation \cite{Gell-Mann:1968hlm}. 

The mass gap between current quarks at large momenta and constituent
quarks at low momenta can be explained by the spontaneous breaking of  chiral symmetry~\cite{Nambu:1960tm,Goldstone:1961eq}.
This mechanism dynamically generates a constituent mass through the interactions of quarks and gluons.
It adds roughly $\sim 350$ MeV to the quark mass of each quark flavor,
which is a subleading or even negligible effect for heavy charm and bottom quarks 
while for light $u$ and $d$  quarks it is the  dominant contribution.
An important consequence is the large mass gap between chiral partners: The masses of the
nucleon with spin-parity $J^P=1/2^+$ and its chiral partner $N(1535)\nicefrac{1}{2}^-$ with $J^P=1/2^-$ differ by about 600~MeV. 
Since quark models provide no insight how this large mass difference is generated,
we should take a step back and ask: How much can we infer about the baryon spectrum
from QCD?

\subsection{What is a baryon in QCD? }\label{sec:whatisabaryon}

\begin{figure}[b]
	\centering
	\includegraphics[width=0.6\columnwidth]{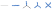}
	\caption{Elementary propagators and interactions in QCD.}
	\label{fig:tree-level}
	\vspace{-2mm}
\end{figure}

As far as we know today, the strong interaction is fully 
described by the QCD Lagrangian $\mL$ or
its classical action $S$:
\begin{eqnarray}\label{qcd-l}
	S = \int d^4x\,\mL\,, \quad
	\mL = \conjg\psi\left(i\slashed{\p} + g\slashed{A} - \mathsf{M} \right) \psi - \frac{1}{4} F_{\mu\nu}^a F^{\mu\nu}_a\,.
\end{eqnarray}
The quark fields $\psi_{\alpha,i,f}(x)$ are Dirac spinors $(\alpha = 1 \dots 4)$ and vectors in color  $(i=1\dots 3)$ and flavor space $(f=1\dots N_f)$.
$\mathsf{M} = \text{diag}(m_u, m_d, m_s, \dots)$ is the quark mass matrix in flavor space.
$F_{\mu\nu}^a$ are the components of the field-strength tensor 
\begin{eqnarray}\label{Fmunu}
	F_{\mu\nu} = \p_\mu A_\nu - \p_\nu A_\mu -ig[A_\mu,A_\nu]\,, \qquad 
    A^\mu = \sum_{a=1}^8 A^\mu_a\,\mathsf{t}_a\,, \qquad 
    F^{\mu\nu} = \sum_{a=1}^8 F^{\mu\nu}_a\,\mathsf{t}_a\,,
\end{eqnarray} 
where $A_a^\mu$ are the eight gluon fields and $\mathsf{t}_a$ the generators of $SU(3)_c$.
Diagrammatically, the action is described by the Feynman diagrams in Fig.~\ref{fig:tree-level}: 
Quarks and gluons interact with a coupling strength $g$,
and due to the non-Abelian nature of the gauge group $SU(3)_c$ the elementary interactions consist of a quark-gluon vertex
as well as three- and four-gluon interactions.
What else can the Lagrangian tell us?

The local $SU(3)_c$ gauge invariance of the action does not tell us much about the baryon spectrum
apart from the fact that hadrons must be color singlets.
When combining three quarks, which transform under the fundamental $\mathbf{3}$ representation of $SU(3)_c$,
to $\mathbf{3}\otimes\mathbf{3}\otimes\mathbf{3} = \mathbf{10}_S \oplus \mathbf{8}_{M_A} \oplus \mathbf{8}_{M_S} \oplus \mathbf{1}_A$,
baryons must belong to the totally antisymmetric color-singlet representation $\mathbf{1}_A$.
Furthermore, the action is invariant under the Poincaré group, which consists of translations, rotations and boosts.
The Poincaré group has two Casimir operators labelling the states: the mass $M$ and the total angular momentum or spin $J$.
Concerning the invariance under the discrete symmetries parity, charge conjugation and time reversal,
the combination $CPT$ is always conserved.
Because  charge conjugation transforms baryons into antibaryons, 
only parity induces a conserved quantum number $P$.
Therefore, so far we arrived at $M$ and $J^P$ as quantum numbers labeling the states.

There are  further global transformations of the quark fields, which
we exemplify for $N_f = 3$ with three flavors $u$, $d$, $s$:
\begin{itemize}

\item
The group $U(1)_V$ consists of global phase transformations $\psi' = e^{i\varepsilon}\psi$
which leave $\mL$ invariant. The conserved quantum number is the baryon number $B= \frac{1}{3}(n_u + n_d + n_s)$,
where $n_q$ is the number of quarks minus antiquarks in a state. Baryons carry $B=1$ (so do also pentaquarks).
In strong interactions also $n_u, n_d$ and $ n_s$ are conserved.

\item
The group $SU(N_f)_V$ is defined by the transformation $\psi' = e^{i\varepsilon}\psi$, where $\varepsilon = \sum_a \varepsilon_a \mathsf{t}_a$ and
$\mathsf{t}_a$ with $a=1 \dots N_f^2-1$ are the $SU(N)$ generators. It induces vector currents and corresponding charges
\begin{eqnarray}
	V_a^\mu = \conjg\psi \gamma^\mu \mathsf{t}_a \psi\,, \qquad
	\p_\mu V_a^\mu = i\conjg\psi \left[ \mathsf{M}, \mathsf{t}_a \right] \psi\,, \qquad Q_a^V(t) = \int d^3x\,\psi^\dag \mathsf{t}_a \psi\,.
\end{eqnarray}
For $N_f=3$ this is the usual $SU(3)_f$ flavor symmetry, which by the Noether theorem is preserved if the divergences of the currents vanish.
This can only work if $\mathsf{M} = m$, i.e., all quark masses are equal and the commutator is zero.
On the other hand,  the diagonal $SU(3)_f$ generators $\mathsf{t}_3$ and $\mathsf{t}_8$ commute with $\mathsf{M}$ and are  always conserved.
The corresponding quantum numbers are the isospin-3 component $I_3 = \frac{1}{2}\,(n_u-n_d)$ and the hypercharge $Y= \frac{1}{3} (n_u + n_d - 2n_s)$,
which are therefore  still good quantum numbers to label the states even if the flavor symmetry is broken.

\item
The transformation $\psi' = e^{i\gamma_5\varepsilon}\psi$, again with $\varepsilon = \sum_{a} \varepsilon_a \mathsf{t}_a$,
define $SU(N_f)_A$  which induces axialvector currents
\begin{eqnarray}\label{ax-currents}
	\tilde A_a^\mu = \conjg\psi \gamma^\mu\gamma_5 \mathsf{t}_a \psi\,, \qquad
	\p_\mu \tilde A_a^\mu = i\conjg\psi \left\{ \mathsf{M}, \mathsf{t}_a \right\} \gamma_5 \psi
\end{eqnarray}
and  charges $Q_a^A(t)$. We used a tilde to distinguish the axialvector currents $\tilde A^\mu_a$ corresponding to $SU(N_f)_A$  
from the gluon fields $A^\mu_a$ corresponding to $SU(3)_c$ in Eq.~\eqref{Fmunu}. 
Chiral symmetry is the combination $SU(N_f)_V \times SU(N_f)_A \simeq SU(N_f)_L \times SU(N_f)_R$.
The axialvector symmetry is classically conserved only in the chiral limit $\mathsf{M} = 0$ where the anticommutator in Eq.~\eqref{ax-currents} vanishes.
However, it is spontaneously broken in the quantum field theory by the quark-gluon dynamics,
which is therefore called dynamical chiral symmetry breaking. As mentioned above, this property has important consequences for the
hadron spectrum and will be discussed in more detail in Sec.~\ref{sec:fm}.  

\item
The axial $U(1)_A$ symmetry defined by $\psi' = e^{i \gamma_5 \varepsilon}\psi$, where $\varepsilon$ is   a number,
is classically also only conserved for $\mathsf{M}=0$. However, this symmetry does not survive the quantization of QCD and is anomalously broken.
\end{itemize}
All in all, the quantum numbers to label baryons ($B=1$) with three quark flavors are the mass $M$, 
spin and parity $J^P$, the isospin $I_3$ and the hypercharge $Y$.

So what is a baryon from the standpoint of QCD? To answer this,
we must turn to the  quantum field theory (QFT). The Lagrangian~\eqref{qcd-l} only gives
us the building blocks of the theory and their elementary interactions, i.e., the 
basic construction rules of QCD. The dynamical content of the theory emerges in the QFT,
which is defined by  the partition function $Z$:
\begin{eqnarray}\label{path-int-0}
	Z = \int  \mD A\,\mD \psi  \,\mD\conjg{\psi} \,e^{iS}\,.
\end{eqnarray}
%{\bl Can you define $\mD A$? You use $A, A^\mu, A^\mu _a$.}
Here, the path integral measure $\mD A\,\mD \psi  \,\mD\conjg{\psi}$ amounts to a statistical average over all possible quark and gluon
field configurations. 
The central quantities in QCD are the  $n$-point correlation functions of the form
\begin{eqnarray}\label{corr-fct}
	G(x_1, \dots, x_n) = \langle 0 | \mathsf{T}\,\phi(x_1) \dots \phi(x_n) | 0 \rangle\,,
\end{eqnarray}
where $\phi \in \{ A^\mu,   \psi,   \conjg{\psi} \}$ is a shorthand for the quark and gluon fields
and $\mathsf{T}$ denotes time ordering.
Abbreviating the combinations of fields by $\mO = \phi(x_1) \dots \phi(x_n)$ and the respective correlation
function by $\langle \mO \rangle = G(x_1, \dots, x_n)$, they can be computed from the path integral via
\begin{eqnarray}\label{path-int-2}
	\langle \mO \rangle = \frac{1}{Z} \int  \mD A\,\mD \psi  \,\mD\conjg{\psi} \,e^{iS}\,\mO\,.
\end{eqnarray}
The generation of the field ensembles that enter in the path integral, and the calculation of such expectation values,
is conceptually straightforward (although technically demanding) and the basis of lattice QCD. 
Another way to obtain the correlation functions is to perform functional derivatives
of the partition function after adding appropriate source terms to the action.
This, in turn, entails that the partition function can 
be reconstructed from the set of all correlation functions. 
This is why the correlation functions play such a central role in the QFT:
If we knew all  (infinitely many)
correlation functions of the theory, we would have solved QCD!

\begin{figure}
	\centering
	\includegraphics[width=1\textwidth]{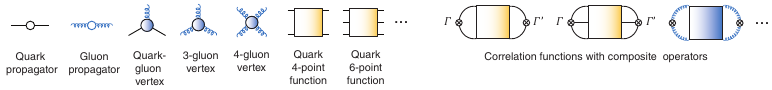}
	\caption{Some $n$-point correlation functions in QCD.}
	\label{fig:cfs}
\end{figure}

Fig.~\ref{fig:cfs} shows some of the simplest correlation functions:
the elementary quark propagator from the Lagrangian turns into the full (`dressed') quark propagator
$S_{\alpha\beta}(x) = \langle \conjg{\psi}_\alpha(x)\,\psi_\beta(0) \rangle$,
%\begin{eqnarray}
%	S_{\alpha\beta}(x) = \langle 0 | \mathsf{T} \,\conjg{\psi}_\alpha(x)\,\psi_\beta(0) | 0 \rangle\,,
%\end{eqnarray}
the elementary gluon propagator turns into the dressed gluon propagator $D^{\mu\nu}(x) = \langle A^\mu(x) A^\nu(0)\rangle$,
the vertices become dressed vertices and so on. In addition, there
are new correlation functions which do not have an elementary counterpart,
such as a quark-antiquark ($q\bar{q}$) four-point function, a three-quark correlation function, etc.
Moreover, the fields in Eq.~\eqref{corr-fct} do not need to be elementary fields but can also be composite: For example, by contraction
with appropriate Dirac-flavor-color matrices, the three-quark correlator turns into a nucleon current correlator $\langle  N(x)\, \bar{N}(0) \rangle$.
This is especially useful for constructing gauge-invariant correlation functions: While the correlation functions
with open quark or gluon legs depend on the gauge, hadronic current correlators are by construction gauge-invariant
and thus in principle measurable.

%So once again: What is a baryon?
An important relation in QFT  says that physical states produce poles in correlation functions
(see e.g.~\cite{Weinberg:1995mt} for a proof).
Suppose we split the $n$ coordinates in Eq.~\eqref{corr-fct} into two sets $\{x_i\}$ and $\{x_j\}$  with $i=1\dots N$. % and $j=1\dots N'$.
Then, by inserting the completeness relation of the state space,  
each asymptotic one-particle state $|\lambda\rangle$ with onshell momentum $P$ and mass $m_\lambda$ (such that $P^2 = m_\lambda^2$) 
produces a pole in any correlation function that is compatible with the quantum numbers of the state:
\begin{align}\label{poles}
	G(x_1, \dots x_n) &= \int \!\!  \frac{d^4P}{(2\pi)^4}\,e^{-iPz}\, \frac{i\Psi(\{x_i\},P)\,\Psi^\dag(\{x_j\},P)}{P^2 - m_\lambda^2 + i\epsilon} +  \text{finite}\,.
\end{align}
The remainder refers to terms that are finite at the pole position $P^2 = m_\lambda^2$.
The residue at the pole defines the Bethe-Salpeter wave function (BSWF),
\begin{eqnarray}\label{bswf}
	\Psi(\{x_i\},P) = \langle 0 | \mathsf{T}\,\phi(x_1) \dots \phi(x_N) | \lambda \rangle\,,
\end{eqnarray}
which is the overlap of the vacuum and the one-particle state.
Here, an overall coordinate $z$ has been factored out by translation invariance, % and only contributes to the phase.
so the BSWF depends only on $N-1$ coordinates.
$\Psi$ is the QFT analogue of a quantum-mechanical wave function since it carries the information about the hadron,
although it does not have a direct probability interpretation (more on that in Sec.~\ref{sec:fm}). Note also that the BSWF is not an observable as it is  gauge- and  renormalization-point dependent.

\begin{figure}
	\centering
	\includegraphics[width=0.95\columnwidth]{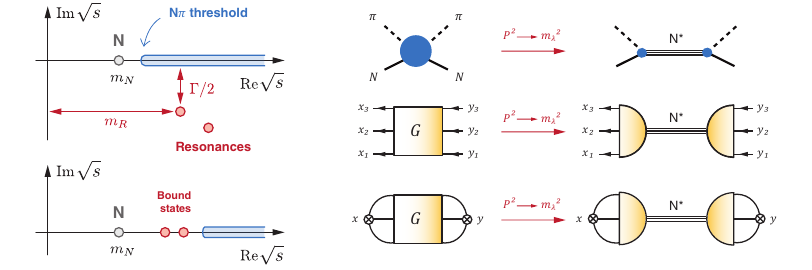}
	\caption{Left: Poles and cuts in the complex energy plane of a correlation function. 
             By tuning certain control parameters, resonances may turn into bound states and vice versa.
         Right: Baryon poles appear in any correlation function that can produce them, such as the $N\pi$ scattering amplitude,
		the gauge-dependent $qqq$ correlation function or the respective gauge-invariant current correlator. }
	\label{fig:poles}
	\vspace{0mm}
\end{figure}

With this we can finally answer our question: What is a baryon? 
A \textit{stable} baryon is a one-particle state $|\lambda\rangle$ with baryon number $B=1$, which creates
poles in those correlation functions that are compatible with its quantum numbers.
A baryon \textit{resonance}, on the other hand, is an unstable baryon above a threshold which can decay. Because it is not an asymptotic state $|\lambda\rangle$ 
the definition~\eqref{bswf} does not apply, but since a resonance
also produces a pole,
its BSWF can still be  defined as the pole residue of the correlation function at the respective pole position like in Eq.~\eqref{poles}.
In turn, these poles no longer appear on the real axis but in the complex plane (i.e., complex masses $m_\lambda$)
on higher Riemann sheets, since the first sheet must be free of singularities.
This goes in parallel with the development of branch cuts corresponding to the possible decay channels, 
 which come from the multiparticle continuum.
The situation is illustrated in the left of Fig.~\ref{fig:poles}:
In the $J^P = \frac{1}{2}^+$ channel, the `one-particle state'  is the nucleon ground state, the lowest multiparticle threshold is $N\pi$,
and the resonances decaying into $N\pi$ lie in the complex plane on the second Riemann sheet.
With certain control parameters one can  convert resonances into bound states and vice versa:
In model calculations one can switch off the $N\pi$ interactions and turn  the resonances  into bound states;
in lattice calculations one can tune the current-quark masses to make the pion heavy and push the branch cuts further to the right,
so that the resonances also become bound.

In any case, the same gauge-invariant spectrum must appear in any correlation function 
that is compatible with the  quantum numbers of the baryon.
This is  Occam's razor at work:
What can be measured experimentally are cross sections for particular scattering amplitudes, whose singularities define the spectrum,
and what can be calculated theoretically are the respective correlation functions. %, and the set of all correlation functions define QCD.
%Eq.~\eqref{poles} is the central formula for extracting baryon properties from QCD, as illustrated in the right of Fig.~\ref{fig:poles}.
%First, 
As illustrated in the right of Fig.~\ref{fig:poles}, Eq.~\eqref{poles} applies to any scattering amplitude that admits baryon poles, 
such as $N\pi$ scattering or pion electroproduction, in which case the residues are the $N\pi$ pion couplings or  electrocouplings.
This is how baryons are measured experimentally, and it is the starting point for EFTs 
 when employing multichannel scattering equations. 
Eq.~\eqref{poles} also applies to the (gauge-dependent)  three-quark correlator, 
in which case the pole residues are the (gauge-dependent) three-quark BSWFs of the baryon in question. 
This is the starting point for functional methods, where  %  discussed in Sec.~\ref{sec:fm}
the physical spectrum is extracted from   the quark and gluon correlation functions.
Finally, Eq.~\eqref{poles} applies to the (gauge-invariant) two-point current correlators
with composite operators, like those shown in Fig.~\ref{fig:cfs}, which follow from the contraction with appropriate Dirac-flavor-color tensors.
In this case the pole residues are the integrated BSWFs which are then also gauge invariant.
This is the basis of lattice QCD calculations, where  gauge-invariant current correlators
are computed from the path integral~\eqref{path-int-2}.  %  discussed in Sec.~\ref{sec:lattice}

Is some given baryon resonance more like a three-quark state or a meson-baryon dominated state?
A priori, QFT does not directly tell us that:
From Fig.~\ref{fig:poles}, a nucleon resonance can be thought of as an $N\pi$ state, but also as a three-quark state ($qqq$), or a five-quark state ($qqqq\bar{q}$) etc.,
because all those correlation function must produce the same spectrum. 
For example, any  dynamically generated `additional' baryon resonance that 
arises from meson-baryon interactions in an EFT must  also have a three-quark core (unless some  quantum number prevents it),
so these two options are not binary.
Vice versa, a pole in the $qqq$ correlation function emerging from quark-gluon interactions 
must also appear in the $qqqq\bar q$ and other correlation functions, but the three-quark component might be small.

The relevant question is therefore: How much \textit{overlap} does a state have
with a given configuration? This is defined by the respective BSWF.
If the $qqq$ overlap is small and the $qqqq\bar q$ overlap large, it implies a dominant $qqqq\bar q$ (or $N\pi$) structure;
if the $qqq$ overlap is large it implies a three-quark dominance.
In light-quark baryon spectroscopy, the $\Lambda(1405)$ is the primary candidate for having a dynamically generated additional state,
which would then translate into two poles that couple  more strongly to  $qqq$ or $qqqq\bar q$~\cite{Meissner:2020khl,Mai:2020ltx,Mai:2022eur,Guo:2023wes}.
Going further, a deeper question is: What is the reason for the existence of a particular resonance?
What are the interactions that produce a $qqq$-like or molecular behavior?
Is there a molecular force that binds $N$ and $\pi$, and this bound state may also couple to $qqq$?
Or is there a $qqq$ bound state that acquires a molecular component?
Microscopic calculations that can extract such information using different approaches are underway and will presumably
shed  more light on these questions in the future.
In addition, much can be learned from the structure of baryons as measured in electroproduction,
which we will discuss in Sec.~\ref{Structure}.

In the following subsections, we will give a brief overview of some of the predominant methods
used to compute the baryon spectrum. Rather than delving into the practical details of each method,
our focus will be on global features: What insights can these methods offer about the baryon spectrum, 
and how do they enhance our  understanding?

\subsection{Effective field theories}	\label{sec:eft}

Effective field theories (EFTs) for describing hadron-hadron interactions have found widespread applications over the last decades, 
see e.g.~\cite{Guo:2017jvc,Oller:2019opk,Mai:2022eur,Doring:2025sgb} for recent reviews. 
%They are now frequently employed in combination with lattice calculations and amplitude analyses.
%In the following we only sketch some basic relations and refer to the reviews~\cite{a,a,Mai} for details.
The primary quantities of interest are hadronic scattering amplitudes like in the top right row of Fig.~\ref{fig:poles}.
These contain the hadron poles matching the quantum numbers allowed by the initial and final states,
which allows one to extract the hadron spectrum from the pole positions.
Generally, the analytic structure of scattering amplitudes is  constrained
by the S-matrix principles of  unitarity, analyticity and crossing symmetry.
The analytic structure is  determined entirely by \textit{physical} particles: 
bound states and resonances manifest as poles, while particles in the loop diagrams generate branch cuts. 
By contrast, confinement implies that quarks and gluons must conspire in such a way that all direct evidence of their existence is erased
from the singularity structure of physical scattering amplitudes.
Although conceptually inevitable, the elimination of these gauge-dependent 
degrees of freedom from gauge-invariant amplitudes is quite astonishing 
and a central theme of quark-hadron duality~\cite{Melnitchouk:2005zr}.

\begin{wrapfigure}[13]{r}{8.5cm}
		\centering
         \vspace{-5mm}
		\includegraphics[width=0.95\linewidth]{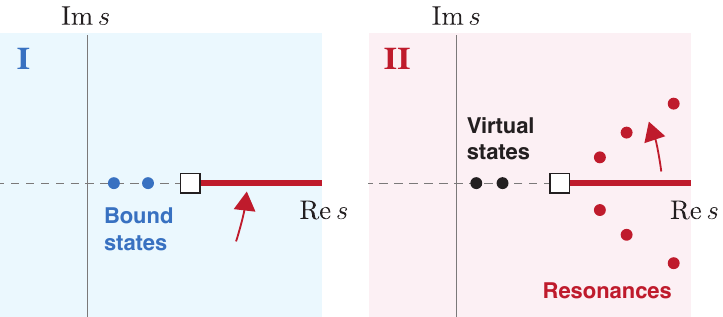}
		\caption{Possible analytic structure of a scattering amplitude with a single branch point ($\Box$) in the complex $s$ plane.
			         The scattering amplitude is analytic when crossing from the first to the second sheet (indicated by the arrow).
			    }
		\label{fig:cuts-0}
%		\vspace{4mm}
	\end{wrapfigure}

For illustration, we consider a
$2\to 2$ scattering process of identical particles with mass $m$, and we only allow for 2-body intermediate states.
The invariant matrix element $\mM(s,z)$ then depends on two Lorentz-invariant variables $s$ and $z$, where in the CM frame
$\sqrt{s}$ takes the meaning of the CM energy and $z$ is the cosine of the CM scattering angle.
Its expansion in partial waves $f_l(s)$ reads
\begin{eqnarray}\label{partial-waves}
	\mM(s,z) = \sum_{l=0}^\infty (2l+1)\,f_l(s)\,P_l(z)\,,
\end{eqnarray}
where  $P_l(s)$ are the Legendre polynomials.
The typical singularity structure of the partial waves $f_l(s)$ in the complex $s$ plane is sketched in Fig.~\ref{fig:cuts-0} for the situation
with only one cut.
Bound-state poles can appear  on the real axis below the threshold $\sqrt{s} = 2m$. The threshold marks the onset of the two-particle continuum,
which creates a branch cut and thus an imaginary part for the scattering amplitude along the cut.
Analyticity implies that no singularities can appear on the first sheet (the physical sheet).
Thus, resonance poles must  appear on the unphysical second sheet, and the pole position corresponds to the mass and width of the resonance (Fig.~\ref{fig:poles}).
In addition, virtual states are possible poles on the real axis of the second sheet below the threshold.

\begin{figure}[!b]
	\centering
	\includegraphics[width=0.8\textwidth]{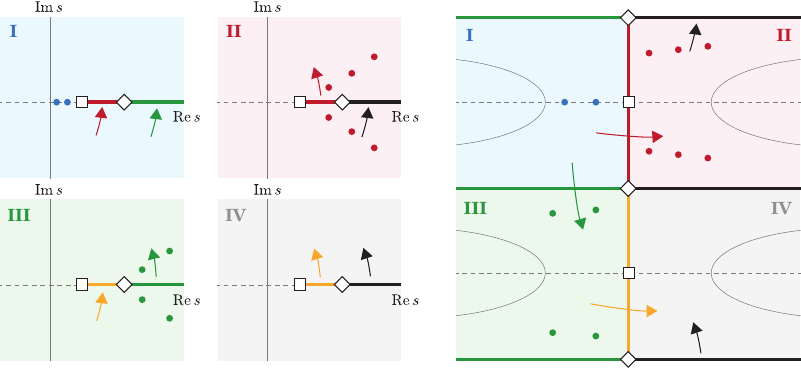}
	\caption{Left: Sheet structure of a scattering amplitude with two branch points ($\Box$ and $\Diamond$) and four Riemann sheets.
		The amplitude is analytic when moving through the sheets, as indicated by the arrows.
        This becomes obvious when unfolding the sheet structure (right). }
	\label{fig:cuts-2}
	\vspace{0mm}
\end{figure}

In practice the situation can be much more complicated than this simple sketch.
For example, the case with two branch points is shown in Fig.~\ref{fig:cuts-2}.
The four Riemann sheets in the left panel can be mapped to a torus~\cite{Mai:2022eur} in  the right panel:
The bottom and top borders of the rectangle are identified, and the left and right borders correspond to $|s|=\infty$.
Crossing over between the four sheets is now an analytic operation.

The crucial relation behind these properties is unitarity for the S matrix, $S^\dag S  = 1$.
With $S=1+iT$, this
turns into $T-T^\dag = i\, T^\dag T$ for the T matrix, which entails 
\begin{eqnarray}\label{unitarity}
	 \text{Im} \,f_l(s) = \tau(s) \,|f_l(s)|^2 
\end{eqnarray}
for the partial waves in Eq.~\eqref{partial-waves}. Here,
$\tau(s)$ is a kinematic function that vanishes at the threshold $\sqrt{s} = 2m$,
\begin{eqnarray}
	\tau(s) = \frac{1}{16\pi} \sqrt{\frac{s-4m^2}{s}} = \frac{1}{16\pi} \frac{2|\vect{q}|}{\sqrt{s}}\,,
\end{eqnarray}
and $\vect{q}$ is the three-momentum in the CM frame (see also~\cite{Mai:2025wjb} for a pedagogical introduction). 
Unitarity thus constrains the imaginary parts of the partial waves $f_l(s)$. This
can be expressed through the phase shift $\delta_l(s)$:
Writing $f_l(s) = R_l(s)\,e^{i\delta_l(s)}$ above the threshold $\sqrt{s} > 2m$, Eq.~\eqref{unitarity} implies $R_l = \frac{1}{\tau} \sin\delta_l$ and hence
\begin{eqnarray}\label{unitarity-2}
	f_l = \frac{1}{\tau}\,e^{i\delta_l} \sin\delta_l = \frac{1}{\tau}\,\frac{1}{\cot\delta_l - i} \qquad \Rightarrow \qquad
	 \text{Re}\,\frac{1}{f_l} = \tau\,\cot\delta_l\,, \quad \text{Im}\,\frac{1}{f_l} = -\tau\,. %\,\Theta(\sqrt{s}-2m)\,,
\end{eqnarray}
Again, these relations are valid above the threshold, whereas below the threshold the scattering amplitude is real.
%where the step function emphasizes that the scattering amplitude is real below the threshold.
With $\text{Im}\,f_l^{-1}= -\tau$,
another way to parametrize the partial-wave amplitudes 
is the so-called $K$ matrix. Here one writes $\text{Re}\,f_l^{-1}= K_l^{-1}$,
such that the real quantity $K_l(s)$ is related to the phase shift by
$\tan \delta_l = \tau K_l$. This yields
\begin{eqnarray}\label{K-matrix}
	  f_l^{-1} = K_l^{-1} - i\tau  \quad \Leftrightarrow \quad f_l = K_l + K_l \,i\tau\,f_l  \,,
\end{eqnarray}
%with $K_l(s)$ being purely real. The
and the resulting expression satisfies the unitarity relation~\eqref{unitarity} irrespective of the form of $K_l$ because
\begin{equation*}
	  f_l = \frac{K_l}{1-i\tau K_l} = \frac{K_l\, (1+i\tau K_l)}{1+\tau^2 K_l^2}\,,   \qquad  |f_l|^2 = \frac{K_l^2}{1+\tau^2 K_l^2}\,.
\end{equation*}
Note that the normalization conventions for $f_l$ and $K_l$ can differ in the literature.

 \begin{figure}[t]
 	\centering
     \vspace{-0mm}
 	\includegraphics[width=0.38\linewidth]{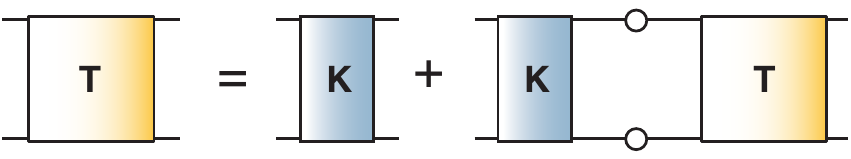}
 	\caption{Two-body Bethe-Salpeter equation.}
 	\label{fig:bse}
 	\vspace{0mm}
 \end{figure}

The unitarity relation~\eqref{unitarity} can also be derived from the scattering equation or 
inhomogeneous Bethe-Salpeter equation (BSE) for the scattering amplitude,
\begin{eqnarray}\label{bse-1}
	 \mathbf{T} = \mathbf{K} + \mathbf{K}\,\mathbf{G}_0\,\mathbf{T} \quad \Leftrightarrow \quad \mathbf{T}^{-1} = \mathbf{K}^{-1} - \mathbf{G}_0\,,
\end{eqnarray}
which is illustrated in Fig.~\ref{fig:bse} for the two-body case with a single channel.
In this symbolic notation the multiplications represent four-momentum integrations,
so this is an integral equation which by iteration
generates all possible diagrams for the scattering amplitude $\mathbf{T}$.
The two-particle irreducible kernel $\mathbf{K}$ contains all diagrams  that do not fall apart by cutting two horizontal lines,
because those are already generated by the iteration.
$\mathbf{G}_0$ is the product of the two propagators. %two-body propagator. 
For BSEs with hadronic degrees of freedom, the propagator lines are usually treated as free propagators
with onshell poles, since only onshell hadrons are well-defined concepts (and accessible experimentally) and offshell extensions are ambiguous.
The three-dimensional or nonrelativistic version of the scattering equation is known as the Lippmann-Schwinger equation,
but to keep the following discussion general we will collectively refer to these types of equations as BSEs.

Note that by subtracting the BSEs slightly above ($\mathbf{T}_+$) and   below  ($\mathbf{T}_-$) 
the cut in the variable $s$, one obtains
 \begin{eqnarray} \label{BSE-unitarity}
	  \mathbf{T}_+ - \mathbf{T}_- = \mathbf{T}_+ (\mathbf{T}_-^{-1} - \mathbf{T}_+^{-1}) \mathbf{T}_-  
		 = \mathbf{T}_+ (\mathbf{G}_{0+} - \mathbf{G}_{0-}) \mathbf{T}_- \,.
\end{eqnarray}
 The kernel $\mathbf{K}$ is real along the cut, so its difference vanishes.
 The l.h.s. in Eq.~\eqref{BSE-unitarity} is proportional to the imaginary part of $\mathbf{T}$ along the cut,
 while the difference of the onshell propagator poles in $\mathbf{G}_0$ on the r.h.s.
  induces $\delta$-functions. 
Working this out for the partial waves yields again Eq.~\eqref{unitarity}.
Thus, the BSE automatically satisfies unitarity!

 It is also often useful to perform a split $\mathbf{G}_0 = \mathbf{G}_1 + \mathbf{G}_2$
 and define a quantity %$\mathbf{K}_1$ by
 $\mathbf{K}_1^{-1} = \mathbf{K}^{-1} - \mathbf{G}_1$ $\Leftrightarrow$ $\mathbf{K}_1 = \mathbf{K} + \mathbf{K}\,\mathbf{G}_1\,\mathbf{K}_1$,
 so that Eq.~\eqref{bse-1} turns into
\begin{eqnarray}\label{bse-2}
	\mathbf{T}^{-1} = \mathbf{K}_1^{-1} - \mathbf{G}_2  \quad \Leftrightarrow \quad \mathbf{T} = \mathbf{K}_1 + \mathbf{K}_1\,\mathbf{G}_2\,\mathbf{T}  \,.
\end{eqnarray}
The BSE is then a two-step process, where one first determines $\mathbf{K}_1$ from $\mathbf{K}$ and $\mathbf{G}_1$
and afterwards $\mathbf{T}$ from $\mathbf{K}_1$ and $\mathbf{G}_2$.
This is especially useful when the split refers to the sum of a principal-value integral $\mathbf{G}_1$ and an onshell
pole contribution $\mathbf{G}_2$, because Eq.~\eqref{bse-2} can then be integrated analytically. This reproduces Eq.~\eqref{K-matrix},
i.e., the $K$-matrix amplitudes $K_l(s)$ are the partial waves corresponding to $\mathbf{K}_1$.
It is also a frequently employed approximation to neglect $\mathbf{G}_1$.

\begin{figure}[!t]
    \vspace{-2mm}
	\centering
	\includegraphics[width=0.7\columnwidth]{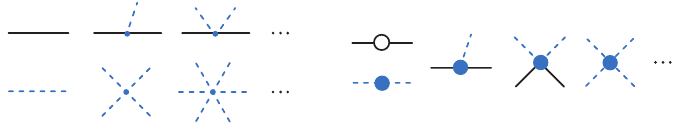}
	\caption{Left: Elementary propagators and interactions in chiral perturbation theory. Solid lines represent nucleons and dashed lines pions.
		Right: Some $n$-point correlation functions.}
	\label{fig:chpt}
	\vspace{-3mm}
\end{figure}

 How is the kernel $\mathbf{K}$ determined, which contains the interactions between the particles and is the input in the equations?
 Since the particles appearing in the loops in Fig.~\eqref{fig:bse} are hadrons,
 these interactions cannot be the ones between quarks and gluons but should rather be derived from an effective Lagrangian
 containing physical fields like nucleons, pions, etc.
 This leads to EFTs, whose prime example is chiral perturbation theory (ChPT)~\cite{Gasser:1983yg,Gasser:1984gg,Weinberg:1978kz}. 
 In the two-flavor case, the ChPT Lagrangian is expressed in terms of pions which  couple to nucleons;
 the extension to three flavors is straightforward. 
 The idea behind ChPT is to be agnostic about the microscopic interactions and include all possible terms compatible with the symmetries of QCD --
 in particular chiral symmetry --
 which leads to a systematic expansion in powers of derivatives and pion masses.
 The elementary Feynman diagrams are shown in Fig.~\ref{fig:chpt} together with the 
 full correlation functions like  the nucleon and pion propagators, 
  $N\pi$ and $\pi\pi$ scattering amplitudes, etc., 
 which can be calculated through perturbative loop expansions.
 The price to pay is that the Lagrangian is non-renormalizable and contains infinitely many terms with infinitely many free parameters.
 However, the derivative couplings ensure that higher loop diagrams in the perturbative expansion are suppressed at low momenta and small pion masses.
 The renormalization can then be done at each  order and requires only a finite number of parameters at each order, the low-energy constants,
 which must be determined from elsewhere, e.g., from experiment or lattice QCD.
 In this way, ChPT is a low-energy EFT and one can make a range of predictions at low momenta and small pion masses,
 e.g. for the nucleon mass as a function of the pion mass, the $N\pi$ scattering amplitude,
 electroweak processes and so on, see~\cite{Meissner:1993ah,Ecker:1994gg,Pich:1995bw,Bernard:1995dp,Scherer:2002tk,Bernard:2006gx,Bijnens:2006zp,Geng:2013xn} for reviews.

On the other hand, because ChPT amounts to perturbative loop expansions it cannot describe bound states and resonances. 
One strategy to deal with this is to include these extra fields ($\rho$ and $\omega$ mesons,  the $\Delta$ resonance,
Roper resonance, $\Lambda(1405)$, etc.) explicitly in the Lagrangian and hence in the perturbative expansions, see e.g.~\cite{Meissner:1987ge,Ecker:1989yg,Jenkins:1991es,Hemmert:1997ye,Lutz:2001yb,Pascalutsa:2002pi,Bernard:2003xf,Kolomeitsev:2003kt,Pascalutsa:2006up,Djukanovic:2009zn,Gegelia:2016xcw} and references therein.
Another strategy is to unitarize ChPT by employing BSEs~\cite{Oller:1997ng,Oller:1997ti,Oller:1998hw,Nieves:1999bx,Bruns:2010sv,Inoue:2001ip}.
A BSE  automatically produces the unitary cuts when the hadrons in the loops  go onshell, but 
because it is genuinely non-perturbative it can also dynamically generate bound states and resonances.
This non-perturbative nature can be understood by  analogy with the geometric series:
The series expansion $f(x) = 1 + x + x^2 + x^3 + \dots$
%\begin{eqnarray}
%	f(x) = 1 + x + x^2 + x^3 + \dots
%\end{eqnarray}
determines the function $f(x) = 1/(1-x)$ only for $|x| < 1$, whereas
the `nonperturbative' equation
\begin{eqnarray}\label{geom-series}
		f(x) = 1 + x f(x) = 1 + x + x^2 f(x)  = 1 + x + x^2 +x^3 f(x) = \dots \qquad \Leftrightarrow \qquad f(x)^{-1} = 1-x
\end{eqnarray}
where $f(x)$ also appears  on the right,
is always exact irrespective of how large $x$ is.
Similarly, for small couplings one can expand Eq.~\eqref{bse-1} in $\mathbf{T} = \mathbf{K} + \mathbf{K}\,\mathbf{G}_0 \,\mathbf{K} + \dots$,
which does not generate any  poles, while the original nonperturbative equation 
generates poles if the condition `$\mathbf{K}\,\mathbf{G}_0 = 1$' is satisfied for  some value of $\sqrt{s}= m_\lambda$.

A shortcoming of the BSE in the form of Eq.~\eqref{bse-1}, which is only iterated in the $s$ channel, 
is that it does not automatically satisfy crossing symmetry.
While it generates the correct $s$-channel cut for $s > 4m^2$ (the unitary or right-hand cut),
a unitarization also in the crossed channels would produce further cuts in the Mandelstam variables $t$ and $u$,
which translates to $s < 0$ and is the so-called left-hand cut. 
For this reason, other unitarization methods based on dispersion relations 
such as the $N/D$ approach~\cite{Chew:1960iv,Oller:1998zr,Meissner:1999vr},
the inverse amplitude method~\cite{Truong:1988zp,Pelaez:2006nj,GomezNicola:2007qj,Pelaez:2010fj} or the Roy-Steiner equations~\cite{Hoferichter:2015hva,Hoferichter:2015tha,RuizdeElvira:2017stg,Hoferichter:2023mgy}
have been frequently employed, see e.g. the reviews~\cite{Pelaez:2010fj,Hoferichter:2015hva,Mai:2022eur} for  details and references.

The application of BSEs with kernels taken from EFTs,
with (multi-)pion and  possibly other hadron exchanges, has found widespread applications in hadron and nuclear physics.
 Although we  exemplified it for the two-body scattering amplitude in a single channel,
 the approach can be generalized to dynamical coupled-channel equations where
 the external particles  create different intermediate states such as  $N\pi$, $\Lambda K$, $\Sigma K$, $N\gamma$, etc.,
 see~\cite{Doring:2025sgb} for a recent review. 
 Coupled-channel models have become indispensable tools for amplitude analyses to extract experimental data,
 especially in view of  extracting the photo- and electrocouplings of light baryons (see also the discussion in Sec.~\ref{PWA-DCC}).
 Examples are the ANL-Osaka approach~\cite{Sato:1996gk,Matsuyama:2006rp,Julia-Diaz:2007qtz,Suzuki:2009nj,Kamano:2013iva},
 the Jülich-Bonn-Washington approach~\cite{Doring:2010ap,Ronchen:2012eg,Ronchen:2014cna,Wang:2022osj,Mai:2021vsw,Wang:2024byt,Doring:2025sgb},
 and various K-matrix formulations like the MAID~\cite{Tiator:2011pw,Drechsel:2007if,Drechsel:1998hk}, SAID~\cite{Arndt:2002xv,Workman:2012jf}, 
 JLab~\cite{Aznauryan:2011qj}, KSU~\cite{Manley:1992yb,Shrestha:2012ep}, Giessen~\cite{Penner:2002ma,Penner:2002md,Shklyar:2006xw,Shklyar:2012js} and Bonn-Gatchina approaches~\cite{Anisovich:2005tf,Anisovich:2010mks,Anisovich:2011fc,Anisovich:2017bsk,Sarantsev:2025lik}.

\begin{figure}[!t]
	\centering
	\includegraphics[width=1\textwidth]{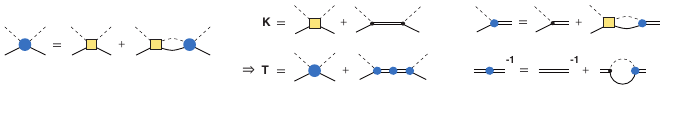}
    \vspace{-14mm}
	\caption{When splitting the kernel of the scattering equation in
    a non-resonant and a resonant part, this leads to the same splitting for the scattering matrix
    and allows one to extract the resonance dressing effects.	   \vspace{-2mm}}
	\label{fig:coupled-channel}
\end{figure}

 Dynamical coupled-channel models also allow one to estimate the meson-baryon dressing effects of nucleon resonances
 and extract their pole trajectories in the complex plane, see e.g.~\cite{Suzuki:2009nj,Guo:2023wes}.
 As an example, the procedure employed in the ANL-Osaka approach is illustrated in Fig.~\ref{fig:coupled-channel} for a single-channel $N\pi$ scattering equation.
 If one  pulls out the resonant $s$-channel $N^\ast$ contributions from the kernel explicitly, 
 then also the T-matrix turns into  the sum of a non-resonant and a resonant part. 
 The former satisfies the scattering equation with the non-resonant kernel, while
  the dressing effects for the resonance coupling and propagator are calculated separately.
  In this way one can study the effects of meson-baryon interactions:
When the full T-matrix is fitted to the electroproduction data, switching off the dressing effects 
for the masses and transition form factors provides an estimate for the `quark core' of the resonance.

A central theme and relevant aspect for our purposes
is the notion of composite hadrons and dynamically generated resonances.
On physical grounds this refers to hadrons that are made of other hadrons,
like the deuteron is a shallow bound state just below the proton-neutron threshold.
In the meson sector, longstanding candidates for hadronic molecules are the light scalar mesons, the $\sigma/f_0(500)$, 
$\kappa/K^*_0(700)$, $a_0(980)$ and $f_0(980)$~\cite{Pelaez:2010fj}.
Especially in the exotic heavy-meson sector there are now several well-established candidates for hadronic molecules~\cite{Guo:2017jvc}.
On technical grounds, the realization of this phenomenon in EFTs is that `pre-existing' hadrons, 
namely those appearing in the effective Lagrangian,
can create new bound states or resonances by the strong channel couplings.
For example, the light scalar mesons arise from the $\pi\pi$, $K\pi$ and $K\bar{K}$ couplings~\cite{Oller:1997ti,Oller:1998hw,Oller:1997ng},
and some states in the baryon sector from $N\pi$, $\Lambda K$, \dots interactions~\cite{Inoue:2001ip,Bruns:2010sv}.
In the light and strange baryon sector, the prime candidates for dynamically generated resonances are the $\Lambda(1405)$, the $N(1440)$ Roper resonance, and several other states~\cite{Meissner:2020khl,Mai:2020ltx,Mai:2022eur,Guo:2023wes,Wang:2023snv}.
For example, the authors of Refs.~\cite{Inoue:2001ip,Bruns:2010sv} argue that also the $N(1535)$ and $N(1650)$
can be  dynamically generated, see e.g.~\cite{Li:2023pjx,Molina:2023jov} for related discussions.

Sometimes it is argued that
if  a baryon resonance is dynamically generated 
it cannot be simultaneously a three-quark state, 
or at least its $qqq$ component should be small compared to $qqqq\bar{q}$ and higher multiquark components.
Although this conclusion seems well  justified in a number of cases,
on  general grounds one should  advocate caution:
Unless some quantum number prevents it,
a baryon resonance must appear in any $qqq$, $qqqq\bar{q}$, \dots correlation function, 
so there \textit{is} no dynamically generated baryon resonance without a three-quark core.
Furthermore, one should also keep in mind that a BSE formulated in hadronic degrees of freedom does not know anything about quarks, 
so  \textit{all it  can do} is create resonances dynamically, at least as long as the interactions allow it. 
For example, the $\pi\pi$, $K\bar{K}$, \dots interactions
generate the light scalar mesons 
but at the same time also the $\rho$ and $K^\ast$, whose $q\bar{q}$ nature is  undisputed.
Unless one can  somehow map the quark-gluon interactions in QCD to the hadronic interactions in EFTs,
these are just two complementary toolboxes.
Statements about compositeness 
require  QCD calculations, e.g., like lattice QCD and functional methods 
where one can calculate the overlaps with $(qqq)$, $(qqq)(q\bar{q})$, \dots operators dynamically, or 
additional information such as the proximity to thresholds, decay patterns, line shapes, pole trajectories,
large $N_c$ arguments, Regge phenomenology or unusual properties of elastic and transition form factors,
see~\cite{Pelaez:2010fj,Guo:2017jvc,Mai:2022eur} and references therein.

Finally, a fruitful strategy which has gained traction over the last years is to combine EFTs with lattice QCD.
Here the term `EFT' does not necessarily refer to any specific effective theory derived from a Lagrangian,
but rather to the general concept of employing hadronic scattering equations to satisfy unitarity.
For example, this is needed to map finite-volume lattice data
to  infinite-volume scattering amplitudes in order to describe states above  decay thresholds via the Lüscher method~\cite{Luscher:1986pf,Luscher:1990ux}. 
Similarly, one can use BSEs to analytically continue  lattice results 
for  scattering amplitudes into the complex plane to determine the resonance pole locations.
These aspects are further discussed in Sec.~\ref{sec:lattice}.

\subsection{Quark models}
\label{pheno}

Let us now adopt a different perspective and discuss baryon spectroscopy from microscopic approaches using quark and gluon degrees of freedom.
The baseline for this are non-relativistic and relativistic constituent quark models, see~\cite{Hey:1982aj,Capstick:2000qj,Klempt:2009pi,Richard:2012xw,Crede:2013kia,Giannini:2015zia,Thiel:2022xtb,Gross:2022hyw} for reviews.
The  idea is to write down  
a quantum-mechanical Hamiltonian for a system of $n$ valence quarks and/or antiquarks,
in the non-relativistic version\\[-1ex]
\begin{eqnarray} 
  H = \sum_{i=1}^n \left(m_i + \frac{\vect{p}_i^2}{2m_i}\right) + \sum_{i<j} V(\vect{r}_{ij})  \,,  
\end{eqnarray}
where the quarks carry constituent masses $m_i$ and
the potential is the sum of pairwise interactions between the quarks with $\vect{r}_{ij} = \vect{r}_{i} - \vect{r}_{j}$.
The solution of the Schrödinger equation $H\Psi_\lambda = E_\lambda\Psi_\lambda$
%\begin{eqnarray}\label{SGE}
%	H\Psi_\lambda = E_\lambda\Psi_\lambda  
%\end{eqnarray}
then determines the wave function $\Psi_\lambda$ of a hadron $|\lambda\rangle$ and its mass $E_\lambda = m_\lambda$ in the rest frame.  
The  wave functions admit a direct probability interpretation through their modulus,
and matrix elements are calculated via expectation values of operators with those wave functions.

Although we already mentioned some concrete models in Sec.~\ref{Models},
%Before turning to  concrete models, 
let us first discuss a few general features of the interquark potential $V(\vect{r}_{ij})$.
%Although there is no unique construction, 
A useful starting point~\cite{DeRujula:1975qlm} is the Breit-Fermi interaction, i.e.,
the nonrelativistic expansion of the one-gluon exchange amplitude between quark and (anti)quark in QCD  
analogous to the one-photon exchange  between electron and proton in the hydrogen atom:\\[-1ex]
\begin{eqnarray}\label{breit-fermi}
	V(\vect{r}) = C \alpha_s \left[ -\frac{1}{r} + V_\text{ss}(\vect{r}) + V_\text{so}(\vect{r}) + \dots\right].
\end{eqnarray}
Besides the Coulomb  and other spin-independent terms, it contains spin-dependent forces:
The hyperfine potential $V_\text{ss}(\vect{r})$  arises from the color-magnetic interaction and
is responsible for the dominant mass splittings between states with different total quark spin $S$, e.g.
between the pseudoscalar ($S=0$) and vector mesons ($S=1$) or the ground-state octet ($S=\nicefrac{1}{2}$) and decuplet ($S=\nicefrac{3}{2}$) baryons.
The spin-orbit interaction $V_\text{so}(\vect{r})$ does not contribute to orbital ground states ($L=0$) but gives rise to splittings for states
with different $J$, e.g. for the charmonium states $\left\{ \chi_{c0}, \chi_{c1}, \chi_{c2} \right\}$ with $J^{PC} = \{0, 1, 2\}^{++}$.
%For baryons, spin-orbit interactions are often neglected to improve the fits for the spectrum.  
By contrast, in the hydrogen atom the spin-orbit force beats the spin-spin interaction (due to $m_e \ll m_p$) and gives rise
to the atomic fine structure effects.

The color factor $C$ in Eq.~\eqref{breit-fermi} depends on the $SU(3)_c$ representation of the $q\bar{q}$ or $qq$ system, with
$C=4/3$ for the singlet representation $\mathbf{1}$, $C=2/3$ for the (anti-) triplet representations $\mathbf{3}$ and $\overline{\mathbf{3}}$,
and $C < 0$ for   higher-dimensional representations ($\mathbf{6}$, $\overline{\mathbf{6}}$, $\mathbf{8}$, \dots).
Thus, the interaction is maximally attractive for color-singlet mesons,
whereas for two quarks with $\mathbf{3}\otimes \mathbf{3} = \overline{\mathbf{3}} \oplus \mathbf{6}$
it is  attractive in the color-antitriplet diquark channel but repulsive in the sextet channel.
This also underlines the role of diquarks in the baryon spectrum: with $\mathbf{3}\otimes \mathbf{3}\otimes \mathbf{3} = (\overline{\mathbf{3}} \oplus \mathbf{6})\otimes \mathbf{3}$,
color-singlet baryons can be obtained by combining quarks with antitriplet diquarks.

Quark confinement is usually implemented by adding 
a linearly rising potential $V_\text{conf} = \sigma r$ by hand to Eq.~\eqref{breit-fermi}, where $\sigma$ is the string tension.
This is motivated by Regge phenomenology and the observed mass spectra 
as well as lattice calculations of the Wilson loop for infinitely heavy quarks~\cite{Bali:2000gf}.
The dynamical origin of confinement is still under debate and several (not mutually exclusive)
mechanisms have been proposed, e.g.,
center vortices in the path integral~\cite{Greensite:2003bk}, 
ghost dominance in the infrared~\cite{Alkofer:2000wg},  
%the emergence of a gluon mass gap~\cite{a},
or the formation of color-electric flux tubes
and condensation of color-magnetic monopoles in the dual superconductor picture~\cite{Kondo:2014sta}. 
A hypothesis with interesting consequences for the electroweak sector
is that confinement may be nothing more than gauge invariance~\cite{Maas:2017wzi}: Quarks and gluons depend on the
gauge, but experimentally we can only measure gauge-invariant hadrons.

\begin{figure*}[!t]
\setlength{\unitlength}{0.45mm}
\linethickness{0.3mm}
\begin{picture}(450.00,180.00)
\put(20.00,00.00){\line(1,0){05.00}}
\put(20.00,170.00){\line(1,0){05.00}}
\multiput(20,0)(40,0){10}{\line(0,1){170}} 
\multiput(65,0)(40,0){8}{\line(0,1){170}} 
\multiput(25,0)(40,0){9}{\line(1,0){35}} 
\multiput(25,170)(40,0){9}{\line(1,0){35}} 
\multiput(25,0)(40,0){9}{\line(1,0){35}} 
\multiput(25,170)(40,0){9}{\line(1,0){35}} 

\multiput(20,-25)(40,0){10}{\line(0,1){20}} 
\multiput(65,-25)(40,0){8}{\line(0,1){20}} 
\multiput(25,-5)(40,0){9}{\line(1,0){35}} 
\multiput(25,-25)(40,0){9}{\line(1,0){35}} 
\put(20.00,-05.00){\line(1,0){05.00}}
\put(20.00,-25.00){\line(1,0){05.00}}
\put(57.00,160.00){\line(1,0){3.00}}
\put(57.00,150.00){\line(1,0){3.00}}
\put(57.00,140.00){\line(1,0){3.00}}
\put(55.00,130.00){\line(1,0){5.00}}
\put(35.00,130.00){\makebox(10.00,2.00)[l]{\bf\large\boldmath 2.0}}
\put(35.00,80.00){\makebox(10.00,2.00)[l]{\bf\large\boldmath 1.5}}
\put(35.00,30.00){\makebox(10.00,2.00)[l]{\bf\large\boldmath 1.0}}
\put(55.00,080.00){\line(1,0){5.00}}
\put(55.00,030.00){\line(1,0){5.00}}
\put(57.00,120.00){\line(1,0){3.00}}
\put(57.00,110.00){\line(1,0){3.00}}
\put(57.00,100.00){\line(1,0){3.00}}
\put(57.00,090.00){\line(1,0){3.00}}
\put(57.00,070.00){\line(1,0){3.00}}
\put(57.00,060.00){\line(1,0){3.00}}
\put(57.00,050.00){\line(1,0){3.00}}
\put(57.00,040.00){\line(1,0){3.00}}
\put(57.00,020.00){\line(1,0){3.00}}
\put(57.00,010.00){\line(1,0){3.00}}
\put(26.00,155.00){\makebox(10.00,2.00)[l]{\bf\large\boldmath GeV}}
\put(26.00,-15.00){\makebox(10.00,2.00)[l]{\bf\large\boldmath\hspace{2mm} $J^P$}}
\put(73.00,-15.00){\makebox(10.00,2.00)[l]{\bf\large\boldmath $N\frac12^+$}}
\put(113.00,-15.00){\makebox(10.00,2.00)[l]{\bf\large\boldmath $N\frac32^+$}}
\put(153.00,-15.00){\makebox(10.00,2.00)[l]{\bf\large\boldmath $N\frac52^+$}}
\put(193.00,-15.00){\makebox(10.00,2.00)[l]{\bf\large\boldmath $N\frac72^+$}}
\put(233.00,-15.00){\makebox(10.00,2.00)[l]{\bf\large\boldmath $N\frac12^-$}}
\put(273.00,-15.00){\makebox(10.00,2.00)[l]{\bf\large\boldmath $N\frac32^-$}}
\put(313.00,-15.00){\makebox(10.00,2.00)[l]{\bf\large\boldmath $N\frac52^-$}}
\put(353.00,-15.00){\makebox(10.00,2.00)[l]{\bf\large\boldmath $N\frac72^-$}}
\linethickness{0.6mm}

                                                                                % Data on N^* begin here:

                                                        %Capstick and Isgur Capstick:1985xss
                                                        %1/2^+: 
\put(66.00,26.00){\bl\line(1,0){5.00}}
\put(66.00,84.00){\bl\line(1,0){5.00}}
\put(66.00,107.00){\bl\line(1,0){5.00}}
\put(66.00,118.00){\bl\line(1,0){5.00}}
\put(66.00,127.50){\bl\line(1,0){5.00}}
\put(66.00,136.50){\color{bl2}\line(1,0){5.00}}
%Capstick and Isgur 3/2^+: Capstick:1985xss
\put(106.00,109.50){\bl\line(1,0){5.00}}
\put(106.00,117.00){\bl\line(1,0){5.00}}
\put(106.00,121.00){\bl\line(1,0){5.00}}
\put(106.00,125.00){\bl\line(1,0){5.00}}
\put(106.00,133.50){\bl\line(1,0){5.00}}
%Capstick and Isgur 5/2^+: Capstick:1985xss
\put(146.00,107.00){\bl\line(1,0){5.00}}
\put(146.00,128.00){\bl\line(1,0){5.00}}
%Capstick and Isgur 7/2^+: Capstick:1985xss
\put(186.00,130.00){\bl\line(1,0){5.00}}

%Glozman, Plessas, Vargas 1/2^+: Glozman:1997ag
\put(73.00,24.00){\mg\line(1,0){5.00}}
\put(73.00,76.00){\mg\line(1,0){5.00}}
\put(73.00,108.00){\mg\line(1,0){5.00}}
%Glozman, Plessas, Vargas 3/2^+: Glozman:1997ag
\put(113.00,102.50){\mg\line(1,0){5.00}}
%Glozman, Plessas, Vargas 5/2^+: Glozman:1997ag
\put(153.00,103.50){\mg\line(1,0){5.00}}

%Giannini, Santopinto, Vassallo 1/2^+: Giannini:2001kb
\put(80.00,24.00){\br\line(1,0){5.00}}
\put(80.00,76.30){\br\line(1,0){5.00}}
\put(80.00,105.20){\br\line(1,0){5.00}}
\put(80.00,112.80){\br\line(1,0){5.00}}
\put(80.00,119.40){\br\line(1,0){5.00}}
\put(80.00,123.80){\color{br2}\line(1,0){5.00}}
%Giannini, Santopinto, Vassallo 3/2^+: Giannini:2001kb
\put(120.00,94.80){\br\line(1,0){5.00}}
\put(120.00,111.60){\br\line(1,0){5.00}}
\put(120.00,119.40){\br\line(1,0){5.00}}
\put(120.00,123.90){\br\line(1,0){5.00}}
\put(120.00,133.40){\br\line(1,0){5.00}}
%Giannini, Santopinto, Vassallo 5/2^+: Giannini:2001kb
\put(160.00,98.00){\br\line(1,0){5.00}}
\put(160.00,113.30){\br\line(1,0){5.00}}
\put(160.00,134.60){\br\line(1,0){5.00}}
%Giannini, Santopinto, Vassallo 7/2^+: Giannini:2001kb
\put(200.00,123.90){\br\line(1,0){5.00}}

%Löring, Metsch, Petry 1/2^+: Loring:2001kx
\put(87.00,24.00){\line(1,0){5.00}}              
\put(87.00,24.00){\line(1,0){5.00}}
\put(87.00,81.80){\line(1,0){5.00}}
\put(87.00,102.90){\line(1,0){5.00}}
\put(87.00,125.00){\line(1,0){5.00}}
\put(87.00,126.60){\line(1,0){5.00}}
\put(87.00,130.90){\color{grey}\line(1,0){5.00}}
\put(87.00,147.40){\color{grey}\line(1,0){5.00}}
\put(87.00,149.90){\color{grey}\line(1,0){5.00}}
\put(87.00,166.10){\color{grey}\line(1,0){5.00}}
                                       %Löring, Metsch, Petry 3/2^+: Loring:2001kx
\put(127.00,98.80){\line(1,0){5.00}}
\put(127.00,119.90){\line(1,0){5.00}}
\put(127.00,123.60){\line(1,0){5.00}}
\put(127.00,126.90){\line(1,0){5.00}}
\put(127.00,131.30){\line(1,0){5.00}}
\put(127.00,141.70){\color{grey}\line(1,0){5.00}}
\put(127.00,156.90){\color{grey}\line(1,0){5.00}}
\put(127.00,161.20){\color{grey}\line(1,0){5.00}}
%Löring, Metsch, Petry 5/2^+: Loring:2001kx
\put(167.00,102.30){\line(1,0){5.00}}
\put(167.00,123.40){\line(1,0){5.00}}
\put(167.00,125.90){\line(1,0){5.00}}
\put(167.00,142.00){\color{grey}\line(1,0){5.00}}
\put(167.00,159.60){\color{grey}\line(1,0){5.00}}
\put(167.00,164.40){\color{grey}\line(1,0){5.00}}
%Löring, Metsch, Petry 7/2^+: Loring:2001kx
\put(207.00,128.90){\line(1,0){6.00}}
\put(207.00,149.00){\color{grey}\line(1,0){5.00}}
\put(207.00,166.50){\color{grey}\line(1,0){5.00}}
\put(207.00,169.90){\color{grey}\line(1,0){5.00}}

%Capstick and Isgur 1/2^-: Capstick:1985xss
\put(226.00,76.00){\bl\line(1,0){5.00}}
\put(226.00,83.50){\bl\line(1,0){5.00}}
\put(226.00,124.50){\color{bl2}\line(1,0){5.00}}
\put(226.00,133.00){\color{bl2}\line(1,0){5.00}}
\put(226.00,137.00){\color{bl2}\line(1,0){5.00}}
\put(226.00,144.50){\color{bl2}\line(1,0){5.00}}
\put(226.00,149.50){\color{bl2}\line(1,0){5.00}}
%Capstick and Isgur 3/2^-: Capstick:1985xss
\put(266.00,79.50){\bl\line(1,0){5.00}}
\put(266.00,92.50){\bl\line(1,0){5.00}}
\put(266.00,126.00){\color{bl2}\line(1,0){5.00}}
\put(266.00,135.50){\color{bl2}\line(1,0){5.00}}
\put(266.00,139.50){\color{bl2}\line(1,0){5.00}}
\put(266.00,146.50){\color{bl2}\line(1,0){5.00}}
\put(266.00,148.00){\color{bl2}\line(1,0){5.00}}
%Capstick and Isgur 5/2^-: Capstick:1985xss
\put(306.00,93.00){\bl\line(1,0){5.00}}
\put(306.00,138.00){\color{bl2}\line(1,0){5.00}}
\put(306.00,139.50){\color{bl2}\line(1,0){5.00}}
\put(306.00,148.00){\color{bl2}\line(1,0){5.00}}
%Capstick and Isgur 7/2^-: Capstick:1985xss
\put(346.00,139.00){\color{bl2}\line(1,0){5.00}}

%Glozman, Plessas, Vargas 1/2^-: Glozman:1997ag
\put(233.00,81.50){\mg\line(1,0){5.00}}
\put(233.00,95.00){\mg\line(1,0){5.00}}
%Glozman, Plessas, Vargas 3/2^-: Glozman:1997ag
\put(273.00,81.50){\mg\line(1,0){5.00}}
\put(273.00,95.00){\mg\line(1,0){5.00}}
%Glozman, Plessas, Vargas 5/2^-: Glozman:1997ag
\put(313.00,95.00){\mg\line(1,0){5.00}}

%Giannini, Santopinto, Vassallo 1/2^-: Giannini:2001kb
\put(240.00,82.40){\br\line(1,0){5.00}}
\put(240.00,98.80){\br\line(1,0){5.00}}
\put(240.00,116.10){\color{br2}\line(1,0){5.00}}
\put(240.00,130.80){\color{br2}\line(1,0){5.00}}
%Giannini, Santopinto, Vassallo 3/2^-: Giannini:2001kb
\put(280.00,82.40){\br\line(1,0){5.00}}
\put(280.00,99.20){\br\line(1,0){5.00}}
\put(280.00,116.00){\color{br2}\line(1,0){5.00}}
\put(280.00,130.80){\color{br2}\line(1,0){5.00}}
%Giannini, Santopinto, Vassallo 5/2^-: Giannini:2001kb
\put(320.00,96.80){\br\line(1,0){5.00}}
\put(320.00,128.40){\color{br2}\line(1,0){5.00}}

%Löring, Metsch, Petry 1/2^-: Loring:2001kx
\put(247.00,73.50){\line(1,0){6.00}}
\put(247.00,96.00){\line(1,0){6.00}}
\put(247.00,120.10){\color{grey}\line(1,0){5.00}}
\put(247.00,121.80){\color{grey}\line(1,0){5.00}}
\put(247.00,145.30){\color{grey}\line(1,0){5.00}}
\put(247.00,148.50){\color{grey}\line(1,0){5.00}}
\put(247.00,148.40){\color{grey}\line(1,0){5.00}}
\put(247.00,153.20){\color{grey}\line(1,0){5.00}}
\put(247.00,154.20){\color{grey}\line(1,0){5.00}}
%Löring, Metsch, Petry 3/2^-: Loring:2001kx
\put(287.00,77.60){\line(1,0){6.00}}
\put(287.00,90.60){\line(1,0){6.00}}
\put(287.00,122.60){\color{grey}\line(1,0){5.00}}
\put(287.00,137.90){\color{grey}\line(1,0){5.00}}
\put(287.00,144.30){\color{grey}\line(1,0){5.00}}
\put(287.00,147.70){\color{grey}\line(1,0){5.00}}
\put(287.00,150.30){\color{grey}\line(1,0){5.00}}
\put(287.00,154.70){\color{grey}\line(1,0){5.00}}
%Löring, Metsch, Petry 5/2^-: Loring:2001kx
\put(327.00,95.50){\line(1,0){6.00}}
\put(327.00,127.00){\color{grey}\line(1,0){5.00}}
\put(327.00,140.40){\color{grey}\line(1,0){5.00}}
\put(327.00,144.70){\color{grey}\line(1,0){5.00}}
\put(327.00,151.70){\color{grey}\line(1,0){5.00}}
\put(327.00,152.50){\color{grey}\line(1,0){5.00}}
\put(327.00,153.30){\color{grey}\line(1,0){5.00}}
\put(327.00,160.30){\color{grey}\line(1,0){5.00}}
%Löring, Metsch, Petry 7/2^-: Loring:2001kx
\put(367.00,131.50){\color{grey}\line(1,0){5.00}}
\put(367.00,147.10){\color{grey}\line(1,0){5.00}}
\put(367.00,152.90){\color{grey}\line(1,0){5.00}}
\put(367.00,154.70){\color{grey}\line(1,0){5.00}}
\put(367.00,157.90){\color{grey}\line(1,0){5.00}}

                                                      %Ronninger and Metsch  Ronniger:2011td   
                                                      %1/2^+
\put(94.00,24.50){\gr\line(1,0){5.00}} 
\put(94.00,74.00){\gr\line(1,0){5.00}} 
\put(94.00,100.90){\gr\line(1,0){5.00}} 
\put(94.00,119.30){\gr\line(1,0){5.00}} 
\put(94.00,126.70){\gr\line(1,0){5.00}} 
\put(94.00,136.30){\grl\line(1,0){5.00}} 
\put(94.00,139.00){\grl\line(1,0){5.00}} 
\put(94.00,149.00){\grl\line(1,0){5.00}} 
\put(94.00,158.00){\grl\line(1,0){5.00}} 
                                                      %3/2^+
\put(134.00,100.30){\gr\line(1,0){5.00}} 
\put(134.00,112.50){\gr\line(1,0){5.00}} 
\put(134.00,124.50){\gr\line(1,0){5.00}} 
\put(134.00,126.60){\gr\line(1,0){5.00}} 
\put(134.00,134.90){\gr\line(1,0){5.00}} 
\put(134.00,136.10){\grl\line(1,0){5.00}} 
\put(134.00,148.00){\grl\line(1,0){5.00}} 
\put(134.00,152.50){\grl\line(1,0){5.00}} 
\put(134.00,156.00){\grl\line(1,0){5.00}} 
\put(134.00,158.80){\grl\line(1,0){5.00}} 
                                                      %5/2^+
\put(174.00,106.10){\gr\line(1,0){5.00}} 
\put(174.00,123.00){\gr\line(1,0){5.00}} 
\put(174.00,128.30){\grl\line(1,0){5.00}} 
\put(174.00,138.90){\grl\line(1,0){5.00}} 
\put(174.00,154.20){\grl\line(1,0){5.00}} 
\put(174.00,155.90){\grl\line(1,0){5.00}} 
\put(174.00,163.50){\grl\line(1,0){5.00}} 
\put(174.00,165.50){\grl\line(1,0){5.00}} 
                                                      %7/2^+
\put(214.00,130.10){\gr\line(1,0){5.00}} 
\put(214.00,143.90){\grl\line(1,0){5.00}} 
\put(214.00,160.60){\grl\line(1,0){5.00}} 
\put(214.00,163.60){\grl\line(1,0){5.00}} 
\put(214.00,164.30){\grl\line(1,0){5.00}} 
\put(214.00,166.50){\grl\line(1,0){5.00}} 
                                                      %1/2^-
\put(254.00,78.40){\gr\line(1,0){5.00}} 
\put(254.00,97.20){\gr\line(1,0){5.00}} 
\put(254.00,115.10){\grl\line(1,0){5.00}} 
\put(254.00,118.10){\grl\line(1,0){5.00}} 
\put(254.00,138.90){\grl\line(1,0){5.00}} 
\put(254.00,144.80){\grl\line(1,0){5.00}} 
\put(254.00,149.20){\grl\line(1,0){5.00}} 
\put(254.00,151.90){\grl\line(1,0){5.00}} 
\put(254.00,154.80){\grl\line(1,0){5.00}} 
\put(254.00,155.30){\grl\line(1,0){5.00}} 
                                                       %3/2^-
\put(294.00,83.40){\gr\line(1,0){5.00}} 
\put(294.00,98.50){\gr\line(1,0){5.00}} 
\put(294.00,115.30){\grl\line(1,0){5.00}} 
\put(294.00,123.40){\grl\line(1,0){5.00}} 
\put(294.00,137.30){\grl\line(1,0){5.00}} 
\put(294.00,139.10){\grl\line(1,0){5.00}} 
\put(294.00,143.70){\grl\line(1,0){5.00}} 
\put(294.00,148.00){\grl\line(1,0){5.00}} 
\put(294.00,151.00){\grl\line(1,0){5.00}} 
\put(294.00,152.00){\grl\line(1,0){5.00}} 

                                                     %5/2^-
\put(334.00,96.70){\gr\line(1,0){5.00}} 
\put(334.00,124.50){\grl\line(1,0){5.00}} 
\put(334.00,130.70){\grl\line(1,0){5.00}} 
\put(334.00,141.60){\grl\line(1,0){5.00}} 
\put(334.00,146.70){\grl\line(1,0){5.00}} 
\put(334.00,149.60){\grl\line(1,0){5.00}} 
\put(334.00,155.30){\grl\line(1,0){5.00}} 
\put(334.00,157.30){\grl\line(1,0){5.00}} 
\put(334.00,159.20){\grl\line(1,0){5.00}} 
\put(334.00,163.20){\grl\line(1,0){5.00}} 
                                                     %7/2^-
\put(374.00,131.10){\gr\line(1,0){5.00}} 
\put(374.00,143.00){\grl\line(1,0){5.00}} 
\put(374.00,148.40){\grl\line(1,0){5.00}} 
\put(374.00,156.90){\grl\line(1,0){5.00}} 
\put(374.00,162.00){\grl\line(1,0){5.00}} 
\put(374.00,163.80){\grl\line(1,0){5.00}}

                                                    %data 1/2^+
\linethickness{0.2mm}
\put(66.00,24.00){\line(1,0){34.00}}
\put(66,71.0){\framebox(33,6)[*]{}}
\put(66,98.0){\framebox(33,6)[*]{}}
\put(66,113.0){\framebox(33,10)[*]{}}
\put(66,135.0){\framebox(33,10)[*]{}}
\put(66,157.0){\cy\framebox(33,13)[*]{}}
%data 3/2^+
\put(106,98.0){\framebox(33,7)[*]{}}
\put(107,119.0){\framebox(31,6)[*]{}}
\put(107,131.5){\cy\framebox(31,5)[*]{}}
%data 5/2^+
\put(146,98.0){\framebox(33,1.0)[*]{}}
\put(146,122.0){\rd\framebox(33,5)[*]{}}
\put(146,131.0){\rd\framebox(33,8)[*]{}}
%data 7/2^+
\put(186,125.0){\framebox(33,7.0)[*]{}}
%data 1/2^-
\put(226,81.5){\framebox(33,3)[*]{}}
\put(226,93.5){\framebox(33,3)[*]{}}
\put(226,117.0){\framebox(33,5.0)[*]{}}
%data 3/2^-
\put(266,81.0){\framebox(33,1)[*]{}}
\put(266,95.0){\framebox(33,10.0)[*]{}}
\put(266,115.0){\framebox(33,7.0)[*]{}}
\put(266,136.0){\framebox(33,10.0)[*]{}}
%data 5/2^-
\put(306,96.5){\framebox(33,1.5)[*]{}}
\put(306,133.0){\framebox(33,17.0)[*]{}}
%data 7/2^-
\put(346,144.0){\framebox(33,8.0)[*]{}}
\end{picture}
\vspace{10mm}

%\end{figure*}
%\begin{figure*}[t]
\setlength{\unitlength}{0.45mm}
\linethickness{0.3mm}
%begin frame
\begin{picture}(450.00,180.00)
\put(20.00,00.00){\line(1,0){05.00}}
\put(20.00,170.00){\line(1,0){05.00}}
\multiput(20,0)(40,0){10}{\line(0,1){170}} 
\multiput(65,0)(40,0){8}{\line(0,1){170}} 
\multiput(25,0)(40,0){9}{\line(1,0){35}} 
\multiput(25,170)(40,0){9}{\line(1,0){35}} 
\multiput(25,0)(40,0){9}{\line(1,0){35}} 
\multiput(25,170)(40,0){9}{\line(1,0){35}} 

\multiput(20,-25)(40,0){10}{\line(0,1){20}} 
\multiput(65,-25)(40,0){8}{\line(0,1){20}} 
\multiput(25,-5)(40,0){9}{\line(1,0){35}} 
\multiput(25,-25)(40,0){9}{\line(1,0){35}} 
\put(20.00,-05.00){\line(1,0){05.00}}
\put(20.00,-25.00){\line(1,0){05.00}}
\put(57.00,160.00){\line(1,0){3.00}}
\put(57.00,150.00){\line(1,0){3.00}}
\put(57.00,140.00){\line(1,0){3.00}}
\put(55.00,130.00){\line(1,0){5.00}}
\put(35.00,130.00){\makebox(10.00,2.00)[l]{\bf\large\boldmath 2.0}}
\put(35.00,80.00){\makebox(10.00,2.00)[l]{\bf\large\boldmath 1.5}}
\put(35.00,30.00){\makebox(10.00,2.00)[l]{\bf\large\boldmath 1.0}}
\put(55.00,080.00){\line(1,0){5.00}}
\put(55.00,030.00){\line(1,0){5.00}}
\put(57.00,120.00){\line(1,0){3.00}}
\put(57.00,110.00){\line(1,0){3.00}}
\put(57.00,100.00){\line(1,0){3.00}}
\put(57.00,090.00){\line(1,0){3.00}}
\put(57.00,070.00){\line(1,0){3.00}}
\put(57.00,060.00){\line(1,0){3.00}}
\put(57.00,050.00){\line(1,0){3.00}}
\put(57.00,040.00){\line(1,0){3.00}}
\put(57.00,020.00){\line(1,0){3.00}}
\put(57.00,010.00){\line(1,0){3.00}}
\put(26.00,155.00){\makebox(10.00,2.00)[l]{\bf\large\boldmath GeV}}
\put(26.00,-15.00){\makebox(10.00,2.00)[l]{\bf\large\boldmath\hspace{2mm} $J^P$}}
\put(73.00,-15.00){\makebox(10.00,2.00)[l]{\bf\large\boldmath $\Delta\frac12^+$}}
\put(113.00,-15.00){\makebox(10.00,2.00)[l]{\bf\large\boldmath $\Delta\frac32^+$}}
\put(153.00,-15.00){\makebox(10.00,2.00)[l]{\bf\large\boldmath $\Delta\frac52^+$}}
\put(193.00,-15.00){\makebox(10.00,2.00)[l]{\bf\large\boldmath $\Delta\frac72^+$}}
\put(233.00,-15.00){\makebox(10.00,2.00)[l]{\bf\large\boldmath $\Delta\frac12^-$}}
\put(273.00,-15.00){\makebox(10.00,2.00)[l]{\bf\large\boldmath $\Delta\frac32^-$}}
\put(313.00,-15.00){\makebox(10.00,2.00)[l]{\bf\large\boldmath $\Delta\frac52^-$}}
\put(353.00,-15.00){\makebox(10.00,2.00)[l]{\bf\large\boldmath $\Delta\frac72^-$}}
%end frame

 \linethickness{0.6mm}
                                                    % Data on Delta^* begin here:
%%%%%%%%%%%%%%%%%%%%%%%%%%%%%%%%%%%%%%%%                                                     
                                                     %Capstick and Isgur Capstick:1985xss
                                                     % 1/2^+:         
\put(66.00,113.50){\color{bl2}\line(1,0){5.00}}
\put(66.00,117.50){\bl\line(1,0){5.00}}
                                                     % 3/2^+:
\put(106.00,53.20){\bl\line(1,0){5.00}}
\put(106.00,109.50){\bl\line(1,0){5.00}}
\put(106.00,121.50){\bl\line(1,0){5.00}}
\put(106.00,128.50){\bl\line(1,0){5.00}}
                                                     %5/2^+: 
\put(146.00,121.00){\bl\line(1,0){5.00}}
\put(146.00,129.00){\bl\line(1,0){5.00}}
                                                     % 7/2^+:  
\put(186.00,124.00){\bl\line(1,0){5.00}}

                                                     %Bijker, Iachello,Leviatan,  Bijker:1994yr
                                                     %1/2^+
\put(73.00,108.60){\mg\line(1,0){5.00}}
\put(73.00,120.90){\mg\line(1,0){5.00}}
                                                
                                                     %3/2^+:
\put(113.00,53.20){\mg\line(1,0){5.00}}
\put(113.00,94.90){\mg\line(1,0){5.00}}
\put(113.00,120.90){\mg\line(1,0){5.00}}
\put(113.00,124.50){\mg\line(1,0){5.00}}
                                                     %5/2^+:
  \put(153.00,120.90){\mg\line(1,0){5.00}}
  \put(153.00,124.50){\mg\line(1,0){5.00}}
                                                     %7/2^+:
  \put(193.00,120.90){\mg\line(1,0){5.00}}
                                                       %Giannini, Santopinto, Vassallo Giannini:2001kb
                                                     %1/2^+: 
\put(80.00,120.00){\br\line(1,0){5.00}}
\put(80.00,125.30){\br\line(1,0){5.00}}
                                                     % 3/2^+: 
\put(120.00,53.20){\br\line(1,0){5.00}}
\put(120.00,102.70){\br\line(1,0){5.00}} 
\put(120.00,122.10){\br\line(1,0){5.00}}
\put(120.00,122.50){\br\line(1,0){5.00}}
\put(120.00,133.49){\br\line(1,0){5.00}}
                                                     % 5/2^+: 
\put(160.00,120.10){\br\line(1,0){5.00}}
\put(160.00,125.60){\br\line(1,0){5.00}}
                                                     % 7/2^+: 
\put(200.00,125.50){\br\line(1,0){5.00}}

                                                     %Löring, Metsch, Petry Loring:2001kx
                                                     %1/2^+: 
\put(87.00,116.60){\line(1,0){5.00}}           
\put(87.00,120.60){\line(1,0){5.00}}
\put(87.00,157.80){\color{grey}\line(1,0){5.00}}
\put(87.00,166.50){\color{grey}\line(1,0){5.00}}

                                                     %3/2^+: 
\put(127.00,56.00){\line(1,0){5.00}}                          
\put(127.00,111.00){\line(1,0){5.00}}
\put(127.00,117.10){\line(1,0){5.00}}
\put(127.00,125.00){\line(1,0){5.00}}
\put(127.00,155.00){\color{grey}\line(1,0){5.00}}
\put(127.00,158.20){\color{grey}\line(1,0){5.00}}
\put(127.00,164.20){\color{grey}\line(1,0){5.00}}
                                                      % 5/2^+:
\put(167.00,119.70){\line(1,0){5.00}}
\put(167.00,128.50){\line(1,0){5.00}}
\put(167.00,160.50){\color{grey}\line(1,0){5.00}}
\put(167.00,165.80){\color{grey}\line(1,0){5.00}}
                                                     % 7/2^+:
\put(207.00,125.60){\line(1,0){5.00}}
\put(207.00,149.00){\color{grey}\line(1,0){5.00}}
\put(207.00,164.00){\color{grey}\line(1,0){5.00}}

                                                       %Capstick and Isgur  Capstick:1985xss
                                                       %1/2^-:
\put(226.00,85.50){\bl\line(1,0){5.00}}
\put(226.00,133.50){\color{bl2}\line(1,0){5.00}}
\put(226.00,144.00){\color{bl2}\line(1,0){5.00}}
                                                       % 3/2^-: 
\put(266.00,92.00){\bl\line(1,0){5.00}}
\put(266.00,128.00){\color{bl2}\line(1,0){5.00}}
\put(266.00,144.50){\color{bl2}\line(1,0){5.00}}
                                                      % 5/2^-: 
\put(306.00,145.50){\color{bl2}\line(1,0){5.00}}
\put(306.00,146.50){\color{bl2}\line(1,0){5.00}}

                                                      %Bijker, Iachello,Leviatan,  Bijker:1994yr

                                                     %1/2^-: 
\put(233.00,94.90){\mg\line(1,0){5.00}} 
\put(233.00,127.70){\mg\line(1,0){5.00}} 
                                                     % 3/2^-: 
\put(273.00,94.90){\mg\line(1,0){5.00}}
\put(273.00,124.50){\mg\line(1,0){5.00}}
\put(273.00,127.70){\mg\line(1,0){5.00}} 
                                                     % 5/2^-: 
\put(313.00,124.50){\mg\line(1,0){5.00}}

                                                     %Giannini, Santopinto, Vassallo Giannini:2001kb
                                                     %1/2^-: 
\put(240.00,87.30){\br\line(1,0){5.00}}
\put(240.00,121.00){\color{br2}\line(1,0){5.00}}
                                                      % 3/2^-: 
\put(280.00,87.30){\br\line(1,0){5.00}}
\put(280.00,120.00){\color{br2}\line(1,0){5.00}}

                                                      %Löring, Metsch, Petry Loring:2001kx
                                                      %1/2^-: 
\put(247.00,95.40){\line(1,0){5.00}}
\put(247.00,140.00){\color{grey}\line(1,0){5.00}}
\put(247.00,144.10){\color{grey}\line(1,0){5.00}}
\put(247.00,150.10){\color{grey}\line(1,0){5.00}}
                                                      % 3/2^-:
\put(287.00,92.80){\line(1,0){5.00}}
\put(287.00,138.90){\color{grey}\line(1,0){5.00}}
\put(287.00,145.60){\color{grey}\line(1,0){5.00}}
\put(287.00,147.00){\color{grey}\line(1,0){5.00}}
\put(287.00,151.80){\color{grey}\line(1,0){5.00}}
\put(287.00,156.00){\color{grey}\line(1,0){5.00}}
                                                       % 5/2^-: 
\put(327.00,147.00){\color{grey}\line(1,0){5.00}}
\put(327.00,148.70){\color{grey}\line(1,0){5.00}}
\put(327.00,151.00){\color{grey}\line(1,0){5.00}}
\put(327.00,159.00){\color{grey}\line(1,0){5.00}}
                                                       % 7/2^-: 
\put(367.00,148.10){\color{grey}\line(1,0){5.00}}
\put(367.00,152.00){\color{grey}\line(1,0){5.00}}

                                                     %Ronninger and Metsch  Ronniger:2011td   
                                                      %1/2^+
\put(94.00,105.30){\gr\line(1,0){5.00}} 
\put(94.00,118.80){\gr\line(1,0){5.00}} 
\put(94.00,136.30){\grl\line(1,0){5.00}} 
\put(94.00,154.60){\grl\line(1,0){5.00}} 
\put(94.00,164.60){\grl\line(1,0){5.00}} 
                                                    %3/2^+                       
\put(134.00,53.10){\gr\line(1,0){5.00}} 
\put(134.00,88.90){\gr\line(1,0){5.00}} 
\put(134.00,119.40){\gr\line(1,0){5.00}} 
\put(134.00,122.50){\gr\line(1,0){5.00}} 
\put(134.00,129.10){\grl\line(1,0){5.00}} 
\put(134.00,148.00){\grl\line(1,0){5.00}} 
\put(134.00,154.00){\grl\line(1,0){5.00}} 
\put(134.00,155.50){\grl\line(1,0){5.00}} 
\put(134.00,157.00){\grl\line(1,0){5.00}} 
\put(134.00,164.00){\grl\line(1,0){5.00}} 
                                                    %5/2^+                       
\put(174.00,119.70){\gr\line(1,0){5.00}} 
\put(174.00,126.10){\gr\line(1,0){5.00}} 
\put(174.00,150.10){\grl\line(1,0){5.00}} 
\put(174.00,153.80){\grl\line(1,0){5.00}} 
\put(174.00,156.10){\grl\line(1,0){5.00}} 
                                                    %7/2^+                       
\put(214.00,123.60){\gr\line(1,0){5.00}} 
\put(214.00,151.90){\grl\line(1,0){5.00}} 
\put(214.00,160.00){\grl\line(1,0){5.00}} 
\put(214.00,164.60){\grl\line(1,0){5.00}} 
\put(214.00,169.30){\grl\line(1,0){5.00}} 

                                                      %Ronninger and Metsch  Ronniger:2011td   
                                                      %1/2^-
\put(254.00,93.40){\gr\line(1,0){5.00}}
\put(254.00,123.20){\grl\line(1,0){5.00}}
\put(254.00,134.00){\grl\line(1,0){5.00}}
\put(254.00,140.90){\grl\line(1,0){5.00}}
\put(254.00,154.60){\grl\line(1,0){5.00}}
\put(254.00,162.20){\grl\line(1,0){5.00}}
                                                      %3/2^-
\put(294.00,89.10){\gr\line(1,0){5.00}}
\put(294.00,117.10){\grl\line(1,0){5.00}}
\put(294.00,124.40){\grl\line(1,0){5.00}}
\put(294.00,141.40){\grl\line(1,0){5.00}}
\put(294.00,147.90){\grl\line(1,0){5.00}}
\put(294.00,149.40){\grl\line(1,0){5.00}}
\put(294.00,152.50){\grl\line(1,0){5.00}}
\put(294.00,155.50){\grl\line(1,0){5.00}}
\put(294.00,168.60){\grl\line(1,0){5.00}}
                                                      %5/2^-
\put(334.00,131.20){\grl\line(1,0){5.00}}
\put(334.00,143.60){\grl\line(1,0){5.00}}
\put(334.00,146.50){\grl\line(1,0){5.00}}
\put(334.00,155.70){\grl\line(1,0){5.00}}
\put(334.00,159.60){\grl\line(1,0){5.00}}
\put(334.00,168.80){\grl\line(1,0){5.00}}
                                                      %7/2^-
\put(374.00,142.60){\grl\line(1,0){5.00}}
\put(374.00,149.80){\grl\line(1,0){5.00}}
\put(374.00,169.10){\grl\line(1,0){5.00}}

%data 1/2^+
\linethickness{0.2mm}
\put(66,106.0){\color{grey}\framebox(33,6.0)[*]{}}
\put(66,115.0){\framebox(33,10.0)[*]{}}
%data 3/2^+
\linethickness{0.2mm}
\put(106.00,53.20){\bl\line(1,0){34.00}}
\put(106,80.0){\framebox(33,14.0)[*]{}}
\put(106,115.0){\framebox(33,10.0)[*]{}}
%data 5/2^+
\put(146,115.5){\framebox(33,5.5)[*]{}}
\put(146,126.0){\color{grey}\framebox(33,16)[*]{}}
%data 7/2^+
\put(186,121.5){\framebox(33,3.5)[*]{}}
%data 1/2^-
\put(226,89.0){\framebox(33,4)[*]{}}
\put(226,113.0){\framebox(33,7)[*]{}}
\put(226,135.0){\color{grey}\framebox(33,20)[*]{}}
%data 3/2^-
\put(266,99.0){\framebox(33,4.0)[*]{}}
\put(266,124.0){\framebox(33,12.0)[*]{}}  %from PDG
\put(266,146.5){\rd\framebox(33,8.0)[*]{}}  %from PDG
%data 52^-
\put(306,115.0){\framebox(33,10.0)[*]{}}
\put(306,147.5){\rd\framebox(33,6.0)[*]{}}
%data 7/2^-
\put(346,145.0){\framebox(33,10.0)[*]{}}
\end{picture}
\vspace{10mm}
\caption{\label{NSTAR}(Color online) Masses of nucleon (top) and $\Delta$ resonances (bottom) from quark model calculations.
         From left to right: Relativized one-gluon exchange by Capstick and Isgur~\cite{Capstick:1986ter}, 
         the algebraic model by Bijker, Iachello and Leviathan~\cite{Bijker:1994yr}, 
         the hypercentral quark model by Giannini, Santopinto and Vassallo~\cite{Giannini:2001kb}, 
          the relativistic Bonn model with instanton-induced interactions by Löring, Metsch and Petry~\cite{Loring:2001kx}, 
         and its extension including Goldstone-boson exchange by Ronniger and Metsch~\cite{Ronniger:2012xp}. 
         The low-lying resonances in the first and second excitation shells are shown in darker colors
         and those in the third and fourth shells in lighter colors. 
         The boxes give the experimental results for the Breit-Wigner masses
         with the uncertainty ranges from the RPP~\cite{ParticleDataGroup:2024cfk}. 
         If no uncertainty estimate is given, or for new
         resonances, the experimental results from Refs.~\cite{BES:2009ufh,BESIII:2012ssm,CLAS:2024iir,Sarantsev:2025lik} are shown.
}
\vspace{2mm}
\end{figure*}
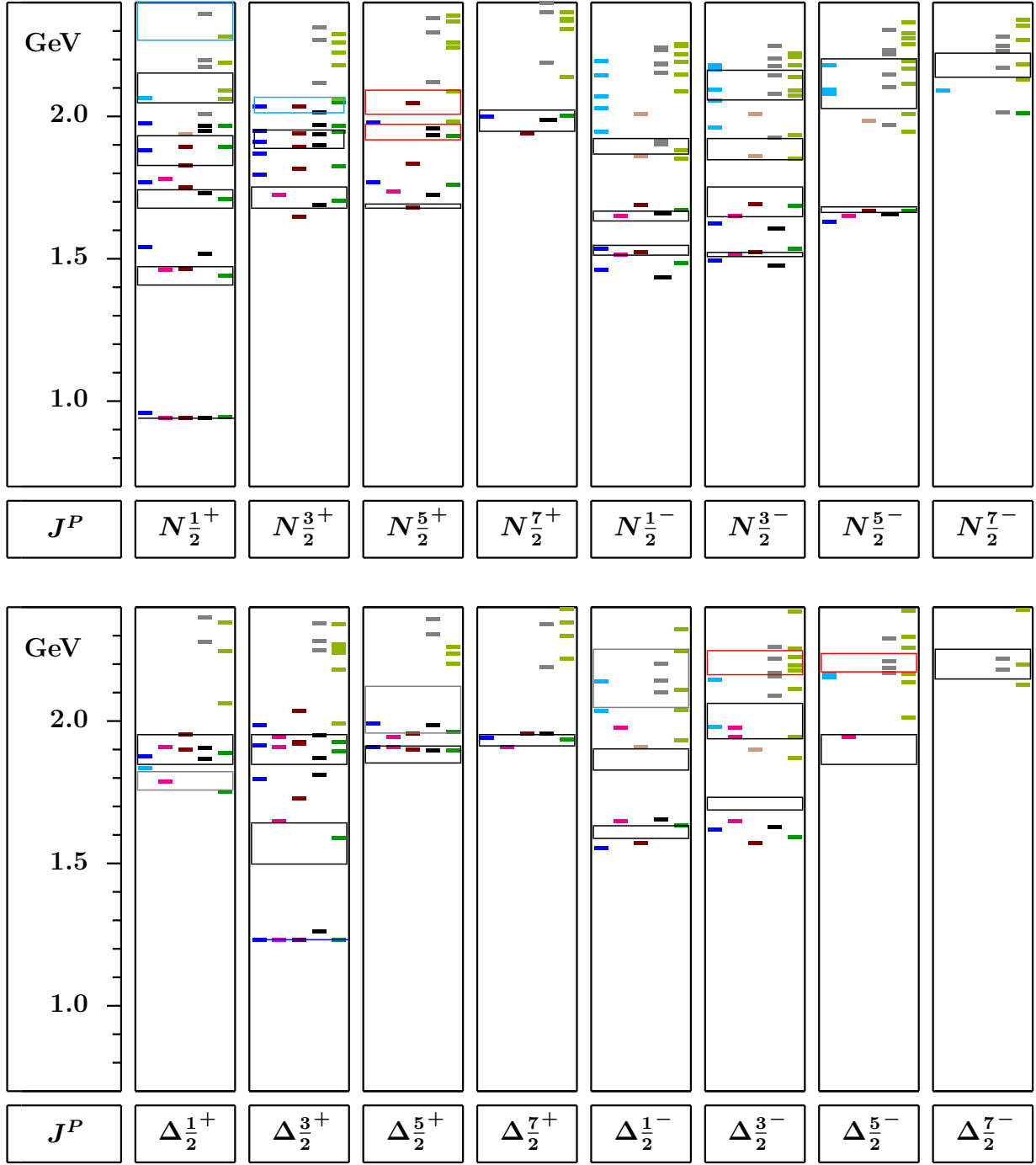

Regarding light baryon spectroscopy, a range of phenomenologically successful models have been employed over the years.
In addition to the confining potential, in these models the constituent quarks interact via `residual' interactions 
based on different mechanisms, which we briefly discuss in the following.
The results of some of these models are compared with the experimental status in Fig.~\ref{NSTAR}.

Various quark models based on one-gluon exchange (OGE) interactions have been studied in the literature.
The prototype model by de Rujula, Georgi and Glashow with spin-dependent forces
reproduces the phenomenological Gell-Mann-Okubo relations for ground states~\cite{DeRujula:1975qlm}.
A particularly influential model is the Isgur-Karl model~\cite{Isgur:1977ef,Isgur:1979be,Isgur:1978xb,Isgur:1978xi,Isgur:1978xj,Isgur:1978wd},  % Isgur:1979be
which implements a confining harmonic oscillator potential together with an anharmonic perturbation and a hyperfine interaction.
We already discussed the construction for baryons in Sec.~\ref{Construction}:
After removing the center-of-mass motion, one arrives at two independent spherical harmonic oscillators
for the coordinates $\rho$ and $\lambda$. 
This leads to the band quantum number N and the pattern in Fig.~\ref{fig:Regge-N} (right), 
where the mass splittings inside the $SU(6)$ multiplets are
due to the hyperfine interaction and the splittings between the multiplets come from the anharmonic perturbation.
The Isgur-Karl model describes the light and strange baryon
spectrum reasonably well but also predicts more states than were observed at that time, 
which is the problem of the \textit{missing resonances}.
Capstick and Isgur used a relativized version of the model to fit the  light and strange baryon spectrum 
with a consistent parameter set~\cite{Capstick:1986ter,Capstick:2000qj}. 
Here the $N(1440)$ Roper resonance turns out to be heavier than the parity partner of the nucleon, the $N(1535)$.
It  was also found that many of the missing states couple only weakly to $N\pi$,
which initiated photo- and electroproduction experiments around the world.

 The number of states in the spectrum can be reduced by assuming diquark clustering in baryons~\cite{Lichtenberg:1969sxc,Anselmino:1992vg,Santopinto:2004hw,Ebert:2007nw,Santopinto:2014opa,Barabanov:2020jvn}.
 For a pointlike diquark one of the oscillators in Eq.~\eqref{jacobi-coordinates} is frozen,
 which reduces the number of excitations compared to the three-quark model.
 Because the total diquark wave function must be antisymmetric and its color part is already antisymmetric,
 for ground-state diquarks with a symmetric spatial wave function also the spin-flavor part must be symmetric.
 The spin-flavor $SU(6)$ diquark combination $\bm{6}\otimes \bm{6} = \bm{21}_S \oplus \bm{15}_A$ then only leaves
  the $\bm{21}$-plet, whose decomposition $\bm{21} = {^1}\bar{\bm{3}} \oplus {^3}\bm{6}$ implies
 a scalar (`good') diquark with spin $S=0$ and an axialvector (`bad') diquark with $S=1$.
 Combining this with the remaining quark gives $\bm{21}\otimes\bm{6} = \bm{56}\oplus \bm{70}$,
 whereas the multiplets arising from $\bm{15}\otimes\bm{6} = \bm{20}\oplus \bm{70}$ are absent.
 On the other hand, since both $\bm{70}$-plets are needed to construct a totally antisymmetric
 baryon wave function, this also means that many states in the diquark model do
 not belong to irreducible representations of $SU(6)$; i.e., static diquarks that cannot exchange
 their roles also break the permutation symmetry.
 The simplest possibility of pointlike diquarks is also disfavored by
 comparison with experiment~\cite{Nikonov:2007br}.
 However, this does not prevent the formation of non-pointlike diquarks as two-body clusters that
 permanently exchange their roles through quark exchange,
 and there are indeed many indications for  strong quark-quark correlations inside baryons~\cite{Barabanov:2020jvn}.
 
A different interaction mechanism is Goldstone-boson exchange (GBE)~\cite{Glozman:1995fu,Glozman:1996wq}, 
which emphasizes the role of chiral symmetry and the flavor dependence of the baryon spectrum. 
 Here, the hyperfine interaction arises not from one-gluon exchange
 but from the exchange of octet pseudoscalar mesons between the quarks.
 This mechanism successfully predicts the correct level orderings of positive- and negative-parity nucleon excitations,
 such as the $N(1440)$ being lighter than the $N(1535)$.
 The relativistic version of the model provides a fair description of the spectrum 
  and especially also various  baryon form factors  in the spacelike region~\cite{Glozman:1997ag,Glozman:2001zc,Melde:2008yr,Melde:2007zz,Melde:2008dg,Plessas:2015mpa}. 
  This is attributed to the importance of relativity, since in non-relativistic approaches form factors  
  acquire relativistic recoil corrections at nonzero $Q^2$ which already affect their charge radii.
  A notable shortcoming here is the $\Lambda(1405)$, which comes out too high in mass,
  but this is a common problem among constituent-quark models.

The hypercentral constituent quark model~\cite{Ferraris:1995ui,Giannini:2001kb,Giannini:2005ks,DeSanctis:2005kt,Santopinto:2012nq,Giannini:2015zia} 
works with the hyperspherical coordinates $x=\sqrt{\rho^2+\lambda^2}$ and $\xi=\arctan\frac\rho\lambda$, 
where $\rho$ and $\lambda$ are the variables from Eq.~\eqref{jacobi-coordinates}, with $x$ representing the size
of the state and $\xi$ its deformation. 
The potential is then expressed in these coordinates
and the baryon wave functions are expanded in hyperspherical harmonics.
The nonrelativistic version of the model with Coulomb, confinement, hyperfine and flavor-dependent interactions
yields a good description of the spectrum and electromagnetic transition amplitudes~\cite{Giannini:2015zia,Santopinto:2012nq}.
It also predicts more states than expected from experiment,
and the flavor-dependent interaction is necessary to get the correct level ordering for the Roper resonance.
The relativistic version of the model~\cite{DeSanctis:2005kt} also describes the elastic nucleon form factors well.

Comprehensive calculations of the baryon spectrum have been achieved in the relativistic Bonn model
employing instanton-induced interactions (I.I.I.)~\cite{Loring:2001kv,Loring:2001kx,Loring:2001ky,Metscha:2008gkf}.
Here one starts from the three-body Bethe-Salpeter or Faddeev equation discussed in Sec.~\ref{sec:fm} (see Fig.~\ref{fig:faddeev}),
except one employs free constituent quark propagators and the kernels are reduced to  instantaneous interactions.
This simplifies the structure of the equation to a Salpeter equation, 
which enables a similar treatment as in quark potential models. 
Quark confinement is implemented through a linearly rising three-quark string potential.
Instead of the hyperfine interaction from one-gluon exchange, the authors employ
a $qq$ interaction motivated by instanton effects, 
which accounts for the dominant mass splittings in the spectrum.
A later extension~\cite{Ronniger:2011td,Ronniger:2012xp} also includes a flavor-dependent interaction similar to Goldstone-boson exchanges.
With this setup, the light and strange baryon spectrum has been calculated up to high $J$ values.
The model with flavor-dependent interactions 
correctly predicts the level ordering between positive- and negative-parity baryons,
but the missing resonances problem remains.
The model also gives a good description of various spacelike baryon elastic and transition form factors~\cite{Merten:2002nz}.

Various other phenomenologically very successful models have been employed to describe the baryon excitation spectrum,
including the `algebraic' or collective models~\cite{Bijker:1994yr,Bijker:1995ii,Bijker:2000gq},
where the residual interactions are described by  a sum of spin-spin and spin-flavor operators,
and large $N_c$ models~\cite{Goity:2003ab,Matagne:2004pm}.
With the observations of exotic hadrons in the heavy-quark sector,
quark models have also found widespread applications in the description of multiquark states in recent years, see e.g.~\cite{Brambilla:2019esw,Liu:2019zoy,Dong:2021bvy,Dong:2021juy,Chen:2022asf,Burns:2022uiv,Klempt:2022zwo,Liu:2024uxn} for reviews. 

Fig.~\ref{NSTAR} shows the spectra of $N^*$ and $\Delta^*$ resonances calculated in
different models in comparison to the experimental results. 
Instead of the real part of the pole position, 
we compare to the Breit-Wigner masses from the RPP~\cite{ParticleDataGroup:2024cfk}
since the parameters of the calculations were chosen to reproduce those. 
The gross features of the low-lying states are well reproduced by all models. 
Below 1.7 GeV there are five negative-parity nucleon resonances and two $\Delta$ resonances, 
corresponding to the first shell in Fig.~\ref{fig:missing-resonances}.
In the positive-parity sector there is one low-mass state in the nucleon sector  with $J^P=\nicefrac12^+$ (the Roper) and one in
the $\Delta$ sector  with $J^P=\nicefrac32^+$. %, whose mass varies considerably between the models.
Concerning the higher-lying positive parity states in the second shell, 
there are three states with $J^P=\nicefrac12^+$, $\nicefrac32^+$ and $\nicefrac52^+$ around 1700 MeV in the nucleon sector 
and one $\Delta$ state, which may be identified with the (one-star) $\Delta(1750)\nicefrac12^+$.
In this region, the masses already vary considerably between the models.
Above 1.9 GeV, numerous states are expected but only few have been found experimentally.

The theoretical assumptions underlying these predictions are very different, and 
some deficiencies can be related to the underlying assumptions. The deficiencies can be seen
inspecting the predicted masses of resonances in a few partial waves, see Table~\ref{tab:onehalf}. 
For example, like many other models, the relativized OGE mechanism~\cite{Capstick:1986ter}
has difficulties in explaining the low masses of the $N(1440)\nicefrac12^+$ and $\Delta(1600)\nicefrac32^+$.
The  mass difference between the Roper resonance and the negative-parity $N(1535)\nicefrac12^-$ is $+90(34)$~MeV in experiment
but $-60$~MeV in the calculation, and the difference between the 
$\Delta(1700)\nicefrac32^-$ and $\Delta(1600)\nicefrac32^+$ is +$140(73)$~MeV but $-175$~MeV   in ~\cite{Capstick:1986ter}.
The GBE model~\cite{Glozman:1997ag} with flavor-dependent interactions brings the mass of the Roper resonance below that of the $N(1535)\nicefrac12^-$,
however at the price of a somewhat mistuned fine structure: The  mass separation between the $N(1650)\nicefrac12^-$ and  $N(1535)\nicefrac12^-$ 
is $85(21)$~MeV compared to the calculated value of 254~MeV, while the difference between the $\Delta(1700)\nicefrac32^-$ 
and $\Delta(1600)\nicefrac32^+$ mass is $140(72)$~MeV but $-73$~MeV in the calculation.
This implies that neither OGE nor GBE alone are sufficient to explain the spectrum.

\begin{figure}
\vspace{-2mm}
\begin{minipage}[r]{.55\linewidth}
    \captionof{table}{Comparison of calculated masses of spin-$\nicefrac12$ nucleon and spin-$\nicefrac32$ $\Delta$ resonances
    with RPP values. The models are characterized by their interaction: One-gluon exchange (OGE)~\cite{Capstick:1986ter}, 
    Goldstone-boson exchange (GBE)~\cite{Glozman:1997ag},
    instanton-induced interactions (I.I.I.) without~\cite{Loring:2001kx} or with additional Goldstone-boson 
    exchange~\cite{Ronniger:2012xp}, the hypercentral constituent-quark model (HQM)~\cite{Giannini:2001kb},
    diquark (DQ) models~\cite{Santopinto:2004hw,Santopinto:2014opa}, and the algebraic model (AM) of Ref.~\cite{Bijker:1994yr}.
    n.c. stands for not calculated.
      }
    \label{tab:onehalf}
\footnotesize
\renewcommand{\arraystretch}{1.2}
    \centering
    \begin{tabular}{|cccccccccc|}
    \hline \rule{-1mm}{0.4cm}
               \hspace{-3mm}&\hspace{-3mm}RPP
               \hspace{-3mm}&\hspace{-3mm}OGE
               \hspace{-3mm}&\hspace{-3mm}GBE
               \hspace{-3mm}&\hspace{-3mm}I.I.I.
               \hspace{-3mm}&\hspace{-3mm}I.I.I.
               \hspace{-3mm}&\hspace{-3mm}HQM
               \hspace{-3mm}&\hspace{-3mm}DQ
               \hspace{-3mm}&\hspace{-3mm}DQ
               \hspace{-3mm}&\hspace{-3mm}AM \\
               \hspace{-3mm}&\hspace{-3mm}\cite{ParticleDataGroup:2024cfk} 
               \hspace{-3mm}&\hspace{-3mm}\cite{Capstick:1986ter}
               \hspace{-3mm}&\hspace{-3mm}\cite{Glozman:1997ag}
               \hspace{-3mm}&\hspace{-3mm}\cite{Loring:2001kx}
               \hspace{-3mm}&\hspace{-3mm}\cite{Ronniger:2012xp}
               \hspace{-3mm}&\hspace{-3mm}\cite{Giannini:2001kb}
               \hspace{-3mm}&\hspace{-3mm}\cite{Santopinto:2004hw}
               \hspace{-3mm}&\hspace{-3mm}\cite{Santopinto:2014opa}
               \hspace{-3mm}&\hspace{-3mm}\cite{Bijker:1994yr}
               \\[1mm] \hline \rule{-1mm}{0.35cm}
$N(1440)$      \hspace{-3mm}&\hspace{-3mm} 1440\er30  \hspace{-3mm}&\hspace{-3mm} 1540 \hspace{-3mm}&\hspace{-3mm} 1462 \hspace{-3mm}&\hspace{-3mm} 1518 \hspace{-3mm}&\hspace{-3mm} 1440 \hspace{-3mm}&\hspace{-3mm} 1463 \hspace{-3mm}&\hspace{-3mm} 1543 \hspace{-3mm}&\hspace{-3mm} 1511 \hspace{-3mm}&\hspace{-3mm} 1440 \\ 
$N(1535)$      \hspace{-3mm}&\hspace{-3mm} 1530\er15  \hspace{-3mm}&\hspace{-3mm} 1460 \hspace{-3mm}&\hspace{-3mm} 1522 \hspace{-3mm}&\hspace{-3mm} 1435 \hspace{-3mm}&\hspace{-3mm} 1484 \hspace{-3mm}&\hspace{-3mm} 1524 \hspace{-3mm}&\hspace{-3mm} 1538 \hspace{-3mm}&\hspace{-3mm} 1537 \hspace{-3mm}&\hspace{-3mm} 1566 \\       
$\Delta(1600)$ \hspace{-3mm}&\hspace{-3mm} 1570\er70  \hspace{-3mm}&\hspace{-3mm} 1795 \hspace{-3mm}&\hspace{-3mm} 1725 \hspace{-3mm}&\hspace{-3mm} 1866 \hspace{-3mm}&\hspace{-3mm} 1589 \hspace{-3mm}&\hspace{-3mm} 1727 \hspace{-3mm}&\hspace{-3mm} 1714 \hspace{-3mm}&\hspace{-3mm} 1689 \hspace{-3mm}&\hspace{-3mm} 1646 \\
$\Delta(1700)$ \hspace{-3mm}&\hspace{-3mm} 1710\er20  \hspace{-3mm}&\hspace{-3mm} 1620 \hspace{-3mm}&\hspace{-3mm} 1652 \hspace{-3mm}&\hspace{-3mm} 1628 \hspace{-3mm}&\hspace{-3mm} 1591 \hspace{-3mm}&\hspace{-3mm} 1573 \hspace{-3mm}&\hspace{-3mm} 1673 \hspace{-3mm}&\hspace{-3mm} 1625 \hspace{-3mm}&\hspace{-3mm} 1649 \\
$N(1710)$      \hspace{-3mm}&\hspace{-3mm} 1710\er30  \hspace{-3mm}&\hspace{-3mm} 1770 \hspace{-3mm}&\hspace{-3mm} 1650 \hspace{-3mm}&\hspace{-3mm} 1729 \hspace{-3mm}&\hspace{-3mm} 1709 \hspace{-3mm}&\hspace{-3mm} 1752 \hspace{-3mm}&\hspace{-3mm} 1640 \hspace{-3mm}&\hspace{-3mm} 1766 \hspace{-3mm}&\hspace{-3mm} 1710 \\
$N(1650)$      \hspace{-3mm}&\hspace{-3mm} 1615\er15  \hspace{-3mm}&\hspace{-3mm} 1535 \hspace{-3mm}&\hspace{-3mm} 1776 \hspace{-3mm}&\hspace{-3mm} 1660 \hspace{-3mm}&\hspace{-3mm} 1672 \hspace{-3mm}&\hspace{-3mm} 1692 \hspace{-3mm}&\hspace{-3mm} 1675 \hspace{-3mm}&\hspace{-3mm} 1625 \hspace{-3mm}&\hspace{-3mm} 1680 \\
$N(1880)$      \hspace{-3mm}&\hspace{-3mm} 1880\er50  \hspace{-3mm}&\hspace{-3mm} 1880 \hspace{-3mm}&\hspace{-3mm} n.c. \hspace{-3mm}&\hspace{-3mm} 1950 \hspace{-3mm}&\hspace{-3mm} 1893 \hspace{-3mm}&\hspace{-3mm} 1828 \hspace{-3mm}&\hspace{-3mm} n.c. \hspace{-3mm}&\hspace{-3mm} 1890 \hspace{-3mm}&\hspace{-3mm} 1972 \\
$N(1895)$      \hspace{-3mm}&\hspace{-3mm} 1895\er15  \hspace{-3mm}&\hspace{-3mm} 1945 \hspace{-3mm}&\hspace{-3mm} n.c. \hspace{-3mm}&\hspace{-3mm} 1901 \hspace{-3mm}&\hspace{-3mm} 1851 \hspace{-3mm}&\hspace{-3mm} 1861 \hspace{-3mm}&\hspace{-3mm} n.c. \hspace{-3mm}&\hspace{-3mm} 1888 \hspace{-3mm}&\hspace{-3mm} 1972 \\
$\Delta(1920)$ \hspace{-3mm}&\hspace{-3mm} 1920\er50  \hspace{-3mm}&\hspace{-3mm} 1915 \hspace{-3mm}&\hspace{-3mm} n.c. \hspace{-3mm}&\hspace{-3mm} 1906 \hspace{-3mm}&\hspace{-3mm} 1894 \hspace{-3mm}&\hspace{-3mm} 1921 \hspace{-3mm}&\hspace{-3mm} 1930 \hspace{-3mm}&\hspace{-3mm} 2042 \hspace{-3mm}&\hspace{-3mm} 1909 \\
$\Delta(1940)$ \hspace{-3mm}&\hspace{-3mm} 1950\er50  \hspace{-3mm}&\hspace{-3mm} 2080 \hspace{-3mm}&\hspace{-3mm} n.c. \hspace{-3mm}&\hspace{-3mm} 2100 \hspace{-3mm}&\hspace{-3mm} 1871 \hspace{-3mm}&\hspace{-3mm} 1900 \hspace{-3mm}&\hspace{-3mm} 2003 \hspace{-3mm}&\hspace{-3mm} 1910 \hspace{-3mm}&\hspace{-3mm} 1945 
 \\[1mm] \hline
 \end{tabular}
 \vspace{5mm}
               \captionof{table}{Right: Fit of the $N^\ast$ and $\Delta^\ast$ c.o.g. masses up to the second excitation shell using Eq.~\eqref{baryon-fit-iii}, 
               cf. Fig.~\ref{fig:missing-resonances}. Masses are given in MeV.}
               \label{tab:iii}
\end{minipage} 
\hspace{5mm}
             \begin{minipage}[r]{.4\linewidth}
                \centering
                \footnotesize
                \setlength\tabcolsep{2mm}
                \renewcommand{\arraystretch}{1.0}
                \begin{tabular}{     | c  | c | c | c | c | c | }  \hline  \rule{-1mm}{0.35cm}

                    N  & $(\mathbf{D},L^P,S)$ & $M_\text{c.o.g.}$ &$x_\text{conf}$ & $I_\text{sym}$  & $M_\text{fit}$       \\[1mm] \hline\hline  \rule{-1mm}{0.35cm}

                    0   &  $N(\mathbf{56},0^+,\nicefrac{1}{2})$       & 939 & 0 & $\nicefrac{1}{2}$  & $939$ \\
                        &  $\Delta(\mathbf{56},0^+,\nicefrac{3}{2})$  & $1210(1)$   &  0 & 0 & $1210$  \\[1mm] \hline\hline  \rule{-1mm}{0.35cm}

                    1  &  $N(\mathbf{70},1^-,\nicefrac{1}{2})$ & $1503(4)$ & 0 & $\nicefrac{1}{4}$  & $1572$   \\ 
                       &  $N(\mathbf{70},1^-,\nicefrac{3}{2})$ & $1690(24)$ & 0 & $0$  & $1662$   \\ 
                       &  $\Delta(\mathbf{70},1^-,\nicefrac{1}{2})$ & $1656(14)$ & 0 & $0$  & $1662$   \\[1mm] \hline\hline  \rule{-1mm}{0.35cm} 

                    2  &  $N(\mathbf{56},0^+,\nicefrac{1}{2})$      & $1366(3)$  & 1 & $\nicefrac{1}{2}$  & $1372$   \\ 
                       &  $\Delta(\mathbf{56},0^+,\nicefrac{3}{2})$ & $1550(15)$ & 1 & 0  & $1570$   \\[1mm] \cline{2-6}  \rule{-1mm}{0.35cm} 
                                           
                       &  $N(\mathbf{70},0^+,\nicefrac{1}{2})$      & $1696(10)$ & $\nicefrac{1}{2}$ & $\nicefrac{1}{4}$  & $1724$   \\  
                       &  $N(\mathbf{70},0^+,\nicefrac{3}{2})$      & $1725(30)$ & $\nicefrac{1}{2}$ & 0  & $1801$   \\  
                       &  $\Delta(\mathbf{70},0^+,\nicefrac{1}{2})$ & $1770(30)$ & $\nicefrac{1}{2}$ & 0  & $1801$    \\[1mm] \cline{2-6}  \rule{-1mm}{0.35cm} 
                                           
                       &  $N(\mathbf{20},1^+,\nicefrac{1}{2})$      &            & 0 & 0  & $2016$  \\[1mm] \cline{2-6}   \rule{-1mm}{0.35cm} 
                                                                                        
                       &  $N(\mathbf{56},2^+,\nicefrac{1}{2})$      & $1689(13)$ & $\nicefrac{2}{5}$ & $\nicefrac{1}{2}$  & $1696$   \\ 
                       &  $\Delta(\mathbf{56},2^+,\nicefrac{3}{2})$ & $1859(14)$ & $\nicefrac{2}{5}$ & 0  & $1850$   \\[1mm] \cline{2-6}  \rule{-1mm}{0.35cm} 
                                           
                       &  $N(\mathbf{70},2^+,\nicefrac{1}{2})$      & $1891(25)$ & $\nicefrac{1}{5}$ & $\nicefrac{1}{4}$  & $1883$   \\  
                       &  $N(\mathbf{70},2^+,\nicefrac{3}{2})$      & $1978(39)$ & $\nicefrac{1}{5}$ & 0  & $1935$   \\  
                       &  $\Delta(\mathbf{70},2^+,\nicefrac{1}{2})$ & $2040(80)$ & $\nicefrac{1}{5}$ & 0  & $1935$  \\[1mm] \hline\hline  
                       \rule{-1mm}{0.35cm} 
                       
                    3  &  $N(\mathbf{56},1^-,\nicefrac{1}{2})$      & $1868(23)$  & 1 & 0  & $1858$   \\ 
                       &  $\Delta(\mathbf{56},1^-,\nicefrac{3}{2})$ & $1858(63)$ & 1 & 0  & $1858$   \\[1mm] \cline{2-6}  \rule{-1mm}{0.35cm} 
                       &  $N(\mathbf{70},3^-,\nicefrac{1}{2})$      & $2081(50)$  & 1 & $\nicefrac{1}{4}$  & $2087$   \\ 
                       &  $N(\mathbf{70},3^-,\nicefrac{3}{2})$      & $2198(50)$  & 1 & 0  & $2155$   \\ 
                       &  $\Delta(\mathbf{70},3^-,\nicefrac{3}{2})$ & $2144(26)$ & 1 & 0  & $2155$   \\[1mm] \hline  
                                           
                \end{tabular}
                \renewcommand{\arraystretch}{1.0}
\end{minipage}
\vspace{-2mm}
        \end{figure}

The relativistic Bonn model with instanton-induced interactions~\cite{Loring:2001kx}
solves the problem with the Roper resonance by lowering the baryon masses with 
a large component of scalar ({\it good}) diquarks; however, the 
$\Delta(1600)\nicefrac32^+$ does not have scalar diquarks and 
still lies $296(70)$~MeV above its experimental value.
Combining the short-range instanton interaction with long-range GBE
results in a spectrum that shows reasonable agreement with the data~\cite{Ronniger:2012xp}, however at the expense of additional parameters.
Finally, the approaches~\cite{Giannini:2001kb,Santopinto:2004hw,Santopinto:2014opa,Bijker:1994yr} do not employ
specific quark-quark interactions but use operators describing spin-spin and flavor correlations;
here the differences  between calculations and data are
always below 100~MeV but no insight is gained into the internal dynamics of resonances.

For comparison, in Table~\ref{tab:iii} we use the simple empirical mass formula \\[-1ex]
\begin{eqnarray}\label{baryon-fit-iii}
   M = \sqrt{M_\Delta^2 + \text{N}\,\omega -  x_\text{conf} \,A   -  I_\text{sym}\,\delta  - D/2 } 
\end{eqnarray}
to fit the light baryon spectrum (c.o.g. masses) up to the third shell. 
The formula makes minimal assumptions and has five parameters: the  $\Delta(1232)\nicefrac32^+$ mass,
the $\Delta(1232)\nicefrac32^+ - N(939)$ hyperfine splitting~$\delta$, 
the level spacing $\omega$ whose coefficient is the shell number N,  a confinement term with parameter $A$,
and a parameter $D$ which is only active in the third shell.
The values of $x_\text{conf}$ are extracted from Fig.~\ref{fig:Regge-N}\,d. 
The effect of instanton-induced interaction is taken into account by a mass shift proportional to the fraction of 
`good' diquarks in the wave function, which
are the  components highlighted in red in Fig.~\ref{fig:flavor-wfs}. This entails
$I_\text{sym} = \frac{1}{2}$ for $\psi_1$, $I_\text{sym} = \frac{1}{4}$ for $\psi_3$ and zero  
for all other $\psi_i$, so this term only contributes to $N^\ast$ resonances with spin $S=\frac{1}{2}$
and the fully antisymmetric quark pair in $S$-wave. (Hence it is not applicable for the
$N(\mathbf{56},1^-,\nicefrac{1}{2})$ doublet in the third shell.)
The square root is chosen to yield a better fit.
Numerically,  $\omega=1.3$~GeV$^2$ and $A=1.6$~GeV$^2$ are used,
and the term  $D=0.8$~GeV$^2$ is added for the third shell.
It is curious that the agreement with the data is even better 
than in the calculations discussed above; in particular, no flavor-dependent
interaction term is required.

So what have we learned about the baryon spectrum? 
Many quark models describe the spectrum reasonably well and several of them
also  the $Q^2$ dependence of various form factors and electroproduction amplitudes.   
In fact, different models describe the spectrum \textit{similarly} well: 
one-gluon exchange, Goldstone-boson exchange,
instanton-induced interactions, the hypercentral model, all capture the essential aspects of the light and strange baryon spectrum.
This suggests that the spectrum is not overly sensitive to the specific form of the interactions between the quarks,
except that flavor-dependent interactions appear necessary for the correct level ordering
of baryons with opposite parity.
Because adding more interaction terms like in Eq.~\eqref{breit-fermi} introduces new  parameters,
some information may also be absorbed in the model parameters
when fitting the spectrum: 
The Capstick-Isgur model has 13 parameters~\cite{Capstick:1986ter} and 
the Bonn model 7--10~\cite{Loring:2001kx,Ronniger:2011td}, 
with similar numbers  for the Goldstone-boson exchange and  hypercentral models depending on the setup~\cite{Plessas:2015mpa,Giannini:2015zia}.
%20(?) parameters for the hypercentral model~\cite{a}, etc.
Common challenges across the different models include the $\Lambda(1405)$, which indicates the absence of meson-baryon effects,
and the missing resonances problem -- while new experiments and improved partial-wave analyses have
 added new states to the RPP, the spectrum as of today
is still much sparser than quark models predict.

Given that the spectrum alone does not seem to suffice for discriminating between models, a
more stringent test of the underlying dynamics are elastic and transition form factors. 
Here, relativity turns out to be crucial to get the correct magnitude and $Q^2$ dependence of the form factors. 
But this is also not sufficient, as
the spacelike ($Q^2 > 0$) and timelike ($Q^2 < 0$) properties are intertwined~\cite{Denig:2012by,Eichmann:2016yit,Ramalho:2023hqd,Lin:2021umz}. 
For instance, electromagnetic form factors
exhibit vector-meson poles in the timelike region which significantly impact the spacelike properties,
as is evident in dispersive calculations of form factors~\cite{Denig:2012by,Lin:2021umz,Alvarado:2023loi}.
These effects must be dynamically  generated from the quark level, which many quark models discussed above would not achieve. 
Is it then satisfactory if the model describes  spacelike experimental data well, even if it does not produce timelike vector-meson poles whose effects are felt throughout the spacelike region?
In other words, should one  prioritize a pointwise description of spacelike  data  
or genuine insights into the underlying dynamics? 
In this context one should highlight the covariant spectator model, which
successfully connects the spacelike and timelike regions and
has been explored for a range of elastic and transition 
form factors~\cite{Gross:2006fg,Ramalho:2008dp,Ramalho:2012ng,Ramalho:2015qna,Ramalho:2016zgc,Ramalho:2020nwk}, see~\cite{Ramalho:2023hqd} for a review.

In general, the relativistic nature of light baryons tells us that 
neither non-relativistic nor relativistic quantum mechanics is sufficient for a complete description; 
instead, one must employ relativistic QFT.
Unfortunately, translating the Schrödinger equation to relativistic
QFT is not straightforward. 
Special relativity forces us to abandon the concept of a wave function with a probability interpretation.
It also calls the spin-flavor $SU(6)$ classification into question, 
because only the mass $M$ and spin $J$ are Casimirs of the Poincaré group, while $L$ and $S$ can mix in different frames and are no longer good quantum numbers 
for labeling states.
Particle creation and annihilation makes the Hilbert space infinite-dimensional, and
renormalization  is highly non-trivial in a Hamiltonian description. % (see e.g.~\cite{a}). 
% As an alternative, a  promising strategy for accessing the spectrum is to extract 
In QFT, the spectrum can be accessed by extracting
hadron poles from QCD's correlation functions,  
which requires an understanding of how those correlation functions are generated to begin with.  
One way to deal with this is lattice QCD discussed in Sec.~\ref{sec:lattice}; 
however, its language is quite different from quark models which can make a meaningful dialogue difficult.
Another option are the functional methods discussed in Sec.~\ref{sec:fm}, 
which also address all of the above points but share sufficient similarities to
provide a possible avenue for connecting quark models with QCD.

\subsection{Light-front holography}

A different approach to calculate the baryon spectrum is light-front holographic QCD~\cite{Brodsky:2008pg,Brodsky:2014yha}.
Here the idea is to convert the QCD Lagrangian into a light-front Hamiltonian $H_\text{LF}$ such
that the spectrum and light-front wave functions are obtained from the eigenvalue equation $H_\text{LF}\,\psi = M^2 \psi$~\cite{Brodsky:1997de}.
Because the diagonalization of this Hamiltonian is a difficult task, a practical simplification is to employ
a semiclassical approximation and reduce the equation to a Schrödinger equation  where  the dynamics is encoded in an effective potential.
It turns out that this equation has the same structure as the eigenvalue equations in AdS (Anti-de-Sitter) space,
which is a spacetime with a  constant negative curvature.
This is the basis of light-front holographic models.

The underlying idea is the holographic principle~\cite{Bekenstein:1973ur,Hawking:1975vcx,tHooft:1993dmi,Susskind:1994vu},
which is based on black hole thermodynamics and postulates that a gravitational system in a given volume is equivalent 
to a non-gravitational system on the boundary of that volume.
A concrete realization of this principle is the AdS/CFT correspondence~\cite{Maldacena:1997re} between gravity on a $(d+1)$-dimensional
AdS spacetime and a conformal field theory (CFT) living on the $d$-dimensional boundary of that spacetime.
In particular, the correspondence implies that a strong coupling in the quantum gauge theory corresponds to a weak coupling
in the classical gravitational theory (its holographic dual), which enables calculations of observables in nonperturbative regimes that are otherwise not accessible.

Concerning QCD, however, so far no holographic dual is known.
Among other reasons, QCD is not a conformal theory: Although the classical QCD Lagrangian for massless quarks is scale invariant,
the regularization of the quantum field theory breaks this scale invariance. % and eventually induces a gluon mass gap.
At low energies, QCD has mass gaps, confinement and dynamical chiral symmetry breaking, which are all hallmarks of a non-conformal theory.
One strategy is then to employ a bottom-up  approach named AdS/QCD, where 
one searches for a higher-dimensional classical gravitational theory that encodes basic aspects of QCD.
Starting from a five-dimensional AdS background, one can break conformal symmetry 
e.g. by introducing an infrared cutoff (`hard wall'~\cite{Polchinski:2001tt}) to mimic the confinement of hadronic modes in a box, 
or by a dilaton background (`soft wall'~\cite{Karch:2006pv}) to model the running coupling or chiral symmetry breaking.
Although not derived from first principles, this is a simple and flexible approach that captures
qualitative features of confinement and the hadron spectrum.
For example,  the soft-wall model leads to the linear Regge trajectories observed in the hadron mass spectrum.

In light-front holography, the dilaton profile in the AdS wave equation is determined by the mechanism proposed by de Alfaro, Fubini and Furlan~\cite{deAlfaro:1976vlx},
which also settles the form of the potential in the light-front Schrödinger equation.
The resulting equation can  be solved analytically to return the hadron masses and their light-front wave functions. 
For mesons, this leads to the mass formula 
\begin{eqnarray}\label{ads-mesons}
  M^2 = 2\lambda \left( J + L + 2n \right) = M_\rho^2 \left( J + L + 2n \right),
\end{eqnarray}
where  $J$ is the total angular momentum of the state,
$L$ its orbital angular momentum and $n$ the radial quantum number.  
$\lambda$ is the coefficient of the dilaton field and plays the role of the dynamically generated mass scale in QCD,
which can be traded for the $\rho$-meson mass $M_\rho$
given that the $\rho$ meson corresponds to $J=1$, $L=0$ and $n=0$.
Note that Eq.~\eqref{ads-mesons} also implies linear Regge trajectories as well as a massless pion ($J=L=n=0$).
Reviews of meson phenomenology using AdS/QCD can be found, e.g., in~\cite{deTeramond:2011aml,Brodsky:2014yha,Kim:2011ey}.

The analogous mass formula for baryons reads
\begin{eqnarray}\label{holo-mass-formula}
   M^2 = 4\lambda\,(1+n+\nu) = M_N^2\,(1+n+\nu)\,,
\end{eqnarray}
where we traded the parameter $\lambda$ for the nucleon mass $M_N$, 
and $\nu$ is related  to the minimal light-front orbital angular momentum of the baryon.
Although this equation does not know about  baryon multiplets, 
one can combine it with the nonrelativistic  spin-flavor $SU(6)$ classification discussed in Sec.~\ref{sec:su(6)}.
Restricting ourselves to flavor octets and decuplets, the $\mathbf{56}$-plet contains  nucleons with three-quark spin $S=\nicefrac{1}{2}$ and 
$\Delta$ baryons with $S=\nicefrac{3}{2}$, while the $\mathbf{70}$-plet consists of nucleons with $S=\nicefrac{1}{2}$ and $S=\nicefrac{3}{2}$
and $\Delta$ baryons with $S=\nicefrac{1}{2}$ (cf.~Fig.~\ref{fig:flavor-wfs}).
In Ref.~\cite{Brodsky:2014yha} it is further assumed that the $\mathbf{56}$-plet only admits baryons with positive parity and even values of $L$, while
the $\mathbf{70}$-plet only admits baryons with negative parity and odd values of $L$.
This leads to a different scheme compared to Fig.~\ref{fig:missing-resonances},  
whose $\mathbf{70}$-plet entries with $L^P = 0^+$ and $2^+$ would no longer appear here,
so one also needs to  reassign the physical states to different $SU(6)$ multiplets. 
The shell number N is also no longer available since the radial quantum number $n$ has taken its role.
By comparing  Regge trajectories, one is then led to the phenomenological identification $\nu = L + S/2 - P/4$,
where $P$ is the parity of the baryon, $S$ its intrinsic spin and $L$ its orbital angular momentum. 
Written out explicitly, this entails
\begin{eqnarray}\label{AdS-SU6-v1}
   \begin{array}{ll}
    N(\mathbf{56},L^+,\nicefrac{1}{2}) \quad &\leftrightarrow \quad \nu = L\,, \\
    \Delta(\mathbf{56},L^+,\nicefrac{3}{2}) \quad &\leftrightarrow \quad \nu = L + \frac{1}{2}\,,
    \end{array}
    \qquad
    \begin{array}{ll}
    \left\{ 
    \displaystyle N(\mathbf{70},L^-,\nicefrac{1}{2}) \atop
    \displaystyle \Delta(\mathbf{70},L^-,\nicefrac{1}{2})\right\} 
    &\leftrightarrow \quad \nu = L + \frac{1}{2}\,, \\
   \; \;\;N(\mathbf{70},L^-,\nicefrac{3}{2}) \quad &\leftrightarrow \quad \nu = L+1\,,
   \end{array}
\end{eqnarray}
with $L$ even (odd) for positive (negative) parity.
Therefore, in contrast to the meson spectrum, light-front holography predicts that the baryon spectrum does not depend on $J$, 
which also entails that there is no spin-orbit coupling.

\begin{figure}[!t]
	\centering
	\includegraphics[width=0.96\textwidth]{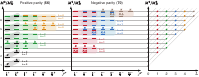}
	\caption{Light baryon spectrum in the light-front holographic model discussed in Ref.~\cite{Brodsky:2014yha}.
    The left panel shows the positive-parity spectrum, the center panel the negative-parity spectrum and the right panel the corresponding Regge trajectories. }
	\label{fig:ads-1}
	\vspace{3mm}
\end{figure}

The resulting spectrum is illustrated in Fig.~\ref{fig:ads-1}. 
The left panel shows the situation for  $L=0$, 2, 4, $\dots$  corresponding to positive parity.
A given value of $L$ admits ground states ($n=0$) and an infinite tower of radial excitations ($n>0$).
Because positive parity is tied to the $\mathbf{56}$-plet, for any combination $(n,L)$ 
one obtains two nucleons with three-quark spin $S=\nicefrac{1}{2}$
and four $\Delta$ baryons with $S=\nicefrac{3}{2}$ (except for $L=0$ which allows  only one $N$ and one $\Delta$).
This results in an alternating mass pattern between nucleons and $\Delta$ baryons, which are equally spaced in the squared mass.
The center panel in Fig.~\ref{fig:ads-1} shows the analogous situation for  $L=1$, 3, 5, $\dots$, which corresponds to negative parity and the $\mathbf{70}$-plet.
In this case, each combination $(n,L)$ returns
two nucleons with $S=\nicefrac{1}{2}$, four nucleons  with $S=\nicefrac{3}{2}$, and two $\Delta$ baryons with $S=\nicefrac{1}{2}$
(except for $L=1$ which admits only three nucleons with $S=\nicefrac{3}{2}$).
Because the mass formula~\eqref{holo-mass-formula} has the form $M^2 \sim n + L + const.$, this naturally leads to Regge trajectories 
displayed in the right panel of Fig.~\ref{fig:ads-1}, where each point in the $(L,M^2)$ plane may contain several states with different total angular momentum $J$.

How well does  Eq.~\eqref{holo-mass-formula} describe the experimental baryon spectrum?
Clearly, this depends on the mapping of the $SU(6)$ multiplets to the $(n,L)$ values.
For the low-lying multiplets the identification is straightforward
and identical to our earlier assignment in Fig.~\ref{fig:missing-resonances}:
$(n,L) = (0,0)$ are the nucleon and $\Delta$ ground states and
$(1,0)$ their first radial excitations;
$(0,1)$ are the negative-parity excitations from the first harmonic oscillator shell;
and $(0,2)$ can be identified with the $N(\mathbf{56},2^+,\nicefrac{1}{2})$ and $\Delta(\mathbf{56},2^+,\nicefrac{3}{2})$ baryons
in the second shell. These systematics are indeed visible in the experimental spectrum as seen in Fig.~\ref{fig:neg-parity}.
However, for the higher-lying states the situation is less clear.
In the positive-parity sector one could identify the $N(1710)\nicefrac{1}{2}^+$ with the second radial excitation in $(n,L) =(2,0)$,
but this leaves no corresponding $\Delta$ state since we already used the $\Delta(1920)\nicefrac{3}{2}^+$ in $(0,2)$.
Concerning $(n,L) =(1,2)$,  one could assign the two  nucleons in $N(\mathbf{70},2^+,\nicefrac{1}{2})$ in the second oscillator shell,
but this leaves  four missing $\Delta$ baryons  while the four nucleons in $N(\mathbf{70},2^+,\nicefrac{3}{2})$ are unaccounted for.
The latter include the (two-star) $N(1990)\nicefrac{7}{2}^+$, while the lowest-lying $N \nicefrac{7}{2}^+$ state in Fig.~\ref{fig:ads-1} only appears 
high  in the spectrum for $L=4$.
The negative-parity spectrum has similar issues: $(n,L) =(1,1)$ consists of a nucleon doublet, a $\Delta$ doublet and a nucleon triplet, 
but experimentally one has a nucleon doublet and a $\Delta$ triplet (the lowest states in the third oscillator shell).
The latter contains the (three-star) $\Delta(1930)\nicefrac{5}{2}^-$ but the lowest  $\Delta \nicefrac{5}{2}^-$ state in Fig.~\ref{fig:ads-1} only enters with  $L=3$.
Taken together, it appears that Eq.~\eqref{holo-mass-formula} with the assumptions made above is too restrictive to explain the higher-lying baryon excitation spectrum.

On the other hand, since the interpretation of the parameter $\nu$ in Eq.~\eqref{holo-mass-formula} is  flexible,
these assignments are not the only options. 
If one relaxes the assumption that the \textbf{56}-plet (\textbf{70}-plet) corresponds only to positive (negative) parity with even (odd) $L$, the agreement with the spectrum can be improved.
For example, one can combine the $SU(6)$ scheme from Fig.~\ref{fig:missing-resonances} 
with the mass formula~\eqref{holo-mass-formula} and assign to each multiplet a value for $\nu$  as follows:
\begin{eqnarray}\label{AdS-SU6}
   N(\mathbf{56},L^P,\nicefrac{1}{2})  \; \leftrightarrow \; \nu = l\,, \qquad
   \left\{\Delta(\mathbf{56},L^P,\nicefrac{3}{2}) \atop
          N(\mathbf{70},L^P,\nicefrac{1}{2})
   \right\} \; \leftrightarrow \; \nu = l + \frac{1}{2}\,, \qquad
   \left\{ N(\mathbf{70},L^P,\nicefrac{3}{2}) \atop
          \Delta(\mathbf{70},L^P,\nicefrac{1}{2}) 
    \right\} \; \leftrightarrow \;  \nu = l + 1\,.
\end{eqnarray}

%where we ignored the \textbf{20}-plets.
Here, $n=n_\rho + n_\lambda$ and $l=l_\rho + l_\lambda$ are the harmonic-oscillator eigenvalues from Fig.~\ref{fig:flavor-wfs}.
These are technically not good quantum numbers in the oscillator assignment; e.g., in the $\textbf{70}$-plet with $L^P = 0^+$ for $\text{N}=2$
the mixed-symmetric and mixed-antisymmetric components of the doublet carry different values of $(n,l)$,
or in the $\textbf{56}$-plet with $L^P = 1^-$ for $\text{N}=3$ the singlet wave functions are sums of terms with different $(n,l)$.
In these cases we take the weighted average  of  the squared mass values~\eqref{holo-mass-formula} for the individual $(n,l)$ components.

\begin{figure} 
\centering
	\includegraphics[width=0.97\linewidth]{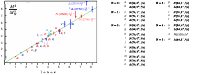}
\caption{\label{fig:Regge-AdS} Squared masses of nucleon and $\Delta$ resonances according to the AdS/QCD mass formula~\eqref{holo-mass-formula}
combined with the assignments in Eq.~\eqref{AdS-SU6}.
The bars are the center-of-gravity masses of the multiplets, see Fig.~\ref{fig:missing-resonances} and the discussion in Sec.~\ref{sec-beyond-3rd-shell} for the assignments of the $SU(6)$ multiplets
to the experimentally known baryon resonances.}
\end{figure}

The result is shown in Fig.~\ref{fig:Regge-AdS}, where most states fall on a single line. 
Even though the assignments in Eq.~\eqref{AdS-SU6} are ad-hoc,  it is  remarkable that almost all known light baryon resonances can be accommodated in this scheme.
For the $N(\mathbf{70},L^P,\nicefrac{3}{2})$ and $\Delta(\mathbf{70},L^P,\nicefrac{1}{2})$ multiplets with $\nu=l+1$, Eq.~\eqref{AdS-SU6} entails
\bq
\label{ads-13}
\frac{M^2}{M_N^2} =  n+l+2 
=  \left(n_\rho+\frac12\right) + \left(n_\lambda+\frac12\right) + \left(l_\rho+\frac12\right) + \left(l_\lambda+\frac12\right),
\eq
such that the total mass gets contributions from the four oscillators and their zero-point fluctuations.
For the $\Delta(\mathbf{56},L^P,\nicefrac{3}{2})$ and $N(\mathbf{70},L^P,\nicefrac{1}{2})$ multiplets with $\nu = l + 1/2$ 
the r.h.s. above is lowered by $1/2$, and the best binding
is achieved for  $N(\mathbf{56},L^P,\nicefrac{1}{2})$ where this value is lowered by 1.
We note that this scheme is also compatible with the observation that oscillator eigenvalues grow with the combination $\text{N}=2n+l$,
since Eq.~\eqref{AdS-SU6} can be written as
\begin{eqnarray}
   \frac{M^2}{M_N^2} = \left\{ \begin{array}{rl} 1 + l + \kappa &\quad n=0\,, \\[-1mm] N + \kappa &\quad n = 1 
   \end{array}\right.
\end{eqnarray}
with $\kappa=0$, $1/2$ or $1$ for the three respective options in Eq.~\eqref{AdS-SU6}.
The connection between AdS/QCD and the non-relativistic three-oscillator model is, however, a speculation and not covered by the holographic approach to the spectrum~\cite{Guy:2025pc}.

In general, light-front holography is a powerful tool to provide analytic insight into QCD phenomenology.
As discussed above, it naturally produces Regge trajectories and gives overall good agreement with experimental data.
Remarkably, the masses of baryons, the Regge slope and the hyperfine splitting
are all described by one parameter, pointing at an intimate connection between these quantities.
Regarding hadron phenomenology, apart from the baryon spectrum light-front holography has been applied to calculate distribution amplitudes, form factors, electromagnetic transition amplitudes 
and a range of other quantities, see e.g.~\cite{Polchinski:2001tt,Katz:2005ir,Brodsky:2006uqa,Boschi-Filho:2005xct,Brodsky:2007hb,Forkel:2007cm,Brodsky:2008pf,Colangelo:2008us,Branz:2010ub,deTeramond:2011aml,Li:2013oda,Brodsky:2014yha,Leutgeb:2019gbz,Aliberti:2025beg} and references therein.

\subsection{Functional methods}\label{sec:fm}

Progress in calculating the baryon spectrum has also been made with functional methods.
Here the central quantities of interest are QCD's $n$-point correlation functions
 in Fig.~\ref{fig:cfs}. As mentioned in Sec.~\ref{sec:whatisabaryon}, 
they can be  obtained through functional derivatives of the QCD partition function~\eqref{path-int-0}. 
Taking the quantum expectation values of the classical (Klein-Gordon, Maxwell, Dirac) equations of motion
yields the Dyson-Schwinger equations (DSEs), which relate different $n$-point functions with each other~\cite{Roberts:1994dr,Alkofer:2000wg,Fischer:2006ub,Bashir:2012fs,Cloet:2013jya,Eichmann:2016yit,Huber:2018ned,Fischer:2018sdj}. 
Similar systems of equations can be derived using the functional renormalization group (FRG)~\cite{Berges:2000ew,Pawlowski:2005xe,Dupuis:2020fhh}.
Therefore, in principle functional methods are ab-initio methods just like lattice QFT:
while on the lattice one calculates the correlation functions directly from the path integral,
the strategy with functional methods is to calculate them  \textit{from each other} using equations derived from the path integral.

\begin{figure}
 	\centering
 	\includegraphics[width=0.96\textwidth]{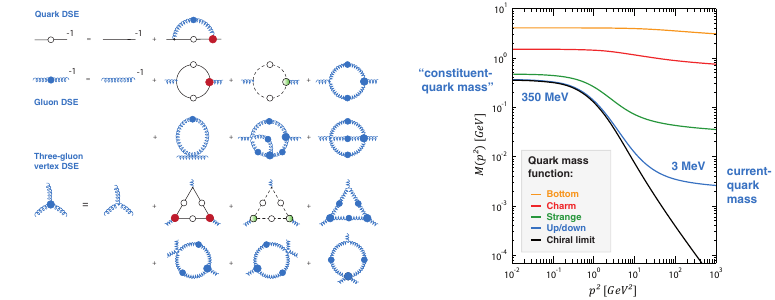}  
 	\caption{(color online) Left: Dyson-Schwinger equations for the quark and gluon propagator and three-gluon vertex. 
      The solid and dashed lines represent quarks and ghosts,
 	  the springs are gluons. Two-loop diagrams in the three-gluon  DSE are omitted.
      Right: Quark mass function for different quark flavors obtained from the quark DSE~\cite{Eichmann:2016yit}.}
 	\label{fig:dses}
 	\vspace{-0mm}
 \end{figure}

As  examples, the left plot in Fig.~\ref{fig:dses} shows the DSEs for the quark and gluon propagator and the three-gluon vertex.
Like  usual Feynman diagrams, each loop represents a four-momentum integration. Because the quantities on the left appear again
on the right inside the integrals, these are integral equations which can be solved recursively. 
In this process, the classical propagators and vertices from the Lagrangian get `dressed' and acquire structure.
For example, to solve the quark DSE one needs to know the gluon propagator and quark-gluon vertex.
These satisfy their own DSEs, which depend again on the quark propagator but also on higher $n$-point functions.
This yields an infinite tower of equations, which for practical calculations
must be truncated. % at some point. 
The truncations employed in the literature so far cover a broad landscape ranging  
from simple models to ab-initio  calculations without any model input.
For example, one could absorb all information from the gluon sector into an effective interaction
and solve the quark DSE as a standalone equation. 
Or one could also solve the gluon DSE and  
set all three- and four-point functions to their classical values,
or calculate all two- and three-point functions while neglecting four-point functions, etc.
Therefore,  systematically enlarging these equations  should improve the precision,
as is already visible in the existing DSEs and  FRG calculations~\cite{Cyrol:2017ewj,Huber:2018ned,Gao:2021wun,Eichmann:2021zuv,Fu:2022uow,Ihssen:2024miv}; e.g.,
recent glueball calculations in Yang-Mills theory (i.e., QCD without quarks) are in good agreement  with lattice results~\cite{Huber:2020ngt,Huber:2021yfy,Huber:2025kwy}. 
Correlation functions with quark and gluon legs are  gauge-dependent, so
 the strategy is  to choose the best-suited gauge for calculating gauge-invariant observables --
 in analogy to choosing spherical coordinates to compute the  energy levels of the hydrogen atom in quantum mechanics. 
  Most calculations so far have employed the Landau gauge as it preserves manifest Lorentz invariance 
  and  is a fixed point under the renormalization group.

One advantage of functional methods is that they facilitate a `hands-on' understanding of QCD's nonperturbative phenomena.
For example, a key feature  of QCD is dynamical mass generation:
The sum of the light quark masses only amounts to 1\% of the proton mass, so the rest must  somehow be generated in QCD.
This is a direct outcome of the quark DSE:
The calculated quark masses run with the momentum scale (right panel in Fig.~\ref{fig:dses}), so that for low momenta the current-quark masses (which are  inputs to QCD)
turn into large constituent-quark masses around 350 MeV. 
For heavier flavors the effect is eventually washed out.
This is a  nonperturbative effect due to the dynamical breaking of chiral symmetry,
which cannot be generated at any order
in perturbation theory (see e.g.~\cite{Eichmann:2025wgs} for a pedagogical discussion).
The resulting mass function at low momenta is  proportional to the quark condensate and sets the scale 
for the pion decay constant as well as the masses of the nucleon and other hadrons. 
 The dynamical generation of a quark mass function is also seen in lattice QCD~\cite{Bowman:2005vx,Oliveira:2018lln}, where it arises from center vortices~\cite{Biddle:2023lod,Kamleh:2023gho}. 
  Note that the influence of such topological gluon field configurations on the $n$-point correlation functions is not directly accessible through DSEs, which are derived from the partition function where the gluon fields are already integrated over -- as mentioned above, in contrast to lattice calculations,  DSEs determine the correlation functions from \textit{each other}.  
Another robust prediction from both functional methods and lattice calculations is
the disappearance of the massless pole in the gluon propagator, 
i.e., the gluon develops a mass gap, see~\cite{Alkofer:2000wg,Sternbeck:2005tk,Aguilar:2006gr,Braun:2007bx,Fischer:2008uz,Binosi:2009qm,Cyrol:2017ewj,Huber:2018ned,Falcao:2020vyr,Eichmann:2021zuv,Ferreira:2023fva}
and references therein.

%vonSmekal:1997ohs, Fischer:2002hna,Fischer:2006ub,

\begin{figure}
	\centering
	\includegraphics[width=1\textwidth]{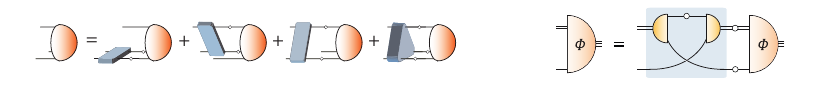}
	\caption{Left: Covariant Faddeev equation for a baryon. 
             Right: Quark-diquark BSE for a baryon. Single lines with circles represent quarks and double lines diquarks.}
	\label{fig:faddeev}
	\vspace{-1mm}
\end{figure}

Hadrons appear as poles  
in QCD's correlation functions, like the $qqq$ correlation function in the second row of Fig.~\ref{fig:poles}.
An efficient way to extract the mass of a hadron, like e.g. a nucleon resonance, is again the Bethe-Salpeter equation (BSE)
discussed in Sec.~\ref{sec:eft},
but now formulated in terms of quarks and gluons.
One writes down a scattering equation
$\mathbf{T} = \mathbf{K} + \mathbf{K}\,\mathbf{G}_0\,\mathbf{T}$,
where $\mathbf{T}$ is now the connected and amputated part of the $qqq$ correlation function (i.e., without external quark propagators). 
$\mathbf{G}_0$ is the product of  three dressed quark propagators which depend on the quark mass function in Fig.~\ref{fig:dses},
so the quarks are also no longer onshell particles with a simple mass pole and a three-dimensional reduction is no longer possible.
Because $\mathbf{T}$ is amputated, its residue at a given hadron pole is the hadron's Bethe-Salpeter amplitude $\mathbf{\Gamma}$  
defined through  $\Psi = \mathbf{G}_0\,\mathbf{\Gamma}$, where $\Psi$ is the Bethe-Salpeter wave function (BSWF) from Eq.~\eqref{poles}.
Comparing the residues on both sides of the scattering equation then yields the homogeneous BSE at the pole, 
\begin{eqnarray}\label{hom-bse}
	 \mathbf\Gamma = \mathbf{K}\,\mathbf{G}_0\,\mathbf\Gamma\,.
\end{eqnarray}
For a three-quark system this is the covariant Faddeev equation  in the left of Fig.~\ref{fig:faddeev}.
It is formally still an exact equation, where the kernel $\mathbf{K}$ is the sum of all two- and three-body irreducible interactions.
(Note that crossing symmetry and the resulting left-hand cuts are not an issue  for the homogeneous BSE because one compares the residues of $s$-channel poles.)
In practice the equation is solved by calculating the eigenvalue spectrum
of the equation $\mathbf{K}\,\mathbf{G}_0\,\mathbf\Gamma = \lambda\, \mathbf\Gamma$ as a function of the total momentum $P^2$;
if an eigenvalue satisfies $\lambda_i(P^2= m_i^2) = 1$, one can read off the hadron mass $m_i$.

In this sense, the homogeneous BSE takes over the role of the Schrödinger equation from quantum mechanics.
The BSWF of a hadron with given quantum numbers, defined as  the residue of the hadron pole,
 is  the QFT analogue of the quantum-mechanical wave function except it does not have a probability interpretation.
 Its general structure is given by $\Psi = \Psi_\text{dyn} \otimes \Psi_\text{flavor}\otimes \Psi_\text{color}$, 
where the dynamical part represents the full Lorentz group including the spin.
The antisymmetric color and the flavor parts are identical to those in the quark model, 
% Eq.~\eqref{total-wf-generic}, \EK{Should this be
% Table~\ref{tab:baryons-flavor-2}?
% }
 but the dynamical part $\Psi_\text{dyn}$ is richer and not just the product of spatial and spin wave functions.
In the case of a baryon, it assumes the following form in momentum space:
 \begin{eqnarray}\label{faddeev-wf}
 	  \Psi_\text{dyn}(p,q,P)_{\alpha\beta\gamma\delta} = \sum_{i} f_i(p^2, q^2, p\cdot q, p\cdot P, q\cdot P) 
 	                                                     \, \tau_i(p,q,P)_{\alpha\beta\gamma\delta}\,.
 \end{eqnarray}
 The wave function has three Dirac indices for the quarks and one for the baryon, and
 depending on the spin $J$ there may also be further Lorentz indices.
 It depends on three independent momenta, e.g. two relative Jacobi momenta $p$ and $q$ and the total momentum $P$.
 The wave function can  be expanded in a Lorentz-covariant tensor basis,
 which yields 64 linearly independent tensors $\tau_i(p,q,P)$ for $J=1/2$ states  and 128 for $J=3/2$ states~\cite{Eichmann:2009qa,Sanchis-Alepuz:2011egq}.
 From three independent momenta one can form six Lorentz invariants, one of them being onshell ($P^2 = m^2$),
 so the dressing functions $f_i$ depend on five momentum variables.

\begin{figure*}
	\centering
	\includegraphics[width=0.96\textwidth]{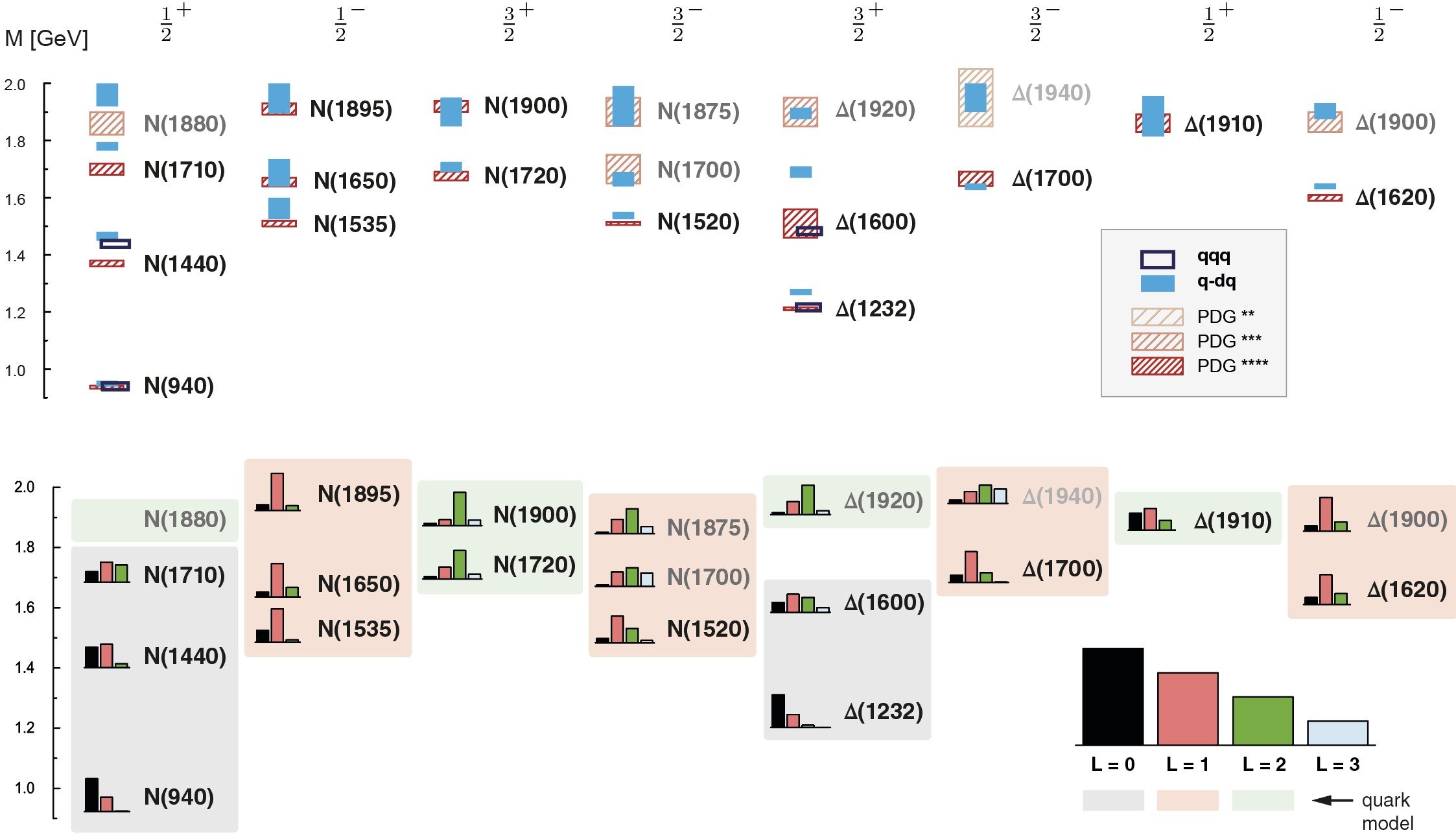}
	\caption{(color online) Top: Light baryon spectrum from the three-quark and quark-diquark calculations for different $J^P$ quantum numbers~\cite{Eichmann:2016hgl,Eichmann:2016nsu}, 
		compared to the RPP masses (real parts of pole positions). %~\cite{ParticleDataGroup:2020ssz}.
		Bottom: Orbital angular momentum contributions to each baryon from the quark-diquark calculation. The bars sum up to 100\%, and the light-colored boxes drawn around them
		are the quark-model values for $L$ from Fig.~\ref{fig:missing-resonances}.
%        {\rd Is this the angular momentum between a quark and a diquark, all with current quark masses?}
        }
	\label{fig:spectrum-qdq}
\end{figure*}

 What are the new effects compared to the quark-model wave functions  discussed in Sec.~\ref{Construction}?
 One can  arrange the $\tau_i$ in Eq.~\eqref{faddeev-wf}  into eigenstates of the three-quark spin $S$ and orbital angular momentum $L$ 
 in the baryon's rest frame, 
 which results in tensors with $L=0$ ($s$ waves), $L=1$ ($p$ waves), $L=2$ ($d$ waves), and so on~\cite{Eichmann:2016yit}.
 These components mix together in the BSWF to produce a baryon with definite $J^P$.
For example, the nucleon  flavor wave function is a doublet $\mD_f$,
 so one can  cast Eq.~\eqref{faddeev-wf}   %the $f_i \,\tau_i$ in 
 in permutation-group doublets $\mD_i$ to obtain a fully symmetric wave function 
 $\Psi = \sum_{i=1}^{64} \mD_i\cdot \mD_f$ (modulo color).
 One of these components is $\psi_1 = (\mS_o\,\mD_s) \cdot \mD_f$ from Fig.~\ref{fig:flavor-wfs},
 which we earlier identified with `the' nucleon wave function.
 However, there are other $L=0$ contributions which are similarly important, 
and relativistic $p$ waves with $L=1$ 
  contribute another substantial fraction of about 33\% to the nucleon's normalization~\cite{Eichmann:2011vu}, as we will discuss below in connection with Fig.~\ref{fig:spectrum-qdq}.
 Relativity is therefore important for light baryons,
 and relativistic effects even persist up to bottom baryons~\cite{Qin:2018dqp}.

The homogeneous BSE~\eqref{hom-bse} is equally applicable for bound states and resonances, just like Eq.~\eqref{poles} holds in both cases.
To extract a resonance mass, however, one must analytically continue the eigenvalues to unphysical Riemann sheets, which can be  technically challenging~\cite{Eichmann:2019dts,Santowsky:2020pwd}.
The situation  is also more involved than with EFTs,
where the particles in the loops are hadrons and so the multihadron cuts are automatic.
In Fig.~\ref{fig:faddeev}, the particles in the loops are quarks and gluons, so the BSE must simultaneously  create hadron poles
dynamically from the quark level but also the multihadron  cuts, which create the Riemann sheet structure to allow for resonances,
all while avoiding decay channels for quarks and gluons as required by confinement. 
Such a mechanism has been  demonstrated  in Ref.~\cite{Eichmann:2015cra}:
A four-quark BSE  implementing gluon exchanges
dynamically creates intermediate meson poles, and these
automatically also generate meson-meson cuts
which turn a four-quark ($qq\bar{q}\bar{q}$) state into a resonance.

   What is the kernel $\mathbf{K}$? 
   This is where the information from QCD's $n$-point function enters.
   The leading term in the irreducible three-body kernel in Fig.~\ref{fig:faddeev},
   which is the three-gluon vertex connecting the three quarks,
   vanishes by the color traces. 
   This hints at quark-quark correlations as the dominant forces in baryons.
   For the two-body kernels, most  studies so far have used the rainbow-ladder truncation, %,Maris:2005tt,Hilger:2014nma,Rojas:2014aka,Fischer:2014xha},
   where the BSE kernel turns into an effective gluon exchange~\cite{Maris:1997tm,Maris:1999nt,Qin:2011dd}.
   This preserves chiral symmetry and its dynamical breaking pattern:
    The quark DSE dynamically generates quark masses like in Fig.~\ref{fig:dses},
    and the pseudoscalar mesons calculated from their BSEs are the massless Goldstone bosons in the chiral limit.
    In the light-meson sector, rainbow-ladder works  well for pseudoscalar and vector mesons, whereas scalar and axialvector mesons are too strongly bound.
    Going beyond rainbow-ladder by  implementing higher $n$-point functions %and thus the information from the gluon sector   
    is computationally demanding,
    but  efforts have been made and improved the meson spectrum~\cite{Chang:2011ei,Williams:2015cvx,Eichmann:2016yit,Huber:2020ngt}.

  Concerning the baryon spectrum, one strategy has been to
  solve the three-body equation in  rainbow-ladder  without further approximations~\cite{Eichmann:2009qa,Eichmann:2011vu,Sanchis-Alepuz:2011egq,Sanchis-Alepuz:2014sca,Qin:2018dqp,Qin:2019hgk,Yao:2024ixu}.
  The other is to simplify the three-body equation to a quark-diquark BSE shown in the right panel of Fig.~\ref{fig:faddeev}, 
  where gluons no longer appear and the interaction becomes a ping-pong quark exchange,
  with  quarks forming (non-pointlike) diquarks in the process~\cite{Oettel:1998bk,Eichmann:2008ef,Eichmann:2016hgl,Eichmann:2016yit,Chen:2017pse,Chen:2019fzn,Barabanov:2020jvn,Liu:2022ndb}. 
  Fig.~\ref{fig:spectrum-qdq} shows the resulting light baryon spectrum~\cite{Eichmann:2016hgl}, which is very similar in both cases.
  In particular, the nucleon mass comes out at $940$ MeV in both approaches without any tuning.
  Diquark clustering is  a good way to think about baryons!
  
The lightest diquarks are the scalar diquarks 
with masses  $\sim 0.8$ GeV (the `good diquarks' in the literature),
which are the parity partners of the pseudoscalar mesons,
followed by the axialvector (`bad') diquarks at $\sim 1$ GeV as  partners of the vector mesons,
follwed by pseudoscalar and vector (`ugly') diquarks as partners of  scalar and axialvector mesons, and so on.
The fact that scalar and axialvector mesons are too strongly bound in rainbow-ladder also translates to their diquark partners
and the resulting baryons. 
These are the orbital excitations in the quark model, i.e., nucleons with $J^P = 1/2^-$, $3/2^\pm$ and $\Delta$ baryons with $J^P = 3/2^-$, $1/2^\pm$,
which come out too light compared to experiment.
This happens both in the three-quark and quark-diquark approach, even though the former does not know about diquarks.
If the strength in the pseudoscalar and vector diquark channels is adjusted to the meson spectrum to simulate  beyond-rainbow-ladder effects, 
one finds a 1:1 correspondence with experiment for the resulting baryon spectrum, as shown in Fig.~\ref{fig:spectrum-qdq}.
No parameters related to baryons have been introduced: 
The light current-quark mass is fixed by the pion mass and  the extra diquark parameter by the axialvector meson masses.
   The effective quark-gluon interaction depends on a scale and a shape parameter, where
   the former is fixed by the  pion decay constant $f_\pi= 92$ MeV but effectively drops out
   when plotting $M/f_\pi$, which  leaves Fig.~\ref{fig:spectrum-qdq} unchanged,
   and the latter has a negligible effect on the spectrum.

An interesting observation in Fig.~\ref{fig:spectrum-qdq} is the correct level ordering between the 
Roper resonance $N(1440)\nicefrac{1}{2}^+$ and the nucleon's parity partner
$N(1535)\nicefrac{1}{2}^-$. This has not so much to do with the Roper resonance but rather with the 
$N(1535)$, which in contrast to the Roper is very sensitive to the higher-lying (`ugly') pseudoscalar and vector diquarks.
In the rainbow-ladder calculation the $N(1535)$ mass comes out too low, along with many other states
in the spectrum. If the strength of these `ugly' diquarks is reduced, 
the $N(1535)$ mass moves up, overtakes the Roper resonance
and ends up in the proximity of the mass of the $N(1650)$  (which is also not sensitive to the higher-lying diquarks).
Corresponding plots can be found in Ref.~\cite{Barabanov:2020jvn}.
In the three-body calculation,  where diquarks do not  appear explicitly, 
this means that the $N(1535)$ is very sensitive to effects beyond rainbow-ladder and, therefore,
gluon exchange alone is  not sufficient to describe the $N(1535)$ well.
This is  analogous to calculations in the meson sector, where the scalar and axialvector mesons receive large repulsive corrections
beyond rainbow-ladder~\cite{Chang:2011ei,Williams:2015cvx}.  
Note also that the observed overall agreement with experiment in Fig.~\ref{fig:spectrum-qdq} is not in conflict with
the importance of meson-baryon interactions, which are 
obviously necessary to generate a resonance mechanism.
Unlike for higher multiquark states, such a mechanism is not automatic in the three-quark system  
and would enter through more advanced truncations with quark and gluon correlation functions that produce internal meson and baryon poles.
The ratio $M/f_\pi$ (which is essentially what is shown in Fig.~\ref{fig:spectrum-qdq}) would also remain unchanged if meson-baryon interactions 
affected the masses and $f_\pi$ proportionally. %, as it happens in several models~\cite{a}.
In any case, as expected from our discussion in Sec.~\ref{sec:whatisabaryon}, 
these results imply that all  nucleon excitations below 2 GeV have a three-quark core.

It is also  interesting to study the internal composition of these baryon excitations~\cite{Eichmann:2016hgl,Eichmann:2016nsu,Chen:2017pse,Liu:2022nku,Liu:2022ndb}. % in terms of their orbital angular momentum contributions.
The bars in the lower panel of Fig.~\ref{fig:spectrum-qdq} show the orbital angular momentum contributions to each baryon's normalization.
These come from the various tensors contributing to the wave functions discussed in connection with Eq.~\eqref{faddeev-wf},
i.e., each baryon is a mixture of different $L$ components in the rest frame.
For comparison, the light-colored boxes show the quark-model values for $L$ in the harmonic oscillator assignment according to Fig.~\ref{fig:missing-resonances}, 
where each baryon carries a specific value of $L$.
For the majority of states one can see clear traces of the nonrelativistic quark model,
e.g., the nucleon states with $J^P = 1/2^-$ are dominated by $L=1$ and those with $J^P=3/2^+$ by $L=2$. 
But there are also clear deviations:
The nucleon and $\Delta(1232)$ ground states are dominated by $L=0$  but have substantial (relativistic) $L=1$ contributions,
and for several states the nonrelativistic assignment does not work at all.

As will be discussed in Sec.~\ref{Structure}, further information on the internal structure of nucleon resonances
can be extracted from nucleon-to-resonance transition form factors.
With functional methods, these have been studied in a variety of different approximations
ranging from contact interactions~\cite{Segovia:2013rca,Raya:2021pyr,Yin:2021uom,Albino:2025fcp,Albino:2025bnr},  % ,
the quark-diquark approach either with model input~\cite{Segovia:2015hra,Segovia:2016zyc,Chen:2018nsg,Lu:2019bjs,Chen:2023zhh} or  
calculated rainbow-ladder ingredients~\cite{Eichmann:2011vu}, or in the three-quark approach~\cite{Sanchis-Alepuz:2017mir}.
Like for the spectrum, the three-quark and quark-diquark calculations
show good overall agreement, see the reviews~\cite{Eichmann:2016yit,Barabanov:2020jvn}.
A typical common feature is missing strength at low $Q^2$, which in some cases can be clearly  attributed to 
missing meson-cloud effects, e.g. for the nucleon elastic electromagnetic form factors: In that case, the meson effects 
are enhanced in the isovector combination but cancel  in the isoscalar one, and the isoscalar anomalous magnetic moment
from the three-body equation is in agreement with the experimental result $\kappa_s = -0.12$~\cite{Eichmann:2011vu}.
In other cases, discrepancies that are sometimes attributed to the meson cloud  can be explained by the 
relativistic $L=1$ components in the baryon wave functions, like for the $R_{EM}$ ratio in the $N\to\Delta\gamma$ transition~\cite{Eichmann:2011aa}.
For the Roper resonance, the transition form factor $F_2$ shows the zero crossing expected from a radial excitation,
which  points towards an underlying three-quark composition~\cite{Segovia:2015hra}.
Especially important in the timelike region ($Q^2 < 0$) are the vector-meson poles, which are part of the dressed quark-photon vertex and 
automatically generated from the quark-gluon interactions if the vertex is calculated self-consistently~\cite{Maris:1999bh}. More advanced truncations
that are currently being explored also capture the $\rho\to\pi\pi$ resonance mechanism which turns the $\rho$ meson into a resonance~\cite{Miramontes:2019mco,Miramontes:2021xgn}.

Similar plots as Fig.~\ref{fig:spectrum-qdq} can  be drawn for hyperons~\cite{Fischer:2017cte,Eichmann:2018adq} 
and heavy baryons~\cite{Qin:2018dqp,Qin:2019hgk,Torcato:2023ijg}.
Especially interesting are charmed baryons such as $\Sigma_c$ and $\Lambda_c$, which consist of two light quarks and one charm quark. 
  Denoting up and down by $n$, it is often assumed that charmed baryons form `planetary' systems made of a light $(nn)$ diquark 
  that `orbits' around the heavier charm quark.
  However, the Faddeev calculations do not support this picture~\cite{Yin:2019bxe,Torcato:2023ijg}. 
  Rather, the $(nc)n$ configurations are dominant and contribute about $60\%$ to the norm of the state,
  while the $(nn)c$ configurations contribute only $40\%$. Therefore, a charmed baryon actually spends most of its time in a $(nc)n$ configuration.

To summarize, functional methods provide a systematically improvable approach to compute hadron properties from nonperturbative QCD.
Although they are limited by the necessity of employing truncations, the studies so far cover a broad spectrum
ranging from QCD-based models to systematic approaches whose ingredients are  calculated self-consistently. 
The main takeaways regarding the composition of baryons can be summarized as follows: 
(i) Spontaneous chiral symmetry breaking and the associated dynamical mass generation
are essential features to explain the light baryon spectrum;  (ii) two-body correlations in the form of diquark clusters
are important for the binding of baryons; (iii) relativity is a central ingredient of the light baryon spectrum 
and induces effects that are not present in nonrelativistic quark models; and (iv) meson-baryon effects
can play an important role in the low-$Q^2$ behavior of nucleon-to-resonance transition form factors.

    \begin{figure}
	\centering
 \includegraphics[width=0.42\columnwidth]{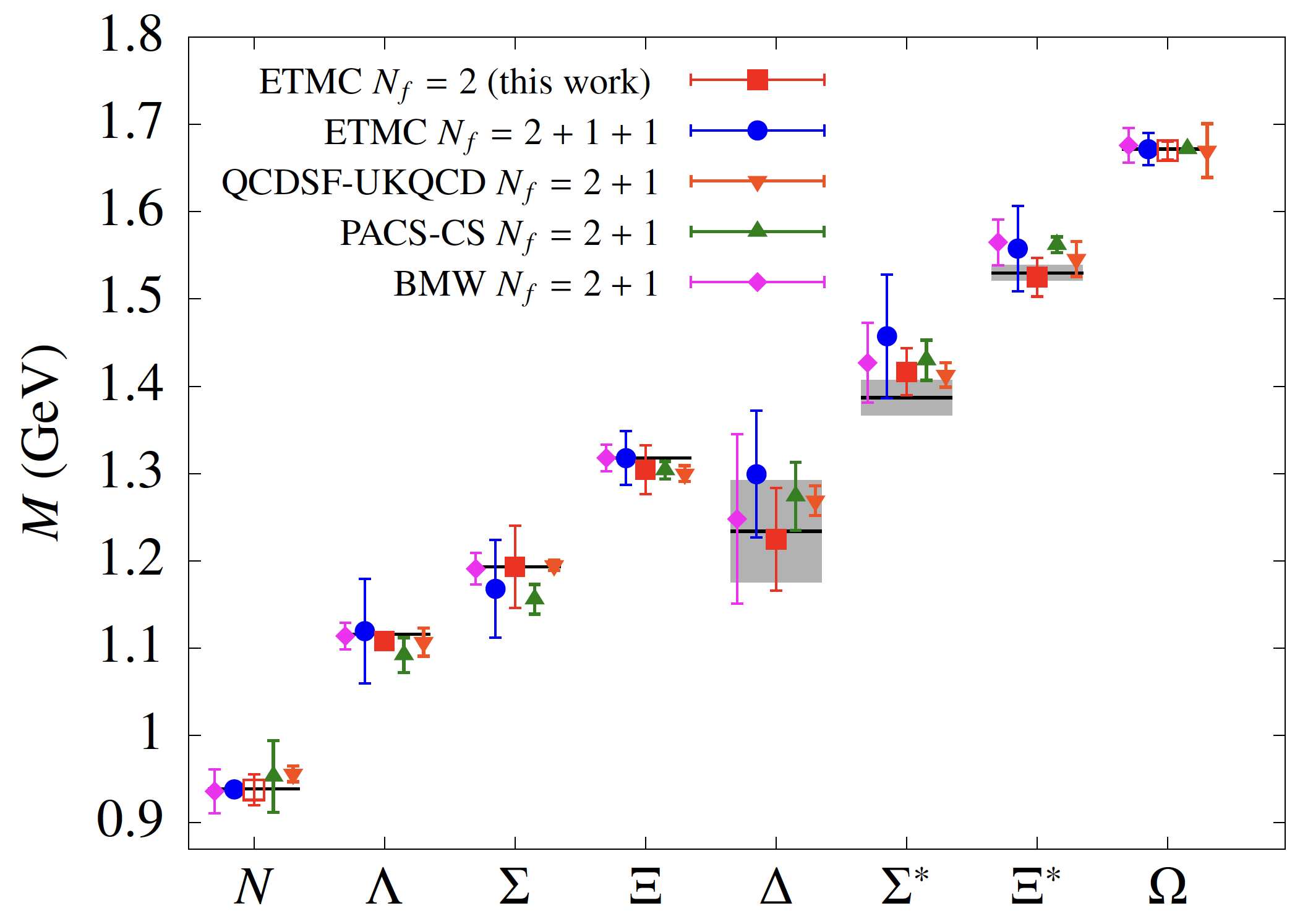} \quad  
    \includegraphics[width=0.54\columnwidth]{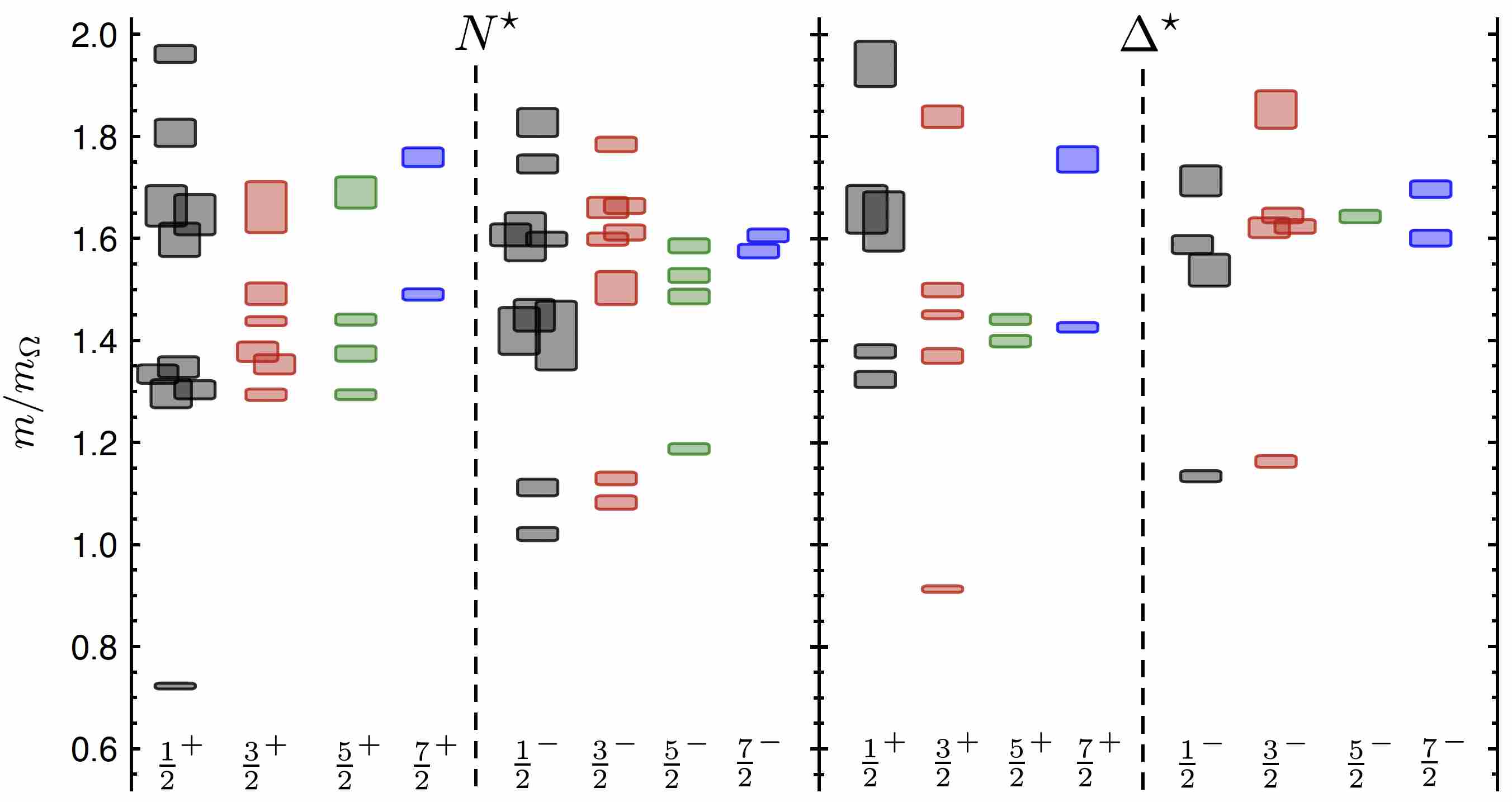}
    \caption{(color online) Left: Comparison of baryon octet and decuplet masses from different lattice calculations~\cite{PACS-CS:2008bkb,BMW:2008jgk,Bietenholz:2011qq,Alexandrou:2014sha,Alexandrou:2017xwd}.
Plot  from Ref.~\cite{Alexandrou:2017xwd}.
Right: Lattice finite-volume energy levels for $N^*$ and $\Delta^*$ excitations  at a pion mass of 396~MeV~\cite{Edwards:2011jj}. 
     }
    \label{fig:lattice}
    \vspace{3mm}
\end{figure} 

 \subsection{Baryons in lattice QCD}\label{sec:lattice}

Lattice QCD is a nonperturbative ab-initio method because in principle, apart from supplying the current-quark masses and setting a scale, the only information needed
to compute baryon observables is the QCD action~\cite{Gattringer:2010zz,Briceno:2017max}.
The basic idea in lattice QFT is to discretize Euclidean spacetime 
and consider a four-dimensional box with lattice spacing $a$, length $L$ and appropriate boundary conditions.
Correlation functions inside that box are then calculated through the Euclidean version of the path integral in Eq.~\eqref{path-int-2}.
The path integral over the quark and antiquark fields can be rewritten as a fermion determinant,
so that the remaining integral amounts to  statistically sampling over many gauge (i.e., gluon) field configurations.
The typical workflow of lattice calculations  consists of three steps: 
generating the gauge-field ensembles that enter in the path integral;
`measuring' the correlation functions $\langle \mO \rangle$ on those ensembles; and 
extracting physical observables from the correlation functions.
The last part comes with several complications
mainly induced by the finite volume ($L < \infty$), the non-zero lattice spacing ($a>0$), and the difficulties
in going to physical pion masses due to the accumulating statistical noise. 

The starting point for extracting the baryon spectrum on the lattice is
again Fig.~\ref{fig:poles}.
To begin with, suppose we start with some (gauge-dependent) correlation function that permits baryon poles,
like the $qqq$ correlation function in the second row on the right.
By contracting it with appropriate structures in Dirac, color and flavor space,
we arrive at a (gauge-invariant) two-point correlator which is shown in the third row.
This two-point function only depends on one variable $z=x-y$ 
and can be interpreted as a baryon-baryon correlator $G(z) = \langle \mathcal{O}_i(x)\,\bar{\mathcal{O}}_j(y) \rangle$,
where $\mathcal{O}_i(x)$ is some gauge-invariant combination of three quark operators.
In Euclidean momentum space, this correlator depends on the total baryon momentum $P$
and inherits the poles from the original correlation function at $P^2 = -m_\lambda^2$:
\begin{eqnarray}\label{lattice-poles}
	G(P) \stackrel{P^2 \to -m_\lambda^2}{\longlonglongrightarrow} \frac{R_\lambda}{P^2 + m_\lambda^2} = \frac{R_\lambda}{P_4^2 + E_\lambda^2}\,,
\end{eqnarray}
where $E_\lambda^2 =  \vect{P}^2 + m_\lambda^2$ and
the overlap $R_\lambda$ is  the product of the contracted Bethe-Salpeter wave functions.
When taking a Fourier transform in $P_4$, a pole in momentum space turns into an exponential decay
in Euclidean time $z_4 = \tau$:
\begin{eqnarray}
	G(\tau,\vect{P}) \stackrel{P^2 \to -m_\lambda^2}{\longlonglongrightarrow} R_\lambda \int \frac{dP_4}{2\pi}  \frac{e^{iP_4 z_4}}{P_4^2 + E_\lambda^2} 
	= \frac{R_\lambda}{2E_\lambda} e^{-E_\lambda \tau}\,.
\end{eqnarray}
Setting $\vect{P}=0$, one can then extract the mass $m_\lambda$ at large Euclidean times $\tau$.
In general Eq.~\eqref{lattice-poles} will be a sum of poles, but for large $\tau$
the ground state with the lowest-lying mass  dominates the sum.

While the above equations  sketch the basic ideas,  a clean extraction of hadronic states from lattice QCD
requires overcoming significant technical challenges. 
Over the years, various methods have been developed to address these issues and reduce the computational cost,
including distillation~\cite{HadronSpectrum:2009krc,Morningstar:2011ka}, noise reduction techniques~\cite{Bali:2009hu},
the inclusion of large operator bases also with multihadron operators 
to solve generalized eigenvalue problems, 
and matching the irreducible representations of the cubic group
to physical $J^P$ quantum numbers; see e.g.~\cite{Dudek:2010wm,Thomas:2011rh,Dudek:2012gj,Edwards:2012fx} for  details. 
From the perspective of Fig.~\ref{fig:poles}, the inclusion of multihadron and gluonic operators 
for solving generalized eigenvalue problems of two-point functions  $\langle \mathcal{O}_i\,\bar{\mathcal{O}}_j\rangle$
amounts to finding a given hadron pole
in a matrix of $qqq$, $(qqq)(q\bar{q})$, $(qqq)g$ \dots correlation functions,
which therefore also provides insight into their multiquark and gluonic admixtures. 
In this way, for instance, the existence of hybrid baryons with significant gluonic components has been demonstrated~\cite{Dudek:2012ag}.

Lattice calculations of stable hadrons are well established by now, see e.g.~\cite{PACS-CS:2008bkb,BMW:2008jgk,Bietenholz:2011qq,Lin:2011ti,Fodor:2012gf,Engel:2013ig,Alexandrou:2014sha,Alexandrou:2017xwd}. 
In the absence of electroweak interactions, the ground-state octet baryons $N$, $\Lambda$, $\Sigma$, $\Xi$ and the decuplet $\Omega$ baryon
are stable under the strong interaction. 
Thanks to improved algorithms, calculations of their masses have also become possible at physical pion masses as shown in the left panel of Fig.~\ref{fig:lattice}.
This also allows one  to  investigate precision effects like isospin breaking and electromagnetic corrections to the masses~\cite{Blum:2010ym,BMW:2014pzb,Horsley:2015eaa,NPLQCD:2020ozd}.
The FLAG (Flavour Lattice Averaging Group) review~\cite{FlavourLatticeAveragingGroupFLAG:2021npn,FlavourLatticeAveragingGroupFLAG:2024oxs}
provides a comprehensive meta-analysis of results for several  hadronic quantities ensuring strict quality criteria 
such as continuum extrapolations, finite-volume corrections and renormalization procedures.

\begin{figure}[!t]
    \centering
	\includegraphics[width=0.49\columnwidth]{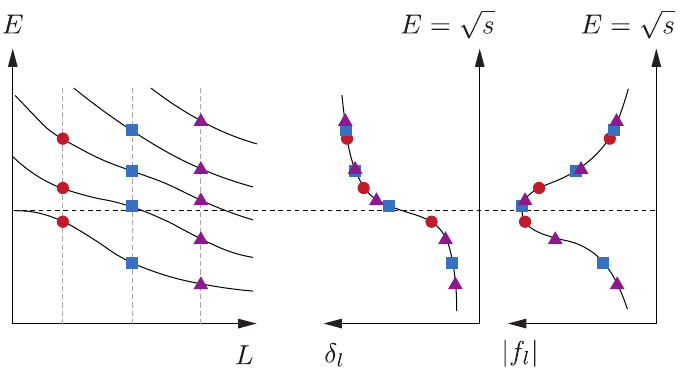} \quad\quad
    \includegraphics[width=0.40\columnwidth]{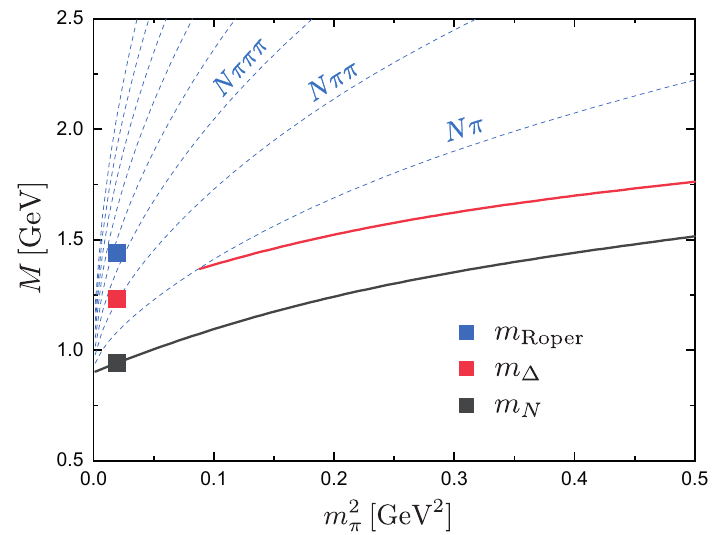}   
    \vspace{2mm}
	\caption{(color online) Left and center: Sketch of the Lüscher method. From the know\-ledge of the finite-volume spectrum
		at different lattice volumes one can reconstruct the phase shifts and hence the scattering amplitude.
        Right: Sketch of the $N\pi$, $N\pi\pi$, \dots thresholds (dashed) plotted over the squared pion mass.
 		The squares are the experimental values for the nucleon, $\Delta$ and Roper mass at the physical pion mass.
 		The solid curves are parametrizations of lattice results for $m_N$ and $m_\Delta$ (see~\cite{Eichmann:2016yit} for references).}
	\label{fig:luescher}
\vspace{0mm}
\end{figure}

In contrast to stable ground states, the extraction of
unstable hadrons above thresholds is substantially more difficult on the lattice.
The reason for this is the finite lattice volume $L$, which does not admit branch cuts, Riemann sheets or resonance poles in scattering amplitudes~\cite{Maiani:1990ca}.
Instead, the energy levels calculated on the lattice correspond to discrete poles on the real axis,
which are the finite-volume analogues of the branch cuts in the infinite volume, i.e., the multiparticle continuum.
As such, the energy levels above thresholds also cannot be directly interpreted as physical states.

One way to deal with this  is 
the Lüscher formalism~\cite{Luscher:1986pf,Luscher:1990ux} including its various 
generalizations~\cite{Rummukainen:1995vs,Kim:2005gf,He:2005ey,Bernard:2008ax,Leskovec:2012gb,Fu:2011xz,Hansen:2012tf,Briceno:2012yi,Guo:2012hv,Briceno:2014oea,Agadjanov:2016mao}, 
see~\cite{Briceno:2017max,Hansen:2019nir,Mai:2021lwb} for reviews. The basic idea is best explained with the help of a Bethe-Salpeter equation.
Consider an elastic  $2\to 2$ scattering amplitude such as $N\pi \to N\pi$ (or a coupled system of such scattering amplitudes),
which creates intermediate poles for the nucleon and its excitations.
Like in Eq.~\eqref{bse-1}, we write down the generic BSE for this amplitude, once in the infinite volume and once in a finite volume 
(the latter with subscript $L$):
\begin{eqnarray}
		\mathbf{T}^{-1} = \mathbf{K}^{-1} - \mathbf{G}_0\,, \qquad
		\mathbf{T}_L^{-1} = \mathbf{K}_L^{-1} - \mathbf{G}_{0L}\,.
\end{eqnarray}
$\mathbf{T}$ is the scattering matrix in the infinite volume, which contains the physical poles.
$\mathbf{T}_L$ is its finite-volume analogue, which is still well-defined as a correlation function, with poles at the positions $\sqrt{s} = E_i(L)$ that can be calculated on the lattice.
$\mathbf{G}_0$ and $\mathbf{G}_{0L}$ are the respective two-body propagators. 
Now, these equations would not be very useful if we needed to know the kernels $\mathbf{K}$ and $\mathbf{K}_L$.
The crucial observation is that below three-particle thresholds the volume corrections to $\mathbf{K}$ are  exponentially  suppressed with $L$, 
so for large enough volumes one can set $\mathbf{K} \approx \mathbf{K}_L$ to arrive at
\begin{eqnarray}
	\mathbf{T}^{-1} - \mathbf{T}_L^{-1} = -(\mathbf{G}_0 - \mathbf{G}_{0L}) =: -\mathbf{F}\,.
\end{eqnarray}
The known function $\mathbf{F}$ is the difference of the two-body propagators.
This also implies that $\mathbf{T}_L$ does not reproduce $\mathbf{T}$ in the limit $L\to\infty$ due to the non-vanishing term $\mathbf{F}$. Instead,
for each  volume $L$ one then obtains the condition $\mathbf{T}^{-1} + \mathbf{F} = \mathbf{T}_L^{-1}$,
which must have zeros at $\sqrt{s} = E_i(L)$.
After multiplication with $\mathbf{T}$ and $\mathbf{F}^{-1}$, this turns into
\begin{eqnarray}\label{luescher}
    \det\,(\mathbf{T} + \mathbf{F}^{-1}) = 0
    \quad \text{at} \quad E = E_i(L)\,,
\end{eqnarray}
which allows one to extract the infinite-volume scattering matrix $\mathbf{T}$ 
at the values $\sqrt{s} = E_i(L)$. 
The procedure is sketched in Fig.~\ref{fig:luescher}: The left panel
illustrates the finite-volume energy levels obtained at different lattice volumes,
and the center panels show the reconstructed phase shift $\delta_l$ and corresponding scattering amplitude $|f_l|$.
There is no 1:1 correspondence between the finite-volume energy levels
and physical resonances, because the energies simply map to different values of the scattering amplitude. 
One can still infer information from finite-volume lattice spectra above thresholds;
e.g., typical signatures for resonances are avoided level crossings in $E_i(L)$, and if
the energy levels cluster in a narrow region it may hint at a nearby resonance.
In general, however, a direct comparison with the experimental spectrum is only applicable for 
stable states below the lowest strong decay threshold.

Over the past decade, large progress has been made in lattice QCD calculations of hadron masses above two-particle thresholds;
see~\cite{Briceno:2017max,Padmanath:2018zqw,Detmold:2019ghl,Brambilla:2019esw,Bulava:2022ovd,Mai:2022eur} for overviews.
Most practical applications of the Lüscher method to date have focused on the meson sector. 
The case of elastic scattering applies to resonances with a single decay channel,
such as the $\rho$ and $K^\ast$ which decay almost exclusively into $\pi\pi$ and $K\pi$, respectively. 
Other resonances like the $\sigma$ and $\kappa$ mesons~\cite{Briceno:2016mjc,Guo:2018zss,Rodas:2023nec} 
and a number of states in the heavy-meson spectrum, including  exotic candidates~\cite{Mohler:2012na,Prelovsek:2013cra,Mohler:2013rwa,Bicudo:2015vta,Francis:2016hui,Moir:2016srx,Bali:2017pdv,Cheung:2017tnt,Junnarkar:2018twb,Padmanath:2022cvl}, have also been studied by various lattice collaborations.
Since most resonances are inelastic and couple to multiple decay channels, 
also coupled-channel generalizations of the Lüscher formalism to inelastic scattering 
have been applied to a variety of systems~\cite{Prelovsek:2014swa,Dudek:2014qha,Wilson:2015dqa,Dudek:2016cru,Briceno:2017qmb,JPAC:2018zyd,Woss:2020ayi,Prelovsek:2020eiw}.
These investigations enable the determination of resonance pole positions in the complex plane 
and provide insight into their dependence on the current-quark masses. %, or equivalently, the pion mass.

The difficulty is that each additional threshold complicates the situation substantially.
For unstable states above three-particle thresholds, the corresponding BSE
turns into a genuine three-body problem.
The generalization of the Lüscher condition
to states above three-particle thresholds has  been worked out by several 
groups~\cite{Hansen:2014eka,Hansen:2015zga,Hammer:2017uqm,Hammer:2017kms,Mai:2017bge,Mai:2018djl,Jackura:2019bmu,Briceno:2019muc,Blanton:2020gha,Blanton:2020jnm}, 
see~\cite{Hansen:2019nir,Mai:2021lwb} for reviews.
This has  found various recent applications in the meson sector, 
e.g., for $\pi\pi\pi$ and $KKK$ systems~\cite{Blanton:2019igq,Horz:2019rrn,Blanton:2019vdk,Culver:2019vvu,Muller:2020wjo}, from where one can
determine the pole positions of the $a_1$(1260) and $\omega(782)$ resonances~\cite{Mai:2021nul,Yan:2024gwp}.

Such progress will be  especially important for the baryon excitation spectrum,
where many resonances couple strongly to three- or more particle channels.
%For example, the Roper resonance $N^\ast(1440)$ has a significant three-particle ($N\pi\pi$) decay mode.
Lattice calculations of baryon excitations %, see e.g.~\cite{Bulava:2010yg,Edwards:2011jj,Dudek:2012ag,Edwards:2012fx,Engel:2013ig}, 
are often  performed  at larger pion masses to reduce the numerical cost.
As sketched in the right panel of  Fig.~\ref{fig:luescher},
because the nucleon mass  increases with the pion mass, the $N\pi$, $N\pi\pi$, \dots thresholds also increase with $m_\pi$.
Therefore, states that are resonances at the physical pion mass may eventually become bound states as $m_\pi$ grows.
In that case, a lattice computation is straightforward and the energy levels directly correspond to the masses of the states. 
For example, the right panel in Fig.~\ref{fig:lattice} shows the finite-volume energy levels for the light baryon excitation spectrum 
at a pion mass of about 400 MeV~\cite{Edwards:2011jj}.
In these units the lowest $N\pi$ threshold corresponds to $m/m_\Omega \sim 1$, 
which is close to the lowest level in the $J^P = \nicefrac{1}{2}^-$ channel that  eventually becomes the $N(1535)\nicefrac{1}{2}^-$. 
However, at the physical pion mass more and more thresholds open up, as shown in  Fig.~\ref{fig:luescher}:
Already the $\Delta$ baryon is close to the $N\pi\pi$ threshold,
and the situation becomes progressively severe for higher-lying resonances above multiple thresholds --
in these cases one would need generalizations of the Lüscher condition to four-body, five-body equations and so on.

Compared to the meson sector, baryon resonances have been studied far less extensively on the lattice due to their increased computational complexity.
The $\Delta$ resonance has been investigated in elastic $N\pi$ scattering~\cite{Andersen:2017una,Silvi:2021uya,Bulava:2022vpq}, which is well justified as it decays almost exclusively into $N\pi$.
The situation becomes more intricate in the negative-parity $J^P = \nicefrac{1}{2}^-$ channel~\cite{Lang:2012db},
where several two- and also three-body decay channels are involved.
This is particularly  true for the Roper resonance, which is  observed experimentally in both $N\pi$ and $N\pi\pi$ final states.
On the lattice, the Roper has remained elusive for several reasons.
First, the finite-volume energy levels in this channel vary significantly between different lattice studies, 
see~\cite{Liu:2016rwa,Padmanath:2018zqw} for reviews and Fig.~\ref{Roper-hQCD} in Sec.~\ref{Structure}.
Notably, only calculations using chiral fermions yield a low-lying nucleon excitation 
in the energy range of the Roper resonance~\cite{Mathur:2003zf,Liu:2016rwa}, 
but the discrepancies may be traced back to diagrams encoding multihadron effects~\cite{xQCD:2019jke}.
Second, an elastic $N\pi$ scattering study using the Lüscher formalism~\cite{Lang:2016hnn} 
found no signal for  the Roper resonance, which 
implies that % $N\pi$ scattering alone is insufficient to produce a low-lying Roper and 
three-body analyses including the $N\pi\pi$ channel are essential.
Third, by combining lattice calculations with Hamiltonian effective field theory, 
Refs.~\cite{Liu:2016uzk,Kiratidis:2016hda,Wu:2017qve,Owa:2025mep} conclude that
the Roper  is a dynamically generated resonance arising from coupled-channel effects between 
$N\pi$, $N\sigma$, $N\eta$ and $\Delta\pi$, with negligible contributions from any underlying `bare' state.
Taken together, these findings highlight  the central role of multihadron dynamics 
in understanding the Roper resonance.

In Sec.~\ref{Structure} we will contrast these findings with the electromagnetic transition form factors of the Roper measured in electroproduction,
which show the characteristic zero crossing expected from a first radial excitation.
Transition form factors are matrix elements of the type $\langle R | J^\mu | H \rangle$, where $H$ is a stable hadron like the nucleon
and $R$ a resonance which can decay into two (or more) hadrons via $R\to H_1 H_2 \dots$.
The corresponding lattice formalism~\cite{Lellouch:2000pv,Agadjanov:2014kha,Briceno:2014uqa,Briceno:2015csa,Agadjanov:2016fbd,Briceno:2021xlc}
is similar to spectroscopy calculations, i.e., 
one looks for pole positions in the matrix elements $\langle H_1  H_2 \dots | J^\mu | H \rangle$, 
which can be related to their finite-volume analogues  on the lattice. 
Such matrix elements have been explored in the meson sector, e.g., for the process $\pi\gamma^\ast \to \rho \to \pi\pi$~\cite{Briceno:2015dca,Briceno:2016kkp,Alexandrou:2018jbt}, and it would be very interesting to extend them to the baryon sector.

Concerning the $\Lambda(1405)$, most  lattice calculations until recently were limited to determining the lowest finite-volume energy level using three-quark operators.
Based on a form factor analysis, the authors of Ref.~\cite{Hall:2014uca} argue that this state is primarily a $\bar{K}N$ molecule.
Recently, a  lattice study of coupled-channel $\pi\Sigma - \bar{K}N$ scattering at 
$m_\pi = 200$ MeV  has found two poles for the $\Lambda(1405)$~\cite{BaryonScatteringBaSc:2023zvt,BaryonScatteringBaSc:2023ori},
which is consistent with analyses in chiral EFT~\cite{Meissner:2020khl,Mai:2020ltx,Mai:2022eur,Guo:2023wes}.

To summarize, even though lattice calculations for baryon resonances remain highly challenging,
progress is being made in connecting finite-volume spectra to physical scattering amplitudes through multihadron dynamics. 
The Lüscher method and its extensions have  stimulated  
a productive interplay between lattice QCD, effective field theory and amplitude analyses~\cite{JPAC:2021rxu,Mai:2022eur}.
In parallel, other formalisms to extract infinite volume scattering information from the lattice such as 
the HALQCD method~\cite{Ishii:2006ec,Aoki:2009ji,Ishii:2012ssm,Aoki:2012tk}, the %aforementioned
finite-volume Hamiltonian EFT~\cite{Hall:2013qba,Wu:2014vma},   
the optical potential method~\cite{Agadjanov:2016mao}, and the determination of spectral densities from Euclidean correlators~\cite{Bulava:2019kbi,Bulava:2023mjc,Patella:2024cto}
are being actively investigated.  
Thanks to improved algorithms, lattice calculations near or  at the physical pion mass have now become feasible,
and the treatment of multi-particle thresholds will be a key ingredient for a reliable determination of the baryon excitation spectrum.

\newpage

\clearpage

\section{\label{Structure}Structure of Excited States}
\begin{figure}[bht]
\begin{minipage}{.54\linewidth}
\includegraphics[width=\linewidth]{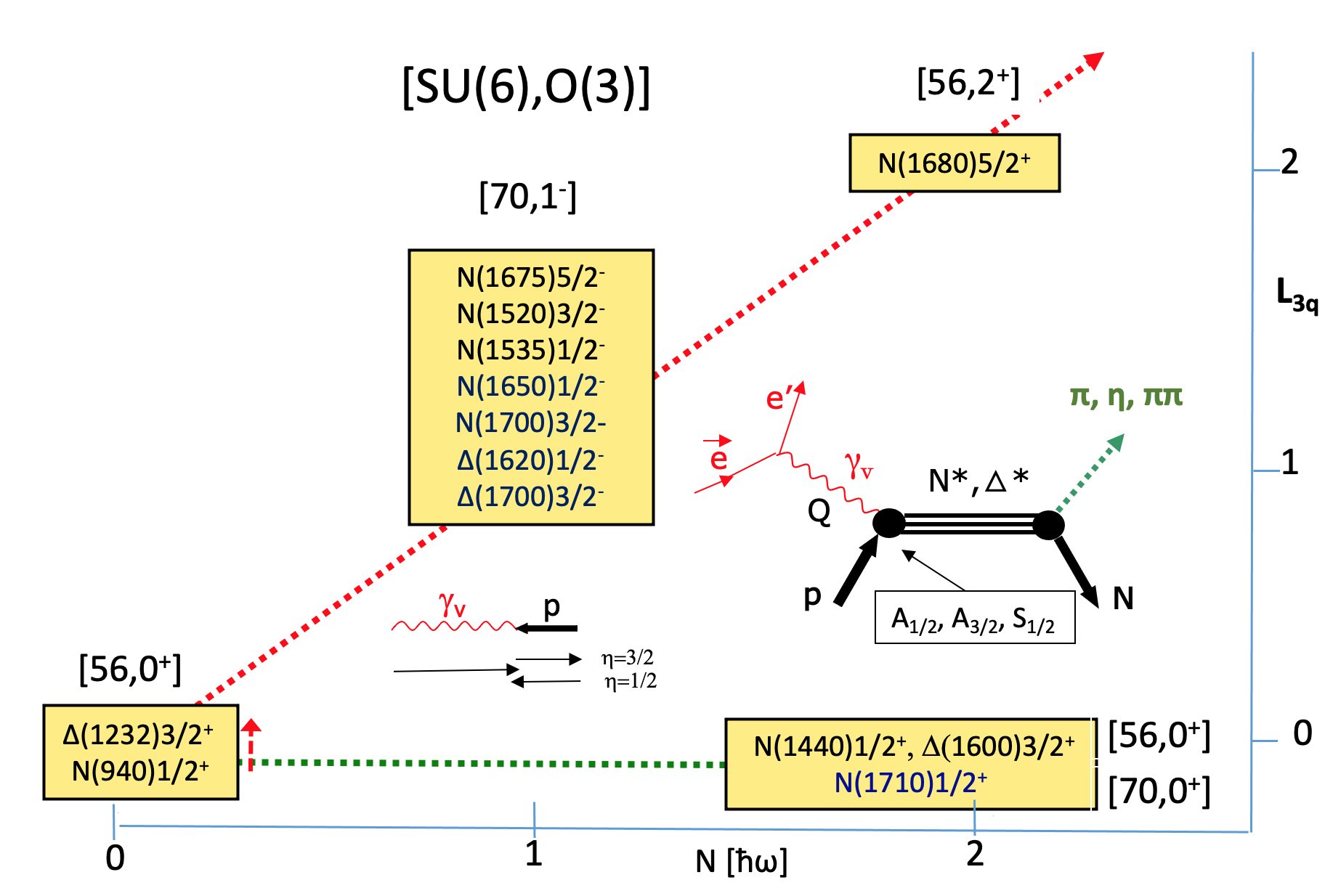}
\caption{Excited states of the proton that have been studied in electroproduction and photoproduction to determine their resonance transition amplitudes or transition form factors  and will be discussed in this review. The states are shown with respect to their assignments to supermultiplets in the quark model with $U(6)\otimes O(3)$ symmetry. The insert shows the s-channel diagram for the resonance transition from the proton to the excited state. The transition is  described by two (for spin-1/2 states) or three (for higher spins) electrocoupling transition amplitudes. Graphics from Ref.~\cite{Burkert:2019kxy}.}
\label{SU6}
\end{minipage}
\hspace{5mm}
\begin{minipage}{.45\linewidth}
\includegraphics[width=1\linewidth]{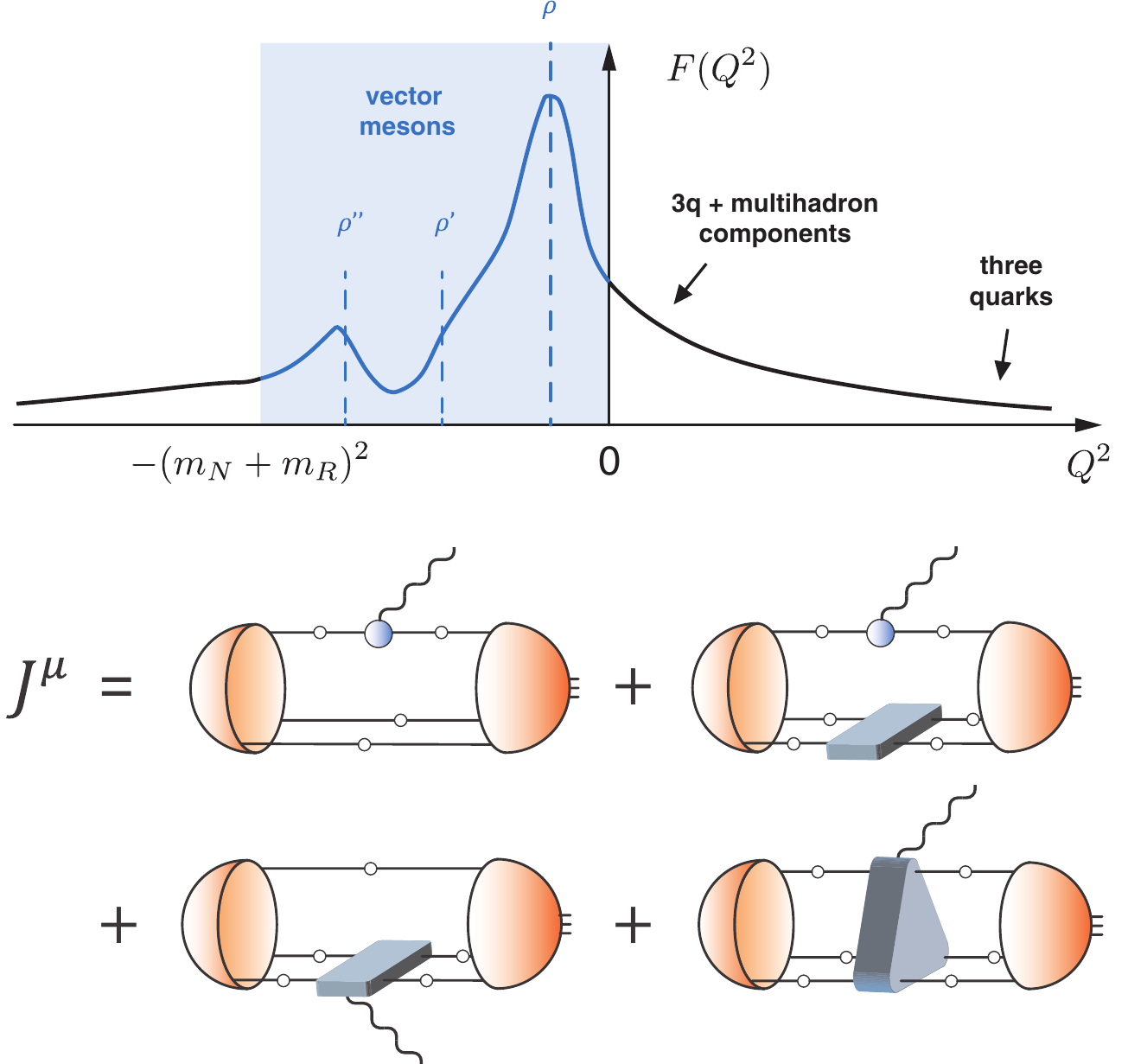}
 \caption{\footnotesize Top: Sketch of a nucleon-to-resonance transition form factor at timelike and spacelike values of $Q^2$.
  Bottom: Microscopic decomposition of the respective current matrix element, where the photon can couple to one, two or all three quarks. }
\label{fig:ffs}    
\end{minipage}
\vspace{-3mm}
\end{figure}

The availability of electron accelerators generating high beam currents, such as CEBAF at Jefferson Lab in the US and MAMI at Mainz in Germany, has enabled precise studies of the internal structure of excited $N^*$ and  $\Delta^*$ states employing s-channel 
resonance excitation in large ranges of the photon virtuality $Q^2$. This allows one to access the degrees of freedom relevant in the resonance 
excitation as a function of the distance scale probed.
Several excited states, shown in Fig.~\ref{SU6} assigned to their primary $SU(6) \otimes O(3)$ supermultiplets, have been studied this way, mostly with CLAS in Hall B.

Most of the resonance couplings have been extracted from the production of single pseudoscalar mesons. In electroproduction, there are six complex helicity amplitudes, requiring a minimum of 11 independent measurements for a complete  model-independent determination of the amplitudes (with the exception of an overall phase that cannot be determined). 
Measurements of isospin amplitudes 
require additional measurements. 
Following this, the complex amplitudes would need to be subjected to analyses of their 
phase motions to determine resonance masses on the (real) energy axis, or poles in the (complex) energy plane.        
Fortunately, in the lower mass range, a variety of constraints can be applied to limit the number of unknowns when fitting the cross sections and polarization observables.   
These include the masses of states quite well known from hadronic processes or from meson photoproduction. Also, the number of possible
angular momenta is limited to $l_\pi \le 3$ in the examples discussed in the following. Additional constraints come from the Watson theorem~\cite{Watson:1954uc}, which 
relates the electromagnetic phases to the hadronic ones, and from dispersion relations, which assume the imaginary parts of the amplitude is  
given by the resonance contribution, and the real parts are determined through dispersion integrals and additional non-resonant pole terms. Other approaches
use unitary isobar models, which parameterize all known resonances and background terms and unitarize the full amplitudes in a K-matrix
procedure.  In the following, we show results based on both approaches, where additional systematic uncertainties have been derived from the differences in the two procedures.   

\subsection{General remarks} 

Electromagnetic transition form factors are interesting examples for highlighting the duality between hadronic and quark-based descriptions.
On the one hand, we know from dispersion relations that the knowledge of the imaginary part of a form factor at timelike momentum transfer ($Q^2 < 0$)
is sufficient to reconstruct the form factor for any value of $Q^2$, including spacelike values ($Q^2 > 0$).
As sketched in Fig.~\ref{fig:ffs}, for small and intermediate values of $q^2 = -Q^2$ the timelike behavior is dominated by vector-meson bumps,
which originate from their respective poles on higher Riemann sheets, and the associated multihadron cuts like $\pi\pi$, $K\bar{K}$ etc.
For larger $q^2$ the hadronic structure is washed out and eventually disappears in the multiparticle continuum. 
The region below the two-baryon threshold $q^2 < (m_N + m_R)^2$, where $m_N$ is the nucleon mass and $m_R$ the mass of the resonance, 
can be probed via dilepton decays, while the region above the threshold is experimentally
accessible in $e^+e^-$ scattering.

On the other hand,  these form factors can also be calculated microscopically from the diagrams shown in the bottom of Fig.~\ref{fig:ffs}.
This follows from coupling a photon to the $qqq\to qqq$ scattering amplitude and reading off the residue at the ($N$, $N^\ast$) double pole~\cite{Eichmann:2016yit}.
Since no assumptions are made, this is also an exact expression but now formulated in terms of quarks.
Nevertheless, it still contains all hadronic effects: The timelike vector-meson poles have their origin in the quark-photon vertex, which describes the 
coupling of the photon to a quark. The meson-baryon intermediate states, on the other hand, can only appear in the last diagram where the photon
couples to all three quarks, because only this object can create intermediate meson and baryon poles and the corresponding cuts.
Note also that all possible gluon topologies are contained in these diagrams, e.g., gluon exchanges between the quarks are absorbed in the wave functions,
gluon dressings of the photon coupling in the quark-photon vertex, and more complicated topologies are absorbed in the two- and three-body kernels.
In this way, we arrived at a description of the form factor in terms of three valence quarks, which is \textit{also} valid at any $Q^2$.

\begin{figure}[!t]
\begin{minipage}{.45\linewidth}
\includegraphics[width=1\linewidth]{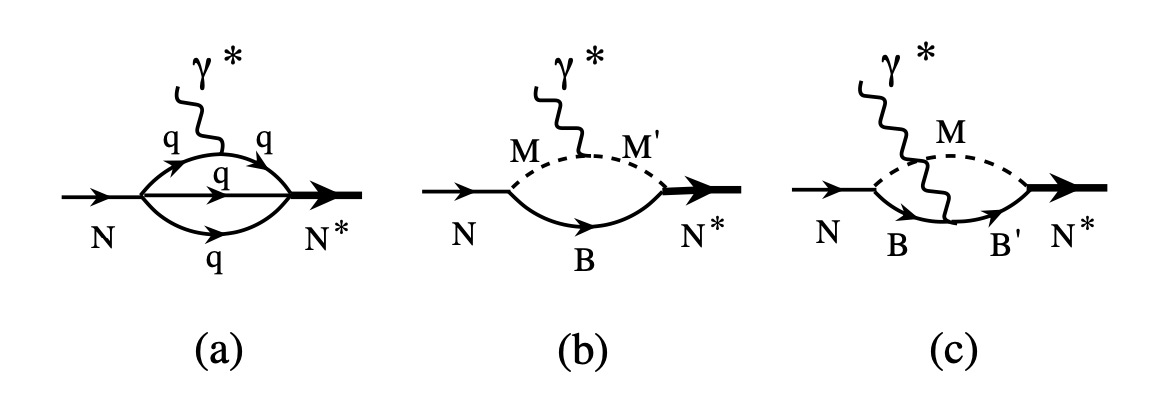}
\caption{Contributions to the analysis model, the LF RQM (a), and meson-baryon contributions 
(b) and (c)).\vspace{-3mm}}
\label{LFRQM}
\hspace{-0mm}\caption{(color online) The running quark mass from Dyson-Schwinger equations and lattice QCD for different current quark masses and in the chiral limit~\cite{Bhagwat:2003vw}.}
\label{Quark-mass}
\end{minipage}
\hspace{5mm}\begin{minipage}{.45\linewidth}
\includegraphics[width=1\linewidth]{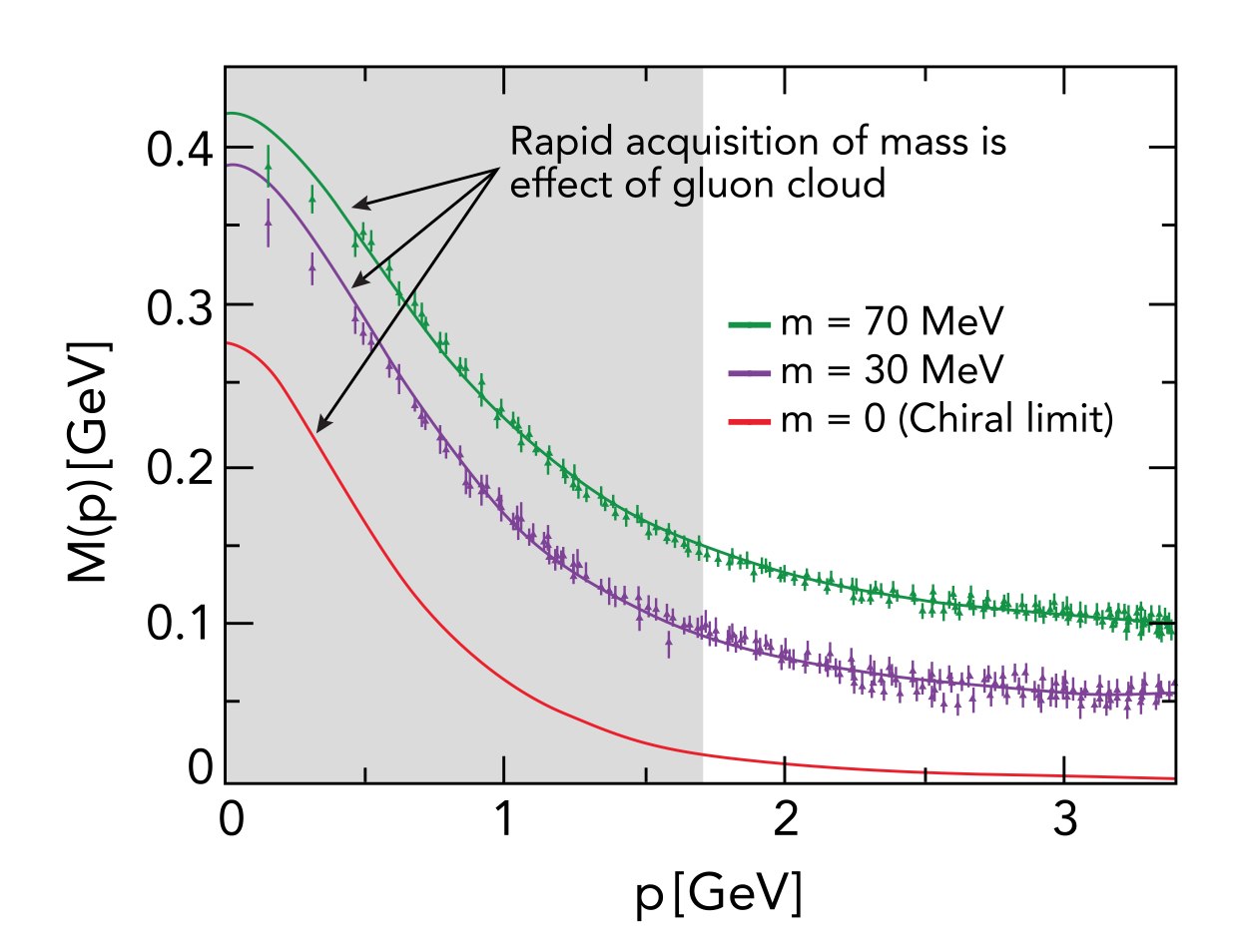}
\end{minipage}
\end{figure}

Therefore, in practice it is less a matter of principle and rather a matter of efficiency 
how to tune the slide control between `quark effects' and `hadronic effects'.
Even though in principle all hadronic effects are contained in the diagrams in Fig.~\ref{fig:ffs}, it is usually more efficient to work with
explicit couplings to mesons and baryons like those shown in Fig.~\ref{LFRQM}. Furthermore, models usually simplify the
diagrams in  Fig.~\ref{fig:ffs} to various degrees, e.g. by keeping only the first diagram, 
or by dropping the vector-meson poles in the quark-photon vertex, or using non-relativistic ansätze for the baryon wave functions, etc.
At large $Q^2$, where all these effects are no longer  relevant, one truly probes the `three-quark core'. The fact that the quark struck by the photon still needs to communicate with the other two quarks via gluons then leads to quark counting rules
which determine the asymptotic behavior of the form factors~\cite{Brodsky:1973kr,Brodsky:1974vy,Carlson:1985mm}.
At lower $Q^2$, the form factors become  more and more sensitive to the multihadron or meson-baryon (MB) components.
Finally, the form factors have a smooth transition to the timelike region, which is dominated by the vector-meson poles.
In the following we adopt this `microscopic' point of view, combined with the available electroproduction data,
 to interpret our current knowledge
on the spacelike transition form factors and corresponding helicity amplitudes.

\subsection{Comparison with state-of-the-art quark-based approaches}

In this section we compare the results on electroproduction of several low-mass states 
with different microscopic approaches using quark degrees of freedom,
including light-cone sum rules (LCSR)~\cite{Anikin:2015ita}, the Dyson-Schwinger (DSE) approach ~\cite{Segovia:2016zyc,Eichmann:2016yit}, the covariant spectator model~\cite{Ramalho:2011ae,Ramalho:2023hqd}, and lattice QCD. 
Our baseline for comparisons will be the light-front relativistic quark model (LF RQM) from Refs.~\cite{Aznauryan:2007ja,Aznauryan:2012ec}.
%In several cases 
%There are also calculations from other microscopic approaches %with traceable links to QCD, 
Where available we also compare with models for dynamically generated  resonances~\cite{Jido:2007sm}. 
Further details and discussions can also be found in the comprehensive review~\cite{Ramalho:2023hqd}.

 The LF RQM ingredients, shown in Fig.~\ref{LFRQM}, are tuned to the elastic electromagnetic form factors by combining the three-quark and  $|\pi N \rangle$  contributions to the nucleon wave function of the form 
\begin{equation}
 |N \rangle= A\,|3q\rangle + B\,|\pi N\rangle. 
 \label{decomp}
\end{equation}
with $A = 0.95$ and $B= 0.313$. 
The portions of the three-quark and pion contributions were determined in a light-front cloudy bag model in the fit to the elastic electromagnetic form factors~\cite{Miller:2002ig};
the condition $F_1{}_p (0) = 1$ provides the normalization.

The electroexcitation of resonances on the proton was investigated in the LF RQM~\cite{Aznauryan:2007ja,Aznaurian:1982qc} for the resonances of the multiplets $(\textbf{56},0^+_0)$, $(\textbf{56},0^+_2)$, and $(\textbf{70},1^- _1)$. In~Refs.~\cite{Aznauryan:2012ec,Aznauryan:2015zta,Aznauryan:2017nkz,Aznauryan:2016wwm} 
the parameters of the LF RQM were specified via description of the nucleon electromagnetic form factors up to $Q^2$ = 16~GeV$^2$ by combining the three-quark and  pion-cloud contributions in the LF dynamics in Eq.~(\ref{decomp})   
and employing a running quark mass as a function of $Q^2$~\cite{Aznauryan:2017nkz,Aznauryan:2012ba}. The inclusion of a running quark mass in the LF RQM is motivated by  calculations of the quark mass function using the quark DSE~\cite{Bhagwat:2003vw} and  lattice QCD~\cite{Bowman:2005vx} as displayed in Fig.~\ref{Quark-mass}. 
For the $\Delta(1232)\nicefrac{3}{2}^+$, $N(1440)\nicefrac{1}{2}^+$, $N(1520)\nicefrac{3}{2}^-$,  $N(1535)\nicefrac{1}{2}^-$, the weights of the $|3q \rangle$ 
component in the expansion $|N^*\rangle= c_{N^*} |3q \rangle + ... $
were found from experimental data on $\gamma^* p \to \pi^+ n$ assuming that at $Q^2 > 3$~GeV$^2$ these resonance transitions are determined  by  $|3q \rangle$ only, while at lower $Q^2$ contributions from meson-baryon terms become significant. This has  been especially verified in the transitions $\gamma^\ast p\to  N(1535)\nicefrac{1}{2}^-$ and $\gamma^\ast p\to N(1440)\nicefrac{1}{2}^+$, where in both cases several different model calculations coincide in their $Q^2$ dependence.

\begin{figure}[!t]
\includegraphics[width=0.61\linewidth]{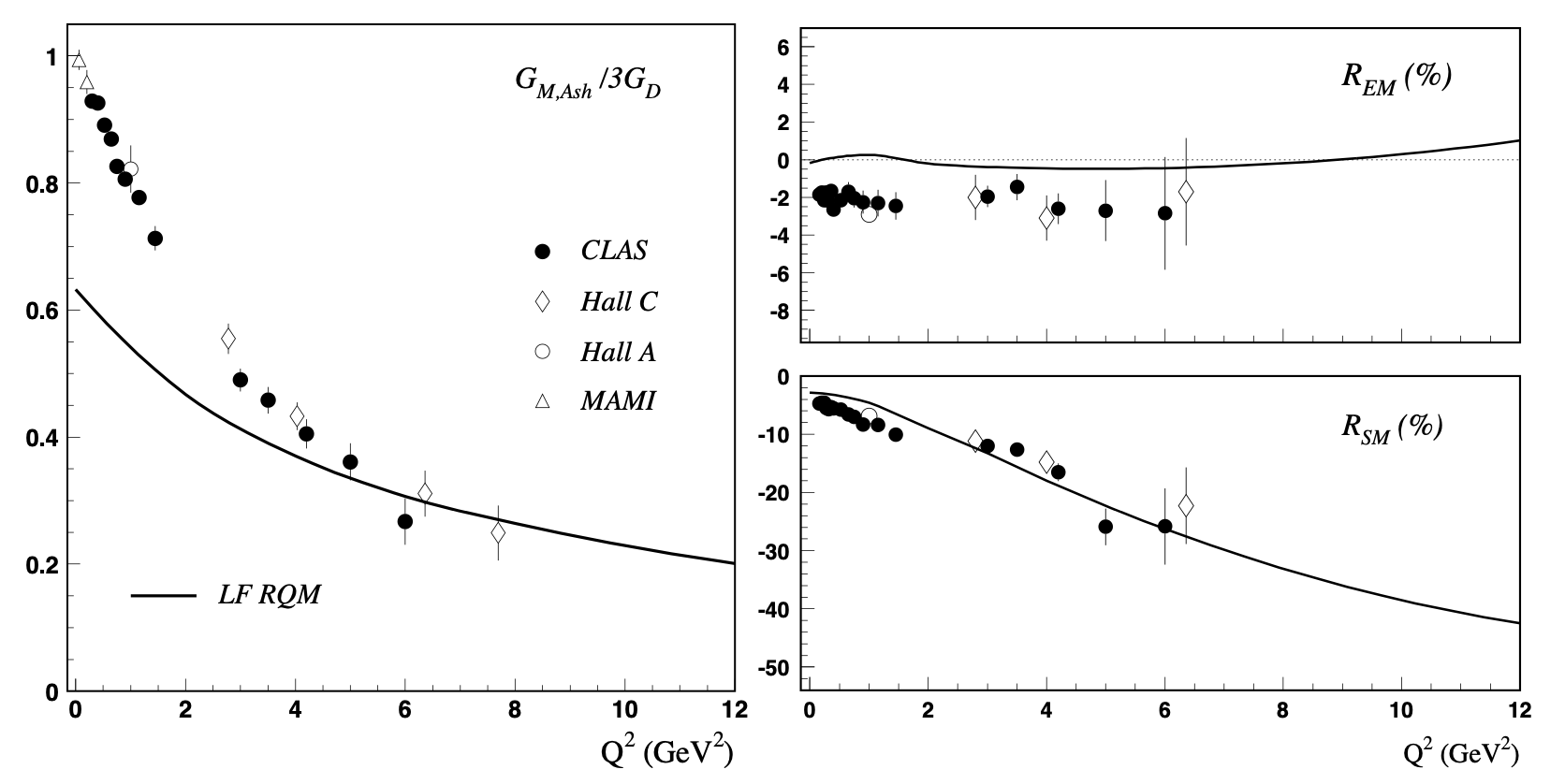}\hspace{1mm}
\includegraphics[width=0.38\textwidth]{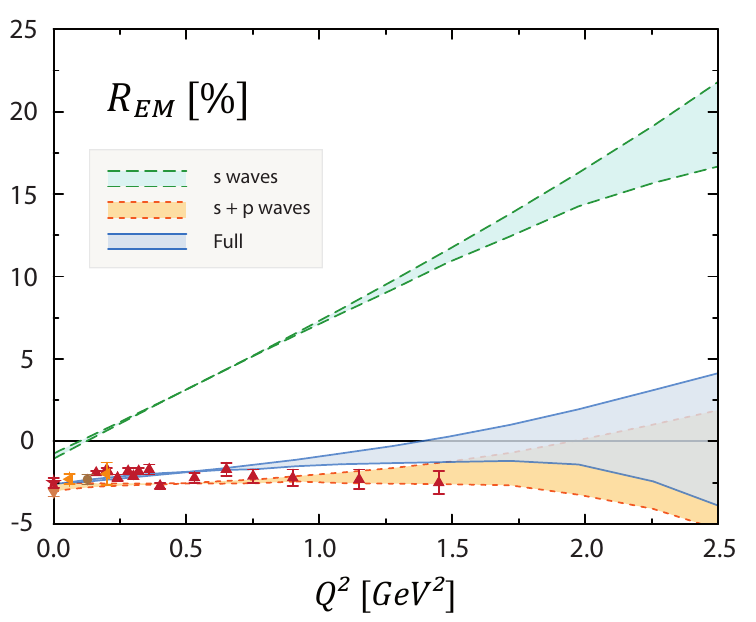}
\caption{\small \small Left: Magnetic dipole form factor $G_M^*(Q^2)$ of the $\gamma^* p \to  \Delta(1232)$ transition
(here shown in the Ash convention~\cite{Ash:1967rlw} and normalized to a dipole). Center:
The ratios of the electric ($R_{EM}$) and scalar ($R_{SM}$) to magnetic multipole transition amplitudes. The data are from Refs.~\cite{CLAS:2001cbm,CLAS:2006ezq,Kelly:2005jy,Frolov:1998pw,Beck:1999ge} and the solid line is the prediction of the LF RQM~\cite{Aznauryan:2012ec}.   Right: DSE results for the $R_{EM}$ ratio;
the full result is compared to the one obtained by  including only $s$ waves (or $s$ and $p$ waves)  in the nucleon and $\Delta$ wave functions~\cite{Eichmann:2011aa}.}
\label{Delta-GM-REM-RSM}
\end{figure}

\subsubsection{\bf The $\gamma^* N\to \Delta(1232)$ transition}

\begin{figure}
\centering
\begin{minipage}{.52\linewidth}
\includegraphics[width=\textwidth]{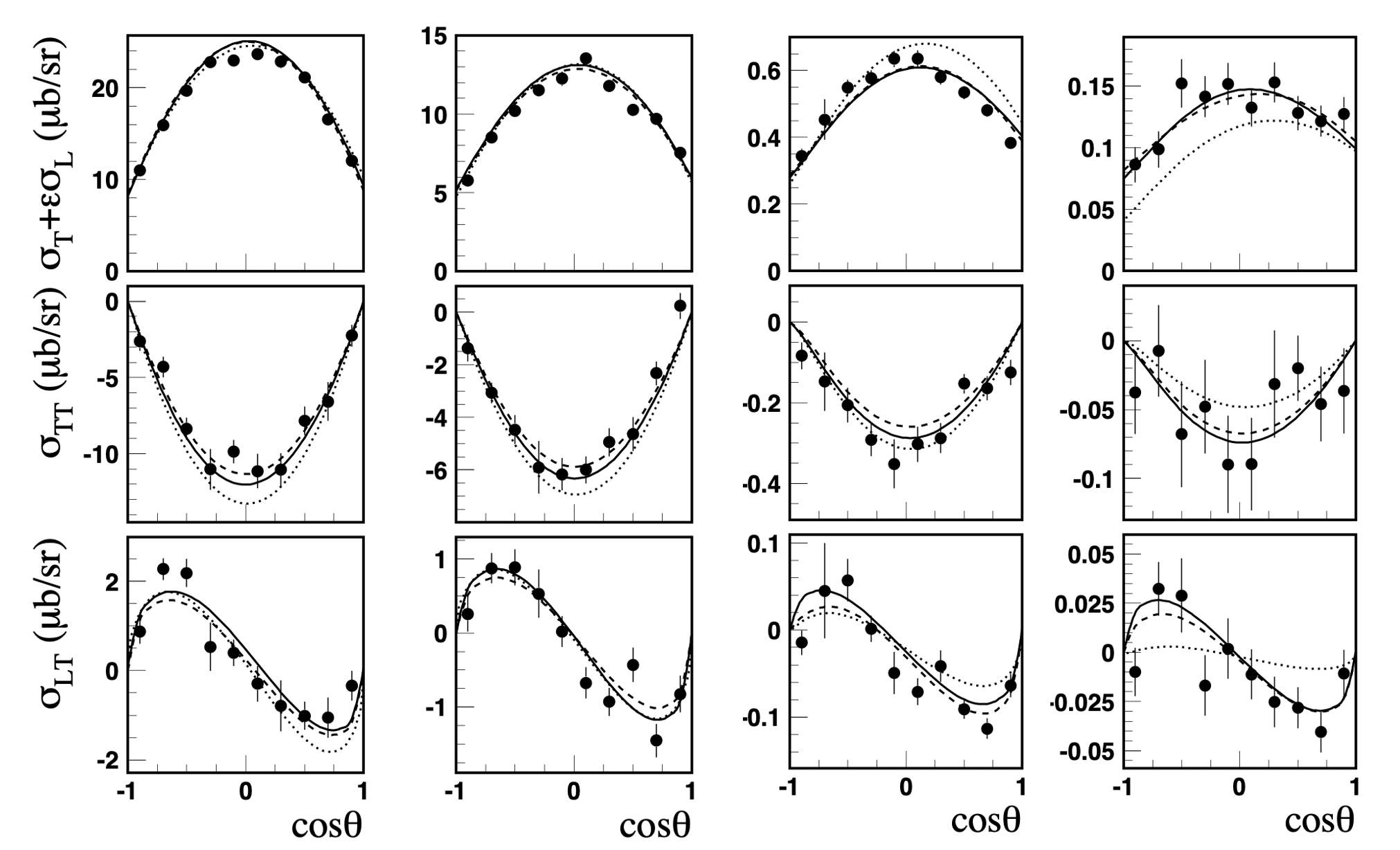}
\caption{\small Samples of response functions extracted from the azimuthal $\phi$-angle dependence of the differential cross section in the $\gamma^*p \to p\pi^0$ channel in the region of the $\Delta(1232)\nicefrac{3}{2}^+$ resonance at different values of $Q^2$, from left to right: $Q^2 = 0.4, 1.45, 3.0, 5.0$~GeV$^2$~\cite{Aznauryan:2011qj}. The solid curve corresponds to the analysis of the data using the dispersion relation approach. The dashed curves uses the unitary isobar model approach, and the dotted curve is the prediction of the MAID2007 model~\cite{Drechsel:2007if}.} 
\label{totalcrs}
\end{minipage}
\hspace{3mm}\begin{minipage}{.40\linewidth}
\centering
\includegraphics[width=0.72\linewidth]{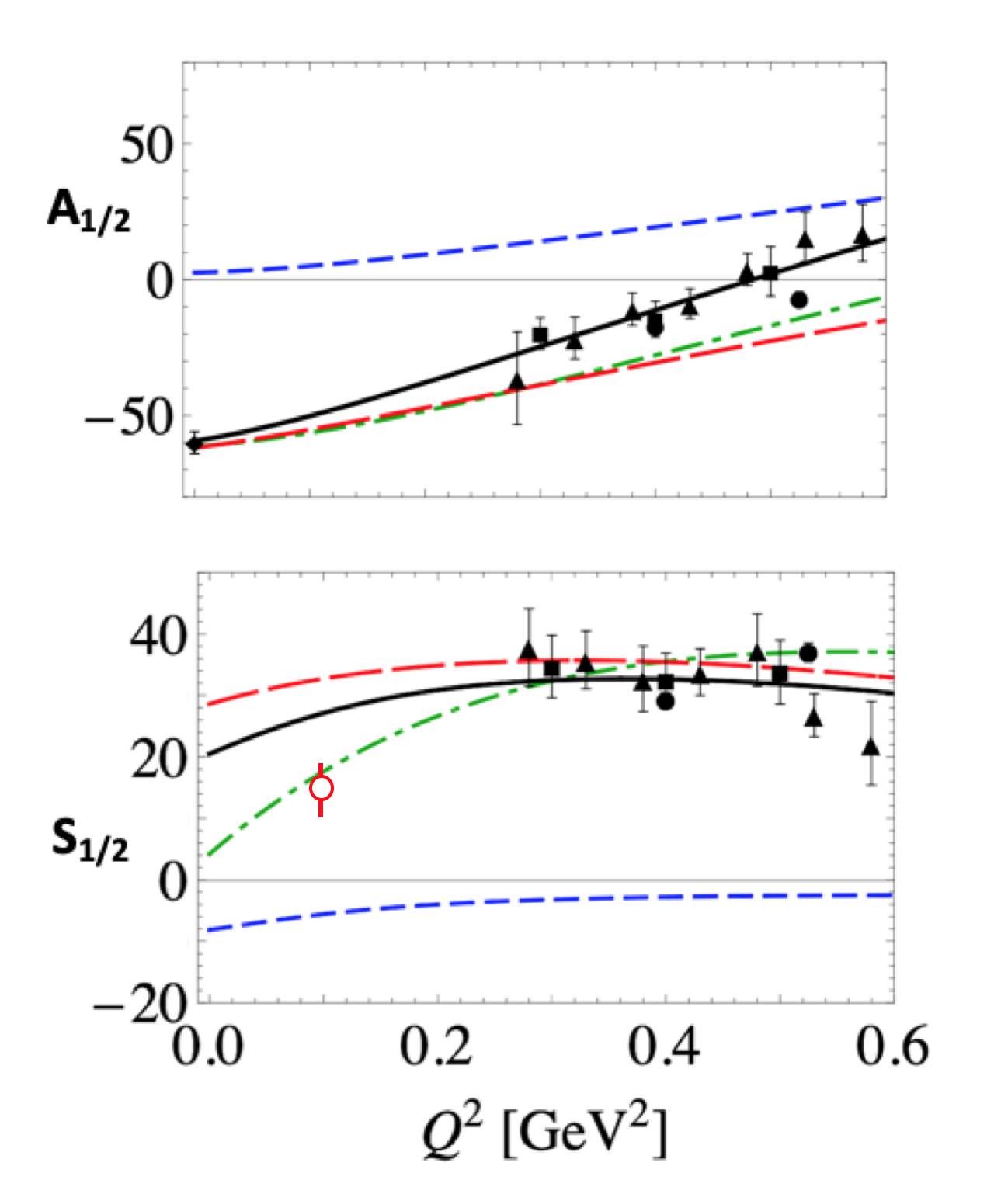}
\caption{The $\gamma^\ast p\to N(1440)$ helicity transition amplitudes $A_{1/2}$ and $S_{1/2}$
in units of $10^{-3}$~GeV$^{-1/2}$. Helicity amplitudes at low $Q^2$;
the data are from ~\cite{CLAS:2012wxw,CLAS:2009ces,Drechsel:2007if}. The open red circle at $Q^2 \approx 0.1$~GeV$^2$ 
is the result of an analysis of $ep\to ep\pi^0$ data from MAMI~\cite{Stajner:2017fmh}.
The data are compared to EFT calculations~\cite{Bauer:2014cqa}, where
the solid (black) lines are full results, the long-dashed (red) lines  
tree contributions and the short-dashed (blue) lines are the loop terms. The dashed-dotted (green) lines are from an 
empirical parametrization.}
\end{minipage}
\end{figure}
As an example, we first discuss the lowest-mass and best studied of all states, the $\Delta(1232)\nicefrac{3}{2}^+$. 
It is most prominently seen both in $\pi N \to \pi N$ elastic scattering and in photoproduction $\gamma N \to N \pi$.
The electromagnetic transition of the ground state proton to the  $\Delta(1232)\nicefrac32^+$ has been experimentally studied in great detail 
and is known in a large range of $Q^2$~\cite{CLAS:2006ezq,CLAS:2001cbm,Frolov:1998pw}. 
Figure~\ref{totalcrs} shows samples of the response functions in the mass range of the $\Delta(1232)\nicefrac{3}{2}^+$.

In the quark model the $\gamma^\ast p \to \Delta^+$ transition is predominantly a single spin flip of one of the valence quarks corresponding to a magnetic dipole transition. 
At small $Q^2$, the data are therefore explained by a dominant magnetic transition from the  nucleon ground state to the $\Delta(1232)\nicefrac{3}{2}^+$ excited state. 
Additional contributions are expected to come from small $d$-wave components in both the nucleon and the $\Delta(1232)\nicefrac{3}{2}^+$ wave functions, which lead to electric and scalar quadrupole transitions. %The resulting three transition form factors are shown in Fig.~\ref{Delta-GM-REM-RSM}.
The experimental status on the magnetic dipole,  electric quadrupole and scalar quadrupole transitions 
of $\gamma^* p\to \pi^0 p$ is shown in the left and center panels of Fig.~\ref{Delta-GM-REM-RSM}. 
Note that the electric and scalar quadrupole transitions are plotted as ratios, which remain in the few-percent range at small $Q^2$.

A well-known shortcoming of quark models and quark-based approaches is that
at the real photon point ($Q^2=0$), the quark contributions do not account for the full magnitude of the transition amplitudes 
but only to  65 \dots 75\% depending on the approach, 
as  seen in Fig.~\ref{Delta-GM-REM-RSM}. 
The remainder can be attributed to meson-baryon contributions, also referred to as the `pion cloud',
 which are not modeled in either of the calculations.   
On the other hand, at $Q^2 \ge 2 \dots 3$~GeV$^2$ 
%the  quark models show good agreement with the data;
%e.g., 
the LF~RQM with a momentum-dependent constituent quark mass and the DSE results~\cite{Eichmann:2011aa,Segovia:2016zyc,Sanchis-Alepuz:2017mir}  account for nearly the full strength of the experimental data. 
This illustrates that the meson-baryon contributions occupy the peripheral, large distance space probed at low $Q^2$, while at large $Q^2$, i.e. short distances, the quark core accounts for the full resonance response. Without a quark core, the transition amplitudes would fall off much faster with $Q^2$ compared to a pure three-quark system.

Also interesting is 
the electric quadrupole ratio, whose experimental value at low $Q^2$ is $R_{EM}  \approx -0.02.$
There has been a longstanding prediction of asymptotic pQCD  that \mbox{$R_{EM} \to +1$ at $Q^2 \to \infty$}~\cite{Carlson:1985mm}. 
However, the quadrupole ratio $R_{EM}$ shows no sign of departing significantly from its value at $Q^2=0$, even at the highest $Q^2 \approx 6.5$~GeV$^2$. 
As seen in Fig.~\ref{Delta-GM-REM-RSM}, the LF RQM calculation also barely departs from $R_{EM}=0$  and remains near zero at all $Q^2 > 2$~GeV$^2$. 
The holographic QCD model~\cite{Grigoryan:2009pp}, on the other hand, predicts a zero-crossing of $R_{EM}$ near $Q^2 = 0.6$~GeV$^2$ followed by a steep rise to $R_{EM} \to +100\%$ at $Q^2 \to \infty$. 
This would indicate that the negative constant value shown by the data could be due to meson-baryon contributions which are not included in the theoretical models.  
However, the DSE result shown in the right panel of Fig.~\ref{Delta-GM-REM-RSM} offers a different explanation~\cite{Eichmann:2011aa}:
As discussed in Sec.~\ref{sec:fm}, the relativistic nucleon and $\Delta$ wave functions have significant $p$-wave components,
which are absent in the non-relativistic quark model. If they are neglected, $R_{EM}$ starts out close to zero
and sharply rises towards the asymptotic value; if they are included, the result is able to explain the data.
This illustrates that relativistic effects for light baryons have observable consequences, and 
not every discrepancy between model calculations and experimental data is necessarily a pion-cloud effect. 

The $R_{SM}$ data show a strong trend of increasing  negative values at larger $Q^2$.
The prediction of a rise in magnitude of $R_{SM} \to -100\%$ at $Q^2 \to \infty$ is supported by the trend of the data and also
 consistent with the LF RQM.

\begin{figure}
\centering
\begin{minipage}{.74\linewidth}
\includegraphics[width=\linewidth]{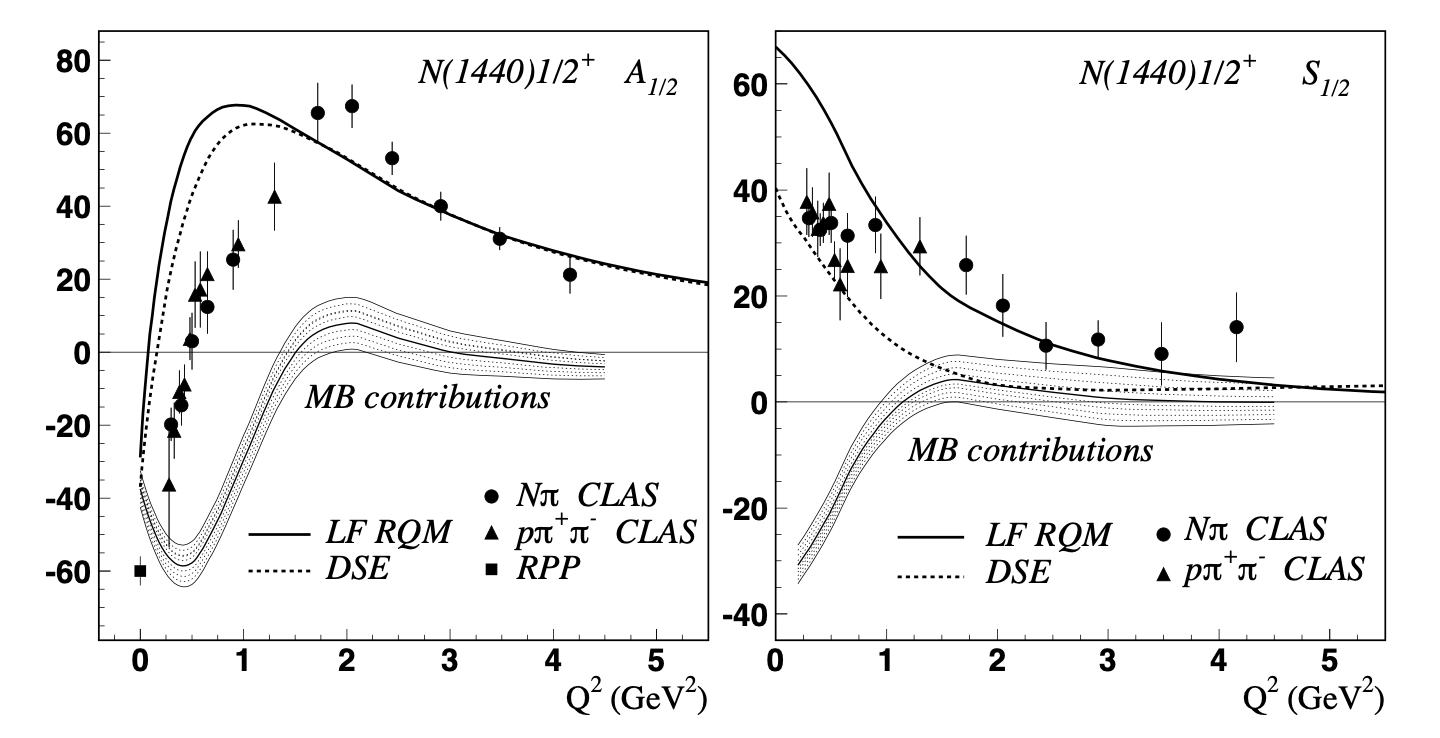}
\end{minipage}
\begin{minipage}{.24\linewidth}
\includegraphics[width=1\linewidth]{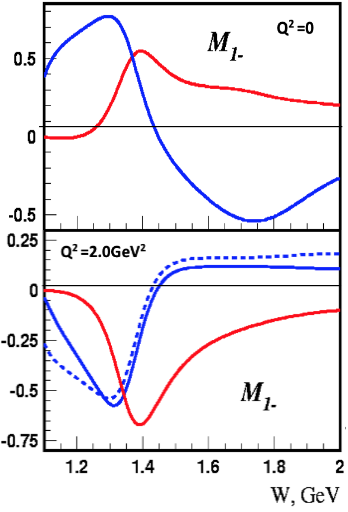}
\end{minipage}
\caption{\small Left and center: The $\gamma^\ast p\to N(1440)$ helicity transition amplitudes $A_{1/2}$ and $S_{1/2}$
in units of $10^{-3}$~GeV$^{-1/2}$. $A_{1/2}$ and $S_{1/2}$ over the full $Q^2$ range. 
The solid curves are from a light-front relativistic quark model~\cite{Aznauryan:2012ec} and the dashed curves are DSE results~\cite{Segovia:2015hra}.
 Data are from the CLAS collaboration~\cite{CLAS:2012wxw,CLAS:2009ces}. 
 Right:  Transition multipole amplitude $M_{1-}$ in the $N(1440)\nicefrac{1}{2}^+$ mass region for two values 
$Q^2 = 0$ and $Q^2 = 2$~GeV$^2$. One can see the dramatic switch from Im~$M_{1-} > 0$ (corresponding to $A_{1/2} < 0$) 
at the photon point to a large negative Im~$M_{1-}$ amplitude at $Q^2=2$ GeV$^2$. The solid lines relate to the 
unitary isobar model and the dashed lines to the fixed-t dispersion relation approach~\cite{Aznauryan:2007ja} and Inna Aznauryan, private communications.}
\label{p11} 
\end{figure}

\subsubsection{\bf The $\gamma^*p\to  N(1440)\nicefrac{1}{2}^+$ transition and the Roper resonance}
\label{Sec:Roper}
The Roper resonance $N(1440)\nicefrac{1}{2}^+$, discovered in 1964~\cite{Roper:1964zza} in a phase shift analysis of 
elastic $\pi N$ scattering data, has been differently interpreted for half a century. A detailed review of the history 
and current status of the this resonance is presented in a colloquium-style 
article published in Review of Modern Physics~\cite{Burkert:2017djo}. Earlier model descriptions, such as the Isgur-Karl model which describes the nucleon as a system of three constituent quarks in a confining potential and a one-gluon exchange contribution leading to a magnetic hyperfine splitting of states~\cite{Isgur:1978xj,Isgur:1977ef}, and the relativized version of Capstick~\cite{Capstick:1986ter}, have popularized the models that became the basis for many further developments and variations, e.g. the light front relativistic quark model and the hypercentral quark model~\cite{Giannini:2003xx}. Other models were developed in parallel. The cloudy bag  model~\cite{Thomas:1981vc} describes the nucleon as a bag of three constituent quarks surrounded by a cloud of pions. It has been mostly applied to nucleon resonance excitations in real photoproduction,  $Q^2=0$~\cite{Bermuth:1988ms,Thomas:1981vc}, with some success in the description of the $\Delta(1232)\nicefrac{3}{2}^+$ and the Roper resonance transitions. 

In quark models, the Roper resonance is the 
first radial excitation of the nucleon ground state with a predicted mass that is considerably higher than the measured Breit-Wigner mass of $\approx 1440$~MeV. 
The wrong level ordering compared to the parity partner of the nucleon, the $N(1535)\nicefrac12^-$,
was one of the major obstacles for a quark-model interpretation of the Roper resonance 
and the main motivation to search for a different understanding of this resonance. 
Another problem with the quark-model interpretation of the Roper resonance was the sign of the transition form 
factor $A_{1/2}(Q^2=0)$, predicted in the non-relativistic quark models as large and positive, while experimental analyses showed a negative value. 

These discrepancies resulted in different interpretations of the  Roper resonance. 
In Ref.~\cite{Krehl:1999km}, a coupled-channel meson exchange model for pion-nucleon scattering
was derived from a chirally symmetric Lagrangian, and the $\pi N$ phase shifts and inelasticities
were fitted. No genuine three-quark resonance was needed, and the
Roper resonance was described by meson-baryon dynamics alone. These findings were confirmed in
Ref.~\cite{Ronchen:2018ury} where the Roper  resonance was listed as a ``dynamically generated
resonance". In subsequent publications \cite{Sekihara:2021eah,Wang:2023snv}, the Roper
resonance was decomposed into a large meson-baryon (MB) and a small $qqq$ component.   

Is the $N(1440)\nicefrac{1}{2}^+$ a dynamically generated resonance or a three-quark baryon with
a MB cloud? This question can be  resolved with electroproduction data 
from CLAS at Jefferson Lab and the parallel development of theoretical approaches.
The internal structure of the Roper resonance is revealed by measurements 
of the  $\gamma^\ast p\to N(1440)\nicefrac{1}{2}^+$ helicity amplitudes  $A_{1/2}$ and $S_{1/2}$.
The CLAS collaboration reported these amplitudes over a wide range in 
$Q^2 \sim 0.25 \dots 4.2$ GeV$^2$~\cite{CLAS:2012wxw,CLAS:2009ces}, see Fig.~\ref{p11}.
New electroproduction data on the Roper yield an additional value at very small $Q^2$ in the amplitude $S_{1/2}$
shown in the left panel of Fig.~\ref{p11}~\cite{Stajner:2017fmh}. Instead of the helicity amplitudes, which  satisfy kinematic constraints
at timelike values $Q^2< 0$, it is sometimes more convenient to parametrize the data in terms of two transition 
form factors $F_1$ and $F_2$ which are free of kinematic constraints, see e.g.~Refs.~\cite{Tiator:2011pw,Eichmann:2018ytt} for  explicit relations.

The striking feature visible in the data is the rapid transition of the transverse amplitude $A_{1/2}$ from a large negative magnitude to 
a large positive one as a function  of $Q^2$. Translated to the transition form factors, this implies a zero crossing 
in $F_2$, % at $Q^2 \sim 0.5 \dots 0.8$ GeV$^2$, 
which is a typical signature of a first radial excitation.
Such a behavior is also found in quark-based approaches like the DSE approach~\cite{Segovia:2015hra},   
the covariant spectator model~\cite{Ramalho:2017pyc},  holographic 
models~\cite{deTeramond:2011qp} and the LF RQM with momentum-dependent quark masses~\cite{Aznauryan:2012ec}. 
In the DSE approach, the Roper resonance appears as the first radial excitation of the nucleon with a mass above the $N(1535)\nicefrac12^-$,
see  Fig.~\ref{fig:spectrum-qdq}, although this does not exclude  MB contributions which may still be important~\cite{Eichmann:2016yit}.
They are especially visible in the helicity amplitudes:
At high $Q^2$, the process is dominated by the three-quark 
structure of the $N(1440)\nicefrac{1}{2}^+$, and the calculated amplitudes agree with the data for the 
highest $Q^2$ values. At low $Q^2$ the Roper resonance is no 
longer well described through its quark structure, which indicates the presence of non-quark core effects. 
If one (tentatively) interprets the difference between the predicted curves and the 
data  as MB effects, then these contributions must be large 
below $Q^2 \sim 1$ GeV$^2$.
%at distances above $1/Q = 0.4$/GeV  or 0.5\,fm.  
For the Roper resonance such contributions have been described successfully in dynamical 
meson-baryon models~\cite{Obukhovsky:2011sc} and in effective field theory~\cite{Bauer:2014cqa}.

The rapid transition of the transverse amplitude $A_{1/2}$ 
can also be seen in the partial wave multipole 
amplitude $M_{1-}$ in the right of Fig.~\ref{p11}. It also shows that the resonance character of 
the Roper $N(1440)\nicefrac{1}{2}^+$ is more prominently visible at  high $Q^2$ 
in both the real and  imaginary parts of the amplitude. This is likely a consequence of the 
virtual photon coupling directly the quark core of the nucleon, where the contribution of the 
meson-baryon cloud is much reduced.

\begin{figure}[!t]
\vspace{-1mm}
\hspace{15mm}
\includegraphics[width=0.74\linewidth]{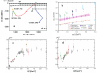}
% \raisebox{-3mm}{\includegraphics[width=0.528\linewidth]{Roper-mass-motion-VB.png}} 
% \hspace{-8mm}\includegraphics[width=0.421\linewidth]{roper-xqcd-vb.png} \vspace{-5mm}\\
% \includegraphics[width=0.726\linewidth]{Roper-F1-F2-LQCD-VB.png}
\caption{\label{Roper-hQCD}\small 
a:  Trajectories of the evolution of the $J^P=\nicefrac12^+$ resonance. 
Three poles evolve from one  bare $N^*$ state  at  1763~MeV~\cite{Suzuki:2009nj}.
b: Lattice results for the nucleon and  Roper masses (adapted from Ref.~\cite{Liu:2016rwa}).
Most data points for the Roper resonance lie above the $N\pi$ threshold.
c and d: Lattice results for the $\gamma^\ast p\to N(1440)$ proton to Roper transition form factors $F_1$ 
and $F_2$  for pion masses 390~MeV (red squares), 450~MeV (orange triangles) and
875~MeV (green circles) compared to data (open circles)~\cite{Lin:2011da}. 
}
\end{figure}

There are also efforts to investigate the Roper resonance using lattice QCD.
As discussed in Sec.~\ref{sec:lattice}, the main difficulty is the fact that the  Roper resonance
lies above several thresholds, where the coupling to $N\pi\pi$ is especially important~\cite{Lang:2016hnn}.
This requires generalizations of the Lüscher formalism, which are  under active current development~\cite{Hansen:2019nir}.
Such caveats must be kept in mind when interpreting lattice results like those shown in Fig.~\ref{Roper-hQCD}\,b,
where the Roper resonance %and $N(1535)\nicefrac12^-$ are stable bound states 
is a stable bound state only at large pion masses (corresponding to large current-quark masses).
The bottom panels in Fig.~\ref{Roper-hQCD} show lattice results for the $\gamma^\ast p\to N(1440)$ proton to Roper transition form factors $F_1$  
and $F_2$, which are compatible with the experimental data and also display a zero crossing in $F_2$~\cite{Lin:2011da}.

Although the zero crossing in the transition form factor $F_2$  suggests an interpretation of the Roper resonance as the first radial
excitation of the nucleon,
the fact that it strongly couples to meson-baryon channels at physical pion masses also allows for different interpretations~\cite{Wu:2017qve,Virgili:2019shg}.
Combining lattice QCD and Hamiltonian EFT, the authors
of Ref.~\cite{Leinweber:2024psf} argue that the 2s radial excitation of the nucleon should have
a mass of 2~GeV, while the Roper resonance is
generated by rescattering effects in the coupled channels of $\pi N$, $\sigma N$,
and $\pi\Delta(1232)$. In a similar spirit,
the coupled-channel analysis of Ref.~\cite{Suzuki:2009nj} 
assumes one ``bare" pole at 1763 MeV, which couples to $\pi N$, $\eta N$, $\pi\Delta$, $\rho N$ and $\sigma N$ 
and thereby generates three further poles (see Fig.~\ref{Roper-hQCD}\,a).
One is interpreted as the Roper resonance and another one as the $N(1710)\nicefrac12^+$,
suggesting that both resonances originate from 
the same excited nucleon state. This possibility is discussed in more detail in Sec.~\ref{sub:2nd}.

Here we would like to emphasize again that, from the point of view of quantum field theory, the
 options ``three-quark state'' or ``dynamically generated resonance'' are not exclusive.
As we argued in Sec.~\ref{sec:eft}, if a baryon pole appears in the $qqq$ correlation function, it must also appear
in the $qqqq\bar{q}$, $qqqq\bar{q}q\bar{q}$ and any other correlation function that can produce baryon poles.
Unless some quantum number prevents it,
there \textit{is} no dynamically generated resonance without an accompanying quark core. 
This implies that there is also no contradiction between the Roper resonance being the radial excitation of the nucleon
while at the same time  having a large admixture of five- or even seven-quark components.
A similar example in the meson sector is the famous $\chi_{c1}(3872)$, 
which couples to $c\bar{c}$ and $c\bar{c}q\bar{q}$ and is therefore simultaneously a  $c\bar{c}$ excitation
and a four-quark state~\cite{Prelovsek:2013cra,Padmanath:2015era,Guo:2017jvc}.

Supported by an extensive amount of single-pion electroproduction data, covering the full phase space in the pion polar and azimuthal center-of-mass angles, and accompanied by  theoretical calculations, we can thus summarize our current understanding of the $N(1440)\nicefrac{1}{2}^+$ state as follows: 
(i) The Roper resonance is, at heart, the first radial excitation of the nucleon. Its three-quark core plays a role in determining the system's properties at all length scales, but exerts a dominant influence on probes with $Q^2 > m_N^2$, where $m_N$ is the nucleon mass.
(ii) The core is augmented by a meson cloud, which can be  understood as an admixture of higher multiquark components 
        and manifests itself as a dynamical coupling between several meson-baryon channels.
        It reduces the Roper's  mass, %by $\approx 20\%$, 
        thereby solving the mass problem that was such a puzzle in constituent quark model treatments. 
        At low $Q^2$ it  contributes an amount to the electroproduction transition form factors whose magnitude may be comparable to the three-quark core but vanishes rapidly as $Q^2$ is increased beyond $m_N^2$. 
        These admixtures  will be further quantified by upcoming theory calculations.

Similar conclusions about the nature of the Roper resonance have been reached in different approaches analyzing the $Q^2$ dependence of electroproduction data~\cite{Karapetyan:2023kzs,Kaewsnod:2021nfw,Ramalho:2010js,Ramalho:2023hqd}. 
As stated in the conclusions of~\cite{Burkert:2017djo}: "The fifty years of 
experience with the Roper resonance have delivered lessons that 
cannot be emphasized too strongly. Namely, in attempting to predict and explain the QCD spectrum, one must fully consider the impact of meson-baryon final state interactions and the coupling between channels and states that they generate, and look beyond merely locating the poles in the S-matrix, which themselves reveal little structural information, to also consider the $Q^2$ dependencies of the residues, which serve as a penetrating scale-dependent probe of resonance composition." \\[-4ex]  

\begin{figure}[!t]
\centering
\includegraphics[width=0.8\linewidth]{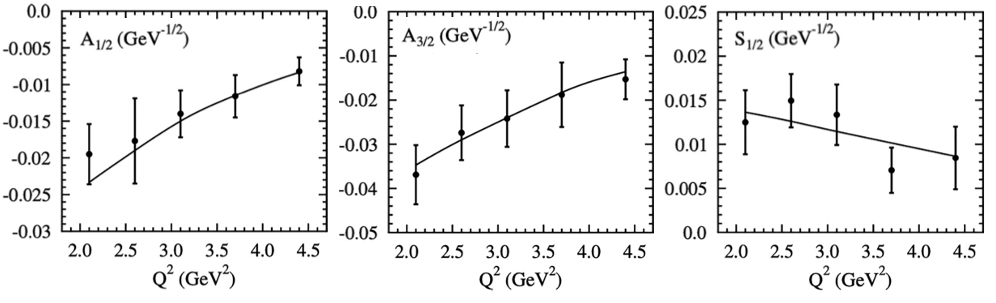} 
\caption{\small Helicity transition amplitudes $A_{1/2}$, $A_{3/2}$ and $S_{1/2}$ for the transition $\gamma^* p \to \Delta(1600)\nicefrac{3}{2}^+$. The data are from \cite{Mokeev:2023zhq}, the curves  are from\,\cite{Lu:2019bjs}.}
\vspace{-0.9cm}
\label{fig:Delta-1600}
\end{figure}

\subsubsection{\bf The $\gamma^*p\to \Delta(1600)\nicefrac{3}{2}^+$ transition}
\label{Delta-1600}

The $\Delta(1600)\nicefrac{3}{2}^+$ resonance is considered the first radial excitation of the ground state $\Delta(1232)\nicefrac{3}{2}^+$. While the resonance has been listed in the Review of Particle Properties for over three decades, it has been fully accepted as a 4-star resonance only in recent years. Because of its small coupling to the $N\pi$ channel, the resonance is hardly visible in $\pi N\to \pi N$ elastic scattering.

The recent results from the CLAS collaboration in $\gamma^* p \to p\pi^+\pi^-$ electro-production\,\cite{Mokeev:2023zhq} have changed the situation significantly and enabled the first extraction of the $Q^2$ dependence of the transition amplitudes in the range $2<Q^2<4.5$~GeV$^2$ as shown in Fig.~\ref{fig:Delta-1600}. The theory curves are based on the DSE approach using a dynamical quark-diquark approximation to the relativistic three-body bound state problem. It provides an excellent fit to all three helicity amplitudes and establishes the resonance as a solid three-quark radial excitation. The lack of electroproduction data at low $Q^2$ prohibits an evaluation of possible meson-baryon contributions in the lower $Q^2$ range. It is worth noting that the theory predictions existed before the data became available.% \\[-4ex]

An interesting question regarding the $\Delta(1600)\nicefrac32^+$ as a radial excitation of the $\Delta(1232)\nicefrac32^+$ is whether it behaves similarly to the $N(1440)\nicefrac12^+$ as a radial excitation of the nucleon ground state, i.e., whether it shows a similar zero-crossing of its helicity amplitudes. This has been studied within the LF RQM~\cite{Aznauryan:2016wwm}.
In this work the configuration mixings that follow from QCD-inspired inter-quark
forces on the results for the electroexcitation of the $\Delta(1232)\nicefrac{3}{2}^+$, 
$N(1440)\nicefrac{1}{2}^+$, and $\Delta(1600)\nicefrac{3}{2}^+$ obtained earlier in the 
light-front relativistic quark model without configuration mixing~\cite{Aznauryan:2015zta}.
In the case of the $N(1440)\nicefrac{1}{2}^+$, the configuration mixing does improve the overall
agreement with the data while not significantly changing the zero crossing of the 
transverse transition amplitude $A_{1/2}$. For the $\gamma^\ast N \to \Delta(1600)\nicefrac{3}{2}^+$ transition, 
configuration mixing changes very strongly the results obtained earlier for the $N$ and $\Delta(1600)\nicefrac{3}{2}^+$ 
taken as pure states in the multiplets $(\textbf{56},0^+)$ and $(\textbf{56}^\prime,0^+)$. While in the case without 
 mixing both amplitudes $A_{1/2}$ and $A_{3/2}$ are predicted to change sign at 
$Q^2 \approx 0.25$ GeV$^2$, both remain negative when mixing is taken into account, as also measured at the real photon point, with a smooth falloff in magnitude with increasing $Q^2$. 
The predicted behavior is consistent with the experimental data obtained at $Q^2 > 2~$GeV$^2$ and shown in Fig.~\ref{fig:Delta-1600}.

%\vfill\eject
\subsubsection{The $\gamma^*p\to  N(1520)\nicefrac{3}{2}^-$ transition and switch of its helicity structure}
\label{N1520}
\begin{figure}[!t]
\centering
\begin{tabular}{cc}
\includegraphics[width=0.74\linewidth]{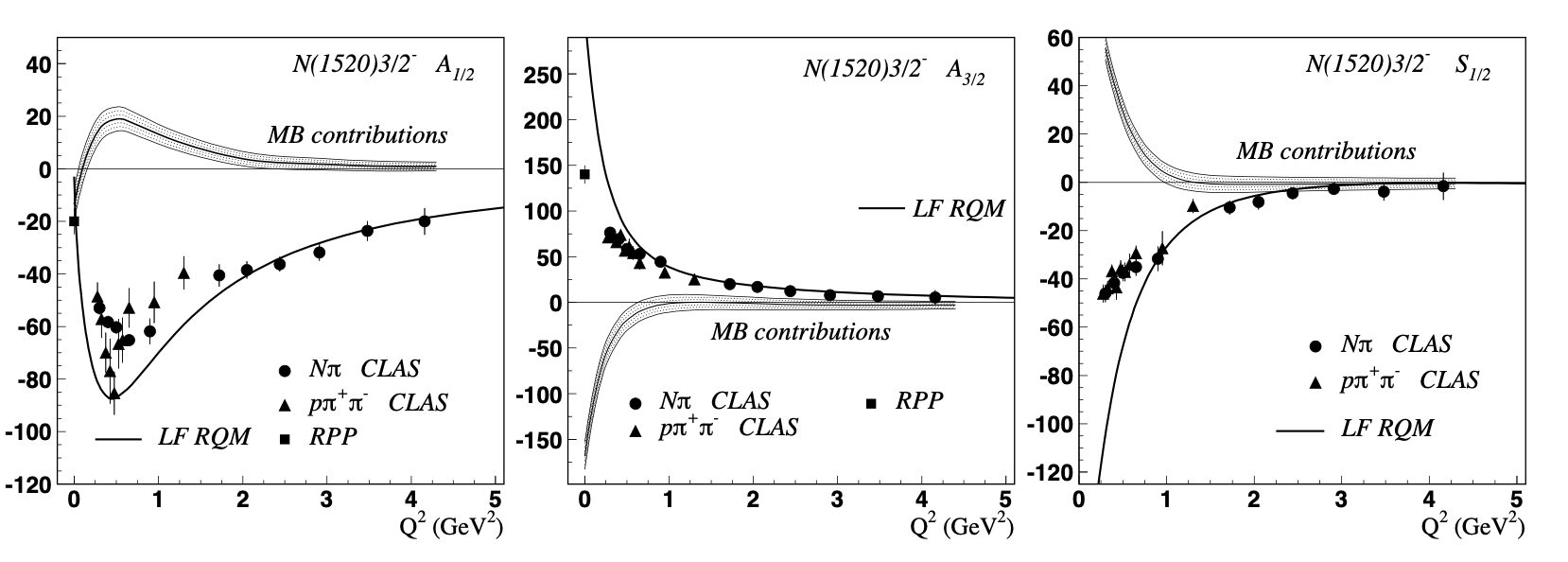}&
\hspace{-6mm}\raisebox{1mm}{\includegraphics[width=0.255\linewidth]{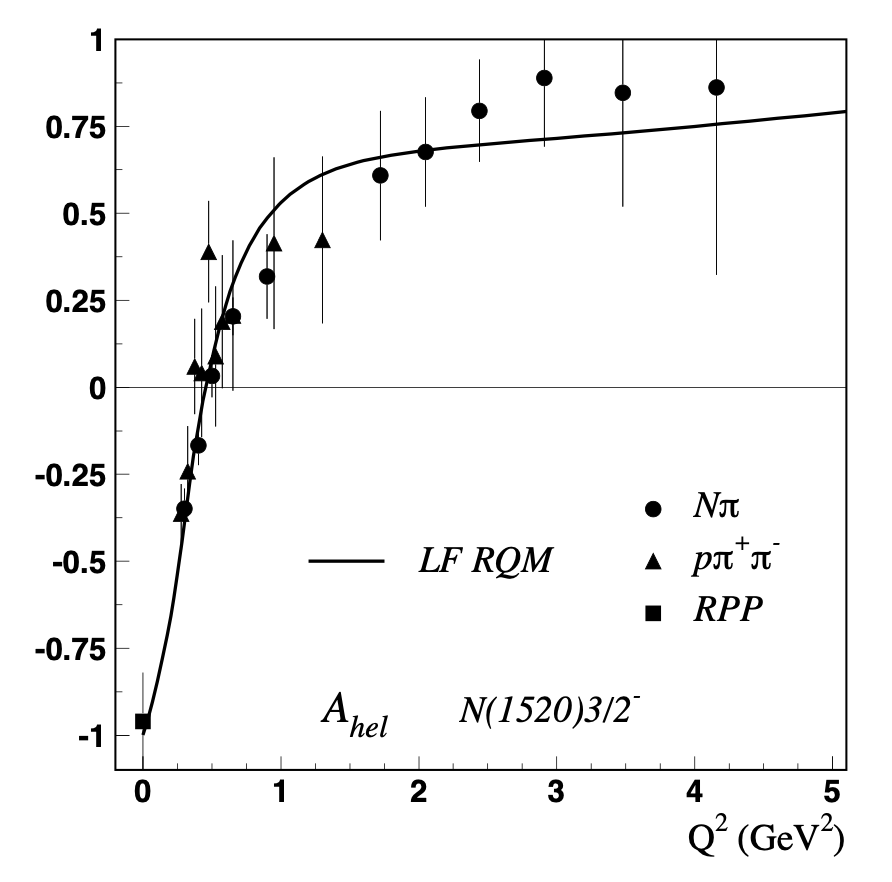}}
\end{tabular}
\vspace{-3mm}
\caption{\small The $\gamma^\ast p\to N(1520)$  resonance helicity transition amplitudes $A_{1/2}$, $A_{3/2}$, $S_{1/2}$ 
and  helicity transition asymmetry $A_{hel}$. Data and LF RQM results from \cite{CLAS:2009ces}.}
\label{fig:N1520}
\end{figure}

The $N(1520)\nicefrac{3}{2}^-$ state is the lowest excited nucleon resonance with $J^P=\nicefrac{3}{2}^-$. Its helicity structure is defined by the $Q^2$ dependence of the two transverse transition amplitudes $A_{1/2}$ and $A_{3/2}$, which are shown in Fig.~\ref{fig:N1520}. A particularly interesting feature of this state is that at the real-photon point $A_{3/2}$ is strongly dominant, while $A_{1/2}$ is very small. However, at increasing $Q^2$, $A_{1/2}$ becomes dominant, while $A_{3/2}$ drops rapidly. Meson-baryon contributions are small at $Q^2 > 1.0 \dots 1.5$~GeV$^2$ making the quark core excitation the dominant contribution.  
At $Q^2 > 2~$~GeV$^2$, $A_{1/2}$ fully dominates the process.
The helicity asymmetry $A_{hel} = (A_{1/2}^2 - A_{3/2}^2)/(A_{1/2}^2 + A_{3/2}^2)$ shown in the right panel of Fig.~\ref{fig:N1520}  illustrates this rapid change in the helicity structure. 
Expressed in terms of constraint-free form factors, 
the effect can be explained by the fact that two form factors have the same asymptotic falloff ($\propto 1/Q^6$) 
and cancel in the helicity amplitude $A_{3/2}$ but add up in $A_{1/2}$~\cite{Eichmann:2018ytt}.
Within the constituent-quark model picture, this behavior can be understood from the quark wave function, see Eq.~\eqref{N1520-wf} below, 
which on the one hand allows for $A_{1/2} = 0$ at $\vect{Q}^2 = \alpha^2$, 
where $\alpha$ is the harmonic oscillator constant, and on the other hand causes a rapid rise of the ratio of $A_{1/2}/A_{3/2}$ with increasing $\vect{Q^2}$.

   A similar behavior is expected for the $N(1680)\nicefrac{5}{2}^+$ with a somewhat slower change in the helicity structure with $Q^2$ as seen in the comparison in Eq.~\eqref{N1520-wf}. 
A somewhat analogous analysis has been conducted by Ramalho and Pena~\cite{Ramalho:2013mxa} within a covariant spectator quark-diquark model that assumes additional pion cloud contributions. A significant difference to the approach adopted here is that the $A_{3/2}$ helicity amplitude does not have a quark core contribution but coincides with the pion cloud contribution. 
Therefore, their results cannot be compared directly with the approach taken here.

\subsubsection{\boldmath  \label{N1535} The $\gamma^*p \to N(1535)\nicefrac{1}{2}^+$ transition} 

\begin{figure}
%\vspace{-10mm}
\centering
\begin{minipage}{.38\linewidth}
\includegraphics[width=1\linewidth]{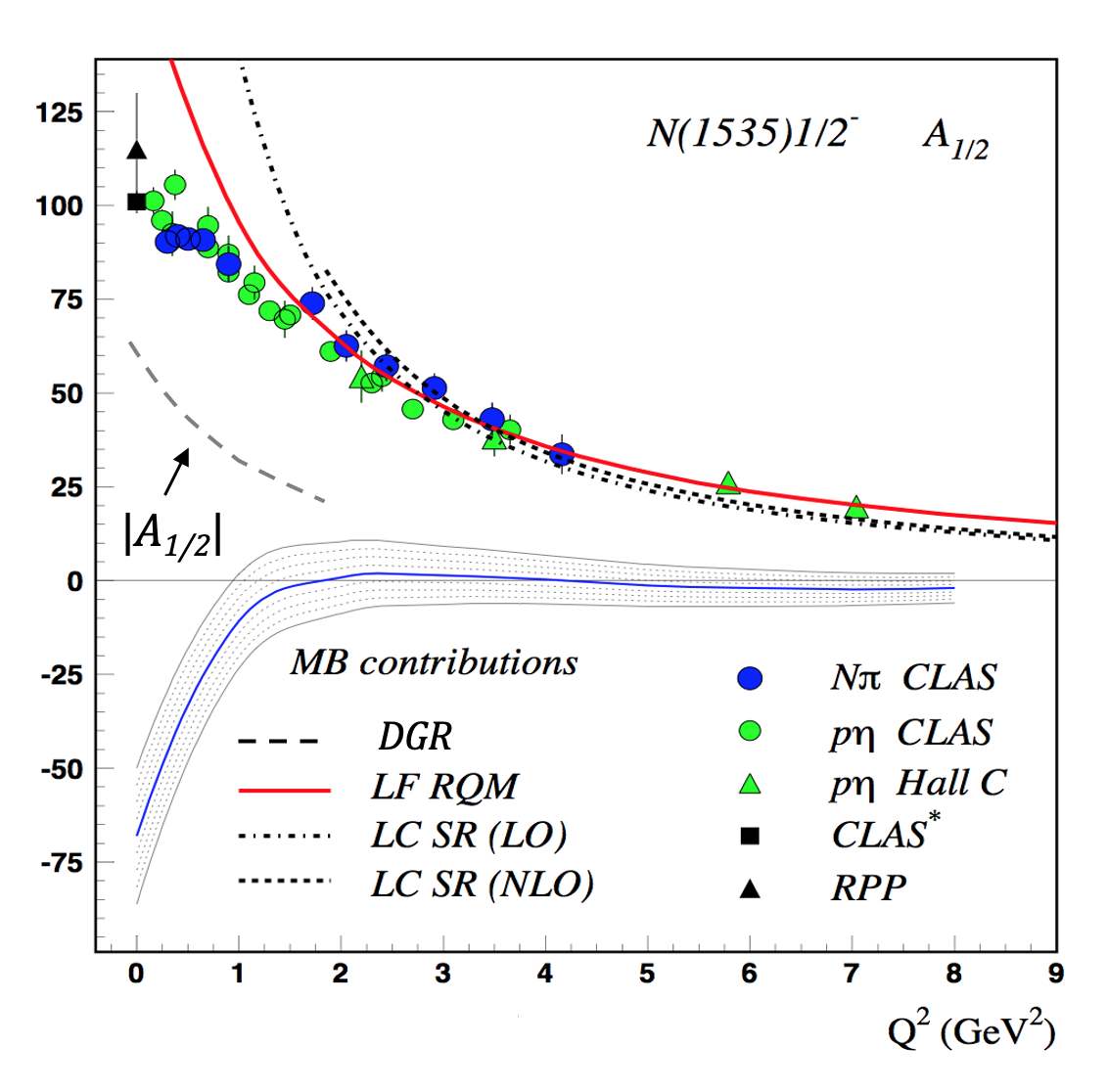}
\vspace{-10mm}
\caption{\small (Color online) $A_{1/2}$ transition amplitude. The  data include both $N\pi$ and $p\eta$ in the final state. The dashed line is the result of assuming $N(1535)$ is a dynamically generated resonance (DGR) and applying a form factor that is external to the model.~\cite{Jido:2007sm}}
\label{fig:N1535}
\end{minipage}
\hspace{4mm}\begin{minipage}{.58\linewidth}
\includegraphics[width=1.03\linewidth]{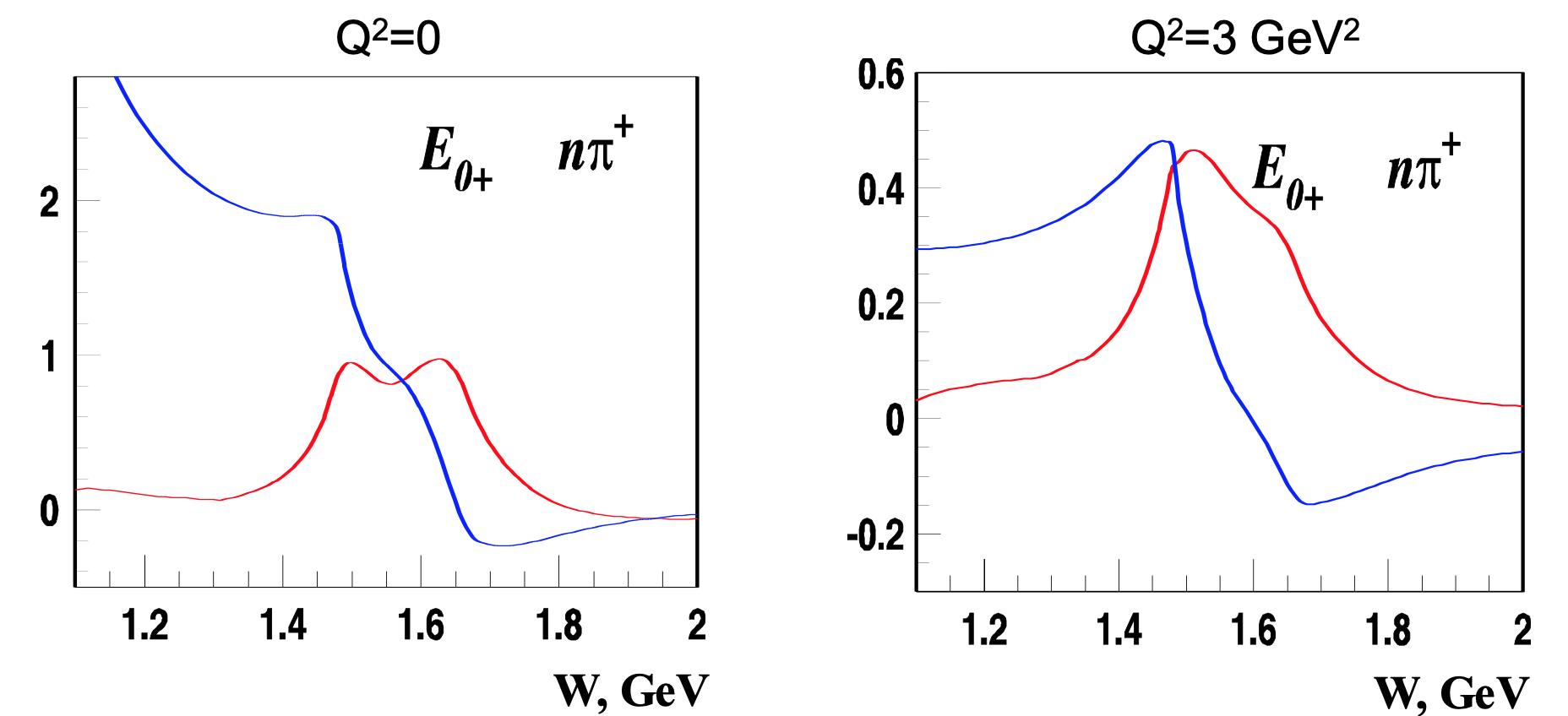}
\caption{Transition multipole $E_{0+}$ in the $N(1535)\nicefrac{1}{2}^-$ and $N(1650)\nicefrac{1}{2}^-$ mass region for $Q^2 = 0$ and $Q^2 = 3$~GeV$^2$. 
The blue (red) lines relate to the real (imaginary) part of the amplitude. 
The imaginary part shows two resonances corresponding to $N(1535)\nicefrac{1}{2}^-$ and $N(1650)\nicefrac{1}{2}^-$. 
At $Q^2=0$ the real part shows large non-resonant $S$-wave production near threshold, while at $Q^2=3$~GeV$^2$ both the real and imaginary 
parts are consistent with dominant two-resonance contributions and much reduced non-resonant strength.
The higher mass resonance appears to drop faster with $Q^2$ than the lower mass one, remaining visible just as a shoulder near 1650~MeV. Figure credit:  I. Aznauryan (private communication).}
\label{E0+}
\end{minipage}
\end{figure}

The $N(1535)\nicefrac{1}{2}^+$ is the parity partner of the ground-state nucleon, with the same spin $J=1/2$ but with opposite parity. Its quark content requires an orbital $L=1$ excitation in the transition 
from the proton. In the $SU(6)\otimes O(3)$ symmetry scheme, the state is a member of the 
$(\textbf{70},1^-)$ supermultiplet  and couples nearly equally to the $N\pi$ and  $N\eta$ final states. It has therefore been probed using both decay channels in $ep \to e p \eta$ and $ep \to eN\pi$, which are shown in Fig.~\ref{fig:N1535}. 
For isospin $I=1/2$ nucleon states, the coupling to the charged $\pi^+n$ channel is preferred over $\pi^0p$ owing to the Clebsch-Gordon coefficients. 
%{\color{blue} VB: Should we include the E0+ multipole located in Figures-VB:E0+.pdf, similar to what we did for the Roper? It shows the two 1/2- resonances at 1535 and 1650, and similar effects to the Roper at higher $Q^2$ (clearer resonance contributions).  }
 The $A_{1/2}$ helicity amplitude for the $\gamma^* p\to N(1535)\nicefrac{1}{2}^-$ resonance excitation  
shown in Fig.~\ref{fig:N1535} is known over the widest range in $Q^2$ of all nucleon states for which resonance transition form factors have been measured as part of the broad experimental program at JLab. We therefore may draw more definitive conclusions compared to some other states. 
For this resonance, as well as for $N(1440)\nicefrac{1}{2}^+$, advanced relativistic quark model calculations~\cite{Aznauryan:2015zta}, DSE calculations~\cite{Segovia:2015hra} and 
Light Cone sum rule results~\cite{Anikin:2015ita} are available, employing QCD-based modeling of the excitation of the quark core for the first time.

The transverse transition amplitude $A_{1/2}$ of $N(1535)\nicefrac{1}{2}^{-}$ is a prime example of the power of meson electroproduction to unravel the internal structure of the resonance transition. The nature of the $N(1535)\nicefrac{1}{2}^{-}$ resonance has been  discussed as a dynamically generated resonance, e.g.~\cite{Kaiser:1995cy,Bruns:2010sv}. The electroproduction data discussed here reveal structural aspects of the state and its nature that support a different interpretation. The transition form factor $A_{1/2}$ of the state, shown in Fig.~\ref{fig:N1535}, is quantitatively reproduced over a large range in $Q^2$ by two alternative approaches, the LF RQM and the LCSR. Both calculations are based on the assumption of the presence of a three-quark core. Note that there is a significant deviation from the quark calculations essentially only at $Q^2 <1 \dots 2$~GeV$^2$ for both three-quark core related calculations. The difference to the data is highlighted for the LF RQM calculation as the shaded area in Fig.~\ref{fig:N1535}, which may be assigned to the presence of non-quark contributions. 
We also note that the results of a multipole analysis shown in Fig.~\ref{E0+} qualitatively confirms the conclusions drawn from the discussion above. 
Attempts to compute the transition form factors within a strictly dynamical model, shown as the DGR labeled curve, have not succeeded in explaining the data in any part of the $Q^2$ range covered~\cite{Jido:2007sm}. However, in this case only the modulos of $A_{1/2}$ is shown in the DGR curve, and the phase of the amplitude was adjusted to obtain a value > 0. We may therefore change the sign of that contribution, which then would make the DGR line closely resemble the shaded MB contributions, both in sign and in absolute value at small $Q^2$. We may then conclude that the DGR result is dominated by the meson-baryon cloud contribution, which is not included in either of the three-quark core related calculations. Small meson-baryon contributions are also consistent with~\cite{Sekihara:2015gvw} who found negligible compositeness for the $N(1535)\nicefrac{1}{2}^-$ state. More generally, Ref.~\cite{Kinugawa:2024kwb}
found that near-threshold resonances have small composite fractions, which applies to the $N(1535)\nicefrac{1}{2}^-$ with a pole position of 1510--1520~MeV, just above the $\eta$-production threshold.    

A caveat arises   when translating the helicity amplitudes to constraint-free transition form factors $F_1$ and $F_2$.
While many models reproduce the qualitative behavior of $F_1$, they struggle to describe $F_2$ which drops unusually fast~\cite{Ramalho:2023hqd}.
In the EBAC analysis~\cite{Julia-Diaz:2009dnz} as well as the covariant spectator model~\cite{Ramalho:2020nwk} the `bare' contribution to $F_2$ even has  an opposite sign, which  points towards substantial meson cloud contributions. On the other hand, models that include such contributions describe the data well~\cite{Golli:2011jk,Gutsche:2019yoo}.

Owing to the availability of two independent approaches in computing the electroproduction amplitudes of $\gamma^\ast p\to N(1535)\nicefrac{1}{2}^-$, using either three-quark related degrees of freedom, and meson-baryon degrees of freedom, we come to the conclusion that in electro-excitation both of these approaches are needed to come to a more complete picture of the electromagnetic excitation of nucleon resonances. The long ongoing dispute of quark-based nucleon resonances versus dynamically generated meson-baryon resonances could be resolved for at least some of the resonances if the character of the electromagnetic probe is taken into account. At low $Q^2$, i.e. long-range electroproduction, and at high $Q^2$, i.e. short-range electroproduction, we probe essentially different parts of the resonance's spatial structure: peripheral in case of the former, and short distance behavior in case of the latter. The peripheral low $Q^2$ meson production reveals the dynamical features of the state, whereas high $Q^2$ electroproduction reveals the structure of the quark core.

\subsubsection{\boldmath  The $\gamma^*p\to N(1675)\nicefrac{5}{2}^-$ transition: Revealing the meson-baryon contributions}

In previous discussions we have concluded that modern quark-based models, tuned to reproduce the high-$Q^2$ behavior of the electroproduction transition amplitudes $A_{1/2}$, $A_{3/2}$, fail to reproduce quantitatively the behavior in $Q^2$ ranges below 1 \dots 2~GeV$^2$. In this range meson-baryon degrees of freedom provide significant strength to the resonance excitation depending on the specific character of the excited state.
For the Roper resonance, the  effective field theory approach of Ref.~\cite{Bauer:2014cqa} reproduces the transition amplitudes at $Q^2 < 0.6$~GeV$^2$, 
whereas quark-based approaches such as the LF RQM and the DSE and LCSR calculations do not reproduce the transition amplitudes quantitatively.
Nevertheless, the LF RQM and  DSE approaches predict features like the change of sign of the $A_{1/2}$ amplitude as seen in Fig.~\ref{p11}, but at somewhat lower $Q^2$.

\begin{figure}[!t]
\hspace{-4mm}
\includegraphics[width=1.05\linewidth]{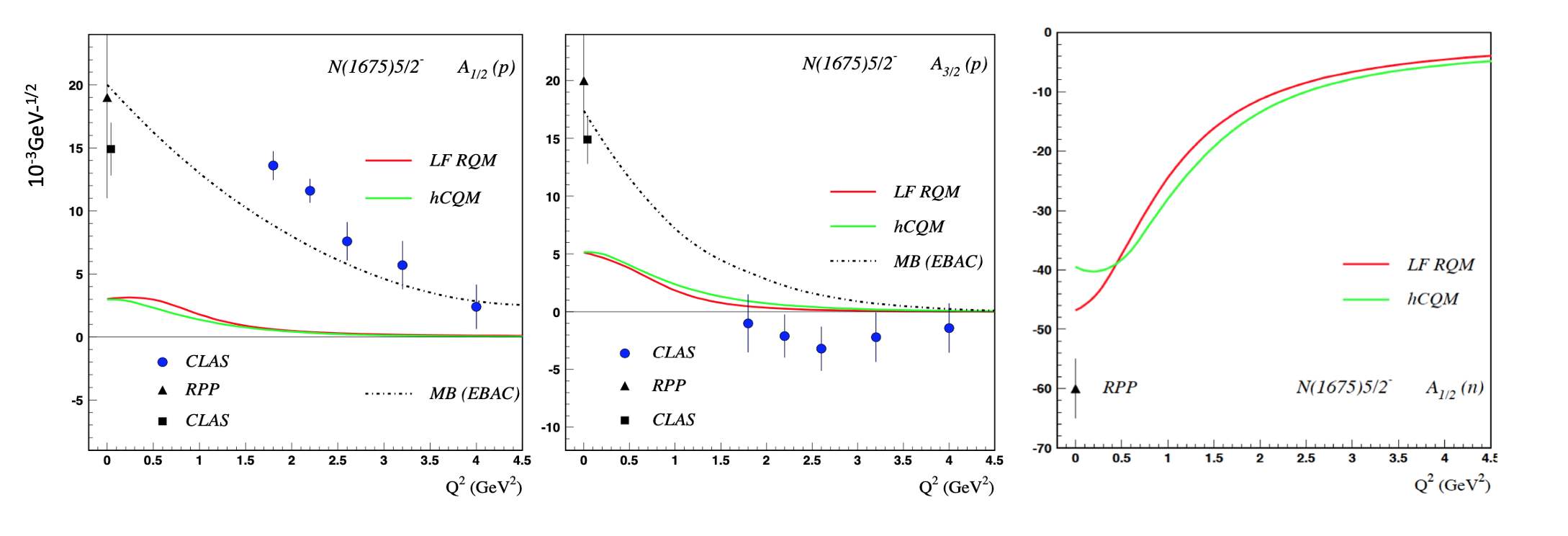} 
\vspace{-8mm}
\caption{\small (Color online)  Left and center: Transverse  helicity amplitudes $A_{1/2}$ and $A_{3/2}$ for $\gamma^* p\to N(1675)\nicefrac{5}{2}^-$ (in %units of
 $10^{-3}$~GeV$^{-1/2}$). Right: The amplitude $A_{1/2}$ on neutron target. The solid curves are from the LF RQM and the hypercentral constituent quark model (hCQM)~\cite{Santopinto:2012nq}.  The dashed-dotted curves are absolute values of the predicted meson-baryon contributions from the dynamical coupled-channel approach of Ref.~\cite{Julia-Diaz:2007mae}. The solid circles are the amplitudes extracted from CLAS pion electroproduction data~\cite{CLAS:2009ces}. The solid square at $Q^2 = 0$ is the RPP estimate~\cite{ParticleDataGroup:2016lqr}. Note that, as an effect of the selection rule, the projected quark model amplitude $A_{1/2}$ of the neutron at the photon point is one order of magnitude larger than the one on the proton. The non-quark contributions for both amplitudes $A_{1/2}(0)$ and $A_{3/2}(0)$  are $+15\pm7\times 10^{-3}$GeV$^{-1/2}$ for the proton, and $-15\pm7\times 10^{-3}$GeV$^{-1/2}$ for the neutron.} 
\label{N1675}
\end{figure}

In the case of the $N(1675)\nicefrac{5}{2}^-$, we can actually measure the meson-baryon contributions directly without having to separate them from the quark core contribution. This is due to the suppression of the quark contributions in the excitation of this resonance when probed on proton targets. 
Fig.~\ref{N1675} shows our current knowledge of the transverse helicity amplitudes $A_{1/2}(Q^2)$ and $A_{3/2}(Q^2)$ for a proton target, compared to LF RQM~\cite{Aznauryan:2017nkz} and  
hypercentral constituent-quark model~\cite{Santopinto:2012nq} calculations. The specific quark transition for a $J^P = \nicefrac{5}{2}^-$ state, belonging to the 
$[SU(6)\otimes O(3)] = (\mathbf{70}, 1^-)$ supermultiplet configuration, in the non-relativistic approximation prohibits the transition from the
 proton in a single quark transition.

This suppression, known as the Moorhouse selection rule~\cite{Moorhouse:1966jn}, is valid for the transverse transition 
amplitudes $A_{1/2}$ and $A_{3/2}$ at all $Q^2$. 
%It should be noted that this selection rule 
It applies to the 
transition from a proton target but not from the neutron, which is consistent with the data at the $Q^2=0$ point, where data from both proton and neutron targets exist. Modern relativistic quark models that go beyond 
single-quark transitions quantitatively confirm  this suppression, resulting in very 
small amplitudes from protons but large ones from neutrons. Furthermore, a direct computation of the 
hadronic contribution to the transition from protons in a coupled-channel approach confirms this (Fig.~\ref{N1675}).  
The measured helicity amplitudes off the protons are almost  
exclusively due to MB contributions, as the dynamical coupled-channel (DCC) calculation indicates (dashed line). 
The close correlation of the 
DCC calculation and the measured data for the case when quark contributions are nearly absent 
supports the phenomenological description %of the helicity amplitudes 
in terms of a three-quark core  
that dominates at high $Q^2$ and MB contributions that can make important contributions at lower $Q^2$. 
From the direct measurement of the MB contributions on the proton, and from measurements on the neutron at the photon point subtracted of the LF RQM predictions, one finds that for this resonance the MB contributions are of isovector nature, as shown in Fig.~\ref{N1675-iso}.

%\newpage
\subsubsection{\bf \boldmath The $\gamma^*p\to N(1680)\nicefrac{5}{2}^+$ transition}

In the quark model, the $N(1680)\nicefrac{5}{2}^+$ is part of the $(\mathbf{56},2^+)$ supermultiplet with spin 
$S=\nicefrac{1}{2}$ and orbital angular momentum $L=2$. 
It belongs to the same Regge trajectory as the $N(1520){\nicefrac{3}{2}^-}$,
which is part of the  $(\mathbf{70},1^-)$ supermultiplet and also has three-quark spin $S=\nicefrac{1}{2}$
but one unit lower angular momentum $L=1$.
The non-relativistic quark model with a harmonic 
oscillator potential predicts a very similar helicity structure for the ratios of $A_{1/2}$ and $A_{3/2}$ of the $N(1520)\nicefrac{3}{2}^-$ and the $N(1680)\nicefrac{5}{2}^+$~\cite{Copley:1969qn}, indicating that the two resonance transitions should have a very similar helicity structure of the ratios
(see, however,  the comments in~\cite{Ono:1976mx}): 
\begin{equation}
\label{N1520-wf}
 \hspace{-6mm} N(1520): \quad \frac{A_{1/2}}{A_{3/2}} = \frac{1}{\sqrt{3}}  \left( {\frac{\vect Q^2}{\alpha^2}}- 1 \right), \qquad 
 N(1680): \quad \frac{A_{1/2}}{A_{3/2}} = -\frac{1}{2\sqrt{2}}  \left (\frac{\vect Q^2}{\alpha^2}- 2 \right ).
\end{equation} 
Because of this similarity, we  discuss the two resonance transitions together as we expect the same nature and a similar $Q^2$ dependence of the transition amplitudes for both states. The authors of Ref.~\cite{Wang:2024jns,Wang:2023snv} consider the $N(1520)\nicefrac{3}{2}^-$ to be predominantly a composite meson-baryon resonance, while the $N(1680)\nicefrac{5}{2}^+$ is considered predominantly an elementary three-quark resonance. In this case, we should expect a different $Q^2$ behavior of their respective helicity amplitudes.

\begin{figure}
\hspace{4mm}\begin{minipage}{.40\linewidth}
\centering
\includegraphics[width=0.6\linewidth]{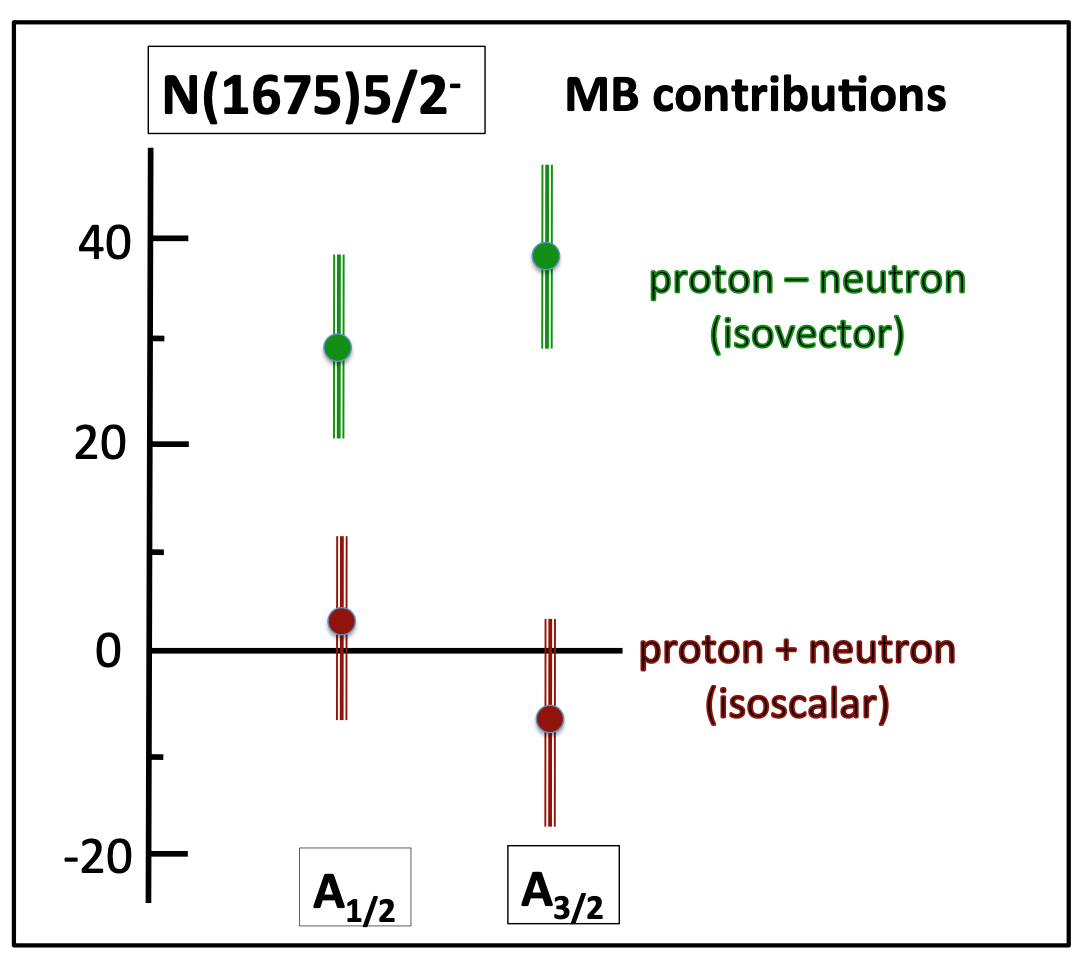}
\caption{\small Isovector and isoscalar part of  $A_{1/2}(0)$ and $A_{3/2}(0)$ amplitudes for the $N(1675)\nicefrac{5}{2}^-$~\cite{Aznauryan:2014xea}.} 
\label{N1675-iso}
\caption{\small Electroproduction transition amplitudes $A_{1/2}$, $A_{3/2}$ and $S_{1/2}$ for the $\gamma^\ast p\to N(1520)\nicefrac{3}{2}^-$ (top) and $\gamma^\ast p\to N(1680)\nicefrac{5}{2}^+$ (bottom) transitions. The curves are to guide the eyes. Source: \textit{Baryons reviews: $N$ and $\Delta$ Resonances} in~\cite{Workman:2022ynf}.} 
\label{N1520-N1680}
\end{minipage}
\hspace{2mm}\begin{minipage}{.57\linewidth}
\centering
\includegraphics[width=\linewidth]{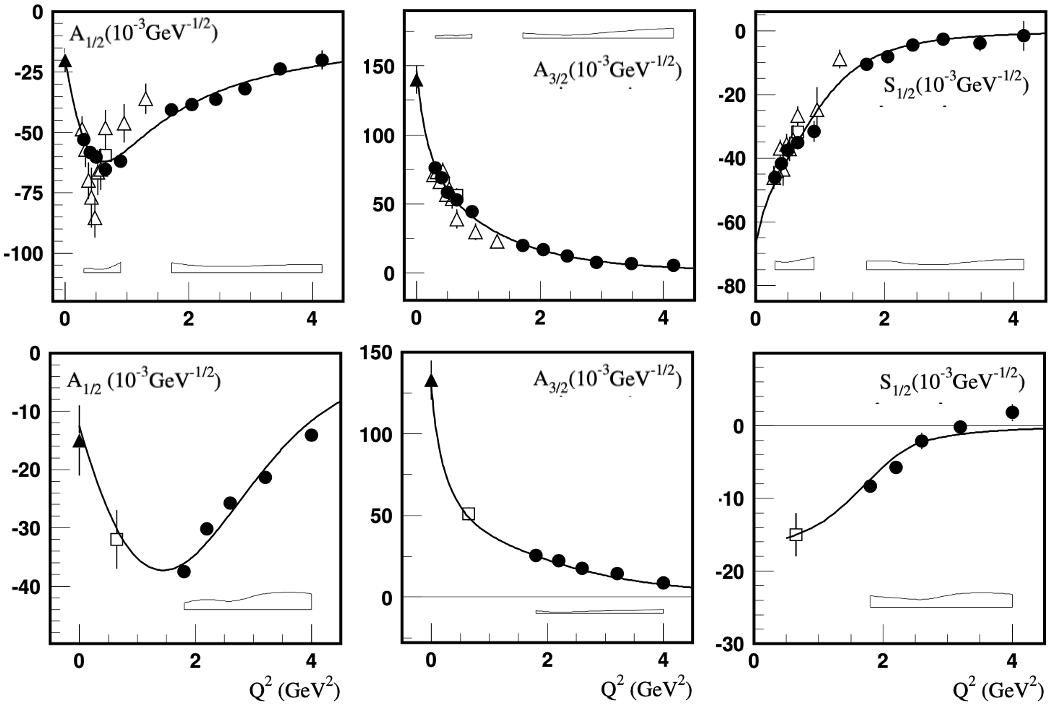}
\end{minipage}
\vspace{-0mm}
\end{figure}

The experimental results of the individual amplitudes for both states are displayed in Fig.~\ref{N1520-N1680} for ease of comparison. Both states show very similar $Q^2$ dependencies: The $A_{1/2}$ amplitudes are close to zero at the real photon point, strongly rise with a negative magnitude at small $Q^2$, and slowly fall to become the dominant contribution high $Q^2$. The $A_{3/2}$ amplitudes are both dominant at the photon point and drop rapidly  with $Q^2$. The $S_{1/2}$ amplitudes are both negative at the photon point and drop rapidly in magnitude with $Q^2$. 
%In both cases, $A_{1/2}$ becomes the dominant contribution at high $Q^2$, consistent with the general prediction of asymptotic QCD at $Q^2 \to \infty$ in (\ref{pQCD}).    
Concerning the nature of both states, the similarity of their $Q^2$ dependence strongly indicates that they both at their core are three-quark states, 
just like all the other resonances we have discussed so far.  

There are a number of nucleon and $\Delta$ excited states where electroproduction data for their transition amplitudes exist but are limited to narrow $Q^2$ ranges, e.g. $N(1650)\nicefrac{1}{2}^-$,  $N(1710)\nicefrac{1}{2}^+$,  $N(1880)\nicefrac{1}{2}^+$, $\Delta(1620)\nicefrac{1}{2}^-$, and $\Delta(1700)\nicefrac{3}{2}^-$.  Due to the limited coverage in $Q^2$~\cite{Ramalho:2023hqd}, these states are  difficult to interpret in terms of the quark core and meson-baryon contributions to the excitation strength.

\subsubsection{\label{sub:2nd}\boldmath  The $\gamma^*p\to N(1710)\nicefrac{1}{2}^+$ transition}

The $N(1710)\nicefrac{1}{2}^+$ is the second excitation in the $J^P = \nicefrac{1}{2}^+$ nucleon channel, i.e.,
above the Roper resonance. This state is interpreted differently in the literature:
In the AdS/QCD approach~\cite{Brodsky:2014yha} it   %  of Brodsky and collaborators
is the second radial excitation of the nucleon, while in Ref.~\cite{Suzuki:2009nj} it was
interpreted as dynamically generated resonance emerging from coupled-channel dynamics, jointly with a two-pole structure
at the Roper mass. In Sec.~\ref{Global} we discussed this resonance as 
a companion of the $N(1440)\nicefrac{1}{2}^+$, which has  one node but a different spatial distribution.

\begin{figure}[!b]
\centering
\begin{tabular}{ccc}
 \hspace{-2mm}   \includegraphics[width=0.33\linewidth,height=0.30\linewidth]{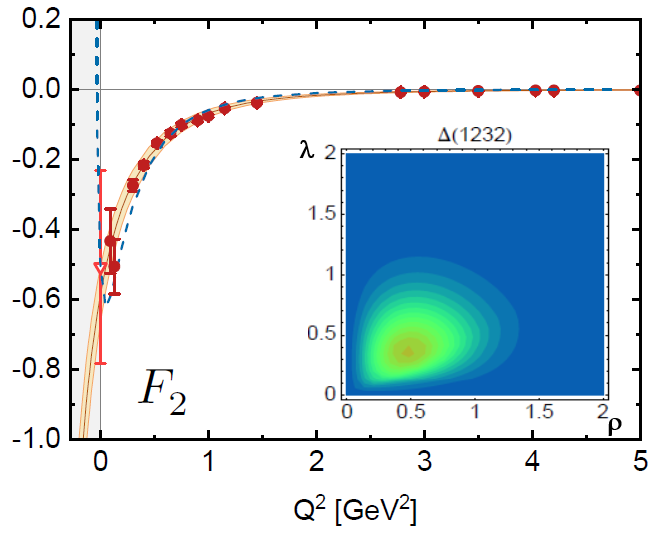}&
\hspace{-4mm}    \includegraphics[width=0.33\linewidth,height=0.30\linewidth]{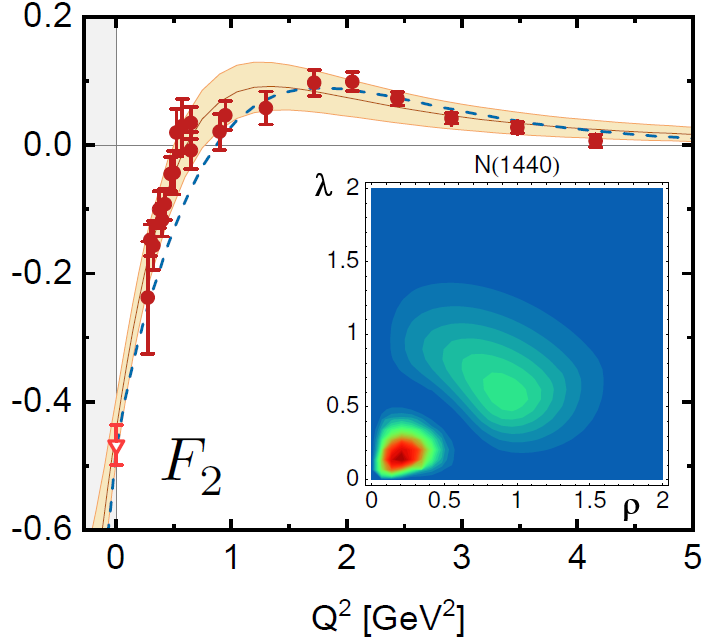}&
\hspace{-4mm}    \includegraphics[width=0.33\linewidth]{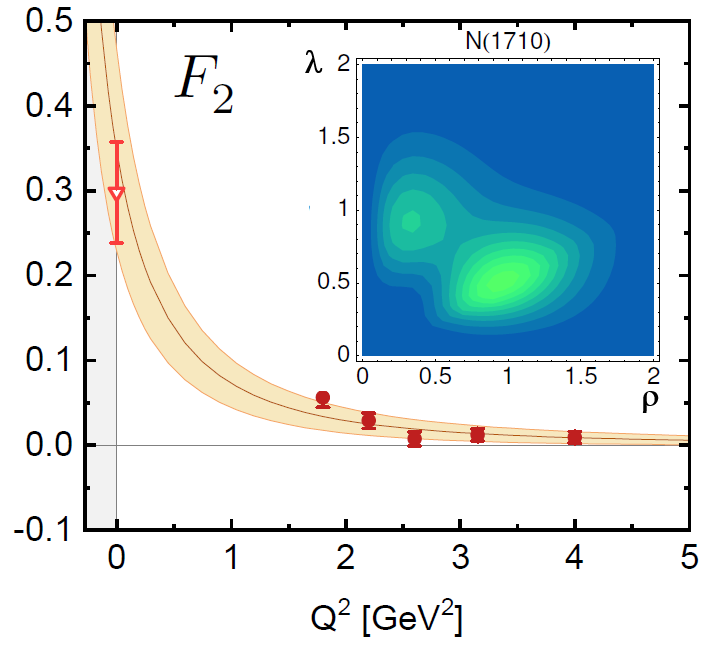}\vspace{-2mm}
 \end{tabular}
    \caption{\label{fig:2radials} Transition form factor $F_2(Q^2)$ for the $\Delta(1232)\nicefrac32^+$, $N(1440)\nicefrac12^+$  
    and $N(1710)\nicefrac12^+$.
    The data are from Refs.~\cite{CLAS:2012wxw,CLAS:2014fml}, the band is from~\cite{Eichmann:2018ytt} and the dashed curve
    from \cite{Drechsel:2007if}. The quark-model spatial probability distributions 
    from \cite{Melde:2008yr} are shown as insets.
    }
\end{figure}

The insets in Fig.~\ref{fig:2radials} shows quark-model calculations of these distributions for the $\Delta(1232)\nicefrac{3}{2}^+,$
$N(1440)\nicefrac{1}{2}^+$ and $N(1710)\nicefrac{1}{2}^+$ as functions of the variables $\rho$ and $\lambda$ (in fm). 
The $\Delta(1232)\nicefrac{3}{2}^+$ density distributions shows no significant structure. 
The  $N(1440)\nicefrac{1}{2}^+$ density distribution shows a minimum as a function of $\rho^2+\lambda^2$, which is related to
the zero crossing in the form factor $F_2$. The $N(1710)\nicefrac{1}{2}^+$ density distribution, on the other hand,  
shows a minimum as a function of $\rho^2-\lambda^2$. 
This is consistent with the data for $F_2$, which  do not show any zero crossings as one might expect for a second radial excitation.
Therefore, the $N(1710)\nicefrac{1}{2}^+$ seems to be incompatible with an interpretation
as a $3s$ state, although higher statistics in the data would certainly be desirable.  
These observations also do not seem to support the interpretation in~\cite{Suzuki:2009nj},
where a parent state is supposed to generate three molecular states; if these states share the $qqq$ content 
of the bare state, they should also have similar transition  form factors in electroproduction 
which is not the case.
A study of a $qqq$ contribution to
the three molecular states within the model of Ref.~\cite{Suzuki:2009nj} could help  resolve this issue.

\subsection{What have we learned on the nature of the excited states from electroproduction?} 

{The preceding discussions of the $\Delta(1232)\nicefrac{3}{2}^+$, $\Delta(1600)\nicefrac{3}{2}^+$, $N(1440)\nicefrac{1}{2}^+$, $N(1520)\nicefrac{3}{2}^-$, and $N(1535)\nicefrac{1}{2}^-$ electroproduction transition amplitudes show clear evidence of quark-core contributions, particularly at $Q^2 >2$~GeV$^2$ where they nearly coincide within uncertainties with the quark-based approaches of the LFRQM, DSE and LCSR calculations. 
Furthermore, for the $N(1675)\nicefrac{5}{2}^-$ state we find a quantitative estimate of the non-quark contribution to the proton transition amplitudes $A_{1/2}(Q^2)$ and $A_{3/2}(Q^2)$ owing to the suppression of the quark contribution. For the neutron an estimate at the real photon point leads to results that are similar to the proton albeit with opposite sign. It indicates the isovector character of these contributions.}

\begin{figure}[!t]
\vspace{3mm}
\centering
\includegraphics[width=1\textwidth]{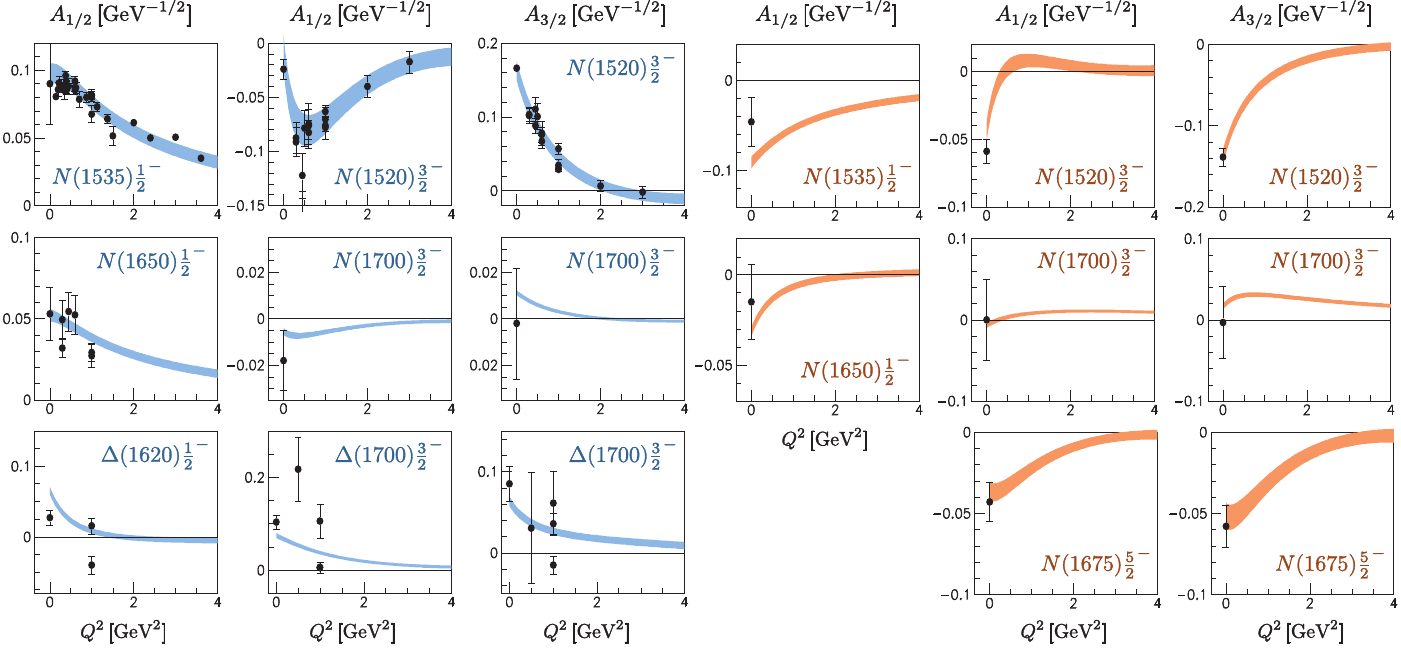}
\caption{Single-quark transition model (SQTM) amplitudes for the transitions from the ground-state $(\mathbf{56},0^+)$ to the $(\mathbf{70},1^-)$ supermultiplet for protons (left, blue) and neutrons (right, orange)~\cite{Burkert:2002zz,Aznauryan:2011qj}. The markers are results of the data analysis. The colored bands for the top three amplitudes are parameterizations fitted to the data, whereas the bands in all other panels are SQTM predictions using those three as inputs. For the two $S$-wave states $N(1535)\nicefrac{1}{2}^-$ and $N(1650)\nicefrac{1}{2}^-$ a mixing angle of $31^o$ is used, and for the two $D$-wave states $N(1520)\nicefrac{3}{2}^-$ and $N(1700)\nicefrac{3}{2}^-$ the mixing angle of $6^o$ is applied. Note that these graphs do not include more recent data covering higher $Q^2$  results. }
\label{SQTM-proton}
\end{figure}

 It is also interesting that the amplitudes are qualitatively compatible with the radial and orbital excitation structure expected from their assignments within the $SU(6) \times O(3)$ three-quark representation. This can be seen in Fig.~\ref{SQTM-proton}, 
where the results from the algebraic single-quark transition model (SQTM)~\cite{Hey:1974qe,Cottingham:1978za,Burkert:2002zz} for the transition amplitudes of the members of the $(SU(6), L^P)$ = $(\mathbf{70},1^-)$ supermultiplet are plotted.
In this model, only single-quark transitions to the excited states are considered. After fitting three of the amplitudes to the data,  the remaining 16 transverse helicity amplitudes for all states in $(\mathbf{70},1^-)$ are determined, including transitions on neutrons. 
For resonances with sufficient reach in $Q^2$ the agreement is surprisingly good. For the neutrons, only data at the photon point are available, but they also show a surprisingly good correspondence.
 
In general, all  resonances we discussed in this section exhibit significant contributions in the lower $Q^2$ range that {\bl may}  be described in hadronic models as meson-baryon contributions. Here the transition $\gamma^* p \to N(1675)\nicefrac{5}{2}^-$ is particularly interesting as it is suppressed on proton targets via a selection rule, and we get a direct measurement of the meson-baryon contributions to the excitation of this resonance.
% As demonstrated in the case of $\gamma^\ast p\to N(1675)\nicefrac{5}{2}^-$, the meson-baryon contributions can be directly accessed and 
These are found to be small at $Q^2 > 1.5$~GeV$^2$ for $A_{3/2}$ and at $Q^2 > 3$ GeV$^2$ for $A_{1/2}$, while both contributions, quark core and meson-baryon, are present in the case of neutron target $\gamma^\ast n \to N(1675)\nicefrac{5}{2}^-$ at the photon point.

The experimental data used in Fig.~\ref{p11} and Figs.~\ref{fig:Delta-1600}--\ref{N1520-N1680} were recently also analyzed by the J\"ulich-Bonn-Washington (JBW) group ~\cite{Wang:2024byt}. The group  employs a dynamical coupled channel approach. The results are presented as the $Q^2$ dependence of the imaginary and real part of the helicity transition amplitudes $A_{1/2},~ A_{3/2}$ and $S_{1/2}$. The real parts compare generally well with the results of the data analysis summarized in ~\cite{Aznauryan:2011qj} with the exception of the $A_{1/2}$ amplitude of $N(1440)1/2^+$ at very small $Q^2$, where the zero crossing is observed at different $Q^2$. This is an interesting discrepancy and should be studied further; it does, however, not affect the conclusions we draw in this article.   

The Argonne-JLab-Osaka  group has determined some of the resonance transition amplitudes employing a dynamical coupled channel approach~\cite{Suzuki:2010yn}. They extract transition amplitudes for two Roper-like states at pole masses just 8 MeV apart from each other. If one adds the real parts of the two amplitudes, they compare qualitatively well with the results we discuss in this article. Similar observations apply for the $N(1520)\nicefrac{3}{2}^-$, where the results compare  quantitatively well in the limited $Q^2$ range that was analyzed by the group.

\begin{figure}
\centering
\hspace{2mm}\begin{minipage}{.51\linewidth}
\centering
\includegraphics[width=0.84\linewidth]{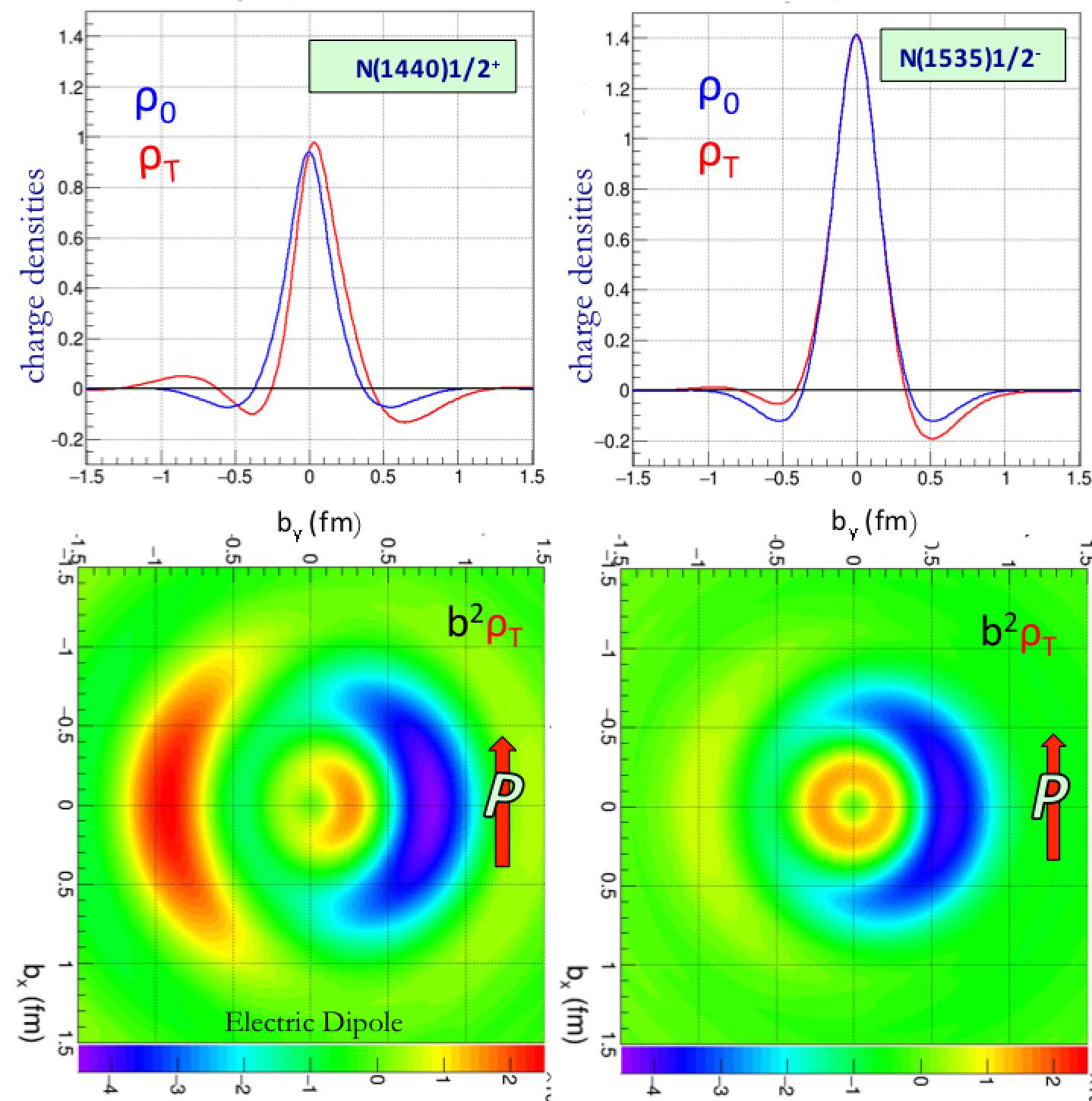}
\end{minipage}
\begin{minipage}{.47\linewidth}
\centering
\includegraphics[width=0.6\linewidth]{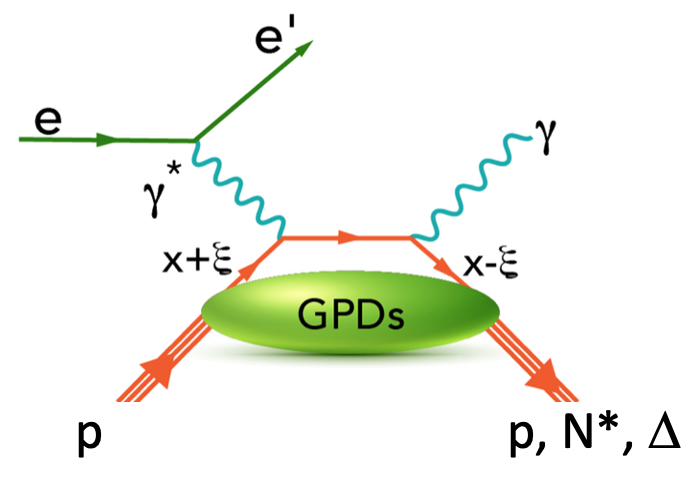}
\caption{(color online) Deeply virtual Compton scattering of a virtual photon from a quark inside the proton target, with a high energy real photon  emitted from the  quark. The proton or an excited baryon ($N^*$, $\Delta$) remains in the final state. }
\label{dvcs}
\caption{(color online) Left panels: $N(1440)$, top: projection of charge densities on $b_y$, bottom: transition charge densities when the proton  is spin polarized along $b_x$. Right panels: same for $N(1535)$. Note that the densities are scaled with 
$b^2$ to emphasize the outer wings. Color code:negative charge is blue, positive charge is red. Note that all scales are the same for ease of comparison~\cite{Burkert:2018oyl}. Graphics credit: F.X. Girod.}
\label{charge_densities}
\end{minipage}
\end{figure}

The knowledge of helicity amplitudes in a large  $Q^2$ range also allows for the determination of transition charge densities on the light cone in transverse impact parameter space ($b_x, b_y$)~\cite{Carlson:2007xd,Tiator:2008kd}. 
A comparison of the $N(1440)\nicefrac{1}{2}^+$ and $N(1535)\nicefrac{1}{2}^-$ light-front densities is shown in Figure~\ref{charge_densities}. There are clear differences in the charge transition densities between the two states. The Roper  has a softer positive core and a wider negative outer cloud than the $N(1535)\nicefrac{1}{2}^-$ and develops a larger shift in $b_y$ when the proton is polarized along the $b_x$ axis. These results are qualitatively consistent with the interpretation of a more extended spatial charge distribution from the widely distributed meson-baryon contributions to the Roper spatial wave function compared to the much narrower meson-baryon contributions to the $\gamma^\ast p\to N(1535)$ transition. The more narrowly shaped core contribution in the latter case also indicates that quark core contributions are strongly dominant over any meson-baryon contributions. 

Despite the very significant progress made in recent years in further establishing the light-quark baryon spectrum and exploring the internal structure of excited states and the relationship to QCD~\cite{Carman:2020qmb,Proceedings:2020fyd}, much remains to be done. A large amount of precision data already collected needs to be included in the multi-channel analysis frameworks, and polarization data are still to be analyzed. 
There are approved proposals to study resonance excitations at much higher $Q^2$ and with higher precision at Jefferson Lab with CLAS12~\cite{CLAS:2022kta,Burkert:2018nvj}, which may begin to reveal the transition from a dressed quark core to a bare quark core at short distances. First attempts have been made on extending these coupled-channel analysis frameworks to the electroproduction data as well~\cite{Mai:2023cbp,Wang:2024byt}.

During the past decade, eight baryon states in the mass range from 1.85 to 2.15~GeV have been discovered, or evidence for the existence of states has been significantly strengthened. Some of these states are in 
the mass range and have $J^{PC}$ quantum numbers that could have significant contributions of gluonic components. Such ``hybrid'' states are in fact predicted in lattice QCD~\cite{Dudek:2012ag}. These states appear with the same quantum numbers as ordinary quark excitations and could possibly be isolated from ordinary states due to the $Q^2$ dependence of their helicity amplitudes~\cite{Li:1991yba}, which is expected to be quite different from ordinary 3-quark excitation. The study of hybrid baryon excitations then requires new electroproduction data, especially at low $Q^2$~\cite{Lanza:2021ayj} with different final states and masses above 2~GeV.

A novel way of studying the structure of the proton  has been experimentally pioneered in recent years and resulted in the first determination of the mechanical properties of the proton, its internal distribution of pressure and shear stress~\cite{Burkert:2018bqq}. This remarkable result is based on the similarity of the graviton coupling, a spin-2 
interaction, with the coupling of two spin-1 photons in the exclusive process of deeply virtual Compton scattering (DVCS). 
For a colloquial introduction in this topic, see ~\cite{Burkert:2023wzr}. If energy is transferred to the ground-state proton, a highly energetic photon may be released while the proton remains in its ground state, in which case the process probes the generalized parton distributions (GPDs) of the 
nucleon. If an intermediate nucleon excitation is formed, which then decays into a nucleon and a pion or some other final state in addition to the photon, 
the process probes the $NN^*$ transition GPDs from the nucleon to the intermediate resonance. 
Recently, the process $ep\to e\pi^-\Delta^{++}$, which is sensitive to the $N \to \Delta$ transition GPDs, has been measured~\cite{CLAS:2023akb},  see~\cite{Diehl:2024bmd} for a review. While this research is still in its very early stage, much more progress can be expected in the coming years. It is worth mentioning that the DVCS process in Fig.~\ref{dvcs} is a hard, deep inelastic process, where the final-state photon has energies $E_\gamma \ge 2-3$~GeV, and the reaction  provides access to the transition GPDs of the resonance's quark core.

Where does one go from here on the theory side?
Despite the substantial progress discussed so far, the theoretical description of baryon resonances remains at an early stage. 
As we have seen, current modelling allows one to separate the physics into a three-quark core at high $Q^2$ and meson-baryon interactions at low $Q^2$, with various approaches combining these components in a suitable framework.
In this section we  focused on  recent results for nucleon resonance transition amplitudes and their interpretation within different approaches, including light-front relativistic quark models with momentum-dependent constituent quark masses, 
light-cone sum rules~\cite{Braun:2009jy},
and functional methods based on Dyson-Schwinger equations~\cite{Roberts:2007ji,Eichmann:2016yit}; for further examples see~\cite{Ramalho:2023hqd} and references therein. Typically, 
such calculations describe the transition form factors at $Q^2 \gtrsim  2$~GeV$^2$, while at lower $Q^2$ values hadronic degrees of freedom must be included. These may account for significant parts of the total resonance excitation strength.

To move beyond models toward a first-principles description of baryon resonances, one could add several more items to the wish list. A systematic approach should (i) ideally start from the QCD Lagrangian, whose only parameters are the current-quark masses; (ii) be fully relativistic; (iii) reproduce the baryon spectrum, including a proper treatment of resonances; (iv) be systematically improvable; (v) ensure that transition amplitudes satisfy all low-energy constraints, as well as electromagnetic gauge invariance and thus transversality; (vi) allow analytic continuations to timelike momenta to generate vector-meson poles on the appropriate Riemann sheets; (vii) treat transition amplitudes as complex quantities, as they represent residues of resonance poles; (viii) and include all relevant contributions, from quark core to meson-baryon effects.
One candidate for such an approach are functional methods, which are starting to evolve from QCD modelling towards ab-initio calculations.
The prime candidate, however, is lattice QCD.
For the lowest-mass states, the $\Delta(1232)\nicefrac{3}{2}^+$ and Roper $N(1440)\nicefrac{1}{2}^+$, lattice calculations have been carried out about a decade ago which are consistent with the data within large uncertainties. 
Following recent progress in the meson sector, lattice QCD is now beginning to tackle resonances above thresholds in the baryon sector as well.

Naturally, the price to pay in a first-principles approach is that the distinction between a ``quark core'' and ``meson-baryon interactions’’ eventually starts to blur: A black box whose only inputs are the current-quark masses and a scale adjusted to experiment does not know how to disentangle these contributions in a clean and model-independent way. However, the same can be said about nature as well. This underscores the importance of combining first-principles approaches with QCD modelling, effective  theories and data-driven analyses to achieve a comprehensive understanding of the baryon excitation spectrum.

\clearpage

\section{\label{Out}Specific Questions in Baryon Spectroscopy}
\subsection{Hybrids, pentaquarks, dynamically generated, molecular, and exotic states}

\subsubsection{\boldmath Wording}

Naming conventions in hadron spectroscopy often reflect the perspective of the authors. Here we outline the terminology used in this work to classify different types of states.

Resonances are unstable particles above thresholds that decay into specific two- or multihadron final states. 
For example, the $\Delta(1232)\nicefrac{3}{2}^+$ is above the $N\pi$ threshold and decays almost exclusively into $N\pi$. 
The $\Delta(1232)\nicefrac{3}{2}^+$ wave function thus contains a $N\pi$ component~\cite{Pascalutsa:2006up} 
but is dominantly a three-quark state. 
The Roper resonance $N(1440)\nicefrac{1}{2}^+$ decays into $N\pi$, $N\sigma$, and $\Delta(1232)\nicefrac{3}{2}^+\pi$. 
Within effective field theory frameworks, 
the $N(1440)\nicefrac{1}{2}^+$ can be dynamically generated from meson-baryon interactions, and no three-quark component is required.
When this mechanism dominates, the resonance is often referred to as a hadronic molecule.
A baryon that arises dynamically from meson-baryon interactions must, at the quark level, contain at least four quarks and one antiquark. If such a state is narrow, it is sometimes classified as a pentaquark.

A state is called exotic if either its $J^{PC}$ quantum numbers or its minimal quark content ($q\bar{q}$ for mesons or $qqq$ for baryons) are not compatible with the conventional quark model. Hadrons in which the gluonic field contributes excitation energy are referred to as hybrids. Only mesonic hybrids can have quantum-number-exotic $J^{PC}$ combinations, i.e., values that are forbidden for  $q\bar{q}$ configurations in the quark model. Finally, states that are not part of the traditional quark model spectrum but also do not possess exotic quantum numbers are sometimes termed cryptoexotic.

\subsubsection{\boldmath``Quark core'' vs. meson-baryon interactions}

In the previous sections, we explored a range of approaches to understanding the nucleon and its excitations. 
Yet no single method
works well on its own: effective field theories  need low-energy constants from other sources or 
from lattice QCD.
Phenomenological models depend on guidance from ab-initio calculations. Functional methods benefit from lattice 
results to improve their truncations. Lattice QCD, in turn, relies on effective field theories to connect
finite-volume spectra to physical resonances.
And of course, all of these approaches depend on experimental input, just as experimental analyses 
require theory input.
A comprehensive understanding of  the baryon spectrum thus calls for an interdisciplinary effort.

A key challenge is to disentangle the interplay between the intrinsic quark core and the surrounding meson 
cloud. A useful conceptual tool in this context is Weinberg's compositeness criterion~\cite{Weinberg:1965zz}, 
which relates observable scattering parameters to the internal structure of a bound state. Specifically, 
Weinberg established a connection between the scattering length $a$, the effective range $r$, 
and the probability $Z$ of finding a compact, or ``elementary", component within the wave function
of a hadronic bound-state:
\begin{eqnarray}
a \approx -\frac{2}{\gamma}\frac{1-Z}{2-Z} \phantom{zzzzzzzzz}
r \approx -\frac{1}{\gamma}\frac{Z}{1-Z}.\label{StW}
\end{eqnarray}
where $\gamma=\sqrt{2\mu E_B}$, with $\mu$ denoting the reduced mass and $E_B$ the 
binding energy. Note that the sign of $a$ depends on the chosen convention.
When a resonance lies near the mass of a conventional hadron, the effective range expansion in Eq.~\eqref{StW} becomes less reliable, and care must be taken in its application.

The compositeness criterion has been generalized to account for coupled-channel dynamics~\cite{Matuschek:2020gqe},
though the interpretation of the effective range in such systems remains nontrivial and warrants special
attention~\cite{Baru:2021ldu}. In the present context, we do not elaborate on these refinements, but 
instead highlight recent results obtained within the J\"ulich-Bonn model~\cite{Wang:2023snv}. This study 
tested multiple definitions of compositeness and analyzed 13 light-quark baryon resonances. For eight 
of these states, reasonably consistent results were found.  
Four states, the $N(1440)\nicefrac12^+$, $N(1520)\nicefrac32^-$, $N(1535)\nicefrac12^-$,
and $N(1710)\nicefrac12^+$, ``have the chance of being composite". For example, 
the $N(1440)\nicefrac12^+$ appears to have an approximately 50\% probability of being ``elementary", with
the remaining strength  arising from molecular components distributed among its 
dominant decay channels. Obviously, it is the ``elementary"
part that is responsible for the node in the wave function found in electroproduction, with 
$N\pi$, $\Delta(1232)\nicefrac32^+\pi$, and $N\sigma$ forming the meson cloud.
These findings support a nuanced interpretation in which a dynamically generated resonance may still contain a genuine three-quark core, dressed by meson-baryon components. In this view, the $N(1440)\nicefrac12^+$
is not dynamically generated in nature but can be dynamically generated 
within a suitable theoretical framework.

As of today, the picture for the $N(1440)\nicefrac{1}{2}^+$ seems reasonably consistent: it is frequently described as a dynamically generated state, as in Ref.~\cite{Wang:2023snv}, where it includes a three-quark core, and in electroproduction studies, where it emerges as a quark-model state dressed by a meson cloud. 
In contrast, conflicting interpretations exist for the $N(1520)\nicefrac{3}{2}^-$. In Ref.~\cite{Wang:2023snv}, this resonance is characterized as dominantly ($\sim 75\%$) molecular in nature. However, electroproduction data reveal it as a compact object with minimal meson-cloud contributions.

An important question remains unanswered: what is the underlying reason for the existence of the $N(1440)\nicefrac{1}{2}^+$, or more generally, of dynamically generated states? Are resonances like the $N(1440)\nicefrac{1}{2}^+$ molecular in nature, coupling to a three-quark core via the annihilation of a $q\bar{q}$ pair? Or does a pole, originating from the strong interactions among three quarks, often referred to as a CDD pole in the literature~\cite{Castillejo:1955ed}, induce the attraction between the decay products?

While this may resemble a chicken-and-egg problem, the implications are substantial. If the strong force is the primary mechanism behind the existence of resonances, then the $\Delta\pi$, $N\rho$, and $N\sigma$ interactions should all resonate at the same mass. Conversely, if molecular interactions are the main driver, a $J^P = \nicefrac{1}{2}^+$ state at 1763~MeV coupling to $\Delta\pi$, $N\rho$, and $N\sigma$ could generate three distinct resonances~\cite{Suzuki:2009nj}.
This latter approach was recently criticized in Ref.~\cite{Owa:2025mep} for not offering ``a meaningful physical understanding of the origin of resonances.” Moreover, it does not fully account for the electroproduction data on the $N(1440)\nicefrac{1}{2}^+$, see Sec.~\ref{Sec:Roper}.

\begin{figure}[!t]
\centering
\begin{minipage}{.55\linewidth}
  \centering\vspace{0mm}
  \captionof{table}{\label{Tab:p1}$P_c$ and $P_{cs}$ states observed by LHCb.\vspace{-2mm}}
\footnotesize
\renewcommand{\arraystretch}{1.5}
\hspace{-5mm}\begin{tabular}{cccccc}
\hline\hline
  state   & mass (MeV) &\multicolumn{2}{c}{threshold (MeV)} & width (MeV) & Ref.\\\hline
  $P_c(4312)$  & $4311.9\pm 0.7^{+6.8} _{-0.6}$ &$\Sigma_c ^+\bar D^{0}$&  4317.5 &$9.8\pm 2.7^{+3.7} _{-4.5}$&\cite{LHCb:2019kea}\\
  $P_c(4337)$  & $4337^{+7} _{-4}{} ^{+2} _{-2}$ &$\Sigma_c ^0 \bar D^{+}$&  4323.4 &$9.8\pm 2.7^{+3.7} _{-4.5}$&\cite{LHCb:2021chn}\\
  $P_c(4380)^\dagger$  & $4380 \pm 8\pm28$ &$\Sigma_c ^0\bar D^{+}$&  4323.4 &$205\pm90$&\cite{LHCb:2016ztz}\\
  $P_c(4440)$  & $4440.3\pm 1.3^{+4.1} _{-4.7}$ &$\Sigma_c ^+\bar D^{*0}$&4459.5 &$20.6\pm 4.9^{+8.7} _{-10.1}$&\cite{LHCb:2019kea}\\
  $P_c(4457)$  & $4457.3\pm 0.6^{+4.1} _{-1.7}$ &$\Sigma_c ^0\bar D^{*+}$&4464.0  &$6.4\pm 2.0^{+5.7} _{-1.9}$&\cite{LHCb:2019kea}\\
  $P_{cs}(4338)$  & $4338.2\pm 0.7\pm0.4$  &$\Xi_c ^0 D^{+}$&  4340.1&$7.0\pm 1.2\pm1.3$&\cite{LHCb:2022ogu}\\
  $P_{cs}(4459)$  & $4458.8\pm 2.9^{+4.7} _{-1.1}$ &$\Xi_c ^{'+}\bar D^{0}$& 4443.0  &$17.3\pm 6.5^{+8.0} _{-5.7}$&\cite{LHCb:2020jpq}\\
  \hline\hline\\[-4ex]
  \multicolumn{5}{l}{$\dagger$~Not confirmed in \cite{LHCb:2019kea}}
\end{tabular}
\end{minipage}
\hspace{4mm}\begin{minipage}{.4\linewidth}
\centering
   \includegraphics[width=1\linewidth]{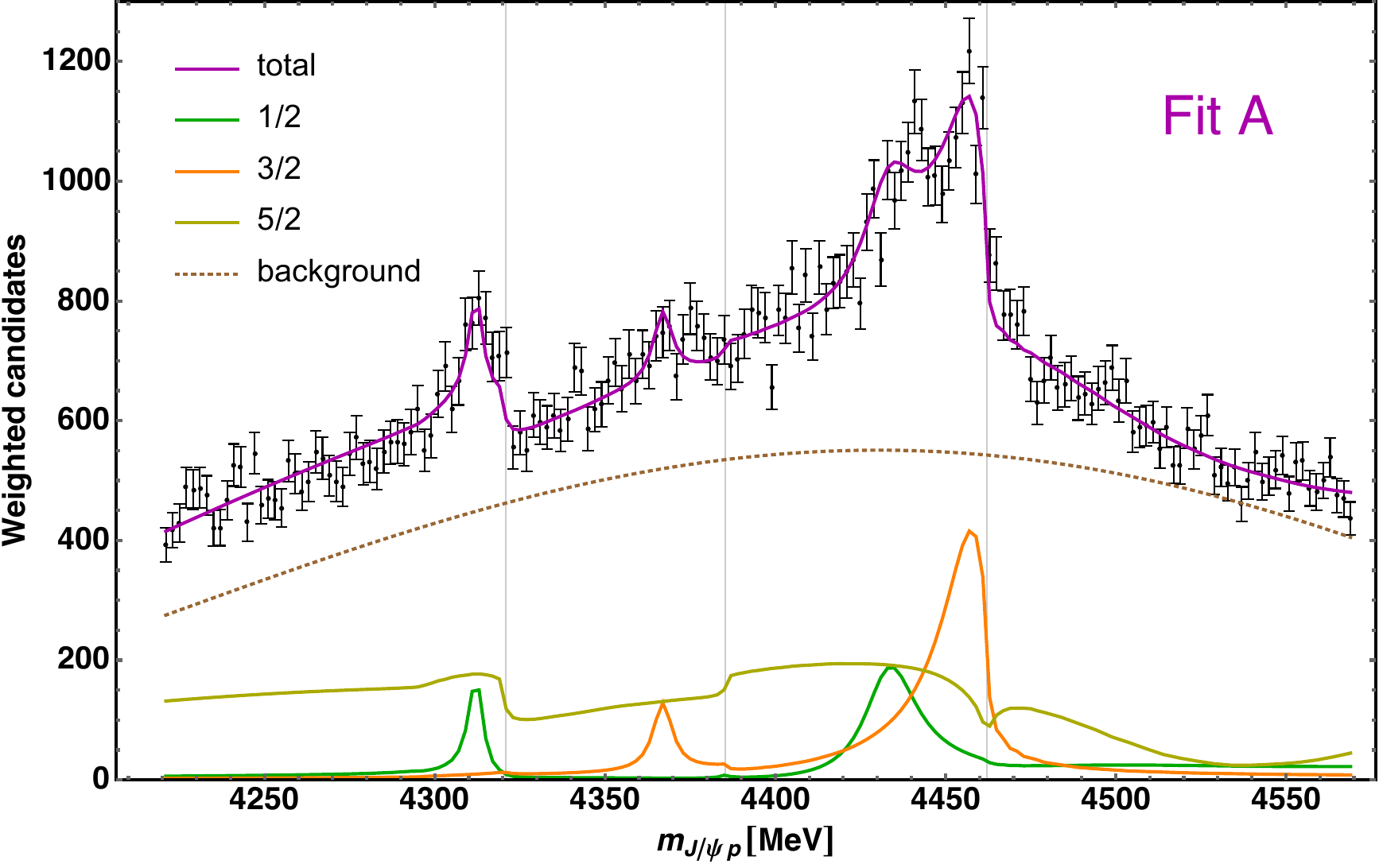}
 \end{minipage}
  \caption{LHCb data on \mbox{$\Lambda_b\to J/\psi pK^-$} \cite{LHCb:2019kea} with fit from Ref.~\cite{Shen:2024nck} using a hadronic molecular picture. The vertical dashed lines are thresholds for $\Sigma_c\bar D$, $\Sigma_c ^*\bar D$, and $\Sigma_c\bar D^{*}$.}
   \label{fig:LHCb}
   \end{figure}

\subsubsection{\boldmath  Heavy pentaquarks}

The discovery of narrow peaks in the $J/\psi\,p$ invariant mass spectrum observed in 
$\Lambda_b \to J/\psi\,p\,K^-$ decays~\cite{LHCb:2015yax,LHCb:2016ztz} provided clear 
evidence for states with a minimal quark content beyond that of the conventional quark model. 
The decay of an excited three-quark nucleon into $J/\psi\,p$ is highly suppressed by the 
OZI rule; thus, the minimal quark content of the observed narrow peaks must be $c\bar{c}uud$.
The most recent results on $J/\psi\,p$ resonances in $\Lambda_b \to J/\psi\,p\,K^-$ decays were reported in Ref.~\cite{LHCb:2019kea}. An additional state was observed in $\Lambda_b \to J/\psi\,p\,\bar{p}$~\cite{LHCb:2021chn}. Peaks in the $J/\psi\,\Lambda$ invariant mass spectrum have also been seen in the decays $\Xi_b^- \to J/\psi\,\Lambda\,K^-$~\cite{LHCb:2020jpq} and $B^- \to J/\psi\,\Lambda\,\bar{p}$~\cite{LHCb:2022ogu}, indicating a minimal quark content of $c\bar{c}uds$.
Table~\ref{Tab:p1} summarizes the current experimental results on heavy pentaquark candidates.

There are numerous interpretations of these structures. Frequently, the observed peaks are described as loosely bound molecular states composed of charmed baryons and anti-charmed mesons. In Fig.~\ref{fig:LHCb}, we show the LHCb data on $\Lambda_b \to J/\psi\,p\,K^-$~\cite{LHCb:2019kea}, along with a fit from Ref.~\cite{Shen:2024nck} based on a hadronic molecular model. 
The small structure near 4380~MeV, originally proposed as a broad resonance in Refs.~\cite{LHCb:2015yax,LHCb:2016ztz} but not confirmed in later LHCb data~\cite{LHCb:2019kea}, is now described by a pole at $(4366.5 - i\,4.7)$~MeV. Two alternative fits yield slightly different values.
Some of these states may originate from charmonium bound to light-quark baryons. Alternatively, they could arise from threshold singularities due to final-state rescattering, which leads to a logarithmic branch point in the amplitude. Other interpretations suggest compact states formed from triquark--diquark configurations, or systems in which a $c\bar{c}$ core is surrounded by light quarks. QCD sum rules have also been employed to describe their properties.
Several comprehensive reviews on tetraquarks, pentaquarks, and their possible interpretations are available; see  
Refs.~\cite{Brambilla:2019esw,Liu:2019zoy,Dong:2021bvy,Dong:2021juy,Chen:2022asf,Burns:2022uiv,Klempt:2022zwo,Liu:2024uxn}.

\subsubsection{\boldmath Light-quark pentaquarks}

In 1997, Diakonov, Petrov and Polyakov predicted the existence of an anti-decuplet of exotic light-quark pentaquarks based on the chiral soliton model, a theoretical framework in which baryons emerge as topological solitons of the pion field rather than as three-quark states~\cite{Diakonov:1997mm}. 
Translated into quark-model 
language, this implies that baryons do not consist solely of three quarks, but also include 
sea quarks. When $q\bar{q}$ pairs are assembled in light chiral fields, their production 
incurs little additional energy, provided that their flavor content matches the quantum numbers of the baryon.
The lightest state in the anti-decuplet, denoted $\Theta^+$, was predicted to have strangeness $S=+1$, spin-parity $J=\nicefrac{1}{2}^+$, a mass of 1530~MeV, and to be narrow in width. Its minimal quark content is given by $\Theta^+ = uud\,d\bar{s}$, identifying it as an exotic pentaquark. The $\Theta^+$ should be accompanied by a nucleon-like isospin doublet, three $\Sigma$-like states, and a quartet of $\Xi$ states, among which $\Xi^{--}$ and $\Xi^{+}$ are exotic.

The 2003 announcement of the discovery of the $\Theta^+(1540)$ in the reaction $\gamma\,n \to K^+ K^- n$ on a carbon target~\cite{LEPS:2003wug} generated tremendous excitement in both the medium-energy and high-energy physics communities. Within less than a year, the initial observation was supported by similar results from ten other experiments. Shortly thereafter, a peak in the $\Xi^- \pi^-$ invariant mass spectrum was reported by the NA49 collaboration at CERN and interpreted as the $\Xi^{--}(1862)$ pentaquark~\cite{NA49:2003fxh}. One year later, data from the GRAAL experiment revealed a structure in the $N\eta$ invariant mass distribution, which was interpreted as a $N(1680)$ pentaquark~\cite{GRAAL:2004ndn}.
In the context of these pentaquark searches, two additional candidates were proposed that were not predicted by the chiral soliton model: a doubly charged isovector companion of the $\Theta^+$, and a narrow anti-charmed baryon denoted $\Theta_c$.

However, the $\Theta^+(1540)$ signals reported by various experiments consisted of only a few dozen 
events each and were observed above substantial backgrounds, with statistical significances 
typically in the range of $5$--$7\sigma$. Numerous selection cuts had to be applied to extract 
the signals, necessitating a ``blind" analysis which, in most cases, was not undertaken. Moreover, 
the spread in the reported mass values, beyond the quoted uncertainties, further undermined the 
credibility of the multiple observations. The states $\Xi^{--}(1862)$, $\Theta^{++}$, and 
$\Theta_c$ were each observed with moderate significance and only in a single experiment.
High-statistics experiments and reanalyses were subsequently performed with the aim of confirming 
or refuting the existence of these pentaquark candidates. As a result, the evidence for the 
$\Theta^+(1540)$ and $\Xi^{--}(1862)$ gradually faded. While the presence of a peak-like structure 
at 1685~MeV in the $N\eta$ invariant mass distribution is undisputed, this feature finds a 
natural explanation as an interference effect between known $N^*$ resonances, most likely 
$N(1535)$ and $N(1650)$. There is no need to introduce a new, narrow state to account for this observation.

Detailed discussions of the experiments that claimed evidence for pentaquark candidates, as well 
as those that reported null results, can be found in a number of reviews~\cite{Klempt:2004yz,Kabana:2005tp,Danilov:2005kt,Burkert:2005ft,Danilov:2008uxa,Hicks:2012zz,Liu:2014yva,%
Strakovsky:2024ppo,Kim:2024tae,Praszalowicz:2024mji,Praszalowicz:2024zsy}. The Particle Data Group ultimately concluded that the non-existence of pentaquarks appears compelling~\cite{ParticleDataGroup:2006fqo,ParticleDataGroup:2008zun}.
While the chiral soliton model offered an elegant and appealing picture of the baryon structure, it did not withstand experimental scrutiny.

\subsection{\label{sec:1405}The \texorpdfstring{$\Lambda(1405)$}\ \ and its debated two-pole structure}

The $\Lambda(1405)$ resonance was first observed in a bubble chamber experiment at the Lawrence Radiation Laboratory in 1961~\cite{Alston:1961zzd}. Tripp {\it et al.}~\cite{Tripp:1968ukt} determined the relative signs of $\bar{K} N \to \pi\Sigma$ transition amplitudes for the $\Sigma(1385)$, $\Lambda(1405)$, and $\Lambda(1520)$, identifying the $\Lambda(1405)$ as primarily an $SU(3)_f$ singlet state.
In Ref.~\cite{Hemingway:1984pz}, the $\Lambda(1405)$ was investigated in the production and decay chain $K^-p \to \pi^-\Sigma^+(1670)\nicefrac{3}{2}^-$, $\Sigma^+(1670)\nicefrac{3}{2}^- \to \Lambda(1405)\pi^+$, $\Lambda(1405) \to \Sigma^\pm\pi^\mp$. The sign of the transition amplitude enables a determination of the $\Lambda(1405)$ $SU(3)_f$ structure \cite{Klempt:2020klf}.
Later experimental studies mainly focused on measuring the line shape of $\Lambda(1405)$ \cite{Zychor:2007mm,Esmaili:2009rf,Moriya:2009mx,Moriya:2010zz,HADES:2012csk,CLAS:2013rjt,CLAS:2013zie,% 
Ahn:2010zzb,Siebenson:2013rpa,Ren:2015bsa,J-PARCE31:2022plu,BGOOD:2021sog,Wickramaarachchi:2022mhi} 
or confirming its spin and parity \cite{CLAS:2014tbc}.
Recent investigations of the reactions $pp \to \Sigma^-\pi^+ p K^+$\cite{Siebenson:2013rpa} and $\gamma p \to K^+\Sigma^0\pi^0$ \cite{BGOOD:2021sog,Wickramaarachchi:2022mhi} provided evidence for a two-pole structure of the invariant 
mass distribution $\Sigma\pi$ in the $\Lambda(1405)$ region.

Concerning its theoretical interpretations, Dalitz and collaborators
already predicted the existence of the
$\Lambda(1405)$ as a quasibound molecular 
state of the $\bar{K}N$ system \cite{Dalitz:1960du,Dalitz:1967fp}
two years before its discovery 
based on the $\bar K N$ scattering lengths~\cite{Dalitz:1959dn}.
Later, Kaiser, Waas and Weise constructed an effective potential from a
chiral Lagrangian, and the $\Lambda(1405)$ emerged as a quasi-bound state in the $\bar KN$ and
$\pi\Sigma$ coupled-channel system \cite{Kaiser:1995eg,Kaiser:1996js}. 
In quark models, the $\Lambda(1405)$ and $\Lambda(1520)$ are interpreted as $qqq$
resonances where one of the quarks is excited to the $p$ state, forming a spin-doublet of
states with a dominant $SU(3)_f$ singlet structure~\cite{Isgur:1978xj}. This is also the
interpretation in later quark-model calculations of baryon 
resonances~\cite{Capstick:1986ter,Glozman:1997ag,Loring:2001ky}. However, such interpretations need not necessarily conflict;  
resonances may have a dual nature. For example,
the Roper resonance $N(1440)\nicefrac12 ^+$ has an unexpectedly low mass of
about 100~MeV below the $N(1535)\nicefrac12^-$, while
quark models predict the
mass of the nucleon's first radial excitation above its parity partner.
This discrepancy led to non-$qqq$ interpretations of
the Roper resonance, and a large number of papers identified it as a molecular state from very
early times \cite{Bhargava:1969jt} to recent publications \cite{Wang:2023snv} (although the $N(1440)\nicefrac12 ^+$ has, in the
latter publication, a significant $qqq$ contribution). In electroproduction, the Roper resonance reveals a
$qqq$ core surrounded by a meson cloud~\cite{Burkert:2017djo}.
Hence, interpretations of states as dynamically generated resonances and as
quark-model states can co-exist without necessarily being in conflict.

\begin{figure}
\centering
\begin{minipage}{.43\linewidth}
\centering 
\includegraphics[width=\linewidth]{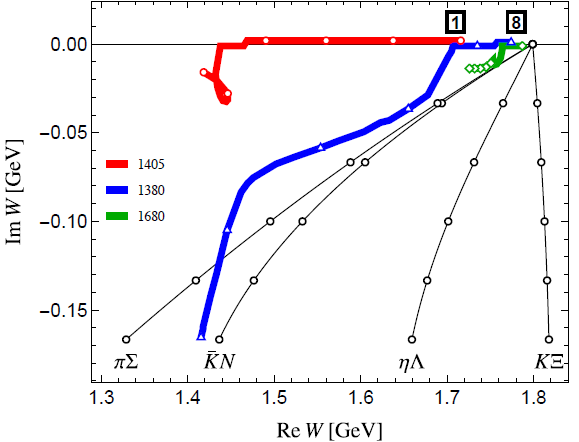}\vspace{-0mm}
    \caption{(color online) Motion of  poles by changing the $SU(3)_f$ breaking
  parameter $x$. In the $SU(3)_f$ limit ($x=0$), the two poles belong to the $SU(3)_f$ 
  singlet {\bf 1} and octet {\bf 8}.    The blue, red, green lines denote the 
  $\Lambda(1380)$, $\Lambda(1405)$ and $\Lambda(1680)$.  The various meson-baryon thresholds are  
   shown as  black solid lines. Reproduced from Ref.~\cite{Guo:2023wes}. \EPJA}
\label{fig:su3traj}
\end{minipage}
\hspace{5mm}
\begin{minipage}{.5\linewidth}
\captionof{table}{\label{octet}Octet fraction of low-mass negative-parity $\Lambda$ hyperons.
\vspace{-2mm}}
\footnotesize
\centering\begin{tabular}{lccccc}
\hline\hline
&\cite{Oller:2000fj}&\phantom{cc}\cite{Jido:2003cb}\phantom{cc}&\phantom{cc}&\phantom{c}\cite{Isgur:1978xj}&\cite{Loring:2001ky}\phantom{c}\\
\hline
$\Lambda(1380)\frac12^-$&  8\% & 47\%\ &&   -    & -\\
$\Lambda(1420)\frac12^-$& 76\% & 69\%\ &&   -    & - \\
$\Lambda(1405)\frac12^-$&  -   &   -   &&19\%&\phantom{c}  29.7\%\phantom{c}\\
$\Lambda(1520)\frac32^-$&  -   &    -  &&17\% &  21.2\%\\
$\Lambda(1670)\frac12^-$& 77\% & 88\%\ &&85\%& 71.5\%\\
$\Lambda(1690)\frac32^-$&   -  &     - &&84\%&  78.5\%\\
\hline\hline
\end{tabular}

\vspace{0mm}

\captionof{table}{\label{GMO}
Test of the GMO formula for octet baryons.
The $\Xi(2020)$ resonance has spin $J=\nicefrac52$, $P=+1$ is assumed.  $\delta$ is the 
mismatch parameter defined in Eq.~(\ref{missm}). The GMO 
formula is well satisfied for the ground states and the $J^P=\nicefrac32^-$
and $\nicefrac52^+$ excitations. Assuming GMO for  $J^P=\nicefrac12^-$, 
the $\Xi$ mass is predicted for the the mass of the $\Lambda$ octet state 
in the quark model (1680~MeV) 
and for the chiral unitary approach (1418~MeV). \vspace{1mm}
}
\centering
\renewcommand{\arraystretch}{1.2}
\footnotesize
\begin{tabular}{cccccc}
\hline\hline
$J^P$      &$M_N$    & $M_\Lambda$ & $M_\Sigma$ & $M_\Xi$ & $\delta$ \\\hline
$\nicefrac12^+$& 938.9   & 1115.7      & 1193.2     & 1318.3  & -6.4\\
$\nicefrac32^-$& 1510(5) & 1690(10)    & 1665(10)   & 1823(5) & -14(16)\\
$\nicefrac52^+$&1670(10) & 1818(7)     & 1900(15)   & 2025(5) & 7(25)\\\hline
$\nicefrac12^-$&1510(10) & 1680(8)     & 1674(4)    & 1841(16)& QM\\ 
$\nicefrac12^-$&1510(10) & 1418       &  1674(4)    & 1454 (!)& ChUA\\ 
\hline\hline
\end{tabular}

\end{minipage}
\end{figure}

Oller and Mei\ss ner studied the S-wave kaon–nucleon interactions for strangeness $S = -1$ in a  
relativistic chiral unitary approach based on coupled channels \cite{Oller:2000fj}. The construction of
meson-baryon potentials compatible with chiral symmetry generated
two poles in the $\Lambda(1405)$ region. This exciting
finding initiated a large number of further investigations
\cite{Oset:2001cn,Lutz:2001yb,Jido:2002zk,Garcia-Recio:2002yxy,Jido:2003cb,Oller:2006jw,%
Doring:2010rd,Haidenbauer:2010ch,%Cieply:2011nq,Doring:2011ip,Ikeda:2011pi,Mai:2012dt,Guo:2012vv,%
Mai:2014xna,Feijoo:2015yja,Cieply:2015pwa,Cieply:2016jby,Liu:2016wxq,Ramos:2016odk,Kamiya:2016jqc,
Miyahara:2018onh,Bruns:2021krp,Lu:2022hwm,Sadasivan:2022srs,Guo:2023wes}, and 
several reviews have highlighted the nature of the $\Lambda(1405)$~\cite{Hyodo:2011ur,Guo:2017jvc,Hyodo:2020czb,Meissner:2020khl,Mai:2020ltx,Nieves:2024dcz}.  
The resulting three poles $\Lambda(1380)$, $\Lambda(1405)$, and $\Lambda(1680)$ are 
suggested to be the singlet and the two octet isoscalar states, as
expected from the decomposition $8\otimes8=1\oplus 8\oplus 8\oplus 10\oplus \overline{10}\oplus 27$. 
Figure~\ref{fig:su3traj}
shows how low-mass strange baryons with $J^P=\nicefrac12^-$ emerge when the
$SU(3)_f$ breaking coefficient is tuned from 0 (the $SU(3)_f$ limit) to 1 (the physical point),
see also Ref.~\cite{Zhuang:2024udv}.

The structure of the $\Lambda(1405)$ has also been intensely studied in lattice QCD. 
Several investigations found a spectrum with a flavor-singlet state 
and two octet states with masses compatible with the conventional picture, i.e. 
with a single pole at about 1405~MeV. The $\Lambda(1405)$ was mostly interpreted
as a molecular bound state of an antikaon and a nucleon, in some cases with 
overlap with local three-quark operators \cite{Melnitchouk:2002eg,Nemoto:2003ft,Burch:2006cc,%
Takahashi:2009ik,Edwards:2012fx,Engel:2013ig,Hall:2014uca,Hall:2016kou}.
These masses only refer to the lowest finite-volume energy level, see the discussion below and in~\cite{BaryonScatteringBaSc:2023ori,BaryonScatteringBaSc:2023zvt}. 
The authors of~\cite{MartinezTorres:2012yi} devised an analysis strategy to obtain
the two $\Lambda(1405)$ poles with sufficient precision to distinguish them; see
also Ref.~\cite{Molina:2015uqp}.  Pavao {\it et al.}~\cite{Pavao:2020zle} argued that most 
lattice QCD studies so far have relied on three-quark operators to generate the physical states. Hence, signals
corresponding to the meson-baryon scattering states, appearing in the finite-volume effective
theory calculations, have not yet been observed. Using a  pion mass of 200~MeV and single-baryon and meson-baryon operators, lattice QCD 
computations of the coupled channel $\pi\Sigma -\bar KN$ scattering amplitudes revealed
a two-pole structure~\cite{BaryonScatteringBaSc:2023ori,BaryonScatteringBaSc:2023zvt} consistent with 
the approach based on unitary chiral perturbation theory~\cite{Guo:2023wes}.

The two-pole structure of the $\Lambda(1405)$ is, however, not undisputed.
Recent conventional multichannel analyses treat the $\Lambda(1405)$ as single 
pole~\cite{Matveev:2019igl,Zhang:2013cua,Zhang:2013sva,Fernandez-Ramirez:2015tfa,Kamano:2014zba,%
Kamano:2015hxa,Sarantsev:2019xxm, Fernandez-Ramirez:2015fbq,Anisovich:2020lec}. In \cite{Anisovich:2020lec}, the
$\Lambda(1405)$ was observed as a flavor-singlet state. Introducing a second pole in the analysis, the fit
converged and the low-mass pole became the flavor-singlet and the high-mass pole the
flavor-octet state.
% , consistent with the findings discussed above. 
From now
on, we speak of $\Lambda(1405)$ to denote the observed structure or the single pole, while the 
two poles are described as $\Lambda(1380)$ and $\Lambda(1420)$.

Table~\ref{octet} shows the octet fraction of low-mass negative-parity $\Lambda^*$ resonances from
two determinations within the chiral unitary approach~\cite{Oller:2000fj,Jido:2003cb}
(both results are given in Ref.~\cite{Jido:2003cb}) 
and from two quark-model calculations~\cite{Isgur:1978xj,Loring:2001ky}.  In the chiral unitary approach, the complex of $\Lambda(1380)$ and
$\Lambda(1420)$ contains one full octet state. In quark models, the octet fraction in $\Lambda(1405)$ is small and
comparable to $\Lambda(1520)$, so they are both predominantly singlets.
The additional octet state arising from the two-pole structure creates tension with the Gell-Mann-Okubo (GMO) formula, which connects 
the octet $\Lambda^*$ and  $\Sigma^*$ sector 
to the $N^*$ and $\Xi^*$ octet family via $SU(3)_f$. Expressed by a mismatch parameter $\delta$, it reads
\begin{equation}
 \delta = \frac12(M_N + M_\Xi) - \frac14(3M_\Lambda+M_\Sigma) \stackrel{!}{\approx} 0\,.
 \label{missm}
\end{equation}
Table~\ref{GMO} lists the pole
masses of baryon resonances and the mismatch parameter $\delta$ assuming the custom quark model assignments. The GMO formula is well satisfied for the
ground states and the $\nicefrac32^-$ states. It suggests that, for instance, the $\Lambda(1690)$ mass
would gradually decrease to 1510~MeV when the $s$-quark mass is tuned to 
the $d$-quark mass. 
Only a few $\Xi$ resonances with known spin and parity are identified; 
in the $\nicefrac52^+$ sector, a $\Xi(2020)$ resonance is 
listed in the RPP with spin $J=\nicefrac52$ but unknown parity. Assuming $P=+1$, the GMO formula is well satisfied. No $\Xi$ state with $J^P=\nicefrac12^-$ 
is known. Using the GMO formula and assuming that the lowest-mass octet state is $\Lambda(1690)$,
the mass of the $\Xi$ state with $J^P=\nicefrac12^-$ can be predicted and is close to that of
its $\nicefrac32^-$ spin partner.
With $M_\Lambda=1418$~MeV as determined in Ref.~\cite{Guo:2023wes} and 
$M_\Sigma=1674$~MeV from the RPP, the corresponding
$\Xi$ mass would be 1454~MeV, which is only 140~MeV above the ground state. This is in conflict with the observation that the lowest-mass negative-parity state is typically found more than 500~MeV
above the ground state. 
Thus, based on the GMO formula, the scenario suggested in~\cite{Jido:2002zk}
cannot be straightforwardly integrated into the baryon spectrum. It would require a low-mass 
$\Sigma^*$ at about 1400~MeV (see Ref.~\cite{Belle:2022ywa,Wang:2024jyk,Lin:2025pyk}) and a $N^*$ below 1300~MeV, both
with $J^P=\nicefrac12^-$.

\begin{figure}
\centering
\begin{minipage}{.32\linewidth}
\vspace{-3mm}
       \captionof{table}{Branching fractions for  $\psi(3686)\to$ $\Lambda\bar\Sigma^0\pi^0+c.c.$ decays.\vspace{-1mm}}
    \label{1405-br}
   \centering\footnotesize
   \renewcommand{\arraystretch}{1.5}
    \begin{tabular}{cc}
    \hline\hline
          Decay &${\cal B}\times 10^{-7}$\\\hline
    \hspace{-4mm}      $\psi(3686)\to\bar\Lambda\Lambda(1690)\frac32^-$+c.c.    &47\er9\er8 \\
    \hspace{-4mm}      $\psi(3686)\to\bar\Lambda\Lambda(1670)\frac12^-$+c.c.    &33\er7\er9 \\
    \hspace{-4mm}      $\psi(3686)\to\bar\Lambda\Lambda(1520)\frac32^-$+c.c.    &20\er5\er5 \\
    \hspace{-4mm}     \hspace{2mm} $\psi(3686)\to\bar\Lambda\Lambda(1420)\frac12^-$+c.c. \hspace{-1mm}\multirow{2}{*}{\Large$\}$}&\multirow{2}{*}{38\er4\er10} \\
    \hspace{-4mm}      $\psi(3686)\to\bar\Lambda\Lambda(1380)\frac12^-$+c.c.    & \\
    \hline\hline
    \end{tabular}
\end{minipage}
\hspace{10mm}
\begin{minipage}{.37\linewidth}
    \centering
\includegraphics[width=\linewidth]{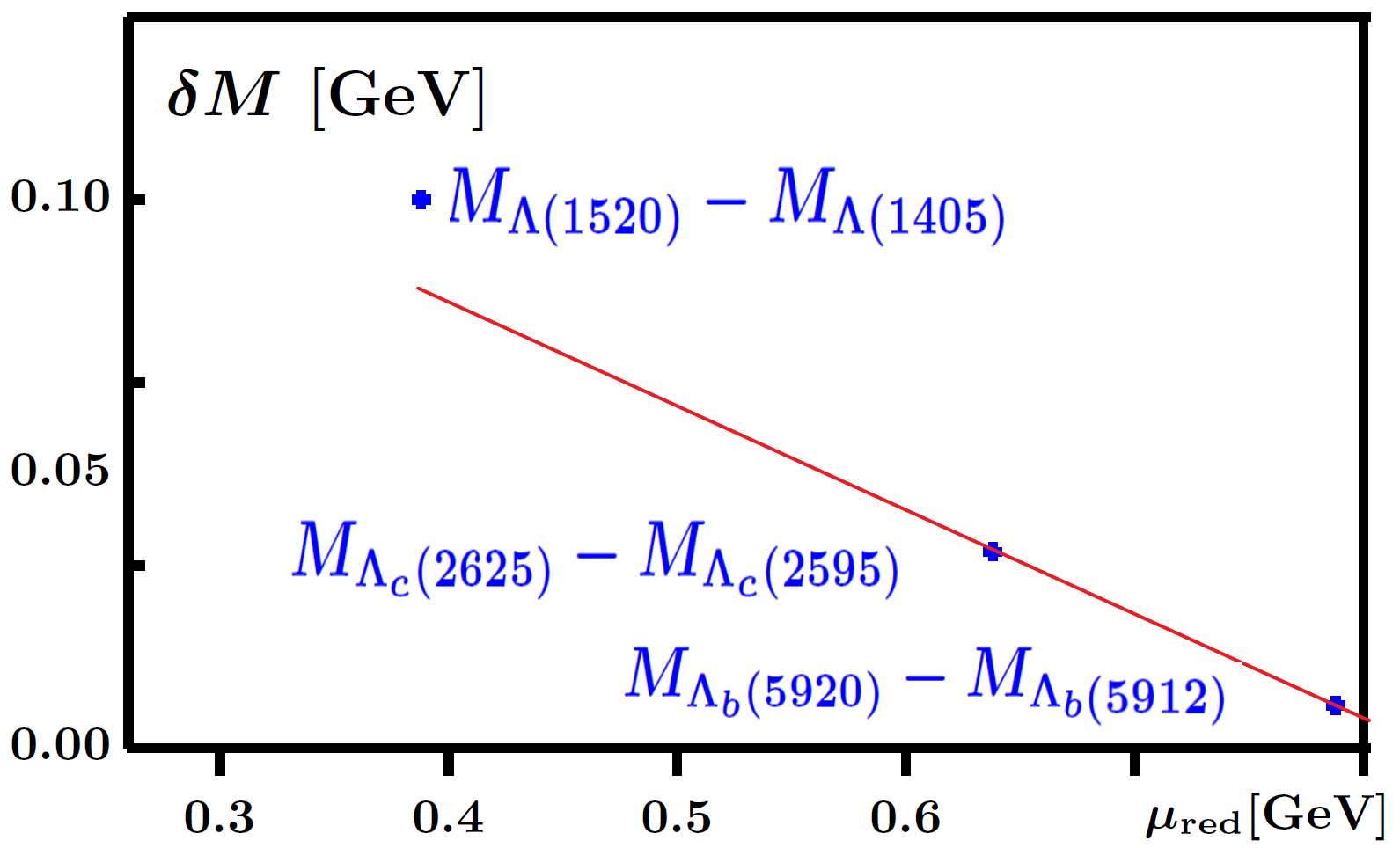}
\end{minipage}
\hspace{1mm}
\begin{minipage}{.20\linewidth}
\phantom{zzz}\\[3ex]
      \caption{Mass difference between $\nicefrac32^-$ and  $\nicefrac12^-$ 
    baryons with fully antisymmetric wave functions as a function of the reduced mass.}
    \label{fig:4plet}
\end{minipage}
    \end{figure}

Another argument comes from the following observation.
The BESIII collaboration studied the reaction $\psi(3686)\to\Lambda\bar\Sigma^0\pi^0+c.c.$~\cite{BESIII:2024jgy}
and found a significant $\Lambda(1405)$ contribution. It was described applying a chiral dynamics model and
a Flatté-like formula for the parametrization, which are both able to describe the data. The former
assumes two poles whose relative strength is fixed in the model, and only one yield is
given in~\cite{BESIII:2024jgy}. Now, because the $\psi(3686)$ is an $SU(3)$ singlet state and the $\overline{\Lambda}$
an $SU(3)$ octet, $\Lambda^*$ resonances are produced in this process because of their octet component.
Table~\ref{1405-br} presents a selection of branching fractions for $\psi(3686)\to\Lambda\bar\Sigma^0\pi^0+c.c.$ decays.
The yields of $J^P=\nicefrac12^-$ and $\nicefrac32^-$ states allow one to predict their  octet
components. 
For example, the $\Lambda(1690)\nicefrac32^-$ has a slightly higher production rate 
compared to the $\Lambda(1670)\nicefrac12^-$ and thus possibly a larger octet component, although both effects are statistically insignificant. We assume that
states with $J^P=\nicefrac12^-$ are neither dynamically favored nor suppressed compared to states with $J^P=\nicefrac32^-$. 
Concerning the $\Lambda(1405)$,
the quark model predicts a production
strength that is larger compared to the $\Lambda(1520)$ by a factor 1.1~\cite{Isgur:1978xj} to 1.4~\cite{Loring:2001ky},
while in the chiral unitary approach one should  expect a higher yield of the two poles $\Lambda(1380)$ and $\Lambda(1420)$
compared to the flavor-singlet $\Lambda(1520)$  by a factor 4~\cite{Oller:2000fj} to 6~\cite{Jido:2003cb}.   
By contrast, in Table~\ref{1405-br} the combined yield of $\Lambda(1380)$ plus $\Lambda(1420)$ is larger compared to
the $\Lambda(1520)$ only by a factor 1.9\er0.9. This is compatible with the one-pole hypothesis
but only incompatible with the two-pole scenario by $2.3\sigma$. A higher precision in the branching ratio
may help to distinguish between the two views.

The $\Lambda(1405)\nicefrac12^-$ and $\Lambda(1520)\nicefrac32^-$ have a remarkably low mass, even below
the $N(1535)\nicefrac12^-$ and $N(1520)\nicefrac32^-$. In quark models, these states are
interpreted as flavor singlet baryons. The three pairs of quarks one can form within the resonances
are all three antisymmetric with respect to their exchange, and there are three pairs of {\it good}  diquarks, 
explaining at least partially their low masses. 
In $SU(4)$, three additional pairs of low-mass negative-parity flavor-singlet states are expected.
These are collected in Fig.~\ref{fig:4plet} (left) for charm and bottom baryons.
On the other hand, 
the mass splitting between the  $\Lambda(1405)$ and $\Lambda(1520)$ is unexpectedly large. In Fig.~\ref{fig:4plet} (right)
we show the  mass difference of resonances with $\nicefrac32^-$ and $\nicefrac12^-$ as a function of the 
reduced mass of the baryon. Indeed, the $\Lambda(1520) - \Lambda(1405)$ mass difference is larger than expected; 
however, the deviation from the expected value is small.

Can further experiments help us  understand the structure of $\Lambda(1405)$? 
First, precise line shapes may reveal the two-pole structure of the $\Lambda(1405)$ and, indeed, there
are indications for a two-pole structure in the $\Lambda(1405)$ region \cite{BGOOD:2021sog,Siebenson:2013rpa,Wickramaarachchi:2022mhi}.
Second, electroproduction may help to elucidate the structure of the $\Lambda(1405)$ region. Are both
states, $\Lambda(1350)$ and $\Lambda(1430)$, seen at large momentum transfer? Do they have a
hard $qqq$ core? 
Third, the one-pole or two-pole structure of $\Lambda(1405)$ leads to
different predictions for the mass spectra of
$\chi_{c0,c2}$ decays to
$(\Sigma\pi)(\overline{\Sigma}\pi)$. The branching ratio for
$\chi_{c0}$\,/\,$\chi_{c2}$ decays to $\Lambda(1520)\overline{\Lambda}(1520)$
is (3.1\er1.2)$\cdot 10^{-4}$\,/\,(4.7\er1.5)~\cite{Wang:2011kti}, 
and we can expect the branching for decays to $\Lambda(1405)\overline{\Lambda}(1405)$ to be of similar magnitude. Experimentally, at least
one neutral decay mode of $\Lambda(1405)$ or
$\overline{\Lambda}(1405)$ should be chosen to avoid
the contamination from $\Sigma(1385)$ or 
$\overline\Sigma(1385)$.
In the chiral unitary approach, the high- and low-mass parts of the
$(\overline{\Sigma}\pi)$ invariant mass spectrum contain different fractions of
$SU(3)$ singlet and octet, so the recoiling $(\Sigma\pi)$ 
mass distribution should be different for the high- and low-mass parts. The low-mass (high-mass) region of the invariant mass spectrum for $(\Sigma\pi)$ should recoil against the low-mass (high-mass) region 
in the invariant mass spectrum for $(\overline{\Sigma}\pi)$. If there is only one pole,
the shape of the $(\overline{\Sigma}\pi)$ invariant mass spectrum should not depend on this cut.

There is overwhelming evidence from calculations based on the chiral unitary approach
and from recent lattice QCD calculations for the existence of two poles in the 1350 -- 1430~MeV region.
Nevertheless, it seems difficult to accommodate these two states in the general baryon spectrum. A singlet
state is expected but a low-mass octet state is not.
There is one perhaps remote possibility: In Ref.~\cite{Sadasivan:2022srs}, the two poles
are found at $[1.355(16) - i 0.038(14)]$ GeV and $[1.430(6) - i 0.023(4)]$ GeV,
and the widths are compatible at  $1\sigma$. In Ref.~\cite{Guo:2023wes}, the two poles are at $[1.4154 -  i 0.1657]$ GeV and $[1.4179 - i 0.0157]$ GeV,
and the masses are separated by just 2.5~MeV (uncertainties are not provided). The masses and widths of the 
two states vary significantly when different approximations are used. Should masses and widths both
be compatible, the single pole expected in quark models would be represented by two compatible poles 
in the molecular picture. The single pole would be responsible for the attractive interaction between $\bar K$ and $N$
and between $\Sigma$ and $\pi$. In the chiral unitary approach, the attractive interactions lead to
two poles, but they would represent only one particle.

\subsection{\label{hybrid}Hybrid resonances and cascade decays}

The question of the existence of hybrid baryons, referred to at the time as ``hermaphrodite baryons", 
dates back to the early 1980s~\cite{Barnes:1977hg,Barnes:1982fj,Golowich:1982kx}. It was argued 
that, analogous to hybrid mesons, hybrid baryons should also exist. In quark models, the 
light $u$ and $d$ quarks, which have current masses of only a few MeV, acquire constituent 
masses of approximately 350~MeV. Within this constituent quark framework, gluons can be 
treated as effective constituents, with constituent gluon masses typically ranging 
between 600~MeV and 1~GeV. This leads to a novel spectroscopy, characterized by the 
emergence of additional $SU(6)$ multiplets, and hence a greater number of baryon states 
than predicted by quark excitations alone.

\paragraph{Is the Roper a hybrid?}
The presence of additional multiplets arising from gluonic degrees of freedom offers a potential route to identifying hybrid baryons. One strategy involves searching for an overpopulation of states with specific quantum numbers in the nucleon sector, such as $N\nicefrac{1}{2}^+$ and $N\nicefrac{3}{2}^+$. The Roper resonance $N(1440)\nicefrac{1}{2}^+$ was long considered a strong candidate for the lowest-lying hybrid nucleon~\cite{Barnes:1982fj}, particularly due to difficulties in explaining its anomalously low mass as a simple radial excitation of the nucleon ground state. Additionally, its electromagnetic transition amplitude $A_{1/2}(0)$ is large and negative, in stark contrast to the large positive value predicted by non-relativistic constituent quark models at the photon point.

\begin{figure}
    \begin{minipage}[c]{0.27\textwidth}
    \vspace{5mm}
\hspace{-1mm}\includegraphics[width=1.05\linewidth,height=0.7\linewidth]{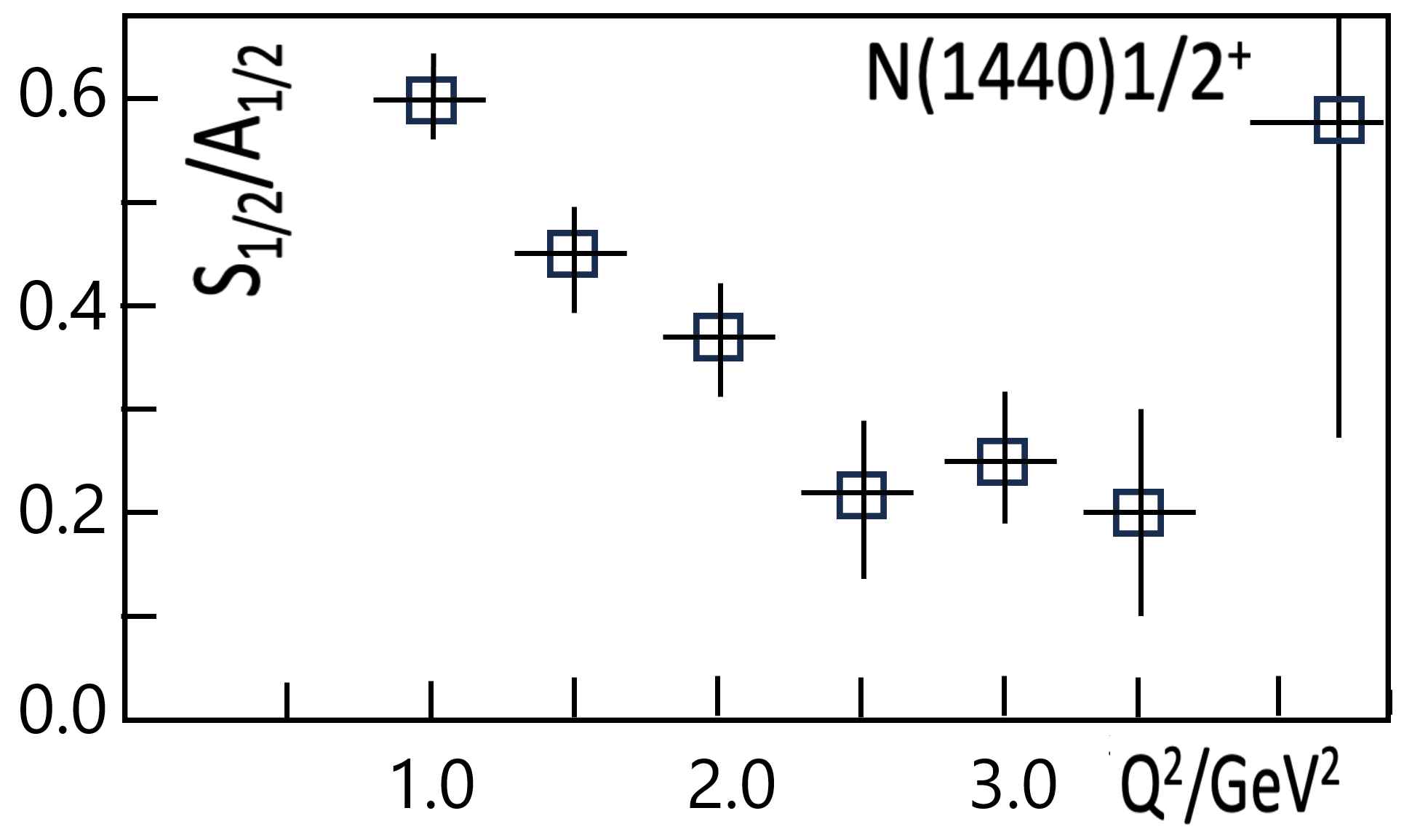}
\caption{The ratio $S_{1/2}/A_{1/2}$ of the scalar to the transverse helicity amplitudes of the $N(1440)\nicefrac{1}{2}^+$ Roper resonance. For a pure hybrid state the ratio should be $\approx 0$. Values taken from  $A_{1/2}$ and $S_{1/2}$ in Fig.~\ref{p11}. }
\label{Roper-S12-A12}
 \end{minipage}
\hspace{1.5mm} \begin{minipage}[c]{0.18\textwidth}
\includegraphics[width=1.08\linewidth]{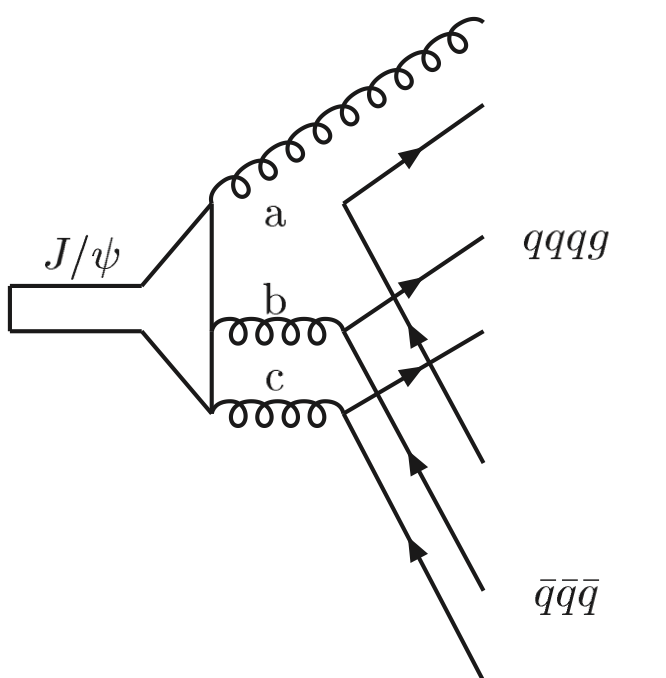}
\caption{Diagram for hybrid baryon production through $J/\psi$ decay. Adopted from \cite{Ping:2004wz}.}
\label{j-psi-hybrid}
\end{minipage}
\hspace{1.5mm}\begin{minipage}[c]{0.22\textwidth}
    \centering
\raisebox{10mm}{\includegraphics[width=\linewidth]{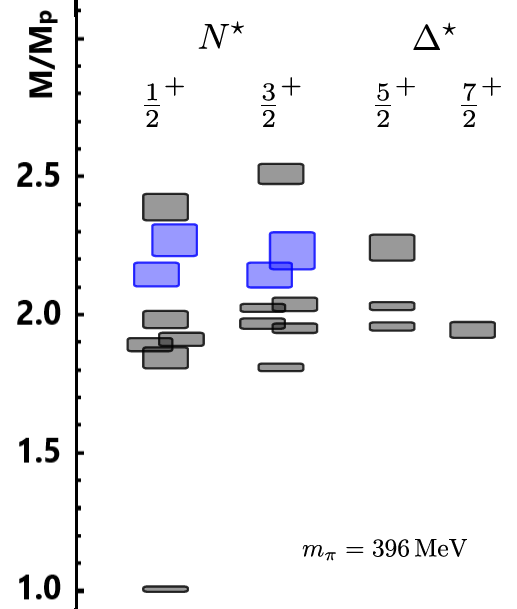}}
 \end{minipage}    
\hspace{1.5mm}\begin{minipage}[c]{0.27\textwidth}
\captionof{table}{\label{tab:Decayall}Branching ratios (in \%) for decays of the $N(2100)\nicefrac12^+$~\cite{PMahlberg} (preliminary). Their sum is 107\er12. 
}
\renewcommand{\arraystretch}{1.4}
{\footnotesize
\begin{tabular}{|c|c|c|c|} \hline
\hspace{-2mm}$N\pi$\hspace{-2mm}&\hspace{-2mm}$N\eta$\hspace{-2mm}&\hspace{-2mm}$N\eta'$\hspace{-2mm}&\hspace{-1mm}$\Lambda K$\\
\hspace{-2mm}17\er5\hspace{-2mm}&\hspace{-2mm}9\er3\hspace{-2mm}&\hspace{-2mm}4\er2   \hspace{-2mm}&\hspace{-1mm}$<1$\\\hline
\hspace{-2mm}$\Sigma K$\hspace{-2mm}&\hspace{-2mm} $N\omega$~\cite{Denisenko:2016ugz}\hspace{-2mm}&
\hspace{-2mm}$\Delta\pi$\hspace{-2mm}&\hspace{-1mm}$N\rho$\\
\hspace{-2mm}8\er3\hspace{-2mm}&\hspace{-2mm}10\er 4\hspace{-2mm}&\hspace{-2mm}6\er2\hspace{-2mm}&\hspace{-1mm}15\er4\\
\hline 
\hspace{-2mm} $N_{1535}\pi$\hspace{-2mm}&\hspace{-2mm} $N_{1710}\pi$\hspace{-2mm}&\hspace{-2mm} $N_{1720}\pi$\hspace{-1mm}&
\hspace{-2mm}$N\sigma$\\
\hspace{-2mm}1\er1\hspace{-2mm}&\hspace{-2mm}5\er3 \hspace{-2mm}&\hspace{-2mm} 9\er4\hspace{-2mm}&\hspace{-1mm}23\er5\\
\hline
\end{tabular}}
\vspace{15mm}
\renewcommand{\arraystretch}{1.0}
\end{minipage} \\[-6ex]
\phantom{zz}
%\end{figure}
%\begin{figure}
\phantom{zzzzzzzzzzzzzzzzzzzzzzzzzzzzzzzzzzzzzzzzzzzzzzzzzzzzzzzzzz}
\begin{minipage}[c]{0.48\textwidth}
\caption{(color online) Mass spectra of $N^*$ and $\Delta^*$ baryons. The mass scale is made to reproduce 
    the masses of the nucleon and of $\Delta(1950)\nicefrac72^+$. Baryons with a large hybrid content are
    highlighted in blue. Adapted from~\cite{Dudek:2012ag}.  }
    \label{hybrid-lattice}
\end{minipage}\\[-4ex]
\end{figure}

However, it was shown in Ref.~\cite{Li:1991yba} that if the Roper  were a pure hybrid baryon, the helicity amplitude $A_{1/2}(Q^2)$ would decrease rapidly in magnitude with increasing $Q^2$. In contrast, quark models that interpret the Roper as the first radial excitation of the nucleon~\cite{Abdullah:1971aq,LeYaouanc:1972ju} predict that its helicity amplitudes become dominant over those of the $N(1535)\nicefrac{1}{2}^-$, $N(1520)\nicefrac{3}{2}^-$, and even $\Delta(1232)\nicefrac{3}{2}^+$ as $Q^2$ increases, whereas a hybrid interpretation would imply further suppression.
The most recent data, shown in Fig.~\ref{p11}, rule out the hybrid interpretation of the Roper resonance. Instead, they confirm that its core is best described as a radial excitation of the nucleon’s three-quark system, with significant contributions from the surrounding meson-baryon cloud, which becomes apparent at small virtuality $Q^2$ of the exchanged photon. Furthermore, the scalar helicity amplitude $S_{1/2}(Q^2)$ is predicted to be very small for a hybrid state, which contradicts what is shown in Fig.~\ref{Roper-S12-A12}, where the ratio $S_{1/2}/A_{1/2}$ is sizeable throughout the $Q^2$ range covered by the data.

Another search for a hybrid Roper-like state has been studied through the decays of $J/\psi$~\cite{Ping:2004wz}, 
produced in $e^+e^-$ collisions. The hybrid baryon may be produced in the decay $J/\psi \to \bar{p} + |Gqqq\rangle$, 
as depicted in Fig.~\ref{j-psi-hybrid}. However, due to the $J/\psi$ mass of 3.1~GeV, only hybrid baryons with
masses below 2.1~GeV can be generated. Decays of $\psi(2S)$ would increase 
the accessible mass range for hybrid searches.

\paragraph{Can electroproduction identify hybrid baryons at high masses?}

More recent lattice QCD predictions place the masses of the lowest hybrid baryons with $J^P = \nicefrac{1}{2}^+$ above 2 GeV. The answer to the question above becomes more complex compared to the case of a possible low-mass hybrid Roper resonance near 1.5 GeV. At such high masses, individual resonances can no longer be clearly observed as enhancements in the mass spectrum, as they overlap with other states. Furthermore, hybrid states have the same quantum numbers as ordinary $qqq$ states in the quark model and may mix with them.
One may hope, however, that the heavy constituent gluon suppresses strong mixing with ordinary states, as suggested by lattice QCD projections. In that case, hybrid baryons may exhibit $Q^2$-dependent features similar to those predicted for the low-mass hybrid Roper. If so, a possible strategy may proceed as follows:
(i) Search for all $J^P = \nicefrac{1}{2}^+$ states in the relevant mass range and compare with predictions from $qqq$ systems.
(ii) Measure transition amplitudes as functions of $Q^2$ to identify deviations from expectations based on $qqq$ and $Gqqq$ models -- particularly a rapid drop in the transverse amplitude $A_{1/2}$ and near-zero scalar amplitude $S_{1/2}$.
(iii) Compare the decay pattern of the hybrid candidate with that of other baryons.

%\subsubsection{\label{2100-hybrid}\boldmath Is $N(2100)\nicefrac{1}{2}^+$ a hybrid?}

\paragraph{Is the $N(2100)\nicefrac{1}{2}^+$ a hybrid?} Fig.~\ref{hybrid-lattice} shows the lattice spectrum of  
``ordinary” $N^*$ baryons with spin-parity $\nicefrac{1}{2}^+$ and $\nicefrac{3}{2}^+$,  
$\Delta^*$ baryons with $\nicefrac{5}{2}^+$ and $\nicefrac{7}{2}^+$~\cite{Edwards:2011jj}, 
and hybrid baryons~\cite{Dudek:2012ag}. Since lattice calculations use large quark masses, 
resulting in baryon masses that are systematically too high, we normalized the results
to the nucleon mass. 
Although the finite-volume spectrum does not directly translate to physical states, it suggests the existence of baryons with large hybrid components in the higher-lying spectrum.
Experimentally, the known states in the $I(J^P)=\nicefrac{1}{2}(\nicefrac{1}{2}^+)$ channel include the $N(1440)\nicefrac{1}{2}^+$, $N(1710)\nicefrac{1}{2}^+$, $N(1880)\nicefrac{1}{2}^+$, and $N(2100)\nicefrac{1}{2}^+$.
Quark models~\cite{Isgur:1978xj,Loring:2001ky} predict four $N^*$ states 
with these quantum numbers, with the highest-mass state belonging to the \textbf{20}-plet. Therefore, the
$N(2100)\nicefrac{1}{2}^+$ could be interpreted as a member of the \textbf{20}-plet, 
or a member of the fourth excitation shell, or also as a hybrid baryon. 

These possibilities can be  explored by analyzing 
cascade decays, as studied by the CBELSA/TAPS collaboration~\cite{CBELSATAPS:2022uad,CBELSATAPS:2014wvh}.
Figure~\ref{fig:dalitz} shows the Dalitz plot for the reaction $\gamma p \to p\pi^0\pi^0$ in the invariant mass range from 1975 to 2225~MeV. Clear bands are observed, demonstrating the significance of cascade decays via intermediate isobars. The most prominent contributions arise from the $\Delta(1232)\nicefrac{3}{2}^+$, $N(1520)\nicefrac{3}{2}^-$, and $N(1680)\nicefrac{5}{2}^+$. Interestingly, Refs.~\cite{CBELSATAPS:2022uad,CBELSATAPS:2014wvh} report markedly different decay patterns for the positive-parity $N^*$ and $\Delta^*$ 
resonances with masses around 1900~MeV. The four 
$N^*$ resonances
$N(1880)\nicefrac12^+$, $N(1965)\nicefrac32^+$, $N(2000)\nicefrac52^+$, $N(1990)\nicefrac72^+$
likely form a \textbf{70}-plet (cf.~Fig.~\ref{fig:missing-resonances}) and decay with an average branching fraction of $(34 \pm 6)$\% into $N\pi$ and 
$\Delta\pi$, and with a branching fraction of $(21 \pm 5)$\% into orbitally excited states such as $N(1520)\nicefrac{3}{2}^-\pi$, $N(1535)\nicefrac{1}{2}^-\pi$, and $N\sigma$.
In contrast, the four $\Delta^*$ resonances
$\Delta(1910)\nicefrac12^+$, $\Delta(1920)\nicefrac32^+$, $\Delta(1905)\nicefrac52^+$, $\Delta(1950)\nicefrac72^+$
are members of a \textbf{56}-plet and have an average decay branching fraction of $(44 \pm 7)$\% into $N\pi$ 
or $\Delta\pi$. Their decay into orbitally excited states is significantly smaller with only $(5 \pm 2)$\%~\cite{CBELSATAPS:2022uad}.
This different behavior can be explained by their different quark-model wave functions: The four positive-parity $\Delta^*$ states have a 
symmetric  spatial wave function $\phi_{n_\rho n_\lambda, l_\rho l_\lambda} \sim \phi_{00,20}+\phi_{00,02}$,
where either the $\rho$ or  $\lambda$ oscillator is excited to $l_\rho=2$ or $l_\lambda=2$, so these states can release their excitation energy in a single step.
The four positive-parity $N^*$ states in the \textbf{70}-plet have a mixed-symmetric component
$\phi_{00,20}-\phi_{00,02}$, but also a mixed-antisymmetric component 
$\phi_{00,11}$
where both oscillators are excited to $l_\rho = l_\lambda = 1$. If one of the oscillators de-excites, the other remains excited; thus, a second de-excitation step is needed, as illustrated in Fig.~\ref{fig:dalitz}, 
which may lead to the observed cascade decays.

Now, if the $N(2100)\nicefrac12^+$ is a member of the \textbf{20}-plet in the second excitation shell (see Fig.~\ref{fig:Regge-N}), its wave function $\phi_{00,11}$ is entirely antisymmetric, with both oscillators excited simultaneously to $l_\rho=l_\lambda=1$. Single-step excitation and de-excitation processes are suppressed, although a small mixing with other $J^P=\nicefrac12^+$ states could allow for a weak production rate. The dominant decay mode should then proceed via a cascade, and we expect the $S$-wave decay $N(2100)\nicefrac12^+\to N(1535)\nicefrac12^-\,\pi$ to be dominant.

On the other hand, 
the Bonn model of baryon resonances predicts the lowest-mass $J^P=\nicefrac12^+$ state in the fourth excitation shell 
to be “Roper-like” with a mass of 2009~MeV~\cite{Loring:2001kx}, making it a plausible candidate for an identification with the $N(2100)\nicefrac12^+$. Its wave function should then contain a large contribution from the harmonic-oscillator wave function
\begin{equation}
\phi_{n_\rho n_\lambda, l_\rho l_\lambda} =
\frac{1}{\sqrt{6}}\,\phi_{20,00} +\frac{\sqrt{5}}{3}\,\phi_{11,00} -\frac{1}{3}\,\phi_{00,22} +\frac{1}{\sqrt{6}}\,\phi_{02,00}\,.
\end{equation}
The $\phi_{20,00}$ and $\phi_{02,00}$ components allow for decays in a single step
and are likely responsible for the decays into two $S$-wave baryons. With a probability of 2/3, both oscillators 
are simultaneously excited. For the $\phi_{11,00}$ component,
we may expect decays into radially excited states like $N(1440)\nicefrac12^+$ or $N(1710)\nicefrac12^+$; 
alternatively, two pions could be emitted coherently, forming a $\sigma$. The $\phi_{00,22}$ 
component could emit, in a $P$-wave decay, a pion recoiling against a $N(1720)\nicefrac32^+$ 
to which the orbital angular momentum $l=2$ of the second oscillator is transferred. There is no 
orbital angular momentum $l_\rho$ or $l_\lambda=1$ in the initial-state wave function, and hence no
$N (2100)\nicefrac12^+\to N(1535)\nicefrac12^-\,\pi$ decays are expected. 

Finally, if the $N(2100)\nicefrac12^+$ is a hybrid, it
likely decays via the breaking of the excited string, leading to intrinsic orbital angular momentum 
in one of the decay products. This decay mechanism was originally proposed for mesons~\cite{Isgur:1985vy,Kokoski:1985is} 
and is likely applicable to the decays of hybrid baryons as well. Indeed, for the $\pi_1(1600)$, the channels 
$f_1(1285)\pi$~\cite{E852:2004gpn} and $b_1(1235)\pi$~\cite{E852:2004rfa} are observed to be strong decay modes, 
although $\rho\pi$~\cite{COMPASS:2018uzl}, $\eta'\pi$, and $f_1(1285)\pi$~\cite{E852:2004gpn} are established 
decay channels as well. We may thus expect some two-body decays, but a significant fraction should proceed into 
orbitally excited states; in particular the $S$-wave decay into $N(1535)\nicefrac12^-$ and a pion should be 
strong. Radially excited states are not expected to appear in such cascade decays.

Table~\ref{tab:Decayall} shows the 
$N(2100)\nicefrac12^+$ branching ratios for twelve decay modes~\cite{PMahlberg}. 
The $P$-wave decays into a nucleon or $\Delta(1232)$ plus 
a pseudoscalar meson are significant; in addition, there are contributions from $N\rho$ and $N\omega$. Only a small 
decay fraction is observed for the $S$-wave decay into $N(1535)\pi$ with intrinsic orbital angular momentum. The 
decay into $N(1520)\pi$ via $D$-wave is not observed. The $N\sigma$ decay (via $S$-wave) is likely due to coherent 
pion emission from the two oscillators.
Both as a hybrid state or as a member of a \textbf{20}-plet, the $N(2100)\nicefrac12^+$ should have a large decay fraction into 
$N(1535)\nicefrac12^-$ and a pion, for which only a tight upper limit is observed. 
The absence of this decay mode 
makes interpretations of the $N(2100)\nicefrac12^+$ as a hybrid baryon or as a member of a \textbf{20}-plet unlikely. Hence, we 
conclude that the $N(2100)\nicefrac12^+$ is likely a Roper-like state in the fourth excitation shell, as predicted 
in the Bonn model~\cite{Loring:2001kx}. This interpretation finds support in the observation of $N(2100)\nicefrac12^+$
decays into $N(1710)\nicefrac12^+\,\pi$ and $N(1720)\nicefrac32^+\,\pi$, even  
these decays contribute only with a significance of about 2$\sigma$.

\subsection{Diquark models and  ``missing resonances''}
In the high-mass spectrum, $\Delta$ resonances often have intrinsic spin $S=3/2$
forming the Regge trajectory $\Delta(1232)\nicefrac32^+$, $\Delta(1950)\nicefrac72^+$, $\Delta(2420)1\nicefrac12^+$, $\Delta(2950)1\nicefrac52^+$.
Nucleons have intrinsic quark spin $S=1/2$ and form the Regge trajectory $N(940)$, $N(1680)\nicefrac52^+$, $N((2220)\nicefrac92^+$, $N((2700)1\nicefrac32^+$.
The slope of these trajectories is similar to the slope observed in leading meson trajectories like
$a_2(1320)$, $a_4(2040)$, $a_6(2450)$. The forces between a quark and an antiquark are identical to those
between a a quark and a diquark (within a three-quark state). 

The similarity of the Regge trajectories suggests a quark-diquark structure of baryon resonances. 
There are different interpretations of the term diquark. In models with a frozen diquark, only 
excitations involving a quasi-stable diquark are allowed. In $SU(6)$, positive-parity diquark excitations 
belong to a \textbf{56}-plet, while negative-parity diquark excitations belong to a \textbf{70}-plet. 
Baryons with $P=+1$ in a \textbf{70}-plet indicate broken symmetry~\cite{Lichtenberg:1969sxc}. 
States belonging to the \textbf{20}-plet are strictly forbidden in frozen diquark models. The role of 
diquarks in spectroscopy, in baryon decays, and in the reaction dynamics of baryons is reviewed in Ref.~\cite{Anselmino:1992vg} and, more recently, in Ref.~\cite{Barabanov:2020jvn}.

Before results from photoproduction experiments were available, only a few resonances had been established, receiving three- or four-star ratings. The first excitation shell was complete, but in the second shell many resonances were missing. Only four $N^*$ resonances with three- or four-star ratings could be assigned to the second shell, whereas the symmetric quark model with three dynamical quarks predicted 13. 
Why were so many resonances missing? Could this reduction be due to frozen diquarks? In that case, 
only three resonances would be expected: $N(1440)\nicefrac12^+$, $N(1720)\nicefrac32^+$, and $N(1680)\nicefrac52^+$. Alternatively, a small $N\pi$ coupling could be responsible, as it enters 
quadratically in $\pi N$ scattering. There was hope that the $\gamma N$ coupling might be large, 
allowing the missing resonances to be observed in photoproduction.

Indeed, 18 $N^*$ and $\Delta^*$ resonances have now been suggested that can be assigned to the second 
excitation shell, 12 of which have 3* or 4* status. Models with a frozen diquark predict only eight states. 
Due to these new findings from photoproduction experiments, the hypothesis of frozen diquarks 
can be ruled out. In addition, we have seen that both oscillators of the three-particle 
dynamics contribute to the excitation spectrum.

\subsection{Chiral symmetry restoration}

The $u$ and $d$ quarks are nearly massless, yet nucleons have a mass of 939~MeV even though they are 
composed of three valence quarks.
This mass cannot be explained by the quark masses alone. In the limit of massless quarks, QCD exhibits chiral symmetry, which implies that left- and right-handed quarks are independent. In such a scenario, hadrons would come in ``parity doublets", i.e., pairs of states with identical masses but opposite parity. 
This is not observed experimentally: for instance, the $\rho(770)$ meson with spin-parity $J^P=1^-$ and its presumed chiral partner $a_1(1260)$ with $J^P=1^+$ differ by nearly 500 MeV. Similarly, the nucleon ($J^P=\nicefrac12^+$) and the $N(1535)$ resonance ($J^P=\nicefrac12^-$) differ by about 600 MeV. These mass splittings signal that chiral symmetry is spontaneously broken in QCD, and this dynamical breaking enables the generation of most of the hadron masses even in the absence of quark masses.

Glozman observed that in the baryon spectrum, especially at higher masses, approximate parity doublets appear~\cite{Glozman:1999tk,Glozman:2002kq}. He pointed out several pairs of nucleon and $\Delta$ resonances with the same spin but opposite parity and nearly degenerate masses, such as:
\bc
\mbox{$N(2220)\frac{9}{2}^+ - N(2250)\frac{9}{2}^-$;\phantom{zzzzzzz}
$N(1990)\frac{7}{2}^+ - N(2190)\frac{7}{2}^-$;\phantom{zzzzzzz}
$N(2000)\frac{5}{2}^+ - N(2200)\frac{5}{2}^-$;}
 \mbox{$N(1900)\frac{3}{2}^+ - N(2080)\frac{3}{2}^-$;\phantom{zzzzzzz}
$\Delta(2300)\frac{9}{2}^+ - \Delta(2400)\frac{9}{2}^-$;\phantom{zzzzzzz}
$\Delta(1950)\frac{7}{2}^+ - \Delta(2200)\frac{7}{2}^-$;}
 \mbox{$\Delta(1905)\frac{5}{2}^+ - \Delta(1930)\frac{5}{2}^-$;\phantom{zzzzzzz} 
$\Delta(1920)\frac{3}{2}^+ - \Delta(1940)\frac{3}{2}^-$;\phantom{zzzzzzz}
$\Delta(1910)\frac{1}{2}^+ - \Delta(1900)\frac{1}{2}^-$.} 
\ec
Subsequent discoveries of new resonances have lent further support to this idea. For example, 
the $N(1900)\nicefrac{3}{2}^+$ is close in mass to the $N(1875)\nicefrac{3}{2}^-$, and the mass of the
$N(1880)\nicefrac{1}{2}^+$ is close to that of the $N(1895)\nicefrac{1}{2}^-$.
The concept of parity doubling was later extended to the meson sector as well~\cite{Glozman:2002jf}. In a comprehensive review~\cite{Glozman:2007ek}, the idea of chiral symmetry restoration was further expanded to include partial restoration of $U(1)_A$ symmetry. In this scenario, not only are positive- and negative-parity states expected to become approximately degenerate at high masses, but also $N^*$ and $\Delta^*$ resonances with the same spin $J$ may converge, reflecting a higher degree of symmetry in the baryon spectrum.

\begin{figure}[thb]
\bc
    \begin{minipage}[c]{0.35\textwidth}
\includegraphics[width=\linewidth]{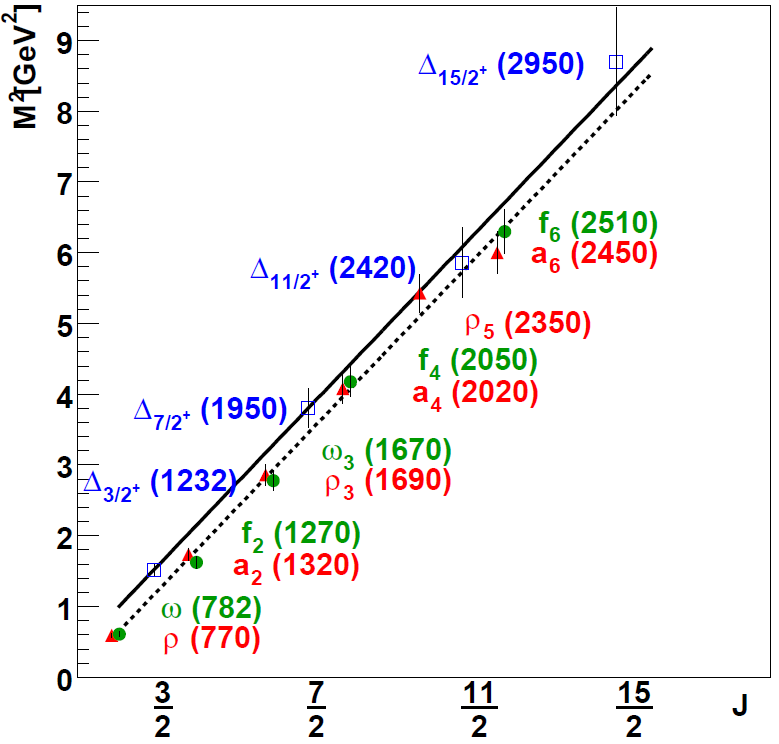}
\caption{(color online) Squared masses of $\Delta^*$ resonances with $J=L+S$ (stretched states)
    in comparison to squared meson masses~\cite{Klempt:2012fy}. \EPJA}
    \label{fig:enter-label}
 \vspace{30mm}
 \end{minipage}
 \hspace{5mm}
   \begin{minipage}[c]{0.35\textwidth}
\includegraphics[width=\linewidth]{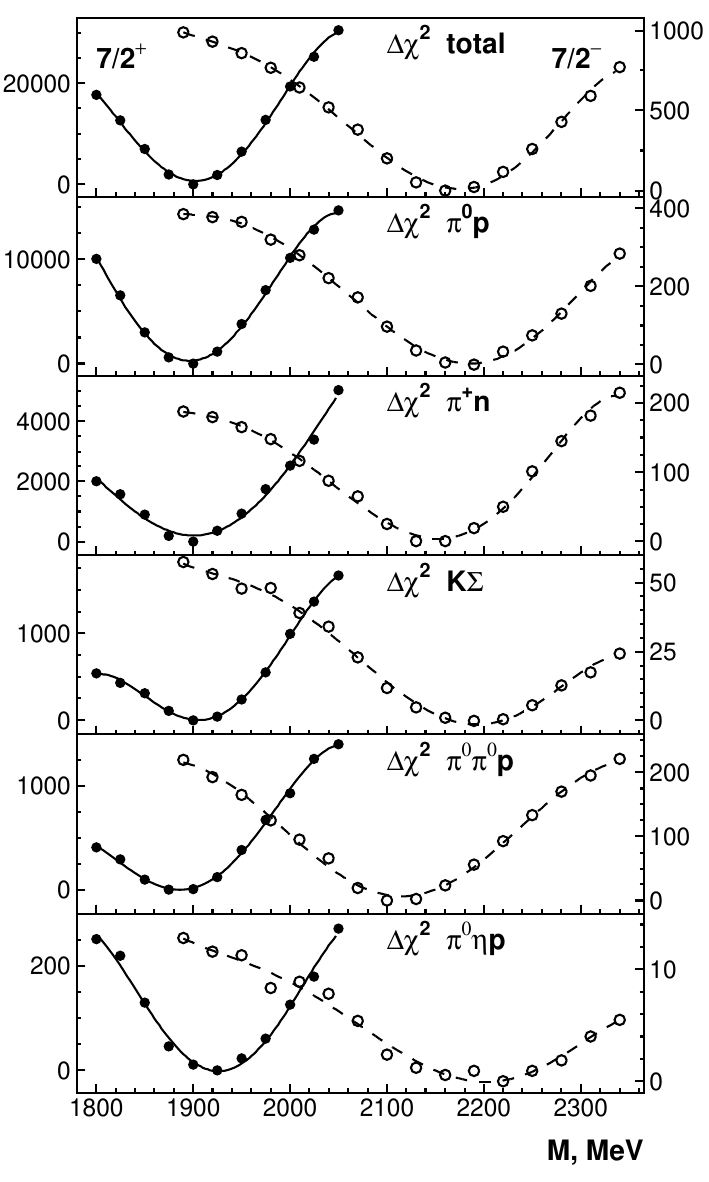}
\end{minipage}
\vspace{-27mm}
\ec
\begin{minipage}[t]{0.48\textwidth}
\caption{\label{scan-g37}The pseudo-$\chi^2$ of the fit to a large body of pion- and photo-produced reactions as a function
of the mass of $\Delta(1950)\nicefrac72^+$ (solid points) or $\Delta(2200)\nicefrac72^-$ (open circles). The scale on the left (right) abscissa refers to the 
$\nicefrac72^+$ ($\nicefrac72^-$) partial wave. The curves are to guide the eye. The Figure is taken from \cite{Anisovich:2011sv}.}
\phantom{zzzzzzzzzzzzzzzz}
\end{minipage}\end{figure}

This interpretation of parity partners as evidence for chiral symmetry restoration
was questioned in Ref.~\cite{Klempt:2002tt}. An empirical mass formula for baryon resonances~\cite{Klempt:2002vp}
suggested that the masses of light baryon resonances do not follow a $M^2 \propto l + 2n$
behavior, but rather $M \propto l + n$ (where $l = l_\rho + l_\lambda$ and $n = n_\rho + n_\lambda$).
Thus, the quartet of positive-parity states $\Delta(1910)\nicefrac{1}{2}^+ \dots \Delta(1950)\nicefrac{7}{2}^+$,
with a leading $l = 2$ and $n = 0$ contribution, and the triplet of negative-parity states
$\Delta(1900)\nicefrac{1}{2}^-, \Delta(1930)\nicefrac{3}{2}^-, \Delta(1940)\nicefrac{5}{2}^-$,
with leading $l = 1$ and $n = 1$, are expected to be mass-degenerate. However, in this scheme,
no $J^P = \nicefrac{7}{2}^-$ state is expected at this mass. Such a state would require $l = 3$
and is expected at about 2150~MeV.
At the time, the $\Delta(2200)\nicefrac72^-$ had a 1* rating, and uncertainties of about 80~MeV were assigned to the
mass determination. Also, further resonances with the predicted properties could be missing (like in the case of  $N(1900)\nicefrac32^+$
where Glozman suggested the $N(2080)\nicefrac32^-$ as parity partner, while later the $N(1875)\nicefrac32^-$ -- in principle a perfect parity partner -- was discovered \cite{Anisovich:2011sv}).

Data on photoproduction enabled a more refined study of the $I=3/2$, $J^P = \nicefrac{7}{2}^-$ channel. 
In a coupled-channel analysis~\cite{Anisovich:2015gia},
the masses of possible $\Delta^*$ resonances with $J^P = \nicefrac{7}{2}^+$ and $J^P = \nicefrac{7}{2}^-$ 
were scanned over a wide mass range. The results of a fit to an extensive data set on pion- and 
photo-induced reactions are shown in Fig.~\ref{scan-g37}. The scans reveal clear minima,
from which Breit-Wigner masses of $(1917 \pm 4)$ and $(2176 \pm 40)$~MeV were derived.
The small coupling of the $\Delta(2200)\nicefrac{7}{2}^-$ to $\pi N$ (branching fraction $(3.3 \pm 1.5)$\%)
explains why it was so difficult to observe it in $\pi N$ elastic scattering.
The observed $\Delta(2200)\nicefrac{7}{2}^-$ mass supports quark models and AdS/QCD,
but conflicts with models in which chiral symmetry is restored in meson and baryon resonances below 2~GeV.
In summary, while these experimental findings are not in conflict with the concept of parity doublets in general, they disfavor their identification with the low-lying excitations in the baryon spectrum.

\clearpage

\section{\label{Summary}Summary and Outlook}
What have we learned from photo- and electroproduction experiments conducted over the past two or three decades? 
First and foremost, the number of known $N^*$ resonances has significantly increased, and many new decay modes 
have been identified. The earlier $\pi N$ elastic scattering data allowed three major research groups — in Karlsruhe, 
at Carnegie Mellon, and at GWU — to extract $N^*$ resonances up to 1.75~GeV and a few states along the leading 
Regge trajectories. However, not a single $N^*$ resonance received a three- or four-star rating within the 
critical 1.75 to 2.15~GeV mass range. This lack of observed states, together with the similarities between 
meson and baryon Regge trajectories, led to the hypothesis that light-quark baryon resonances may consist of 
quasi-stable diquarks and a third quark that carries the excitation. 
Today, nine states in this mass range have either been discovered or confirmed by new data 
from ELSA in Bonn, JLab in Newport News (VA), MAMI in Mainz, and other laboratories. Notably, several positive-parity $N^*$ resonances have now been established that are 
inconsistent with static diquark models. These resonances often decay via cascades such as 
$N(1900)\nicefrac32^+\to N(1520)\nicefrac32^-\pi$, which serves as direct evidence of three-body 
dynamics in the parent nucleon resonance.

Despite these advances, several resonances predicted by quark models in the second 
shell are still missing. In some instances, such as the $\Delta(1750)\nicefrac12^+$, 
evidence for their existence is sparse. This resonance is listed in the RPP with a one-star rating
at a pole mass of about 1730~MeV and an essentially unknown width. Another resonance in the same partial wave, 
the $\Delta(1910)\nicefrac12^+$, has $M_{\rm pole}\sim 1850$~MeV and $\Gamma_{\rm pole}\sim 350$~MeV; 
yet no analysis has claimed the presence of two separate states. 
Interestingly, functional methods based on QCD's $n$-point correlation functions predict only one resonance in 
this partial wave, contrasting with quark models that expect two states below 2~GeV. The latest Bonn-Gatchina 
fit includes two poles, but without conclusive evidence that both are necessary. A firm decision on 
the (non-)existence of the $\Delta(1750)\nicefrac12^+$ is needed.

It is worth noting that in the late 1970s, $\Delta^*$ resonances were reported with masses 
as high as 4.1~GeV, spins up to $J=\nicefrac{21}{2}$ and orbital angular momenta up to $L=9$~\cite{Hendry:1978cd}.
In the RPP, these states are listed under the $\Delta(3000)$ region.
Relative to the known Regge trajectory of $\Delta^*$s, these high-spin resonances have masses 
that are too high. Can the gluon string 
connecting a quark and a diquark persist over such large 
distances for long enough to sustain a resonance? If so, does the associated Regge trajectory begin to curve in the far-distance
confinement region? Investigating this would likely require a $\pi^+$ beam — to 
produce $\Delta^{++}$ baryons via interactions with a proton target — with momenta reaching up to 10~GeV.

The baryon excitation spectra and the density of states play an important role in the early phase of 
the Universe and in the interpretation
of heavy-ion collisions. Details of the freeze-out process define the particle yields and the net-particle
fluctuations. For an acceptable description of the process, the inclusion of the resonances found in the recent decades is mandatory.

Quark models have been developed to help explain the structure of excited states.
A description of resonances through a linear confining potential combined with 
residual interactions among quarks, such as effective one-gluon exchange, pseudoscalar meson exchange, 
or instanton-induced forces,
captures many essential features.
Yet, none of them 
provides a fully satisfactory account both of the $N^*$ and $\Delta^*$ mass spectrum and  their (transition) form factors in the spacelike and timelike regions, and 
they
may be overall too simplistic in view of the rich underlying dynamical phenomena and the multihadron components of many states. Still, the simplest model — 
the nonrelativistic model by Isgur and Karl — remains a useful framework for classifying resonances. 

AdS/QCD calculations aim to reproduce the essential features of QCD. A new interpretation of an 
AdS/QCD mass formula successfully matches the masses of all known $N^*$ and $\Delta^*$ resonances. 
Remarkably, the nucleon mass, Regge slope, and $\Delta(1232)\nicefrac32^+ - N$ hyperfine splitting 
can all be described using a single parameter such as the nucleon mass.

In quark models, the $u$ and $d$ quark masses are taken to be around 350~MeV, even though the actual 
current-quark masses entering the QCD Lagrangian are only a few MeV. Advancements with functional methods based on Dyson-Schwinger equations, which are coupled integral equations for the quark and gluon $n$-point correlation functions, yield dynamical running quark masses via chiral 
symmetry breaking. 
They also provide insight in the forces
that bind baryons: Two-body interactions appear to be dominant, and even though in principle the kernels contain all possible diagrams (gluon exchange, three-gluon vertices, crossed-gluon exchange and so on), many of these contributions can be modelled by  effective gluon exchange. In some exploratory studies, higher-order interactions and effective pion exchanges have been considered as well.
In general, so far these calculations have been limited to low-mass and low-spin baryon
excitations up to 2~GeV and $J^P=1/2^\pm$ and $3/2^\pm$. The results show good agreement with the experimental spectrum and also with electromagnetic transition form factors (or helicity 
transition amplitudes) at sufficiently high $Q^2$.
The tools to describe the multihadron components of baryons are under development, and
work on higher-spin states is ongoing.

Much progress has also been made in lattice QCD. Calculations of stable baryons at physical quark masses have become possible and pave the way for detailed structure investigations of the nucleon. Concerning resonances, in the meson sector the mapping of finite-volume energy levels to infinite-volume scattering amplitudes and resonance poles in the complex plane is now well-established. Similar efforts are beginning to take shape in the baryon sector, which opens up the exciting prospect of first-principles calculations of baryon resonance properties and transition form factors in the near future.

Effective field theories (EFTs) have become a powerful tool for interpreting new data. 
Rather than resolving the quark substructure of baryons, they focus on the dynamics of hadronic interactions between mesons and baryons. In fitting the data, some resonance poles must be included explicitly, 
while others emerge dynamically from the interactions. Dynamically generated states are often 
considered unrelated to $qqq$ spectroscopy. The Roper resonance $N(1440)\nicefrac12^+$, for instance, 
was long thought to be a dynamically generated state with no three-quark core. However, electroproduction 
experiments have resolved this issue, confirming that the $N(1440)\nicefrac12^+$ is indeed a genuine 
three-quark excited state at its core, albeit with a substantial meson cloud. 
The so-called Weinberg criterion — based on meson-baryon scattering amplitudes at zero momentum 
transfer — provides a way to estimate the degree to which hadrons are composite. Based  on
this criterium, the $N(1440)\nicefrac12^+$ is characterized as a dynamically 
generated resonance with a notable three-quark core, and electroproduction data and EFTs now
agree that the Roper resonance has a $qqq$ core and a significant meson cloud. In other cases, 
results remain contradictory. For example, 
the $N(1520)\nicefrac32^-$ is sometimes described as a predominantly composite meson-baryon state, yet 
electroproduction data prove it has a clear three-quark structure with little meson cloud involvement.

One of the central unresolved issues concerns the very origin of resonances. While the interquark 
forces clearly account for the nucleon and the $\Delta(1232)\nicefrac32^+$, some resonances might 
arise from molecular interactions between hadrons. 
The deuteron, for example, is well understood as 
a proton-neutron bound state stabilized by meson-exchange forces. Could some light-quark baryons 
follow a similar pattern? Are molecular interactions between mesons and baryons the primary source 
for the existence of certain resonances, even if they possess a three-quark component due to $q\bar q$ annihilation? 
Or do interquark forces create the resonance, with the pole subsequently generating the attractive 
molecular interaction? The $\Lambda(1405)$ may help to elucidate this question. The two-pole nature of 
the $\Lambda(1405)$ -- decaying into $NK$ and $\Sigma\pi$ -- is a significant discovery based on EFTs. 
Both decay modes offer limited phase space, and the attractive interactions in both 
channels lead to the presence of two poles contributing to the $\Lambda(1405)$ mass distribution. 
Still, some analyses manage to fit the data using only a single resonance compatible with a
three-quark structure. Precision measurements of the $\Lambda(1405)$ line shape, electroproduction of $\Lambda(1405)$,
and new BESIII experiments are essential to resolve this question.

Several accelerators and colliders are still in operation, and new insight into the nucleon, its structure and its excitations 
are expected. 
{The doubling of the CEBAF accelerator from 6 GeV to 12 GeV maximum energy has opened up the kinematic space in hadron physics significantly. It extends both the mass reach in the photoproduction of excited baryons, and the reach in photon virtuality $Q^2$ in the electroproduction of mesons. The former program may be part of the GlueX experiment in Hall D ~\cite{GlueX:2020idb}, while the latter enables probing the excited baryons at significantly shorter distances and should lead to more quantifiable results in identifying the effective degrees of freedom. This is a focus of the newly constructed CLAS12 spectrometer~\cite{Burkert:2020akg} in Hall B.  Much of the data that may be of interest to readers of this review have already been taken and are in the analysis stage. Another program that can be more rigorously pursued with CLAS12 is the study of nucleon mass creation using excited baryon states~\cite{Achenbach:2025kfx}. }

At the low-energy end, electron-proton scattering experiments are in the planning at MESA in Mainz, Germany,  
aiming at precise measurements of electroweak coupling constants like the Weinberg angle, and searches for light dark matter
will be pursued~\cite{Schlimme:2024eky}. The role of gluons in hadron spectroscopy has been a central topic since the advent of QCD. Gluonic degrees 
of freedom represent a challenging and often debated aspect of meson spectroscopy, while baryonic hybrids 
are even more complex. If hybrid baryons indeed have high masses, as suggested by lattice calculations, 
mixing with conventional $qqq$ states may occur. However, the presence of a heavy constituent gluon could 
suppress such mixing. At JLab, an energy upgrade to 22~GeV is planned \cite{Accardi:2023chb} to perform
high-statistics electroproduction experiments over a broad range of momentum 
transfers $Q^2$ providing crucial insights into gluonic components within baryon resonance wave functions. The considerable higher $Q^2$ range that is achievable at this energy will significantly benefit the nucleon resonance program in probing the emerging hadron mass program~\cite{Achenbach:2025kfx}.   
In addition, cascade decays could help illuminate this issue. At high $Q^2$, backgrounds from $t$- and 
$u$-channel exchanges are reduced, offering a cleaner environment for future studies of the baryon 
resonance spectrum.

In this review we have only briefly touched upon hyperon spectroscopy, i.e., on the $\Lambda$, $\Sigma$, $\Xi$, 
and $\Omega$ baryons. An important question is how well SU(3) symmetry relations hold in these systems. 
The $\Lambda^*$ spectrum includes the enigmatic $\Lambda(1405)$ complex; its structure deserves further
studies and little is known about the higher-mass singlet $\Lambda^*$'s. These hyperons and 
$\Sigma^*$'s will be topical themes at JLab and the new Cluster of Excellence ``Color meets flavor" in Bonn. 
The hyperon spectra are also important components of the FAIR program at GSI (Germany). Their study will benefit from the 
narrowing of resonance widths with increasing strangeness. 

At high energies, structure functions will remain a topical theme at JLab, at the electron-ion collider at the Brookhaven National Laboratory presently in the construction stage, and at the planned electron-ion collider in China. 
Gravity-like
interactions are still 
in a very early stage but show great promise as a novel way of probing the strong interaction in hadrons and 
light nuclei. Gravity as a probe was theoretically studied already in the 1960's~\cite{Kobzarev:1962wt,Pagels:1966zza} but remained dormant for four decades. 
% because of the extreme weakness of gravitation. 
Maxim Polyakov~\cite{Polyakov:2002yz} considered using  deeply virtual Compton scattering as a way to mimic gravity 
% at a $10^{39}$ times stronger interaction to 
 and probe the gravitational structure of the strong force. 
Following the first successful experimental application of this technique in 2018 with the ground-state proton, the field developed rapidly and has expanded to probe gravitational properties of nucleon resonances through the energy-momentum tensor~\cite{Kim:2024hhd,Azizi:2020jog,Goharipour:2024atx}.  

Considerable progress has been achieved arising from photo- and electroproduction off nucleons, and new insight has been gained in our understanding how baryons emerge from nearly massless quarks. 
Yet, as is often the case in science, new discoveries bring new questions. Fortunately, the tools to explore them are already in place.

%\vfill
\section*{Acknowledgements}
We would like to thank John Bulava, Maxim Mai, Ulf-G. Mei\ss ner, Bernard Metsch, Viktor Mokeev, Raquel Molina,
Deborah Rönchen, Andrey Sarantsev, 
Hartmut Schmieden, Ulrike Thoma, Guy de Teramond, Lothar Tiator, and Wolfram Weise for fruitful 
discussions and valuable comments to the manuscript. 
The work of V. Burkert was supported by the US Department of Energy under contract DE-AC-06OR23177.  
GE acknowledges support from the Austrian Science Fund FWF under grant number
10.55776/PAT2089624. This work contributes to the aims of the USDOE ExoHad Topical Collaboration, contract
DE-SC0023598.

%\newpage
%\phantom{zz}
\bibliography{Baryon}% Produces the bibliography via BibTeX.

\begin{thebibliography}{1000}
\expandafter\ifx\csname url\endcsname\relax
  \def\url#1{\texttt{#1}}\fi
\expandafter\ifx\csname urlprefix\endcsname\relax\def\urlprefix{URL }\fi
\expandafter\ifx\csname href\endcsname\relax
  \def\href#1#2{#2} \def\path#1{#1}\fi

\bibitem{Gross:2022hyw}
\protect{Gross, F., Klempt, E. (eds.)}, et~al., {50 Years of Quantum Chromodynamics}, Eur. Phys. J. C 83 (2023) 1--636.
\newblock \href {http://arxiv.org/abs/2212.11107} {\path{arXiv:2212.11107}}.

\bibitem{Aliberti:2025beg}
R.~Aliberti, et~al., {The anomalous magnetic moment of the muon in the Standard Model: an update} (5 2025).
\newblock \href {http://arxiv.org/abs/2505.21476} {\path{arXiv:2505.21476}}.

\bibitem{ParticleDataGroup:2024cfk}
S.~Navas, et~al., {Review of particle physics}, Phys. Rev. D 110~(3) (2024) 030001.
\newblock \href {https://doi.org/10.1103/PhysRevD.110.030001} {\path{doi:10.1103/PhysRevD.110.030001}}.

\bibitem{Deur:2022msf}
A.~Deur, V.~Burkert, J.~P. Chen, W.~Korsch, {Experimental determination of the QCD effective charge $\alpha_{g_1}(Q)$}, Particles 5 (2022) 171.
\newblock \href {http://arxiv.org/abs/2205.01169} {\path{arXiv:2205.01169}}, \href {https://doi.org/10.3390/particles5020015} {\path{doi:10.3390/particles5020015}}.

\bibitem{Gell-Mann:1964ewy}
M.~Gell-Mann, {A Schematic Model of Baryons and Mesons}, Phys. Lett. 8 (1964) 214--215.
\newblock \href {https://doi.org/10.1016/S0031-9163(64)92001-3} {\path{doi:10.1016/S0031-9163(64)92001-3}}.

\bibitem{Zweig:1964jf}
G.~Zweig, {An SU(3) model for strong interaction symmetry and its breaking. Version 2}, Nonantum, MA : Hadronic Press, 1964, pp. 22--101.

\bibitem{Isgur:2000ad}
N.~Isgur, {Why N*'s are important}, in: V.~Burkert, L.~Elouadrhiri, J.~Kelly, R.~M. (eds.) (Eds.), NSTAR2000: Excited nucleons and hadronic structure, Newport News, USA, 2000, pp. 422--445.
\newblock \href {http://arxiv.org/abs/nucl-th/0007008} {\path{arXiv:nucl-th/0007008}}.

\bibitem{Hey:1982aj}
A.~J.~G. Hey, R.~L. Kelly, {Baryon spectroscopy}, Phys. Rept. 96 (1983) 71.
\newblock \href {https://doi.org/10.1016/0370-1573(83)90114-X} {\path{doi:10.1016/0370-1573(83)90114-X}}.

\bibitem{Burkert:2004sk}
V.~D. Burkert, T.~S.~H. Lee, {Electromagnetic meson production in the nucleon resonance region}, Int. J. Mod. Phys. E 13 (2004) 1035--1112.
\newblock \href {http://arxiv.org/abs/nucl-ex/0407020} {\path{arXiv:nucl-ex/0407020}}, \href {https://doi.org/10.1142/S0218301304002545} {\path{doi:10.1142/S0218301304002545}}.

\bibitem{Klempt:2009pi}
E.~Klempt, J.-M. Richard, {Baryon spectroscopy}, Rev. Mod. Phys. 82 (2010) 1095--1153.
\newblock \href {http://arxiv.org/abs/0901.2055} {\path{arXiv:0901.2055}}, \href {https://doi.org/10.1103/RevModPhys.82.1095} {\path{doi:10.1103/RevModPhys.82.1095}}.

\bibitem{Crede:2013kia}
V.~Crede, W.~Roberts, {Progress towards understanding baryon resonances}, Rept. Prog. Phys. 76 (2013) 076301.
\newblock \href {http://arxiv.org/abs/1302.7299} {\path{arXiv:1302.7299}}, \href {https://doi.org/10.1088/0034-4885/76/7/076301} {\path{doi:10.1088/0034-4885/76/7/076301}}.

\bibitem{Ireland:2019uwn}
D.~G. Ireland, E.~Pasyuk, I.~Strakovsky, {Photoproduction Reactions and Non-Strange Baryon Spectroscopy}, Prog. Part. Nucl. Phys. 111 (2020) 103752.
\newblock \href {http://arxiv.org/abs/1906.04228} {\path{arXiv:1906.04228}}, \href {https://doi.org/10.1016/j.ppnp.2019.103752} {\path{doi:10.1016/j.ppnp.2019.103752}}.

\bibitem{Eichmann:2022zxn}
G.~Eichmann, {Theory Introduction to Baryon Spectroscopy}, Few Body Syst. 63~(3) (2022) 57.
\newblock \href {http://arxiv.org/abs/2202.13378} {\path{arXiv:2202.13378}}, \href {https://doi.org/10.1007/s00601-022-01756-y} {\path{doi:10.1007/s00601-022-01756-y}}.

\bibitem{Thiel:2022xtb}
A.~Thiel, F.~Afzal, Y.~Wunderlich, {Light Baryon Spectroscopy}, Prog. Part. Nucl. Phys. 125 (2022) 103949.
\newblock \href {http://arxiv.org/abs/2202.05055} {\path{arXiv:2202.05055}}, \href {https://doi.org/10.1016/j.ppnp.2022.103949} {\path{doi:10.1016/j.ppnp.2022.103949}}.

\bibitem{Mai:2022eur}
M.~Mai, U.-G. Mei\ss{}ner, C.~Urbach, {Towards a theory of hadron resonances}, Phys. Rept. 1001 (2023) 1--66.
\newblock \href {http://arxiv.org/abs/2206.01477} {\path{arXiv:2206.01477}}, \href {https://doi.org/10.1016/j.physrep.2022.11.005} {\path{doi:10.1016/j.physrep.2022.11.005}}.

\bibitem{Doring:2025sgb}
M.~D\"oring, J.~Haidenbauer, M.~Mai, T.~Sato, {Dynamical coupled-channel models for hadron dynamics} (5 2025).
\newblock \href {http://arxiv.org/abs/2505.02745} {\path{arXiv:2505.02745}}.

\bibitem{Aznauryan:2011qj}
I.~G. Aznauryan, V.~D. Burkert, {Electroexcitation of nucleon resonances}, Prog. Part. Nucl. Phys. 67 (2012) 1--54.
\newblock \href {http://arxiv.org/abs/1109.1720} {\path{arXiv:1109.1720}}, \href {https://doi.org/10.1016/j.ppnp.2011.08.001} {\path{doi:10.1016/j.ppnp.2011.08.001}}.

\bibitem{Burkert:2022ioj}
V.~D. Burkert, {Nucleon resonances and transition form factors}, Eur. Phys. J. C 83, in: \cite{Gross:2022hyw} (12 2022).
\newblock \href {http://arxiv.org/abs/2212.08980} {\path{arXiv:2212.08980}}.

\bibitem{Carman:2020qmb}
D.~S. Carman, K.~Joo, V.~I. Mokeev, {Strong QCD Insights from Excited Nucleon Structure Studies with CLAS and CLAS12}, Few Body Syst. 61~(3) (2020) 29.

\bibitem{Ramalho:2023hqd}
G.~Ramalho, M.~T. Pe\~na, {Electromagnetic transition form factors of baryon resonances}, Prog. Part. Nucl. Phys. 136 (2024) 104097.
\newblock \href {http://arxiv.org/abs/2306.13900} {\path{arXiv:2306.13900}}, \href {https://doi.org/10.1016/j.ppnp.2024.104097} {\path{doi:10.1016/j.ppnp.2024.104097}}.

\bibitem{Achenbach:2025kfx}
P.~Achenbach, et~al., {Electroexcitation of Nucleon Resonances and the Emergence of Hadron Mass} (5 2025).
\newblock \href {http://arxiv.org/abs/2505.23550} {\path{arXiv:2505.23550}}.

\bibitem{Hohler:1977em}
G.~Höhler, K.~H. Augenstein, E.~Pietarinen, H.~M. Staudenmaier, Physics Data: Handbook of pion-nucleon scattering. Supplement 1: Differential Cross-Sections for $\pi^- p \to \pi^- p$., Published by Fachinformationszentrum, Karlsruhe 1977, Zaed 1-2, 1977.

\bibitem{Hohler:1979yr}
G.~Höhler, F.~Kaiser, R.~Koch, E.~Pietarinen, {Physics Data: Handbook of pion-nucleon scattering}, Phys. Daten 12N1 (1979) 1.

\bibitem{Koch:1980ay}
R.~Koch, E.~Pietarinen, {Low-Energy $\pi N$ Partial Wave Analysis}, Nucl. Phys. A 336 (1980) 331--346.
\newblock \href {https://doi.org/10.1016/0375-9474(80)90214-6} {\path{doi:10.1016/0375-9474(80)90214-6}}.

\bibitem{Koch:1983ht}
R.~Koch, M.~Hutt, {A Partial Wave Dispersion Relation Analysis of Pion - Nucleon Scattering Amplitudes}, Z. Phys. C 19 (1983) 119.
\newblock \href {https://doi.org/10.1007/BF01571772} {\path{doi:10.1007/BF01571772}}.

\bibitem{Hohler:1984ux}
G.~Höhler, {Numerical data and functional relationships in science and technology. Group I: Nuclear and Particle Physics. Vol. 9: Elastic and Charge Echange Scattering of Elementary Particles. B: Pion Nucleon Scattering. Part 2: Methods and results}, Springer (1983), 601 P. Berlin, Germany ( Landolt- Boernstein. New Series, I/9B2, H. Schopper, (Ed.)), 1984.

\bibitem{Koch:1985bp}
R.~Koch, {Improved $\pi N$ Partial Waves Consistent With Analyticity and Unitarity}, Z. Phys. C 29 (1985) 597.
\newblock \href {https://doi.org/10.1007/BF01560295} {\path{doi:10.1007/BF01560295}}.

\bibitem{Koch:1985bn}
R.~Koch, {A Calculation of Low-Energy $\pi N$ Partial Waves Based on Fixed t Analyticity}, Nucl. Phys. A 448 (1986) 707--731.
\newblock \href {https://doi.org/10.1016/0375-9474(86)90438-0} {\path{doi:10.1016/0375-9474(86)90438-0}}.

\bibitem{Hohler:1993lbk}
G.~Höhler, {Determination of $\pi N$ resonance pole parameters}, PiN Newslett. 1993~(9) (1993) 1--36.

\bibitem{Kelly:1979uf}
R.~L. Kelly, R.~E. Cutkosky, {Amalgamation of meson - nucleon scattering data}, Phys. Rev. D20 (1979) 2782.
\newblock \href {https://doi.org/10.1103/PhysRevD.20.2782} {\path{doi:10.1103/PhysRevD.20.2782}}.

\bibitem{gwdac}
\url{https://gwdac.phys.gwu.edu/analysis/pin_analysis.html}.

\bibitem{Alekseev:2005zr}
I.~G. Alekseev, et~al., {Measurement of the spin rotation parameter A in the elastic pion-proton scattering at 1.43-GeV/c}, Eur. Phys. J. C 45 (2006) 383--386.
\newblock \href {http://arxiv.org/abs/hep-ex/0510010} {\path{arXiv:hep-ex/0510010}}, \href {https://doi.org/10.1140/epjc/s2005-02454-y} {\path{doi:10.1140/epjc/s2005-02454-y}}.

\bibitem{Alekseev:1996gs}
I.~G. Alekseev, et~al., {Influence of spin rotation measurements on partial wave analyses of elastic pion - nucleon scattering}, Phys. Rev. C 55 (1997) 2049--2053.
\newblock \href {http://arxiv.org/abs/nucl-th/9608043} {\path{arXiv:nucl-th/9608043}}, \href {https://doi.org/10.1103/PhysRevC.55.2049} {\path{doi:10.1103/PhysRevC.55.2049}}.

\bibitem{ITEP-PNPI:2000gqp}
I.~G. Alekseev, et~al., {Measurements of spin rotation parameter A in pion proton elastic scattering at 1.62-GeV/c}, Phys. Lett. B 485 (2000) 32--36.
\newblock \href {http://arxiv.org/abs/hep-ex/0004025} {\path{arXiv:hep-ex/0004025}}, \href {https://doi.org/10.1016/S0370-2693(00)00653-5} {\path{doi:10.1016/S0370-2693(00)00653-5}}.

\bibitem{ITEP-PNPI:2008cmv}
I.~G. Alekseev, et~al., {Backward asymmetry measurements in the elastic pion-proton scattering at resonance energies}, Eur. Phys. J. A 39 (2009) 163--168.
\newblock \href {http://arxiv.org/abs/0810.1143} {\path{arXiv:0810.1143}}, \href {https://doi.org/10.1140/epja/i2008-10709-0} {\path{doi:10.1140/epja/i2008-10709-0}}.

\bibitem{Herndon:1974xd}
D.~Herndon, et~al., {A Partial - Wave Analysis of the Reaction $\pi N \to \pi \pi N$ in the Center-Of-Mass Energy Range 1300-MeV - 2000-MeV}, Phys. Rev. D 11 (1975) 3183.
\newblock \href {https://doi.org/10.1103/PhysRevD.11.3183} {\path{doi:10.1103/PhysRevD.11.3183}}.

\bibitem{Longacre:1975qj}
R.~S. Longacre, J.~Dolbeau, {Comparison Between $\gamma n \to \pi n$ and $\pi n \to \rho n$ in the Resonance Region Up to $\sqrt s$ = 1.75-GeV via the Vector Dominance Model}, Phys. Lett. B 58 (1975) 455--458.
\newblock \href {https://doi.org/10.1016/0370-2693(75)90588-2} {\path{doi:10.1016/0370-2693(75)90588-2}}.

\bibitem{Longacre:1976ja}
R.~S. Longacre, J.~Dolbeau, {K-Matrix Fits to $\pi n \to n \pi$ and $\pi n \to n \pi \pi$ in the Resonance Region $\sqrt s$ = 1380-MeV to 1740-MeV}, Nucl. Phys. B 122 (1977) 493--524.
\newblock \href {https://doi.org/10.1016/0550-3213(77)90142-0} {\path{doi:10.1016/0550-3213(77)90142-0}}.

\bibitem{Longacre:1977ga}
R.~S. Longacre, et~al., {K-Matrix Fits to $\pi n \to n \pi$ and $\pi n \to n \pi \pi$ in the Resonance Region $\sqrt s$ = 1.3-GeV to 2.0-GeV}, Phys. Rev. D 17 (1978) 1795.
\newblock \href {https://doi.org/10.1103/PhysRevD.17.1795} {\path{doi:10.1103/PhysRevD.17.1795}}.

\bibitem{Novoseller:1977xt}
D.~E. Novoseller, {Partial Wave Analysis Including $\pi$ Exchange for $\pi n \to n \pi \pi$ in the Center-of-Mass Energy Range 1.65-GeV-1.97-GeV}, Nucl. Phys. B 137 (1978) 445--508.
\newblock \href {https://doi.org/10.1016/0550-3213(78)90325-5} {\path{doi:10.1016/0550-3213(78)90325-5}}.

\bibitem{Novoseller:1977xs}
D.~E. Novoseller, {Analysis of Decay of $\pi n$ Resonances Into $\pi \pi n$ Channels}, Nucl. Phys. B 137 (1978) 509--520.
\newblock \href {https://doi.org/10.1016/0550-3213(78)90326-7} {\path{doi:10.1016/0550-3213(78)90326-7}}.

\bibitem{Manley:1984zs}
D.~M. Manley, {Isospin Analysis of Low-energy $\pi N \to \pi \pi N$ Data and Chiral Symmetry Breaking}, Phys. Rev. D 30 (1984) 536--540.
\newblock \href {https://doi.org/10.1103/PhysRevD.30.536} {\path{doi:10.1103/PhysRevD.30.536}}.

\bibitem{Manley:1992yb}
D.~M. Manley, E.~M. Saleski, {Multichannel resonance parametrization of $\pi N$ scattering amplitudes}, Phys. Rev. D45 (1992) 4002--4033.
\newblock \href {https://doi.org/10.1103/PhysRevD.45.4002} {\path{doi:10.1103/PhysRevD.45.4002}}.

\bibitem{CrystalBall:2004qln}
S.~Prakhov, et~al., {Measurement of $\pi^- p \to \pi^0 \pi^0 n$ from threshold to $p_{\pi^-}= 750$-MeV/c}, Phys. Rev. C 69 (2004) 045202.
\newblock \href {https://doi.org/10.1103/PhysRevC.69.045202} {\path{doi:10.1103/PhysRevC.69.045202}}.

\bibitem{HADES:2020kce}
J.~Adamczewski-Musch, et~al., {Two-pion production in the second resonance region in ${\pi}^-p$ collisions with the High-Acceptance Di-Electron Spectrometer (HADES)}, Phys. Rev. C 102~(2) (2020) 024001.
\newblock \href {http://arxiv.org/abs/2004.08265} {\path{arXiv:2004.08265}}, \href {https://doi.org/10.1103/PhysRevC.102.024001} {\path{doi:10.1103/PhysRevC.102.024001}}.

\bibitem{Candlin:1984av}
D.~J. Candlin, et~al., {A measurement of $\pi^+ p$ backward elastic differential cross sections from 1.282-GeV/c to 2.472-GeV/c}, Nucl. Phys. B 244 (1984) 23--56.
\newblock \href {https://doi.org/10.1016/0550-3213(84)90180-9} {\path{doi:10.1016/0550-3213(84)90180-9}}.

\bibitem{Edinburgh-Rutherford-Westfield:1983qsm}
D.~J. Candlin, et~al., {An Energy Dependent Partial Wave Analysis of $\pi^+ p \to K^+ \Sigma^+$ Between Threshold and 2.35-{GeV}}, Nucl. Phys. B 238 (1984) 477--491.
\newblock \href {https://doi.org/10.1016/0550-3213(84)90332-8} {\path{doi:10.1016/0550-3213(84)90332-8}}.

\bibitem{Candlin:1988pn}
D.~J. Candlin, et~al., {Measurement of the Spin Rotation Parameter, $\beta$, in the Reaction $\pi^+ p \to K^+ \Sigma^+$ at 1.60-{GeV}/$c$ and 1.88-{GeV}/$c$}, Nucl. Phys. B 311 (1989) 613--629.
\newblock \href {https://doi.org/10.1016/0550-3213(89)90170-3} {\path{doi:10.1016/0550-3213(89)90170-3}}.

\bibitem{Matveev:2019igl}
M.~Matveev, et~al., {Hyperon I: Partial-wave amplitudes for K$^{-}p$ scattering}, Eur. Phys. J. A 55~(10) (2019) 179.
\newblock \href {http://arxiv.org/abs/1907.03645} {\path{arXiv:1907.03645}}, \href {https://doi.org/10.1140/epja/i2019-12878-y} {\path{doi:10.1140/epja/i2019-12878-y}}.

\bibitem{Althoff:1963hfa}
K.~Althoff, H.~Fischer, W.~Paul, {Photoproduction of positive pions on hydrogen between 200 and 450 MeV}, Z. Phys. 175~(1) (1963) 19--33.
\newblock \href {https://doi.org/10.1007/BF01375391} {\path{doi:10.1007/BF01375391}}.

\bibitem{Althoff:1968twa}
K.~H. Althoff, et~al., {The 2.5 GeV electron synchrotron of the University of Bonn}, Nucl. Instrum. Meth. 61 (1968) 1--30.
\newblock \href {https://doi.org/10.1016/0029-554X(68)90443-6} {\path{doi:10.1016/0029-554X(68)90443-6}}.

\bibitem{Mango:1969ww}
S.~Mango, O.~Runolfsson, M.~Borghini, {A butanol polarized proton target}, Nucl. Instrum. Meth. 72 (1969) 45--50.
\newblock \href {https://doi.org/10.1016/0029-554X(69)90263-8} {\path{doi:10.1016/0029-554X(69)90263-8}}.

\bibitem{Althoff:1971zza}
K.~H. Althoff, et~al., {Experience with a polarized proton target at the bonn 2.5 gev synchrotron.}, in: G.~Shapiro (Ed.), 2nd International Conference on Polarized Targets, 30 August - 2 September 1971. Berkeley, CA, USA, Vol. 710830, 1971, pp. 205--208.

\bibitem{Baum:1969es}
G.~Baum, U.~Koch, {A source of polarized electrons}, Nucl. Instrum. Meth. 71 (1969) 189--195.
\newblock \href {https://doi.org/10.1016/0029-554X(69)90011-1} {\path{doi:10.1016/0029-554X(69)90011-1}}.

\bibitem{Hillert:2006yb}
W.~Hillert, {The Bonn electron stretcher accelerator ELSA: Past and future}, Eur. Phys. J. A 28S1 (2006) 139--148.
\newblock \href {https://doi.org/10.1140/epja/i2006-09-015-4} {\path{doi:10.1140/epja/i2006-09-015-4}}.

\bibitem{Brefeld:1984iv}
W.~Brefeld, et~al., {Measurement of the Polarization Degree of Accelerated Polarized Electrons at the 2.5-{GeV} Synchrotron in Bonn for Energies Between 0.85-{GeV} and 2-{GeV}}, Nucl. Instrum. Meth. A 228 (1985) 228.
\newblock \href {https://doi.org/10.1016/0168-9002(85)90264-5} {\path{doi:10.1016/0168-9002(85)90264-5}}.

\bibitem{Husmann:1988vg}
D.~Husmann, W.~J. Schwille, {ELSA - the New Electron Stretcher Ring in Bonn. (In German)}, Phys. Bl. 44 (1988) 40--44.

\bibitem{Dutz:1996uc}
H.~Dutz, et~al., {Photoproduction of positive pions from polarized protons}, Nucl. Phys. A 601 (1996) 319--332.
\newblock \href {https://doi.org/10.1016/0375-9474(96)00008-5} {\path{doi:10.1016/0375-9474(96)00008-5}}.

\bibitem{Dutz:2004zz}
H.~Dutz, et~al., {Experimental Check of the Gerasimov-Drell-Hearn Sum Rule for H-1}, Phys. Rev. Lett. 93 (2004) 032003.
\newblock \href {https://doi.org/10.1103/PhysRevLett.93.032003} {\path{doi:10.1103/PhysRevLett.93.032003}}.

\bibitem{Hoffmann-Rothe:1997jmq}
P.~Hoffmann-Rothe, et~al., {Break up and coherent photoproduction of eta mesons on the deuteron}, Phys. Rev. Lett. 78 (1997) 4697--4700.
\newblock \href {https://doi.org/10.1103/PhysRevLett.78.4697} {\path{doi:10.1103/PhysRevLett.78.4697}}.

\bibitem{Bock:1998rk}
A.~Bock, et~al., {Measurement of the target asymmetry of eta and pi0 photoproduction on the proton}, Phys. Rev. Lett. 81 (1998) 534--537.
\newblock \href {https://doi.org/10.1103/PhysRevLett.81.534} {\path{doi:10.1103/PhysRevLett.81.534}}.

\bibitem{GDH:2003xhc}
H.~Dutz, et~al., {First measurement of the Gerasimov-Drell-Hearn sum rule for H-1 from 0.7-GeV to 1.8-GeV at ELSA}, Phys. Rev. Lett. 91 (2003) 192001.
\newblock \href {https://doi.org/10.1103/PhysRevLett.91.192001} {\path{doi:10.1103/PhysRevLett.91.192001}}.

\bibitem{Naumann:2003vf}
J.~Naumann, et~al., {A photon tagging system for the GDH experiment at ELSA}, Nucl. Instrum. Meth. A 498 (2003) 211--219.
\newblock \href {https://doi.org/10.1016/S0168-9002(02)02147-2} {\path{doi:10.1016/S0168-9002(02)02147-2}}.

\bibitem{GDH:2005noz}
H.~Dutz, et~al., {Measurement of helicity-dependent photoabsorption cross sections on the neutron from 815-MeV to 1825-MeV}, Phys. Rev. Lett. 94 (2005) 162001.
\newblock \href {https://doi.org/10.1103/PhysRevLett.94.162001} {\path{doi:10.1103/PhysRevLett.94.162001}}.

\bibitem{Gothe:2001yjy}
R.~W. Gothe, {Pion Electroproduction at Elsa}, in: {2nd International Workshop on the Physics of Excited Nucleons}, 2001, pp. 19--26.
\newblock \href {https://doi.org/10.1142/9789812810878_0003} {\path{doi:10.1142/9789812810878_0003}}.

\bibitem{Schwille:1994vg}
W.~J. Schwille, et~al., {Design and construction of the SAPHIR detector}, Nucl. Instrum. Meth. A 344 (1994) 470--486.
\newblock \href {https://doi.org/10.1016/0168-9002(94)90868-0} {\path{doi:10.1016/0168-9002(94)90868-0}}.

\bibitem{Bockhorst:1994jf}
M.~Bockhorst, et~al., {Measurement of $\gamma p \to \Lambda$ and $\gamma p \to K^+ \Sigma^0$ at photon energies up to 1.47-GeV}, Z. Phys. C 63 (1994) 37--47.
\newblock \href {https://doi.org/10.1007/BF01577542} {\path{doi:10.1007/BF01577542}}.

\bibitem{Mirazita:1997yp}
M.~Mirazita, et~al., {Total hadronic photoabsorption on carbon and lead in the shadowing threshold region}, Phys. Lett. B 407 (1997) 225--228.
\newblock \href {https://doi.org/10.1016/S0370-2693(97)00764-8} {\path{doi:10.1016/S0370-2693(97)00764-8}}.

\bibitem{Bennhold:1997mg}
C.~Bennhold, et~al., {$K^0 \Sigma^+$ photoproduction with SAPHIR}, Nucl. Phys. A 639 (1998) 209--212.
\newblock \href {http://arxiv.org/abs/nucl-th/9711048} {\path{arXiv:nucl-th/9711048}}, \href {https://doi.org/10.1016/S0375-9474(98)00275-9} {\path{doi:10.1016/S0375-9474(98)00275-9}}.

\bibitem{SAPHIR:1998noz}
R.~Plötzke, et~al., {Photoproduction of $\eta'$ mesons with the $4\pi$-detector SAPHIR}, Phys. Lett. B 444 (1998) 555--562.
\newblock \href {https://doi.org/10.1016/S0370-2693(98)01444-0} {\path{doi:10.1016/S0370-2693(98)01444-0}}.

\bibitem{SAPHIR:1998fev}
M.~Q. Tran, et~al., {Measurement of $\gamma p \to K^+ \Lambda$ and $\gamma p \to K^+ \Sigma^0$ at photon energies up to 2-GeV}, Phys. Lett. B 445 (1998) 20--26.
\newblock \href {https://doi.org/10.1016/S0370-2693(98)01393-8} {\path{doi:10.1016/S0370-2693(98)01393-8}}.

\bibitem{Muccifora:1998ct}
V.~Muccifora, et~al., {Photoabsorption on nuclei in the shadowing threshold region}, Phys. Rev. C 60 (1999) 064616.
\newblock \href {http://arxiv.org/abs/nucl-ex/9810015} {\path{arXiv:nucl-ex/9810015}}, \href {https://doi.org/10.1103/PhysRevC.60.064616} {\path{doi:10.1103/PhysRevC.60.064616}}.

\bibitem{SAPHIR:1999wfu}
S.~Goers, et~al., {Measurement of $\gamma p \to K^0 \Sigma^+$ at photon energies up to 1.55-GeV}, Phys. Lett. B 464 (1999) 331--338.
\newblock \href {https://doi.org/10.1016/S0370-2693(99)01031-X} {\path{doi:10.1016/S0370-2693(99)01031-X}}.

\bibitem{Barth:2001cb}
J.~Barth, et~al., {New results on $\eta$ and $\eta'$ photoproduction with SAPHIR at ELSA}, Nucl. Phys. A 691 (2001) 374--380.
\newblock \href {https://doi.org/10.1016/S0375-9474(01)01059-4} {\path{doi:10.1016/S0375-9474(01)01059-4}}.

\bibitem{Barth:2003kv}
J.~Barth, et~al., {Low-energy of photoproduciton of $\omega$-mesons}, Eur. Phys. J. A 18 (2003) 117--127.
\newblock \href {https://doi.org/10.1140/epja/i2003-10061-y} {\path{doi:10.1140/epja/i2003-10061-y}}.

\bibitem{Barth:2003bq}
J.~Barth, et~al., {Low-energy photoproduction of $\Phi$ mesons}, Eur. Phys. J. A 17 (2003) 269--274.
\newblock \href {https://doi.org/10.1140/epja/i2002-10154-1} {\path{doi:10.1140/epja/i2002-10154-1}}.

\bibitem{SAPHIR:2003lnh}
J.~Barth, et~al., {Evidence for the positive strangeness pentaquark $\Theta^+$ in photoproduction with the SAPHIR detector at ELSA}, Phys. Lett. B 572 (2003) 127--132.
\newblock \href {http://arxiv.org/abs/hep-ex/0307083} {\path{arXiv:hep-ex/0307083}}, \href {https://doi.org/10.1016/j.physletb.2003.08.019} {\path{doi:10.1016/j.physletb.2003.08.019}}.

\bibitem{Glander:2003jw}
K.~H. Glander, et~al., {Measurement of $\gamma p \to K^+ \Lambda$ and $\gamma p \to K^+ \Sigma^0$ at photon energies up to 2.6-GeV}, Eur. Phys. J. A 19 (2004) 251--273.
\newblock \href {http://arxiv.org/abs/nucl-ex/0308025} {\path{arXiv:nucl-ex/0308025}}, \href {https://doi.org/10.1140/epja/i2003-10119-x} {\path{doi:10.1140/epja/i2003-10119-x}}.

\bibitem{Wu:2005wf}
C.~Wu, et~al., {Photoproduction of $\rho^0$ mesons and $\Delta$-baryons in the reaction $\gamma p \to p \pi^+ \pi^-$ at energies up to $\sqrt s = 2.6$-GeV}, Eur. Phys. J. A 23 (2005) 317--344.
\newblock \href {https://doi.org/10.1140/epja/i2004-10093-9} {\path{doi:10.1140/epja/i2004-10093-9}}.

\bibitem{Lawall:2005np}
R.~Lawall, et~al., {Measurement of the reaction $\gamma p \to K^0 \Sigma^+$ at photon energies up to 2.6-GeV}, Eur. Phys. J. A 24 (2005) 275--286.
\newblock \href {http://arxiv.org/abs/nucl-ex/0504014} {\path{arXiv:nucl-ex/0504014}}, \href {https://doi.org/10.1140/epja/i2005-10002-x} {\path{doi:10.1140/epja/i2005-10002-x}}.

\bibitem{Wieland:2010run}
F.~W. Wieland, et~al., {Measurement of the reaction $\gamma p \to K^ + \Lambda(1520)$ at photon energies up to 2.65 GeV}, Eur. Phys. J. A 47 (2011) 47, [Erratum: Eur.Phys.J.A 47, 133 (2011)].
\newblock \href {http://arxiv.org/abs/1011.0822} {\path{arXiv:1011.0822}}, \href {https://doi.org/10.1140/epja/i2011-11047-x} {\path{doi:10.1140/epja/i2011-11047-x}}.

\bibitem{CrystalBarrel:1992qav}
E.~Aker, et~al., {The Crystal Barrel spectrometer at LEAR}, Nucl. Instrum. Meth. A 321 (1992) 69--108.
\newblock \href {https://doi.org/10.1016/0168-9002(92)90379-I} {\path{doi:10.1016/0168-9002(92)90379-I}}.

\bibitem{Novotny:1991ht}
R.~Novotny, {The BaF-2 photon spectrometer TAPS}, IEEE Trans. Nucl. Sci. 38 (1991) 379--385.
\newblock \href {https://doi.org/10.1109/23.289329} {\path{doi:10.1109/23.289329}}.

\bibitem{CB-ELSA:2003rxy}
V.~Crede, et~al., {Photoproduction of eta mesons off protons for 0.75-GeV $< E_\gamma < 3$-GeV}, Phys. Rev. Lett. 94 (2005) 012004.
\newblock \href {http://arxiv.org/abs/hep-ex/0311045} {\path{arXiv:hep-ex/0311045}}, \href {https://doi.org/10.1103/PhysRevLett.94.012004} {\path{doi:10.1103/PhysRevLett.94.012004}}.

\bibitem{CB-ELSA:2004sqg}
O.~Bartholomy, et~al., {Neutral pion photoproduction off protons in the energy range 0.3-GeV $< E_\gamma < 3$-GeV}, Phys. Rev. Lett. 94 (2005) 012003.
\newblock \href {http://arxiv.org/abs/hep-ex/0407022} {\path{arXiv:hep-ex/0407022}}, \href {https://doi.org/10.1103/PhysRevLett.94.012003} {\path{doi:10.1103/PhysRevLett.94.012003}}.

\bibitem{CBELSATAPS:2005iwc}
D.~Trnka, et~al., {First observation of in-medium modifications of the $\omega$ meson}, Phys. Rev. Lett. 94 (2005) 192303.
\newblock \href {http://arxiv.org/abs/nucl-ex/0504010} {\path{arXiv:nucl-ex/0504010}}, \href {https://doi.org/10.1103/PhysRevLett.94.192303} {\path{doi:10.1103/PhysRevLett.94.192303}}.

\bibitem{CB-ELSA:2007htf}
O.~Bartholomy, et~al., {Photoproduction of $\eta$-mesons off protons}, Eur. Phys. J. A 33 (2007) 133--146.
\newblock \href {https://doi.org/10.1140/epja/i2007-10455-9} {\path{doi:10.1140/epja/i2007-10455-9}}.

\bibitem{CB-ELSA:2007wep}
H.~van Pee, et~al., {Photoproduction of $\pi^0$-mesons off protons from the $\Delta(1232)$ region to $E_\gamma = 3$-GeV}, Eur. Phys. J. A 31 (2007) 61--77.
\newblock \href {http://arxiv.org/abs/0704.1776} {\path{arXiv:0704.1776}}, \href {https://doi.org/10.1140/epja/i2006-10160-3} {\path{doi:10.1140/epja/i2006-10160-3}}.

\bibitem{CBELSATAPS:2007oqn}
R.~Castelijns, et~al., {Nucleon resonance decay by the $K^0 \Sigma^+$ channel}, Eur. Phys. J. A 35 (2008) 39--45.
\newblock \href {http://arxiv.org/abs/nucl-ex/0702033} {\path{arXiv:nucl-ex/0702033}}, \href {https://doi.org/10.1140/epja/i2007-10529-8} {\path{doi:10.1140/epja/i2007-10529-8}}.

\bibitem{CBELSA:2007vce}
D.~Elsner, et~al., {Measurement of the beam asymmetry in $\eta$-photoproduction off the proton}, Eur. Phys. J. A 33 (2007) 147--155.
\newblock \href {http://arxiv.org/abs/nucl-ex/0702032} {\path{arXiv:nucl-ex/0702032}}, \href {https://doi.org/10.1140/epja/i2007-10447-9} {\path{doi:10.1140/epja/i2007-10447-9}}.

\bibitem{Thoma:2007bm}
U.~Thoma, et~al., {$N^*$ and $\Delta^*$ decays into $N \pi^0 \pi^0$}, Phys. Lett. B 659 (2008) 87--93.
\newblock \href {http://arxiv.org/abs/0707.3592} {\path{arXiv:0707.3592}}, \href {https://doi.org/10.1016/j.physletb.2007.11.054} {\path{doi:10.1016/j.physletb.2007.11.054}}.

\bibitem{Sarantsev:2007aa}
A.~V. Sarantsev, et~al., {New results on the Roper resonance and the $P_{11}$ partial wave}, Phys. Lett. B 659 (2008) 94--100.
\newblock \href {http://arxiv.org/abs/0707.3591} {\path{arXiv:0707.3591}}, \href {https://doi.org/10.1016/j.physletb.2007.11.055} {\path{doi:10.1016/j.physletb.2007.11.055}}.

\bibitem{CB-ELSA:2007xbv}
I.~Horn, et~al., Evidence for a parity doublet \protect{$\Delta (1920) P_{33}$} and \protect{$\Delta (1940) D_{33}$} from $\gamma p \to p \pi^0 \eta$, Phys. Rev. Lett. 101 (2008) 202002.
\newblock \href {http://arxiv.org/abs/0711.1138} {\path{arXiv:0711.1138}}, \href {https://doi.org/10.1103/PhysRevLett.101.202002} {\path{doi:10.1103/PhysRevLett.101.202002}}.

\bibitem{CBELSA:2008wwq}
E.~Gutz, et~al., {Measurement of the beam asymmetry $\Sigma$ in $\pi\eta$ production off the proton with the CBELSA/TAPS experiment}, Eur. Phys. J. A 35 (2008) 291--293.
\newblock \href {https://doi.org/10.1140/epja/i2008-10566-9} {\path{doi:10.1140/epja/i2008-10566-9}}.

\bibitem{CBELSATAPS:2019ylw}
J.~M\"uller, et~al., {New data on $\vec{\gamma} \vec{p}\rightarrow \eta p$ with polarized photons and protons and their implications for $N^* \to N\eta$ decays}, Phys. Lett. B 803 (2020) 135323.
\newblock \href {http://arxiv.org/abs/1909.08464} {\path{arXiv:1909.08464}}, \href {https://doi.org/10.1016/j.physletb.2020.135323} {\path{doi:10.1016/j.physletb.2020.135323}}.

\bibitem{CBELSATAPS:2020cwk}
F.~Afzal, et~al., {Observation of the p\ensuremath{\eta}' Cusp in the New Precise Beam Asymmetry \ensuremath{\Sigma} Data for \ensuremath{\gamma}p\textrightarrow{}p\ensuremath{\eta}}, Phys. Rev. Lett. 125~(15) (2020) 152002.
\newblock \href {http://arxiv.org/abs/2009.06248} {\path{arXiv:2009.06248}}, \href {https://doi.org/10.1103/PhysRevLett.125.152002} {\path{doi:10.1103/PhysRevLett.125.152002}}.

\bibitem{CBELSATAPS:2021osa}
V.~Metag, et~al., {Observation of a structure in the $M_{p\eta }$ invariant mass distribution near 1700 ${MeV}/ {c}^2$ in the $\gamma {p \rightarrow p \pi ^0 \eta } $ reaction}, Eur. Phys. J. A 57~(12) (2021) 325.
\newblock \href {http://arxiv.org/abs/2110.05155} {\path{arXiv:2110.05155}}, \href {https://doi.org/10.1140/epja/s10050-021-00634-1} {\path{doi:10.1140/epja/s10050-021-00634-1}}.

\bibitem{CBELSATAPS:2022uad}
T.~Seifen, et~al., {Polarization observables in double neutral pion photoproduction}, Eur. Phys. J. A.\,\,Accepted for publication (2025).
\newblock \href {http://arxiv.org/abs/2207.01981} {\path{arXiv:2207.01981}}.

\bibitem{Ehrenberg:1972tg}
H.~Ehrenberg, et~al., {Die Elektronenstreu-Apparatur am Mainzer 300 MeV-Elektronen-Linearbeschleuniger}, Nucl. Instrum. Meth. 105 (1972) 253--263.
\newblock \href {https://doi.org/10.1016/0029-554X(72)90566-6} {\path{doi:10.1016/0029-554X(72)90566-6}}.

\bibitem{Herminghaus:1983nv}
H.~Herminghaus, et~al., {Status Report on the Normal Conducting cw Racetrack Microtron Cascade MAMI}, IEEE Trans. Nucl. Sci. 30 (1983) 3274--3278.
\newblock \href {https://doi.org/10.1109/TNS.1983.4336636} {\path{doi:10.1109/TNS.1983.4336636}}.

\bibitem{Euteneuer:1992qe}
H.~Euteneuer, et~al., {Experience with the 855-MeV RTM-Cascade MAMI}, Conf. Proc. C 920324 (1992) 418--420.

\bibitem{Kaiser:2008zza}
K.~H. Kaiser, et~al., {The 1.5-GeV harmonic double-sided microtron at Mainz University}, Nucl. Instrum. Meth. A 593 (2008) 159--170.
\newblock \href {https://doi.org/10.1016/j.nima.2008.05.018} {\path{doi:10.1016/j.nima.2008.05.018}}.

\bibitem{Chan:1978ck}
Y.~Chan, et~al., {Design and performance of a modularized NaI(Tl) detector (the Crystal Ball prototype)}, IEEE Trans. Nucl. Sci. 25 (1978) 333--339.
\newblock \href {https://doi.org/10.1109/TNS.1978.4329326} {\path{doi:10.1109/TNS.1978.4329326}}.

\bibitem{CrystalBallatMAMI:2008cye}
M.~Unverzagt, et~al., {Determination of the Dalitz plot parameter $\alpha$ for the decay $\eta \to 3 \pi^0$ with the Crystal Ball at MAMI-B}, Eur. Phys. J. A 39 (2009) 169--177.
\newblock \href {http://arxiv.org/abs/0812.3324} {\path{arXiv:0812.3324}}, \href {https://doi.org/10.1140/epja/i2008-10710-7} {\path{doi:10.1140/epja/i2008-10710-7}}.

\bibitem{CrystalBallatMAMI:2009lze}
V.~L. Kashevarov, et~al., {Photoproduction of $\pi^0 \eta$ on protons and the $\Delta(1700)D_{33}$ resonance}, Eur. Phys. J. A 42 (2009) 141--149.
\newblock \href {http://arxiv.org/abs/0901.3888} {\path{arXiv:0901.3888}}, \href {https://doi.org/10.1140/epja/i2009-10868-4} {\path{doi:10.1140/epja/i2009-10868-4}}.

\bibitem{Starostin:2009zz}
A.~Starostin, et~al., {Search for the charge-conjugation-forbidden decay $\omega \to \eta\pi^0$}, Phys. Rev. C 79 (2009) 065201.
\newblock \href {https://doi.org/10.1103/PhysRevC.79.065201} {\path{doi:10.1103/PhysRevC.79.065201}}.

\bibitem{CrystalBallatMAMI:2009iym}
D.~Krambrich, et~al., {Beam-Helicity Asymmetries in Double Pion Photoproduction off the Proton}, Phys. Rev. Lett. 103 (2009) 052002.
\newblock \href {http://arxiv.org/abs/0907.0358} {\path{arXiv:0907.0358}}, \href {https://doi.org/10.1103/PhysRevLett.103.052002} {\path{doi:10.1103/PhysRevLett.103.052002}}.

\bibitem{Schumann:2010js}
S.~Schumann, et~al., {Radiative $\pi^0$ photoproduction on protons in the $\Delta^+(1232)$ region}, Eur. Phys. J. A 43 (2010) 269--282.
\newblock \href {http://arxiv.org/abs/1001.3626} {\path{arXiv:1001.3626}}, \href {https://doi.org/10.1140/epja/i2010-10925-y} {\path{doi:10.1140/epja/i2010-10925-y}}.

\bibitem{CrystalBallatMAMI:2010slt}
E.~F. McNicoll, et~al., {Study of the $\gamma p \to \eta p$ reaction with the Crystal Ball detector at the Mainz Microtron (MAMI-C)}, Phys. Rev. C 82 (2010) 035208, [Erratum: Phys.Rev.C 84, 029901 (2011)].
\newblock \href {http://arxiv.org/abs/1007.0777} {\path{arXiv:1007.0777}}, \href {https://doi.org/10.1103/PhysRevC.84.029901} {\path{doi:10.1103/PhysRevC.84.029901}}.

\bibitem{Zehr:2012tj}
F.~Zehr, et~al., {Photoproduction of $\pi^0\pi^0$ and $\pi^0\pi^\pm$ pairs off the proton from threshold to the second resonance region}, Eur. Phys. J. A 48 (2012) 98.
\newblock \href {http://arxiv.org/abs/1207.2361} {\path{arXiv:1207.2361}}, \href {https://doi.org/10.1140/epja/i2012-12098-1} {\path{doi:10.1140/epja/i2012-12098-1}}.

\bibitem{A2:2012lnr}
D.~Hornidge, et~al., {Accurate Test of Chiral Dynamics in the $\gamma p\to \pi^0 p$ Reaction}, Phys. Rev. Lett. 111~(6) (2013) 062004.
\newblock \href {http://arxiv.org/abs/1211.5495} {\path{arXiv:1211.5495}}, \href {https://doi.org/10.1103/PhysRevLett.111.062004} {\path{doi:10.1103/PhysRevLett.111.062004}}.

\bibitem{Thiel:2013cea}
M.~Thiel, et~al., {In-medium modifications of the $\omega$ meson near the production threshold}, Eur. Phys. J. A 49 (2013) 132.
\newblock \href {https://doi.org/10.1140/epja/i2013-13132-6} {\path{doi:10.1140/epja/i2013-13132-6}}.

\bibitem{Robinson:2013sxa}
J.~Robinson, et~al., {Measurements of $^{12}C(\overline{\gamma},pp)$ photon asymmetries for $E_{\gamma}$ = 200-450 MeV}, Eur. Phys. J. A 49 (2013) 65.
\newblock \href {https://doi.org/10.1140/epja/i2013-13065-0} {\path{doi:10.1140/epja/i2013-13065-0}}.

\bibitem{CrystalBall:2013oow}
Y.~Maghrbi, et~al., {Coherent photoproduction of $\pi^0$- and $\eta$-mesons off $^{7}$Li}, Eur. Phys. J. A 49 (2013) 38.
\newblock \href {http://arxiv.org/abs/1303.2254} {\path{arXiv:1303.2254}}, \href {https://doi.org/10.1140/epja/i2013-13038-3} {\path{doi:10.1140/epja/i2013-13038-3}}.

\bibitem{Oberle:2013kvb}
M.~Oberle, et~al., {Measurement of the beam-helicity asymmetry $I^{\odot}$ in the photoproduction of $\pi^0$-pairs off the proton and off the neutron}, Phys. Lett. B 721 (2013) 237--243.
\newblock \href {http://arxiv.org/abs/1304.1919} {\path{arXiv:1304.1919}}, \href {https://doi.org/10.1016/j.physletb.2013.03.021} {\path{doi:10.1016/j.physletb.2013.03.021}}.

\bibitem{Maghrbi:2013xqd}
Y.~Maghrbi, et~al., {Double pion photoproduction off nuclei \textendash{} Are there effects beyond final-state interaction?}, Phys. Lett. B 722 (2013) 69--75.
\newblock \href {http://arxiv.org/abs/1304.1918} {\path{arXiv:1304.1918}}, \href {https://doi.org/10.1016/j.physletb.2013.04.014} {\path{doi:10.1016/j.physletb.2013.04.014}}.

\bibitem{A2:2013cqk}
P.~Aguar-Bartolome, et~al., {Measurement of the $\gamma p \to K^{0} \Sigma^{+}$ reaction with the Crystal Ball/TAPS detectors at the Mainz Microtron}, Phys. Rev. C 88~(4) (2013) 044601.
\newblock \href {http://arxiv.org/abs/1306.1243} {\path{arXiv:1306.1243}}, \href {https://doi.org/10.1103/PhysRevC.88.044601} {\path{doi:10.1103/PhysRevC.88.044601}}.

\bibitem{AguarBartolome:2013mga}
P.~Aguar~Bartolome, et~al., {First measurement of the helicity dependence of $^{3}$He photoreactions in the $\Delta$(1232) resonance region}, Phys. Lett. B 723 (2013) 71--77.
\newblock \href {https://doi.org/10.1016/j.physletb.2013.04.057} {\path{doi:10.1016/j.physletb.2013.04.057}}.

\bibitem{CrystalBallatMAMI:2013iig}
T.~C. Jude, et~al., {$K^+\Lambda$ and $K^+\Sigma^0$ photoproduction with fine center-of-mass energy resolution}, Phys. Lett. B 735 (2014) 112--118.
\newblock \href {http://arxiv.org/abs/1308.5659} {\path{arXiv:1308.5659}}, \href {https://doi.org/10.1016/j.physletb.2014.06.015} {\path{doi:10.1016/j.physletb.2014.06.015}}.

\bibitem{A2:2013wad}
P.~Aguar-Bartolome, et~al., {New determination of the $\eta$ transition form factor in the Dalitz decay $\eta \to e^+e^{-} \gamma$ with the Crystal Ball/TAPS detectors at the Mainz Microtron}, Phys. Rev. C 89~(4) (2014) 044608.
\newblock \href {http://arxiv.org/abs/1309.5648} {\path{arXiv:1309.5648}}, \href {https://doi.org/10.1103/PhysRevC.89.044608} {\path{doi:10.1103/PhysRevC.89.044608}}.

\bibitem{Sikora:2013vfa}
M.~H. Sikora, et~al., {Measurement of the $^{1}H$($\vec{\gamma}$, $\vec{p}$)$\pi^{0}$ reaction using a novel nucleon spin polarimeter}, Phys. Rev. Lett. 112~(2) (2014) 022501.
\newblock \href {http://arxiv.org/abs/1309.7897} {\path{arXiv:1309.7897}}, \href {https://doi.org/10.1103/PhysRevLett.112.022501} {\path{doi:10.1103/PhysRevLett.112.022501}}.

\bibitem{A2:2013tbo}
D.~Werthm\"uller, et~al., {Narrow Structure in the Excitation Function of \ensuremath{\eta} Photoproduction off the Neutron}, Phys. Rev. Lett. 111~(23) (2013) 232001.
\newblock \href {http://arxiv.org/abs/1311.2781} {\path{arXiv:1311.2781}}, \href {https://doi.org/10.1103/PhysRevLett.111.232001} {\path{doi:10.1103/PhysRevLett.111.232001}}.

\bibitem{A2:2014bam}
S.~Costanza, et~al., {Helicity dependence of the $\gamma^3He \to \pi X$ reactions in the $\Delta(1232)$ resonance region}, Eur. Phys. J. A 50~(11) (2014) 173.
\newblock \href {https://doi.org/10.1140/epja/i2014-14173-y} {\path{doi:10.1140/epja/i2014-14173-y}}.

\bibitem{A2:2015mqi}
M.~Martemianov, et~al., {A new measurement of the neutron detection efficiency for the NaI Crystal Ball detector}, JINST 10~(04) (2015) T04001.
\newblock \href {http://arxiv.org/abs/1502.07317} {\path{arXiv:1502.07317}}, \href {https://doi.org/10.1088/1748-0221/10/04/T04001} {\path{doi:10.1088/1748-0221/10/04/T04001}}.

\bibitem{A2:2015ocx}
J.~R.~M. Annand, et~al., {First measurement of target and beam-target asymmetries in the $\gamma p \to \pi^0 \eta p$ reaction}, Phys. Rev. C 91~(5) (2015) 055208.
\newblock \href {http://arxiv.org/abs/1505.07963} {\path{arXiv:1505.07963}}, \href {https://doi.org/10.1103/PhysRevC.91.055208} {\path{doi:10.1103/PhysRevC.91.055208}}.

\bibitem{A2:2015mhs}
P.~Adlarson, et~al., {Measurement of $\pi^0$ photoproduction on the proton at MAMI C}, Phys. Rev. C 92~(2) (2015) 024617.
\newblock \href {http://arxiv.org/abs/1506.08849} {\path{arXiv:1506.08849}}, \href {https://doi.org/10.1103/PhysRevC.92.024617} {\path{doi:10.1103/PhysRevC.92.024617}}.

\bibitem{MAINZ-A2:2016iua}
S.~Gardner, et~al., {Photon asymmetry measurements of $\overrightarrow{\gamma}p \rightarrow \pi^{0} p$ for $E_{\gamma}=$ 320-650 MeV}, Eur. Phys. J. A 52~(11) (2016) 333.
\newblock \href {http://arxiv.org/abs/1606.07930} {\path{arXiv:1606.07930}}, \href {https://doi.org/10.1140/epja/i2016-16333-5} {\path{doi:10.1140/epja/i2016-16333-5}}.

\bibitem{A2:2016vzp}
A.~K\"aser, et~al., {Photoproduction of $\eta$ $\pi$ pairs off nucleons and deuterons}, Eur. Phys. J. A 52~(9) (2016) 272.
\newblock \href {http://arxiv.org/abs/1608.07475} {\path{arXiv:1608.07475}}, \href {https://doi.org/10.1140/epja/i2016-16272-1} {\path{doi:10.1140/epja/i2016-16272-1}}.

\bibitem{Adlarson:2016hpp}
P.~Adlarson, et~al., {Measurement of the $\omega \to \pi^0 e^+e^-$ and $\eta \to e^+e^- \gamma$ Dalitz decays with the A2 setup at MAMI}, Phys. Rev. C 95~(3) (2017) 035208.
\newblock \href {http://arxiv.org/abs/1609.04503} {\path{arXiv:1609.04503}}, \href {https://doi.org/10.1103/PhysRevC.95.035208} {\path{doi:10.1103/PhysRevC.95.035208}}.

\bibitem{A2:2016bij}
L.~Witthauer, et~al., {Insight into the Narrow Structure in \ensuremath{\eta} Photoproduction on the Neutron from Helicity-Dependent Cross Sections}, Phys. Rev. Lett. 117~(13) (2016) 132502.
\newblock \href {http://arxiv.org/abs/1702.01408} {\path{arXiv:1702.01408}}, \href {https://doi.org/10.1103/PhysRevLett.117.132502} {\path{doi:10.1103/PhysRevLett.117.132502}}.

\bibitem{A2:2016nio}
V.~Sokhoyan, et~al., {Determination of the scalar polarizabilities of the proton using beam asymmetry $\Sigma_{3}$ in Compton scattering}, Eur. Phys. J. A 53~(1) (2017) 14.
\newblock \href {http://arxiv.org/abs/1611.03769} {\path{arXiv:1611.03769}}, \href {https://doi.org/10.1140/epja/i2017-12203-0} {\path{doi:10.1140/epja/i2017-12203-0}}.

\bibitem{A2:2016sjm}
P.~Adlarson, et~al., {Measurement of the $\pi^{0}\to e^{+}e^{-}\gamma$ Dalitz decay at the Mainz Microtron}, Phys. Rev. C 95~(2) (2017) 025202.
\newblock \href {http://arxiv.org/abs/1611.04739} {\path{arXiv:1611.04739}}, \href {https://doi.org/10.1103/PhysRevC.95.025202} {\path{doi:10.1103/PhysRevC.95.025202}}.

\bibitem{A2:2017gwp}
V.~L. Kashevarov, et~al., {Study of \ensuremath{\eta} and \ensuremath{\eta}' Photoproduction at MAMI}, Phys. Rev. Lett. 118~(21) (2017) 212001.
\newblock \href {http://arxiv.org/abs/1701.04809} {\path{arXiv:1701.04809}}, \href {https://doi.org/10.1103/PhysRevLett.118.212001} {\path{doi:10.1103/PhysRevLett.118.212001}}.

\bibitem{A2:2017auj}
L.~Witthauer, et~al., {Helicity-dependent cross sections and double-polarization observable E in \ensuremath{\eta} photoproduction from quasifree protons and neutrons}, Phys. Rev. C 95~(5) (2017) 055201.
\newblock \href {http://arxiv.org/abs/1704.00649} {\path{arXiv:1704.00649}}, \href {https://doi.org/10.1103/PhysRevC.95.055201} {\path{doi:10.1103/PhysRevC.95.055201}}.

\bibitem{A2:2018doh}
C.~S. Akondi, et~al., {Experimental study of the $\gamma p\rightarrow K^0\Sigma^+$, $\gamma n\rightarrow K^0\Lambda$, and $\gamma n\rightarrow K^0 \Sigma^0$ reactions at the Mainz Microtron}, Eur. Phys. J. A 55~(11) (2019) 202.
\newblock \href {http://arxiv.org/abs/1811.05547} {\path{arXiv:1811.05547}}, \href {https://doi.org/10.1140/epja/i2019-12924-x} {\path{doi:10.1140/epja/i2019-12924-x}}.

\bibitem{A2:2019fvn}
V.~Sokhoyan, et~al., {Measurement of the beam-helicity asymmetry in photoproduction of $\pi^{0}\eta$ pairs on carbon, aluminum, and lead}, Phys. Lett. B 802 (2020) 135243.
\newblock \href {http://arxiv.org/abs/1907.00232} {\path{arXiv:1907.00232}}, \href {https://doi.org/10.1016/j.physletb.2020.135243} {\path{doi:10.1016/j.physletb.2020.135243}}.

\bibitem{A2:2019yud}
W.~J. Briscoe, et~al., {Cross section for $\gamma n \to \pi^0 n$ at the Mainz A2 experiment}, Phys. Rev. C 100~(6) (2019) 065205.
\newblock \href {http://arxiv.org/abs/1908.02730} {\path{arXiv:1908.02730}}, \href {https://doi.org/10.1103/PhysRevC.100.065205} {\path{doi:10.1103/PhysRevC.100.065205}}.

\bibitem{A2:2019bqm}
D.~Paudyal, et~al., {Extracting the spin polarizabilities of the proton by measurement of Compton double-polarization observables}, Phys. Rev. C 102~(3) (2020) 035205.
\newblock \href {http://arxiv.org/abs/1909.02032} {\path{arXiv:1909.02032}}, \href {https://doi.org/10.1103/PhysRevC.102.035205} {\path{doi:10.1103/PhysRevC.102.035205}}.

\bibitem{A2:2019arr}
M.~Bashkanov, et~al., {Signatures of the $d^*(2380)$ Hexaquark in d($\gamma$,$p\vec{n}$)}, Phys. Rev. Lett. 124~(13) (2020) 132001.
\newblock \href {http://arxiv.org/abs/1911.08309} {\path{arXiv:1911.08309}}, \href {https://doi.org/10.1103/PhysRevLett.124.132001} {\path{doi:10.1103/PhysRevLett.124.132001}}.

\bibitem{Dieterle:2020vug}
M.~Dieterle, et~al., {Helicity-Dependent Cross Sections for the Photoproduction of \ensuremath{\pi}$^0$ Pairs from Nucleons}, Phys. Rev. Lett. 125~(6) (2020) 062001.
\newblock \href {http://arxiv.org/abs/2007.06079} {\path{arXiv:2007.06079}}, \href {https://doi.org/10.1103/PhysRevLett.125.062001} {\path{doi:10.1103/PhysRevLett.125.062001}}.

\bibitem{A2CollaborationatMAMI:2021vfy}
E.~Mornacchi, et~al., {Measurement of Compton Scattering at MAMI for the Extraction of the Electric and Magnetic Polarizabilities of the Proton}, Phys. Rev. Lett. 128~(13) (2022) 132503.
\newblock \href {http://arxiv.org/abs/2110.15691} {\path{arXiv:2110.15691}}, \href {https://doi.org/10.1103/PhysRevLett.128.132503} {\path{doi:10.1103/PhysRevLett.128.132503}}.

\bibitem{A2:2022ycp}
F.~Cividini, et~al., {Measurement of the helicity dependence for single $\pi ^{0}$ photoproduction from the deuteron}, Eur. Phys. J. A 58~(6) (2022) 113.
\newblock \href {http://arxiv.org/abs/2203.00535} {\path{arXiv:2203.00535}}, \href {https://doi.org/10.1140/epja/s10050-022-00760-4} {\path{doi:10.1140/epja/s10050-022-00760-4}}.

\bibitem{A2:2022kkx}
M.~Bashkanov, et~al., {First measurement of polarisation transfer Cx'n in deuteron photodisintegration}, Phys. Lett. B 844 (2023) 138080.
\newblock \href {http://arxiv.org/abs/2206.12299} {\path{arXiv:2206.12299}}, \href {https://doi.org/10.1016/j.physletb.2023.138080} {\path{doi:10.1016/j.physletb.2023.138080}}.

\bibitem{A2:2022ipx}
S.~Garni, et~al., {Target and beam-target asymmetries for the $\gamma p \to \pi^0 \pi^0 p$ reaction} (7 2022).
\newblock \href {http://arxiv.org/abs/2207.14079} {\path{arXiv:2207.14079}}.

\bibitem{Mornacchi:2023oir}
E.~Mornacchi, et~al., {Evaluation of the E2/M1 ratio in the $N\to \Delta(1232)$ transition from the $ \vec{\gamma} \vec{p} \to p \pi^0 $ reaction}, Phys. Rev. C 109~(5) (2024) 055201.
\newblock \href {http://arxiv.org/abs/2312.08211} {\path{arXiv:2312.08211}}.

\bibitem{CLAS:2003umf}
B.~A. Mecking, et~al., {The CEBAF Large Acceptance Spectrometer (CLAS)}, Nucl. Instrum. Meth. A 503 (2003) 513--553.
\newblock \href {https://doi.org/10.1016/S0168-9002(03)01001-5} {\path{doi:10.1016/S0168-9002(03)01001-5}}.

\bibitem{CLAS:2023ddn}
C.~W. Kim, et~al., {Measurement of the helicity asymmetry ${\mathbb {E}}$ for the $\vec {\gamma }\vec {p} \rightarrow p \pi ^0$ reaction in the resonance region: The~CLAS~Collaboration}, Eur. Phys. J. A 59~(9) (2023) 217.
\newblock \href {http://arxiv.org/abs/2305.08616} {\path{arXiv:2305.08616}}, \href {https://doi.org/10.1140/epja/s10050-023-01123-3} {\path{doi:10.1140/epja/s10050-023-01123-3}}.

\bibitem{CLAS:2021jhm}
S.~Moran, et~al., {Measurement of charged-pion production in deep-inelastic scattering off nuclei with the CLAS detector}, Phys. Rev. C 105~(1) (2022) 015201.
\newblock \href {http://arxiv.org/abs/2109.09951} {\path{arXiv:2109.09951}}, \href {https://doi.org/10.1103/PhysRevC.105.015201} {\path{doi:10.1103/PhysRevC.105.015201}}.

\bibitem{CLAS:2021hex}
N.~Zachariou, et~al., {Beam-spin asymmetry $\boldsymbol{\Sigma}$ for $\Sigma^-$ hyperon photoproduction off the neutron}, Phys. Lett. B 827 (2022) 136985.
\newblock \href {http://arxiv.org/abs/2106.13957} {\path{arXiv:2106.13957}}, \href {https://doi.org/10.1016/j.physletb.2022.136985} {\path{doi:10.1016/j.physletb.2022.136985}}.

\bibitem{CLAS:2021osv}
U.~Shrestha, et~al., {Differential cross sections for $\Lambda(1520)$ using photoproduction at CLAS}, Phys. Rev. C 103~(2) (2021) 025206.
\newblock \href {http://arxiv.org/abs/2101.06134} {\path{arXiv:2101.06134}}, \href {https://doi.org/10.1103/PhysRevC.103.025206} {\path{doi:10.1103/PhysRevC.103.025206}}.

\bibitem{CLAS:2020ngl}
M.~Carver, et~al., {Photoproduction of the $f_2(1270)$ meson using the CLAS detector}, Phys. Rev. Lett. 126~(8) (2021) 082002.
\newblock \href {http://arxiv.org/abs/2010.16006} {\path{arXiv:2010.16006}}, \href {https://doi.org/10.1103/PhysRevLett.126.082002} {\path{doi:10.1103/PhysRevLett.126.082002}}.

\bibitem{CLAS:2020rdz}
A.~Celentano, et~al., {First measurement of direct photoproduction of the $a_2(1320)^0$ meson on the proton}, Phys. Rev. C 102~(3) (2020) 032201.
\newblock \href {http://arxiv.org/abs/2004.05359} {\path{arXiv:2004.05359}}, \href {https://doi.org/10.1103/PhysRevC.102.032201} {\path{doi:10.1103/PhysRevC.102.032201}}.

\bibitem{CLAS:2020spy}
N.~Zachariou, et~al., {Beam target helicity asymmetry $E$ in $K^+$ $\Sigma^-$ photoproduction on the neutron}, Phys. Lett. B 808 (2020) 135662.
\newblock \href {https://doi.org/10.1016/j.physletb.2020.135662} {\path{doi:10.1016/j.physletb.2020.135662}}.

\bibitem{Mokeev:2020hhu}
V.~I. Mokeev, et~al., {Evidence for the $N'(1720)3/2^+$ Nucleon Resonance from Combined Studies of CLAS $\pi^+\pi^-p$ Photo- and Electroproduction Data}, Phys. Lett. B 805 (2020) 135457.
\newblock \href {http://arxiv.org/abs/2004.13531} {\path{arXiv:2004.13531}}, \href {https://doi.org/10.1016/j.physletb.2020.135457} {\path{doi:10.1016/j.physletb.2020.135457}}.

\bibitem{CLAS:2018mmb}
P.~Roy, et~al., {First Measurements of the Double-Polarization Observables F , P , and H in \ensuremath{\omega} Photoproduction off Transversely Polarized Protons in the N* Resonance Region}, Phys. Rev. Lett. 122~(16) (2019) 162301.
\newblock \href {http://arxiv.org/abs/1812.02106} {\path{arXiv:1812.02106}}, \href {https://doi.org/10.1103/PhysRevLett.122.162301} {\path{doi:10.1103/PhysRevLett.122.162301}}.

\bibitem{CLAS:2018kvn}
J.~T. Goetz, et~al., {Study of $\Xi^*$ Photoproduction from Threshold to $W = 3.3$ GeV}, Phys. Rev. C 98~(6) (2018) 062201.
\newblock \href {http://arxiv.org/abs/1809.00074} {\path{arXiv:1809.00074}}, \href {https://doi.org/10.1103/PhysRevC.98.062201} {\path{doi:10.1103/PhysRevC.98.062201}}.

\bibitem{CLAS:2018azo}
S.~Lombardo, et~al., {Photoproduction of $K^+K^-$ meson pairs on the proton}, Phys. Rev. D 98~(5) (2018) 052009.
\newblock \href {http://arxiv.org/abs/1808.01918} {\path{arXiv:1808.01918}}, \href {https://doi.org/10.1103/PhysRevD.98.052009} {\path{doi:10.1103/PhysRevD.98.052009}}.

\bibitem{CLAS:2018avi}
T.~Chetry, et~al., {Differential cross section for $\gamma d \rightarrow \omega d$ using CLAS at Jefferson Lab}, Phys. Lett. B 782 (2018) 646--651.
\newblock \href {http://arxiv.org/abs/1802.06746} {\path{arXiv:1802.06746}}, \href {https://doi.org/10.1016/j.physletb.2018.06.003} {\path{doi:10.1016/j.physletb.2018.06.003}}.

\bibitem{CLAS:2018drk}
E.~Golovatch, et~al., {First results on nucleon resonance photocouplings from the $\gamma p \to \pi^+\pi^-p$ reaction}, Phys. Lett. B 788 (2019) 371--379.
\newblock \href {http://arxiv.org/abs/1806.01767} {\path{arXiv:1806.01767}}, \href {https://doi.org/10.1016/j.physletb.2018.10.013} {\path{doi:10.1016/j.physletb.2018.10.013}}.

\bibitem{CLAS:2018gxz}
D.~H. Ho, et~al., {Beam-target helicity asymmetry $E$ in $K^{0}\Lambda$ and $K^{0}\Sigma^0$ photoproduction on the neutron}, Phys. Rev. C 98~(4) (2018) 045205.
\newblock \href {http://arxiv.org/abs/1805.04561} {\path{arXiv:1805.04561}}, \href {https://doi.org/10.1103/PhysRevC.98.045205} {\path{doi:10.1103/PhysRevC.98.045205}}.

\bibitem{CLAS:2018xbd}
J.~Bono, et~al., {First measurement of $\Xi^-$ polarization in photoproduction}, Phys. Lett. B 783 (2018) 280--286.
\newblock \href {http://arxiv.org/abs/1804.04564} {\path{arXiv:1804.04564}}, \href {https://doi.org/10.1016/j.physletb.2018.07.004} {\path{doi:10.1016/j.physletb.2018.07.004}}.

\bibitem{Anisovich:2018yoo}
A.~V. Anisovich, et~al., {Proton-$\eta^\prime$ interactions at threshold}, Phys. Lett. B 785 (2018) 626--630.
\newblock \href {http://arxiv.org/abs/1803.06814} {\path{arXiv:1803.06814}}, \href {https://doi.org/10.1016/j.physletb.2018.06.034} {\path{doi:10.1016/j.physletb.2018.06.034}}.

\bibitem{CLAS:2017kyf}
M.~C. Kunkel, et~al., {Exclusive photoproduction of $\pi^0$ up to large values of Mandelstam variables $s, t$ and $u$ with CLAS}, Phys. Rev. C 98~(1) (2018) 015207.
\newblock \href {http://arxiv.org/abs/1712.10314} {\path{arXiv:1712.10314}}, \href {https://doi.org/10.1103/PhysRevC.98.015207} {\path{doi:10.1103/PhysRevC.98.015207}}.

\bibitem{CLAS:2017vxx}
S.~Chandavar, et~al., {Double $K_S^0$ Photoproduction off the Proton at CLAS}, Phys. Rev. C 97~(2) (2018) 025203.
\newblock \href {http://arxiv.org/abs/1712.02184} {\path{arXiv:1712.02184}}, \href {https://doi.org/10.1103/PhysRevC.97.025203} {\path{doi:10.1103/PhysRevC.97.025203}}.

\bibitem{CLAS:2017jrx}
P.~Roy, et~al., {Measurement of the beam asymmetry $\Sigma$ and the target asymmetry $T$ in the photoproduction of $\omega$ mesons off the proton using CLAS at Jefferson Laboratory}, Phys. Rev. C 97~(5) (2018) 055202.
\newblock \href {http://arxiv.org/abs/1711.05176} {\path{arXiv:1711.05176}}, \href {https://doi.org/10.1103/PhysRevC.97.055202} {\path{doi:10.1103/PhysRevC.97.055202}}.

\bibitem{CLAS:2017sgi}
A.~V. Anisovich, et~al., {Differential cross sections and polarization observables from CLAS $K$* photoproduction and the search for new $N$* states}, Phys. Lett. B 771 (2017) 142--150.
\newblock \href {https://doi.org/10.1016/j.physletb.2017.05.029} {\path{doi:10.1016/j.physletb.2017.05.029}}.

\bibitem{Anisovich:2017pox}
A.~V. Anisovich, et~al., {$N^*\to N \eta^\prime$ decays from photoproduction of $\eta^\prime$-mesons off protons}, Phys. Lett. B 772 (2017) 247--252.
\newblock \href {http://arxiv.org/abs/1706.05144} {\path{arXiv:1706.05144}}, \href {https://doi.org/10.1016/j.physletb.2017.06.052} {\path{doi:10.1016/j.physletb.2017.06.052}}.

\bibitem{Anisovich:2015gia}
A.~V. Anisovich, et~al., {Evidence for $\Delta(2200)7/2^-$ from photoproduction and consequence for chiral-symmetry restoration at high mass}, Phys. Lett. B 766 (2017) 357--361.
\newblock \href {http://arxiv.org/abs/1503.05774} {\path{arXiv:1503.05774}}, \href {https://doi.org/10.1016/j.physletb.2016.12.014} {\path{doi:10.1016/j.physletb.2016.12.014}}.

\bibitem{Anisovich:2017ygb}
A.~V. Anisovich, et~al., {$N^*$ resonances from $K\Lambda$ amplitudes in sliced bins in energy}, Eur. Phys. J. A 53~(12) (2017) 242.
\newblock \href {http://arxiv.org/abs/1712.07537} {\path{arXiv:1712.07537}}, \href {https://doi.org/10.1140/epja/i2017-12443-x} {\path{doi:10.1140/epja/i2017-12443-x}}.

\bibitem{Anisovich:2017xqg}
A.~V. Anisovich, V.~Burkert, E.~Klempt, V.~A. Nikonov, A.~V. Sarantsev, U.~Thoma, {Scrutinizing the evidence for N(1685)}, Phys. Rev. C 95~(3) (2017) 035211.
\newblock \href {http://arxiv.org/abs/1701.06387} {\path{arXiv:1701.06387}}, \href {https://doi.org/10.1103/PhysRevC.95.035211} {\path{doi:10.1103/PhysRevC.95.035211}}.

\bibitem{CLAS:2017gsu}
N.~Compton, et~al., {Measurement of the differential and total cross sections of the ${\gamma}d{\rightarrow}{K}^{0}\mathrm{{\Lambda}}(p)$ reaction within the resonance region}, Phys. Rev. C 96~(6) (2017) 065201.
\newblock \href {http://arxiv.org/abs/1706.04748} {\path{arXiv:1706.04748}}, \href {https://doi.org/10.1103/PhysRevC.96.065201} {\path{doi:10.1103/PhysRevC.96.065201}}.

\bibitem{CLAS:2017dco}
P.~T. Mattione, et~al., {Differential cross section measurements for $\gamma n\rightarrow{\pi}^{-}p$ above the first nucleon resonance region}, Phys. Rev. C 96~(3) (2017) 035204.
\newblock \href {http://arxiv.org/abs/1706.01963} {\path{arXiv:1706.01963}}, \href {https://doi.org/10.1103/PhysRevC.96.035204} {\path{doi:10.1103/PhysRevC.96.035204}}.

\bibitem{CLAS:2017kua}
D.~Ho, et~al., {Beam-Target Helicity Asymmetry for $\vec{\gamma} \vec{n} \rightarrow \pi^- p$ in the $N^*$ Resonance Region}, Phys. Rev. Lett. 118~(24) (2017) 242002.
\newblock \href {http://arxiv.org/abs/1705.04713} {\path{arXiv:1705.04713}}, \href {https://doi.org/10.1103/PhysRevLett.118.242002} {\path{doi:10.1103/PhysRevLett.118.242002}}.

\bibitem{CLAS:2017yjv}
Z.~Akbar, et~al., {Measurement of the helicity asymmetry $E$ in $\omega\to\pi^+\pi^-\pi^0$ photoproduction}, Phys. Rev. C 96~(6) (2017) 065209.
\newblock \href {http://arxiv.org/abs/1708.02608} {\path{arXiv:1708.02608}}, \href {https://doi.org/10.1103/PhysRevC.96.065209} {\path{doi:10.1103/PhysRevC.96.065209}}.

\bibitem{CLAS:2017rxe}
P.~Collins, et~al., {Photon beam asymmetry $\Sigma$ for $\eta$ and $\eta^\prime$ photoproduction from the proton}, Phys. Lett. B 771 (2017) 213--221.
\newblock \href {http://arxiv.org/abs/1703.00433} {\path{arXiv:1703.00433}}, \href {https://doi.org/10.1016/j.physletb.2017.05.045} {\path{doi:10.1016/j.physletb.2017.05.045}}.

\bibitem{CLAS:2016zjy}
R.~Dickson, et~al., {Photoproduction of the $f_1(1285)$ Meson}, Phys. Rev. C 93~(6) (2016) 065202.
\newblock \href {http://arxiv.org/abs/1604.07425} {\path{arXiv:1604.07425}}, \href {https://doi.org/10.1103/PhysRevC.93.065202} {\path{doi:10.1103/PhysRevC.93.065202}}.

\bibitem{CLAS:2016wrl}
C.~A. Paterson, et~al., {Photoproduction of $\Lambda$ and $\Sigma^0$ hyperons using linearly polarized photons}, Phys. Rev. C 93~(6) (2016) 065201.
\newblock \href {http://arxiv.org/abs/1603.06492} {\path{arXiv:1603.06492}}, \href {https://doi.org/10.1103/PhysRevC.93.065201} {\path{doi:10.1103/PhysRevC.93.065201}}.

\bibitem{CLAS:2015pjm}
I.~Senderovich, et~al., {First measurement of the helicity asymmetry $E$ in $\eta$ photoproduction on the proton}, Phys. Lett. B 755 (2016) 64--69.
\newblock \href {http://arxiv.org/abs/1507.00325} {\path{arXiv:1507.00325}}, \href {https://doi.org/10.1016/j.physletb.2016.01.044} {\path{doi:10.1016/j.physletb.2016.01.044}}.

\bibitem{CLAS:2015ykk}
S.~Strauch, et~al., {First Measurement of the Polarization Observable E in the $\vec p(\vec \gamma,\pi^+)n$ Reaction up to 2.25 GeV}, Phys. Lett. B 750 (2015) 53--58.
\newblock \href {http://arxiv.org/abs/1503.05163} {\path{arXiv:1503.05163}}, \href {https://doi.org/10.1016/j.physletb.2015.08.053} {\path{doi:10.1016/j.physletb.2015.08.053}}.

\bibitem{CLAS:2014tbc}
K.~Moriya, et~al., {Spin and parity measurement of the $\Lambda(1405)$ baryon}, Phys. Rev. Lett. 112~(8) (2014) 082004.
\newblock \href {http://arxiv.org/abs/1402.2296} {\path{arXiv:1402.2296}}, \href {https://doi.org/10.1103/PhysRevLett.112.082004} {\path{doi:10.1103/PhysRevLett.112.082004}}.

\bibitem{Anisovich:2014yza}
A.~V. Anisovich, et~al., {Energy-independent PWA of the reaction $\gamma p\to K^+\Lambda$}, Eur. Phys. J. A 50 (2014) 129.
\newblock \href {http://arxiv.org/abs/1404.4587} {\path{arXiv:1404.4587}}, \href {https://doi.org/10.1140/epja/i2014-14129-3} {\path{doi:10.1140/epja/i2014-14129-3}}.

\bibitem{CLAS:2013pcs}
M.~Dugger, et~al., {Beam asymmetry $\Sigma$ for $\pi^+$ and $\pi^0$ photoproduction on the proton for photon energies from 1.102 to 1.862 GeV}, Phys. Rev. C 88~(6) (2013) 065203, [Addendum: Phys.Rev.C 89, 029901 (2014)].
\newblock \href {http://arxiv.org/abs/1308.4028} {\path{arXiv:1308.4028}}, \href {https://doi.org/10.1103/PhysRevC.88.065203} {\path{doi:10.1103/PhysRevC.88.065203}}.

\bibitem{CLAS:2013jlg}
H.~Seraydaryan, et~al., {$\phi$-meson photoproduction on Hydrogen in the neutral decay mode}, Phys. Rev. C 89~(5) (2014) 055206.
\newblock \href {http://arxiv.org/abs/1308.1363} {\path{arXiv:1308.1363}}, \href {https://doi.org/10.1103/PhysRevC.89.055206} {\path{doi:10.1103/PhysRevC.89.055206}}.

\bibitem{CLAS:2013rxx}
K.~Moriya, et~al., {Differential Photoproduction Cross Sections of the $\Sigma^0(1385)$, $\Lambda(1405)$, and $\Lambda(1520)$}, Phys. Rev. C 88 (2013) 045201, [Addendum: Phys.Rev.C 88, 049902 (2013)].
\newblock \href {http://arxiv.org/abs/1305.6776} {\path{arXiv:1305.6776}}, \href {https://doi.org/10.1103/PhysRevC.88.045201} {\path{doi:10.1103/PhysRevC.88.045201}}.

\bibitem{CLAS:2013qgi}
W.~Tang, et~al., {Cross sections for the $\gamma p \to K^{*+}\Lambda$ and $\gamma p \to K^{*+}\Sigma^0$ reactions measured at CLAS}, Phys. Rev. C 87~(6) (2013) 065204.
\newblock \href {http://arxiv.org/abs/1303.2615} {\path{arXiv:1303.2615}}, \href {https://doi.org/10.1103/PhysRevC.87.065204} {\path{doi:10.1103/PhysRevC.87.065204}}.

\bibitem{CLAS:2013owj}
C.~S. Nepali, et~al., {Transverse polarization of $\Sigma^+$(1189) in photoproduction on a hydrogen target in CLAS}, Phys. Rev. C 87~(4) (2013) 045206.
\newblock \href {http://arxiv.org/abs/1302.0322} {\path{arXiv:1302.0322}}, \href {https://doi.org/10.1103/PhysRevC.87.045206} {\path{doi:10.1103/PhysRevC.87.045206}}.

\bibitem{Anisovich:2013jya}
A.~V. Anisovich, et~al., {Helicity amplitudes for photoexcitation of nucleon resonances off neutrons}, Eur. Phys. J. A 49 (2013) 67.
\newblock \href {http://arxiv.org/abs/1304.2177} {\path{arXiv:1304.2177}}, \href {https://doi.org/10.1140/epja/i2013-13067-x} {\path{doi:10.1140/epja/i2013-13067-x}}.

\bibitem{CLAS:2013rjt}
K.~Moriya, et~al., {Measurement of the \ensuremath{\Sigma}\ensuremath{\pi} photoproduction line shapes near the \ensuremath{\Lambda}(1405)}, Phys. Rev. C 87~(3) (2013) 035206.
\newblock \href {http://arxiv.org/abs/1301.5000} {\path{arXiv:1301.5000}}, \href {https://doi.org/10.1103/PhysRevC.87.035206} {\path{doi:10.1103/PhysRevC.87.035206}}.

\bibitem{Afanasev:2012fh}
A.~Afanasev, et~al., {Photoproduction of the Very Strangest Baryons on a Proton Target in CLAS12}, unpublished note (5 2012).

\bibitem{Price:2004hr}
J.~W. Price, J.~Ducote, J.~Goetz, B.~M.~K. Nefkens, {Photoproduction of the doubly strange $\Xi$ hyperons}, Nucl. Phys. A 754 (2005) 272--280.
\newblock \href {http://arxiv.org/abs/nucl-ex/0402006} {\path{arXiv:nucl-ex/0402006}}, \href {https://doi.org/10.1016/j.nuclphysa.2005.02.075} {\path{doi:10.1016/j.nuclphysa.2005.02.075}}.

\bibitem{CLAS:2011aa}
H.~Egiyan, et~al., {Upper limits for the photoproduction cross section for the $\Phi^{--}(1860)$ pentaquark state off the deuteron}, Phys. Rev. C 85 (2012) 015205.
\newblock \href {http://arxiv.org/abs/1109.1238} {\path{arXiv:1109.1238}}, \href {https://doi.org/10.1103/PhysRevC.85.015205} {\path{doi:10.1103/PhysRevC.85.015205}}.

\bibitem{CLAS:2010gcv}
R.~Nasseripour, et~al., {Coherent Photoproduction of $\pi^{+}$ from $^3$He}, Phys. Rev. C 83 (2011) 034001.
\newblock \href {http://arxiv.org/abs/1011.0073} {\path{arXiv:1011.0073}}, \href {https://doi.org/10.1103/PhysRevC.83.034001} {\path{doi:10.1103/PhysRevC.83.034001}}.

\bibitem{CLAS:2010aen}
B.~Dey, et~al., {Differential cross sections and recoil polarizations for the reaction $\gamma p \to K^{+} \Sigma^{0}$}, Phys. Rev. C 82 (2010) 025202.
\newblock \href {http://arxiv.org/abs/1006.0374} {\path{arXiv:1006.0374}}, \href {https://doi.org/10.1103/PhysRevC.82.025202} {\path{doi:10.1103/PhysRevC.82.025202}}.

\bibitem{Qian:2010rr}
X.~Qian, et~al., {Near-threshold Photoproduction of Phi Mesons from Deuterium}, Phys. Lett. B 696 (2011) 338--342.
\newblock \href {http://arxiv.org/abs/1011.1305} {\path{arXiv:1011.1305}}, \href {https://doi.org/10.1016/j.physletb.2010.12.065} {\path{doi:10.1016/j.physletb.2010.12.065}}.

\bibitem{CLAS:2009rdi}
M.~E. McCracken, et~al., {Differential cross section and recoil polarization measurements for the $\gamma p \to K^+ \Lambda$ reaction using CLAS at Jefferson Lab}, Phys. Rev. C 81 (2010) 025201.
\newblock \href {http://arxiv.org/abs/0912.4274} {\path{arXiv:0912.4274}}, \href {https://doi.org/10.1103/PhysRevC.81.025201} {\path{doi:10.1103/PhysRevC.81.025201}}.

\bibitem{CLAS:2009ngd}
M.~Battaglieri, et~al., {Photoproduction of $\pi^+ \pi^-$ meson pairs on the proton}, Phys. Rev. D 80 (2009) 072005.
\newblock \href {http://arxiv.org/abs/0907.1021} {\path{arXiv:0907.1021}}, \href {https://doi.org/10.1103/PhysRevD.80.072005} {\path{doi:10.1103/PhysRevD.80.072005}}.

\bibitem{CLAS:2009tyz}
M.~Dugger, et~al., {$\pi^+$ photoproduction on the proton for photon energies from 0.725 to 2.875-GeV}, Phys. Rev. C 79 (2009) 065206.
\newblock \href {http://arxiv.org/abs/0903.1110} {\path{arXiv:0903.1110}}, \href {https://doi.org/10.1103/PhysRevC.79.065206} {\path{doi:10.1103/PhysRevC.79.065206}}.

\bibitem{CLAS:2009wde}
M.~Williams, et~al., {Differential cross s ections for the reactions $\gamma p \to p \eta$ and $\gamma p \to p \eta^\prime$}, Phys. Rev. C 80 (2009) 045213.
\newblock \href {http://arxiv.org/abs/0909.0616} {\path{arXiv:0909.0616}}, \href {https://doi.org/10.1103/PhysRevC.80.045213} {\path{doi:10.1103/PhysRevC.80.045213}}.

\bibitem{CLAS:2009hpc}
M.~Williams, et~al., {Differential cross sections and spin density matrix elements for the reaction $\gamma p \to p \omega$}, Phys. Rev. C 80 (2009) 065208.
\newblock \href {http://arxiv.org/abs/0908.2910} {\path{arXiv:0908.2910}}, \href {https://doi.org/10.1103/PhysRevC.80.065208} {\path{doi:10.1103/PhysRevC.80.065208}}.

\bibitem{Chen:2009sda}
W.~Chen, et~al., {A Measurement of the differential cross section for the reaction $\gamma n \to \pi^- p$ from deuterium}, Phys. Rev. Lett. 103 (2009) 012301.
\newblock \href {http://arxiv.org/abs/0903.1260} {\path{arXiv:0903.1260}}, \href {https://doi.org/10.1103/PhysRevLett.103.012301} {\path{doi:10.1103/PhysRevLett.103.012301}}.

\bibitem{CLAS:2008ycy}
M.~Battaglieri, et~al., {First measurement of direct $f_0(980)$ photoproduction on the proton}, Phys. Rev. Lett. 102 (2009) 102001.
\newblock \href {http://arxiv.org/abs/0811.1681} {\path{arXiv:0811.1681}}, \href {https://doi.org/10.1103/PhysRevLett.102.102001} {\path{doi:10.1103/PhysRevLett.102.102001}}.

\bibitem{Dugger:2007bt}
M.~Dugger, et~al., {pi0 photoproduction on the proton for photon energies from 0.675 to 2.875-GeV}, Phys. Rev. C 76 (2007) 025211.
\newblock \href {http://arxiv.org/abs/0705.0816} {\path{arXiv:0705.0816}}, \href {https://doi.org/10.1103/PhysRevC.76.025211} {\path{doi:10.1103/PhysRevC.76.025211}}.

\bibitem{CLAS:2007xhu}
T.~Mibe, et~al., {First measurement of coherent phi-meson photoproduction on deuteron at low energies}, Phys. Rev. C 76 (2007) 052202.
\newblock \href {http://arxiv.org/abs/nucl-ex/0703013} {\path{arXiv:nucl-ex/0703013}}, \href {https://doi.org/10.1103/PhysRevC.76.052202} {\path{doi:10.1103/PhysRevC.76.052202}}.

\bibitem{Guo:2007dw}
L.~Guo, et~al., {Cascade production in the reactions $\gamma p \to K^+ K^+ (X)$ and $\gamma p \to K^+ K^+ \pi^- (X)$}, Phys. Rev. C 76 (2007) 025208.
\newblock \href {http://arxiv.org/abs/nucl-ex/0702027} {\path{arXiv:nucl-ex/0702027}}, \href {https://doi.org/10.1103/PhysRevC.76.025208} {\path{doi:10.1103/PhysRevC.76.025208}}.

\bibitem{CLAS:2007kab}
I.~Hleiqawi, et~al., {Cross-sections for the $\gamma p\to K^{*0} \Sigma^+$ reaction at $E_\gamma = 1.7$\,GeV - 3.0\,GeV}, Phys. Rev. C 75 (2007) 042201, [Erratum: Phys.Rev.C 76, 039905 (2007)].
\newblock \href {http://arxiv.org/abs/nucl-ex/0701036} {\path{arXiv:nucl-ex/0701036}}, \href {https://doi.org/10.1103/PhysRevC.76.039905} {\path{doi:10.1103/PhysRevC.76.039905}}.

\bibitem{CLAS:2006pde}
R.~K. Bradford, et~al., {First measurement of beam-recoil observables C(x) and C(z) in hyperon photoproduction}, Phys. Rev. C 75 (2007) 035205.
\newblock \href {http://arxiv.org/abs/nucl-ex/0611034} {\path{arXiv:nucl-ex/0611034}}, \href {https://doi.org/10.1103/PhysRevC.75.035205} {\path{doi:10.1103/PhysRevC.75.035205}}.

\bibitem{CLAS:2005koo}
M.~Battaglieri, et~al., {Search for $\Theta^+(1540)$ pentaquark in high statistics measurement of $\gamma p \to \bar{K}^0 K^+ n$ at CLAS}, Phys. Rev. Lett. 96 (2006) 042001.
\newblock \href {http://arxiv.org/abs/hep-ex/0510061} {\path{arXiv:hep-ex/0510061}}, \href {https://doi.org/10.1103/PhysRevLett.96.042001} {\path{doi:10.1103/PhysRevLett.96.042001}}.

\bibitem{CLAS:2006czw}
B.~McKinnon, et~al., {Search for the $\Theta^+$ pentaquark in the reaction $\gamma d \to p K^- K^+ n$}, Phys. Rev. Lett. 96 (2006) 212001.
\newblock \href {http://arxiv.org/abs/hep-ex/0603028} {\path{arXiv:hep-ex/0603028}}, \href {https://doi.org/10.1103/PhysRevLett.96.212001} {\path{doi:10.1103/PhysRevLett.96.212001}}.

\bibitem{CLAS:2006rru}
S.~Niccolai, et~al., {Search for the Theta+ pentaquark in the $\gamma d \to \Lambda n K^+$ reaction measured with CLAS}, Phys. Rev. Lett. 97 (2006) 032001.
\newblock \href {http://arxiv.org/abs/hep-ex/0604047} {\path{arXiv:hep-ex/0604047}}, \href {https://doi.org/10.1103/PhysRevLett.97.032001} {\path{doi:10.1103/PhysRevLett.97.032001}}.

\bibitem{CLAS:2006anj}
R.~De~Vita, et~al., {Search for the $\Theta^+$ pentaquark in the reactions $\gamma p \to \bar{K}^0 K^+n$ and $\gamma p -\to \bar{K}^0 K^0p$}, Phys. Rev. D 74 (2006) 032001.
\newblock \href {http://arxiv.org/abs/hep-ex/0606062} {\path{arXiv:hep-ex/0606062}}, \href {https://doi.org/10.1103/PhysRevD.74.032001} {\path{doi:10.1103/PhysRevD.74.032001}}.

\bibitem{CLAS:2005rxe}
M.~Dugger, et~al., {$\eta\prime$ photoproduction on the proton for photon energies from 1.527-GeV to 2.227-GeV}, Phys. Rev. Lett. 96 (2006) 062001, [Erratum: Phys.Rev.Lett. 96, 169905 (2006)].
\newblock \href {http://arxiv.org/abs/nucl-ex/0512019} {\path{arXiv:nucl-ex/0512019}}, \href {https://doi.org/10.1103/PhysRevLett.96.062001} {\path{doi:10.1103/PhysRevLett.96.062001}}.

\bibitem{CLAS:2005oqk}
S.~Strauch, et~al., {Beam-helicity asymmetries in double-charged-pion photoproduction on the proton}, Phys. Rev. Lett. 95 (2005) 162003.
\newblock \href {http://arxiv.org/abs/hep-ex/0508002} {\path{arXiv:hep-ex/0508002}}, \href {https://doi.org/10.1103/PhysRevLett.95.162003} {\path{doi:10.1103/PhysRevLett.95.162003}}.

\bibitem{CLAS:2005lui}
R.~Bradford, et~al., {Differential cross sections for $\gamma + p \to K^+ + Y$ for $\Lambda$ and $\Sigma^0$ hyperons}, Phys. Rev. C 73 (2006) 035202.
\newblock \href {http://arxiv.org/abs/nucl-ex/0509033} {\path{arXiv:nucl-ex/0509033}}, \href {https://doi.org/10.1103/PhysRevC.73.035202} {\path{doi:10.1103/PhysRevC.73.035202}}.

\bibitem{Burkert:2005ft}
V.~D. Burkert, {Have pentaquark states been seen?}, Int. J. Mod. Phys. A 21 (2006) 1764--1777.
\newblock \href {http://arxiv.org/abs/hep-ph/0510309} {\path{arXiv:hep-ph/0510309}}, \href {https://doi.org/10.1142/S0217751X06032745} {\path{doi:10.1142/S0217751X06032745}}.

\bibitem{CLAS:2004gjf}
J.~W. Price, et~al., {Exclusive photoproduction of the cascade $\Xi$ hyperons}, Phys. Rev. C 71 (2005) 058201.
\newblock \href {http://arxiv.org/abs/nucl-ex/0409030} {\path{arXiv:nucl-ex/0409030}}, \href {https://doi.org/10.1103/PhysRevC.71.058201} {\path{doi:10.1103/PhysRevC.71.058201}}.

\bibitem{CLAS:2003wfm}
V.~Kubarovsky, et~al., {Observation of an exotic baryon with S = +1 in photoproduction from the proton}, Phys. Rev. Lett. 92 (2004) 032001, [Erratum: Phys.Rev.Lett. 92, 049902 (2004)].
\newblock \href {http://arxiv.org/abs/hep-ex/0311046} {\path{arXiv:hep-ex/0311046}}, \href {https://doi.org/10.1103/PhysRevLett.92.032001} {\path{doi:10.1103/PhysRevLett.92.032001}}.

\bibitem{CLAS:2003zrd}
J.~W.~C. McNabb, et~al., {Hyperon photoproduction in the nucleon resonance region}, Phys. Rev. C 69 (2004) 042201.
\newblock \href {http://arxiv.org/abs/nucl-ex/0305028} {\path{arXiv:nucl-ex/0305028}}, \href {https://doi.org/10.1103/PhysRevC.69.042201} {\path{doi:10.1103/PhysRevC.69.042201}}.

\bibitem{CLAS:2003yuj}
S.~Stepanyan, et~al., {Observation of an exotic S = +1 baryon in exclusive photoproduction from the deuteron}, Phys. Rev. Lett. 91 (2003) 252001.
\newblock \href {http://arxiv.org/abs/hep-ex/0307018} {\path{arXiv:hep-ex/0307018}}, \href {https://doi.org/10.1103/PhysRevLett.91.252001} {\path{doi:10.1103/PhysRevLett.91.252001}}.

\bibitem{CLAS:2002cdi}
M.~Battaglieri, et~al., {Photoproduction of the $\omega$ meson on the proton at large momentum transfer}, Phys. Rev. Lett. 90 (2003) 022002.
\newblock \href {http://arxiv.org/abs/hep-ex/0210023} {\path{arXiv:hep-ex/0210023}}, \href {https://doi.org/10.1103/PhysRevLett.90.022002} {\path{doi:10.1103/PhysRevLett.90.022002}}.

\bibitem{CLAS:2002rxi}
M.~Dugger, et~al., {Eta photoproduction on the proton for photon energies from 0.75-GeV to 1.95-GeV}, Phys. Rev. Lett. 89 (2002) 222002, [Erratum: Phys.Rev.Lett. 89, 249904 (2002)].
\newblock \href {https://doi.org/10.1103/PhysRevLett.89.222002} {\path{doi:10.1103/PhysRevLett.89.222002}}.

\bibitem{CLAS:2001zxv}
M.~Battaglieri, et~al., {Photoproduction of the $\rho^0$ meson on the proton at large momentum transfer}, Phys. Rev. Lett. 87 (2001) 172002.
\newblock \href {http://arxiv.org/abs/hep-ex/0107028} {\path{arXiv:hep-ex/0107028}}, \href {https://doi.org/10.1103/PhysRevLett.87.172002} {\path{doi:10.1103/PhysRevLett.87.172002}}.

\bibitem{CLAS:2000kid}
E.~Anciant, et~al., {Photoproduction of $\phi(1020)$ mesons on the proton at large momentum transfer}, Phys. Rev. Lett. 85 (2000) 4682--4686.
\newblock \href {http://arxiv.org/abs/hep-ex/0006022} {\path{arXiv:hep-ex/0006022}}, \href {https://doi.org/10.1103/PhysRevLett.85.4682} {\path{doi:10.1103/PhysRevLett.85.4682}}.

\bibitem{Manak:2000ty}
J.~J. Manak, V.~Burkert, F.~Klein, B.~Mecking, A.~Coleman, H.~Funsten, {Electro- and photoproduction of $\omega(783)$ mesons using CLAS at Jefferson Lab}, Nucl. Phys. A 663 (2000) 671--674.
\newblock \href {https://doi.org/10.1016/S0375-9474(99)00713-7} {\path{doi:10.1016/S0375-9474(99)00713-7}}.

\bibitem{JeffersonLabE94014:1998czy}
C.~S. Armstrong, et~al., {Electroproduction of the $S_{11}(1535)$ resonance at high momentum transfer}, Phys. Rev. D 60 (1999) 052004.
\newblock \href {http://arxiv.org/abs/nucl-ex/9811001} {\path{arXiv:nucl-ex/9811001}}, \href {https://doi.org/10.1103/PhysRevD.60.052004} {\path{doi:10.1103/PhysRevD.60.052004}}.

\bibitem{Frolov:1998pw}
V.~V. Frolov, et~al., {Electroproduction of the $\Delta (1232)$ resonance at high momentum transfer}, Phys. Rev. Lett. 82 (1999) 45--48.
\newblock \href {https://doi.org/10.1103/PhysRevLett.82.45} {\path{doi:10.1103/PhysRevLett.82.45}}.

\bibitem{GRAAL:2000gnk}
J.~Ajaka, et~al., {Precise measurement of $\Sigma$ beam asymmetry for positive pion photoproduction on the proton from 550-MeV to 1100-MeV}, Phys. Lett. B 475 (2000) 372--377.
\newblock \href {https://doi.org/10.1016/S0370-2693(99)01396-9} {\path{doi:10.1016/S0370-2693(99)01396-9}}.

\bibitem{GRAAL:2000qng}
F.~Renard, et~al., {Differential cross-section measurement of $\eta$ photoproduction on the proton from threshold to 1100-MeV}, Phys. Lett. B 528 (2002) 215--220.
\newblock \href {http://arxiv.org/abs/hep-ex/0011098} {\path{arXiv:hep-ex/0011098}}, \href {https://doi.org/10.1016/S0370-2693(02)01196-6} {\path{doi:10.1016/S0370-2693(02)01196-6}}.

\bibitem{GRAAL:2002mil}
O.~Bartalini, et~al., {Precise measurement of $\Sigma$ beam asymmetry for positive pion photoproduction on the proton from 800-MeV to 1500-MeV}, Phys. Lett. B 544 (2002) 113--120.
\newblock \href {http://arxiv.org/abs/nucl-ex/0207010} {\path{arXiv:nucl-ex/0207010}}, \href {https://doi.org/10.1016/S0370-2693(02)02467-X} {\path{doi:10.1016/S0370-2693(02)02467-X}}.

\bibitem{Assafiri:2003mv}
Y.~Assafiri, et~al., {Double $\pi^0$ photoproduction on the proton at GRAAL}, Phys. Rev. Lett. 90 (2003) 222001.
\newblock \href {https://doi.org/10.1103/PhysRevLett.90.222001} {\path{doi:10.1103/PhysRevLett.90.222001}}.

\bibitem{Gurzadyan:2004rx}
V.~G. Gurzadyan, et~al., {Probing the light speed anisotropy with respect to the Cosmic Microwave Background radiation dipole}, Mod. Phys. Lett. A 20 (2005) 19--28.
\newblock \href {http://arxiv.org/abs/astro-ph/0410742} {\path{arXiv:astro-ph/0410742}}, \href {https://doi.org/10.1142/S0217732305016294} {\path{doi:10.1142/S0217732305016294}}.

\bibitem{GRAAL:2005mor}
O.~Bartalini, et~al., {Measurement of $\pi^0$ photoproduction on the proton from 550-MeV to 1500-MeV at GRAAL}, Eur. Phys. J. A 26 (2005) 399--419.
\newblock \href {https://doi.org/10.1140/epja/i2005-10191-2} {\path{doi:10.1140/epja/i2005-10191-2}}.

\bibitem{Lleres:2007tx}
A.~Lleres, et~al., {Polarization observable measurements for $\gamma p \to K^+ \Lambda$ and $\gamma p \to K^+ \Sigma^0$ for energies up to 1.5-GeV}, Eur. Phys. J. A 31 (2007) 79--93.
\newblock \href {https://doi.org/10.1140/epja/i2006-10167-8} {\path{doi:10.1140/epja/i2006-10167-8}}.

\bibitem{GRAAL:2007gsc}
O.~Bartalini, et~al., {Measurement of $\eta$ photoproduction on the proton from threshold to 1500-MeV}, Eur. Phys. J. A 33 (2007) 169--184.
\newblock \href {http://arxiv.org/abs/0707.1385} {\path{arXiv:0707.1385}}, \href {https://doi.org/10.1140/epja/i2007-10439-9} {\path{doi:10.1140/epja/i2007-10439-9}}.

\bibitem{Bartalini:2008zza}
O.~Bartalini, et~al., {Measurement of the total photoabsorption cross section on a proton in the energy range 600-MeV - 1500-MeV at the GRAAL}, Phys. Atom. Nucl. 71 (2008) 75--82.
\newblock \href {https://doi.org/10.1007/s11450-008-1008-9} {\path{doi:10.1007/s11450-008-1008-9}}.

\bibitem{GRAAL:2013tzy}
V.~Vegna, et~al., {Measurement of the $\Sigma$ beam asymmetry for the $\omega$ photo-production off the proton and the neutron at GRAAL}, Phys. Rev. C 91~(6) (2015) 065207.
\newblock \href {http://arxiv.org/abs/1306.5943} {\path{arXiv:1306.5943}}, \href {https://doi.org/10.1103/PhysRevC.91.065207} {\path{doi:10.1103/PhysRevC.91.065207}}.

\bibitem{LeviSandri:2014uhc}
P.~Levi~Sandri, et~al., {First Measurement of the $\Sigma$ Beam Asymmetry in $\eta^{\prime}$ Photoproduction off the Proton near Threshold}, Eur. Phys. J. A 51~(7) (2015) 77.
\newblock \href {http://arxiv.org/abs/1407.6991} {\path{arXiv:1407.6991}}, \href {https://doi.org/10.1140/epja/i2015-15077-0} {\path{doi:10.1140/epja/i2015-15077-0}}.

\bibitem{Schmieden:2018yjy}
H.~Schmieden, {First Results from the BGO--OD Experiment at ELSA}, Few Body Syst. 59~(6) (2018) 135.
\newblock \href {https://doi.org/10.1007/s00601-018-1453-0} {\path{doi:10.1007/s00601-018-1453-0}}.

\bibitem{BGO-OD:2019utx}
S.~Alef, et~al., {The BGOOD experimental setup at ELSA}, Eur. Phys. J. A 56~(4) (2020) 104.
\newblock \href {http://arxiv.org/abs/1910.11939} {\path{arXiv:1910.11939}}, \href {https://doi.org/10.1140/epja/s10050-020-00107-x} {\path{doi:10.1140/epja/s10050-020-00107-x}}.

\bibitem{Jude:2020byj}
T.~C. Jude, et~al., {Observation of a cusp-like structure in the $\gamma p\to K^+\Sigma^0$ cross section at forward angles and low momentum transfer}, Phys. Lett. B 820 (2021) 136559.
\newblock \href {http://arxiv.org/abs/2006.12437} {\path{arXiv:2006.12437}}, \href {https://doi.org/10.1016/j.physletb.2021.136559} {\path{doi:10.1016/j.physletb.2021.136559}}.

\bibitem{Alef:2020yul}
S.~Alef, et~al., {$K^+\Lambda$ photoproduction at forward angles and low momentum transfer}, Eur. Phys. J. A 57~(2) (2021) 80.
\newblock \href {http://arxiv.org/abs/2006.12350} {\path{arXiv:2006.12350}}, \href {https://doi.org/10.1140/epja/s10050-021-00392-0} {\path{doi:10.1140/epja/s10050-021-00392-0}}.

\bibitem{BGOOD:2021sog}
G.~Scheluchin, et~al., {Photoproduction of $K^+\Lambda(1405)\to K^+\pi^0\Sigma^0$ extending to forward angles and low momentum transfer}, Phys. Lett. B 833 (2022) 137375.
\newblock \href {http://arxiv.org/abs/2108.12235} {\path{arXiv:2108.12235}}, \href {https://doi.org/10.1016/j.physletb.2022.137375} {\path{doi:10.1016/j.physletb.2022.137375}}.

\bibitem{BGOOD:2021oxp}
K.~Kohl, et~al., {Measurement of the $\gamma n\rightarrow K^0\Sigma ^0$ differential cross section over the $K^*$ threshold}, Eur. Phys. J. A 59~(11) (2023) 254.
\newblock \href {http://arxiv.org/abs/2108.13319} {\path{arXiv:2108.13319}}, \href {https://doi.org/10.1140/epja/s10050-023-01133-1} {\path{doi:10.1140/epja/s10050-023-01133-1}}.

\bibitem{Jude:2022atd}
T.~C. Jude, et~al., {Evidence of a dibaryon spectrum in coherent $\pi^0\pi^0p$ photoproduction at forward deuteron angles}, Phys. Lett. B 832 (2022) 137277.
\newblock \href {http://arxiv.org/abs/2202.08594} {\path{arXiv:2202.08594}}, \href {https://doi.org/10.1016/j.physletb.2022.137277} {\path{doi:10.1016/j.physletb.2022.137277}}.

\bibitem{Figueiredo:2024zmh}
A.~J.~C. Figueiredo, et~al., {Coherent $\pi^0\eta d$ photoproduction at forward deuteron angles measured at BGOOD} (5 2024).
\newblock \href {http://arxiv.org/abs/2405.09392} {\path{arXiv:2405.09392}}.

\bibitem{Rosanowski:2024rww}
E.~O. Rosanowski, et~al., {$K^+\Lambda(1520)$ photoproduction at forward angles near threshold with the BGOOD experiment} (6 2024).
\newblock \href {http://arxiv.org/abs/2406.01121} {\path{arXiv:2406.01121}}.

\bibitem{Nakano:2000ku}
T.~Nakano, et~al., {Experiment at SPring-8}, Nucl. Phys. A 670 (2000) 332--339.
\newblock \href {https://doi.org/10.1016/S0375-9474(00)00124-X} {\path{doi:10.1016/S0375-9474(00)00124-X}}.

\bibitem{Nakano:2001xp}
T.~Nakano, et~al., {Multi-GeV laser electron photon project at SPring-8}, Nucl. Phys. A 684 (2001) 71--79.
\newblock \href {https://doi.org/10.1016/S0375-9474(01)00490-0} {\path{doi:10.1016/S0375-9474(01)00490-0}}.

\bibitem{Muramatsu:2021bpl}
N.~Muramatsu, et~al., {SPring-8 LEPS2 beamline: A facility to produce a multi-GeV photon beam via laser Compton scattering}, Nucl. Instrum. Meth. A 1033 (2022) 166677.
\newblock \href {http://arxiv.org/abs/2112.07832} {\path{arXiv:2112.07832}}, \href {https://doi.org/10.1016/j.nima.2022.166677} {\path{doi:10.1016/j.nima.2022.166677}}.

\bibitem{Matsumura:2003mw}
T.~Matsumura, et~al., {2$\pi^0$ photoproduction experiment at SPring-8}, Nucl. Phys. A 721 (2003) 723--726.
\newblock \href {https://doi.org/10.1016/S0375-9474(03)01166-7} {\path{doi:10.1016/S0375-9474(03)01166-7}}.

\bibitem{LEPS:2003wug}
T.~Nakano, et~al., {Evidence for a narrow S = +1 baryon resonance in photoproduction from the neutron}, Phys. Rev. Lett. 91 (2003) 012002.
\newblock \href {http://arxiv.org/abs/hep-ex/0301020} {\path{arXiv:hep-ex/0301020}}, \href {https://doi.org/10.1103/PhysRevLett.91.012002} {\path{doi:10.1103/PhysRevLett.91.012002}}.

\bibitem{LEPS:2003buk}
R.~G.~T. Zegers, et~al., {Beam polarization asymmetries for the $p(\gamma, K^+) \Lambda$ and $p(\gamma, K^+) \Sigma^0$ reactions at $E_\gamma$ = 1.5 Gev - 2.4 GeV}, Phys. Rev. Lett. 91 (2003) 092001.
\newblock \href {http://arxiv.org/abs/nucl-ex/0302005} {\path{arXiv:nucl-ex/0302005}}, \href {https://doi.org/10.1103/PhysRevLett.91.092001} {\path{doi:10.1103/PhysRevLett.91.092001}}.

\bibitem{CDF:2003epb}
D.~Acosta, et~al., {Measurement of the mass difference $m(D_s^+) - m(D^+)$ at {CDF} II}, Phys. Rev. D 68 (2003) 072004.
\newblock \href {http://arxiv.org/abs/hep-ex/0310043} {\path{arXiv:hep-ex/0310043}}, \href {https://doi.org/10.1103/PhysRevD.68.072004} {\path{doi:10.1103/PhysRevD.68.072004}}.

\bibitem{Nakano:2004cr}
T.~Nakano, K.~Hicks, {Discovery of the strangeness S = +1 pentaquark}, Mod. Phys. Lett. A 19 (2004) 645--657.
\newblock \href {https://doi.org/10.1142/S0217732304013477} {\path{doi:10.1142/S0217732304013477}}.

\bibitem{Ishikawa:2004id}
T.~Ishikawa, et~al., {$\phi$ photo-production from Li, C, Al, and Cu nuclei at E(gamma) = 1.5-GeV to 2.4-GeV}, Phys. Lett. B 608 (2005) 215--222.
\newblock \href {http://arxiv.org/abs/nucl-ex/0411016} {\path{arXiv:nucl-ex/0411016}}, \href {https://doi.org/10.1016/j.physletb.2005.01.023} {\path{doi:10.1016/j.physletb.2005.01.023}}.

\bibitem{LEPS:2005hax}
T.~Mibe, et~al., {Diffractive $\phi$-meson photoproduction on proton near threshold}, Phys. Rev. Lett. 95 (2005) 182001.
\newblock \href {http://arxiv.org/abs/nucl-ex/0506015} {\path{arXiv:nucl-ex/0506015}}, \href {https://doi.org/10.1103/PhysRevLett.95.182001} {\path{doi:10.1103/PhysRevLett.95.182001}}.

\bibitem{LEPS:2005hji}
M.~Sumihama, et~al., {The $\vec\gamma p \to K^+ \Lambda$ and $\vec\gamma p \to K^+ \Sigma^0$ reactions at forward angles with photon energies from 1.5-GeV to 2.4-GeV}, Phys. Rev. C 73 (2006) 035214.
\newblock \href {http://arxiv.org/abs/hep-ex/0512053} {\path{arXiv:hep-ex/0512053}}, \href {https://doi.org/10.1103/PhysRevC.73.035214} {\path{doi:10.1103/PhysRevC.73.035214}}.

\bibitem{Kohri:2006yx}
H.~Kohri, et~al., {Differential cross section and photon beam asymmetry for the polarized $\gamma n \to K^+ \Sigma^-$ reaction at $E_\gamma$ = 1.5GeV-2.4GeV}, Phys. Rev. Lett. 97 (2006) 082003.
\newblock \href {http://arxiv.org/abs/hep-ex/0602015} {\path{arXiv:hep-ex/0602015}}, \href {https://doi.org/10.1103/PhysRevLett.97.082003} {\path{doi:10.1103/PhysRevLett.97.082003}}.

\bibitem{Chang:2007fc}
W.~C. Chang, et~al., {Forward coherent $\phi$-meson photoproduction from deuterons near threshold}, Phys. Lett. B 658 (2008) 209--215.
\newblock \href {http://arxiv.org/abs/nucl-ex/0703034} {\path{arXiv:nucl-ex/0703034}}, \href {https://doi.org/10.1016/j.physletb.2007.11.009} {\path{doi:10.1016/j.physletb.2007.11.009}}.

\bibitem{LEPS:2007nri}
K.~Hicks, et~al., {Measurement of the $\gamma\vec p \to K^+ \Lambda$ Reaction at Backward Angles}, Phys. Rev. C 76 (2007) 042201.
\newblock \href {http://arxiv.org/abs/0707.4412} {\path{arXiv:0707.4412}}, \href {https://doi.org/10.1103/PhysRevC.76.042201} {\path{doi:10.1103/PhysRevC.76.042201}}.

\bibitem{Sumihama:2007qa}
M.~Sumihama, et~al., {Backward-angle photoproduction of $\pi^0$ mesons on the proton at $E_\gamma = 1.5-2.4$\,GeV}, Phys. Lett. B 657 (2007) 32--37.
\newblock \href {http://arxiv.org/abs/0708.1600} {\path{arXiv:0708.1600}}, \href {https://doi.org/10.1016/j.physletb.2007.10.032} {\path{doi:10.1016/j.physletb.2007.10.032}}.

\bibitem{Sumihama:2007gfo}
M.~Sumihama, et~al., {$K^+$ Photoproduction And $\pi^0$ Photoproduction By Linearly Polarized Photons At Spring-8/Leps}, eConf C070910 (2007) 113.

\bibitem{Miwa:2007xk}
K.~Miwa, et~al., {Search for $\Theta^+$ via $K^+ p \to \pi^+ X$ reaction with a 1.2-GeV/c $K^+$ beam}, Phys. Rev. C 77 (2008) 045203.
\newblock \href {http://arxiv.org/abs/0712.3839} {\path{arXiv:0712.3839}}, \href {https://doi.org/10.1103/PhysRevC.77.045203} {\path{doi:10.1103/PhysRevC.77.045203}}.

\bibitem{Niiyama:2008rt}
M.~Niiyama, et~al., {Photoproduction of $\Lambda(1405)$ and $\Sigma^0(1385)$ on the proton at $E_\gamma = 1.5-2.4$\,GeV}, Phys. Rev. C 78 (2008) 035202.
\newblock \href {http://arxiv.org/abs/0805.4051} {\path{arXiv:0805.4051}}, \href {https://doi.org/10.1103/PhysRevC.78.035202} {\path{doi:10.1103/PhysRevC.78.035202}}.

\bibitem{LEPS:2008ghm}
T.~Nakano, et~al., {Evidence of the $\Theta^+$ in the $\gamma d \to K^+ K^- pn$ reaction}, Phys. Rev. C 79 (2009) 025210.
\newblock \href {http://arxiv.org/abs/0812.1035} {\path{arXiv:0812.1035}}, \href {https://doi.org/10.1103/PhysRevC.79.025210} {\path{doi:10.1103/PhysRevC.79.025210}}.

\bibitem{LEPS:2008azb}
K.~Hicks, et~al., {Cross Sections and Beam Asymmetries for $K^+ \Sigma^{*-}$ photoproduction from the deuteron at $E_\gamma = 1.5$\,GeV - 2.4-GeV}, Phys. Rev. Lett. 102 (2009) 012501.
\newblock \href {http://arxiv.org/abs/0812.0771} {\path{arXiv:0812.0771}}, \href {https://doi.org/10.1103/PhysRevLett.102.012501} {\path{doi:10.1103/PhysRevLett.102.012501}}.

\bibitem{Muramatsu:2009zp}
N.~Muramatsu, et~al., {Near-threshold photoproduction of $\Lambda(1520)$ from protons and deuterons}, Phys. Rev. Lett. 103 (2009) 012001.
\newblock \href {http://arxiv.org/abs/0904.2034} {\path{arXiv:0904.2034}}, \href {https://doi.org/10.1103/PhysRevLett.103.012001} {\path{doi:10.1103/PhysRevLett.103.012001}}.

\bibitem{LEPS:2009isz}
H.~Kohri, et~al., {Near-threshold $\Lambda(1520)$ production by the $\vec\gamma p \to K^+ \Lambda(1520)$ reaction at forward $K^+$ angles}, Phys. Rev. Lett. 104 (2010) 172001.
\newblock \href {http://arxiv.org/abs/0906.0197} {\path{arXiv:0906.0197}}, \href {https://doi.org/10.1103/PhysRevLett.104.172001} {\path{doi:10.1103/PhysRevLett.104.172001}}.

\bibitem{LEPS:2009nuw}
W.~C. Chang, et~al., {Measurement of the incoherent $\gamma d \to \phi p n$ photoproduction near threshold}, Phys. Lett. B 684 (2010) 6--10.
\newblock \href {http://arxiv.org/abs/0907.1705} {\path{arXiv:0907.1705}}, \href {https://doi.org/10.1016/j.physletb.2009.12.051} {\path{doi:10.1016/j.physletb.2009.12.051}}.

\bibitem{LEPS:2009pib}
M.~Sumihama, et~al., {Backward-angle $\eta$ photoproduction from protons at $E_\gamma = 1.6 - 2.4$\,GeV}, Phys. Rev. C 80 (2009) 052201.
\newblock \href {http://arxiv.org/abs/0910.0900} {\path{arXiv:0910.0900}}, \href {https://doi.org/10.1103/PhysRevC.80.052201} {\path{doi:10.1103/PhysRevC.80.052201}}.

\bibitem{LEPS:2010ovn}
W.~C. Chang, et~al., {Measurement of Spin-Density Matrix Elements for $\phi$-Meson Photoproduction from Protons and Deuterons Near Threshold}, Phys. Rev. C 82 (2010) 015205.
\newblock \href {http://arxiv.org/abs/1006.4197} {\path{arXiv:1006.4197}}, \href {https://doi.org/10.1103/PhysRevC.82.015205} {\path{doi:10.1103/PhysRevC.82.015205}}.

\bibitem{Hwang:2012zza}
S.~H. Hwang, et~al., {Spin-Density Matrix Elements for $\gamma p \to K^{*0} \Sigma^+$ at $E_\gamma=1.85-3.0$ GeV with Evidence for the $\kappa(800)$ Meson Exchange}, Phys. Rev. Lett. 108 (2012) 092001.
\newblock \href {https://doi.org/10.1103/PhysRevLett.108.092001} {\path{doi:10.1103/PhysRevLett.108.092001}}.

\bibitem{Hwang:2013usa}
S.~Hwang, K.~Hicks, J.~K. Ahn, T.~Nakano, {$K^{*0} \Sigma^+$ Photoproduction with Evidence for the Kappa Meson Exchange at SPring-8/LEPS}, Few Body Syst. 54 (2013) 1037--1041.
\newblock \href {https://doi.org/10.1007/s00601-013-0650-0} {\path{doi:10.1007/s00601-013-0650-0}}.

\bibitem{Morino:2013raa}
Y.~Morino, et~al., {Backward-angle photoproduction of $\omega$ and $\eta'$ mesons from protons in the photon energy range from 1.5 to 3.0 GeV}, PTEP 2015~(1) (2015) 013D01.
\newblock \href {http://arxiv.org/abs/1306.3031} {\path{arXiv:1306.3031}}, \href {https://doi.org/10.1093/ptep/ptu167} {\path{doi:10.1093/ptep/ptu167}}.

\bibitem{LEPS:2013dqu}
A.~O. Tokiyasu, et~al., {Search for $K^-pp$ bound state via $\gamma d \rightarrow K^+ \pi^-X$ reaction at $E_\gamma=1.5-2.4$ GeV}, Phys. Lett. B 728 (2014) 616--621.
\newblock \href {http://arxiv.org/abs/1306.5320} {\path{arXiv:1306.5320}}, \href {https://doi.org/10.1016/j.physletb.2013.12.039} {\path{doi:10.1016/j.physletb.2013.12.039}}.

\bibitem{LEPS:2016ljn}
S.~Y. Ryu, et~al., {Interference effect between $\phi$ and $\Lambda(1520)$ production channels in the $\gamma p \rightarrow K^+K^-p$ reaction near threshold}, Phys. Rev. Lett. 116~(23) (2016) 232001.
\newblock \href {http://arxiv.org/abs/1603.00236} {\path{arXiv:1603.00236}}, \href {https://doi.org/10.1103/PhysRevLett.116.232001} {\path{doi:10.1103/PhysRevLett.116.232001}}.

\bibitem{LEPS:2017jqw}
H.~Kohri, et~al., {Differential cross section and photon beam asymmetry for the $\overrightarrow \gamma p \to \pi^+$ n reaction at forward $\pi^+$ angles at E$_\gamma$ =1.5-2.95 GeV}, Phys. Rev. C 97~(1) (2018) 015205.
\newblock \href {http://arxiv.org/abs/1708.09574} {\path{arXiv:1708.09574}}, \href {https://doi.org/10.1103/PhysRevC.97.015205} {\path{doi:10.1103/PhysRevC.97.015205}}.

\bibitem{LEPS:2017vas}
K.~Mizutani, et~al., {$\phi$ photoproduction on the proton at $E_{\gamma}$ = 1.5 - 2.9 GeV}, Phys. Rev. C 96~(6) (2017) 062201.
\newblock \href {http://arxiv.org/abs/1710.00169} {\path{arXiv:1710.00169}}, \href {https://doi.org/10.1103/PhysRevC.96.062201} {\path{doi:10.1103/PhysRevC.96.062201}}.

\bibitem{LEPS:2017nqz}
T.~Hiraiwa, et~al., {First measurement of coherent $\phi$-meson photoproduction from $^4He$ near threshold}, Phys. Rev. C 97~(3) (2018) 035208.
\newblock \href {http://arxiv.org/abs/1711.01095} {\path{arXiv:1711.01095}}, \href {https://doi.org/10.1103/PhysRevC.97.035208} {\path{doi:10.1103/PhysRevC.97.035208}}.

\bibitem{LEPS:2017pzl}
S.~H. Shiu, et~al., {Photoproduction of $\Lambda$ and $\Sigma^{0}$ hyperons off protons with linearly polarized photons at $E_{\gamma} = 1.5-3.0$ GeV}, Phys. Rev. C 97~(1) (2018) 015208.
\newblock \href {http://arxiv.org/abs/1711.04996} {\path{arXiv:1711.04996}}, \href {https://doi.org/10.1103/PhysRevC.97.015208} {\path{doi:10.1103/PhysRevC.97.015208}}.

\bibitem{LEPS:2018pbi}
H.~Kohri, et~al., {Differential cross section and photon-beam asymmetry for the $\gamma p \to \pi^- \Delta^{++}$(1232) reaction at forward $\pi^-$ angles for $E_\gamma$=1.5-2.95 GeV}, Phys. Rev. Lett. 120~(20) (2018) 202004.
\newblock \href {http://arxiv.org/abs/1801.07900} {\path{arXiv:1801.07900}}, \href {https://doi.org/10.1103/PhysRevLett.120.202004} {\path{doi:10.1103/PhysRevLett.120.202004}}.

\bibitem{LEPS2:2019bek}
N.~Muramatsu, et~al., {Measurement of neutral pion photoproduction off the proton with the large acceptance electromagnetic calorimeter BGOegg}, Phys. Rev. C 100~(5) (2019) 055202.
\newblock \href {https://doi.org/10.1103/PhysRevC.100.055202} {\path{doi:10.1103/PhysRevC.100.055202}}.

\bibitem{LEPS2BGOegg:2020cth}
N.~Tomida, et~al., {Search for $\eta'$ bound nuclei in the $^{12}{\rm C}(\gamma,p)$ reaction with simultaneous detection of decay products}, Phys. Rev. Lett. 124~(20) (2020) 202501.
\newblock \href {http://arxiv.org/abs/2005.03449} {\path{arXiv:2005.03449}}, \href {https://doi.org/10.1103/PhysRevLett.124.202501} {\path{doi:10.1103/PhysRevLett.124.202501}}.

\bibitem{LEPS2:2020ttk}
N.~Muramatsu, et~al., {Differential cross sections, photon beam asymmetries, and spin density matrix elements of $\omega$ photoproduction off the proton at $E_\gamma$=1.3 - 2.4 GeV}, Phys. Rev. C 102~(2) (2020) 025201.
\newblock \href {https://doi.org/10.1103/PhysRevC.102.025201} {\path{doi:10.1103/PhysRevC.102.025201}}.

\bibitem{LEPS2BGOegg:2022dop}
T.~Hashimoto, et~al., {Differential cross sections and photon beam asymmetries of $\eta$ photoproduction on the proton at $E_\gamma$ = 1.3 - 2.4 GeV}, Phys. Rev. C 106~(3) (2022) 035201.
\newblock \href {http://arxiv.org/abs/2202.13688} {\path{arXiv:2202.13688}}, \href {https://doi.org/10.1103/PhysRevC.106.035201} {\path{doi:10.1103/PhysRevC.106.035201}}.

\bibitem{LEPS2BGOegg:2023ssr}
N.~Muramatsu, et~al., {First measurement of differential cross sections and photon beam asymmetries for photoproduction of the $f_0(980)$ meson decaying into $\pi^0\pi^0$ at $E_\gamma < 2.4$\,GeV}, Phys. Rev. C 107~(4) (2023) L042201.
\newblock \href {https://doi.org/10.1103/PhysRevC.107.L042201} {\path{doi:10.1103/PhysRevC.107.L042201}}.

\bibitem{Anisovich:2017bsk}
A.~V. Anisovich, et~al., {Strong evidence for nucleon resonances near 1900\textbackslash{},MeV}, Phys. Rev. Lett. 119~(6) (2017) 062004.
\newblock \href {http://arxiv.org/abs/1712.07549} {\path{arXiv:1712.07549}}, \href {https://doi.org/10.1103/PhysRevLett.119.062004} {\path{doi:10.1103/PhysRevLett.119.062004}}.

\bibitem{ParticleDataGroup:2022pth}
R.~L. Workman, et~al., {Review of Particle Physics}, PTEP 2022 (2022) 083C01.
\newblock \href {https://doi.org/10.1093/ptep/ptac097} {\path{doi:10.1093/ptep/ptac097}}.

\bibitem{Stoler:1993yk}
P.~Stoler, {Baryon form-factors at high $Q^2$ and the transition to perturbative QCD}, Phys. Rept. 226 (1993) 103--171.
\newblock \href {https://doi.org/10.1016/0370-1573(93)90088-U} {\path{doi:10.1016/0370-1573(93)90088-U}}.

\bibitem{Aznauryan:2012ba}
I.~G. Aznauryan, et~al., {Studies of Nucleon Resonance Structure in Exclusive Meson Electroproduction}, Int. J. Mod. Phys. E 22 (2013) 1330015.
\newblock \href {https://doi.org/10.1142/S0218301313300154} {\path{doi:10.1142/S0218301313300154}}.

\bibitem{Proceedings:2020fyd}
S.~J. Brodsky, et~al., {Strong QCD from Hadron Structure Experiments}: {Newport News, VA, USA, November 4-8, 2019}, Int. J. Mod. Phys. E 29~(08) (2020) 2030006.
\newblock \href {http://arxiv.org/abs/2006.06802} {\path{arXiv:2006.06802}}, \href {https://doi.org/10.1142/S0218301320300064} {\path{doi:10.1142/S0218301320300064}}.

\bibitem{CLAS:2023mfc}
I.~A. Skorodumina, et~al., {Double-pion electroproduction off protons in deuterium: Quasifree cross sections~and final-state interactions}, Phys. Rev. C 109~(6) (2024) 065205.
\newblock \href {http://arxiv.org/abs/2308.13962} {\path{arXiv:2308.13962}}, \href {https://doi.org/10.1103/PhysRevC.109.065205} {\path{doi:10.1103/PhysRevC.109.065205}}.

\bibitem{Mokeev:2023zhq}
V.~I. Mokeev, et~al., {First Results on Nucleon Resonance Electroexcitation Amplitudes from $ep\to e'\pi^+\pi^-p'$ Cross Sections at $W$ = 1.4-1.7 GeV and $Q^2$ = 2.0-5.0 GeV$^2$}, Phys. Rev. C 108~(2) (2023) 025204.
\newblock \href {http://arxiv.org/abs/2306.13777} {\path{arXiv:2306.13777}}, \href {https://doi.org/10.1103/PhysRevC.108.025204} {\path{doi:10.1103/PhysRevC.108.025204}}.

\bibitem{CLAS:2022iqy}
S.~Diehl, et~al., {A multidimensional study of the structure function ratio \ensuremath{\sigma}LT'/\ensuremath{\sigma}0 from hard exclusive \ensuremath{\pi}+ electro-production off protons in the GPD regime}, Phys. Lett. B 839 (2023) 137761.
\newblock \href {http://arxiv.org/abs/2210.14557} {\path{arXiv:2210.14557}}, \href {https://doi.org/10.1016/j.physletb.2023.137761} {\path{doi:10.1016/j.physletb.2023.137761}}.

\bibitem{CLAS:2022kta}
Y.~Tian, et~al., {Exclusive $\pi^{-}$ Electroproduction off the Neutron in Deuterium in the Resonance Region}, Phys. Rev. C 107~(1) (2023) 015201.
\newblock \href {http://arxiv.org/abs/2203.16785} {\path{arXiv:2203.16785}}, \href {https://doi.org/10.1103/PhysRevC.107.015201} {\path{doi:10.1103/PhysRevC.107.015201}}.

\bibitem{CLAS:2021cvy}
E.~L. Isupov, et~al., {Polarized structure function $\sigma_{LT'}$ from $\pi^0 p$ electroproduction data in the resonance region at 0.4GeV$^2$\ensuremath{<}Q$^2$\ensuremath{<}1.0GeV$^2$}, Phys. Rev. C 105~(2) (2022) L022201.
\newblock \href {http://arxiv.org/abs/2112.07732} {\path{arXiv:2112.07732}}, \href {https://doi.org/10.1103/PhysRevC.105.L022201} {\path{doi:10.1103/PhysRevC.105.L022201}}.

\bibitem{CLAS:2021ovm}
R.~Dupr\'e, et~al., {Measurement of deeply virtual Compton scattering off $^{4}\mathrm{He}$ with the CEBAF Large Acceptance Spectrometer at Jefferson Lab}, Phys. Rev. C 104~(2) (2021) 025203.
\newblock \href {http://arxiv.org/abs/2102.07419} {\path{arXiv:2102.07419}}, \href {https://doi.org/10.1103/PhysRevC.104.025203} {\path{doi:10.1103/PhysRevC.104.025203}}.

\bibitem{CLAS:2019cpp}
N.~Markov, et~al., {Exclusive ${\pi^{0}p}$ electroproduction off protons in the resonance region at photon virtualities 0.4 GeV${^{2}}$ ${\leq~ Q^{2} \leq~1}$ GeV${^{2}}$}, Phys. Rev. C 101~(1) (2020) 015208.
\newblock \href {http://arxiv.org/abs/1907.11974} {\path{arXiv:1907.11974}}, \href {https://doi.org/10.1103/PhysRevC.101.015208} {\path{doi:10.1103/PhysRevC.101.015208}}.

\bibitem{CLAS:2019uzc}
B.~Zhao, et~al., {Measurement of the beam spin asymmetry of $\overrightarrow e p \to e'p' \eta$ in the deep-inelastic regime with CLAS}, Phys. Lett. B 789 (2019) 426--431.
\newblock \href {https://doi.org/10.1016/j.physletb.2018.12.065} {\path{doi:10.1016/j.physletb.2018.12.065}}.

\bibitem{CLAS:2018fon}
G.~V. Fedotov, et~al., {Measurements of the $\gamma_{v} p \rightarrow p' \pi^{+} \pi^{-}$ cross section with the CLAS detector for $0.4$ GeV$^{2}$ $< Q^{2} <$ $1.0$ GeV$^{2}$ and $1.3$ GeV $< W <$ $1.825$ GeV}, Phys. Rev. C 98~(2) (2018) 025203.
\newblock \href {http://arxiv.org/abs/1804.05136} {\path{arXiv:1804.05136}}, \href {https://doi.org/10.1103/PhysRevC.98.025203} {\path{doi:10.1103/PhysRevC.98.025203}}.

\bibitem{Aznauryan:2018okk}
I.~G. Aznauryan, V.~D. Burkert, {Electroexcitation of nucleon resonances in a light-front relativistic quark model}, Few Body Syst. 59~(5) (2018) 98.
\newblock \href {http://arxiv.org/abs/1805.10785} {\path{arXiv:1805.10785}}, \href {https://doi.org/10.1007/s00601-018-1415-6} {\path{doi:10.1007/s00601-018-1415-6}}.

\bibitem{Aznauryan:2017nkz}
I.~G. Aznauryan, V.~Burkert, {Electroexcitation of nucleon resonances of the [$70,1^-$] multiplet in a light-front relativistic quark model}, Phys. Rev. C 95~(6) (2017) 065207.
\newblock \href {http://arxiv.org/abs/1703.01751} {\path{arXiv:1703.01751}}, \href {https://doi.org/10.1103/PhysRevC.95.065207} {\path{doi:10.1103/PhysRevC.95.065207}}.

\bibitem{Aznauryan:2016wwm}
I.~G. Aznauryan, V.~D. Burkert, {Configuration mixings and light-front relativistic quark model predictions for the electroexcitation of the $\Delta(1232)3/2^+$, $N(1440)1/2^+$, and $\Delta(1600)3/2^+$}, unpublished note (3 2016).
\newblock \href {http://arxiv.org/abs/1603.06692} {\path{arXiv:1603.06692}}.

\bibitem{CLAS:2017rgp}
K.~Park, et~al., {Hard exclusive pion electroproduction at backward angles with CLAS}, Phys. Lett. B 780 (2018) 340--345.
\newblock \href {http://arxiv.org/abs/1711.08486} {\path{arXiv:1711.08486}}, \href {https://doi.org/10.1016/j.physletb.2018.03.026} {\path{doi:10.1016/j.physletb.2018.03.026}}.

\bibitem{CLAS:2017fja}
E.~L. Isupov, et~al., {Measurements of $e p \to e' \pi^+ \pi^- p'$ Cross Sections with CLAS at $1.40$ Gev $< W < 2.0$ GeV and $2.0$ GeV$^2$ $< Q^2 < 5.0$ GeV$^2$}, Phys. Rev. C 96~(2) (2017) 025209.
\newblock \href {http://arxiv.org/abs/1705.01901} {\path{arXiv:1705.01901}}, \href {https://doi.org/10.1103/PhysRevC.96.025209} {\path{doi:10.1103/PhysRevC.96.025209}}.

\bibitem{CLAS:2017jjr}
I.~Bedlinskiy, et~al., {Exclusive $\eta$ electroproduction at $W$\ensuremath{>}2 GeV with CLAS and transversity generalized parton distributions}, Phys. Rev. C 95~(3) (2017) 035202.
\newblock \href {http://arxiv.org/abs/1703.06982} {\path{arXiv:1703.06982}}, \href {https://doi.org/10.1103/PhysRevC.95.035202} {\path{doi:10.1103/PhysRevC.95.035202}}.

\bibitem{CLAS:2016tqs}
P.~E. Bosted, et~al., {Target and beam-target spin asymmetries in exclusive pion electroproduction for $Q^2>1$ GeV$^2$. II. $e p \rightarrow e \pi^0 p$}, Phys. Rev. C 95~(3) (2017) 035207.
\newblock \href {http://arxiv.org/abs/1611.04987} {\path{arXiv:1611.04987}}, \href {https://doi.org/10.1103/PhysRevC.95.035207} {\path{doi:10.1103/PhysRevC.95.035207}}.

\bibitem{CLAS:2016cks}
M.~Mayer, et~al., {Beam-target double-spin asymmetry in quasielastic electron scattering off the deuteron with CLAS}, Phys. Rev. C 95~(2) (2017) 024005.
\newblock \href {http://arxiv.org/abs/1610.06109} {\path{arXiv:1610.06109}}, \href {https://doi.org/10.1103/PhysRevC.95.024005} {\path{doi:10.1103/PhysRevC.95.024005}}.

\bibitem{CLAS:2016dlz}
P.~E. Bosted, et~al., {Target and Beam-Target Spin Asymmetries in Exclusive Pion Electroproduction for $Q^2>1$ GeV$^2$. I. $e p \rightarrow e \pi^+ n$}, Phys. Rev. C 95~(3) (2017) 035206.
\newblock \href {http://arxiv.org/abs/1607.07518} {\path{arXiv:1607.07518}}, \href {https://doi.org/10.1103/PhysRevC.95.035206} {\path{doi:10.1103/PhysRevC.95.035206}}.

\bibitem{CLAS:2016qxj}
X.~Zheng, et~al., {Measurement of Target and Double-spin Asymmetries for the $\vec e\vec p\to e\pi^+ (n)$ Reaction in the Nucleon Resonance Region at Low $Q^2$}, Phys. Rev. C 94~(4) (2016) 045206.
\newblock \href {http://arxiv.org/abs/1607.03924} {\path{arXiv:1607.03924}}, \href {https://doi.org/10.1103/PhysRevC.94.045206} {\path{doi:10.1103/PhysRevC.94.045206}}.

\bibitem{CLAS:2016ikd}
P.~E. Bosted, et~al., {Target and beam-target spin asymmetries in exclusive $\pi^+$ and $\pi^-$ electroproduction with 1.6- to 5.7-GeV electrons}, Phys. Rev. C 94~(5) (2016) 055201.
\newblock \href {http://arxiv.org/abs/1604.04350} {\path{arXiv:1604.04350}}, \href {https://doi.org/10.1103/PhysRevC.94.055201} {\path{doi:10.1103/PhysRevC.94.055201}}.

\bibitem{Kim:2015pkf}
A.~Kim, et~al., {Target and double spin asymmetries of deeply virtual $\pi^0$ production with a longitudinally polarized proton target and CLAS}, Phys. Lett. B 768 (2017) 168--173.
\newblock \href {http://arxiv.org/abs/1511.03338} {\path{arXiv:1511.03338}}, \href {https://doi.org/10.1016/j.physletb.2017.02.032} {\path{doi:10.1016/j.physletb.2017.02.032}}.

\bibitem{Skorodumina:2015ccu}
I.~A. Skorodumina, V.~D. Burkert, E.~N. Golovach, R.~W. Gothe, E.~L. Isupov, B.~S. Ishkhanov, V.~I. Mokeev, G.~V. Fedotov, {Nucleon resonances in exclusive reactions of photo- and electroproduction of mesons}, Moscow Univ. Phys. Bull. 70~(6) (2015) 429--447.
\newblock \href {https://doi.org/10.3103/S002713491506017X} {\path{doi:10.3103/S002713491506017X}}.

\bibitem{Mokeev:2015lda}
V.~I. Mokeev, et~al., {New Results from the \mbox{Studies} of the $N(1440)\frac{1}{2}^+$, $N(1520)\frac{3}{2}^-$, and $\Delta(1620)\frac{1}{2}^-$ Resonances in Exclusive $ep \to e'p'\pi^+ \pi^-$ Electroproduction with the CLAS Detector}, Phys. Rev. C 93~(2) (2016) 025206.

\bibitem{CLAS:2014fml}
K.~Park, et~al., {Measurements of $ep \to e^\prime \pi^+n$ at W = 1.6 - 2.0 GeV and extraction of nucleon resonance electrocouplings at CLAS}, Phys. Rev. C 91 (2015) 045203.
\newblock \href {http://arxiv.org/abs/1412.0274} {\path{arXiv:1412.0274}}, \href {https://doi.org/10.1103/PhysRevC.91.045203} {\path{doi:10.1103/PhysRevC.91.045203}}.

\bibitem{CLAS:2014jpc}
I.~Bedlinskiy, et~al., {Exclusive ${\pi}^0$ electroproduction at $W>2$ GeV with CLAS}, Phys. Rev. C 90~(2) (2014) 025205, [Addendum: Phys.Rev.C 90, 039901 (2014)].
\newblock \href {http://arxiv.org/abs/1405.0988} {\path{arXiv:1405.0988}}, \href {https://doi.org/10.1103/PhysRevC.90.039901} {\path{doi:10.1103/PhysRevC.90.039901}}.

\bibitem{CLAS:2014udv}
M.~Gabrielyan, et~al., {Induced polarization of $\Lambda(1116)$ in kaon electroproduction}, Phys. Rev. C 90~(3) (2014) 035202.
\newblock \href {http://arxiv.org/abs/1406.4046} {\path{arXiv:1406.4046}}, \href {https://doi.org/10.1103/PhysRevC.90.035202} {\path{doi:10.1103/PhysRevC.90.035202}}.

\bibitem{Aznauryan:2015zta}
I.~G. Aznauryan, V.~D. Burkert, {Electroexcitation of the $\Delta(1232)\frac{3}{2}^+$ and $\Delta(1600)\frac{3}{2}^+$ in a light-front relativistic quark model}, Phys. Rev. C 92~(3) (2015) 035211.

\bibitem{Aznauryan:2014xea}
I.~G. Aznauryan, V.~D. Burkert, {Extracting meson-baryon contributions to the electroexcitation of the $N(1675){\frac{5}{2}}^-$ nucleon resonance}, Phys. Rev. C 92~(1) (2015) 015203.
\newblock \href {http://arxiv.org/abs/1412.1296} {\path{arXiv:1412.1296}}, \href {https://doi.org/10.1103/PhysRevC.92.015203} {\path{doi:10.1103/PhysRevC.92.015203}}.

\bibitem{CLAS:2012qga}
K.~Park, et~al., {Deep exclusive $\pi^+$ electroproduction off the proton at CLAS}, Eur. Phys. J. A 49 (2013) 16.
\newblock \href {http://arxiv.org/abs/1206.2326} {\path{arXiv:1206.2326}}, \href {https://doi.org/10.1140/epja/i2013-13016-9} {\path{doi:10.1140/epja/i2013-13016-9}}.

\bibitem{CLAS:2012wxw}
V.~I. Mokeev, et~al., {Experimental Study of the $P_{11}(1440)$ and $D_{13}(1520)$ resonances from CLAS data on $ep \rightarrow e'\pi^{+} \pi^{-} p'$}, Phys. Rev. C 86 (2012) 035203.
\newblock \href {http://arxiv.org/abs/1205.3948} {\path{arXiv:1205.3948}}, \href {https://doi.org/10.1103/PhysRevC.86.035203} {\path{doi:10.1103/PhysRevC.86.035203}}.

\bibitem{CLAS:2012cna}
I.~Bedlinskiy, et~al., {Measurement of Exclusive $\pi^0$ Electroproduction Structure Functions and their Relationship to Transversity GPDs}, Phys. Rev. Lett. 109 (2012) 112001.
\newblock \href {http://arxiv.org/abs/1206.6355} {\path{arXiv:1206.6355}}, \href {https://doi.org/10.1103/PhysRevLett.109.112001} {\path{doi:10.1103/PhysRevLett.109.112001}}.

\bibitem{Aznauryan:2012ec}
I.~G. Aznauryan, V.~D. Burkert, {Nucleon electromagnetic form factors and electroexcitation of low lying nucleon resonances in a light-front relativistic quark model}, Phys. Rev. C 85 (2012) 055202.

\bibitem{CLAS:2009ces}
I.~G. Aznauryan, et~al., {Electroexcitation of nucleon resonances from CLAS data on single pion electroproduction}, Phys. Rev. C 80 (2009) 055203.
\newblock \href {https://doi.org/10.1103/PhysRevC.80.055203} {\path{doi:10.1103/PhysRevC.80.055203}}.

\bibitem{CLAS:2009sbn}
D.~S. Carman, et~al., {Beam-Recoil Polarization Transfer in the Nucleon Resonance Region in the Exclusive $\vec{e}p\to e' K^+ \vec{\Lambda}$ and $\vec{e}p K^+ \vec{\Sigma^0}$ Reactions at CLAS}, Phys. Rev. C 79 (2009) 065205.
\newblock \href {http://arxiv.org/abs/0904.3246} {\path{arXiv:0904.3246}}, \href {https://doi.org/10.1103/PhysRevC.79.065205} {\path{doi:10.1103/PhysRevC.79.065205}}.

\bibitem{CLAS:2008roe}
I.~G. Aznauryan, et~al., {Electroexcitation of the Roper resonance for $1.7 < Q^2 < 4.5$\,GeV$^2$ in $\vec ep \to e n \pi^+$}, Phys. Rev. C 78 (2008) 045209.
\newblock \href {http://arxiv.org/abs/0804.0447} {\path{arXiv:0804.0447}}, \href {https://doi.org/10.1103/PhysRevC.78.045209} {\path{doi:10.1103/PhysRevC.78.045209}}.

\bibitem{CLAS:2008wls}
A.~S. Biselli, et~al., {First measurement of target and double spin asymmetries for $\vec e \vec p \to e p \pi^0$ in the nucleon resonance region above the $\Delta(1232)$}, Phys. Rev. C 78 (2008) 045204.
\newblock \href {http://arxiv.org/abs/0804.3079} {\path{arXiv:0804.3079}}, \href {https://doi.org/10.1103/PhysRevC.78.045204} {\path{doi:10.1103/PhysRevC.78.045204}}.

\bibitem{CLAS:2008agj}
R.~Nasseripour, et~al., {Polarized Structure Function $\sigma_{LT'}$ for $^1H(\vec e, e' K^+) \Lambda$ in the Nucleon Resonance Region}, Phys. Rev. C 77 (2008) 065208.
\newblock \href {http://arxiv.org/abs/0801.4711} {\path{arXiv:0801.4711}}, \href {https://doi.org/10.1103/PhysRevC.77.065208} {\path{doi:10.1103/PhysRevC.77.065208}}.

\bibitem{CLAS:2007jvi}
R.~De~Masi, et~al., {Measurement of $e p \to e p \pi^0$ beam spin asymmetries above the resonance region}, Phys. Rev. C 77 (2008) 042201.
\newblock \href {http://arxiv.org/abs/0711.4736} {\path{arXiv:0711.4736}}, \href {https://doi.org/10.1103/PhysRevC.77.042201} {\path{doi:10.1103/PhysRevC.77.042201}}.

\bibitem{CLAS:2008ihz}
G.~V. Fedotov, et~al., {Electroproduction of $p \pi^+ \pi^-$ off protons at $0.2 < Q^2 < 0.6$\,GeV$^2$ and $1.3 < W < 1.57$\,GeV with CLAS}, Phys. Rev. C 79 (2009) 015204.
\newblock \href {http://arxiv.org/abs/0809.1562} {\path{arXiv:0809.1562}}, \href {https://doi.org/10.1103/PhysRevC.79.015204} {\path{doi:10.1103/PhysRevC.79.015204}}.

\bibitem{CLAS:2008rpm}
S.~A. Morrow, et~al., {Exclusive $rho^0$ electroproduction on the proton at CLAS}, Eur. Phys. J. A 39 (2009) 5--31.
\newblock \href {http://arxiv.org/abs/0807.3834} {\path{arXiv:0807.3834}}, \href {https://doi.org/10.1140/epja/i2008-10683-5} {\path{doi:10.1140/epja/i2008-10683-5}}.

\bibitem{CLAS:2008cms}
J.~P. Santoro, et~al., {Electroproduction of $\phi(1020)$ mesons at $1.4 < Q^2 < 3.8$\,GeV$^2$ measured with the CLAS spectrometer}, Phys. Rev. C 78 (2008) 025210.
\newblock \href {http://arxiv.org/abs/0803.3537} {\path{arXiv:0803.3537}}, \href {https://doi.org/10.1103/PhysRevC.78.025210} {\path{doi:10.1103/PhysRevC.78.025210}}.

\bibitem{CLAS:2007bvs}
H.~Denizli, et~al., {$Q^2$ dependence of the $S_{11}(1535)$ photocoupling and evidence for a P-wave resonance in $\eta$ electroproduction}, Phys. Rev. C 76 (2007) 015204.
\newblock \href {http://arxiv.org/abs/0704.2546} {\path{arXiv:0704.2546}}, \href {https://doi.org/10.1103/PhysRevC.76.015204} {\path{doi:10.1103/PhysRevC.76.015204}}.

\bibitem{CLAS:2006sjw}
H.~Egiyan, et~al., {Single $\pi^+$ electroproduction on the proton in the first and second resonance regions at 0.25\,GeV$^2 < Q^2 < 0.65$\,GeV$^2$ using CLAS}, Phys. Rev. C 73 (2006) 025204.
\newblock \href {http://arxiv.org/abs/nucl-ex/0601007} {\path{arXiv:nucl-ex/0601007}}, \href {https://doi.org/10.1103/PhysRevC.73.025204} {\path{doi:10.1103/PhysRevC.73.025204}}.

\bibitem{CLAS:2006ogr}
P.~Ambrozewicz, et~al., {Separated structure functions for the exclusive electroproduction of $K^+ \Lambda$ and $K^+ \Sigma^0$ final states}, Phys. Rev. C 75 (2007) 045203.
\newblock \href {http://arxiv.org/abs/hep-ex/0611036} {\path{arXiv:hep-ex/0611036}}, \href {https://doi.org/10.1103/PhysRevC.75.045203} {\path{doi:10.1103/PhysRevC.75.045203}}.

\bibitem{CLAS:2006ezq}
M.~Ungaro, et~al., {Measurement of the $N \to \Delta^+(1232)$ transition at high momentum transfer by $\pi^0$ electroproduction}, Phys. Rev. Lett. 97 (2006) 112003.
\newblock \href {https://doi.org/10.1103/PhysRevLett.97.112003} {\path{doi:10.1103/PhysRevLett.97.112003}}.

\bibitem{Aznauryan:2005tp}
I.~G. Aznauryan, et~al., {Electroexcitation of nucleon resonances at $Q^2 = 0.65$(GeV/c)$^2$ from a combined analysis of single- and double-pion electroproduction data}, Phys. Rev. C 72 (2005) 045201.
\newblock \href {http://arxiv.org/abs/hep-ph/0508057} {\path{arXiv:hep-ph/0508057}}, \href {https://doi.org/10.1103/PhysRevC.72.045201} {\path{doi:10.1103/PhysRevC.72.045201}}.

\bibitem{CLAS:2005vxa}
K.~Joo, et~al., {Measurement of the polarized structure function $\sigma_{LT^\prime}$ for pion electroproduction in the Roper resonance region}, Phys. Rev. C 72 (2005) 058202.
\newblock \href {http://arxiv.org/abs/nucl-ex/0504027} {\path{arXiv:nucl-ex/0504027}}, \href {https://doi.org/10.1103/PhysRevC.72.058202} {\path{doi:10.1103/PhysRevC.72.058202}}.

\bibitem{CLAS:2005nkx}
L.~Morand, et~al., {Deeply virtual and exclusive electroproduction of $\omega$ mesons}, Eur. Phys. J. A 24 (2005) 445--458.
\newblock \href {http://arxiv.org/abs/hep-ex/0504057} {\path{arXiv:hep-ex/0504057}}, \href {https://doi.org/10.1140/epja/i2005-10032-4} {\path{doi:10.1140/epja/i2005-10032-4}}.

\bibitem{CLAS:2004cri}
C.~Hadjidakis, et~al., {Exclusive $\rho^0$ meson electroproduction from hydrogen at CLAS}, Phys. Lett. B 605 (2005) 256--264.
\newblock \href {http://arxiv.org/abs/hep-ex/0408005} {\path{arXiv:hep-ex/0408005}}, \href {https://doi.org/10.1016/j.physletb.2004.11.019} {\path{doi:10.1016/j.physletb.2004.11.019}}.

\bibitem{Aznauryan:2004jd}
I.~G. Aznauryan, et~al., {Electroexcitation of the $P_{33}(1232)$, $P_{11}(1440)$, $D_{13}(1520)$, $S_{11}(1535)$ at $Q^2$ = 0.4 and 0.65 (GeV/c)$^2$}, Phys. Rev. C 71 (2005) 015201.
\newblock \href {http://arxiv.org/abs/nucl-th/0407021} {\path{arXiv:nucl-th/0407021}}, \href {https://doi.org/10.1103/PhysRevC.71.015201} {\path{doi:10.1103/PhysRevC.71.015201}}.

\bibitem{CLAS:2004ncx}
K.~Joo, et~al., {Measurement of the polarized structure function $\sigma_{LT^\prime}$ for $p(\vec e, e^\prime \pi^+)n$ in the $\Delta(1232)$ resonance region}, Phys. Rev. C 70 (2004) 042201.
\newblock \href {http://arxiv.org/abs/nucl-ex/0407013} {\path{arXiv:nucl-ex/0407013}}, \href {https://doi.org/10.1103/PhysRevC.70.042201} {\path{doi:10.1103/PhysRevC.70.042201}}.

\bibitem{CLAS:2002xbv}
M.~Ripani, et~al., {Measurement of $e p \to e^\prime p \pi^+ \pi^-$ and baryon resonance analysis}, Phys. Rev. Lett. 91 (2003) 022002.
\newblock \href {http://arxiv.org/abs/hep-ex/0210054} {\path{arXiv:hep-ex/0210054}}, \href {https://doi.org/10.1103/PhysRevLett.91.022002} {\path{doi:10.1103/PhysRevLett.91.022002}}.

\bibitem{CLAS:2003hro}
A.~S. Biselli, et~al., {Study of $e p \to e p \pi^0$ in the $\Delta(1232)$ mass region using polarization asymmetries}, Phys. Rev. C 68 (2003) 035202.
\newblock \href {http://arxiv.org/abs/nucl-ex/0307004} {\path{arXiv:nucl-ex/0307004}}, \href {https://doi.org/10.1103/PhysRevC.68.035202} {\path{doi:10.1103/PhysRevC.68.035202}}.

\bibitem{CLAS:2002zlc}
D.~S. Carman, et~al., {First measurement of transferred polarization in the exclusive $\vec e p \to e^\prime K^+ \vec\Lambda$ reaction}, Phys. Rev. Lett. 90 (2003) 131804.
\newblock \href {http://arxiv.org/abs/hep-ex/0212014} {\path{arXiv:hep-ex/0212014}}, \href {https://doi.org/10.1103/PhysRevLett.90.131804} {\path{doi:10.1103/PhysRevLett.90.131804}}.

\bibitem{CLAS:2001cbm}
K.~Joo, et~al., {$Q^2$ dependence of quadrupole strength in the $\gamma^* p \to \Delta^+(1232)\to p \pi^0$ transition}, Phys. Rev. Lett. 88 (2002) 122001.
\newblock \href {https://doi.org/10.1103/PhysRevLett.88.122001} {\path{doi:10.1103/PhysRevLett.88.122001}}.

\bibitem{CLAS:2000mbw}
R.~Thompson, et~al., {The $e p \to e^\prime p \eta$ reaction at and above the $S_{11}(1535)$ baryon resonance}, Phys. Rev. Lett. 86 (2001) 1702--1706.
\newblock \href {http://arxiv.org/abs/hep-ex/0011029} {\path{arXiv:hep-ex/0011029}}, \href {https://doi.org/10.1103/PhysRevLett.86.1702} {\path{doi:10.1103/PhysRevLett.86.1702}}.

\bibitem{Breuker:1977vy}
H.~Breuker, et~al., {Forward $\pi^+$ Electroproduction in the First Resonance Region at Four Momentum Transfers $q^2 = 0.15$\,(GeV/c)$^2$ and 0.3\,(GeV/c)$^2$}, Nucl. Phys. B 146 (1978) 285--302.
\newblock \href {https://doi.org/10.1016/0550-3213(78)90069-X} {\path{doi:10.1016/0550-3213(78)90069-X}}.

\bibitem{Breuker:1982nw}
H.~Breuker, et~al., {Electroproduction of $\pi^+$ Mesons at Forward and Backward Direction in the Region of the $D^-$13 (1520) and $^{15}$F (1688) Resonances}, Z. Phys. C 13 (1982) 113.
\newblock \href {https://doi.org/10.1007/BF01547674} {\path{doi:10.1007/BF01547674}}.

\bibitem{Beck:1974wd}
U.~Beck, et~al., {Electroproduction of $\eta$-mesons at the $S_{11}(1535)$ resonance}, Phys. Lett. B 51 (1974) 103--105.
\newblock \href {https://doi.org/10.1016/0370-2693(74)90162-2} {\path{doi:10.1016/0370-2693(74)90162-2}}.

\bibitem{Siddle:1971ug}
R.~Siddle, et~al., {Coincidence $\pi^0$ electroproduction experiments in the first resonance region at momentum transfers of 0.3, 0.45, 0.60, 0.76 GeV/c$^2$}, Nucl. Phys. B 35 (1971) 93--119.
\newblock \href {https://doi.org/10.1016/0550-3213(71)90134-9} {\path{doi:10.1016/0550-3213(71)90134-9}}.

\bibitem{Alder:1975tv}
J.~C. Alder, et~al., {Electroproduction of $\eta$ mesons in the region of the resonance S$_{11}$(1535)}, Nucl. Phys. B 91 (1975) 386--398.
\newblock \href {https://doi.org/10.1016/0550-3213(75)90114-5} {\path{doi:10.1016/0550-3213(75)90114-5}}.

\bibitem{Brasse:1975bg}
F.~W. Brasse, et~al., {Electroproduction of Neutral Pions at Energies Above the Resonance Region}, Phys. Lett. B 58 (1975) 467.
\newblock \href {https://doi.org/10.1016/0370-2693(75)90703-0} {\path{doi:10.1016/0370-2693(75)90703-0}}.

\bibitem{Alder:1975na}
J.~C. Alder, et~al., {Electroproduction of $\pi^+$ Mesons in the Resonance Region}, Nucl. Phys. B 99 (1975) 1--12.
\newblock \href {https://doi.org/10.1016/0550-3213(75)90052-8} {\path{doi:10.1016/0550-3213(75)90052-8}}.

\bibitem{Alder:1975xt}
J.~C. Alder, et~al., {Electroproduction of neutral pions in the resonance region}, Nucl. Phys. B 105 (1976) 253--271.
\newblock \href {https://doi.org/10.1016/0550-3213(76)90266-2} {\path{doi:10.1016/0550-3213(76)90266-2}}.

\bibitem{Brasse:1976bf}
F.~W. Brasse, et~al., {Parametrization of the $q^2$ dependence of $\gamma_V p$ total cross sections in the resonance region}, Nucl. Phys. B 110 (1976) 413--433.
\newblock \href {https://doi.org/10.1016/0550-3213(76)90231-5} {\path{doi:10.1016/0550-3213(76)90231-5}}.

\bibitem{Brasse:1977as}
F.~W. Brasse, et~al., {Separation of $\Sigma_L$ and $\Sigma_T$ in $\eta$-Electroproduction at the Resonance S$_{11}$(1535)}, Nucl. Phys. B 139 (1978) 37--44.
\newblock \href {https://doi.org/10.1016/0550-3213(78)90177-3} {\path{doi:10.1016/0550-3213(78)90177-3}}.

\bibitem{Haidan:1979yqa}
R.~Haidan, {Elektroproduktion pseudoskalarer Mesonen im Resonanzgebiet bei gro\ss{}en Impuls\"ubertr\"agen}, Ph.D. thesis, Hamburg U. (1979).
\newblock \href {https://doi.org/10.3204/PUBDB-2017-12089} {\path{doi:10.3204/PUBDB-2017-12089}}.

\bibitem{Koch:1980ae}
R.~Koch, {The Karlsruhe-Helsinki $\pi N$ elastic partial wave analysis. }In *Toronto 1980, Proceedings, Baryon 1980*, 3-17 (1980).

\bibitem{Hohler:1992ru}
G.~{\protect H\"ohler}, A.~Schulte, {Determination of $\pi N$ resonance pole parameters from speed plots}, PiN Newslett. 7 (1992) 94.

\bibitem{Cutkosky:1979fy}
R.~E. Cutkosky, C.~P. Forsyth, R.~E. Hendrick, R.~L. Kelly, {Pion - nucleon partial wave amplitudes}, Phys. Rev. D20 (1979) 2839.
\newblock \href {https://doi.org/10.1103/PhysRevD.20.2839} {\path{doi:10.1103/PhysRevD.20.2839}}.

\bibitem{Cutkosky:1979zv}
R.~E. Cutkosky, et~al., {Pion - nucleon partial wave analysis}, Phys. Rev. D20 (1979) 2804.
\newblock \href {https://doi.org/10.1103/PhysRevD.20.2804} {\path{doi:10.1103/PhysRevD.20.2804}}.

\bibitem{Cutkosky:1980rh}
R.~E. Cutkosky, et~al., {Pion - Nucleon Partial Wave Analysis}, in: {4th International Conference on Baryon Resonances}, 1980, p.~19.

\bibitem{Arndt:1970wr}
R.~A. Arndt, L.~D. Roper, {Deuteron constraints on low-energy nucleon-nucleon scattering analyses}, Phys. Rev. D1 (1970) 129--133.
\newblock \href {https://doi.org/10.1103/PhysRevD.1.129} {\path{doi:10.1103/PhysRevD.1.129}}.

\bibitem{Arndt:1985vj}
R.~A. Arndt, J.~M. Ford, L.~D. Roper, {Pion - nucleon partial wave analysis to 1100\,MeV}, Phys. Rev. D32 (1985) 1085.
\newblock \href {https://doi.org/10.1103/PhysRevD.32.1085} {\path{doi:10.1103/PhysRevD.32.1085}}.

\bibitem{Arndt:1990bp}
R.~A. Arndt, others., {Pion - nucleon partial wave analysis to 2\,GeV}, Phys. Rev. D43 (1991) 2131--2139.
\newblock \href {https://doi.org/10.1103/PhysRevD.43.2131} {\path{doi:10.1103/PhysRevD.43.2131}}.

\bibitem{Arndt:1995bj}
R.~A. Arndt, I.~I. Strakovsky, R.~L. Workman, M.~M. Pavan, {Updated analysis of $\pi N$ elastic scattering data to 2.1\,GeV: The baryon spectrum}, Phys. Rev. C52 (1995) 2120--2130.
\newblock \href {http://arxiv.org/abs/nucl-th/9505040} {\path{arXiv:nucl-th/9505040}}, \href {https://doi.org/10.1103/PhysRevC.52.2120} {\path{doi:10.1103/PhysRevC.52.2120}}.

\bibitem{Arndt:2003if}
R.~A. Arndt, et~al., {Dispersion relation constrained partial wave analysis of $\pi N$ elastic and $\pi N\to\eta N$ scattering data: The baryon spectrum}, Phys. Rev. C69 (2004) 035213.
\newblock \href {http://arxiv.org/abs/nucl-th/0311089} {\path{arXiv:nucl-th/0311089}}, \href {https://doi.org/10.1103/PhysRevC.69.035213} {\path{doi:10.1103/PhysRevC.69.035213}}.

\bibitem{Arndt:2006bf}
R.~A. Arndt, W.~J. Briscoe, I.~I. Strakovsky, R.~L. Workman, {Extended partial-wave analysis of $\pi N$ scattering data}, Phys. Rev. C74 (2006) 045205.
\newblock \href {http://arxiv.org/abs/nucl-th/0605082} {\path{arXiv:nucl-th/0605082}}, \href {https://doi.org/10.1103/PhysRevC.74.045205} {\path{doi:10.1103/PhysRevC.74.045205}}.

\bibitem{Workman:2008iv}
R.~L. Workman, R.~A. Arndt, M.~W. Paris, {Resonance parameters from K-matrix and T-matrix poles}, Phys. Rev. C 79 (2009) 038201.
\newblock \href {http://arxiv.org/abs/0808.2176} {\path{arXiv:0808.2176}}, \href {https://doi.org/10.1103/PhysRevC.79.038201} {\path{doi:10.1103/PhysRevC.79.038201}}.

\bibitem{Workman:2011vb}
R.~L. Workman, W.~J. Briscoe, M.~W. Paris, I.~I. Strakovsky, {Updated SAID analysis of pion photoproduction data}, Phys. Rev. C 85 (2012) 025201.
\newblock \href {http://arxiv.org/abs/1109.0722} {\path{arXiv:1109.0722}}, \href {https://doi.org/10.1103/PhysRevC.85.025201} {\path{doi:10.1103/PhysRevC.85.025201}}.

\bibitem{Workman:2012jf}
R.~L. Workman, M.~W. Paris, W.~J. Briscoe, I.~I. Strakovsky, {Unified Chew-Mandelstam SAID analysis of pion photoproduction data}, Phys. Rev. C 86 (2012) 015202.
\newblock \href {http://arxiv.org/abs/1202.0845} {\path{arXiv:1202.0845}}, \href {https://doi.org/10.1103/PhysRevC.86.015202} {\path{doi:10.1103/PhysRevC.86.015202}}.

\bibitem{Workman:2012hx}
R.~L. Workman, et~al., {Parameterization dependence of T matrix poles and eigenphases from a fit to $\pi$N elastic scattering data}, Phys. Rev. C 86 (2012) 035202.
\newblock \href {http://arxiv.org/abs/1204.2277} {\path{arXiv:1204.2277}}, \href {https://doi.org/10.1103/PhysRevC.86.035202} {\path{doi:10.1103/PhysRevC.86.035202}}.

\bibitem{Briscoe:2012ni}
W.~J. Briscoe, et~al., {Evaluation of the $\gamma n \to\pi^-p$ differential cross sections in the $\Delta$-isobar region}, Phys. Rev. C 86 (2012) 065207.
\newblock \href {http://arxiv.org/abs/1209.0024} {\path{arXiv:1209.0024}}, \href {https://doi.org/10.1103/PhysRevC.86.065207} {\path{doi:10.1103/PhysRevC.86.065207}}.

\bibitem{Workman:2016ysf}
R.~L. Workman, W.~J. Briscoe, I.~I. Strakovsky, {Partial-Wave Analysis of Nucleon-Nucleon Elastic Scattering Data}, Phys. Rev. C 94~(6) (2016) 065203.
\newblock \href {http://arxiv.org/abs/1609.01741} {\path{arXiv:1609.01741}}, \href {https://doi.org/10.1103/PhysRevC.94.065203} {\path{doi:10.1103/PhysRevC.94.065203}}.

\bibitem{Briscoe:2020qat}
W.~J. Briscoe, et~al., {Threshold $\pi ^-$ photoproduction on the neutron}, Eur. Phys. J. A 56~(8) (2020) 218.
\newblock \href {http://arxiv.org/abs/2004.01742} {\path{arXiv:2004.01742}}, \href {https://doi.org/10.1140/epja/s10050-020-00221-w} {\path{doi:10.1140/epja/s10050-020-00221-w}}.

\bibitem{Briscoe:2021siu}
W.~J. Briscoe, et~al., {On the photoproduction reactions $\gamma d \rightarrow \pi NN$}, Eur. Phys. J. A 58~(2) (2022) 23.
\newblock \href {http://arxiv.org/abs/2112.08150} {\path{arXiv:2112.08150}}, \href {https://doi.org/10.1140/epja/s10050-022-00671-4} {\path{doi:10.1140/epja/s10050-022-00671-4}}.

\bibitem{Strakovsky:2022tvu}
I.~Strakovsky, et~al., {Single-pion contribution to the Gerasimov-Drell-Hearn sum rule and related integrals}, Phys. Rev. C 105~(4) (2022) 045202.
\newblock \href {http://arxiv.org/abs/2201.06495} {\path{arXiv:2201.06495}}, \href {https://doi.org/10.1103/PhysRevC.105.045202} {\path{doi:10.1103/PhysRevC.105.045202}}.

\bibitem{Briscoe:2023gmb}
W.~J. Briscoe, et~al., {Extended SAID partial-wave analysis of pion photoproduction}, Phys. Rev. C 108~(6) (2023) 065205.
\newblock \href {http://arxiv.org/abs/2309.06631} {\path{arXiv:2309.06631}}, \href {https://doi.org/10.1103/PhysRevC.108.065205} {\path{doi:10.1103/PhysRevC.108.065205}}.

\bibitem{Drechsel:1998hk}
D.~Drechsel, O.~Hanstein, S.~S. Kamalov, L.~Tiator, {A Unitary isobar model for pion photoproduction and electroproduction on the proton up to 1-GeV}, Nucl. Phys. A 645 (1999) 145--174.
\newblock \href {http://arxiv.org/abs/nucl-th/9807001} {\path{arXiv:nucl-th/9807001}}, \href {https://doi.org/10.1016/S0375-9474(98)00572-7} {\path{doi:10.1016/S0375-9474(98)00572-7}}.

\bibitem{Kamalov:2000en}
S.~S. Kamalov, et~al., {$\gamma^* N \to \Delta$ transition form-factors: A New analysis of the JLab data on $p (e, e' p) \pi^0$ at $Q^2$ = 2.8-(GeV/c)$^2$ and 4.0-(GeV/c)$^2$)}, Phys. Rev. C 64 (2001) 032201.
\newblock \href {http://arxiv.org/abs/nucl-th/0006068} {\path{arXiv:nucl-th/0006068}}, \href {https://doi.org/10.1103/PhysRevC.64.032201} {\path{doi:10.1103/PhysRevC.64.032201}}.

\bibitem{Tiator:2008kd}
L.~Tiator, M.~Vanderhaeghen, {Empirical transverse charge densities in the nucleon-to-$P_{11}(1440)$ transition}, Phys. Lett. B 672 (2009) 344--348.
\newblock \href {http://arxiv.org/abs/0811.2285} {\path{arXiv:0811.2285}}, \href {https://doi.org/10.1016/j.physletb.2009.01.048} {\path{doi:10.1016/j.physletb.2009.01.048}}.

\bibitem{Tiator:2010rp}
L.~Tiator, et~al., {Singularity structure of the $\pi N$ scattering amplitude in a meson-exchange model up to energies $W \leq 2.0-$GeV}, Phys. Rev. C 82 (2010) 055203.
\newblock \href {http://arxiv.org/abs/1007.2126} {\path{arXiv:1007.2126}}, \href {https://doi.org/10.1103/PhysRevC.82.055203} {\path{doi:10.1103/PhysRevC.82.055203}}.

\bibitem{Stajner:2017fmh}
S.~\v{S}tajner, et~al., {Beam-Recoil Polarization Measurement of $\pi^0$ Electroproduction on the Proton in the Region of the Roper Resonance}, Phys. Rev. Lett. 119~(2) (2017) 022001.
\newblock \href {https://doi.org/10.1103/PhysRevLett.119.022001} {\path{doi:10.1103/PhysRevLett.119.022001}}.

\bibitem{A2atMAMI:2014rwc}
C.~S. Akondi, et~al., {Measurement of the Transverse Target and Beam-Target Asymmetries in \ensuremath{\eta} Meson Photoproduction at MAMI}, Phys. Rev. Lett. 113~(10) (2014) 102001.
\newblock \href {http://arxiv.org/abs/1408.3274} {\path{arXiv:1408.3274}}, \href {https://doi.org/10.1103/PhysRevLett.113.102001} {\path{doi:10.1103/PhysRevLett.113.102001}}.

\bibitem{Drechsel:2007if}
D.~Drechsel, S.~S. Kamalov, L.~Tiator, {Unitary Isobar Model - MAID2007}, Eur. Phys. J. A 34 (2007) 69--97.
\newblock \href {http://arxiv.org/abs/0710.0306} {\path{arXiv:0710.0306}}, \href {https://doi.org/10.1140/epja/i2007-10490-6} {\path{doi:10.1140/epja/i2007-10490-6}}.

\bibitem{Chiang:2001as}
W.-T. Chiang, S.-N. Yang, L.~Tiator, D.~Drechsel, {An Isobar model for $\eta$ photoproduction and electroproduction on the nucleon}, Nucl. Phys. A 700 (2002) 429--453.
\newblock \href {http://arxiv.org/abs/nucl-th/0110034} {\path{arXiv:nucl-th/0110034}}, \href {https://doi.org/10.1016/S0375-9474(01)01325-2} {\path{doi:10.1016/S0375-9474(01)01325-2}}.

\bibitem{Tiator:2018heh}
L.~Tiator, et~al., {$\eta$ and $\eta^\prime$ Photoproduction on the Nucleon with the Isobar Model EtaMAID2018}, Eur. Phys. J. A 54~(12) (2018) 210.
\newblock \href {http://arxiv.org/abs/1807.04525} {\path{arXiv:1807.04525}}, \href {https://doi.org/10.1140/epja/i2018-12643-x} {\path{doi:10.1140/epja/i2018-12643-x}}.

\bibitem{Fix:2005if}
A.~Fix, H.~Arenhövel, {Double pion photoproduction on nucleon and deuteron}, Eur. Phys. J. A 25 (2005) 115--135.
\newblock \href {http://arxiv.org/abs/nucl-th/0503042} {\path{arXiv:nucl-th/0503042}}, \href {https://doi.org/10.1140/epja/i2005-10067-5} {\path{doi:10.1140/epja/i2005-10067-5}}.

\bibitem{Fix:2010nv}
A.~Fix, H.~Arenhövel, {Polarization observables in $\pi^0\eta$-photoproduction on the proton}, Phys. Rev. C 83 (2011) 015503.
\newblock \href {http://arxiv.org/abs/1010.4213} {\path{arXiv:1010.4213}}, \href {https://doi.org/10.1103/PhysRevC.83.015503} {\path{doi:10.1103/PhysRevC.83.015503}}.

\bibitem{Mart:1999ed}
T.~Mart, C.~Bennhold, {Evidence for a missing nucleon resonance in kaon photoproduction}, Phys. Rev. C 61 (2000) 012201.
\newblock \href {http://arxiv.org/abs/nucl-th/9906096} {\path{arXiv:nucl-th/9906096}}, \href {https://doi.org/10.1103/PhysRevC.61.012201} {\path{doi:10.1103/PhysRevC.61.012201}}.

\bibitem{MAIDkph}
\url{https://maid.kph.uni-mainz.de/}.

\bibitem{Manley:1984ih}
D.~M. Manley, {Comment on 'Pion decay widths of $N$ and $\Delta$ baryons.'}, Phys. Rev. D 30 (1984) 240.
\newblock \href {https://doi.org/10.1103/PhysRevD.30.240} {\path{doi:10.1103/PhysRevD.30.240}}.

\bibitem{Manley:1984jz}
D.~M. Manley, R.~A. Arndt, Y.~Goradia, V.~L. Teplitz, {An isobar model partial wave analysis of $\pi N \to\pi\pi N$ in the center-of-mass energy range 1320\,MeV to 1930\,MeV}, Phys. Rev. D30 (1984) 904.
\newblock \href {https://doi.org/10.1103/PhysRevD.30.904} {\path{doi:10.1103/PhysRevD.30.904}}.

\bibitem{Shrestha:2012va}
M.~Shrestha, D.~M.~M. Manley, {Partial-Wave Analysis of $\pi^-p \to \eta n$ and $\pi^-p \to K^0\Lambda$ Reactions}, Phys. Rev. C 86 (2012) 045204.
\newblock \href {http://arxiv.org/abs/1205.5294} {\path{arXiv:1205.5294}}, \href {https://doi.org/10.1103/PhysRevC.86.045204} {\path{doi:10.1103/PhysRevC.86.045204}}.

\bibitem{Hunt:2018tvt}
B.~C. Hunt, D.~M. Manley, {Partial-wave analyses of $\gamma p \rightarrow \eta p$ and $\gamma n \rightarrow \eta n$ using a multichannel framework}, Phys. Rev. C 99~(5) (2019) 055203.
\newblock \href {http://arxiv.org/abs/1804.06031} {\path{arXiv:1804.06031}}, \href {https://doi.org/10.1103/PhysRevC.99.055203} {\path{doi:10.1103/PhysRevC.99.055203}}.

\bibitem{Hunt:2018mrt}
B.~C. Hunt, D.~M. Manley, {Partial-Wave Analysis of $\gamma p \rightarrow K^+ \Lambda$ using a multichannel framework}, Phys. Rev. C 99~(5) (2019) 055204.
\newblock \href {http://arxiv.org/abs/1804.07422} {\path{arXiv:1804.07422}}, \href {https://doi.org/10.1103/PhysRevC.99.055204} {\path{doi:10.1103/PhysRevC.99.055204}}.

\bibitem{Hunt:2018wqz}
B.~C. Hunt, D.~M. Manley, {Updated determination of $N^*$ resonance parameters using a unitary, multichannel formalism}, Phys. Rev. C 99~(5) (2019) 055205.
\newblock \href {http://arxiv.org/abs/1810.13086} {\path{arXiv:1810.13086}}, \href {https://doi.org/10.1103/PhysRevC.99.055205} {\path{doi:10.1103/PhysRevC.99.055205}}.

\bibitem{Penner:2001fu}
G.~Penner, U.~Mosel, {$\pi N \to \omega N$ in a coupled channel approach}, Phys. Rev. C 65 (2002) 055202, [Erratum: Phys.Rev.C 65, 059901 (2002)].
\newblock \href {http://arxiv.org/abs/nucl-th/0111023} {\path{arXiv:nucl-th/0111023}}, \href {https://doi.org/10.1103/PhysRevC.65.059901} {\path{doi:10.1103/PhysRevC.65.059901}}.

\bibitem{Penner:2002ma}
G.~Penner, U.~Mosel, {Vector meson production and nucleon resonance analysis in a coupled channel approach for energies $m_N <\sqrt s < 2$\,GeV. 1. Pion induced results and hadronic parameters}, Phys. Rev. C 66 (2002) 055211.
\newblock \href {http://arxiv.org/abs/nucl-th/0207066} {\path{arXiv:nucl-th/0207066}}, \href {https://doi.org/10.1103/PhysRevC.66.055211} {\path{doi:10.1103/PhysRevC.66.055211}}.

\bibitem{Penner:2002md}
G.~Penner, U.~Mosel, {Vector meson production and nucleon resonance analysis in a coupled channel approach for energies $m_N <\sqrt s < 2$\,GeV. 2. Photon induced results}, Phys. Rev. C 66 (2002) 055212.
\newblock \href {http://arxiv.org/abs/nucl-th/0207069} {\path{arXiv:nucl-th/0207069}}, \href {https://doi.org/10.1103/PhysRevC.66.055212} {\path{doi:10.1103/PhysRevC.66.055212}}.

\bibitem{Shklyar:2004dy}
V.~Shklyar, G.~Penner, U.~Mosel, {Spin 5/2 resonance contributions to the pion induced reactions for energies $sqrt s \leq 2.0$-GeV}, Eur. Phys. J. A 21 (2004) 445--454.
\newblock \href {http://arxiv.org/abs/nucl-th/0403064} {\path{arXiv:nucl-th/0403064}}, \href {https://doi.org/10.1140/epja/i2004-10003-3} {\path{doi:10.1140/epja/i2004-10003-3}}.

\bibitem{Shklyar:2004ba}
V.~Shklyar, H.~Lenske, U.~Mosel, G.~Penner, {Coupled-channel analysis of the omega-meson production in $\pi N$ and $\gamma N$ reactions for c.m. energies up to 2-GeV}, Phys. Rev. C 71 (2005) 055206, [Erratum: Phys.Rev.C 72, 019903 (2005)].
\newblock \href {http://arxiv.org/abs/nucl-th/0412029} {\path{arXiv:nucl-th/0412029}}, \href {https://doi.org/10.1103/PhysRevC.72.019903} {\path{doi:10.1103/PhysRevC.72.019903}}.

\bibitem{Shklyar:2005xg}
V.~Shklyar, H.~Lenske, U.~Mosel, {A Coupled-channel analysis of $K \Lambda$ production in the nucleon resonance region}, Phys. Rev. C 72 (2005) 015210.
\newblock \href {http://arxiv.org/abs/nucl-th/0505010} {\path{arXiv:nucl-th/0505010}}, \href {https://doi.org/10.1103/PhysRevC.72.015210} {\path{doi:10.1103/PhysRevC.72.015210}}.

\bibitem{Shklyar:2006xw}
V.~Shklyar, H.~Lenske, U.~Mosel, {$\eta$-photoproduction in the resonance energy region}, Phys. Lett. B 650 (2007) 172--178.
\newblock \href {http://arxiv.org/abs/nucl-th/0611036} {\path{arXiv:nucl-th/0611036}}, \href {https://doi.org/10.1016/j.physletb.2007.05.005} {\path{doi:10.1016/j.physletb.2007.05.005}}.

\bibitem{Shklyar:2012js}
V.~Shklyar, H.~Lenske, U.~Mosel, {$\eta$-meson production in the resonance-energy region}, Phys. Rev. C 87~(1) (2013) 015201.
\newblock \href {http://arxiv.org/abs/1206.5414} {\path{arXiv:1206.5414}}, \href {https://doi.org/10.1103/PhysRevC.87.015201} {\path{doi:10.1103/PhysRevC.87.015201}}.

\bibitem{Cao:2013psa}
X.~Cao, V.~Shklyar, H.~Lenske, {Coupled-channel analysis of $K \Sigma$ production on the nucleon up to 2.0 GeV}, Phys. Rev. C 88~(5) (2013) 055204.
\newblock \href {http://arxiv.org/abs/1303.2604} {\path{arXiv:1303.2604}}, \href {https://doi.org/10.1103/PhysRevC.88.055204} {\path{doi:10.1103/PhysRevC.88.055204}}.

\bibitem{Shklyar:2014kra}
V.~Shklyar, H.~Lenske, U.~Mosel, {$2\pi$ production in the Giessen coupled-channel model}, Phys. Rev. C 93~(4) (2016) 045206.
\newblock \href {http://arxiv.org/abs/1409.7920} {\path{arXiv:1409.7920}}, \href {https://doi.org/10.1103/PhysRevC.93.045206} {\path{doi:10.1103/PhysRevC.93.045206}}.

\bibitem{Anisovich:2004zz}
A.~Anisovich, E.~Klempt, A.~Sarantsev, U.~Thoma, {Partial wave decomposition of pion and photoproduction amplitudes}, Eur. Phys. J. A 24 (2005) 111--128.
\newblock \href {http://arxiv.org/abs/hep-ph/0407211} {\path{arXiv:hep-ph/0407211}}, \href {https://doi.org/10.1140/epja/i2004-10125-6} {\path{doi:10.1140/epja/i2004-10125-6}}.

\bibitem{Anisovich:2006bc}
A.~V. Anisovich, A.~V. Sarantsev, {Partial decay widths of baryons in the spin-momentum operator expansion method}, Eur. Phys. J. A 30 (2006) 427--441.
\newblock \href {http://arxiv.org/abs/hep-ph/0605135} {\path{arXiv:hep-ph/0605135}}, \href {https://doi.org/10.1140/epja/i2006-10102-1} {\path{doi:10.1140/epja/i2006-10102-1}}.

\bibitem{Burkert:2022bqo}
V.~Burkert, et~al., {Note on the definitions of branching ratios of overlapping resonances}, Phys. Lett. B 844 (2023) 138070.
\newblock \href {http://arxiv.org/abs/2207.08472} {\path{arXiv:2207.08472}}, \href {https://doi.org/10.1016/j.physletb.2023.138070} {\path{doi:10.1016/j.physletb.2023.138070}}.

\bibitem{Sarantsev:2005tg}
A.~V. Sarantsev, et~al., {Decays of baryon resonances into $\Lambda K^+, \Sigma^0 K^+$ and $\Sigma^+ K^0$}, Eur. Phys. J. A 25 (2005) 441--453.
\newblock \href {http://arxiv.org/abs/hep-ex/0506011} {\path{arXiv:hep-ex/0506011}}, \href {https://doi.org/10.1140/epja/i2005-10121-4} {\path{doi:10.1140/epja/i2005-10121-4}}.

\bibitem{Anisovich:2005tf}
A.~V. Anisovich, et~al., {Photoproduction of baryons decaying into $N\pi$ and $N \eta$}, Eur. Phys. J. A 25 (2005) 427--439.
\newblock \href {http://arxiv.org/abs/hep-ex/0506010} {\path{arXiv:hep-ex/0506010}}, \href {https://doi.org/10.1140/epja/i2005-10120-5} {\path{doi:10.1140/epja/i2005-10120-5}}.

\bibitem{Klempt:2006sa}
E.~Klempt, et~al., {Phase motion of baryon resonances}, Eur. Phys. J. A 29 (2006) 307--313.
\newblock \href {https://doi.org/10.1140/epja/i2006-10092-x} {\path{doi:10.1140/epja/i2006-10092-x}}.

\bibitem{Nikonov:2007br}
V.~A. Nikonov, et~al., {Further evidence for $N(1900) P_{13}$ from photoproduction of hyperons}, Phys. Lett. B 662 (2008) 245--251.
\newblock \href {http://arxiv.org/abs/0707.3600} {\path{arXiv:0707.3600}}, \href {https://doi.org/10.1016/j.physletb.2008.03.004} {\path{doi:10.1016/j.physletb.2008.03.004}}.

\bibitem{Anisovich:2007bq}
A.~V. Anisovich, et~al., {Baryon resonances and polarization transfer in hyperon photoproduction}, Eur. Phys. J. A 34 (2007) 243--254.
\newblock \href {http://arxiv.org/abs/0707.3596} {\path{arXiv:0707.3596}}, \href {https://doi.org/10.1140/epja/i2007-10503-6} {\path{doi:10.1140/epja/i2007-10503-6}}.

\bibitem{Anisovich:2008wd}
A.~V. Anisovich, et~al., {Photoproduction of $\eta$ mesons off neutrons from a deuteron target}, Eur. Phys. J. A 41 (2009) 13--24.
\newblock \href {http://arxiv.org/abs/0809.3340} {\path{arXiv:0809.3340}}, \href {https://doi.org/10.1140/epja/i2009-10766-9} {\path{doi:10.1140/epja/i2009-10766-9}}.

\bibitem{Anisovich:2010mks}
A.~V. Anisovich, et~al., {Photoproduction of pions and properties of baryon resonances from a Bonn-Gatchina partial wave analysis}, Eur. Phys. J. A 44 (2010) 203--220.
\newblock \href {http://arxiv.org/abs/0911.5277} {\path{arXiv:0911.5277}}, \href {https://doi.org/10.1140/epja/i2010-10950-x} {\path{doi:10.1140/epja/i2010-10950-x}}.

\bibitem{Anisovich:2010an}
A.~V. Anisovich, et~al., {P-wave excited baryons from pion- and photo-induced hyperon production}, Eur. Phys. J. A 47 (2011) 27.
\newblock \href {http://arxiv.org/abs/1009.4803} {\path{arXiv:1009.4803}}, \href {https://doi.org/10.1140/epja/i2011-11027-2} {\path{doi:10.1140/epja/i2011-11027-2}}.

\bibitem{Anisovich:2011ye}
A.~V. Anisovich, et~al., {Nucleon resonances in the fourth resonance region}, Eur. Phys. J. A 47 (2011) 153.
\newblock \href {http://arxiv.org/abs/1109.0970} {\path{arXiv:1109.0970}}, \href {https://doi.org/10.1140/epja/i2011-11153-9} {\path{doi:10.1140/epja/i2011-11153-9}}.

\bibitem{Anisovich:2011sv}
A.~V. Anisovich, et~al., {Evidence for a negative-parity spin-doublet of nucleon resonances at 1.88\textbackslash{},GeV}, Phys. Lett. B 711 (2012) 162--166.
\newblock \href {http://arxiv.org/abs/1111.6151} {\path{arXiv:1111.6151}}, \href {https://doi.org/10.1016/j.physletb.2012.03.077} {\path{doi:10.1016/j.physletb.2012.03.077}}.

\bibitem{Anisovich:2011su}
A.~V. Anisovich, et~al., {Evidence for a spin-quartet of nucleon resonances at 2\textbackslash{},GeV}, Phys. Lett. B 711 (2012) 167--172.
\newblock \href {http://arxiv.org/abs/1111.6150} {\path{arXiv:1111.6150}}, \href {https://doi.org/10.1016/j.physletb.2012.03.066} {\path{doi:10.1016/j.physletb.2012.03.066}}.

\bibitem{Anisovich:2011fc}
A.~V. Anisovich, et~al., {Properties of baryon resonances from a multichannel partial wave analysis}, Eur. Phys. J. A 48 (2012) 15.
\newblock \href {http://arxiv.org/abs/1112.4937} {\path{arXiv:1112.4937}}, \href {https://doi.org/10.1140/epja/i2012-12015-8} {\path{doi:10.1140/epja/i2012-12015-8}}.

\bibitem{Anisovich:2012ct}
A.~V. Anisovich, et~al., {Pion- and photo-induced transition amplitudes to $\Lambda K$, $\Sigma K$, and $N\eta$}, Eur. Phys. J. A 48 (2012) 88.
\newblock \href {http://arxiv.org/abs/1205.2255} {\path{arXiv:1205.2255}}, \href {https://doi.org/10.1140/epja/i2012-12088-3} {\path{doi:10.1140/epja/i2012-12088-3}}.

\bibitem{Anisovich:2013sva}
A.~V. Anisovich, et~al., {Study of the narrow structure at 1685-MeV in $\gamma p \to \eta p$}, Phys. Lett. B 719 (2013) 89--94.
\newblock \href {https://doi.org/10.1016/j.physletb.2012.09.029} {\path{doi:10.1016/j.physletb.2012.09.029}}.

\bibitem{Anisovich:2013tij}
A.~V. Anisovich, et~al., {Study of ambiguities in $\pi^-p\to \Lambda K^0$ scattering amplitudes}, Eur. Phys. J. A 49 (2013) 121.
\newblock \href {http://arxiv.org/abs/1306.5126} {\path{arXiv:1306.5126}}, \href {https://doi.org/10.1140/epja/i2013-13121-9} {\path{doi:10.1140/epja/i2013-13121-9}}.

\bibitem{Anisovich:2013vpa}
A.~V. Anisovich, et~al., {Sign ambiguity in the K$\Sigma$ channel}, Eur. Phys. J. A 49 (2013) 158.
\newblock \href {http://arxiv.org/abs/1310.3610} {\path{arXiv:1310.3610}}, \href {https://doi.org/10.1140/epja/i2013-13158-8} {\path{doi:10.1140/epja/i2013-13158-8}}.

\bibitem{Anisovich:2015tla}
A.~V. Anisovich, et~al., {Interference phenomena in the $J^P=1/2^-$ wave in $\eta$ photoproduction}, Eur. Phys. J. A 51~(6) (2015) 72.
\newblock \href {http://arxiv.org/abs/1501.02093} {\path{arXiv:1501.02093}}, \href {https://doi.org/10.1140/epja/i2015-15072-5} {\path{doi:10.1140/epja/i2015-15072-5}}.

\bibitem{Denisenko:2016ugz}
I.~Denisenko, et~al., {$N^{\bf *}$ decays to $N\omega$ from new data on $\gamma p\to \omega p$}, Phys. Lett. B 755 (2016) 97--101.
\newblock \href {http://arxiv.org/abs/1601.06092} {\path{arXiv:1601.06092}}, \href {https://doi.org/10.1016/j.physletb.2016.01.061} {\path{doi:10.1016/j.physletb.2016.01.061}}.

\bibitem{Anisovich:2016vzt}
A.~V. Anisovich, et~al., {The impact of new polarization data from Bonn, Mainz and Jefferson Laboratory on $\gamma p \rightarrow \pi N$ multipoles}, Eur. Phys. J. A 52~(9) (2016) 284.
\newblock \href {http://arxiv.org/abs/1604.05704} {\path{arXiv:1604.05704}}, \href {https://doi.org/10.1140/epja/i2016-16284-9} {\path{doi:10.1140/epja/i2016-16284-9}}.

\bibitem{Anisovich:2017afs}
A.~V. Anisovich, et~al., {Neutron helicity amplitudes}, Phys. Rev. C 96~(5) (2017) 055202.
\newblock \href {https://doi.org/10.1103/PhysRevC.96.055202} {\path{doi:10.1103/PhysRevC.96.055202}}.

\bibitem{pwahiskp}
\url{https://pwa.hiskp.uni-bonn.de/}.

\bibitem{Sato:1996gk}
T.~Sato, T.-. S.~H. Lee, {Meson exchange model for $\pi N$ scattering and $\gamma N\to \pi N$ reaction}, Phys. Rev. C 54 (1996) 2660--2684.
\newblock \href {https://doi.org/10.1103/PhysRevC.54.2660} {\path{doi:10.1103/PhysRevC.54.2660}}.

\bibitem{Sato:2000jf}
T.~Sato, T.~S.~H. Lee, {Dynamical study of the $\Delta$ excitation in $N (e, e^\prime \pi)$ reactions}, Phys. Rev. C 63 (2001) 055201.
\newblock \href {https://doi.org/10.1103/PhysRevC.63.055201} {\path{doi:10.1103/PhysRevC.63.055201}}.

\bibitem{Julia-Diaz:2006ios}
B.~Julia-Diaz, T.~S.~H. Lee, T.~Sato, L.~C. Smith, {Extraction and Interpretation of $\gamma N \to \Delta$ Form Factors within a Dynamical Model}, Phys. Rev. C 75 (2007) 015205.
\newblock \href {http://arxiv.org/abs/nucl-th/0611033} {\path{arXiv:nucl-th/0611033}}, \href {https://doi.org/10.1103/PhysRevC.75.015205} {\path{doi:10.1103/PhysRevC.75.015205}}.

\bibitem{Julia-Diaz:2007qtz}
B.~Julia-Diaz, T.~S.~H. Lee, A.~Matsuyama, T.~Sato, {Dynamical coupled-channel model of $\pi N$ scattering in the $W \leq 2$\,GeV nucleon resonance region}, Phys. Rev. C 76 (2007) 065201.
\newblock \href {http://arxiv.org/abs/0704.1615} {\path{arXiv:0704.1615}}, \href {https://doi.org/10.1103/PhysRevC.76.065201} {\path{doi:10.1103/PhysRevC.76.065201}}.

\bibitem{Matsuyama:2006rp}
A.~Matsuyama, T.~Sato, T.~S.~H. Lee, {Dynamical coupled-channel model of meson production reactions in the nucleon resonance region}, Phys. Rept. 439 (2007) 193--253.
\newblock \href {http://arxiv.org/abs/nucl-th/0608051} {\path{arXiv:nucl-th/0608051}}, \href {https://doi.org/10.1016/j.physrep.2006.12.003} {\path{doi:10.1016/j.physrep.2006.12.003}}.

\bibitem{Suzuki:2008rp}
N.~Suzuki, T.~Sato, T.~S.~H. Lee, {Extraction of Resonances from Meson-Nucleon Reactions}, Phys. Rev. C 79 (2009) 025205.
\newblock \href {http://arxiv.org/abs/0806.2043} {\path{arXiv:0806.2043}}, \href {https://doi.org/10.1103/PhysRevC.79.025205} {\path{doi:10.1103/PhysRevC.79.025205}}.

\bibitem{Kamano:2008gr}
H.~Kamano, et~al., {Dynamical coupled-channels study of $\pi N \to \pi \pi N$ reactions}, Phys. Rev. C 79 (2009) 025206.
\newblock \href {http://arxiv.org/abs/0807.2273} {\path{arXiv:0807.2273}}, \href {https://doi.org/10.1103/PhysRevC.79.025206} {\path{doi:10.1103/PhysRevC.79.025206}}.

\bibitem{Julia-Diaz:2009dnz}
B.~Julia-Diaz, et~al., {Dynamical coupled-channels analysis of $p(e,e' \pi)N$ reactions}, Phys. Rev. C 80 (2009) 025207.
\newblock \href {http://arxiv.org/abs/0904.1918} {\path{arXiv:0904.1918}}, \href {https://doi.org/10.1103/PhysRevC.80.025207} {\path{doi:10.1103/PhysRevC.80.025207}}.

\bibitem{Suzuki:2009nj}
N.~Suzuki, et~al., {Disentangling the Dynamical Origin of P$_{11}$ Nucleon Resonances}, Phys. Rev. Lett. 104 (2010) 042302.
\newblock \href {http://arxiv.org/abs/0909.1356} {\path{arXiv:0909.1356}}, \href {https://doi.org/10.1103/PhysRevLett.104.042302} {\path{doi:10.1103/PhysRevLett.104.042302}}.

\bibitem{Kamano:2009im}
H.~Kamano, et~al., {Double and single pion photoproduction within a dynamical coupled-channels model}, Phys. Rev. C 80 (2009) 065203.
\newblock \href {http://arxiv.org/abs/0909.1129} {\path{arXiv:0909.1129}}, \href {https://doi.org/10.1103/PhysRevC.80.065203} {\path{doi:10.1103/PhysRevC.80.065203}}.

\bibitem{Sandorfi:2010uv}
A.~M. Sandorfi, S.~Hoblit, H.~Kamano, T.~S.~H. Lee, {Determining pseudoscalar meson photo-production amplitudes from complete experiments}, J. Phys. G 38 (2011) 053001.
\newblock \href {http://arxiv.org/abs/1010.4555} {\path{arXiv:1010.4555}}, \href {https://doi.org/10.1088/0954-3899/38/5/053001} {\path{doi:10.1088/0954-3899/38/5/053001}}.

\bibitem{Suzuki:2010yn}
N.~Suzuki, T.~Sato, T.~S.~H. Lee, {Extraction of Electromagnetic Transition Form Factors for Nucleon Resonances within a Dynamical Coupled-Channels Model}, Phys. Rev. C 82 (2010) 045206.
\newblock \href {http://arxiv.org/abs/1006.2196} {\path{arXiv:1006.2196}}, \href {https://doi.org/10.1103/PhysRevC.82.045206} {\path{doi:10.1103/PhysRevC.82.045206}}.

\bibitem{Kamano:2010ud}
H.~Kamano, S.~X. Nakamura, T.~S.~H. Lee, T.~Sato, {Extraction of $P_{11}$ resonances from $\pi N$ data}, Phys. Rev. C 81 (2010) 065207.
\newblock \href {http://arxiv.org/abs/1001.5083} {\path{arXiv:1001.5083}}, \href {https://doi.org/10.1103/PhysRevC.81.065207} {\path{doi:10.1103/PhysRevC.81.065207}}.

\bibitem{Kamano:2013iva}
H.~Kamano, S.~X. Nakamura, T.~S.~H. Lee, T.~Sato, {Nucleon resonances within a dynamical coupled-channels model of $\pi N$ and $\gamma N$ reactions}, Phys. Rev. C 88~(3) (2013) 035209.
\newblock \href {http://arxiv.org/abs/1305.4351} {\path{arXiv:1305.4351}}, \href {https://doi.org/10.1103/PhysRevC.88.035209} {\path{doi:10.1103/PhysRevC.88.035209}}.

\bibitem{Nakamura:2015rta}
S.~X. Nakamura, H.~Kamano, T.~Sato, {Dynamical coupled-channels model for neutrino-induced meson productions in resonance region}, Phys. Rev. D 92~(7) (2015) 074024.
\newblock \href {http://arxiv.org/abs/1506.03403} {\path{arXiv:1506.03403}}, \href {https://doi.org/10.1103/PhysRevD.92.074024} {\path{doi:10.1103/PhysRevD.92.074024}}.

\bibitem{Kamano:2016bgm}
H.~Kamano, S.~X. Nakamura, T.~S.~H. Lee, T.~Sato, {Isospin decomposition of $\gamma N \to N^*$ transitions within a dynamical coupled-channels model}, Phys. Rev. C 94~(1) (2016) 015201.
\newblock \href {http://arxiv.org/abs/1605.00363} {\path{arXiv:1605.00363}}, \href {https://doi.org/10.1103/PhysRevC.94.015201} {\path{doi:10.1103/PhysRevC.94.015201}}.

\bibitem{Kamano:2019gtm}
H.~Kamano, T.~S.~H. Lee, S.~X. Nakamura, T.~Sato, {The ANL-Osaka Partial-Wave Amplitudes of $\pi N$ and $\gamma N$ Reactions} (9 2019).
\newblock \href {http://arxiv.org/abs/1909.11935} {\path{arXiv:1909.11935}}.

\bibitem{Mai:2021vsw}
M.~Mai, et~al., {J\"ulich-Bonn-Washington model for pion electroproduction multipoles}, Phys. Rev. C 103~(6) (2021) 065204.
\newblock \href {http://arxiv.org/abs/2104.07312} {\path{arXiv:2104.07312}}, \href {https://doi.org/10.1103/PhysRevC.103.065204} {\path{doi:10.1103/PhysRevC.103.065204}}.

\bibitem{Mai:2021aui}
M.~Mai, et~al., {Coupled-channels analysis of pion and $\eta$ electroproduction within the J\"ulich-Bonn-Washington model}, Phys. Rev. C 106~(1) (2022) 015201.
\newblock \href {http://arxiv.org/abs/2111.04774} {\path{arXiv:2111.04774}}, \href {https://doi.org/10.1103/PhysRevC.106.015201} {\path{doi:10.1103/PhysRevC.106.015201}}.

\bibitem{Mai:2023cbp}
M.~Mai, et~al., {Inclusion of $K\Lambda $ electroproduction data in a coupled channel analysis}, Eur. Phys. J. A 59~(12) (2023) 286.
\newblock \href {http://arxiv.org/abs/2307.10051} {\path{arXiv:2307.10051}}, \href {https://doi.org/10.1140/epja/s10050-023-01188-0} {\path{doi:10.1140/epja/s10050-023-01188-0}}.

\bibitem{Doring:2009yv}
M.~Döring, et~al., {Analytic properties of the scattering amplitude and resonances parameters in a meson exchange model}, Nucl. Phys. A 829 (2009) 170--209.
\newblock \href {http://arxiv.org/abs/0903.4337} {\path{arXiv:0903.4337}}, \href {https://doi.org/10.1016/j.nuclphysa.2009.08.010} {\path{doi:10.1016/j.nuclphysa.2009.08.010}}.

\bibitem{Doring:2009bi}
M.~Döring, et~al., {The Role of the background in the extraction of resonance contributions from meson-baryon scattering}, Phys. Lett. B 681 (2009) 26--31.
\newblock \href {http://arxiv.org/abs/0903.1781} {\path{arXiv:0903.1781}}, \href {https://doi.org/10.1016/j.physletb.2009.09.052} {\path{doi:10.1016/j.physletb.2009.09.052}}.

\bibitem{Doring:2009uc}
M.~Döring, K.~Nakayama, {The Phase and pole structure of the $N^*(1535)$ in $\pi N \to \pi N$ and $\gamma N \to \pi N$}, Eur. Phys. J. A 43 (2010) 83--105.
\newblock \href {http://arxiv.org/abs/0906.2949} {\path{arXiv:0906.2949}}, \href {https://doi.org/10.1140/epja/i2009-10892-4} {\path{doi:10.1140/epja/i2009-10892-4}}.

\bibitem{CBELSATAPS:2015tyg}
V.~Sokhoyan, et~al., {Data on $I^s$ and $I^c$ in $\overrightarrow{\gamma}p\to p\pi^0\pi^0$ reveal cascade decays of $N(1900)$ via $N(1520)\pi$}, Phys. Lett. B 746 (2015) 127--131.
\newblock \href {https://doi.org/10.1016/j.physletb.2015.04.063} {\path{doi:10.1016/j.physletb.2015.04.063}}.

\bibitem{Doring:2005bx}
M.~Doring, E.~Oset, D.~Strottman, {Chiral dynamics in the $\gamma p \to \pi^0 \eta p$ and $\gamma p \to \pi^0 K^0 \Sigma^+$ reactions}, Phys. Rev. C 73 (2006) 045209.
\newblock \href {http://arxiv.org/abs/nucl-th/0510015} {\path{arXiv:nucl-th/0510015}}, \href {https://doi.org/10.1103/PhysRevC.73.045209} {\path{doi:10.1103/PhysRevC.73.045209}}.

\bibitem{Doring:2010ap}
M.~Döring, et~al., {The reaction $\pi^+ p \to K^+ \Sigma^+$ in a unitary coupled-channels model}, Nucl. Phys. A 851 (2011) 58--98.
\newblock \href {http://arxiv.org/abs/1009.3781} {\path{arXiv:1009.3781}}, \href {https://doi.org/10.1016/j.nuclphysa.2010.12.010} {\path{doi:10.1016/j.nuclphysa.2010.12.010}}.

\bibitem{Ronchen:2012eg}
D.~Rönchen, et~al., {Coupled-channel dynamics in the reactions $\pi N \to \pi N, \eta N, K\Lambda, K\Sigma$}, Eur. Phys. J. A 49 (2013) 44.
\newblock \href {http://arxiv.org/abs/1211.6998} {\path{arXiv:1211.6998}}, \href {https://doi.org/10.1140/epja/i2013-13044-5} {\path{doi:10.1140/epja/i2013-13044-5}}.

\bibitem{Ronchen:2014cna}
D.~R\"onchen, et~al., {Photocouplings at the Pole from Pion Photoproduction}, Eur. Phys. J. A 50~(6) (2014) 101, [Erratum: Eur.Phys.J.A 51, 63 (2015)].
\newblock \href {http://arxiv.org/abs/1401.0634} {\path{arXiv:1401.0634}}, \href {https://doi.org/10.1140/epja/i2014-14101-3} {\path{doi:10.1140/epja/i2014-14101-3}}.

\bibitem{Ronchen:2015vfa}
D.~R\"onchen, et~al., {Eta photoproduction in a combined analysis of pion- and photon-induced reactions}, Eur. Phys. J. A 51~(6) (2015) 70.
\newblock \href {http://arxiv.org/abs/1504.01643} {\path{arXiv:1504.01643}}, \href {https://doi.org/10.1140/epja/i2015-15070-7} {\path{doi:10.1140/epja/i2015-15070-7}}.

\bibitem{Ronchen:2018ury}
D.~R\"onchen, M.~D\"oring, U.~G. Mei\ss{}ner, {The impact of $K^{+}\Lambda$ photoproduction on the resonance spectrum}, Eur. Phys. J. A 54~(6) (2018) 110.
\newblock \href {http://arxiv.org/abs/1801.10458} {\path{arXiv:1801.10458}}, \href {https://doi.org/10.1140/epja/i2018-12541-3} {\path{doi:10.1140/epja/i2018-12541-3}}.

\bibitem{Ronchen:2022hqk}
D.~R\"onchen, M.~D\"oring, U.-G. Mei\ss{}ner, C.-W. Shen, {Light baryon resonances from a coupled-channel study including ${K\Sigma}$ photoproduction}, Eur. Phys. J. A 58 (2022) 229.
\newblock \href {http://arxiv.org/abs/2208.00089} {\path{arXiv:2208.00089}}, \href {https://doi.org/10.1140/epja/s10050-022-00852-1} {\path{doi:10.1140/epja/s10050-022-00852-1}}.

\bibitem{collaborations.fz-juelich}
\url{https://collaborations.fz-juelich.de/ikp/meson-baryon/juelich_amplitudes.html}.

\bibitem{Svarc:2012pt}
A.~Svarc, M.~Hadzimehmedovic, H.~Osmanovic, J.~Stahov, {A new method for extracting poles from single-channel data based on Laurent expansion of T-matrices with Pietarinen power series representing the non-singular part}, unpublished note (12 2012).
\newblock \href {http://arxiv.org/abs/1212.1295} {\path{arXiv:1212.1295}}.

\bibitem{Svarc:2013laa}
A.~\v{S}varc, others., {Introducing the Pietarinen expansion method into the single-channel pole extraction problem}, Phys. Rev. C 88~(3) (2013) 035206.
\newblock \href {http://arxiv.org/abs/1307.4613} {\path{arXiv:1307.4613}}, \href {https://doi.org/10.1103/PhysRevC.88.035206} {\path{doi:10.1103/PhysRevC.88.035206}}.

\bibitem{Svarc:2014zja}
A.~\v{S}varc, et~al., {Poles of Karlsruhe-Helsinki KH80 and KA84 solutions extracted by using the Laurent-Pietarinen method}, Phys. Rev. C 89~(4) (2014) 045205.
\newblock \href {http://arxiv.org/abs/1401.1947} {\path{arXiv:1401.1947}}, \href {https://doi.org/10.1103/PhysRevC.89.045205} {\path{doi:10.1103/PhysRevC.89.045205}}.

\bibitem{Svarc:2014sqa}
A.~\v{S}varc, et~al., {Pole positions and residues from pion photoproduction using the Laurent-Pietarinen expansion method}, Phys. Rev. C 89~(6) (2014) 065208.
\newblock \href {http://arxiv.org/abs/1404.1544} {\path{arXiv:1404.1544}}, \href {https://doi.org/10.1103/PhysRevC.89.065208} {\path{doi:10.1103/PhysRevC.89.065208}}.

\bibitem{Svarc:2014aga}
A.~\v{S}varc, et~al., {Pole structure from energy-dependent and single-energy fits to GWU-SAID \ensuremath{\pi}N elastic scattering data}, Phys. Rev. C 91~(1) (2015) 015207.
\newblock \href {http://arxiv.org/abs/1405.6474} {\path{arXiv:1405.6474}}, \href {https://doi.org/10.1103/PhysRevC.91.015207} {\path{doi:10.1103/PhysRevC.91.015207}}.

\bibitem{Svarc:2015usk}
A.~\v{S}varc, et~al., {Generalization of the model-independent Laurent-Pietarinen single-channel pole-extraction formalism to multiple channels}, Phys. Lett. B 755 (2016) 452--455.
\newblock \href {http://arxiv.org/abs/1512.07403} {\path{arXiv:1512.07403}}, \href {https://doi.org/10.1016/j.physletb.2016.02.058} {\path{doi:10.1016/j.physletb.2016.02.058}}.

\bibitem{Svarc:2018eol}
A.~\v{S}varc, et~al., {From Experimental Data to Pole Parameters in a Direct Way (Angle Dependent Continuum Ambiguity and Laurent + Pietarinen Expansion)}, Few Body Syst. 59~(5) (2018) 96.
\newblock \href {http://arxiv.org/abs/1806.05675} {\path{arXiv:1806.05675}}, \href {https://doi.org/10.1007/s00601-018-1410-y} {\path{doi:10.1007/s00601-018-1410-y}}.

\bibitem{Svarc:2018aay}
A.~\v{S}varc, et~al., {Role of angle-dependent phase rotations of reaction amplitudes in $\eta$ photoproduction on protons}, Phys. Rev. C 98~(4) (2018) 045206.
\newblock \href {http://arxiv.org/abs/1807.02759} {\path{arXiv:1807.02759}}, \href {https://doi.org/10.1103/PhysRevC.98.045206} {\path{doi:10.1103/PhysRevC.98.045206}}.

\bibitem{Svarc:2020waq}
A.~\v{S}varc, {From Experimental Data to Pole Parameters in a Model Independent Way (Fixed-t Single-Energy Partial Wave Analysis (SE PWA) and Laurent + Pietarinen Model (L+P) for Pole Extraction)}, EPJ Web Conf. 241 (2020) 03005.
\newblock \href {https://doi.org/10.1051/epjconf/202024103005} {\path{doi:10.1051/epjconf/202024103005}}.

\bibitem{Svarc:2020cic}
A.~\v{S}varc, Y.~Wunderlich, L.~Tiator, {Amplitude- and truncated partial-wave analyses combined: A novel, almost theory-independent single-channel method for extracting photoproduction multipoles directly from measured data}, Phys. Rev. C 102 (2020) 064609.
\newblock \href {http://arxiv.org/abs/2008.01355} {\path{arXiv:2008.01355}}, \href {https://doi.org/10.1103/PhysRevC.102.064609} {\path{doi:10.1103/PhysRevC.102.064609}}.

\bibitem{Svarc:2021zkb}
A.~Svarc, {Each single-energy, single-channel partial-wave analysis is inherently model-dependent}, Phys. Rev. C 104~(1) (2021) 014605.
\newblock \href {http://arxiv.org/abs/2103.13052} {\path{arXiv:2103.13052}}, \href {https://doi.org/10.1103/PhysRevC.104.014605} {\path{doi:10.1103/PhysRevC.104.014605}}.

\bibitem{Svarc:2021gcs}
A.~\v{S}varc, Y.~Wunderlich, L.~Tiator, {Application of the single-channel, single-energy amplitude and partial-wave analysis method to $K^+\Lambda$ photoproduction}, Phys. Rev. C 105~(2) (2022) 024614.
\newblock \href {http://arxiv.org/abs/2111.09587} {\path{arXiv:2111.09587}}, \href {https://doi.org/10.1103/PhysRevC.105.024614} {\path{doi:10.1103/PhysRevC.105.024614}}.

\bibitem{Svarc:2022buh}
A.~\v{S}varc, R.~L. Workman, {Laurent+Pietarinen partial-wave analysis}, Phys. Rev. C 108~(1) (2023) 014615.
\newblock \href {http://arxiv.org/abs/2206.05978} {\path{arXiv:2206.05978}}, \href {https://doi.org/10.1103/PhysRevC.108.014615} {\path{doi:10.1103/PhysRevC.108.014615}}.

\bibitem{Wunderlich:2013iga}
Y.~Wunderlich, R.~Beck, L.~Tiator, {The complete-experiment problem of photoproduction of pseudoscalar mesons in a truncated partial-wave analysis}, Phys. Rev. C 89~(5) (2014) 055203.
\newblock \href {http://arxiv.org/abs/1312.0245} {\path{arXiv:1312.0245}}, \href {https://doi.org/10.1103/PhysRevC.89.055203} {\path{doi:10.1103/PhysRevC.89.055203}}.

\bibitem{Workman:2016irf}
R.~L. Workman, L.~Tiator, Y.~Wunderlich, M.~D{\"o}ring, H.~Haberzettl, {Amplitude reconstruction from complete photoproduction experiments and truncated partial-wave expansions}, Phys. Rev. C 95~(1) (2017) 015206.
\newblock \href {http://arxiv.org/abs/1611.04434} {\path{arXiv:1611.04434}}, \href {https://doi.org/10.1103/PhysRevC.95.015206} {\path{doi:10.1103/PhysRevC.95.015206}}.

\bibitem{Fix:2022shn}
A.~Fix, I.~Dementjev, {Ambiguities in a partial-wave analysis of {\ensuremath{\gamma}} p {\textrightarrow} {\ensuremath{\pi}} $^{0}$ p with truncation in total angular momentum}, J. Phys. G 50~(5) (2023) 055107.
\newblock \href {http://arxiv.org/abs/2205.08074} {\path{arXiv:2205.08074}}, \href {https://doi.org/10.1088/1361-6471/acc67a} {\path{doi:10.1088/1361-6471/acc67a}}.

\bibitem{Kroenert:2023ovd}
P.~Kroenert, Y.~Wunderlich, F.~Afzal, A.~Thiel, {Truncated partial-wave analysis for {\ensuremath{\eta}}-photoproduction observables via Bayesian statistics}, Phys. Rev. C 109~(4) (2024) 045206.
\newblock \href {http://arxiv.org/abs/2305.10367} {\path{arXiv:2305.10367}}, \href {https://doi.org/10.1103/PhysRevC.109.045206} {\path{doi:10.1103/PhysRevC.109.045206}}.

\bibitem{CLAS:2024iir}
A.~V. Sarantsev, et~al., {Photoproduction of two charged pions off protons in the resonance region}, Phys. Rev. C 111~(3) (2025) 035203.
\newblock \href {http://arxiv.org/abs/2411.15423} {\path{arXiv:2411.15423}}, \href {https://doi.org/10.1103/PhysRevC.111.035203} {\path{doi:10.1103/PhysRevC.111.035203}}.

\bibitem{Sarantsev:2025lik}
A.~V. Sarantsev, et~al., {Decays of $N^*$ and $\Delta^*$ resonances into $N\rho,~\Delta\pi$, and $Nf_0(500)$}, Phys. Rev. C 112~(1) (2025) 015202.
\newblock \href {http://arxiv.org/abs/2503.16636} {\path{arXiv:2503.16636}}, \href {https://doi.org/10.1103/qfpf-tcs3} {\path{doi:10.1103/qfpf-tcs3}}.

\bibitem{Gopal:1980ur}
G.~P. Gopal, {$S = -1$ Baryons: An experimental review}, in: {4th International Conference on Baryon Resonances}, 1980, baryon 1980, Toronto, Canada, July 14-16, 1980.

\bibitem{Zhang:2013cua}
H.~Zhang, J.~Tulpan, M.~Shrestha, D.~M. Manley, {Partial-wave analysis of $\bar K N$ scattering reactions}, Phys. Rev. C 88~(3) (2013) 035204.
\newblock \href {http://arxiv.org/abs/1305.3598} {\path{arXiv:1305.3598}}, \href {https://doi.org/10.1103/PhysRevC.88.035204} {\path{doi:10.1103/PhysRevC.88.035204}}.

\bibitem{Zhang:2013sva}
H.~Zhang, J.~Tulpan, M.~Shrestha, D.~M. Manley, {Multichannel parametrization of $\bar K N$ scattering amplitudes and extraction of resonance parameters}, Phys. Rev. C 88~(3) (2013) 035205.
\newblock \href {http://arxiv.org/abs/1305.4575} {\path{arXiv:1305.4575}}, \href {https://doi.org/10.1103/PhysRevC.88.035205} {\path{doi:10.1103/PhysRevC.88.035205}}.

\bibitem{Fernandez-Ramirez:2015tfa}
C.~Fernandez-Ramirez, et~al., {Coupled-channel model for $\bar{K}N$ scattering in the resonant region}, Phys. Rev. D 93~(3) (2016) 034029.
\newblock \href {http://arxiv.org/abs/1510.07065} {\path{arXiv:1510.07065}}, \href {https://doi.org/10.1103/PhysRevD.93.034029} {\path{doi:10.1103/PhysRevD.93.034029}}.

\bibitem{Kamano:2014zba}
H.~Kamano, S.~X. Nakamura, T.~S.~H. Lee, T.~Sato, {Dynamical coupled-channels model of K$^-$p reactions: Determination of partial-wave amplitudes}, Phys. Rev. C 90~(6) (2014) 065204.
\newblock \href {http://arxiv.org/abs/1407.6839} {\path{arXiv:1407.6839}}, \href {https://doi.org/10.1103/PhysRevC.90.065204} {\path{doi:10.1103/PhysRevC.90.065204}}.

\bibitem{Kamano:2015hxa}
H.~Kamano, S.~X. Nakamura, T.~S.~H. Lee, T.~Sato, {Dynamical coupled-channels model of $K^- p$ reactions. II. Extraction of $\Lambda^*$ and $\Sigma^*$ hyperon resonances}, Phys. Rev. C 92~(2) (2015) 025205, [Erratum: Phys.Rev.C 95, 049903 (2017)].
\newblock \href {http://arxiv.org/abs/1506.01768} {\path{arXiv:1506.01768}}, \href {https://doi.org/10.1103/PhysRevC.92.025205} {\path{doi:10.1103/PhysRevC.92.025205}}.

\bibitem{Sarantsev:2019xxm}
A.~V. Sarantsev, et~al., {Hyperon II: Properties of excited hyperons}, Eur. Phys. J. A 55~(10) (2019) 180.
\newblock \href {http://arxiv.org/abs/1907.13387} {\path{arXiv:1907.13387}}, \href {https://doi.org/10.1140/epja/i2019-12880-5} {\path{doi:10.1140/epja/i2019-12880-5}}.

\bibitem{CBELSATAPS:2015kka}
V.~Sokhoyan, et~al., {High-statistics study of the reaction $\gamma p\to p\;2\pi^0$}, Eur. Phys. J. A 51~(8) (2015) 95, [Erratum: Eur.Phys.J.A 51, 187 (2015)].
\newblock \href {http://arxiv.org/abs/1507.02488} {\path{arXiv:1507.02488}}, \href {https://doi.org/10.1140/epja/i2015-15187-7} {\path{doi:10.1140/epja/i2015-15187-7}}.

\bibitem{Vrana:1999nt}
T.~P. Vrana, S.~A. Dytman, T.~S.~H. Lee, {Baryon resonance extraction from pi N data using a unitary multichannel model}, Phys. Rept. 328 (2000) 181--236.
\newblock \href {http://arxiv.org/abs/nucl-th/9910012} {\path{arXiv:nucl-th/9910012}}, \href {https://doi.org/10.1016/S0370-1573(99)00108-8} {\path{doi:10.1016/S0370-1573(99)00108-8}}.

\bibitem{ParticleDataGroup:2004fcd}
S.~Eidelman, et~al., {Review of particle physics. Particle Data Group}, Phys. Lett. B 592~(1-4) (2004) 1.
\newblock \href {https://doi.org/10.1016/j.physletb.2004.06.001} {\path{doi:10.1016/j.physletb.2004.06.001}}.

\bibitem{Chatterjee:2017yhp}
S.~Chatterjee, D.~Mishra, B.~Mohanty, S.~Samanta, {Freezeout systematics due to the hadron spectrum}, Phys. Rev. C 96~(5) (2017) 054907.
\newblock \href {http://arxiv.org/abs/1708.08152} {\path{arXiv:1708.08152}}, \href {https://doi.org/10.1103/PhysRevC.96.054907} {\path{doi:10.1103/PhysRevC.96.054907}}.

\bibitem{Walker:1968xu}
R.~L. Walker, {Phenomenological analysis of single pion photoproduction}, Phys. Rev. 182 (1969) 1729--1748.
\newblock \href {https://doi.org/10.1103/PhysRev.182.1729} {\path{doi:10.1103/PhysRev.182.1729}}.

\bibitem{Berends:1967vi}
F.~A. Berends, A.~Donnachie, D.~L. Weaver, {Photoproduction and electroproduction of pions. 1. Dispersion relation theory}, Nucl. Phys. B 4 (1967) 1--53.
\newblock \href {https://doi.org/10.1016/0550-3213(67)90196-4} {\path{doi:10.1016/0550-3213(67)90196-4}}.

\bibitem{Kelly:2005jy}
J.~J. Kelly, et~al., {Recoil polarization measurements for neutral pion electroproduction at $Q^2 = 1$ (GeV/c)$^2$ near the $\Delta$ resonance}, Phys. Rev. C 75 (2007) 025201.
\newblock \href {https://doi.org/10.1103/PhysRevC.75.025201} {\path{doi:10.1103/PhysRevC.75.025201}}.

\bibitem{Tiator:2011pw}
L.~Tiator, D.~Drechsel, S.~S. Kamalov, M.~Vanderhaeghen, {Electromagnetic Excitation of Nucleon Resonances}, Eur. Phys. J. ST 198 (2011) 141--170.
\newblock \href {http://arxiv.org/abs/1109.6745} {\path{arXiv:1109.6745}}, \href {https://doi.org/10.1140/epjst/e2011-01488-9} {\path{doi:10.1140/epjst/e2011-01488-9}}.

\bibitem{Peccei:1969sb}
R.~D. Peccei, {Chiral lagrangian model of single-pion photoproduction}, Phys. Rev. 181 (1969) 1902--1904.
\newblock \href {https://doi.org/10.1103/PhysRev.181.1902} {\path{doi:10.1103/PhysRev.181.1902}}.

\bibitem{Arndt:2002xv}
R.~A. Arndt, W.~J. Briscoe, I.~I. Strakovsky, R.~L. Workman, {Analysis of pion photoproduction data}, Phys. Rev. C 66 (2002) 055213.
\newblock \href {https://doi.org/10.1103/PhysRevC.66.055213} {\path{doi:10.1103/PhysRevC.66.055213}}.

\bibitem{Kamalov:1999hs}
S.~S. Kamalov, S.~N. Yang, {Pion cloud and the $Q^2$ dependence of $\gamma^* N \leftrightarrow \Delta$ transition form-factors}, Phys. Rev. Lett. 83 (1999) 4494--4497.
\newblock \href {https://doi.org/10.1103/PhysRevLett.83.4494} {\path{doi:10.1103/PhysRevLett.83.4494}}.

\bibitem{Wang:2024jns}
Y.-F. Wang, {On the nature of the $N^*$ and $\Delta$ resonances via coupled-channel dynamics}, in: {16th International Conference on Meson-Nucleon Physics and the Structure of the Nucleon}, Vol. 303, 2024, p. 01023.
\newblock \href {http://arxiv.org/abs/2401.01191} {\path{arXiv:2401.01191}}.

\bibitem{Gell-Mann:1957uuj}
M.~Gell-Mann, A.~H. Rosenfeld, {Hyperons and heavy mesons (systematics and decay)}, Ann. Rev. Nucl. Part. Sci. 7 (1957) 407--478.
\newblock \href {https://doi.org/10.1146/annurev.ns.07.120157.002203} {\path{doi:10.1146/annurev.ns.07.120157.002203}}.

\bibitem{Anderson:1952nw}
H.~L. Anderson, E.~Fermi, E.~A. Long, D.~E. Nagle, {Total Cross-sections of Positive Pions in Hydrogen}, Phys. Rev. 85 (1952) 936.
\newblock \href {https://doi.org/10.1103/PhysRev.85.936} {\path{doi:10.1103/PhysRev.85.936}}.

\bibitem{barkas1961data}
W.~Barkas, A.~Rosenfeld, L.~R. Laboratory, U.~A.~E. Commission, \href{https://books.google.de/books?id=oZE1ft4_D6IC}{Data for Elementary-particle Physics}, UCRL (Series), University of California, Lawrence Radiation Laboratory, 1961.
\newline\urlprefix\url{https://books.google.de/books?id=oZE1ft4_D6IC}

\bibitem{Pjerrou:1962oxa}
G.~M. Pjerrou, et~al., {A resonance in the E\ensuremath{\pi} system at 1.53 GeV}, in: {11th International Conference on High-energy Physics}, 1962, pp. 289--290.

\bibitem{Gell-Mann:1962yej}
M.~Gell-Mann, {Symmetries of baryons and mesons}, Phys. Rev. 125 (1962) 1067--1084.
\newblock \href {https://doi.org/10.1103/PhysRev.125.1067} {\path{doi:10.1103/PhysRev.125.1067}}.

\bibitem{Gell-Mann:1964hhf}
M.~Gell-Mann, {The Symmetry group of vector and axial vector currents}, Physics Physique Fizika 1 (1964) 63--75.
\newblock \href {https://doi.org/10.1103/PhysicsPhysiqueFizika.1.63} {\path{doi:10.1103/PhysicsPhysiqueFizika.1.63}}.

\bibitem{Greenberg:1964pe}
O.~W. Greenberg, {Spin and Unitary Spin Independence in a Paraquark Model of Baryons and Mesons}, Phys. Rev. Lett. 13 (1964) 598--602.
\newblock \href {https://doi.org/10.1103/PhysRevLett.13.598} {\path{doi:10.1103/PhysRevLett.13.598}}.

\bibitem{CMS:2025kzt}
A.~Hayrapetyan, et~al., {Observation of a pseudoscalar excess at the top quark pair production threshold} (3 2025).
\newblock \href {http://arxiv.org/abs/2503.22382} {\path{arXiv:2503.22382}}.

\bibitem{Heisenberg:1932aaa}
W.~Heisenberg, H.~Euler, {Über den Bau der Atomkerne}, Z. Phys. 77 (1932) 1--11.
\newblock \href {https://doi.org/doi:10.1007/BF01342433} {\path{doi:doi:10.1007/BF01342433}}.

\bibitem{Neeman:1961jhl}
Y.~Ne'eman, {Derivation of strong interactions from a gauge invariance}, Nucl. Phys. 26 (1961) 222--229.
\newblock \href {https://doi.org/10.1016/0029-5582(61)90134-1} {\path{doi:10.1016/0029-5582(61)90134-1}}.

\bibitem{Gell-Mann:1961omu}
M.~Gell-Mann, {The Eightfold Way: A Theory of strong interaction symmetry} (3 1961).
\newblock \href {https://doi.org/10.2172/4008239} {\path{doi:10.2172/4008239}}.

\bibitem{Okubo:1961jc}
S.~Okubo, {Note on unitary symmetry in strong interactions}, Prog. Theor. Phys. 27 (1962) 949--966.
\newblock \href {https://doi.org/10.1143/PTP.27.949} {\path{doi:10.1143/PTP.27.949}}.

\bibitem{EuropeanMuon:1987isl}
J.~Ashman, et~al., {A Measurement of the Spin Asymmetry and Determination of the Structure Function $g_1$ in Deep Inelastic Muon-Proton Scattering}, Phys. Lett. B 206 (1988) 364.
\newblock \href {https://doi.org/10.1016/0370-2693(88)91523-7} {\path{doi:10.1016/0370-2693(88)91523-7}}.

\bibitem{Ji:2020ena}
X.~Ji, F.~Yuan, Y.~Zhao, {What we know and what we don\textquoteright{}t know about the proton spin after 30 years}, Nature Rev. Phys. 3~(1) (2021) 27--38.
\newblock \href {http://arxiv.org/abs/2009.01291} {\path{arXiv:2009.01291}}, \href {https://doi.org/10.1038/s42254-020-00248-4} {\path{doi:10.1038/s42254-020-00248-4}}.

\bibitem{Dalitz:1965fb}
R.~H. Dalitz, {Quark models for the ''elementary particles''}, in: {Summer School of Theoretical Physics}: {High Energy Physics}, 1965, pp. 251--324.

\bibitem{Dalitz:1966fd}
R.~H. Dalitz, {Resonant states and strong interactions}, in: {Oxford International Conference on Elementary Particles}, 1966, pp. 157--181.

\bibitem{Fritzsch:1972jv}
H.~Fritzsch, M.~Gell-Mann, {Current algebra: Quarks and what else?}, eConf C720906V2 (1972) 135--165.
\newblock \href {http://arxiv.org/abs/hep-ph/0208010} {\path{arXiv:hep-ph/0208010}}.

\bibitem{Fritzsch:1973pi}
H.~Fritzsch, M.~Gell-Mann, H.~Leutwyler, {Advantages of the Color Octet Gluon Picture}, Phys. Lett. B 47 (1973) 365--368.
\newblock \href {https://doi.org/10.1016/0370-2693(73)90625-4} {\path{doi:10.1016/0370-2693(73)90625-4}}.

\bibitem{Gross:1973id}
D.~J. Gross, F.~Wilczek, {Ultraviolet Behavior of Nonabelian Gauge Theories}, Phys. Rev. Lett. 30 (1973) 1343--1346.
\newblock \href {https://doi.org/10.1103/PhysRevLett.30.1343} {\path{doi:10.1103/PhysRevLett.30.1343}}.

\bibitem{Politzer:1973fx}
H.~D. Politzer, {Reliable Perturbative Results for Strong Interactions?}, Phys. Rev. Lett. 30 (1973) 1346--1349.
\newblock \href {https://doi.org/10.1103/PhysRevLett.30.1346} {\path{doi:10.1103/PhysRevLett.30.1346}}.

\bibitem{E598:1974sol}
J.~J. Aubert, et~al., {Experimental Observation of a Heavy Particle $J$}, Phys. Rev. Lett. 33 (1974) 1404--1406.
\newblock \href {https://doi.org/10.1103/PhysRevLett.33.1404} {\path{doi:10.1103/PhysRevLett.33.1404}}.

\bibitem{SLAC-SP-017:1974ind}
J.~E. Augustin, et~al., {Discovery of a Narrow Resonance in $e^+ e^-$ Annihilation}, Phys. Rev. Lett. 33 (1974) 1406--1408.
\newblock \href {https://doi.org/10.1103/PhysRevLett.33.1406} {\path{doi:10.1103/PhysRevLett.33.1406}}.

\bibitem{Abrams:1974yy}
G.~S. Abrams, et~al., {The Discovery of a Second Narrow Resonance in $e^+ e^-$ Annihilation}, Phys. Rev. Lett. 33 (1974) 1453--1455.
\newblock \href {https://doi.org/10.1103/PhysRevLett.33.1453} {\path{doi:10.1103/PhysRevLett.33.1453}}.

\bibitem{Eichten:1974af}
E.~Eichten, K.~Gottfried, T.~Kinoshita, J.~B. Kogut, K.~D. Lane, T.-M. Yan, {The Spectrum of Charmonium}, Phys. Rev. Lett. 34 (1975) 369--372, [Erratum: Phys.Rev.Lett. 36, 1276 (1976)].
\newblock \href {https://doi.org/10.1103/PhysRevLett.34.369} {\path{doi:10.1103/PhysRevLett.34.369}}.

\bibitem{Wilczek:2004im}
F.~Wilczek, {Diquarks as inspiration and as objects}, in: {Deserfest: A Celebration of the Life and Works of Stanley Deser}, 2004, pp. 322--338.
\newblock \href {http://arxiv.org/abs/hep-ph/0409168} {\path{arXiv:hep-ph/0409168}}, \href {https://doi.org/10.1142/9789812775344_0007} {\path{doi:10.1142/9789812775344_0007}}.

\bibitem{Anselmino:1992vg}
M.~Anselmino, et~al., {Diquarks}, Rev. Mod. Phys. 65 (1993) 1199--1234.
\newblock \href {https://doi.org/10.1103/RevModPhys.65.1199} {\path{doi:10.1103/RevModPhys.65.1199}}.

\bibitem{Santopinto:2014opa}
E.~Santopinto, J.~Ferretti, {Strange and nonstrange baryon spectra in the relativistic interacting quark-diquark model with a G\"ursey and Radicati-inspired exchange interaction}, Phys. Rev. C 92~(2) (2015) 025202.
\newblock \href {http://arxiv.org/abs/1412.7571} {\path{arXiv:1412.7571}}, \href {https://doi.org/10.1103/PhysRevC.92.025202} {\path{doi:10.1103/PhysRevC.92.025202}}.

\bibitem{Barabanov:2020jvn}
M.~Y. Barabanov, et~al., {Diquark correlations in hadron physics: Origin, impact and evidence}, Prog. Part. Nucl. Phys. 116 (2021) 103835.
\newblock \href {http://arxiv.org/abs/2008.07630} {\path{arXiv:2008.07630}}, \href {https://doi.org/10.1016/j.ppnp.2020.103835} {\path{doi:10.1016/j.ppnp.2020.103835}}.

\bibitem{Giannini:2001kb}
M.~M. Giannini, E.~Santopinto, A.~Vassallo, {Hypercentral constituent quark model and isospin dependence}, Eur. Phys. J. A 12 (2001) 447--452.
\newblock \href {http://arxiv.org/abs/nucl-th/0111073} {\path{arXiv:nucl-th/0111073}}, \href {https://doi.org/10.1007/s10050-001-8668-y} {\path{doi:10.1007/s10050-001-8668-y}}.

\bibitem{Giannini:2002vp}
M.~M. Giannini, E.~Santopinto, A.~Vassallo, {The hypercentral constituent quark model}, Nucl. Phys. A 699 (2002) 308--311.
\newblock \href {https://doi.org/10.1016/S0375-9474(01)01508-1} {\path{doi:10.1016/S0375-9474(01)01508-1}}.

\bibitem{Giannini:2003xx}
M.~M. Giannini, E.~Santopinto, A.~Vassallo, {An Overview of the hypercentral constituent quark model}, Prog. Part. Nucl. Phys. 50 (2003) 263--272.
\newblock \href {http://arxiv.org/abs/nucl-th/0301017} {\path{arXiv:nucl-th/0301017}}.

\bibitem{DeSanctis:2005kt}
M.~De~Sanctis, M.~M. Giannini, E.~Santopinto, A.~Vassallo, {Electromagnetic form factors and the hypercentral constituent quark model}, Phys. Rev. C 76 (2007) 062201.
\newblock \href {http://arxiv.org/abs/nucl-th/0506033} {\path{arXiv:nucl-th/0506033}}, \href {https://doi.org/10.1103/PhysRevC.76.062201} {\path{doi:10.1103/PhysRevC.76.062201}}.

\bibitem{Santopinto:2010zz}
E.~Santopinto, A.~Vassallo, M.~M. Giannini, M.~De~Sanctis, {High $Q^2$ behavior of the electromagnetic form factors in the relativistic hypercentral constituent quark model}, Phys. Rev. C 82 (2010) 065204.
\newblock \href {http://arxiv.org/abs/1506.08427} {\path{arXiv:1506.08427}}, \href {https://doi.org/10.1103/PhysRevC.82.065204} {\path{doi:10.1103/PhysRevC.82.065204}}.

\bibitem{Santopinto:2012nq}
E.~Santopinto, M.~M. Giannini, {Systematic study of longitudinal and transverse helicity amplitudes \,in \,the \,hypercentral \,constituent \,quark model}, Phys. Rev. C 86 (2012) 065202.

\bibitem{Salehi:2013qga}
N.~Salehi, H.~Hassanabadi, A.~Akbar~Rajabi, {The light and strange baryon spectrum in a non-relativistic hypercentral quark potential model and algebraic framework}, Eur. Phys. J. Plus 128 (2013) 27.
\newblock \href {https://doi.org/10.1140/epjp/i2013-13027-y} {\path{doi:10.1140/epjp/i2013-13027-y}}.

\bibitem{Giannini:2015zia}
M.~M. Giannini, E.~Santopinto, {The hypercentral Constituent Quark Model and its application to baryon properties}, Chin. J. Phys. 53 (2015) 020301.
\newblock \href {http://arxiv.org/abs/1501.03722} {\path{arXiv:1501.03722}}, \href {https://doi.org/10.6122/CJP.20150120} {\path{doi:10.6122/CJP.20150120}}.

\bibitem{Giannini:2016jta}
M.~M. Giannini, {High $Q^2$ Helicity Amplitudes in the Hypercentral Constituent Quark Model}, Few Body Syst. 57~(11) (2016) 1009--1017.
\newblock \href {https://doi.org/10.1007/s00601-016-1142-9} {\path{doi:10.1007/s00601-016-1142-9}}.

\bibitem{Sattari:2019kkd}
F.~Sattari, M.~Aslanzadeh, {Structure of Baryons in a Semi-Relativistic Quark Model}, Braz. J. Phys. 49~(3) (2019) 402--411.
\newblock \href {https://doi.org/10.1007/s13538-019-00649-6} {\path{doi:10.1007/s13538-019-00649-6}}.

\bibitem{Tazimi:2021azh}
N.~Tazimi, P.~Sadeghi~Alavijeh, {Investigation of Baryons in the Hypercentral Quark Model}, Adv. High Energy Phys. 2021 (2021) 7713697.
\newblock \href {https://doi.org/10.1155/2021/7713697} {\path{doi:10.1155/2021/7713697}}.

\bibitem{Isgur:1977ef}
N.~Isgur, G.~Karl, {Hyperfine Interactions in Negative Parity Baryons}, Phys. Lett. B 72 (1977) 109.

\bibitem{Isgur:1978xj}
N.~Isgur, G.~Karl, {P Wave Baryons in the Quark Model}, Phys. Rev. D 18 (1978) 4187.

\bibitem{Isgur:1978xi}
N.~Isgur, G.~Karl, {Symmetry Breaking in Baryons}, Phys. Lett. B 74 (1978) 353--356.
\newblock \href {https://doi.org/10.1016/0370-2693(78)90676-7} {\path{doi:10.1016/0370-2693(78)90676-7}}.

\bibitem{Isgur:1978wd}
N.~Isgur, G.~Karl, {Positive Parity Excited Baryons in a Quark Model with Hyperfine Interactions}, Phys. Rev. D 19 (1979) 2653, [Erratum: Phys.Rev.D 23, 817 (1981)].

\bibitem{Isgur:1978xb}
N.~Isgur, G.~Karl, R.~Koniuk, {Violations of SU(6) Selection Rules from Quark Hyperfine Interactions}, Phys. Rev. Lett. 41 (1978) 1269, [Erratum: Phys.Rev.Lett. 45, 1738 (1980)].
\newblock \href {https://doi.org/10.1103/PhysRevLett.41.1269} {\path{doi:10.1103/PhysRevLett.41.1269}}.

\bibitem{Isgur:1979be}
N.~Isgur, G.~Karl, {Ground State Baryons in a Quark Model with Hyperfine Interactions}, Phys. Rev. D 20 (1979) 1191--1194.
\newblock \href {https://doi.org/10.1103/PhysRevD.20.1191} {\path{doi:10.1103/PhysRevD.20.1191}}.

\bibitem{Isgur:1979ee}
N.~Isgur, G.~Karl, {Ground State Baryon Magnetic Moments}, Phys. Rev. D 21 (1980) 3175.
\newblock \href {https://doi.org/10.1103/PhysRevD.21.3175} {\path{doi:10.1103/PhysRevD.21.3175}}.

\bibitem{Flamm:1982jv}
D.~Flamm, F.~Schoberl, {Introduction to the quark model of elementary particles. Quantum numbers, gauge theories and hadron spectroscopy}, Gordon and Breach, New York, London, Paris, 1982.

\bibitem{Veneziano:1968yb}
G.~Veneziano, {Construction of a crossing - symmetric, Regge behaved amplitude for linearly rising trajectories}, Nuovo Cim. A 57 (1968) 190--197.
\newblock \href {https://doi.org/10.1007/BF02824451} {\path{doi:10.1007/BF02824451}}.

\bibitem{Klempt:2021nuf}
E.~Klempt, {Scalar mesons and the fragmented glueball}, Phys. Lett. B 820 (2021) 136512.
\newblock \href {http://arxiv.org/abs/2104.09922} {\path{arXiv:2104.09922}}, \href {https://doi.org/10.1016/j.physletb.2021.136512} {\path{doi:10.1016/j.physletb.2021.136512}}.

\bibitem{Bowler:1980pfa}
K.~C. Bowler, P.~J. Corvi, A.~J.~G. Hey, P.~D. Jarvis, {Is the $\Delta$ $D_{35} (1925)$ Resonance Evidence for New Baryonic Degrees of Freedom?}, Phys. Rev. Lett. 45 (1980) 97.
\newblock \href {https://doi.org/10.1103/PhysRevLett.45.97} {\path{doi:10.1103/PhysRevLett.45.97}}.

\bibitem{Bowler:1981xh}
K.~C. Bowler, et~al., {The Role of Sp(12,r) in the Harmonic Oscillator Quark Model}, Phys. Rev. D 24 (1981) 197.
\newblock \href {https://doi.org/10.1103/PhysRevD.24.197} {\path{doi:10.1103/PhysRevD.24.197}}.

\bibitem{Burkert:2017djo}
V.~D. Burkert, C.~D. Roberts, {Colloquium : Roper resonance: Toward a solution to the fifty year puzzle}, Rev. Mod. Phys. 91~(1) (2019) 011003.

\bibitem{Nefkens:2005dh}
B.~M.~K. Nefkens, {Ten menus for hadron physics}, Int. J. Mod. Phys. A 20 (2005) 1803--1809.
\newblock \href {https://doi.org/10.1142/S0217751X05023360} {\path{doi:10.1142/S0217751X05023360}}.

\bibitem{Capstick:1986ter}
S.~Capstick, N.~Isgur, {Baryons in a relativized quark model with chromodynamics}, Phys. Rev. D 34~(9) (1986) 2809--2835.

\bibitem{Loring:2001kx}
U.~Löring, B.~C. Metsch, H.~R. Petry, {The Light baryon spectrum in a relativistic quark model with instanton induced quark forces: \,The \,nonstrange \,baryon \,spectrum \,and \,ground states}, Eur. Phys. J. A 10 (2001) 395--446.
\newblock \href {http://arxiv.org/abs/hep-ph/0103289} {\path{arXiv:hep-ph/0103289}}, \href {https://doi.org/10.1007/s100500170105} {\path{doi:10.1007/s100500170105}}.

\bibitem{Khemchandani:2013nma}
K.~P. Khemchandani, A.~Martinez~Torres, H.~Nagahiro, A.~Hosaka, {Role of vector and pseudoscalar mesons in understanding $1/2^- N^*$ and \ensuremath{\Delta} resonances}, Phys. Rev. D 88~(11) (2013) 114016.
\newblock \href {http://arxiv.org/abs/1307.8420} {\path{arXiv:1307.8420}}, \href {https://doi.org/10.1103/PhysRevD.88.114016} {\path{doi:10.1103/PhysRevD.88.114016}}.

\bibitem{Khemchandani:2020exc}
K.~P. Khemchandani, A.~Martinez~Torres, H.~Nagahiro, A.~Hosaka, {Decay properties of $N^*(1895)$}, Phys. Rev. D 103~(1) (2021) 016015.
\newblock \href {http://arxiv.org/abs/2010.04584} {\path{arXiv:2010.04584}}, \href {https://doi.org/10.1103/PhysRevD.103.016015} {\path{doi:10.1103/PhysRevD.103.016015}}.

\bibitem{Klempt:2020bdu}
E.~Klempt, et~al., {$\varLambda $ and $\varSigma $ Excitations and the Quark Model}, Eur. Phys. J. A 56~(10) (2020) 261.
\newblock \href {http://arxiv.org/abs/2007.04232} {\path{arXiv:2007.04232}}, \href {https://doi.org/10.1140/epja/s10050-020-00261-2} {\path{doi:10.1140/epja/s10050-020-00261-2}}.

\bibitem{Burkert:2020akg}
V.~D. Burkert, et~al., {The CLAS12 Spectrometer at Jefferson Laboratory}, Nucl. Instrum. Meth. A 959 (2020) 163419.
\newblock \href {https://doi.org/10.1016/j.nima.2020.163419} {\path{doi:10.1016/j.nima.2020.163419}}.

\bibitem{Ernst:2020hru}
A.~Ernst, {Photoproduction of Cascade Baryons Using the GlueX Detector at Jefferson Laboratory}, Ph.D. thesis, Florida State U., Tallahassee (main), Florida State U. (2020).

\bibitem{PANDA:2020hmi}
F.~J\"ulich, et~al., {Study of excited $\Xi$ baryons with the $\overline{\text{ P }}$ANDA detector}, Eur. Phys. J. A 57~(4) (2021) 149.
\newblock \href {http://arxiv.org/abs/2012.01776} {\path{arXiv:2012.01776}}, \href {https://doi.org/10.1140/epja/s10050-021-00444-5} {\path{doi:10.1140/epja/s10050-021-00444-5}}.

\bibitem{Cluster:2025}
J.~Dingfelder, U.~Thoma, J.~Albrecht, A.~Lenz, Color meets flavor, bonn-Dortmund-Siegen Cluster of Excellence (2025).

\bibitem{Crede:2024hur}
V.~Crede, J.~Yelton, {70 years of hyperon spectroscopy: a review of strange $\Xi$, $\Omega$ baryons, and the spectrum of charmed and bottom baryons}, Rept. Prog. Phys. 87~(10) (2024) 106301.
\newblock \href {http://arxiv.org/abs/2502.08815} {\path{arXiv:2502.08815}}, \href {https://doi.org/10.1088/1361-6633/ad7610} {\path{doi:10.1088/1361-6633/ad7610}}.

\bibitem{Capstick:2000qj}
S.~Capstick, W.~Roberts, {Quark models of baryon masses and decays}, Prog. Part. Nucl. Phys. 45 (2000) S241--S331.
\newblock \href {http://arxiv.org/abs/nucl-th/0008028} {\path{arXiv:nucl-th/0008028}}, \href {https://doi.org/10.1016/S0146-6410(00)00109-5} {\path{doi:10.1016/S0146-6410(00)00109-5}}.

\bibitem{Okubo:1963fa}
S.~Okubo, {Phi meson and unitary symmetry model}, Phys. Lett. 5 (1963) 165--168.
\newblock \href {https://doi.org/10.1016/S0375-9601(63)92548-9} {\path{doi:10.1016/S0375-9601(63)92548-9}}.

\bibitem{Iizuka:1966fk}
J.~Iizuka, {Systematics and phenomenology of meson family}, Prog. Theor. Phys. Suppl. 37 (1966) 21--34.
\newblock \href {https://doi.org/10.1143/PTPS.37.21} {\path{doi:10.1143/PTPS.37.21}}.

\bibitem{Fernandez-Ramirez:2015fbq}
C.~Fernandez-Ramirez, I.~V. Danilkin, V.~Mathieu, A.~P. Szczepaniak, {Understanding the Nature of $\Lambda$(1405) through Regge Physics}, Phys. Rev. D 93~(7) (2016) 074015.
\newblock \href {http://arxiv.org/abs/1512.03136} {\path{arXiv:1512.03136}}, \href {https://doi.org/10.1103/PhysRevD.93.074015} {\path{doi:10.1103/PhysRevD.93.074015}}.

\bibitem{JPAC:2018zjz}
J.~A. Silva-Castro, et~al., {Regge phenomenology of the $N^*$ and $\Delta^*$ poles}, Phys. Rev. D 99~(3) (2019) 034003.
\newblock \href {http://arxiv.org/abs/1809.01954} {\path{arXiv:1809.01954}}, \href {https://doi.org/10.1103/PhysRevD.99.034003} {\path{doi:10.1103/PhysRevD.99.034003}}.

\bibitem{Bijker:1994yr}
R.~Bijker, F.~Iachello, A.~Leviatan, {Algebraic models of hadron structure. 1. Nonstrange baryons}, Annals Phys. 236 (1994) 69--116.
\newblock \href {http://arxiv.org/abs/nucl-th/9402012} {\path{arXiv:nucl-th/9402012}}, \href {https://doi.org/10.1006/aphy.1994.1108} {\path{doi:10.1006/aphy.1994.1108}}.

\bibitem{Bijker:1995ii}
R.~Bijker, F.~Iachello, A.~Leviatan, {Electromagnetic form-factors in a collective model of the nucleon}, Phys. Rev. C 54 (1996) 1935--1953.
\newblock \href {http://arxiv.org/abs/nucl-th/9510001} {\path{arXiv:nucl-th/9510001}}, \href {https://doi.org/10.1103/PhysRevC.54.1935} {\path{doi:10.1103/PhysRevC.54.1935}}.

\bibitem{Bijker:1996tr}
R.~Bijker, F.~Iachello, A.~Leviatan, {Strong decays of nonstrange $q^3$ baryons}, Phys. Rev. D 55 (1997) 2862--2873.
\newblock \href {http://arxiv.org/abs/nucl-th/9608057} {\path{arXiv:nucl-th/9608057}}, \href {https://doi.org/10.1103/PhysRevD.55.2862} {\path{doi:10.1103/PhysRevD.55.2862}}.

\bibitem{Merten:2002nz}
D.~Merten, et~al., {Electroweak form-factors of nonstrange baryons}, Eur. Phys. J. A 14 (2002) 477--489.
\newblock \href {http://arxiv.org/abs/hep-ph/0204024} {\path{arXiv:hep-ph/0204024}}, \href {https://doi.org/10.1140/epja/i2002-10009-9} {\path{doi:10.1140/epja/i2002-10009-9}}.

\bibitem{Metsch:2003ix}
B.~Metsch, U.~Löring, D.~Merten, H.~Petry, {The spectrum and strong decays of baryons in a relativistic quark model}, Eur. Phys. J. A 18 (2003) 189--192.
\newblock \href {https://doi.org/10.1140/epja/i2002-10298-x} {\path{doi:10.1140/epja/i2002-10298-x}}.

\bibitem{Koniuk:1979vy}
R.~Koniuk, N.~Isgur, {Baryon Decays in a Quark Model with Chromodynamics}, Phys. Rev. D 21 (1980) 1868, [Erratum: Phys.Rev.D 23, 818 (1981)].
\newblock \href {https://doi.org/10.1103/PhysRevD.21.1868} {\path{doi:10.1103/PhysRevD.21.1868}}.

\bibitem{Koniuk:1979vw}
R.~Koniuk, N.~Isgur, {Where Have All the Resonances Gone? An Analysis of Baryon Couplings in a Quark Model With Chromodynamics}, Phys. Rev. Lett. 44 (1980) 845.
\newblock \href {https://doi.org/10.1103/PhysRevLett.44.845} {\path{doi:10.1103/PhysRevLett.44.845}}.

\bibitem{Koniuk:1981ej}
R.~Koniuk, {Baryon - Vector Meson Couplings in a Quark Model With Chromodynamics}, Nucl. Phys. B 195 (1982) 452--465.
\newblock \href {https://doi.org/10.1016/0550-3213(82)90005-0} {\path{doi:10.1016/0550-3213(82)90005-0}}.

\bibitem{Capstick:1992th}
S.~Capstick, W.~Roberts, {$N\pi$ decays of baryons in a relativized model}, Phys. Rev. D 47 (1993) 1994--2010.
\newblock \href {https://doi.org/10.1103/PhysRevD.47.1994} {\path{doi:10.1103/PhysRevD.47.1994}}.

\bibitem{Capstick:1993kb}
S.~Capstick, W.~Roberts, {Quasi two-body decays of nonstrange baryons}, Phys. Rev. D 49 (1994) 4570--4586.
\newblock \href {http://arxiv.org/abs/nucl-th/9310030} {\path{arXiv:nucl-th/9310030}}, \href {https://doi.org/10.1103/PhysRevD.49.4570} {\path{doi:10.1103/PhysRevD.49.4570}}.

\bibitem{Capstick:1998uh}
S.~Capstick, W.~Roberts, {Strange decays of nonstrange baryons}, Phys. Rev. D 58 (1998) 074011.
\newblock \href {http://arxiv.org/abs/nucl-th/9804070} {\path{arXiv:nucl-th/9804070}}, \href {https://doi.org/10.1103/PhysRevD.58.074011} {\path{doi:10.1103/PhysRevD.58.074011}}.

\bibitem{Capstick:1998md}
S.~Capstick, W.~Roberts, {New baryons in the $\Delta \eta$ and $\Delta \omega$ channels}, Phys. Rev. D 57 (1998) 4301--4309.
\newblock \href {http://arxiv.org/abs/nucl-th/9708048} {\path{arXiv:nucl-th/9708048}}, \href {https://doi.org/10.1103/PhysRevD.57.4301} {\path{doi:10.1103/PhysRevD.57.4301}}.

\bibitem{Goity:2007ft}
J.~L. Goity, N.~N. Scoccola, {Photo-production of Positive Parity Excited Baryons in the 1/$N_c$ Expansion of QCD}, Phys. Rev. Lett. 99 (2007) 062002.
\newblock \href {http://arxiv.org/abs/hep-ph/0701244} {\path{arXiv:hep-ph/0701244}}, \href {https://doi.org/10.1103/PhysRevLett.99.062002} {\path{doi:10.1103/PhysRevLett.99.062002}}.

\bibitem{Goity:2009wq}
J.~L. Goity, C.~Jayalath, N.~N. Scoccola, {Analysis of 56-plet positive parity baryon decays in the \protect{$1/N_c$} expansion}, Phys. Rev. D 80 (2009) 074027.
\newblock \href {http://arxiv.org/abs/0907.2706} {\path{arXiv:0907.2706}}, \href {https://doi.org/10.1103/PhysRevD.80.074027} {\path{doi:10.1103/PhysRevD.80.074027}}.

\bibitem{Jayalath:2011uc}
C.~Jayalath, J.~L. Goity, E.~Gonzalez~de Urreta, N.~N. Scoccola, {Negative parity baryon decays in the $1/N_c$ expansion}, Phys. Rev. D 84 (2011) 074012.
\newblock \href {http://arxiv.org/abs/1108.2042} {\path{arXiv:1108.2042}}, \href {https://doi.org/10.1103/PhysRevD.84.074012} {\path{doi:10.1103/PhysRevD.84.074012}}.

\bibitem{LeYaouanc:1973ldf}
A.~Le~Yaouanc, L.~Oliver, O.~Pene, J.~C. Raynal, {Naive quark pair creation model and baryon decays}, Phys. Rev. D 9 (1974) 1415--1419.
\newblock \href {https://doi.org/10.1103/PhysRevD.9.1415} {\path{doi:10.1103/PhysRevD.9.1415}}.

\bibitem{LeYaouanc:1974cvx}
A.~Le~Yaouanc, L.~Oliver, O.~Pene, J.~C. Raynal, {Resonant Partial Wave Amplitudes in $\pi + n \to \pi + \pi + n$ According to the Naive Quark Pair Creation Model}, Phys. Rev. D 11 (1975) 1272.
\newblock \href {https://doi.org/10.1103/PhysRevD.11.1272} {\path{doi:10.1103/PhysRevD.11.1272}}.

\bibitem{Stancu:1988gb}
F.~Stancu, P.~Stassart, {Pion Decay of Baryons in a Flux Tube Quark Model}, Phys. Rev. D 38 (1988) 233--237.
\newblock \href {https://doi.org/10.1103/PhysRevD.38.233} {\path{doi:10.1103/PhysRevD.38.233}}.

\bibitem{Stancu:1989iu}
F.~Stancu, P.~Stassart, {Role of the Pion Size and Flux Tube Extension in a Baryon Decay Model}, Phys. Rev. D 39 (1989) 343--346.
\newblock \href {https://doi.org/10.1103/PhysRevD.39.343} {\path{doi:10.1103/PhysRevD.39.343}}.

\bibitem{Stassart:1990zt}
P.~Stassart, F.~Stancu, {$N + \rho$ decay of baryons in a flux tube breaking mechanism}, Phys. Rev. D 42 (1990) 1521--1526.
\newblock \href {https://doi.org/10.1103/PhysRevD.42.1521} {\path{doi:10.1103/PhysRevD.42.1521}}.

\bibitem{Stancu:1993xz}
F.~Stancu, P.~Stassart, {$N + \omega$ decay of baryons in a flux tube breaking mechanism}, Phys. Rev. D 47 (1993) 2140--2142.
\newblock \href {https://doi.org/10.1103/PhysRevD.47.2140} {\path{doi:10.1103/PhysRevD.47.2140}}.

\bibitem{Stassart:1995qf}
P.~Stassart, F.~Stancu, {$\Delta\pi$ decay of baryons in a flux tube breaking mechanism}, Z. Phys. A 351 (1995) 77--82.
\newblock \href {https://doi.org/10.1007/BF01292788} {\path{doi:10.1007/BF01292788}}.

\bibitem{Melde:2005hy}
T.~Melde, W.~Plessas, R.~F. Wagenbrunn, {Covariant calculation of mesonic baryon decays}, Phys. Rev. C 72 (2005) 015207, [Erratum: Phys.Rev.C 74, 069901 (2006)].
\newblock \href {http://arxiv.org/abs/hep-ph/0505198} {\path{arXiv:hep-ph/0505198}}, \href {https://doi.org/10.1103/PhysRevC.74.069901} {\path{doi:10.1103/PhysRevC.74.069901}}.

\bibitem{Faessler:2010zzc}
A.~Faessler, et~al., {Low-lying baryon decays in the $^{3}P_0$ quark model}, J. Phys. G 37~(11) (2010) 115002.
\newblock \href {https://doi.org/10.1088/0954-3899/37/11/115002} {\path{doi:10.1088/0954-3899/37/11/115002}}.

\bibitem{Theussl:2000sj}
L.~Theussl, R.~F. Wagenbrunn, B.~Desplanques, W.~Plessas, {Hadronic decays of $N$ and $\Delta$ resonances in a chiral quark model}, Eur. Phys. J. A 12 (2001) 91--101.
\newblock \href {http://arxiv.org/abs/nucl-th/0010099} {\path{arXiv:nucl-th/0010099}}, \href {https://doi.org/10.1007/s100500170042} {\path{doi:10.1007/s100500170042}}.

\bibitem{Wagenbrunn:2000es}
R.~F. Wagenbrunn, et~al., {Covariant nucleon electromagnetic form-factors from the Goldstone boson exchange quark model}, Phys. Lett. B 511 (2001) 33--39.
\newblock \href {http://arxiv.org/abs/nucl-th/0010048} {\path{arXiv:nucl-th/0010048}}, \href {https://doi.org/10.1016/S0370-2693(01)00622-0} {\path{doi:10.1016/S0370-2693(01)00622-0}}.

\bibitem{Boffi:2001zb}
S.~Boffi, et~al., {Covariant electroweak nucleon form-factors in a chiral constituent quark model}, Eur. Phys. J. A 14 (2002) 17--21.
\newblock \href {http://arxiv.org/abs/hep-ph/0108271} {\path{arXiv:hep-ph/0108271}}, \href {https://doi.org/10.1007/s10050-002-8784-3} {\path{doi:10.1007/s10050-002-8784-3}}.

\bibitem{Melde:2007zz}
T.~Melde, et~al., {Electromagnetic nucleon form factors in instant and point form}, Phys. Rev. D 76 (2007) 074020.
\newblock \href {https://doi.org/10.1103/PhysRevD.76.074020} {\path{doi:10.1103/PhysRevD.76.074020}}.

\bibitem{Sengl:2007yq}
B.~Sengl, T.~Melde, W.~Plessas, {Covariant calculation of strange decays of baryon resonances}, Phys. Rev. D 76 (2007) 054008.
\newblock \href {http://arxiv.org/abs/0705.1642} {\path{arXiv:0705.1642}}, \href {https://doi.org/10.1103/PhysRevD.76.054008} {\path{doi:10.1103/PhysRevD.76.054008}}.

\bibitem{Samios:1974tw}
N.~P. Samios, M.~Goldberg, B.~T. Meadows, {Hadrons and SU(3): a critical review}, Rev. Mod. Phys. 46 (1974) 49--81.
\newblock \href {https://doi.org/10.1103/RevModPhys.46.49} {\path{doi:10.1103/RevModPhys.46.49}}.

\bibitem{Guzey:2005vz}
V.~Guzey, M.~V. Polyakov, {SU(3) systematization of baryons}, unpublished (12, 2005).
\newblock \href {http://arxiv.org/abs/hep-ph/0512355} {\path{arXiv:hep-ph/0512355}}.

\bibitem{CBELSATAPS:2015taz}
A.~Thiel, et~al., {Three-body nature of $N^{\bf *}$ and $\Delta^*$ resonances from sequential decay chains}, Phys. Rev. Lett. 114~(9) (2015) 091803.
\newblock \href {http://arxiv.org/abs/1501.02094} {\path{arXiv:1501.02094}}, \href {https://doi.org/10.1103/PhysRevLett.114.091803} {\path{doi:10.1103/PhysRevLett.114.091803}}.

\bibitem{Morel:2002vk}
D.~Morel, S.~Capstick, {Baryon meson loop effects on the spectrum of nonstrange baryons}, unpublished (4 2002).
\newblock \href {http://arxiv.org/abs/nucl-th/0204014} {\path{arXiv:nucl-th/0204014}}.

\bibitem{Anisovich:1998au}
V.~V. Anisovich, D.~V. Bugg, A.~V. Sarantsev, {Exotic mesons, locking states and their role in the formation of the confinement barrier}, Phys. Rev. D 58 (1998) 111503.
\newblock \href {http://arxiv.org/abs/hep-ph/9802426} {\path{arXiv:hep-ph/9802426}}, \href {https://doi.org/10.1103/PhysRevD.58.111503} {\path{doi:10.1103/PhysRevD.58.111503}}.

\bibitem{Weinberg:1978kz}
S.~Weinberg, {Phenomenological Lagrangians}, Physica A 96~(1-2) (1979) 327--340.
\newblock \href {https://doi.org/10.1016/0378-4371(79)90223-1} {\path{doi:10.1016/0378-4371(79)90223-1}}.

\bibitem{Gell-Mann:1968hlm}
M.~Gell-Mann, R.~J. Oakes, B.~Renner, {Behavior of current divergences under SU(3) x SU(3)}, Phys. Rev. 175 (1968) 2195--2199.
\newblock \href {https://doi.org/10.1103/PhysRev.175.2195} {\path{doi:10.1103/PhysRev.175.2195}}.

\bibitem{Nambu:1960tm}
Y.~Nambu, {Quasiparticles and Gauge Invariance in the Theory of Superconductivity}, Phys. Rev. 117 (1960) 648--663.
\newblock \href {https://doi.org/10.1103/PhysRev.117.648} {\path{doi:10.1103/PhysRev.117.648}}.

\bibitem{Goldstone:1961eq}
J.~Goldstone, {Field Theories with Superconductor Solutions}, Nuovo Cim. 19 (1961) 154--164.
\newblock \href {https://doi.org/10.1007/BF02812722} {\path{doi:10.1007/BF02812722}}.

\bibitem{Weinberg:1995mt}
S.~Weinberg, {The Quantum theory of fields. Vol. 1: Foundations}, Cambridge University Press, 2005.
\newblock \href {https://doi.org/10.1017/CBO9781139644167} {\path{doi:10.1017/CBO9781139644167}}.

\bibitem{Meissner:2020khl}
U.-G. Mei\ss{}ner, {Two-pole structures in QCD: Facts, not fantasy!}, Symmetry 12~(6) (2020) 981.
\newblock \href {http://arxiv.org/abs/2005.06909} {\path{arXiv:2005.06909}}, \href {https://doi.org/10.3390/sym12060981} {\path{doi:10.3390/sym12060981}}.

\bibitem{Mai:2020ltx}
M.~Mai, {Review of the ${\Lambda }$(1405): A curious case of a strangeness resonance}, Eur. Phys. J. ST 230~(6) (2021) 1593--1607.
\newblock \href {http://arxiv.org/abs/2010.00056} {\path{arXiv:2010.00056}}, \href {https://doi.org/10.1140/epjs/s11734-021-00144-7} {\path{doi:10.1140/epjs/s11734-021-00144-7}}.

\bibitem{Guo:2023wes}
F.-K. Guo, Y.~Kamiya, M.~Mai, U.-G. Mei\ss{}ner, {New insights into the nature of the \ensuremath{\Lambda}(1380) and \ensuremath{\Lambda}(1405) resonances away from the SU(3) limit}, Phys. Lett. B 846 (2023) 138264.
\newblock \href {http://arxiv.org/abs/2308.07658} {\path{arXiv:2308.07658}}, \href {https://doi.org/10.1016/j.physletb.2023.138264} {\path{doi:10.1016/j.physletb.2023.138264}}.

\bibitem{Guo:2017jvc}
F.-K. Guo, et~al., {Hadronic molecules}, Rev. Mod. Phys. 90~(1) (2018) 015004, [Erratum: Rev.Mod.Phys. 94, 029901 (2022)].
\newblock \href {http://arxiv.org/abs/1705.00141} {\path{arXiv:1705.00141}}, \href {https://doi.org/10.1103/RevModPhys.90.015004} {\path{doi:10.1103/RevModPhys.90.015004}}.

\bibitem{Oller:2019opk}
J.~A. Oller, {Coupled-channel approach in hadron\textendash{}hadron scattering}, Prog. Part. Nucl. Phys. 110 (2020) 103728.
\newblock \href {http://arxiv.org/abs/1909.00370} {\path{arXiv:1909.00370}}, \href {https://doi.org/10.1016/j.ppnp.2019.103728} {\path{doi:10.1016/j.ppnp.2019.103728}}.

\bibitem{Melnitchouk:2005zr}
W.~Melnitchouk, R.~Ent, C.~Keppel, {Quark-hadron duality in electron scattering}, Phys. Rept. 406 (2005) 127--301.
\newblock \href {http://arxiv.org/abs/hep-ph/0501217} {\path{arXiv:hep-ph/0501217}}, \href {https://doi.org/10.1016/j.physrep.2004.10.004} {\path{doi:10.1016/j.physrep.2004.10.004}}.

\bibitem{Mai:2025wjb}
M.~Mai, {Theory of resonances} (2 2025).
\newblock \href {http://arxiv.org/abs/2502.02654} {\path{arXiv:2502.02654}}.

\bibitem{Gasser:1983yg}
J.~Gasser, H.~Leutwyler, {Chiral Perturbation Theory to One Loop}, Annals Phys. 158 (1984) 142.
\newblock \href {https://doi.org/10.1016/0003-4916(84)90242-2} {\path{doi:10.1016/0003-4916(84)90242-2}}.

\bibitem{Gasser:1984gg}
J.~Gasser, H.~Leutwyler, {Chiral Perturbation Theory: Expansions in the Mass of the Strange Quark}, Nucl. Phys. B 250 (1985) 465--516.
\newblock \href {https://doi.org/10.1016/0550-3213(85)90492-4} {\path{doi:10.1016/0550-3213(85)90492-4}}.

\bibitem{Meissner:1993ah}
U.~G. Mei{\ss}ner, {Recent developments in chiral perturbation theory}, Rept. Prog. Phys. 56 (1993) 903--996.
\newblock \href {http://arxiv.org/abs/hep-ph/9302247} {\path{arXiv:hep-ph/9302247}}, \href {https://doi.org/10.1088/0034-4885/56/8/001} {\path{doi:10.1088/0034-4885/56/8/001}}.

\bibitem{Ecker:1994gg}
G.~Ecker, {Chiral perturbation theory}, Prog. Part. Nucl. Phys. 35 (1995) 1--80.
\newblock \href {http://arxiv.org/abs/hep-ph/9501357} {\path{arXiv:hep-ph/9501357}}, \href {https://doi.org/10.1016/0146-6410(95)00041-G} {\path{doi:10.1016/0146-6410(95)00041-G}}.

\bibitem{Pich:1995bw}
A.~Pich, {Chiral perturbation theory}, Rept. Prog. Phys. 58 (1995) 563--610.
\newblock \href {http://arxiv.org/abs/hep-ph/9502366} {\path{arXiv:hep-ph/9502366}}, \href {https://doi.org/10.1088/0034-4885/58/6/001} {\path{doi:10.1088/0034-4885/58/6/001}}.

\bibitem{Bernard:1995dp}
V.~Bernard, N.~Kaiser, U.-G. Mei{\ss}ner, {Chiral dynamics in nucleons and nuclei}, Int. J. Mod. Phys. E 4 (1995) 193--346.
\newblock \href {http://arxiv.org/abs/hep-ph/9501384} {\path{arXiv:hep-ph/9501384}}, \href {https://doi.org/10.1142/S0218301395000092} {\path{doi:10.1142/S0218301395000092}}.

\bibitem{Scherer:2002tk}
S.~Scherer, {Introduction to chiral perturbation theory}, Adv. Nucl. Phys. 27 (2003) 277.
\newblock \href {http://arxiv.org/abs/hep-ph/0210398} {\path{arXiv:hep-ph/0210398}}.

\bibitem{Bernard:2006gx}
V.~Bernard, U.-G. Mei{\ss}ner, {Chiral perturbation theory}, Ann. Rev. Nucl. Part. Sci. 57 (2007) 33--60.
\newblock \href {http://arxiv.org/abs/hep-ph/0611231} {\path{arXiv:hep-ph/0611231}}, \href {https://doi.org/10.1146/annurev.nucl.56.080805.140449} {\path{doi:10.1146/annurev.nucl.56.080805.140449}}.

\bibitem{Bijnens:2006zp}
J.~Bijnens, {Chiral perturbation theory beyond one loop}, Prog. Part. Nucl. Phys. 58 (2007) 521--586.
\newblock \href {http://arxiv.org/abs/hep-ph/0604043} {\path{arXiv:hep-ph/0604043}}, \href {https://doi.org/10.1016/j.ppnp.2006.08.002} {\path{doi:10.1016/j.ppnp.2006.08.002}}.

\bibitem{Geng:2013xn}
L.~Geng, {Recent developments in SU(3) covariant baryon chiral perturbation theory}, Front. Phys. (Beijing) 8 (2013) 328--348.
\newblock \href {http://arxiv.org/abs/1301.6815} {\path{arXiv:1301.6815}}, \href {https://doi.org/10.1007/s11467-013-0327-7} {\path{doi:10.1007/s11467-013-0327-7}}.

\bibitem{Meissner:1987ge}
U.~G. Mei{\ss}ner, {Low-Energy Hadron Physics from Effective Chiral Lagrangians with Vector Mesons}, Phys. Rept. 161 (1988) 213.
\newblock \href {https://doi.org/10.1016/0370-1573(88)90090-7} {\path{doi:10.1016/0370-1573(88)90090-7}}.

\bibitem{Ecker:1989yg}
G.~Ecker, et~al., {Chiral Lagrangians for Massive Spin 1 Fields}, Phys. Lett. B 223 (1989) 425--432.
\newblock \href {https://doi.org/10.1016/0370-2693(89)91627-4} {\path{doi:10.1016/0370-2693(89)91627-4}}.

\bibitem{Jenkins:1991es}
E.~E. Jenkins, A.~V. Manohar, {Chiral corrections to the baryon axial currents}, Phys. Lett. B 259 (1991) 353--358.
\newblock \href {https://doi.org/10.1016/0370-2693(91)90840-M} {\path{doi:10.1016/0370-2693(91)90840-M}}.

\bibitem{Hemmert:1997ye}
T.~R. Hemmert, B.~R. Holstein, J.~Kambor, {Chiral Lagrangians and $\Delta(1232)$ interactions: Formalism}, J. Phys. G 24 (1998) 1831--1859.
\newblock \href {http://arxiv.org/abs/hep-ph/9712496} {\path{arXiv:hep-ph/9712496}}, \href {https://doi.org/10.1088/0954-3899/24/10/003} {\path{doi:10.1088/0954-3899/24/10/003}}.

\bibitem{Lutz:2001yb}
M.~F.~M. Lutz, E.~E. Kolomeitsev, {Relativistic chiral SU(3) symmetry, large $N_c$ sum rules and meson baryon scattering}, Nucl. Phys. A 700 (2002) 193--308.
\newblock \href {http://arxiv.org/abs/nucl-th/0105042} {\path{arXiv:nucl-th/0105042}}, \href {https://doi.org/10.1016/S0375-9474(01)01312-4} {\path{doi:10.1016/S0375-9474(01)01312-4}}.

\bibitem{Pascalutsa:2002pi}
V.~Pascalutsa, D.~R. Phillips, {Effective theory of the $\Delta(1232)$ in Compton scattering off the nucleon}, Phys. Rev. C 67 (2003) 055202.
\newblock \href {http://arxiv.org/abs/nucl-th/0212024} {\path{arXiv:nucl-th/0212024}}, \href {https://doi.org/10.1103/PhysRevC.67.055202} {\path{doi:10.1103/PhysRevC.67.055202}}.

\bibitem{Bernard:2003xf}
V.~Bernard, T.~R. Hemmert, U.-G. Mei{\ss}ner, {Infrared regularization with spin 3/2 fields}, Phys. Lett. B 565 (2003) 137--145.
\newblock \href {http://arxiv.org/abs/hep-ph/0303198} {\path{arXiv:hep-ph/0303198}}, \href {https://doi.org/10.1016/S0370-2693(03)00538-0} {\path{doi:10.1016/S0370-2693(03)00538-0}}.

\bibitem{Kolomeitsev:2003kt}
E.~E. Kolomeitsev, M.~F.~M. Lutz, {On baryon resonances and chiral symmetry}, Phys. Lett. B 585 (2004) 243--252.
\newblock \href {http://arxiv.org/abs/nucl-th/0305101} {\path{arXiv:nucl-th/0305101}}, \href {https://doi.org/10.1016/j.physletb.2004.01.066} {\path{doi:10.1016/j.physletb.2004.01.066}}.

\bibitem{Pascalutsa:2006up}
V.~Pascalutsa, M.~Vanderhaeghen, S.~N. Yang, {Electromagnetic excitation of the $\Delta(1232)$-resonance}, Phys. Rept. 437 (2007) 125--232.
\newblock \href {http://arxiv.org/abs/hep-ph/0609004} {\path{arXiv:hep-ph/0609004}}, \href {https://doi.org/10.1016/j.physrep.2006.09.006} {\path{doi:10.1016/j.physrep.2006.09.006}}.

\bibitem{Djukanovic:2009zn}
D.~Djukanovic, J.~Gegelia, A.~Keller, S.~Scherer, {Complex-mass renormalization in chiral effective field theory}, Phys. Lett. B 680 (2009) 235--238.
\newblock \href {http://arxiv.org/abs/0902.4347} {\path{arXiv:0902.4347}}, \href {https://doi.org/10.1016/j.physletb.2009.08.068} {\path{doi:10.1016/j.physletb.2009.08.068}}.

\bibitem{Gegelia:2016xcw}
J.~Gegelia, U.-G. Mei\ss{}ner, D.-L. Yao, {The width of the Roper resonance in baryon chiral perturbation theory}, Phys. Lett. B 760 (2016) 736--741.
\newblock \href {http://arxiv.org/abs/1606.04873} {\path{arXiv:1606.04873}}, \href {https://doi.org/10.1016/j.physletb.2016.07.068} {\path{doi:10.1016/j.physletb.2016.07.068}}.

\bibitem{Oller:1997ng}
J.~A. Oller, E.~Oset, J.~R. Pelaez, {Nonperturbative approach to effective chiral Lagrangians and meson interactions}, Phys. Rev. Lett. 80 (1998) 3452--3455.
\newblock \href {http://arxiv.org/abs/hep-ph/9803242} {\path{arXiv:hep-ph/9803242}}, \href {https://doi.org/10.1103/PhysRevLett.80.3452} {\path{doi:10.1103/PhysRevLett.80.3452}}.

\bibitem{Oller:1997ti}
J.~A. Oller, E.~Oset, {Chiral symmetry amplitudes in the S wave isoscalar and isovector channels and the $\sigma$, f$_0$(980), a$_0$(980) scalar mesons}, Nucl. Phys. A 620 (1997) 438--456, [Erratum: Nucl.Phys.A 652, 407--409 (1999)].
\newblock \href {http://arxiv.org/abs/hep-ph/9702314} {\path{arXiv:hep-ph/9702314}}, \href {https://doi.org/10.1016/S0375-9474(97)00160-7} {\path{doi:10.1016/S0375-9474(97)00160-7}}.

\bibitem{Oller:1998hw}
J.~A. Oller, E.~Oset, J.~R. Pelaez, {Meson meson interaction in a nonperturbative chiral approach}, Phys. Rev. D 59 (1999) 074001, [Erratum: Phys.Rev.D 60, 099906 (1999), Erratum: Phys.Rev.D 75, 099903 (2007)].
\newblock \href {http://arxiv.org/abs/hep-ph/9804209} {\path{arXiv:hep-ph/9804209}}, \href {https://doi.org/10.1103/PhysRevD.59.074001} {\path{doi:10.1103/PhysRevD.59.074001}}.

\bibitem{Nieves:1999bx}
J.~Nieves, E.~Ruiz~Arriola, {Bethe-Salpeter approach for unitarized chiral perturbation theory}, Nucl. Phys. A 679 (2000) 57--117.
\newblock \href {http://arxiv.org/abs/hep-ph/9907469} {\path{arXiv:hep-ph/9907469}}, \href {https://doi.org/10.1016/S0375-9474(00)00321-3} {\path{doi:10.1016/S0375-9474(00)00321-3}}.

\bibitem{Bruns:2010sv}
P.~C. Bruns, M.~Mai, U.~G. Mei{\ss}ner, {Chiral dynamics of the S11(1535) and S11(1650) resonances revisited}, Phys. Lett. B 697 (2011) 254--259.
\newblock \href {http://arxiv.org/abs/1012.2233} {\path{arXiv:1012.2233}}, \href {https://doi.org/10.1016/j.physletb.2011.02.008} {\path{doi:10.1016/j.physletb.2011.02.008}}.

\bibitem{Inoue:2001ip}
T.~Inoue, E.~Oset, M.~J. Vicente~Vacas, {Chiral unitary approach to S wave meson baryon scattering in the strangeness S = O sector}, Phys. Rev. C 65 (2002) 035204.
\newblock \href {http://arxiv.org/abs/hep-ph/0110333} {\path{arXiv:hep-ph/0110333}}, \href {https://doi.org/10.1103/PhysRevC.65.035204} {\path{doi:10.1103/PhysRevC.65.035204}}.

\bibitem{Chew:1960iv}
G.~F. Chew, S.~Mandelstam, {Theory of low-energy pion pion interactions}, Phys. Rev. 119 (1960) 467--477.
\newblock \href {https://doi.org/10.1103/PhysRev.119.467} {\path{doi:10.1103/PhysRev.119.467}}.

\bibitem{Oller:1998zr}
J.~A. Oller, E.~Oset, {N/D description of two meson amplitudes and chiral symmetry}, Phys. Rev. D 60 (1999) 074023.
\newblock \href {http://arxiv.org/abs/hep-ph/9809337} {\path{arXiv:hep-ph/9809337}}, \href {https://doi.org/10.1103/PhysRevD.60.074023} {\path{doi:10.1103/PhysRevD.60.074023}}.

\bibitem{Meissner:1999vr}
U.-G. Mei{\ss}ner, J.~A. Oller, {Chiral unitary meson baryon dynamics in the presence of resonances: Elastic pion nucleon scattering}, Nucl. Phys. A 673 (2000) 311--334.
\newblock \href {http://arxiv.org/abs/nucl-th/9912026} {\path{arXiv:nucl-th/9912026}}, \href {https://doi.org/10.1016/S0375-9474(00)00150-0} {\path{doi:10.1016/S0375-9474(00)00150-0}}.

\bibitem{Truong:1988zp}
T.~N. Truong, {Chiral Perturbation Theory and Final State Theorem}, Phys. Rev. Lett. 61 (1988) 2526.
\newblock \href {https://doi.org/10.1103/PhysRevLett.61.2526} {\path{doi:10.1103/PhysRevLett.61.2526}}.

\bibitem{Pelaez:2006nj}
J.~R. Pelaez, G.~Rios, {Nature of the $f_0(600)$ from its $N_c$ dependence at two loops in unitarized Chiral Perturbation Theory}, Phys. Rev. Lett. 97 (2006) 242002.
\newblock \href {http://arxiv.org/abs/hep-ph/0610397} {\path{arXiv:hep-ph/0610397}}, \href {https://doi.org/10.1103/PhysRevLett.97.242002} {\path{doi:10.1103/PhysRevLett.97.242002}}.

\bibitem{GomezNicola:2007qj}
A.~Gomez~Nicola, J.~R. Pelaez, G.~Rios, {The Inverse Amplitude Method and Adler Zeros}, Phys. Rev. D 77 (2008) 056006.
\newblock \href {http://arxiv.org/abs/0712.2763} {\path{arXiv:0712.2763}}, \href {https://doi.org/10.1103/PhysRevD.77.056006} {\path{doi:10.1103/PhysRevD.77.056006}}.

\bibitem{Pelaez:2010fj}
J.~R. Pelaez, G.~Rios, {Chiral extrapolation of light resonances from one and two-loop unitarized Chiral Perturbation Theory versus lattice results}, Phys. Rev. D 82 (2010) 114002.
\newblock \href {http://arxiv.org/abs/1010.6008} {\path{arXiv:1010.6008}}, \href {https://doi.org/10.1103/PhysRevD.82.114002} {\path{doi:10.1103/PhysRevD.82.114002}}.

\bibitem{Hoferichter:2015hva}
M.~Hoferichter, J.~Ruiz~de Elvira, B.~Kubis, U.-G. Mei\ss{}ner, {Roy\textendash{}Steiner-equation analysis of pion\textendash{}nucleon scattering}, Phys. Rept. 625 (2016) 1--88.
\newblock \href {http://arxiv.org/abs/1510.06039} {\path{arXiv:1510.06039}}, \href {https://doi.org/10.1016/j.physrep.2016.02.002} {\path{doi:10.1016/j.physrep.2016.02.002}}.

\bibitem{Hoferichter:2015tha}
M.~Hoferichter, J.~Ruiz~de Elvira, B.~Kubis, U.-G. Mei\ss{}ner, {Matching pion-nucleon Roy-Steiner equations to chiral perturbation theory}, Phys. Rev. Lett. 115~(19) (2015) 192301.
\newblock \href {http://arxiv.org/abs/1507.07552} {\path{arXiv:1507.07552}}, \href {https://doi.org/10.1103/PhysRevLett.115.192301} {\path{doi:10.1103/PhysRevLett.115.192301}}.

\bibitem{RuizdeElvira:2017stg}
J.~Ruiz~de Elvira, M.~Hoferichter, B.~Kubis, U.-G. Mei\ss{}ner, {Extracting the $\sigma$-term from low-energy pion-nucleon scattering}, J. Phys. G 45~(2) (2018) 024001.
\newblock \href {http://arxiv.org/abs/1706.01465} {\path{arXiv:1706.01465}}, \href {https://doi.org/10.1088/1361-6471/aa9422} {\path{doi:10.1088/1361-6471/aa9422}}.

\bibitem{Hoferichter:2023mgy}
M.~Hoferichter, J.~R. de~Elvira, B.~Kubis, U.-G. Mei\ss{}ner, {Nucleon resonance parameters from Roy\textendash{}Steiner equations}, Phys. Lett. B 853 (2024) 138698.
\newblock \href {http://arxiv.org/abs/2312.15015} {\path{arXiv:2312.15015}}, \href {https://doi.org/10.1016/j.physletb.2024.138698} {\path{doi:10.1016/j.physletb.2024.138698}}.

\bibitem{Wang:2022osj}
Y.-F. Wang, et~al., {Reaction $\pi N\to\omega N$ in a dynamical coupled-channel approach}, Phys. Rev. D 106~(9) (2022) 094031.
\newblock \href {http://arxiv.org/abs/2208.03061} {\path{arXiv:2208.03061}}, \href {https://doi.org/10.1103/PhysRevD.106.094031} {\path{doi:10.1103/PhysRevD.106.094031}}.

\bibitem{Wang:2024byt}
Y.-F. Wang, et~al., {Global Data-Driven Determination of Baryon Transition Form Factors}, Phys. Rev. Lett. 133~(10) (2024) 101901.
\newblock \href {http://arxiv.org/abs/2404.17444} {\path{arXiv:2404.17444}}, \href {https://doi.org/10.1103/PhysRevLett.133.101901} {\path{doi:10.1103/PhysRevLett.133.101901}}.

\bibitem{Shrestha:2012ep}
M.~Shrestha, D.~M. Manley, {Multichannel parametrization of $\pi N$ scattering amplitudes and extraction of resonance parameters}, Phys. Rev. C 86 (2012) 055203.
\newblock \href {http://arxiv.org/abs/1208.2710} {\path{arXiv:1208.2710}}, \href {https://doi.org/10.1103/PhysRevC.86.055203} {\path{doi:10.1103/PhysRevC.86.055203}}.

\bibitem{Wang:2023snv}
Y.-F. Wang, U.-G. Mei\ss{}ner, D.~R\"onchen, C.-W. Shen, {Examination of the nature of the N* and \ensuremath{\Delta} resonances via coupled-channels dynamics}, Phys. Rev. C 109~(1) (2024) 015202.
\newblock \href {http://arxiv.org/abs/2307.06799} {\path{arXiv:2307.06799}}, \href {https://doi.org/10.1103/PhysRevC.109.015202} {\path{doi:10.1103/PhysRevC.109.015202}}.

\bibitem{Li:2023pjx}
H.-P. Li, et~al., {Contrasting observables related to the $N^*(1535)$ from the molecular or a genuine structure}, Eur. Phys. J. C 84~(7) (2024) 656.
\newblock \href {http://arxiv.org/abs/2311.14365} {\path{arXiv:2311.14365}}, \href {https://doi.org/10.1140/epjc/s10052-024-13015-x} {\path{doi:10.1140/epjc/s10052-024-13015-x}}.

\bibitem{Molina:2023jov}
R.~Molina, C.-W. Xiao, W.-H. Liang, E.~Oset, {Correlation functions for the N*(1535) and the inverse problem}, Phys. Rev. D 109~(5) (2024) 054002.
\newblock \href {http://arxiv.org/abs/2310.12593} {\path{arXiv:2310.12593}}, \href {https://doi.org/10.1103/PhysRevD.109.054002} {\path{doi:10.1103/PhysRevD.109.054002}}.

\bibitem{Luscher:1986pf}
M.~Lüscher, {Volume Dependence of the Energy Spectrum in Massive Quantum Field Theories. 2. Scattering States}, Commun. Math. Phys. 105 (1986) 153--188.
\newblock \href {https://doi.org/10.1007/BF01211097} {\path{doi:10.1007/BF01211097}}.

\bibitem{Luscher:1990ux}
M.~Lüscher, {Two particle states on a torus and their relation to the scattering matrix}, Nucl. Phys. B 354 (1991) 531--578.
\newblock \href {https://doi.org/10.1016/0550-3213(91)90366-6} {\path{doi:10.1016/0550-3213(91)90366-6}}.

\bibitem{Richard:2012xw}
J.-M. Richard, An introduction to the quark model, in: {Ferrara International School Niccol\`o Cabeo 2012: Hadronic spectroscopy}, 2012.
\newblock \href {http://arxiv.org/abs/1205.4326} {\path{arXiv:1205.4326}}.

\bibitem{DeRujula:1975qlm}
A.~De~Rujula, H.~Georgi, S.~L. Glashow, {Hadron Masses in a Gauge Theory}, Phys. Rev. D 12 (1975) 147--162.
\newblock \href {https://doi.org/10.1103/PhysRevD.12.147} {\path{doi:10.1103/PhysRevD.12.147}}.

\bibitem{Bali:2000gf}
G.~S. Bali, {QCD forces and heavy quark bound states}, Phys. Rept. 343 (2001) 1--136.
\newblock \href {http://arxiv.org/abs/hep-ph/0001312} {\path{arXiv:hep-ph/0001312}}, \href {https://doi.org/10.1016/S0370-1573(00)00079-X} {\path{doi:10.1016/S0370-1573(00)00079-X}}.

\bibitem{Greensite:2003bk}
J.~Greensite, {The Confinement problem in lattice gauge theory}, Prog. Part. Nucl. Phys. 51 (2003) 1.
\newblock \href {http://arxiv.org/abs/hep-lat/0301023} {\path{arXiv:hep-lat/0301023}}, \href {https://doi.org/10.1016/S0146-6410(03)90012-3} {\path{doi:10.1016/S0146-6410(03)90012-3}}.

\bibitem{Alkofer:2000wg}
R.~Alkofer, L.~von Smekal, {The Infrared behavior of QCD Green's functions: Confinement dynamical symmetry breaking, and hadrons as relativistic bound states}, Phys. Rept. 353 (2001) 281.
\newblock \href {http://arxiv.org/abs/hep-ph/0007355} {\path{arXiv:hep-ph/0007355}}, \href {https://doi.org/10.1016/S0370-1573(01)00010-2} {\path{doi:10.1016/S0370-1573(01)00010-2}}.

\bibitem{Kondo:2014sta}
K.-I. Kondo, S.~Kato, A.~Shibata, T.~Shinohara, {Quark confinement: Dual superconductor picture based on a non-Abelian Stokes theorem and reformulations of Yang\textendash{}Mills theory}, Phys. Rept. 579 (2015) 1--226.
\newblock \href {http://arxiv.org/abs/1409.1599} {\path{arXiv:1409.1599}}, \href {https://doi.org/10.1016/j.physrep.2015.03.002} {\path{doi:10.1016/j.physrep.2015.03.002}}.

\bibitem{Maas:2017wzi}
A.~Maas, Brout-englert-higgs physics: From foundations to phenomenology, Prog. Part. Nucl. Phys. 106 (2019) 132--209.
\newblock \href {http://arxiv.org/abs/1712.04721} {\path{arXiv:1712.04721}}, \href {https://doi.org/10.1016/j.ppnp.2019.02.003} {\path{doi:10.1016/j.ppnp.2019.02.003}}.

\bibitem{Lichtenberg:1969sxc}
D.~B. Lichtenberg, {Baryon supermultiplets of SU(6) x O(3) in a quark-diquark model}, Phys. Rev. 178 (1969) 2197--2200.
\newblock \href {https://doi.org/10.1103/PhysRev.178.2197} {\path{doi:10.1103/PhysRev.178.2197}}.

\bibitem{Santopinto:2004hw}
E.~Santopinto, An interacting quark-diquark model of baryons, Phys. Rev. C 72 (2005) 022201.
\newblock \href {http://arxiv.org/abs/hep-ph/0412319} {\path{arXiv:hep-ph/0412319}}, \href {https://doi.org/10.1103/PhysRevC.72.022201} {\path{doi:10.1103/PhysRevC.72.022201}}.

\bibitem{Ebert:2007nw}
D.~Ebert, R.~N. Faustov, V.~O. Galkin, Masses of excited heavy baryons in the relativistic quark model, Phys. Lett. B 659 (2008) 612--620.
\newblock \href {http://arxiv.org/abs/0705.2957} {\path{arXiv:0705.2957}}, \href {https://doi.org/10.1016/j.physletb.2007.11.037} {\path{doi:10.1016/j.physletb.2007.11.037}}.

\bibitem{Glozman:1995fu}
L.~Y. Glozman, D.~O. Riska, {The Spectrum of the nucleons and the strange hyperons and chiral dynamics}, Phys. Rept. 268 (1996) 263--303.
\newblock \href {http://arxiv.org/abs/hep-ph/9505422} {\path{arXiv:hep-ph/9505422}}, \href {https://doi.org/10.1016/0370-1573(95)00062-3} {\path{doi:10.1016/0370-1573(95)00062-3}}.

\bibitem{Glozman:1996wq}
L.~Y. Glozman, Z.~Papp, W.~Plessas, Light baryons in a constituent quark model with chiral dynamics, Phys. Lett. B 381 (1996) 311--316.
\newblock \href {http://arxiv.org/abs/hep-ph/9601353} {\path{arXiv:hep-ph/9601353}}, \href {https://doi.org/10.1016/0370-2693(96)00610-7} {\path{doi:10.1016/0370-2693(96)00610-7}}.

\bibitem{Glozman:1997ag}
L.~Y. Glozman, W.~Plessas, K.~Varga, R.~F. Wagenbrunn, {Unified description of light and strange baryon spectra}, Phys. Rev. D 58 (1998) 094030.
\newblock \href {http://arxiv.org/abs/hep-ph/9706507} {\path{arXiv:hep-ph/9706507}}, \href {https://doi.org/10.1103/PhysRevD.58.094030} {\path{doi:10.1103/PhysRevD.58.094030}}.

\bibitem{Glozman:2001zc}
L.~Y. Glozman, et~al., Covariant axial form-factor of the nucleon in a chiral constituent quark model, Phys. Lett. B 516 (2001) 183--190.
\newblock \href {http://arxiv.org/abs/nucl-th/0105028} {\path{arXiv:nucl-th/0105028}}, \href {https://doi.org/10.1016/S0370-2693(01)00915-7} {\path{doi:10.1016/S0370-2693(01)00915-7}}.

\bibitem{Melde:2008yr}
T.~Melde, W.~Plessas, B.~Sengl, {Quark-Model Identification of Baryon Ground and Resonant States}, Phys. Rev. D 77 (2008) 114002.
\newblock \href {http://arxiv.org/abs/0806.1454} {\path{arXiv:0806.1454}}, \href {https://doi.org/10.1103/PhysRevD.77.114002} {\path{doi:10.1103/PhysRevD.77.114002}}.

\bibitem{Melde:2008dg}
T.~Melde, L.~Canton, W.~Plessas, Structure of meson-baryon interaction vertices, Phys. Rev. Lett. 102 (2009) 132002.
\newblock \href {http://arxiv.org/abs/0811.0277} {\path{arXiv:0811.0277}}, \href {https://doi.org/10.1103/PhysRevLett.102.132002} {\path{doi:10.1103/PhysRevLett.102.132002}}.

\bibitem{Plessas:2015mpa}
W.~Plessas, The constituent-quark model — nowadays, Int. J. Mod. Phys. A 30~(02) (2015) 1530013.
\newblock \href {https://doi.org/10.1142/S0217751X15300136} {\path{doi:10.1142/S0217751X15300136}}.

\bibitem{Ronniger:2012xp}
M.~Ronniger, B.~C. Metsch, {Effects of a spin-flavour dependent interaction on light-flavoured baryon helicity amplitudes}, Eur. Phys. J. A 49 (2013) 8.
\newblock \href {http://arxiv.org/abs/1207.2640} {\path{arXiv:1207.2640}}, \href {https://doi.org/10.1140/epja/i2013-13008-9} {\path{doi:10.1140/epja/i2013-13008-9}}.

\bibitem{BES:2009ufh}
M.~Ablikim, et~al., {Partial wave analysis of $J/\psi \to p \bar p \pi^0$}, Phys. Rev. D 80 (2009) 052004.
\newblock \href {http://arxiv.org/abs/0905.1562} {\path{arXiv:0905.1562}}, \href {https://doi.org/10.1103/PhysRevD.80.052004} {\path{doi:10.1103/PhysRevD.80.052004}}.

\bibitem{BESIII:2012ssm}
M.~Ablikim, et~al., {Observation of two new $N^*$ resonances in the decay $\psi(3686) \rightarrow p\bar{p}\pi^0$}, Phys. Rev. Lett. 110~(2) (2013) 022001.
\newblock \href {http://arxiv.org/abs/1207.0223} {\path{arXiv:1207.0223}}, \href {https://doi.org/10.1103/PhysRevLett.110.022001} {\path{doi:10.1103/PhysRevLett.110.022001}}.

\bibitem{Ferraris:1995ui}
M.~Ferraris, et~al., {A Three body force model for the baryon spectrum}, Phys. Lett. B 364 (1995) 231--238.
\newblock \href {https://doi.org/10.1016/0370-2693(95)01091-2} {\path{doi:10.1016/0370-2693(95)01091-2}}.

\bibitem{Giannini:2005ks}
M.~M. Giannini, E.~Santopinto, A.~Vassallo, A new application of the {G}ursey and {R}adicati mass formula, Eur. Phys. J. A 25 (2005) 241--247.
\newblock \href {http://arxiv.org/abs/nucl-th/0506032} {\path{arXiv:nucl-th/0506032}}, \href {https://doi.org/10.1140/epja/i2005-10113-4} {\path{doi:10.1140/epja/i2005-10113-4}}.

\bibitem{Loring:2001kv}
U.~Löring, K.~Kretzschmar, B.~C. Metsch, H.~R. Petry, {Relativistic quark models of baryons with instantaneous forces: Theoretical background}, Eur. Phys. J. A 10 (2001) 309--346.
\newblock \href {http://arxiv.org/abs/hep-ph/0103287} {\path{arXiv:hep-ph/0103287}}, \href {https://doi.org/10.1007/s100500170117} {\path{doi:10.1007/s100500170117}}.

\bibitem{Loring:2001ky}
U.~Löring, B.~C. Metsch, H.~R. Petry, {The Light baryon spectrum in a relativistic quark model with instanton induced quark forces: The Strange baryon spectrum}, Eur. Phys. J. A 10 (2001) 447--486.
\newblock \href {http://arxiv.org/abs/hep-ph/0103290} {\path{arXiv:hep-ph/0103290}}, \href {https://doi.org/10.1007/s100500170106} {\path{doi:10.1007/s100500170106}}.

\bibitem{Metscha:2008gkf}
B.~Metsch, Quark models, Eur. Phys. J. A 35 (2008) 275--280.
\newblock \href {https://doi.org/10.1140/epja/i2007-10557-4} {\path{doi:10.1140/epja/i2007-10557-4}}.

\bibitem{Ronniger:2011td}
M.~Ronniger, B.~C. Metsch, Effects of a spin-flavour dependent interaction on the baryon mass spectrum, Eur. Phys. J. A 47 (2011) 162.
\newblock \href {http://arxiv.org/abs/1111.3835} {\path{arXiv:1111.3835}}, \href {https://doi.org/10.1140/epja/i2011-11162-8} {\path{doi:10.1140/epja/i2011-11162-8}}.

\bibitem{Bijker:2000gq}
R.~Bijker, F.~Iachello, A.~Leviatan, {Algebraic models of hadron structure. 2. Strange baryons}, Annals Phys. 284 (2000) 89--133.
\newblock \href {http://arxiv.org/abs/nucl-th/0004034} {\path{arXiv:nucl-th/0004034}}, \href {https://doi.org/10.1006/aphy.2000.6064} {\path{doi:10.1006/aphy.2000.6064}}.

\bibitem{Goity:2003ab}
J.~L. Goity, C.~Schat, N.~N. Scoccola, Analysis of the [56,$2^+$] baryon masses in the \protect{$1/N_c$} expansion, Phys. Lett. B 564 (2003) 83--89.
\newblock \href {http://arxiv.org/abs/hep-ph/0304167} {\path{arXiv:hep-ph/0304167}}, \href {https://doi.org/10.1016/S0370-2693(03)00700-7} {\path{doi:10.1016/S0370-2693(03)00700-7}}.

\bibitem{Matagne:2004pm}
N.~Matagne, F.~Stancu, The [56,$4^+$] baryons in the 1/$n_c$ expansion, Phys. Rev. D 71 (2005) 014010.
\newblock \href {http://arxiv.org/abs/hep-ph/0409261} {\path{arXiv:hep-ph/0409261}}, \href {https://doi.org/10.1103/PhysRevD.71.014010} {\path{doi:10.1103/PhysRevD.71.014010}}.

\bibitem{Brambilla:2019esw}
N.~Brambilla, et~al., {The $XYZ$ states: experimental and theoretical status and perspectives}, Phys. Rept. 873 (2020) 1--154.
\newblock \href {http://arxiv.org/abs/1907.07583} {\path{arXiv:1907.07583}}, \href {https://doi.org/10.1016/j.physrep.2020.05.001} {\path{doi:10.1016/j.physrep.2020.05.001}}.

\bibitem{Liu:2019zoy}
Y.-R. Liu, H.-X. Chen, W.~Chen, X.~Liu, S.-L. Zhu, {Pentaquark and Tetraquark states}, Prog. Part. Nucl. Phys. 107 (2019) 237--320.
\newblock \href {http://arxiv.org/abs/1903.11976} {\path{arXiv:1903.11976}}, \href {https://doi.org/10.1016/j.ppnp.2019.04.003} {\path{doi:10.1016/j.ppnp.2019.04.003}}.

\bibitem{Dong:2021bvy}
X.-K. Dong, F.-K. Guo, B.-S. Zou, {A survey of heavy\textendash{}heavy hadronic molecules}, Commun. Theor. Phys. 73~(12) (2021) 125201.
\newblock \href {http://arxiv.org/abs/2108.02673} {\path{arXiv:2108.02673}}, \href {https://doi.org/10.1088/1572-9494/ac27a2} {\path{doi:10.1088/1572-9494/ac27a2}}.

\bibitem{Dong:2021juy}
X.-K. Dong, F.-K. Guo, B.-S. Zou, {A survey of heavy-antiheavy hadronic molecules}, Progr. Phys. 41 (2021) 65--93.
\newblock \href {http://arxiv.org/abs/2101.01021} {\path{arXiv:2101.01021}}, \href {https://doi.org/10.13725/j.cnki.pip.2021.02.001} {\path{doi:10.13725/j.cnki.pip.2021.02.001}}.

\bibitem{Chen:2022asf}
H.-X.~a. Chen, {An updated review of the new hadron states}, Rept. Prog. Phys. 86~(2) (2023) 026201.
\newblock \href {http://arxiv.org/abs/2204.02649} {\path{arXiv:2204.02649}}, \href {https://doi.org/10.1088/1361-6633/aca3b6} {\path{doi:10.1088/1361-6633/aca3b6}}.

\bibitem{Burns:2022uiv}
T.~J. Burns, E.~S. Swanson, {Production of $P_c$ states in \ensuremath{\Lambda}b decays}, Phys. Rev. D 106~(5) (2022) 054029.
\newblock \href {http://arxiv.org/abs/2207.00511} {\path{arXiv:2207.00511}}, \href {https://doi.org/10.1103/PhysRevD.106.054029} {\path{doi:10.1103/PhysRevD.106.054029}}.

\bibitem{Klempt:2022zwo}
E.~Klempt, S.~Neubert, {Heavy-flavor baryons}, Eur. Phys. J. C 83, in: \cite{Gross:2022hyw} (11 2022).
\newblock \href {http://arxiv.org/abs/2211.12897} {\path{arXiv:2211.12897}}.

\bibitem{Liu:2024uxn}
M.-Z. Liu, et~al., {Three ways to decipher the nature of exotic hadrons: Multiplets, three-body hadronic molecules, and correlation functions}, Phys. Rept. 1108 (2025) 1--108.
\newblock \href {http://arxiv.org/abs/2404.06399} {\path{arXiv:2404.06399}}, \href {https://doi.org/10.1016/j.physrep.2024.12.001} {\path{doi:10.1016/j.physrep.2024.12.001}}.

\bibitem{Denig:2012by}
A.~Denig, G.~Salme, {Nucleon Electromagnetic Form Factors in the Timelike Region}, Prog. Part. Nucl. Phys. 68 (2013) 113--157.
\newblock \href {http://arxiv.org/abs/1210.4689} {\path{arXiv:1210.4689}}, \href {https://doi.org/10.1016/j.ppnp.2012.09.005} {\path{doi:10.1016/j.ppnp.2012.09.005}}.

\bibitem{Eichmann:2016yit}
G.~Eichmann, et~al., {Baryons as relativistic three-quark bound states}, Prog. Part. Nucl. Phys. 91 (2016) 1--100.
\newblock \href {http://arxiv.org/abs/1606.09602} {\path{arXiv:1606.09602}}, \href {https://doi.org/10.1016/j.ppnp.2016.07.001} {\path{doi:10.1016/j.ppnp.2016.07.001}}.

\bibitem{Lin:2021umz}
Y.-H. Lin, H.-W. Hammer, U.-G. Mei\ss{}ner, {Dispersion-theoretical analysis of the electromagnetic form factors of the nucleon: Past, present and future}, Eur. Phys. J. A 57~(8) (2021) 255.
\newblock \href {http://arxiv.org/abs/2106.06357} {\path{arXiv:2106.06357}}, \href {https://doi.org/10.1140/epja/s10050-021-00562-0} {\path{doi:10.1140/epja/s10050-021-00562-0}}.

\bibitem{Alvarado:2023loi}
F.~Alvarado, D.~An, L.~Alvarez-Ruso, S.~Leupold, {Light quark mass dependence of nucleon electromagnetic form factors in dispersively modified chiral perturbation theory}, Phys. Rev. D 108~(11) (2023) 114021.
\newblock \href {http://arxiv.org/abs/2310.07796} {\path{arXiv:2310.07796}}, \href {https://doi.org/10.1103/PhysRevD.108.114021} {\path{doi:10.1103/PhysRevD.108.114021}}.

\bibitem{Gross:2006fg}
F.~Gross, G.~Ramalho, M.~T. Pena, {A Pure S-wave covariant model for the nucleon}, Phys. Rev. C 77 (2008) 015202.
\newblock \href {http://arxiv.org/abs/nucl-th/0606029} {\path{arXiv:nucl-th/0606029}}, \href {https://doi.org/10.1103/PhysRevC.77.015202} {\path{doi:10.1103/PhysRevC.77.015202}}.

\bibitem{Ramalho:2008dp}
G.~Ramalho, M.~T. Pena, F.~Gross, {$D$-state effects in the electromagnetic $N \Delta$ transition}, Phys. Rev. D 78 (2008) 114017.
\newblock \href {http://arxiv.org/abs/0810.4126} {\path{arXiv:0810.4126}}, \href {https://doi.org/10.1103/PhysRevD.78.114017} {\path{doi:10.1103/PhysRevD.78.114017}}.

\bibitem{Ramalho:2012ng}
G.~Ramalho, M.~T. Pena, {Timelike $\gamma^* N \to \Delta$ form factors and $\Delta$ Dalitz decay}, Phys. Rev. D 85 (2012) 113014.
\newblock \href {http://arxiv.org/abs/1205.2575} {\path{arXiv:1205.2575}}, \href {https://doi.org/10.1103/PhysRevD.85.113014} {\path{doi:10.1103/PhysRevD.85.113014}}.

\bibitem{Ramalho:2015qna}
G.~Ramalho, et~al., {Role of the pion electromagnetic form factor in the $\Delta(1232) \to \gamma^\ast N$ timelike transition}, Phys. Rev. D 93~(3) (2016) 033004.
\newblock \href {http://arxiv.org/abs/1512.03764} {\path{arXiv:1512.03764}}, \href {https://doi.org/10.1103/PhysRevD.93.033004} {\path{doi:10.1103/PhysRevD.93.033004}}.

\bibitem{Ramalho:2016zgc}
G.~Ramalho, M.~T. Pe\~na, {$\gamma^*N\to N^*(1520)$ form factors in the timelike regime}, Phys. Rev. D 95~(1) (2017) 014003.
\newblock \href {http://arxiv.org/abs/1610.08788} {\path{arXiv:1610.08788}}, \href {https://doi.org/10.1103/PhysRevD.95.014003} {\path{doi:10.1103/PhysRevD.95.014003}}.

\bibitem{Ramalho:2020nwk}
G.~Ramalho, M.~T. Pe\~na, {Covariant model for the Dalitz decay of the $N(1535)$ resonance}, Phys. Rev. D 101~(11) (2020) 114008.
\newblock \href {http://arxiv.org/abs/2003.04850} {\path{arXiv:2003.04850}}, \href {https://doi.org/10.1103/PhysRevD.101.114008} {\path{doi:10.1103/PhysRevD.101.114008}}.

\bibitem{Brodsky:2008pg}
S.~J. Brodsky, G.~F. de~Teramond, {AdS/CFT and Light-Front QCD}, Subnucl. Ser. 45 (2009) 139--183.
\newblock \href {http://arxiv.org/abs/0802.0514} {\path{arXiv:0802.0514}}, \href {https://doi.org/10.1142/9789814293242_0008} {\path{doi:10.1142/9789814293242_0008}}.

\bibitem{Brodsky:2014yha}
S.~J. Brodsky, G.~F. de~Teramond, H.~G. Dosch, J.~Erlich, {Light-Front Holographic QCD and Emerging Confinement}, Phys. Rept. 584 (2015) 1--105.
\newblock \href {http://arxiv.org/abs/1407.8131} {\path{arXiv:1407.8131}}, \href {https://doi.org/10.1016/j.physrep.2015.05.001} {\path{doi:10.1016/j.physrep.2015.05.001}}.

\bibitem{Brodsky:1997de}
S.~J. Brodsky, H.-C. Pauli, S.~S. Pinsky, {Quantum chromodynamics and other field theories on the light cone}, Phys. Rept. 301 (1998) 299--486.
\newblock \href {http://arxiv.org/abs/hep-ph/9705477} {\path{arXiv:hep-ph/9705477}}, \href {https://doi.org/10.1016/S0370-1573(97)00089-6} {\path{doi:10.1016/S0370-1573(97)00089-6}}.

\bibitem{Bekenstein:1973ur}
J.~D. Bekenstein, {Black holes and entropy}, Phys. Rev. D 7 (1973) 2333--2346.
\newblock \href {https://doi.org/10.1103/PhysRevD.7.2333} {\path{doi:10.1103/PhysRevD.7.2333}}.

\bibitem{Hawking:1975vcx}
S.~W. Hawking, {Particle Creation by Black Holes}, Commun. Math. Phys. 43 (1975) 199--220, [Erratum: Commun.Math.Phys. 46, 206 (1976)].
\newblock \href {https://doi.org/10.1007/BF02345020} {\path{doi:10.1007/BF02345020}}.

\bibitem{tHooft:1993dmi}
G.~'t~Hooft, {Dimensional reduction in quantum gravity}, Conf. Proc. C 930308 (1993) 284--296.
\newblock \href {http://arxiv.org/abs/gr-qc/9310026} {\path{arXiv:gr-qc/9310026}}.

\bibitem{Susskind:1994vu}
L.~Susskind, {The World as a hologram}, J. Math. Phys. 36 (1995) 6377--6396.
\newblock \href {http://arxiv.org/abs/hep-th/9409089} {\path{arXiv:hep-th/9409089}}, \href {https://doi.org/10.1063/1.531249} {\path{doi:10.1063/1.531249}}.

\bibitem{Maldacena:1997re}
J.~M. Maldacena, {The Large N limit of superconformal field theories and supergravity}, Adv. Theor. Math. Phys. 2 (1998) 231--252.
\newblock \href {http://arxiv.org/abs/hep-th/9711200} {\path{arXiv:hep-th/9711200}}, \href {https://doi.org/10.4310/ATMP.1998.v2.n2.a1} {\path{doi:10.4310/ATMP.1998.v2.n2.a1}}.

\bibitem{Polchinski:2001tt}
J.~Polchinski, M.~J. Strassler, {Hard scattering and gauge / string duality}, Phys. Rev. Lett. 88 (2002) 031601.
\newblock \href {http://arxiv.org/abs/hep-th/0109174} {\path{arXiv:hep-th/0109174}}, \href {https://doi.org/10.1103/PhysRevLett.88.031601} {\path{doi:10.1103/PhysRevLett.88.031601}}.

\bibitem{Karch:2006pv}
A.~Karch, E.~Katz, D.~T. Son, M.~A. Stephanov, {Linear confinement and AdS/QCD}, Phys. Rev. D 74 (2006) 015005.
\newblock \href {http://arxiv.org/abs/hep-ph/0602229} {\path{arXiv:hep-ph/0602229}}, \href {https://doi.org/10.1103/PhysRevD.74.015005} {\path{doi:10.1103/PhysRevD.74.015005}}.

\bibitem{deAlfaro:1976vlx}
V.~de~Alfaro, S.~Fubini, G.~Furlan, {Conformal Invariance in Quantum Mechanics}, Nuovo Cim. A 34 (1976) 569.
\newblock \href {https://doi.org/10.1007/BF02785666} {\path{doi:10.1007/BF02785666}}.

\bibitem{deTeramond:2011aml}
G.~F. de~Teramond, S.~J. Brodsky, {Hadronic Form Factor Models and Spectroscopy Within the Gauge/Gravity Correspondence}, in: {Ferrara International School Niccol\`o Cabeo 2011}: {Hadronic Physics}, 2011, pp. 54--109.
\newblock \href {http://arxiv.org/abs/1203.4025} {\path{arXiv:1203.4025}}.

\bibitem{Kim:2011ey}
Y.~Kim, D.~Yi, {Holography at work for nuclear and hadron physics}, Adv. High Energy Phys. 2011 (2011) 259025.
\newblock \href {http://arxiv.org/abs/1107.0155} {\path{arXiv:1107.0155}}, \href {https://doi.org/10.1155/2011/259025} {\path{doi:10.1155/2011/259025}}.

\bibitem{Guy:2025pc}
G.~F. de~Téramond, private communication (July 2025).

\bibitem{Katz:2005ir}
E.~Katz, A.~Lewandowski, M.~D. Schwartz, {Tensor mesons in AdS/QCD}, Phys. Rev. D 74 (2006) 086004.
\newblock \href {http://arxiv.org/abs/hep-ph/0510388} {\path{arXiv:hep-ph/0510388}}, \href {https://doi.org/10.1103/PhysRevD.74.086004} {\path{doi:10.1103/PhysRevD.74.086004}}.

\bibitem{Brodsky:2006uqa}
S.~J. Brodsky, G.~F. de~Teramond, {Hadronic spectra and light-front wavefunctions in holographic QCD}, Phys. Rev. Lett. 96 (2006) 201601.
\newblock \href {http://arxiv.org/abs/hep-ph/0602252} {\path{arXiv:hep-ph/0602252}}, \href {https://doi.org/10.1103/PhysRevLett.96.201601} {\path{doi:10.1103/PhysRevLett.96.201601}}.

\bibitem{Boschi-Filho:2005xct}
H.~Boschi-Filho, N.~R.~F. Braga, H.~L. Carrion, {Glueball Regge trajectories from gauge/string duality and the Pomeron}, Phys. Rev. D 73 (2006) 047901.
\newblock \href {http://arxiv.org/abs/hep-th/0507063} {\path{arXiv:hep-th/0507063}}, \href {https://doi.org/10.1103/PhysRevD.73.047901} {\path{doi:10.1103/PhysRevD.73.047901}}.

\bibitem{Brodsky:2007hb}
S.~J. Brodsky, G.~F. de~Teramond, {Light-Front Dynamics and AdS/QCD Correspondence: The Pion Form Factor in the Space- and Time-Like Regions}, Phys. Rev. D 77 (2008) 056007.
\newblock \href {http://arxiv.org/abs/0707.3859} {\path{arXiv:0707.3859}}, \href {https://doi.org/10.1103/PhysRevD.77.056007} {\path{doi:10.1103/PhysRevD.77.056007}}.

\bibitem{Forkel:2007cm}
H.~Forkel, M.~Beyer, T.~Frederico, {Linear square-mass trajectories of radially and orbitally excited hadrons in holographic QCD}, JHEP 07 (2007) 077.
\newblock \href {http://arxiv.org/abs/0705.1857} {\path{arXiv:0705.1857}}, \href {https://doi.org/10.1088/1126-6708/2007/07/077} {\path{doi:10.1088/1126-6708/2007/07/077}}.

\bibitem{Brodsky:2008pf}
S.~J. Brodsky, G.~F. de~Teramond, {Light-Front Dynamics and AdS/QCD Correspondence: Gravitational Form Factors of Composite Hadrons}, Phys. Rev. D 78 (2008) 025032.
\newblock \href {http://arxiv.org/abs/0804.0452} {\path{arXiv:0804.0452}}, \href {https://doi.org/10.1103/PhysRevD.78.025032} {\path{doi:10.1103/PhysRevD.78.025032}}.

\bibitem{Colangelo:2008us}
P.~Colangelo, et~al., {Light scalar mesons in the soft-wall model of AdS/QCD}, Phys. Rev. D 78 (2008) 055009.
\newblock \href {http://arxiv.org/abs/0807.1054} {\path{arXiv:0807.1054}}, \href {https://doi.org/10.1103/PhysRevD.78.055009} {\path{doi:10.1103/PhysRevD.78.055009}}.

\bibitem{Branz:2010ub}
T.~Branz, et~al., {Light and heavy mesons in a soft-wall holographic approach}, Phys. Rev. D 82 (2010) 074022.
\newblock \href {http://arxiv.org/abs/1008.0268} {\path{arXiv:1008.0268}}, \href {https://doi.org/10.1103/PhysRevD.82.074022} {\path{doi:10.1103/PhysRevD.82.074022}}.

\bibitem{Li:2013oda}
D.~Li, M.~Huang, {Dynamical holographic QCD model for glueball and light meson spectra}, JHEP 11 (2013) 088.
\newblock \href {http://arxiv.org/abs/1303.6929} {\path{arXiv:1303.6929}}, \href {https://doi.org/10.1007/JHEP11(2013)088} {\path{doi:10.1007/JHEP11(2013)088}}.

\bibitem{Leutgeb:2019gbz}
J.~Leutgeb, A.~Rebhan, {Axial vector transition form factors in holographic QCD and their contribution to the anomalous magnetic moment of the muon}, Phys. Rev. D 101~(11) (2020) 114015.
\newblock \href {http://arxiv.org/abs/1912.01596} {\path{arXiv:1912.01596}}, \href {https://doi.org/10.1103/PhysRevD.101.114015} {\path{doi:10.1103/PhysRevD.101.114015}}.

\bibitem{Roberts:1994dr}
C.~D. Roberts, A.~G. Williams, {Dyson-Schwinger equations and their application to hadronic physics}, Prog. Part. Nucl. Phys. 33 (1994) 477--575.
\newblock \href {http://arxiv.org/abs/hep-ph/9403224} {\path{arXiv:hep-ph/9403224}}, \href {https://doi.org/10.1016/0146-6410(94)90049-3} {\path{doi:10.1016/0146-6410(94)90049-3}}.

\bibitem{Fischer:2006ub}
C.~S. Fischer, {Infrared properties of QCD from Dyson-Schwinger equations}, J. Phys. G 32 (2006) R253--R291.
\newblock \href {http://arxiv.org/abs/hep-ph/0605173} {\path{arXiv:hep-ph/0605173}}, \href {https://doi.org/10.1088/0954-3899/32/8/R02} {\path{doi:10.1088/0954-3899/32/8/R02}}.

\bibitem{Bashir:2012fs}
A.~Bashir, et~al., {Collective perspective on advances in Dyson-Schwinger Equation QCD}, Commun. Theor. Phys. 58 (2012) 79--134.
\newblock \href {http://arxiv.org/abs/1201.3366} {\path{arXiv:1201.3366}}, \href {https://doi.org/10.1088/0253-6102/58/1/16} {\path{doi:10.1088/0253-6102/58/1/16}}.

\bibitem{Cloet:2013jya}
I.~C. Cloet, C.~D. Roberts, {Explanation and Prediction of Observables using Continuum Strong QCD}, Prog. Part. Nucl. Phys. 77 (2014) 1--69.
\newblock \href {http://arxiv.org/abs/1310.2651} {\path{arXiv:1310.2651}}, \href {https://doi.org/10.1016/j.ppnp.2014.02.001} {\path{doi:10.1016/j.ppnp.2014.02.001}}.

\bibitem{Huber:2018ned}
M.~Q. Huber, {Nonperturbative properties of Yang\textendash{}Mills theories}, Phys. Rept. 879 (2020) 1--92.
\newblock \href {http://arxiv.org/abs/1808.05227} {\path{arXiv:1808.05227}}, \href {https://doi.org/10.1016/j.physrep.2020.04.004} {\path{doi:10.1016/j.physrep.2020.04.004}}.

\bibitem{Fischer:2018sdj}
C.~S. Fischer, {QCD at finite temperature and chemical potential from Dyson\textendash{}Schwinger equations}, Prog. Part. Nucl. Phys. 105 (2019) 1--60.
\newblock \href {http://arxiv.org/abs/1810.12938} {\path{arXiv:1810.12938}}, \href {https://doi.org/10.1016/j.ppnp.2019.01.002} {\path{doi:10.1016/j.ppnp.2019.01.002}}.

\bibitem{Berges:2000ew}
J.~Berges, N.~Tetradis, C.~Wetterich, {Nonperturbative renormalization flow in quantum field theory and statistical physics}, Phys. Rept. 363 (2002) 223--386.
\newblock \href {http://arxiv.org/abs/hep-ph/0005122} {\path{arXiv:hep-ph/0005122}}, \href {https://doi.org/10.1016/S0370-1573(01)00098-9} {\path{doi:10.1016/S0370-1573(01)00098-9}}.

\bibitem{Pawlowski:2005xe}
J.~M. Pawlowski, {Aspects of the functional renormalisation group}, Annals Phys. 322 (2007) 2831--2915.
\newblock \href {http://arxiv.org/abs/hep-th/0512261} {\path{arXiv:hep-th/0512261}}, \href {https://doi.org/10.1016/j.aop.2007.01.007} {\path{doi:10.1016/j.aop.2007.01.007}}.

\bibitem{Dupuis:2020fhh}
N.~Dupuis, et~al., {The nonperturbative functional renormalization group and its applications}, Phys. Rept. 910 (2021) 1--114.
\newblock \href {http://arxiv.org/abs/2006.04853} {\path{arXiv:2006.04853}}, \href {https://doi.org/10.1016/j.physrep.2021.01.001} {\path{doi:10.1016/j.physrep.2021.01.001}}.

\bibitem{Cyrol:2017ewj}
A.~K. Cyrol, M.~Mitter, J.~M. Pawlowski, N.~Strodthoff, {Nonperturbative quark, gluon, and meson correlators of unquenched QCD}, Phys. Rev. D 97~(5) (2018) 054006.
\newblock \href {http://arxiv.org/abs/1706.06326} {\path{arXiv:1706.06326}}, \href {https://doi.org/10.1103/PhysRevD.97.054006} {\path{doi:10.1103/PhysRevD.97.054006}}.

\bibitem{Gao:2021wun}
F.~Gao, J.~Papavassiliou, J.~M. Pawlowski, {Fully coupled functional equations for the quark sector of QCD}, Phys. Rev. D 103~(9) (2021) 094013.
\newblock \href {http://arxiv.org/abs/2102.13053} {\path{arXiv:2102.13053}}, \href {https://doi.org/10.1103/PhysRevD.103.094013} {\path{doi:10.1103/PhysRevD.103.094013}}.

\bibitem{Eichmann:2021zuv}
G.~Eichmann, J.~M. Pawlowski, J.~a.~M. Silva, {Mass generation in Landau-gauge Yang-Mills theory}, Phys. Rev. D 104~(11) (2021) 114016.
\newblock \href {http://arxiv.org/abs/2107.05352} {\path{arXiv:2107.05352}}, \href {https://doi.org/10.1103/PhysRevD.104.114016} {\path{doi:10.1103/PhysRevD.104.114016}}.

\bibitem{Fu:2022uow}
W.-j. Fu, C.~Huang, J.~M. Pawlowski, Y.-y. Tan, {Four-quark scatterings in QCD I}, SciPost Phys. 14~(4) (2023) 069.
\newblock \href {http://arxiv.org/abs/2209.13120} {\path{arXiv:2209.13120}}, \href {https://doi.org/10.21468/SciPostPhys.14.4.069} {\path{doi:10.21468/SciPostPhys.14.4.069}}.

\bibitem{Ihssen:2024miv}
F.~Ihssen, J.~M. Pawlowski, F.~R. Sattler, N.~Wink, {Towards quantitative precision in functional QCD I} (8 2024).
\newblock \href {http://arxiv.org/abs/2408.08413} {\path{arXiv:2408.08413}}.

\bibitem{Huber:2020ngt}
M.~Q. Huber, C.~S. Fischer, H.~Sanchis-Alepuz, Spectrum of scalar and pseudoscalar glueballs from functional methods, Eur. Phys. J. C 80~(11) (2020) 1077.
\newblock \href {http://arxiv.org/abs/2004.00415} {\path{arXiv:2004.00415}}, \href {https://doi.org/10.1140/epjc/s10052-020-08649-6} {\path{doi:10.1140/epjc/s10052-020-08649-6}}.

\bibitem{Huber:2021yfy}
M.~Q. Huber, C.~S. Fischer, H.~Sanchis-Alepuz, {Higher spin glueballs from functional methods}, Eur. Phys. J. C 81~(12) (2021) 1083, [Erratum: Eur.Phys.J.C 82, 38 (2022)].
\newblock \href {http://arxiv.org/abs/2110.09180} {\path{arXiv:2110.09180}}, \href {https://doi.org/10.1140/epjc/s10052-021-09864-5} {\path{doi:10.1140/epjc/s10052-021-09864-5}}.

\bibitem{Huber:2025kwy}
M.~Q. Huber, C.~S. Fischer, H.~Sanchis-Alepuz, {Apparent convergence in functional glueball calculations} (3 2025).
\newblock \href {http://arxiv.org/abs/2503.03821} {\path{arXiv:2503.03821}}.

\bibitem{Eichmann:2025wgs}
G.~Eichmann, {Hadron physics with functional methods} (2025).
\newblock \href {http://arxiv.org/abs/2503.10397} {\path{arXiv:2503.10397}}.

\bibitem{Bowman:2005vx}
P.~O. Bowman, et~al., {Unquenched quark propagator in Landau gauge}, Phys. Rev. D 71 (2005) 054507.
\newblock \href {http://arxiv.org/abs/hep-lat/0501019} {\path{arXiv:hep-lat/0501019}}, \href {https://doi.org/10.1103/PhysRevD.71.054507} {\path{doi:10.1103/PhysRevD.71.054507}}.

\bibitem{Oliveira:2018lln}
O.~Oliveira, P.~J. Silva, J.-I. Skullerud, A.~Sternbeck, {Quark propagator with two flavors of O(a)-improved Wilson fermions}, Phys. Rev. D 99~(9) (2019) 094506.
\newblock \href {http://arxiv.org/abs/1809.02541} {\path{arXiv:1809.02541}}, \href {https://doi.org/10.1103/PhysRevD.99.094506} {\path{doi:10.1103/PhysRevD.99.094506}}.

\bibitem{Biddle:2023lod}
J.~C. Biddle, W.~Kamleh, D.~B. Leinweber, {Center vortex structure in the presence of dynamical fermions}, Phys. Rev. D 107~(9) (2023) 094507.
\newblock \href {http://arxiv.org/abs/2302.05897} {\path{arXiv:2302.05897}}, \href {https://doi.org/10.1103/PhysRevD.107.094507} {\path{doi:10.1103/PhysRevD.107.094507}}.

\bibitem{Kamleh:2023gho}
W.~Kamleh, D.~B. Leinweber, A.~Virgili, {Numerical indication that center vortices drive dynamical mass generation in QCD}, Phys. Rev. D 110~(5) (2024) L051502.
\newblock \href {http://arxiv.org/abs/2305.18690} {\path{arXiv:2305.18690}}, \href {https://doi.org/10.1103/PhysRevD.110.L051502} {\path{doi:10.1103/PhysRevD.110.L051502}}.

\bibitem{Sternbeck:2005tk}
A.~Sternbeck, E.~M. Ilgenfritz, M.~Muller-Preussker, A.~Schiller, {Towards the infrared limit in SU(3) Landau gauge lattice gluodynamics}, Phys. Rev. D 72 (2005) 014507.
\newblock \href {http://arxiv.org/abs/hep-lat/0506007} {\path{arXiv:hep-lat/0506007}}, \href {https://doi.org/10.1103/PhysRevD.72.014507} {\path{doi:10.1103/PhysRevD.72.014507}}.

\bibitem{Aguilar:2006gr}
A.~C. Aguilar, J.~Papavassiliou, {Gluon mass generation in the PT-BFM scheme}, JHEP 12 (2006) 012.
\newblock \href {http://arxiv.org/abs/hep-ph/0610040} {\path{arXiv:hep-ph/0610040}}, \href {https://doi.org/10.1088/1126-6708/2006/12/012} {\path{doi:10.1088/1126-6708/2006/12/012}}.

\bibitem{Braun:2007bx}
J.~Braun, H.~Gies, J.~M. Pawlowski, {Quark Confinement from Color Confinement}, Phys. Lett. B 684 (2010) 262--267.
\newblock \href {http://arxiv.org/abs/0708.2413} {\path{arXiv:0708.2413}}, \href {https://doi.org/10.1016/j.physletb.2010.01.009} {\path{doi:10.1016/j.physletb.2010.01.009}}.

\bibitem{Fischer:2008uz}
C.~S. Fischer, A.~Maas, J.~M. Pawlowski, {On the infrared behavior of Landau gauge Yang-Mills theory}, Annals Phys. 324 (2009) 2408--2437.
\newblock \href {http://arxiv.org/abs/0810.1987} {\path{arXiv:0810.1987}}, \href {https://doi.org/10.1016/j.aop.2009.07.009} {\path{doi:10.1016/j.aop.2009.07.009}}.

\bibitem{Binosi:2009qm}
D.~Binosi, J.~Papavassiliou, {Pinch Technique: Theory and Applications}, Phys. Rept. 479 (2009) 1--152.
\newblock \href {http://arxiv.org/abs/0909.2536} {\path{arXiv:0909.2536}}, \href {https://doi.org/10.1016/j.physrep.2009.05.001} {\path{doi:10.1016/j.physrep.2009.05.001}}.

\bibitem{Falcao:2020vyr}
A.~F. Falc\~ao, O.~Oliveira, P.~J. Silva, {Analytic structure of the lattice Landau gauge gluon and ghost propagators}, Phys. Rev. D 102~(11) (2020) 114518.
\newblock \href {http://arxiv.org/abs/2008.02614} {\path{arXiv:2008.02614}}, \href {https://doi.org/10.1103/PhysRevD.102.114518} {\path{doi:10.1103/PhysRevD.102.114518}}.

\bibitem{Ferreira:2023fva}
M.~N. Ferreira, J.~Papavassiliou, {Gauge Sector Dynamics in QCD}, Particles 6~(1) (2023) 312--363.
\newblock \href {http://arxiv.org/abs/2301.02314} {\path{arXiv:2301.02314}}, \href {https://doi.org/10.3390/particles6010017} {\path{doi:10.3390/particles6010017}}.

\bibitem{Eichmann:2009qa}
G.~Eichmann, R.~Alkofer, A.~Krassnigg, D.~Nicmorus, {N}ucleon mass from a covariant three-quark {F}addeev equation, Phys. Rev. Lett. 104 (2010) 201601.
\newblock \href {http://arxiv.org/abs/0912.2246} {\path{arXiv:0912.2246}}, \href {https://doi.org/10.1103/PhysRevLett.104.201601} {\path{doi:10.1103/PhysRevLett.104.201601}}.

\bibitem{Sanchis-Alepuz:2011egq}
H.~Sanchis-Alepuz, G.~Eichmann, S.~Villalba-Chavez, R.~Alkofer, {$\Delta$ and $\Omega$ masses in a three-quark covariant Faddeev approach}, Phys. Rev. D 84 (2011) 096003.
\newblock \href {http://arxiv.org/abs/1109.0199} {\path{arXiv:1109.0199}}, \href {https://doi.org/10.1103/PhysRevD.84.096003} {\path{doi:10.1103/PhysRevD.84.096003}}.

\bibitem{Eichmann:2011vu}
G.~Eichmann, {N}ucleon electromagnetic form factors from the covariant {F}addeev equation, Phys. Rev. D84 (2011) 014014.
\newblock \href {http://arxiv.org/abs/1104.4505} {\path{arXiv:1104.4505}}, \href {https://doi.org/10.1103/PhysRevD.84.014014} {\path{doi:10.1103/PhysRevD.84.014014}}.

\bibitem{Qin:2018dqp}
\protect{Qin, Si-Xue}, C.~D. Roberts, S.~M. Schmidt, Poincaré-covariant analysis of heavy-quark baryons, Phys. Rev. D 97~(11) (2018) 114017.
\newblock \href {http://arxiv.org/abs/1801.09697} {\path{arXiv:1801.09697}}, \href {https://doi.org/10.1103/PhysRevD.97.114017} {\path{doi:10.1103/PhysRevD.97.114017}}.

\bibitem{Eichmann:2019dts}
G.~Eichmann, P.~Duarte, M.~T. Pe\~na, A.~Stadler, {Scattering amplitudes and contour deformations}, Phys. Rev. D 100~(9) (2019) 094001.
\newblock \href {http://arxiv.org/abs/1907.05402} {\path{arXiv:1907.05402}}, \href {https://doi.org/10.1103/PhysRevD.100.094001} {\path{doi:10.1103/PhysRevD.100.094001}}.

\bibitem{Santowsky:2020pwd}
N.~Santowsky, et~al., {$\sigma$-meson: Four-quark versus two-quark components and decay width in a Bethe-Salpeter approach}, Phys. Rev. D 102~(5) (2020) 056014.
\newblock \href {http://arxiv.org/abs/2007.06495} {\path{arXiv:2007.06495}}, \href {https://doi.org/10.1103/PhysRevD.102.056014} {\path{doi:10.1103/PhysRevD.102.056014}}.

\bibitem{Eichmann:2015cra}
G.~Eichmann, C.~S. Fischer, W.~Heupel, {T}he light scalar mesons as tetraquarks, Phys. Lett. B753 (2016) 282--287.
\newblock \href {http://arxiv.org/abs/1508.07178} {\path{arXiv:1508.07178}}, \href {https://doi.org/10.1016/j.physletb.2015.12.036} {\path{doi:10.1016/j.physletb.2015.12.036}}.

\bibitem{Eichmann:2016hgl}
G.~Eichmann, C.~S. Fischer, H.~Sanchis-Alepuz, {Light baryons and their excitations}, Phys. Rev. D 94~(9) (2016) 094033.
\newblock \href {http://arxiv.org/abs/1607.05748} {\path{arXiv:1607.05748}}, \href {https://doi.org/10.1103/PhysRevD.94.094033} {\path{doi:10.1103/PhysRevD.94.094033}}.

\bibitem{Eichmann:2016nsu}
G.~Eichmann, {More about the light baryon spectrum}, Few Body Syst. 58~(2) (2017) 81.
\newblock \href {http://arxiv.org/abs/1611.10118} {\path{arXiv:1611.10118}}, \href {https://doi.org/10.1007/s00601-016-1200-3} {\path{doi:10.1007/s00601-016-1200-3}}.

\bibitem{Maris:1997tm}
P.~Maris, C.~D. Roberts, $\pi$- and \protect{$K$-meson Bethe-Salpeter} amplitudes, Phys. Rev. C 56 (1997) 3369--3383.
\newblock \href {http://arxiv.org/abs/nucl-th/9708029} {\path{arXiv:nucl-th/9708029}}, \href {https://doi.org/10.1103/PhysRevC.56.3369} {\path{doi:10.1103/PhysRevC.56.3369}}.

\bibitem{Maris:1999nt}
P.~Maris, P.~C. Tandy, \protect{Bethe-Salpeter} study of vector meson masses and decay constants, Phys. Rev. C 60 (1999) 055214.
\newblock \href {http://arxiv.org/abs/nucl-th/9905056} {\path{arXiv:nucl-th/9905056}}, \href {https://doi.org/10.1103/PhysRevC.60.055214} {\path{doi:10.1103/PhysRevC.60.055214}}.

\bibitem{Qin:2011dd}
\protect{Qin, Si-Xue}, et~al., Interaction model for the gap equation, Phys. Rev. C 84 (2011) 042202.
\newblock \href {http://arxiv.org/abs/1108.0603} {\path{arXiv:1108.0603}}, \href {https://doi.org/10.1103/PhysRevC.84.042202} {\path{doi:10.1103/PhysRevC.84.042202}}.

\bibitem{Chang:2011ei}
L.~Chang, C.~D. Roberts, Tracing masses of ground-state light-quark mesons, Phys. Rev. C 85 (2012) 052201.
\newblock \href {http://arxiv.org/abs/1104.4821} {\path{arXiv:1104.4821}}, \href {https://doi.org/10.1103/PhysRevC.85.052201} {\path{doi:10.1103/PhysRevC.85.052201}}.

\bibitem{Williams:2015cvx}
R.~Williams, C.~S. Fischer, W.~Heupel, Light mesons in qcd and unquenching effects from the 3pi effective action, Phys. Rev. D 93~(3) (2016) 034026.
\newblock \href {http://arxiv.org/abs/1512.00455} {\path{arXiv:1512.00455}}, \href {https://doi.org/10.1103/PhysRevD.93.034026} {\path{doi:10.1103/PhysRevD.93.034026}}.

\bibitem{Sanchis-Alepuz:2014sca}
H.~Sanchis-Alepuz, C.~S. Fischer, {Octet and Decuplet masses: a covariant three-body Faddeev calculation}, Phys. Rev. D 90~(9) (2014) 096001.
\newblock \href {http://arxiv.org/abs/1408.5577} {\path{arXiv:1408.5577}}, \href {https://doi.org/10.1103/PhysRevD.90.096001} {\path{doi:10.1103/PhysRevD.90.096001}}.

\bibitem{Qin:2019hgk}
\protect{Qin, Si-Xue}, C.~D. Roberts, S.~M. Schmidt, {Spectrum of light- and heavy-baryons}, Few Body Syst. 60~(2) (2019) 26.
\newblock \href {http://arxiv.org/abs/1902.00026} {\path{arXiv:1902.00026}}, \href {https://doi.org/10.1007/s00601-019-1488-x} {\path{doi:10.1007/s00601-019-1488-x}}.

\bibitem{Yao:2024ixu}
Z.-Q. Yao, et~al., {Nucleon gravitational form factors}, Eur. Phys. J. A 61~(5) (2025) 92.
\newblock \href {http://arxiv.org/abs/2409.15547} {\path{arXiv:2409.15547}}, \href {https://doi.org/10.1140/epja/s10050-025-01557-x} {\path{doi:10.1140/epja/s10050-025-01557-x}}.

\bibitem{Oettel:1998bk}
M.~Oettel, G.~Hellstern, R.~Alkofer, H.~Reinhardt, {Octet and decuplet baryons in a covariant and confining diquark - quark model}, Phys. Rev. C 58 (1998) 2459--2477.
\newblock \href {http://arxiv.org/abs/nucl-th/9805054} {\path{arXiv:nucl-th/9805054}}, \href {https://doi.org/10.1103/PhysRevC.58.2459} {\path{doi:10.1103/PhysRevC.58.2459}}.

\bibitem{Eichmann:2008ef}
G.~Eichmann, et~al., {Toward unifying the description of meson and baryon properties}, Phys. Rev. C 79 (2009) 012202.
\newblock \href {http://arxiv.org/abs/0810.1222} {\path{arXiv:0810.1222}}, \href {https://doi.org/10.1103/PhysRevC.79.012202} {\path{doi:10.1103/PhysRevC.79.012202}}.

\bibitem{Chen:2017pse}
C.~Chen, et~al., {Structure of the nucleon\textquoteright{}s low-lying excitations}, Phys. Rev. D 97~(3) (2018) 034016.
\newblock \href {http://arxiv.org/abs/1711.03142} {\path{arXiv:1711.03142}}, \href {https://doi.org/10.1103/PhysRevD.97.034016} {\path{doi:10.1103/PhysRevD.97.034016}}.

\bibitem{Chen:2019fzn}
C.~Chen, et~al., {Spectrum and structure of octet and decuplet baryons and their positive-parity excitations}, Phys. Rev. D 100~(5) (2019) 054009.
\newblock \href {http://arxiv.org/abs/1901.04305} {\path{arXiv:1901.04305}}, \href {https://doi.org/10.1103/PhysRevD.100.054009} {\path{doi:10.1103/PhysRevD.100.054009}}.

\bibitem{Liu:2022ndb}
L.~Liu, et~al., {Composition of low-lying $J=\frac{3}{2}^\pm\,\Delta$ baryons}, Phys. Rev. D 105~(11) (2022) 114047.
\newblock \href {http://arxiv.org/abs/2203.12083} {\path{arXiv:2203.12083}}, \href {https://doi.org/10.1103/PhysRevD.105.114047} {\path{doi:10.1103/PhysRevD.105.114047}}.

\bibitem{Liu:2022nku}
L.~Liu, C.~Chen, C.~D. Roberts, {Wave functions of ($I$,$J^P$)=($\frac{1}{2}$,$\frac{3}{2}^\mp$) baryons}, Phys. Rev. D 107~(1) (2023) 014002.
\newblock \href {http://arxiv.org/abs/2208.12353} {\path{arXiv:2208.12353}}, \href {https://doi.org/10.1103/PhysRevD.107.014002} {\path{doi:10.1103/PhysRevD.107.014002}}.

\bibitem{Segovia:2013rca}
J.~Segovia, C.~Chen, C.~D. Roberts, S.~Wan, {Insights into the $\gamma^* N \to \Delta$ transition}, Phys. Rev. C 88~(3) (2013) 032201.
\newblock \href {http://arxiv.org/abs/1305.0292} {\path{arXiv:1305.0292}}, \href {https://doi.org/10.1103/PhysRevC.88.032201} {\path{doi:10.1103/PhysRevC.88.032201}}.

\bibitem{Raya:2021pyr}
K.~Raya, et~al., {Dynamical diquarks in the $\gamma^{(\ast)} p\to N(1535)\tfrac{1}{2}^-$ transition}, Eur. Phys. J. A 57~(9) (2021) 266.
\newblock \href {http://arxiv.org/abs/2108.02306} {\path{arXiv:2108.02306}}, \href {https://doi.org/10.1140/epja/s10050-021-00574-w} {\path{doi:10.1140/epja/s10050-021-00574-w}}.

\bibitem{Yin:2021uom}
P.-L. Yin, Z.-F. Cui, C.~D. Roberts, J.~Segovia, {Masses of positive- and negative-parity hadron ground-states, including those with heavy quarks}, Eur. Phys. J. C 81~(4) (2021) 327.
\newblock \href {http://arxiv.org/abs/2102.12568} {\path{arXiv:2102.12568}}, \href {https://doi.org/10.1140/epjc/s10052-021-09097-6} {\path{doi:10.1140/epjc/s10052-021-09097-6}}.

\bibitem{Albino:2025fcp}
L.~Albino, et~al., {Insights into the ${\gamma^{(*)} + N(940)\frac{1}{2}^+ \to \Delta(1700)\frac{3}{2}^{-}}$ transition} (2025).
\newblock \href {http://arxiv.org/abs/2502.06206} {\path{arXiv:2502.06206}}.

\bibitem{Albino:2025bnr}
L.~Albino, et~al., {${\gamma^{(*)} + N(940)\frac{1}{2}^+ \to N(1520)\frac{3}{2}^{-}}$ helicity amplitudes and transition form factors} (4, 2025).
\newblock \href {http://arxiv.org/abs/2504.01770} {\path{arXiv:2504.01770}}.

\bibitem{Segovia:2015hra}
J.~Segovia, et~al., {Completing the picture of the Roper resonance}, Phys. Rev. Lett. 115~(17) (2015) 171801.
\newblock \href {http://arxiv.org/abs/1504.04386} {\path{arXiv:1504.04386}}, \href {https://doi.org/10.1103/PhysRevLett.115.171801} {\path{doi:10.1103/PhysRevLett.115.171801}}.

\bibitem{Segovia:2016zyc}
J.~Segovia, C.~D. Roberts, {Dissecting nucleon transition electromagnetic form factors}, Phys. Rev. C 94~(4) (2016) 042201.
\newblock \href {http://arxiv.org/abs/1607.04405} {\path{arXiv:1607.04405}}, \href {https://doi.org/10.1103/PhysRevC.94.042201} {\path{doi:10.1103/PhysRevC.94.042201}}.

\bibitem{Chen:2018nsg}
C.~Chen, et~al., {Nucleon-to-Roper electromagnetic transition form factors at large $Q^2$}, Phys. Rev. D 99~(3) (2019) 034013.
\newblock \href {http://arxiv.org/abs/1811.08440} {\path{arXiv:1811.08440}}, \href {https://doi.org/10.1103/PhysRevD.99.034013} {\path{doi:10.1103/PhysRevD.99.034013}}.

\bibitem{Lu:2019bjs}
Y.~Lu, et~al., {Transition form factors: $\gamma^\ast + p \to \Delta(1232)$, $\Delta(1600)$}, Phys. Rev. D 100~(3) (2019) 034001.
\newblock \href {http://arxiv.org/abs/1904.03205} {\path{arXiv:1904.03205}}, \href {https://doi.org/10.1103/PhysRevD.100.034001} {\path{doi:10.1103/PhysRevD.100.034001}}.

\bibitem{Chen:2023zhh}
C.~Chen, C.~S. Fischer, C.~D. Roberts, {Nucleon-to-\ensuremath{\Delta} Axial and Pseudoscalar Transition Form Factors}, Phys. Rev. Lett. 133~(13) (2024) 131901.
\newblock \href {http://arxiv.org/abs/2312.13724} {\path{arXiv:2312.13724}}, \href {https://doi.org/10.1103/PhysRevLett.133.131901} {\path{doi:10.1103/PhysRevLett.133.131901}}.

\bibitem{Sanchis-Alepuz:2017mir}
H.~Sanchis-Alepuz, R.~Alkofer, C.~S. Fischer, {Electromagnetic transition form factors of baryons in the space-like momentum region}, Eur. Phys. J. A 54~(3) (2018) 41.
\newblock \href {http://arxiv.org/abs/1707.08463} {\path{arXiv:1707.08463}}, \href {https://doi.org/10.1140/epja/i2018-12465-x} {\path{doi:10.1140/epja/i2018-12465-x}}.

\bibitem{Eichmann:2011aa}
G.~Eichmann, D.~Nicmorus, {Nucleon to $\Delta$ electromagnetic transition in the Dyson-Schwinger approach}, Phys. Rev. D 85 (2012) 093004.
\newblock \href {http://arxiv.org/abs/1112.2232} {\path{arXiv:1112.2232}}, \href {https://doi.org/10.1103/PhysRevD.85.093004} {\path{doi:10.1103/PhysRevD.85.093004}}.

\bibitem{Maris:1999bh}
P.~Maris, P.~C. Tandy, {The Quark photon vertex and the pion charge radius}, Phys. Rev. C 61 (2000) 045202.
\newblock \href {http://arxiv.org/abs/nucl-th/9910033} {\path{arXiv:nucl-th/9910033}}, \href {https://doi.org/10.1103/PhysRevC.61.045202} {\path{doi:10.1103/PhysRevC.61.045202}}.

\bibitem{Miramontes:2019mco}
A.~S. Miramontes, H.~Sanchis-Alepuz, {On the effect of resonances in the quark-photon vertex}, Eur. Phys. J. A 55~(10) (2019) 170.
\newblock \href {http://arxiv.org/abs/1906.06227} {\path{arXiv:1906.06227}}, \href {https://doi.org/10.1140/epja/i2019-12847-6} {\path{doi:10.1140/epja/i2019-12847-6}}.

\bibitem{Miramontes:2021xgn}
A.~S. Miramontes, H.~Sanchis~Alepuz, R.~Alkofer, {Elucidating the effect of intermediate resonances in the quark interaction kernel on the timelike electromagnetic pion form factor}, Phys. Rev. D 103~(11) (2021) 116006.
\newblock \href {http://arxiv.org/abs/2102.12541} {\path{arXiv:2102.12541}}, \href {https://doi.org/10.1103/PhysRevD.103.116006} {\path{doi:10.1103/PhysRevD.103.116006}}.

\bibitem{Fischer:2017cte}
C.~S. Fischer, G.~Eichmann, {Overview of multiquark states}, PoS Hadron2017 (2018) 007.
\newblock \href {https://doi.org/10.22323/1.310.0007} {\path{doi:10.22323/1.310.0007}}.

\bibitem{Eichmann:2018adq}
G.~Eichmann, C.~S. Fischer, {Baryon Structure and Reactions}, Few Body Syst. 60~(1) (2019) 2.
\newblock \href {https://doi.org/10.1007/s00601-018-1469-5} {\path{doi:10.1007/s00601-018-1469-5}}.

\bibitem{Torcato:2023ijg}
A.~Torcato, A.~Arriaga, G.~Eichmann, M.~T. Pe\~na, {Heavy Baryon Spectroscopy in a Quark\textendash{}Diquark Approach}, Few Body Syst. 64~(3) (2023) 45.
\newblock \href {http://arxiv.org/abs/2304.07393} {\path{arXiv:2304.07393}}, \href {https://doi.org/10.1007/s00601-023-01826-9} {\path{doi:10.1007/s00601-023-01826-9}}.

\bibitem{Yin:2019bxe}
P.-L. Yin, et~al., {Masses of ground-state mesons and baryons, including those with heavy quarks}, Phys. Rev. D 100~(3) (2019) 034008.
\newblock \href {http://arxiv.org/abs/1903.00160} {\path{arXiv:1903.00160}}, \href {https://doi.org/10.1103/PhysRevD.100.034008} {\path{doi:10.1103/PhysRevD.100.034008}}.

\bibitem{PACS-CS:2008bkb}
S.~Aoki, et~al., {2+1 Flavor Lattice QCD toward the Physical Point}, Phys. Rev. D 79 (2009) 034503.
\newblock \href {http://arxiv.org/abs/0807.1661} {\path{arXiv:0807.1661}}, \href {https://doi.org/10.1103/PhysRevD.79.034503} {\path{doi:10.1103/PhysRevD.79.034503}}.

\bibitem{BMW:2008jgk}
S.~Durr, et~al., {Ab-Initio Determination of Light Hadron Masses}, Science 322 (2008) 1224--1227.
\newblock \href {http://arxiv.org/abs/0906.3599} {\path{arXiv:0906.3599}}, \href {https://doi.org/10.1126/science.1163233} {\path{doi:10.1126/science.1163233}}.

\bibitem{Bietenholz:2011qq}
W.~Bietenholz, et~al., {Flavour blindness and patterns of flavour symmetry breaking in lattice simulations of up, down and strange quarks}, Phys. Rev. D 84 (2011) 054509.
\newblock \href {http://arxiv.org/abs/1102.5300} {\path{arXiv:1102.5300}}, \href {https://doi.org/10.1103/PhysRevD.84.054509} {\path{doi:10.1103/PhysRevD.84.054509}}.

\bibitem{Alexandrou:2014sha}
C.~Alexandrou, et~al., {Baryon spectrum with $N_f=2+1+1$ twisted mass fermions}, Phys. Rev. D 90~(7) (2014) 074501.
\newblock \href {http://arxiv.org/abs/1406.4310} {\path{arXiv:1406.4310}}, \href {https://doi.org/10.1103/PhysRevD.90.074501} {\path{doi:10.1103/PhysRevD.90.074501}}.

\bibitem{Alexandrou:2017xwd}
C.~Alexandrou, C.~Kallidonis, {Low-lying baryon masses using $N_f=2$ twisted mass clover-improved fermions directly at the physical pion mass}, Phys. Rev. D 96~(3) (2017) 034511.
\newblock \href {http://arxiv.org/abs/1704.02647} {\path{arXiv:1704.02647}}, \href {https://doi.org/10.1103/PhysRevD.96.034511} {\path{doi:10.1103/PhysRevD.96.034511}}.

\bibitem{Edwards:2011jj}
R.~G. Edwards, J.~J. Dudek, D.~G. Richards, S.~J. Wallace, {Excited state baryon spectroscopy from lattice QCD}, Phys. Rev. D 84 (2011) 074508.
\newblock \href {http://arxiv.org/abs/1104.5152} {\path{arXiv:1104.5152}}, \href {https://doi.org/10.1103/PhysRevD.84.074508} {\path{doi:10.1103/PhysRevD.84.074508}}.

\bibitem{Gattringer:2010zz}
C.~Gattringer, C.~B. Lang, {Quantum chromodynamics on the lattice}, Vol. 788, Springer, Berlin, 2010.
\newblock \href {https://doi.org/10.1007/978-3-642-01850-3} {\path{doi:10.1007/978-3-642-01850-3}}.

\bibitem{Briceno:2017max}
R.~A. Briceno, J.~J. Dudek, R.~D. Young, {Scattering processes and resonances from lattice QCD}, Rev. Mod. Phys. 90~(2) (2018) 025001.
\newblock \href {http://arxiv.org/abs/1706.06223} {\path{arXiv:1706.06223}}, \href {https://doi.org/10.1103/RevModPhys.90.025001} {\path{doi:10.1103/RevModPhys.90.025001}}.

\bibitem{HadronSpectrum:2009krc}
M.~Peardon, et~al., {A Novel quark-field creation operator construction for hadronic physics in lattice QCD}, Phys. Rev. D 80 (2009) 054506.
\newblock \href {http://arxiv.org/abs/0905.2160} {\path{arXiv:0905.2160}}, \href {https://doi.org/10.1103/PhysRevD.80.054506} {\path{doi:10.1103/PhysRevD.80.054506}}.

\bibitem{Morningstar:2011ka}
C.~Morningstar, et~al., {Improved stochastic estimation of quark propagation with Laplacian Heaviside smearing in lattice QCD}, Phys. Rev. D 83 (2011) 114505.
\newblock \href {http://arxiv.org/abs/1104.3870} {\path{arXiv:1104.3870}}, \href {https://doi.org/10.1103/PhysRevD.83.114505} {\path{doi:10.1103/PhysRevD.83.114505}}.

\bibitem{Bali:2009hu}
G.~S. Bali, S.~Collins, A.~Schafer, {Effective noise reduction techniques for disconnected loops in Lattice QCD}, Comput. Phys. Commun. 181 (2010) 1570--1583.
\newblock \href {http://arxiv.org/abs/0910.3970} {\path{arXiv:0910.3970}}, \href {https://doi.org/10.1016/j.cpc.2010.05.008} {\path{doi:10.1016/j.cpc.2010.05.008}}.

\bibitem{Dudek:2010wm}
J.~J. Dudek, et~al., {Toward the excited meson spectrum of dynamical QCD}, Phys. Rev. D 82 (2010) 034508.
\newblock \href {http://arxiv.org/abs/1004.4930} {\path{arXiv:1004.4930}}, \href {https://doi.org/10.1103/PhysRevD.82.034508} {\path{doi:10.1103/PhysRevD.82.034508}}.

\bibitem{Thomas:2011rh}
C.~E. Thomas, R.~G. Edwards, J.~J. Dudek, {Helicity operators for mesons in flight on the lattice}, Phys. Rev. D 85 (2012) 014507.
\newblock \href {http://arxiv.org/abs/1107.1930} {\path{arXiv:1107.1930}}, \href {https://doi.org/10.1103/PhysRevD.85.014507} {\path{doi:10.1103/PhysRevD.85.014507}}.

\bibitem{Dudek:2012gj}
J.~J. Dudek, R.~G. Edwards, C.~E. Thomas, {S and D-wave phase shifts in isospin-2 $\pi \pi$ scattering from lattice QCD}, Phys. Rev. D 86 (2012) 034031.
\newblock \href {http://arxiv.org/abs/1203.6041} {\path{arXiv:1203.6041}}, \href {https://doi.org/10.1103/PhysRevD.86.034031} {\path{doi:10.1103/PhysRevD.86.034031}}.

\bibitem{Edwards:2012fx}
R.~G. Edwards, N.~Mathur, D.~G. Richards, S.~J. Wallace, {Flavor structure of the excited baryon spectra from lattice QCD}, Phys. Rev. D 87~(5) (2013) 054506.
\newblock \href {http://arxiv.org/abs/1212.5236} {\path{arXiv:1212.5236}}, \href {https://doi.org/10.1103/PhysRevD.87.054506} {\path{doi:10.1103/PhysRevD.87.054506}}.

\bibitem{Dudek:2012ag}
J.~J. Dudek, R.~G. Edwards, {Hybrid Baryons in QCD}, Phys. Rev. D 85 (2012) 054016.

\bibitem{Lin:2011ti}
H.-W. Lin, {Review of Baryon Spectroscopy in Lattice QCD}, Chin. J. Phys. 49 (2011) 827.
\newblock \href {http://arxiv.org/abs/1106.1608} {\path{arXiv:1106.1608}}.

\bibitem{Fodor:2012gf}
Z.~Fodor, C.~Hoelbling, {Light Hadron Masses from Lattice QCD}, Rev. Mod. Phys. 84 (2012) 449.
\newblock \href {http://arxiv.org/abs/1203.4789} {\path{arXiv:1203.4789}}, \href {https://doi.org/10.1103/RevModPhys.84.449} {\path{doi:10.1103/RevModPhys.84.449}}.

\bibitem{Engel:2013ig}
G.~P. Engel, C.~B. Lang, D.~Mohler, A.~Sch\"afer, {QCD with Two Light Dynamical Chirally Improved Quarks: Baryons}, Phys. Rev. D 87~(7) (2013) 074504.
\newblock \href {http://arxiv.org/abs/1301.4318} {\path{arXiv:1301.4318}}, \href {https://doi.org/10.1103/PhysRevD.87.074504} {\path{doi:10.1103/PhysRevD.87.074504}}.

\bibitem{Blum:2010ym}
T.~Blum, et~al., {Electromagnetic mass splittings of the low lying hadrons and quark masses from 2+1 flavor lattice QCD+QED}, Phys. Rev. D 82 (2010) 094508.
\newblock \href {http://arxiv.org/abs/1006.1311} {\path{arXiv:1006.1311}}, \href {https://doi.org/10.1103/PhysRevD.82.094508} {\path{doi:10.1103/PhysRevD.82.094508}}.

\bibitem{BMW:2014pzb}
S.~Borsanyi, et~al., {Ab initio calculation of the neutron-proton mass difference}, Science 347 (2015) 1452--1455.
\newblock \href {http://arxiv.org/abs/1406.4088} {\path{arXiv:1406.4088}}, \href {https://doi.org/10.1126/science.1257050} {\path{doi:10.1126/science.1257050}}.

\bibitem{Horsley:2015eaa}
R.~Horsley, et~al., {Isospin splittings of meson and baryon masses from three-flavor lattice QCD + QED}, J. Phys. G 43~(10) (2016) 10LT02.
\newblock \href {http://arxiv.org/abs/1508.06401} {\path{arXiv:1508.06401}}, \href {https://doi.org/10.1088/0954-3899/43/10/10LT02} {\path{doi:10.1088/0954-3899/43/10/10LT02}}.

\bibitem{NPLQCD:2020ozd}
S.~R. Beane, et~al., {Charged multihadron systems in lattice QCD+QED}, Phys. Rev. D 103~(5) (2021) 054504.
\newblock \href {http://arxiv.org/abs/2003.12130} {\path{arXiv:2003.12130}}, \href {https://doi.org/10.1103/PhysRevD.103.054504} {\path{doi:10.1103/PhysRevD.103.054504}}.

\bibitem{FlavourLatticeAveragingGroupFLAG:2021npn}
Y.~Aoki, et~al., {FLAG Review 2021}, Eur. Phys. J. C 82~(10) (2022) 869.
\newblock \href {http://arxiv.org/abs/2111.09849} {\path{arXiv:2111.09849}}, \href {https://doi.org/10.1140/epjc/s10052-022-10536-1} {\path{doi:10.1140/epjc/s10052-022-10536-1}}.

\bibitem{FlavourLatticeAveragingGroupFLAG:2024oxs}
Y.~Aoki, et~al., {FLAG Review 2024} (11 2024).
\newblock \href {http://arxiv.org/abs/2411.04268} {\path{arXiv:2411.04268}}.

\bibitem{Maiani:1990ca}
L.~Maiani, M.~Testa, {Final state interactions from Euclidean correlation functions}, Phys. Lett. B 245 (1990) 585--590.
\newblock \href {https://doi.org/10.1016/0370-2693(90)90695-3} {\path{doi:10.1016/0370-2693(90)90695-3}}.

\bibitem{Rummukainen:1995vs}
K.~Rummukainen, S.~A. Gottlieb, {Resonance scattering phase shifts on a nonrest frame lattice}, Nucl. Phys. B 450 (1995) 397--436.
\newblock \href {http://arxiv.org/abs/hep-lat/9503028} {\path{arXiv:hep-lat/9503028}}, \href {https://doi.org/10.1016/0550-3213(95)00313-H} {\path{doi:10.1016/0550-3213(95)00313-H}}.

\bibitem{Kim:2005gf}
C.~h. Kim, C.~T. Sachrajda, S.~R. Sharpe, {Finite-volume effects for two-hadron states in moving frames}, Nucl. Phys. B 727 (2005) 218--243.
\newblock \href {http://arxiv.org/abs/hep-lat/0507006} {\path{arXiv:hep-lat/0507006}}, \href {https://doi.org/10.1016/j.nuclphysb.2005.08.029} {\path{doi:10.1016/j.nuclphysb.2005.08.029}}.

\bibitem{He:2005ey}
S.~He, X.~Feng, C.~Liu, {Two particle states and the S-matrix elements in multi-channel scattering}, JHEP 07 (2005) 011.
\newblock \href {http://arxiv.org/abs/hep-lat/0504019} {\path{arXiv:hep-lat/0504019}}, \href {https://doi.org/10.1088/1126-6708/2005/07/011} {\path{doi:10.1088/1126-6708/2005/07/011}}.

\bibitem{Bernard:2008ax}
V.~Bernard, M.~Lage, U.-G. Meissner, A.~Rusetsky, {Resonance properties from the finite-volume energy spectrum}, JHEP 08 (2008) 024.
\newblock \href {http://arxiv.org/abs/0806.4495} {\path{arXiv:0806.4495}}, \href {https://doi.org/10.1088/1126-6708/2008/08/024} {\path{doi:10.1088/1126-6708/2008/08/024}}.

\bibitem{Leskovec:2012gb}
L.~Leskovec, S.~Prelovsek, {Scattering phase shifts for two particles of different mass and non-zero total momentum in lattice QCD}, Phys. Rev. D 85 (2012) 114507.
\newblock \href {http://arxiv.org/abs/1202.2145} {\path{arXiv:1202.2145}}, \href {https://doi.org/10.1103/PhysRevD.85.114507} {\path{doi:10.1103/PhysRevD.85.114507}}.

\bibitem{Fu:2011xz}
Z.~Fu, {Rummukainen-Gottlieb's formula on two-particle system with different mass}, Phys. Rev. D 85 (2012) 014506.
\newblock \href {http://arxiv.org/abs/1110.0319} {\path{arXiv:1110.0319}}, \href {https://doi.org/10.1103/PhysRevD.85.014506} {\path{doi:10.1103/PhysRevD.85.014506}}.

\bibitem{Hansen:2012tf}
M.~T. Hansen, S.~R. Sharpe, {Multiple-channel generalization of Lellouch-Luscher formula}, Phys. Rev. D 86 (2012) 016007.
\newblock \href {http://arxiv.org/abs/1204.0826} {\path{arXiv:1204.0826}}, \href {https://doi.org/10.1103/PhysRevD.86.016007} {\path{doi:10.1103/PhysRevD.86.016007}}.

\bibitem{Briceno:2012yi}
R.~A. Briceno, Z.~Davoudi, {Moving multichannel systems in a finite volume with application to proton-proton fusion}, Phys. Rev. D 88~(9) (2013) 094507.
\newblock \href {http://arxiv.org/abs/1204.1110} {\path{arXiv:1204.1110}}, \href {https://doi.org/10.1103/PhysRevD.88.094507} {\path{doi:10.1103/PhysRevD.88.094507}}.

\bibitem{Guo:2012hv}
P.~Guo, J.~Dudek, R.~Edwards, A.~P. Szczepaniak, {Coupled-channel scattering on a torus}, Phys. Rev. D 88~(1) (2013) 014501.
\newblock \href {http://arxiv.org/abs/1211.0929} {\path{arXiv:1211.0929}}, \href {https://doi.org/10.1103/PhysRevD.88.014501} {\path{doi:10.1103/PhysRevD.88.014501}}.

\bibitem{Briceno:2014oea}
R.~A. Briceno, {Two-particle multichannel systems in a finite volume with arbitrary spin}, Phys. Rev. D 89~(7) (2014) 074507.
\newblock \href {http://arxiv.org/abs/1401.3312} {\path{arXiv:1401.3312}}, \href {https://doi.org/10.1103/PhysRevD.89.074507} {\path{doi:10.1103/PhysRevD.89.074507}}.

\bibitem{Agadjanov:2016mao}
D.~Agadjanov, et~al., {The Optical Potential on the Lattice}, JHEP 06 (2016) 043.
\newblock \href {http://arxiv.org/abs/1603.07205} {\path{arXiv:1603.07205}}, \href {https://doi.org/10.1007/JHEP06(2016)043} {\path{doi:10.1007/JHEP06(2016)043}}.

\bibitem{Hansen:2019nir}
M.~T. Hansen, S.~R. Sharpe, {Lattice QCD and Three-particle Decays of Resonances}, Ann. Rev. Nucl. Part. Sci. 69 (2019) 65--107.
\newblock \href {http://arxiv.org/abs/1901.00483} {\path{arXiv:1901.00483}}, \href {https://doi.org/10.1146/annurev-nucl-101918-023723} {\path{doi:10.1146/annurev-nucl-101918-023723}}.

\bibitem{Mai:2021lwb}
M.~Mai, M.~D\"oring, A.~Rusetsky, {Multi-particle systems on the lattice and chiral extrapolations: a brief review}, Eur. Phys. J. ST 230~(6) (2021) 1623--1643.
\newblock \href {http://arxiv.org/abs/2103.00577} {\path{arXiv:2103.00577}}, \href {https://doi.org/10.1140/epjs/s11734-021-00146-5} {\path{doi:10.1140/epjs/s11734-021-00146-5}}.

\bibitem{Padmanath:2018zqw}
M.~Padmanath, {Hadron Spectroscopy and Resonances: Review}, PoS LATTICE2018 (2018) 013.
\newblock \href {http://arxiv.org/abs/1905.09651} {\path{arXiv:1905.09651}}, \href {https://doi.org/10.22323/1.334.0013} {\path{doi:10.22323/1.334.0013}}.

\bibitem{Detmold:2019ghl}
W.~Detmold, et~al., {Hadrons and Nuclei}, Eur. Phys. J. A 55~(11) (2019) 193.
\newblock \href {http://arxiv.org/abs/1904.09512} {\path{arXiv:1904.09512}}, \href {https://doi.org/10.1140/epja/i2019-12902-4} {\path{doi:10.1140/epja/i2019-12902-4}}.

\bibitem{Bulava:2022ovd}
J.~Bulava, et~al., {Hadron Spectroscopy with Lattice QCD}, in: {Snowmass 2021}, 2022.
\newblock \href {http://arxiv.org/abs/2203.03230} {\path{arXiv:2203.03230}}.

\bibitem{Briceno:2016mjc}
R.~A. Briceno, J.~J. Dudek, R.~G. Edwards, D.~J. Wilson, {Isoscalar $\pi\pi$ scattering and the $\sigma$ meson resonance from QCD}, Phys. Rev. Lett. 118~(2) (2017) 022002.
\newblock \href {http://arxiv.org/abs/1607.05900} {\path{arXiv:1607.05900}}, \href {https://doi.org/10.1103/PhysRevLett.118.022002} {\path{doi:10.1103/PhysRevLett.118.022002}}.

\bibitem{Guo:2018zss}
D.~Guo, et~al., {Extraction of isoscalar $\pi\pi$ phase-shifts from lattice QCD}, Phys. Rev. D 98~(1) (2018) 014507.
\newblock \href {http://arxiv.org/abs/1803.02897} {\path{arXiv:1803.02897}}, \href {https://doi.org/10.1103/PhysRevD.98.014507} {\path{doi:10.1103/PhysRevD.98.014507}}.

\bibitem{Rodas:2023nec}
A.~Rodas, J.~J. Dudek, R.~G. Edwards, {Determination of crossing-symmetric \ensuremath{\pi}\ensuremath{\pi} scattering amplitudes and the quark mass evolution of the \ensuremath{\sigma} constrained by lattice QCD}, Phys. Rev. D 109~(3) (2024) 034513.
\newblock \href {http://arxiv.org/abs/2304.03762} {\path{arXiv:2304.03762}}, \href {https://doi.org/10.1103/PhysRevD.109.034513} {\path{doi:10.1103/PhysRevD.109.034513}}.

\bibitem{Mohler:2012na}
D.~Mohler, S.~Prelovsek, R.~M. Woloshyn, {$D \pi$ scattering and $D$ meson resonances from lattice QCD}, Phys. Rev. D 87~(3) (2013) 034501.
\newblock \href {http://arxiv.org/abs/1208.4059} {\path{arXiv:1208.4059}}, \href {https://doi.org/10.1103/PhysRevD.87.034501} {\path{doi:10.1103/PhysRevD.87.034501}}.

\bibitem{Prelovsek:2013cra}
S.~Prelovsek, L.~Leskovec, {Evidence for X(3872) from DD* scattering on the lattice}, Phys. Rev. Lett. 111 (2013) 192001.
\newblock \href {http://arxiv.org/abs/1307.5172} {\path{arXiv:1307.5172}}, \href {https://doi.org/10.1103/PhysRevLett.111.192001} {\path{doi:10.1103/PhysRevLett.111.192001}}.

\bibitem{Mohler:2013rwa}
D.~Mohler, et~al., {$D_{s0}^*(2317)$ Meson and $D$-Meson-Kaon Scattering from Lattice QCD}, Phys. Rev. Lett. 111~(22) (2013) 222001.
\newblock \href {http://arxiv.org/abs/1308.3175} {\path{arXiv:1308.3175}}, \href {https://doi.org/10.1103/PhysRevLett.111.222001} {\path{doi:10.1103/PhysRevLett.111.222001}}.

\bibitem{Bicudo:2015vta}
P.~Bicudo, et~al., {Evidence for the existence of $u d \bar{b} \bar{b}$ and the non-existence of $s s \bar{b} \bar{b}$ and $c c \bar{b} \bar{b}$ tetraquarks from lattice QCD}, Phys. Rev. D 92~(1) (2015) 014507.
\newblock \href {http://arxiv.org/abs/1505.00613} {\path{arXiv:1505.00613}}, \href {https://doi.org/10.1103/PhysRevD.92.014507} {\path{doi:10.1103/PhysRevD.92.014507}}.

\bibitem{Francis:2016hui}
A.~Francis, R.~J. Hudspith, R.~Lewis, K.~Maltman, {Lattice Prediction for Deeply Bound Doubly Heavy Tetraquarks}, Phys. Rev. Lett. 118~(14) (2017) 142001.
\newblock \href {http://arxiv.org/abs/1607.05214} {\path{arXiv:1607.05214}}, \href {https://doi.org/10.1103/PhysRevLett.118.142001} {\path{doi:10.1103/PhysRevLett.118.142001}}.

\bibitem{Moir:2016srx}
G.~Moir, et~al., {Coupled-Channel $D\pi$, $D\eta$ and $D_{s}\bar{K}$ Scattering from Lattice QCD}, JHEP 10 (2016) 011.
\newblock \href {http://arxiv.org/abs/1607.07093} {\path{arXiv:1607.07093}}, \href {https://doi.org/10.1007/JHEP10(2016)011} {\path{doi:10.1007/JHEP10(2016)011}}.

\bibitem{Bali:2017pdv}
G.~S. Bali, S.~Collins, A.~Cox, A.~Sch\"afer, {Masses and decay constants of the $D_{s0}^*(2317)$ and $D_{s1}(2460)$ from $N_f=2$ lattice QCD close to the physical point}, Phys. Rev. D 96~(7) (2017) 074501.
\newblock \href {http://arxiv.org/abs/1706.01247} {\path{arXiv:1706.01247}}, \href {https://doi.org/10.1103/PhysRevD.96.074501} {\path{doi:10.1103/PhysRevD.96.074501}}.

\bibitem{Cheung:2017tnt}
G.~K.~C. Cheung, C.~E. Thomas, J.~J. Dudek, R.~G. Edwards, {Tetraquark operators in lattice QCD and exotic flavour states in the charm sector}, JHEP 11 (2017) 033.
\newblock \href {http://arxiv.org/abs/1709.01417} {\path{arXiv:1709.01417}}, \href {https://doi.org/10.1007/JHEP11(2017)033} {\path{doi:10.1007/JHEP11(2017)033}}.

\bibitem{Junnarkar:2018twb}
P.~Junnarkar, N.~Mathur, M.~Padmanath, {Study of doubly heavy tetraquarks in Lattice QCD}, Phys. Rev. D 99~(3) (2019) 034507.
\newblock \href {http://arxiv.org/abs/1810.12285} {\path{arXiv:1810.12285}}, \href {https://doi.org/10.1103/PhysRevD.99.034507} {\path{doi:10.1103/PhysRevD.99.034507}}.

\bibitem{Padmanath:2022cvl}
M.~Padmanath, S.~Prelovsek, {Signature of a Doubly Charm Tetraquark Pole in DD* Scattering on the Lattice}, Phys. Rev. Lett. 129~(3) (2022) 032002.
\newblock \href {http://arxiv.org/abs/2202.10110} {\path{arXiv:2202.10110}}, \href {https://doi.org/10.1103/PhysRevLett.129.032002} {\path{doi:10.1103/PhysRevLett.129.032002}}.

\bibitem{Prelovsek:2014swa}
S.~Prelovsek, C.~B. Lang, L.~Leskovec, D.~Mohler, {Study of the $Z_c^+$ channel using lattice QCD}, Phys. Rev. D 91~(1) (2015) 014504.
\newblock \href {http://arxiv.org/abs/1405.7623} {\path{arXiv:1405.7623}}, \href {https://doi.org/10.1103/PhysRevD.91.014504} {\path{doi:10.1103/PhysRevD.91.014504}}.

\bibitem{Dudek:2014qha}
J.~J. Dudek, R.~G. Edwards, C.~E. Thomas, D.~J. Wilson, {Resonances in coupled $\pi K -\eta K$ scattering from quantum chromodynamics}, Phys. Rev. Lett. 113~(18) (2014) 182001.
\newblock \href {http://arxiv.org/abs/1406.4158} {\path{arXiv:1406.4158}}, \href {https://doi.org/10.1103/PhysRevLett.113.182001} {\path{doi:10.1103/PhysRevLett.113.182001}}.

\bibitem{Wilson:2015dqa}
D.~J. Wilson, et~al., {Coupled $\pi\pi, K\bar{K}$ scattering in $P$-wave and the $\rho$ resonance from lattice QCD}, Phys. Rev. D 92~(9) (2015) 094502.
\newblock \href {http://arxiv.org/abs/1507.02599} {\path{arXiv:1507.02599}}, \href {https://doi.org/10.1103/PhysRevD.92.094502} {\path{doi:10.1103/PhysRevD.92.094502}}.

\bibitem{Dudek:2016cru}
J.~J. Dudek, R.~G. Edwards, D.~J. Wilson, {An $a_0$ resonance in strongly coupled $\pi \eta$, $K\overline{K}$ scattering from lattice QCD}, Phys. Rev. D 93~(9) (2016) 094506.
\newblock \href {http://arxiv.org/abs/1602.05122} {\path{arXiv:1602.05122}}, \href {https://doi.org/10.1103/PhysRevD.93.094506} {\path{doi:10.1103/PhysRevD.93.094506}}.

\bibitem{Briceno:2017qmb}
R.~A. Briceno, J.~J. Dudek, R.~G. Edwards, D.~J. Wilson, {Isoscalar $\pi\pi, K\overline{K}, \eta\eta$ scattering and the $\sigma, f_0, f_2$ mesons from QCD}, Phys. Rev. D 97~(5) (2018) 054513.
\newblock \href {http://arxiv.org/abs/1708.06667} {\path{arXiv:1708.06667}}, \href {https://doi.org/10.1103/PhysRevD.97.054513} {\path{doi:10.1103/PhysRevD.97.054513}}.

\bibitem{JPAC:2018zyd}
A.~Rodas, et~al., {Determination of the pole position of the lightest hybrid meson candidate}, Phys. Rev. Lett. 122~(4) (2019) 042002.
\newblock \href {http://arxiv.org/abs/1810.04171} {\path{arXiv:1810.04171}}, \href {https://doi.org/10.1103/PhysRevLett.122.042002} {\path{doi:10.1103/PhysRevLett.122.042002}}.

\bibitem{Woss:2020ayi}
A.~J. Woss, et~al., {Decays of an exotic $1^{-+}$ hybrid meson resonance in QCD}, Phys. Rev. D 103~(5) (2021) 054502.
\newblock \href {http://arxiv.org/abs/2009.10034} {\path{arXiv:2009.10034}}, \href {https://doi.org/10.1103/PhysRevD.103.054502} {\path{doi:10.1103/PhysRevD.103.054502}}.

\bibitem{Prelovsek:2020eiw}
S.~Prelovsek, et~al., {Charmonium-like resonances with J$^{PC}$ = 0$^{++}$, 2$^{++}$ in coupled $ \mathrm{D}\overline{\mathrm{D}} $, $ {\mathrm{D}}_{\mathrm{s}}{\overline{\mathrm{D}}}_{\mathrm{s}} $ scattering on the lattice}, JHEP 06 (2021) 035.
\newblock \href {http://arxiv.org/abs/2011.02542} {\path{arXiv:2011.02542}}, \href {https://doi.org/10.1007/JHEP06(2021)035} {\path{doi:10.1007/JHEP06(2021)035}}.

\bibitem{Hansen:2014eka}
M.~T. Hansen, S.~R. Sharpe, {Relativistic, model-independent, three-particle quantization condition}, Phys. Rev. D 90~(11) (2014) 116003.
\newblock \href {http://arxiv.org/abs/1408.5933} {\path{arXiv:1408.5933}}, \href {https://doi.org/10.1103/PhysRevD.90.116003} {\path{doi:10.1103/PhysRevD.90.116003}}.

\bibitem{Hansen:2015zga}
M.~T. Hansen, S.~R. Sharpe, {Expressing the three-particle finite-volume spectrum in terms of the three-to-three scattering amplitude}, Phys. Rev. D 92~(11) (2015) 114509.
\newblock \href {http://arxiv.org/abs/1504.04248} {\path{arXiv:1504.04248}}, \href {https://doi.org/10.1103/PhysRevD.92.114509} {\path{doi:10.1103/PhysRevD.92.114509}}.

\bibitem{Hammer:2017uqm}
H.-W. Hammer, J.-Y. Pang, A.~Rusetsky, {Three-particle quantization condition in a finite volume: 1. The role of the three-particle force}, JHEP 09 (2017) 109.
\newblock \href {http://arxiv.org/abs/1706.07700} {\path{arXiv:1706.07700}}, \href {https://doi.org/10.1007/JHEP09(2017)109} {\path{doi:10.1007/JHEP09(2017)109}}.

\bibitem{Hammer:2017kms}
H.~W. Hammer, J.~Y. Pang, A.~Rusetsky, {Three particle quantization condition in a finite volume: 2. general formalism and the analysis of data}, JHEP 10 (2017) 115.
\newblock \href {http://arxiv.org/abs/1707.02176} {\path{arXiv:1707.02176}}, \href {https://doi.org/10.1007/JHEP10(2017)115} {\path{doi:10.1007/JHEP10(2017)115}}.

\bibitem{Mai:2017bge}
M.~Mai, M.~D\"oring, {Three-body Unitarity in the Finite Volume}, Eur. Phys. J. A 53~(12) (2017) 240.
\newblock \href {http://arxiv.org/abs/1709.08222} {\path{arXiv:1709.08222}}, \href {https://doi.org/10.1140/epja/i2017-12440-1} {\path{doi:10.1140/epja/i2017-12440-1}}.

\bibitem{Mai:2018djl}
M.~Mai, M.~Döring, {Finite-Volume Spectrum of $\pi^+\pi^+$ and $\pi^+\pi^+\pi^+$ Systems}, Phys. Rev. Lett. 122~(6) (2019) 062503.
\newblock \href {http://arxiv.org/abs/1807.04746} {\path{arXiv:1807.04746}}, \href {https://doi.org/10.1103/PhysRevLett.122.062503} {\path{doi:10.1103/PhysRevLett.122.062503}}.

\bibitem{Jackura:2019bmu}
A.~W. Jackura, et~al., {Equivalence of three-particle scattering formalisms}, Phys. Rev. D 100~(3) (2019) 034508.
\newblock \href {http://arxiv.org/abs/1905.12007} {\path{arXiv:1905.12007}}, \href {https://doi.org/10.1103/PhysRevD.100.034508} {\path{doi:10.1103/PhysRevD.100.034508}}.

\bibitem{Briceno:2019muc}
R.~A. Brice\~no, M.~T. Hansen, S.~R. Sharpe, A.~P. Szczepaniak, {Unitarity of the infinite-volume three-particle scattering amplitude arising from a finite-volume formalism}, Phys. Rev. D 100~(5) (2019) 054508.
\newblock \href {http://arxiv.org/abs/1905.11188} {\path{arXiv:1905.11188}}, \href {https://doi.org/10.1103/PhysRevD.100.054508} {\path{doi:10.1103/PhysRevD.100.054508}}.

\bibitem{Blanton:2020gha}
T.~D. Blanton, S.~R. Sharpe, {Alternative derivation of the relativistic three-particle quantization condition}, Phys. Rev. D 102~(5) (2020) 054520.
\newblock \href {http://arxiv.org/abs/2007.16188} {\path{arXiv:2007.16188}}, \href {https://doi.org/10.1103/PhysRevD.102.054520} {\path{doi:10.1103/PhysRevD.102.054520}}.

\bibitem{Blanton:2020jnm}
T.~D. Blanton, S.~R. Sharpe, {Equivalence of relativistic three-particle quantization conditions}, Phys. Rev. D 102~(5) (2020) 054515.
\newblock \href {http://arxiv.org/abs/2007.16190} {\path{arXiv:2007.16190}}, \href {https://doi.org/10.1103/PhysRevD.102.054515} {\path{doi:10.1103/PhysRevD.102.054515}}.

\bibitem{Blanton:2019igq}
T.~D. Blanton, F.~Romero-L\'opez, S.~R. Sharpe, {Implementing the three-particle quantization condition including higher partial waves}, JHEP 03 (2019) 106.
\newblock \href {http://arxiv.org/abs/1901.07095} {\path{arXiv:1901.07095}}, \href {https://doi.org/10.1007/JHEP03(2019)106} {\path{doi:10.1007/JHEP03(2019)106}}.

\bibitem{Horz:2019rrn}
B.~H\"orz, A.~Hanlon, {Two- and three-pion finite-volume spectra at maximal isospin from lattice QCD}, Phys. Rev. Lett. 123~(14) (2019) 142002.
\newblock \href {http://arxiv.org/abs/1905.04277} {\path{arXiv:1905.04277}}, \href {https://doi.org/10.1103/PhysRevLett.123.142002} {\path{doi:10.1103/PhysRevLett.123.142002}}.

\bibitem{Blanton:2019vdk}
T.~D. Blanton, F.~Romero-L\'opez, S.~R. Sharpe, {$I=3$ Three-Pion Scattering Amplitude from Lattice QCD}, Phys. Rev. Lett. 124~(3) (2020) 032001.
\newblock \href {http://arxiv.org/abs/1909.02973} {\path{arXiv:1909.02973}}, \href {https://doi.org/10.1103/PhysRevLett.124.032001} {\path{doi:10.1103/PhysRevLett.124.032001}}.

\bibitem{Culver:2019vvu}
C.~Culver, et~al., {Three pion spectrum in the $I=3$ channel from lattice QCD}, Phys. Rev. D 101~(11) (2020) 114507.
\newblock \href {http://arxiv.org/abs/1911.09047} {\path{arXiv:1911.09047}}, \href {https://doi.org/10.1103/PhysRevD.101.114507} {\path{doi:10.1103/PhysRevD.101.114507}}.

\bibitem{Muller:2020wjo}
F.~M\"uller, A.~Rusetsky, {On the three-particle analog of the Lellouch-L\"uscher formula}, JHEP 03 (2021) 152.
\newblock \href {http://arxiv.org/abs/2012.13957} {\path{arXiv:2012.13957}}, \href {https://doi.org/10.1007/JHEP03(2021)152} {\path{doi:10.1007/JHEP03(2021)152}}.

\bibitem{Mai:2021nul}
M.~Mai, et~al., {Three-Body Dynamics of the $a_1(1260)$ Resonance from Lattice QCD}, Phys. Rev. Lett. 127~(22) (2021) 222001.
\newblock \href {http://arxiv.org/abs/2107.03973} {\path{arXiv:2107.03973}}, \href {https://doi.org/10.1103/PhysRevLett.127.222001} {\path{doi:10.1103/PhysRevLett.127.222001}}.

\bibitem{Yan:2024gwp}
H.~Yan, et~al., {\ensuremath{\omega} Meson from Lattice QCD}, Phys. Rev. Lett. 133~(21) (2024) 211906.
\newblock \href {http://arxiv.org/abs/2407.16659} {\path{arXiv:2407.16659}}, \href {https://doi.org/10.1103/PhysRevLett.133.211906} {\path{doi:10.1103/PhysRevLett.133.211906}}.

\bibitem{Andersen:2017una}
C.~W. Andersen, J.~Bulava, B.~H\"orz, C.~Morningstar, {Elastic $I=3/2$ $p$-wave nucleon-pion scattering amplitude and the $\Delta$(1232) resonance from N$_f$=2+1 lattice QCD}, Phys. Rev. D 97~(1) (2018) 014506.
\newblock \href {http://arxiv.org/abs/1710.01557} {\path{arXiv:1710.01557}}, \href {https://doi.org/10.1103/PhysRevD.97.014506} {\path{doi:10.1103/PhysRevD.97.014506}}.

\bibitem{Silvi:2021uya}
G.~Silvi, et~al., {$P$-wave nucleon-pion scattering amplitude in the $\Delta$(1232) channel from lattice QCD}, Phys. Rev. D 103~(9) (2021) 094508.
\newblock \href {http://arxiv.org/abs/2101.00689} {\path{arXiv:2101.00689}}, \href {https://doi.org/10.1103/PhysRevD.103.094508} {\path{doi:10.1103/PhysRevD.103.094508}}.

\bibitem{Bulava:2022vpq}
J.~Bulava, et~al., {Elastic nucleon-pion scattering at $m_{\pi}=200$\,MeV from lattice QCD}, Nucl. Phys. B 987 (2023) 116105.
\newblock \href {http://arxiv.org/abs/2208.03867} {\path{arXiv:2208.03867}}, \href {https://doi.org/10.1016/j.nuclphysb.2023.116105} {\path{doi:10.1016/j.nuclphysb.2023.116105}}.

\bibitem{Lang:2012db}
C.~B. Lang, V.~Verduci, {Scattering in the \ensuremath{\pi}N negative parity channel in lattice QCD}, Phys. Rev. D 87~(5) (2013) 054502.
\newblock \href {http://arxiv.org/abs/1212.5055} {\path{arXiv:1212.5055}}, \href {https://doi.org/10.1103/PhysRevD.87.054502} {\path{doi:10.1103/PhysRevD.87.054502}}.

\bibitem{Liu:2016rwa}
K.-F. Liu, {Baryons and Chiral Symmetry}, Int. J. Mod. Phys. E 26~(01n02) (2017) 1740016.
\newblock \href {http://arxiv.org/abs/1609.02572} {\path{arXiv:1609.02572}}, \href {https://doi.org/10.1142/S021830131740016X} {\path{doi:10.1142/S021830131740016X}}.

\bibitem{Mathur:2003zf}
N.~Mathur, et~al., {Roper resonance and $S_{11}(1535)$ from lattice QCD}, Phys. Lett. B 605 (2005) 137--143.

\bibitem{xQCD:2019jke}
M.~Sun, et~al., {Roper State from Overlap Fermions}, Phys. Rev. D 101~(5) (2020) 054511.
\newblock \href {http://arxiv.org/abs/1911.02635} {\path{arXiv:1911.02635}}, \href {https://doi.org/10.1103/PhysRevD.101.054511} {\path{doi:10.1103/PhysRevD.101.054511}}.

\bibitem{Lang:2016hnn}
C.~B. Lang, L.~Leskovec, M.~Padmanath, S.~Prelovsek, {Pion-nucleon scattering in the Roper channel from lattice QCD}, Phys. Rev. D 95~(1) (2017) 014510.
\newblock \href {http://arxiv.org/abs/1610.01422} {\path{arXiv:1610.01422}}, \href {https://doi.org/10.1103/PhysRevD.95.014510} {\path{doi:10.1103/PhysRevD.95.014510}}.

\bibitem{Liu:2016uzk}
Z.-W. Liu, et~al., {Hamiltonian effective field theory study of the ${N^*(1440)}$ resonance in lattice QCD}, Phys. Rev. D 95~(3) (2017) 034034.
\newblock \href {http://arxiv.org/abs/1607.04536} {\path{arXiv:1607.04536}}, \href {https://doi.org/10.1103/PhysRevD.95.034034} {\path{doi:10.1103/PhysRevD.95.034034}}.

\bibitem{Kiratidis:2016hda}
A.~L. Kiratidis, et~al., {Search for low-lying lattice QCD eigenstates in the Roper regime}, Phys. Rev. D 95~(7) (2017) 074507.
\newblock \href {http://arxiv.org/abs/1608.03051} {\path{arXiv:1608.03051}}, \href {https://doi.org/10.1103/PhysRevD.95.074507} {\path{doi:10.1103/PhysRevD.95.074507}}.

\bibitem{Wu:2017qve}
J.-J. Wu, D.~B. Leinweber, Z.-W. Liu, A.~W. Thomas, {Structure of the Roper Resonance from Lattice QCD Constraints}, Phys. Rev. D 97~(9) (2018) 094509.
\newblock \href {http://arxiv.org/abs/1703.10715} {\path{arXiv:1703.10715}}, \href {https://doi.org/10.1103/PhysRevD.97.094509} {\path{doi:10.1103/PhysRevD.97.094509}}.

\bibitem{Owa:2025mep}
S.~Owa, D.~B. Leinweber, A.~W. Thomas, {Nucleon resonance structure to 2 GeV and the nature of the Roper} (3 2025).
\newblock \href {http://arxiv.org/abs/2503.09945} {\path{arXiv:2503.09945}}.

\bibitem{Lellouch:2000pv}
L.~Lellouch, M.~Luscher, {Weak transition matrix elements from finite volume correlation functions}, Commun. Math. Phys. 219 (2001) 31--44.
\newblock \href {http://arxiv.org/abs/hep-lat/0003023} {\path{arXiv:hep-lat/0003023}}, \href {https://doi.org/10.1007/s002200100410} {\path{doi:10.1007/s002200100410}}.

\bibitem{Agadjanov:2014kha}
A.~Agadjanov, V.~Bernard, U.~G. Mei\ss{}ner, A.~Rusetsky, {A framework for the calculation of the $\Delta N \gamma^\ast$ transition form factors on the lattice}, Nucl. Phys. B 886 (2014) 1199--1222.
\newblock \href {http://arxiv.org/abs/1405.3476} {\path{arXiv:1405.3476}}, \href {https://doi.org/10.1016/j.nuclphysb.2014.07.023} {\path{doi:10.1016/j.nuclphysb.2014.07.023}}.

\bibitem{Briceno:2014uqa}
R.~A. Brice\~no, M.~T. Hansen, A.~Walker-Loud, {Multichannel 1 $\rightarrow$ 2 transition amplitudes in a finite volume}, Phys. Rev. D 91~(3) (2015) 034501.
\newblock \href {http://arxiv.org/abs/1406.5965} {\path{arXiv:1406.5965}}, \href {https://doi.org/10.1103/PhysRevD.91.034501} {\path{doi:10.1103/PhysRevD.91.034501}}.

\bibitem{Briceno:2015csa}
R.~A. Brice\~no, M.~T. Hansen, {Multichannel 0 $\to$ 2 and 1 $\to$ 2 transition amplitudes for arbitrary spin particles in a finite volume}, Phys. Rev. D 92~(7) (2015) 074509.
\newblock \href {http://arxiv.org/abs/1502.04314} {\path{arXiv:1502.04314}}, \href {https://doi.org/10.1103/PhysRevD.92.074509} {\path{doi:10.1103/PhysRevD.92.074509}}.

\bibitem{Agadjanov:2016fbd}
A.~Agadjanov, V.~Bernard, U.-G. Mei\ss{}ner, A.~Rusetsky, {The $B\to K^\ast$ form factors on the lattice}, Nucl. Phys. B 910 (2016) 387--409.
\newblock \href {http://arxiv.org/abs/1605.03386} {\path{arXiv:1605.03386}}, \href {https://doi.org/10.1016/j.nuclphysb.2016.07.005} {\path{doi:10.1016/j.nuclphysb.2016.07.005}}.

\bibitem{Briceno:2021xlc}
R.~A. Brice\~no, J.~J. Dudek, L.~Leskovec, {Constraining $1+\mathcal{J}\to 2$ coupled-channel amplitudes in finite-volume}, Phys. Rev. D 104~(5) (2021) 054509.
\newblock \href {http://arxiv.org/abs/2105.02017} {\path{arXiv:2105.02017}}, \href {https://doi.org/10.1103/PhysRevD.104.054509} {\path{doi:10.1103/PhysRevD.104.054509}}.

\bibitem{Briceno:2015dca}
R.~A. Briceno, et~al., {The resonant $\pi^+\gamma\to\pi^+\pi^0$ amplitude from Quantum Chromodynamics}, Phys. Rev. Lett. 115 (2015) 242001.
\newblock \href {http://arxiv.org/abs/1507.06622} {\path{arXiv:1507.06622}}, \href {https://doi.org/10.1103/PhysRevLett.115.242001} {\path{doi:10.1103/PhysRevLett.115.242001}}.

\bibitem{Briceno:2016kkp}
R.~A. Brice\~no, et~al., {The $\pi\pi\to\pi\gamma^\star$ amplitude and the resonant $\rho\to\pi\gamma^\star$ transition from lattice QCD}, Phys. Rev. D 93~(11) (2016) 114508, [Erratum: Phys.Rev.D 105, 079902 (2022)].
\newblock \href {http://arxiv.org/abs/1604.03530} {\path{arXiv:1604.03530}}, \href {https://doi.org/10.1103/PhysRevD.93.114508} {\path{doi:10.1103/PhysRevD.93.114508}}.

\bibitem{Alexandrou:2018jbt}
C.~Alexandrou, et~al., {$\pi\gamma \to \pi\pi$ transition and the $\rho$ radiative decay width from lattice QCD}, Phys. Rev. D 98~(7) (2018) 074502, [Erratum: Phys.Rev.D 105, 019902 (2022)].
\newblock \href {http://arxiv.org/abs/1807.08357} {\path{arXiv:1807.08357}}, \href {https://doi.org/10.1103/PhysRevD.98.074502} {\path{doi:10.1103/PhysRevD.98.074502}}.

\bibitem{Hall:2014uca}
J.~M.~M. Hall, et~al., {Lattice QCD Evidence that the \ensuremath{\Lambda}(1405) Resonance is an Antikaon-Nucleon Molecule}, Phys. Rev. Lett. 114~(13) (2015) 132002.
\newblock \href {http://arxiv.org/abs/1411.3402} {\path{arXiv:1411.3402}}, \href {https://doi.org/10.1103/PhysRevLett.114.132002} {\path{doi:10.1103/PhysRevLett.114.132002}}.

\bibitem{BaryonScatteringBaSc:2023zvt}
J.~Bulava, et~al., {Two-Pole Nature of the \ensuremath{\Lambda}(1405) resonance from Lattice QCD}, Phys. Rev. Lett. 132~(5) (2024) 051901.
\newblock \href {http://arxiv.org/abs/2307.10413} {\path{arXiv:2307.10413}}, \href {https://doi.org/10.1103/PhysRevLett.132.051901} {\path{doi:10.1103/PhysRevLett.132.051901}}.

\bibitem{BaryonScatteringBaSc:2023ori}
J.~Bulava, et~al., {Lattice QCD study of $\pi\Sigma-K^-N$ scattering and the $\Lambda(1405)$ resonance}, Phys. Rev. D 109~(1) (2024) 014511.
\newblock \href {http://arxiv.org/abs/2307.13471} {\path{arXiv:2307.13471}}, \href {https://doi.org/10.1103/PhysRevD.109.014511} {\path{doi:10.1103/PhysRevD.109.014511}}.

\bibitem{JPAC:2021rxu}
M.~Albaladejo, et~al., {Novel approaches in hadron spectroscopy}, Prog. Part. Nucl. Phys. 127 (2022) 103981.
\newblock \href {http://arxiv.org/abs/2112.13436} {\path{arXiv:2112.13436}}, \href {https://doi.org/10.1016/j.ppnp.2022.103981} {\path{doi:10.1016/j.ppnp.2022.103981}}.

\bibitem{Ishii:2006ec}
N.~Ishii, S.~Aoki, T.~Hatsuda, {The Nuclear Force from Lattice QCD}, Phys. Rev. Lett. 99 (2007) 022001.
\newblock \href {http://arxiv.org/abs/nucl-th/0611096} {\path{arXiv:nucl-th/0611096}}, \href {https://doi.org/10.1103/PhysRevLett.99.022001} {\path{doi:10.1103/PhysRevLett.99.022001}}.

\bibitem{Aoki:2009ji}
S.~Aoki, T.~Hatsuda, N.~Ishii, {Theoretical Foundation of the Nuclear Force in QCD and its applications to Central and Tensor Forces in Quenched Lattice QCD Simulations}, Prog. Theor. Phys. 123 (2010) 89--128.
\newblock \href {http://arxiv.org/abs/0909.5585} {\path{arXiv:0909.5585}}, \href {https://doi.org/10.1143/PTP.123.89} {\path{doi:10.1143/PTP.123.89}}.

\bibitem{Ishii:2012ssm}
N.~Ishii, et~al., {Hadron\textendash{}hadron interactions from imaginary-time Nambu\textendash{}Bethe\textendash{}Salpeter wave function on the lattice}, Phys. Lett. B 712 (2012) 437--441.
\newblock \href {http://arxiv.org/abs/1203.3642} {\path{arXiv:1203.3642}}, \href {https://doi.org/10.1016/j.physletb.2012.04.076} {\path{doi:10.1016/j.physletb.2012.04.076}}.

\bibitem{Aoki:2012tk}
S.~Aoki, et~al., {Lattice QCD approach to Nuclear Physics}, PTEP 2012 (2012) 01A105.
\newblock \href {http://arxiv.org/abs/1206.5088} {\path{arXiv:1206.5088}}, \href {https://doi.org/10.1093/ptep/pts010} {\path{doi:10.1093/ptep/pts010}}.

\bibitem{Hall:2013qba}
J.~M.~M. Hall, et~al., {Finite-volume matrix Hamiltonian model for a $\Delta \to N\pi$ system}, Phys. Rev. D 87~(9) (2013) 094510.
\newblock \href {http://arxiv.org/abs/1303.4157} {\path{arXiv:1303.4157}}, \href {https://doi.org/10.1103/PhysRevD.87.094510} {\path{doi:10.1103/PhysRevD.87.094510}}.

\bibitem{Wu:2014vma}
J.-J. Wu, T.~S.~H. Lee, A.~W. Thomas, R.~D. Young, {Finite-volume Hamiltonian method for coupled-channels interactions in lattice QCD}, Phys. Rev. C 90~(5) (2014) 055206.
\newblock \href {http://arxiv.org/abs/1402.4868} {\path{arXiv:1402.4868}}, \href {https://doi.org/10.1103/PhysRevC.90.055206} {\path{doi:10.1103/PhysRevC.90.055206}}.

\bibitem{Bulava:2019kbi}
J.~Bulava, M.~T. Hansen, {Scattering amplitudes from finite-volume spectral functions}, Phys. Rev. D 100~(3) (2019) 034521.
\newblock \href {http://arxiv.org/abs/1903.11735} {\path{arXiv:1903.11735}}, \href {https://doi.org/10.1103/PhysRevD.100.034521} {\path{doi:10.1103/PhysRevD.100.034521}}.

\bibitem{Bulava:2023mjc}
J.~Bulava, {The spectral reconstruction of inclusive rates}, PoS LATTICE2022 (2023) 231.
\newblock \href {http://arxiv.org/abs/2301.04072} {\path{arXiv:2301.04072}}, \href {https://doi.org/10.22323/1.430.0231} {\path{doi:10.22323/1.430.0231}}.

\bibitem{Patella:2024cto}
A.~Patella, N.~Tantalo, {Scattering amplitudes from Euclidean correlators: Haag-Ruelle theory and approximation formulae}, JHEP 01 (2025) 091.
\newblock \href {http://arxiv.org/abs/2407.02069} {\path{arXiv:2407.02069}}, \href {https://doi.org/10.1007/JHEP01(2025)091} {\path{doi:10.1007/JHEP01(2025)091}}.

\bibitem{Watson:1954uc}
K.~M. Watson, {Some general relations between the photoproduction and scattering of pi mesons}, Phys. Rev. 95 (1954) 228--236.
\newblock \href {https://doi.org/10.1103/PhysRev.95.228} {\path{doi:10.1103/PhysRev.95.228}}.

\bibitem{Burkert:2019kxy}
V.~D. Burkert, {$N^*$ Experiments and what they tell us about Strong QCD Physics}, EPJ Web Conf. 241 (2020) 01004.
\newblock \href {http://arxiv.org/abs/1912.11400} {\path{arXiv:1912.11400}}, \href {https://doi.org/10.1051/epjconf/202024101004} {\path{doi:10.1051/epjconf/202024101004}}.

\bibitem{Brodsky:1973kr}
S.~J. Brodsky, G.~R. Farrar, {Scaling Laws at Large Transverse Momentum}, Phys. Rev. Lett. 31 (1973) 1153--1156.
\newblock \href {https://doi.org/10.1103/PhysRevLett.31.1153} {\path{doi:10.1103/PhysRevLett.31.1153}}.

\bibitem{Brodsky:1974vy}
S.~J. Brodsky, G.~R. Farrar, {Scaling Laws for Large Momentum Transfer Processes}, Phys. Rev. D 11 (1975) 1309.
\newblock \href {https://doi.org/10.1103/PhysRevD.11.1309} {\path{doi:10.1103/PhysRevD.11.1309}}.

\bibitem{Carlson:1985mm}
C.~E. Carlson, {Electromagnetic $N\to\Delta$ transition at high $Q^2$}, Phys. Rev. D 34 (1986) 2704.
\newblock \href {https://doi.org/10.1103/PhysRevD.34.2704} {\path{doi:10.1103/PhysRevD.34.2704}}.

\bibitem{Bhagwat:2003vw}
M.~S. Bhagwat, M.~A. Pichowsky, C.~D. Roberts, P.~C. Tandy, {Analysis of a quenched lattice QCD dressed quark propagator}, Phys. Rev. C 68 (2003) 015203.
\newblock \href {http://arxiv.org/abs/nucl-th/0304003} {\path{arXiv:nucl-th/0304003}}, \href {https://doi.org/10.1103/PhysRevC.68.015203} {\path{doi:10.1103/PhysRevC.68.015203}}.

\bibitem{Anikin:2015ita}
I.~V. Anikin, V.~M. Braun, N.~Offen, {Electroproduction of the $N^*(1535)$ nucleon resonance in QCD}, Phys. Rev. D 92~(1) (2015) 014018.
\newblock \href {http://arxiv.org/abs/1505.05759} {\path{arXiv:1505.05759}}, \href {https://doi.org/10.1103/PhysRevD.92.014018} {\path{doi:10.1103/PhysRevD.92.014018}}.

\bibitem{Ramalho:2011ae}
G.~Ramalho, M.~T. Pena, {A covariant model for the $\gamma N \to N(1535)$ transition at high momentum transfer}, Phys. Rev. D 84 (2011) 033007.
\newblock \href {http://arxiv.org/abs/1105.2223} {\path{arXiv:1105.2223}}, \href {https://doi.org/10.1103/PhysRevD.84.033007} {\path{doi:10.1103/PhysRevD.84.033007}}.

\bibitem{Aznauryan:2007ja}
I.~G. Aznauryan, {Electroexcitation of the Roper resonance in the relativistic quark models}, Phys. Rev. C 76 (2007) 025212.
\newblock \href {http://arxiv.org/abs/nucl-th/0701012} {\path{arXiv:nucl-th/0701012}}, \href {https://doi.org/10.1103/PhysRevC.76.025212} {\path{doi:10.1103/PhysRevC.76.025212}}.

\bibitem{Jido:2007sm}
D.~Jido, M.~Döring, E.~Oset, {Transition form factors of the N*(1535) as a dynamically generated resonance}, Phys. Rev. C 77 (2008) 065207.
\newblock \href {http://arxiv.org/abs/0712.0038} {\path{arXiv:0712.0038}}, \href {https://doi.org/10.1103/PhysRevC.77.065207} {\path{doi:10.1103/PhysRevC.77.065207}}.

\bibitem{Miller:2002ig}
G.~A. Miller, {Light front cloudy bag model: Nucleon electromagnetic form-factors}, Phys. Rev. C 66 (2002) 032201.
\newblock \href {http://arxiv.org/abs/nucl-th/0207007} {\path{arXiv:nucl-th/0207007}}, \href {https://doi.org/10.1103/PhysRevC.66.032201} {\path{doi:10.1103/PhysRevC.66.032201}}.

\bibitem{Aznaurian:1982qc}
I.~G. Aznaurian, A.~S. Bagdasaryan, N.~L. Ter-Isaakian, {Relativistic quark model in the infinite momentum frame and static characteristics of nucleons}, Phys. Lett. B 112 (1982) 393--396.
\newblock \href {https://doi.org/10.1016/0370-2693(82)91076-0} {\path{doi:10.1016/0370-2693(82)91076-0}}.

\bibitem{Ash:1967rlw}
W.~W. Ash, et~al., {Measurement of the $\gamma NN^*$ form factor}, Phys. Lett. B 24 (1967) 165--168.

\bibitem{Beck:1999ge}
R.~Beck, et~al., {Determination of the E2\,/\,M1 ratio in the $\gamma N \to \Delta (1232)$ transition from a simultaneous measurement of $p(\vec\gamma, p)\pi^0$ and $p(\\vecgamma, \pi^+)n$}, Phys. Rev. C 61 (2000) 035204.
\newblock \href {http://arxiv.org/abs/nucl-ex/9908017} {\path{arXiv:nucl-ex/9908017}}, \href {https://doi.org/10.1103/PhysRevC.61.035204} {\path{doi:10.1103/PhysRevC.61.035204}}.

\bibitem{Bauer:2014cqa}
T.~Bauer, S.~Scherer, L.~Tiator, {Electromagnetic transition form factors of the Roper resonance in effective field theory}, Phys. Rev. C 90~(1) (2014) 015201.

\bibitem{Grigoryan:2009pp}
H.~R. Grigoryan, T.~S.~H. Lee, H.-U. Yee, {Electromagnetic Nucleon-to-Delta Transition in Holographic QCD}, Phys. Rev. D 80 (2009) 055006.

\bibitem{Roper:1964zza}
L.~D. Roper, {Evidence for a $P_{11}$ Pion-Nucleon Resonance at 556 MeV}, Phys. Rev. Lett. 12 (1964) 340--342.

\bibitem{Thomas:1981vc}
A.~W. Thomas, S.~Theberge, G.~A. Miller, {The Cloudy Bag Model of the Nucleon}, Phys. Rev. D 24 (1981) 216.

\bibitem{Bermuth:1988ms}
K.~Bermuth, D.~Drechsel, L.~Tiator, J.~B. Seaborn, {Photoproduction of $\Delta$ and Roper Resonances in the Cloudy Bag Model}, Phys. Rev. D 37 (1988) 89--100.

\bibitem{Krehl:1999km}
O.~Krehl, C.~Hanhart, S.~Krewald, J.~Speth, {What is the structure of the Roper resonance?}, Phys. Rev. C 62 (2000) 025207.
\newblock \href {http://arxiv.org/abs/nucl-th/9911080} {\path{arXiv:nucl-th/9911080}}, \href {https://doi.org/10.1103/PhysRevC.62.025207} {\path{doi:10.1103/PhysRevC.62.025207}}.

\bibitem{Sekihara:2021eah}
T.~Sekihara, {Two-body wave functions and compositeness from scattering amplitudes. II. Application to the physical N* and \ensuremath{\Delta}* resonances}, Phys. Rev. C 104~(3) (2021) 035202.
\newblock \href {http://arxiv.org/abs/2104.01962} {\path{arXiv:2104.01962}}, \href {https://doi.org/10.1103/PhysRevC.104.035202} {\path{doi:10.1103/PhysRevC.104.035202}}.

\bibitem{Eichmann:2018ytt}
G.~Eichmann, G.~Ramalho, {Nucleon resonances in Compton scattering}, Phys. Rev. D 98~(9) (2018) 093007.
\newblock \href {http://arxiv.org/abs/1806.04579} {\path{arXiv:1806.04579}}, \href {https://doi.org/10.1103/PhysRevD.98.093007} {\path{doi:10.1103/PhysRevD.98.093007}}.

\bibitem{Ramalho:2017pyc}
G.~Ramalho, D.~Melnikov, {Valence quark contributions for the $\gamma^\ast N \to N(1440)$ form factors from light-front holography}, Phys. Rev. D 97~(3) (2018) 034037.
\newblock \href {http://arxiv.org/abs/1703.03819} {\path{arXiv:1703.03819}}, \href {https://doi.org/10.1103/PhysRevD.97.034037} {\path{doi:10.1103/PhysRevD.97.034037}}.

\bibitem{deTeramond:2011qp}
G.~F. de~Teramond, S.~J. Brodsky, {Excited Baryons in Holographic QCD}, AIP Conf. Proc. 1432~(1) (2012) 168--175.
\newblock \href {http://arxiv.org/abs/1108.0965} {\path{arXiv:1108.0965}}, \href {https://doi.org/10.1063/1.3701207} {\path{doi:10.1063/1.3701207}}.

\bibitem{Obukhovsky:2011sc}
I.~T. Obukhovsky, et~al., {Electroproduction of the Roper resonance on the proton: the role of the three-quark core and the molecular $N\sigma$ component}, Phys. Rev. D 84 (2011) 014004.

\bibitem{Lin:2011da}
H.-W. Lin, S.~D. Cohen, {Roper Properties on the Lattice: An Update}, AIP Conf. Proc. 1432~(1) (2012) 305--308.
\newblock \href {http://arxiv.org/abs/1108.2528} {\path{arXiv:1108.2528}}, \href {https://doi.org/10.1063/1.3701236} {\path{doi:10.1063/1.3701236}}.

\bibitem{Virgili:2019shg}
A.~Virgili, W.~Kamleh, D.~Leinweber, {Role of chiral symmetry in the nucleon excitation spectrum}, Phys. Rev. D 101~(7) (2020) 074504.
\newblock \href {http://arxiv.org/abs/1910.13782} {\path{arXiv:1910.13782}}, \href {https://doi.org/10.1103/PhysRevD.101.074504} {\path{doi:10.1103/PhysRevD.101.074504}}.

\bibitem{Leinweber:2024psf}
D.~B. Leinweber, C.~D. Abell, L.~C. Hockley, W.~Kamleh, Z.-W. Liu, F.~M. Stokes, A.~W. Thomas, J.-J. Wu, {Understanding the nature of baryon resonances}, Nuovo Cim. C 47~(4) (2024) 146.
\newblock \href {http://arxiv.org/abs/2401.04901} {\path{arXiv:2401.04901}}, \href {https://doi.org/10.1393/ncc/i2024-24146-4} {\path{doi:10.1393/ncc/i2024-24146-4}}.

\bibitem{Padmanath:2015era}
M.~Padmanath, C.~B. Lang, S.~Prelovsek, {X(3872) and Y(4140) using diquark-antidiquark operators with lattice QCD}, Phys. Rev. D 92~(3) (2015) 034501.
\newblock \href {http://arxiv.org/abs/1503.03257} {\path{arXiv:1503.03257}}, \href {https://doi.org/10.1103/PhysRevD.92.034501} {\path{doi:10.1103/PhysRevD.92.034501}}.

\bibitem{Karapetyan:2023kzs}
G.~Karapetyan, {Configurational entropy and the N\ensuremath{*}(1440) Roper resonance in QCD}, Annals Phys. 462 (2024) 169612.
\newblock \href {http://arxiv.org/abs/2305.05413} {\path{arXiv:2305.05413}}, \href {https://doi.org/10.1016/j.aop.2024.169612} {\path{doi:10.1016/j.aop.2024.169612}}.

\bibitem{Kaewsnod:2021nfw}
A.~Kaewsnod, et~al., {Study of N(1440) structure via \ensuremath{\gamma}*p\textrightarrow{}N(1440) transition}, Phys. Rev. D 105~(1) (2022) 016008.
\newblock \href {http://arxiv.org/abs/2106.10615} {\path{arXiv:2106.10615}}, \href {https://doi.org/10.1103/PhysRevD.105.016008} {\path{doi:10.1103/PhysRevD.105.016008}}.

\bibitem{Ramalho:2010js}
G.~Ramalho, K.~Tsushima, {Valence quark contributions for the $\gamma N \to P_{11}(1440)$ form factors}, Phys. Rev. D 81 (2010) 074020.
\newblock \href {http://arxiv.org/abs/1002.3386} {\path{arXiv:1002.3386}}, \href {https://doi.org/10.1103/PhysRevD.81.074020} {\path{doi:10.1103/PhysRevD.81.074020}}.

\bibitem{Ramalho:2013mxa}
G.~Ramalho, M.~T. Pe\~na, {$\gamma^\ast N \to N^\ast(1520)$ form factors in the spacelike region}, Phys. Rev. D 89~(9) (2014) 094016.
\newblock \href {http://arxiv.org/abs/1309.0730} {\path{arXiv:1309.0730}}, \href {https://doi.org/10.1103/PhysRevD.89.094016} {\path{doi:10.1103/PhysRevD.89.094016}}.

\bibitem{Kaiser:1995cy}
N.~Kaiser, P.~B. Siegel, W.~Weise, {Chiral dynamics and the $S_{11}(1535)$ nucleon resonance}, Phys. Lett. B 362 (1995) 23--28.
\newblock \href {http://arxiv.org/abs/nucl-th/9507036} {\path{arXiv:nucl-th/9507036}}, \href {https://doi.org/10.1016/0370-2693(95)01203-3} {\path{doi:10.1016/0370-2693(95)01203-3}}.

\bibitem{Sekihara:2015gvw}
T.~Sekihara, T.~Arai, J.~Yamagata-Sekihara, S.~Yasui, {Compositeness of baryonic resonances: Application to the $\Delta$(1232), $N$(1535), and $N$(1650) resonances}, Phys. Rev. C 93~(3) (2016) 035204.
\newblock \href {http://arxiv.org/abs/1511.01200} {\path{arXiv:1511.01200}}, \href {https://doi.org/10.1103/PhysRevC.93.035204} {\path{doi:10.1103/PhysRevC.93.035204}}.

\bibitem{Kinugawa:2024kwb}
T.~Kinugawa, T.~Hyodo, {Compositeness of near-threshold $s$-wave resonances} (3 2024).
\newblock \href {http://arxiv.org/abs/2403.12635} {\path{arXiv:2403.12635}}.

\bibitem{Golli:2011jk}
B.~Golli, S.~Sirca, {A Chiral quark model for meson electro-production in the S11 partial wave}, Eur. Phys. J. A 47 (2011) 61.
\newblock \href {http://arxiv.org/abs/1101.5527} {\path{arXiv:1101.5527}}, \href {https://doi.org/10.1140/epja/i2011-11061-0} {\path{doi:10.1140/epja/i2011-11061-0}}.

\bibitem{Gutsche:2019yoo}
T.~Gutsche, V.~E. Lyubovitskij, I.~Schmidt, {$\gamma N \to N^*(1535)$ transition in soft-wall AdS/QCD}, Phys. Rev. D 101~(3) (2020) 034026.
\newblock \href {http://arxiv.org/abs/1911.00076} {\path{arXiv:1911.00076}}, \href {https://doi.org/10.1103/PhysRevD.101.034026} {\path{doi:10.1103/PhysRevD.101.034026}}.

\bibitem{Julia-Diaz:2007mae}
B.~Julia-Diaz, et~al., {Dynamical coupled-channels effects on pion photoproduction}, Phys. Rev. C 77 (2008) 045205.
\newblock \href {http://arxiv.org/abs/0712.2283} {\path{arXiv:0712.2283}}, \href {https://doi.org/10.1103/PhysRevC.77.045205} {\path{doi:10.1103/PhysRevC.77.045205}}.

\bibitem{ParticleDataGroup:2016lqr}
C.~Patrignani, et~al., {Review of Particle Physics}, Chin. Phys. C 40~(10) (2016) 100001.
\newblock \href {https://doi.org/10.1088/1674-1137/40/10/100001} {\path{doi:10.1088/1674-1137/40/10/100001}}.

\bibitem{Moorhouse:1966jn}
R.~G. Moorhouse, {Photoproduction of $N^*$ resonances in the quark model}, Phys. Rev. Lett. 16 (1966) 772--774.
\newblock \href {https://doi.org/10.1103/PhysRevLett.16.772} {\path{doi:10.1103/PhysRevLett.16.772}}.

\bibitem{Copley:1969qn}
L.~A. Copley, G.~Karl, E.~Obryk, {Backward single pion photoproduction and the symmetric quark model}, Phys. Lett. B 29 (1969) 117--120.
\newblock \href {https://doi.org/10.1016/0370-2693(69)90261-5} {\path{doi:10.1016/0370-2693(69)90261-5}}.

\bibitem{Ono:1976mx}
S.~Ono, {Electromagnetic Properties of Baryons in the Quark Model}, Nucl. Phys. B 107 (1976) 522--534.
\newblock \href {https://doi.org/10.1016/0550-3213(76)90152-8} {\path{doi:10.1016/0550-3213(76)90152-8}}.

\bibitem{Workman:2022ynf}
R.~L. Workman, Others, {Review of Particle Physics}, PTEP 2022 (2022) 083C01.
\newblock \href {https://doi.org/10.1093/ptep/ptac097} {\path{doi:10.1093/ptep/ptac097}}.

\bibitem{Hey:1974qe}
A.~J.~G. Hey, J.~Weyers, {Quarks and the helicity structure of photoproduction amplitudes}, Phys. Lett. B 48 (1974) 69--72.

\bibitem{Cottingham:1978za}
W.~N. Cottingham, I.~H. Dunbar, {Baryon Multipole Moments in the Single Quark Transition Model}, Z. Phys. C 2 (1979) 41.

\bibitem{Burkert:2002zz}
V.~D. Burkert, et~al., {Single quark transition model analysis of electromagnetic nucleon resonance transitions in the [70,1-] supermultiplet}, Phys. Rev. C 67 (2003) 035204.

\bibitem{Carlson:2007xd}
C.~E. Carlson, M.~Vanderhaeghen, {Empirical transverse charge densities in the nucleon and the nucleon-to-Delta transition}, Phys. Rev. Lett. 100 (2008) 032004.

\bibitem{Burkert:2018oyl}
V.~D. Burkert, {N$^*$ Experiments and Their Impact on Strong QCD Physics}, Few Body Syst. 59~(4) (2018) 57.
\newblock \href {http://arxiv.org/abs/1801.10480} {\path{arXiv:1801.10480}}, \href {https://doi.org/10.1007/s00601-018-1378-7} {\path{doi:10.1007/s00601-018-1378-7}}.

\bibitem{Burkert:2018nvj}
V.~D. Burkert, {Jefferson Lab at 12 GeV: The Science Program}, Ann. Rev. Nucl. Part. Sci. 68 (2018) 405--428.

\bibitem{Li:1991yba}
Z.-P. Li, V.~Burkert, Z.-J. Li, {Electroproduction of the Roper resonance as a hybrid state}, Phys. Rev. D 46 (1992) 70--74.

\bibitem{Lanza:2021ayj}
L.~Lanza, A.~D'Angelo, {KY electroproduction at CLAS12}, Nuovo Cim. C 44~(2-3) (2021) 51.

\bibitem{Burkert:2018bqq}
V.~D. Burkert, L.~Elouadrhiri, F.~X. Girod, {The pressure distribution inside the proton}, Nature 557~(7705) (2018) 396--399.

\bibitem{Burkert:2023wzr}
V.~D. Burkert, et~al., {Colloquium: Gravitational form factors of the proton}, Rev. Mod. Phys. 95~(4) (2023) 041002.
\newblock \href {http://arxiv.org/abs/2303.08347} {\path{arXiv:2303.08347}}, \href {https://doi.org/10.1103/RevModPhys.95.041002} {\path{doi:10.1103/RevModPhys.95.041002}}.

\bibitem{CLAS:2023akb}
S.~Diehl, et~al., {First Measurement of Hard Exclusive \ensuremath{\pi}-\ensuremath{\Delta^{++}} Electroproduction Beam-Spin Asymmetries off the Proton}, Phys. Rev. Lett. 131~(2) (2023) 021901.
\newblock \href {http://arxiv.org/abs/2303.11762} {\path{arXiv:2303.11762}}, \href {https://doi.org/10.1103/PhysRevLett.131.021901} {\path{doi:10.1103/PhysRevLett.131.021901}}.

\bibitem{Diehl:2024bmd}
S.~Diehl, et~al., {Exploring Baryon Resonances with Transition Generalized Parton Distributions: Status and Perspectives} (5 2024).
\newblock \href {http://arxiv.org/abs/2405.15386} {\path{arXiv:2405.15386}}.

\bibitem{Braun:2009jy}
V.~M. Braun, et~al., {Electroproduction of the N*(1535) resonance at large momentum transfer}, Phys. Rev. Lett. 103 (2009) 072001.

\bibitem{Roberts:2007ji}
C.~D. Roberts, {Hadron Properties and Dyson-Schwinger Equations}, Prog. Part. Nucl. Phys. 61 (2008) 50--65.

\bibitem{Weinberg:1965zz}
S.~Weinberg, {Evidence That the Deuteron Is Not an Elementary Particle}, Phys. Rev. 137 (1965) B672--B678.
\newblock \href {https://doi.org/10.1103/PhysRev.137.B672} {\path{doi:10.1103/PhysRev.137.B672}}.

\bibitem{Matuschek:2020gqe}
I.~Matuschek, V.~Baru, F.-K. Guo, C.~Hanhart, {On the nature of near-threshold bound and virtual states}, Eur. Phys. J. A 57~(3) (2021) 101.
\newblock \href {http://arxiv.org/abs/2007.05329} {\path{arXiv:2007.05329}}, \href {https://doi.org/10.1140/epja/s10050-021-00413-y} {\path{doi:10.1140/epja/s10050-021-00413-y}}.

\bibitem{Baru:2021ldu}
V.~Baru, et~al., {Effective range expansion for narrow near-threshold resonances}, Phys. Lett. B 833 (2022) 137290.
\newblock \href {http://arxiv.org/abs/2110.07484} {\path{arXiv:2110.07484}}, \href {https://doi.org/10.1016/j.physletb.2022.137290} {\path{doi:10.1016/j.physletb.2022.137290}}.

\bibitem{Castillejo:1955ed}
L.~Castillejo, R.~H. Dalitz, F.~J. Dyson, {Low's scattering equation for the charged and neutral scalar theories}, Phys. Rev. 101 (1956) 453--458.
\newblock \href {https://doi.org/10.1103/PhysRev.101.453} {\path{doi:10.1103/PhysRev.101.453}}.

\bibitem{LHCb:2019kea}
R.~Aaij, et~al., {Observation of a narrow pentaquark state, $P_c(4312)^+$, and of two-peak structure of the $P_c(4450)^+$}, Phys. Rev. Lett. 122~(22) (2019) 222001.
\newblock \href {http://arxiv.org/abs/1904.03947} {\path{arXiv:1904.03947}}, \href {https://doi.org/10.1103/PhysRevLett.122.222001} {\path{doi:10.1103/PhysRevLett.122.222001}}.

\bibitem{LHCb:2021chn}
R.~Aaij, et~al., {Evidence for a new structure in the $J/\psi p$ and $J/\psi \bar{p}$ systems in $B_s^0 \to J/\psi p \bar{p}$ decays}, Phys. Rev. Lett. 128~(6) (2022) 062001.
\newblock \href {http://arxiv.org/abs/2108.04720} {\path{arXiv:2108.04720}}, \href {https://doi.org/10.1103/PhysRevLett.128.062001} {\path{doi:10.1103/PhysRevLett.128.062001}}.

\bibitem{LHCb:2016ztz}
R.~Aaij, et~al., {Model-independent evidence for $J/\psi p$ contributions to $\Lambda_b^0\to J/\psi p K^-$ decays}, Phys. Rev. Lett. 117~(8) (2016) 082002.
\newblock \href {http://arxiv.org/abs/1604.05708} {\path{arXiv:1604.05708}}, \href {https://doi.org/10.1103/PhysRevLett.117.082002} {\path{doi:10.1103/PhysRevLett.117.082002}}.

\bibitem{LHCb:2022ogu}
R.~Aaij, et~al., {Observation of a J/\ensuremath{\psi}\ensuremath{\Lambda} Resonance Consistent with a Strange Pentaquark Candidate in B-\textrightarrow{}J/\ensuremath{\psi}\ensuremath{\Lambda}p\textasciimacron{} Decays}, Phys. Rev. Lett. 131~(3) (2023) 031901.
\newblock \href {http://arxiv.org/abs/2210.10346} {\path{arXiv:2210.10346}}, \href {https://doi.org/10.1103/PhysRevLett.131.031901} {\path{doi:10.1103/PhysRevLett.131.031901}}.

\bibitem{LHCb:2020jpq}
R.~Aaij, et~al., {Evidence of a $J/\psi\Lambda$ structure and observation of excited $\Xi^-$ states in the $\Xi^-_b \to J/\psi\Lambda K^-$ decay}, Sci. Bull. 66 (2021) 1278--1287.
\newblock \href {http://arxiv.org/abs/2012.10380} {\path{arXiv:2012.10380}}, \href {https://doi.org/10.1016/j.scib.2021.02.030} {\path{doi:10.1016/j.scib.2021.02.030}}.

\bibitem{Shen:2024nck}
C.-W. Shen, et~al., {Exploration of the LHCb $P_c$ states and possible resonances in a unitary coupled-channel model}, Eur. Phys. J. C 84~(7) (2024) 764.
\newblock \href {http://arxiv.org/abs/2405.02626} {\path{arXiv:2405.02626}}, \href {https://doi.org/10.1140/epjc/s10052-024-13139-0} {\path{doi:10.1140/epjc/s10052-024-13139-0}}.

\bibitem{LHCb:2015yax}
R.~Aaij, et~al., {Observation of $J/\psi p$ Resonances Consistent with Pentaquark States in $\Lambda_b^0 \to J/\psi K^- p$ Decays}, Phys. Rev. Lett. 115 (2015) 072001.
\newblock \href {http://arxiv.org/abs/1507.03414} {\path{arXiv:1507.03414}}, \href {https://doi.org/10.1103/PhysRevLett.115.072001} {\path{doi:10.1103/PhysRevLett.115.072001}}.

\bibitem{Diakonov:1997mm}
D.~Diakonov, V.~Petrov, M.~V. Polyakov, {Exotic anti-decuplet of baryons: Prediction from chiral solitons}, Z. Phys. A 359 (1997) 305--314.
\newblock \href {http://arxiv.org/abs/hep-ph/9703373} {\path{arXiv:hep-ph/9703373}}, \href {https://doi.org/10.1007/s002180050406} {\path{doi:10.1007/s002180050406}}.

\bibitem{NA49:2003fxh}
C.~Alt, et~al., {Observation of an exotic S = -2, Q = -2 baryon resonance in proton proton collisions at the CERN SPS}, Phys. Rev. Lett. 92 (2004) 042003.
\newblock \href {http://arxiv.org/abs/hep-ex/0310014} {\path{arXiv:hep-ex/0310014}}, \href {https://doi.org/10.1103/PhysRevLett.92.042003} {\path{doi:10.1103/PhysRevLett.92.042003}}.

\bibitem{GRAAL:2004ndn}
V.~Kuznetsov, et~al., {$\eta$ photoproduction off the neutron at GRAAL: Evidence for a resonant structure at W = 1.67-GeV}, in: {4th International Workshop on the Physics of Excited Nucleons}, 2004, pp. 197--203.
\newblock \href {http://arxiv.org/abs/hep-ex/0409032} {\path{arXiv:hep-ex/0409032}}, \href {https://doi.org/10.1142/9789812702272_0022} {\path{doi:10.1142/9789812702272_0022}}.

\bibitem{Klempt:2004yz}
E.~Klempt, {Glueballs, hybrids, pentaquarks: Introduction to hadron spectroscopy and review of selected topics}, in: {18th Annual Hampton University Graduate Studies}, 2004, pp. 1--139.
\newblock \href {http://arxiv.org/abs/hep-ph/0404270} {\path{arXiv:hep-ph/0404270}}.

\bibitem{Kabana:2005tp}
S.~Kabana, {Pentaquarks: Review of the experimental evidence}, J. Phys. G 31 (2005) S1155--S1164.
\newblock \href {http://arxiv.org/abs/hep-ex/0503019} {\path{arXiv:hep-ex/0503019}}, \href {https://doi.org/10.1088/0954-3899/31/6/078} {\path{doi:10.1088/0954-3899/31/6/078}}.

\bibitem{Danilov:2005kt}
M.~Danilov, {Experimental review on pentaquarks}, Frascati Phys. Ser. 39 (2005) 193--209.
\newblock \href {http://arxiv.org/abs/hep-ex/0509012} {\path{arXiv:hep-ex/0509012}}.

\bibitem{Danilov:2008uxa}
M.~Danilov, R.~Mizuk, {Experimental review on pentaquarks}, Phys. Atom. Nucl. 71 (2008) 605--617.
\newblock \href {http://arxiv.org/abs/0704.3531} {\path{arXiv:0704.3531}}, \href {https://doi.org/10.1134/S1063778808040029} {\path{doi:10.1134/S1063778808040029}}.

\bibitem{Hicks:2012zz}
K.~H. Hicks, {On the conundrum of the pentaquark}, Eur. Phys. J. H 37 (2012) 1--31.
\newblock \href {https://doi.org/10.1140/epjh/e2012-20032-0} {\path{doi:10.1140/epjh/e2012-20032-0}}.

\bibitem{Liu:2014yva}
T.~Liu, Y.~Mao, B.-Q. Ma, {Present status on experimental search for pentaquarks}, Int. J. Mod. Phys. A 29~(13) (2014) 1430020.
\newblock \href {http://arxiv.org/abs/1403.4455} {\path{arXiv:1403.4455}}, \href {https://doi.org/10.1142/S0217751X14300208} {\path{doi:10.1142/S0217751X14300208}}.

\bibitem{Strakovsky:2024ppo}
I.~Strakovsky, {History of N(1680)}, Acta Physica Polonica B 56 (2025) 3--A12.
\newblock \href {http://arxiv.org/abs/2405.13749} {\path{arXiv:2405.13749}}.

\bibitem{Kim:2024tae}
H.-C. Kim, {Pentaquarks and Maxim V. Polyakov}, Acta Physica Polonica B 56 (2025) 3--A10.
\newblock \href {http://arxiv.org/abs/2411.13292} {\path{arXiv:2411.13292}}.

\bibitem{Praszalowicz:2024mji}
M.~Praszalowicz, {Odyssey of the elusive $\Theta^+$}, Acta Physica Polonica B 56 (2025) 3--A8.
\newblock \href {http://arxiv.org/abs/2411.08429} {\path{arXiv:2411.08429}}.

\bibitem{Praszalowicz:2024zsy}
M.~Praszalowicz, {20 years of $\Theta^+$}, PoS CORFU2023 (2024) 063.
\newblock \href {http://arxiv.org/abs/2405.09926} {\path{arXiv:2405.09926}}, \href {https://doi.org/10.22323/1.463.0063} {\path{doi:10.22323/1.463.0063}}.

\bibitem{ParticleDataGroup:2006fqo}
W.~M. Yao, et~al., {Review of Particle Physics}, J. Phys. G 33 (2006) 1--1232.
\newblock \href {https://doi.org/10.1088/0954-3899/33/1/001} {\path{doi:10.1088/0954-3899/33/1/001}}.

\bibitem{ParticleDataGroup:2008zun}
C.~Amsler, et~al., {Review of Particle Physics}, Phys. Lett. B 667 (2008) 1--1340.
\newblock \href {https://doi.org/10.1016/j.physletb.2008.07.018} {\path{doi:10.1016/j.physletb.2008.07.018}}.

\bibitem{Alston:1961zzd}
M.~H. Alston, et~al., {Study of Resonances of the $\Sigma-\pi$ System}, Phys. Rev. Lett. 6 (1961) 698--702.
\newblock \href {https://doi.org/10.1103/PhysRevLett.6.698} {\path{doi:10.1103/PhysRevLett.6.698}}.

\bibitem{Tripp:1968ukt}
R.~D. Tripp, R.~O. Bangerter, A.~Barbaro-Galtieri, T.~S. Mast, {Direct evidence for the multiplet assignments of $\Lambda(1520)$ and $\Lambda(1405)$}, Phys. Rev. Lett. 21 (1968) 1721--1724.
\newblock \href {https://doi.org/10.1103/PhysRevLett.21.1721} {\path{doi:10.1103/PhysRevLett.21.1721}}.

\bibitem{Hemingway:1984pz}
R.~J. Hemingway, {Production of $\Lambda(1405)$ in $K^- p$ Reactions at 4.2-{GeV}/$c$}, Nucl. Phys. B 253 (1985) 742--752.
\newblock \href {https://doi.org/10.1016/0550-3213(85)90556-5} {\path{doi:10.1016/0550-3213(85)90556-5}}.

\bibitem{Klempt:2020klf}
E.~Klempt, {Highlights for KLF — a personal view.}, {KLF collaboration meeting, December 2020.} (2020).

\bibitem{Zychor:2007mm}
I.~Zychor, {Studies of the $\Lambda(1405)$ in proton-proton collisions with ANKE at COSY-Julich}, eConf C070910 (2007) 310.
\newblock \href {http://arxiv.org/abs/0711.3084} {\path{arXiv:0711.3084}}.

\bibitem{Esmaili:2009rf}
J.~Esmaili, Y.~Akaishi, T.~Yamazaki, {Resonant formation of $\Lambda(1405)$ by stopped-$K^-$ absorption in deuteron}, Phys. Rev. C 83 (2011) 055207.
\newblock \href {http://arxiv.org/abs/0909.2573} {\path{arXiv:0909.2573}}, \href {https://doi.org/10.1103/PhysRevC.83.055207} {\path{doi:10.1103/PhysRevC.83.055207}}.

\bibitem{Moriya:2009mx}
K.~Moriya, R.~Schumacher, {Properties of the $\Lambda(1405)$ Measured at CLAS}, Nucl. Phys. A 835 (2010) 325--328.
\newblock \href {http://arxiv.org/abs/0911.2705} {\path{arXiv:0911.2705}}, \href {https://doi.org/10.1016/j.nuclphysa.2010.01.210} {\path{doi:10.1016/j.nuclphysa.2010.01.210}}.

\bibitem{Moriya:2010zz}
K.~Moriya, {Photoproduction of the $\Lambda(1405)$ measured at CLAS}, Prog. Theor. Phys. Suppl. 186 (2010) 234--239.
\newblock \href {https://doi.org/10.1143/PTPS.186.234} {\path{doi:10.1143/PTPS.186.234}}.

\bibitem{HADES:2012csk}
G.~Agakishiev, et~al., {Baryonic resonances close to the $\bar{K}N$ threshold: the case of $\Lambda$(1405) in $pp$ collisions}, Phys. Rev. C 87 (2013) 025201.
\newblock \href {http://arxiv.org/abs/1208.0205} {\path{arXiv:1208.0205}}, \href {https://doi.org/10.1103/PhysRevC.87.025201} {\path{doi:10.1103/PhysRevC.87.025201}}.

\bibitem{CLAS:2013zie}
H.~Y. Lu, et~al., {First Observation of the $\Lambda(1405)$ Line Shape in Electroproduction}, Phys. Rev. C 88 (2013) 045202.
\newblock \href {http://arxiv.org/abs/1307.4411} {\path{arXiv:1307.4411}}, \href {https://doi.org/10.1103/PhysRevC.88.045202} {\path{doi:10.1103/PhysRevC.88.045202}}.

\bibitem{Ahn:2010zzb}
J.~K. Ahn, {The nature of the $\Lambda(1405)$ from photoproduction at SPring-8/LEPS}, Nucl. Phys. A 835 (2010) 329--332.
\newblock \href {https://doi.org/10.1016/j.nuclphysa.2010.01.211} {\path{doi:10.1016/j.nuclphysa.2010.01.211}}.

\bibitem{Siebenson:2013rpa}
J.~Siebenson, L.~Fabbietti, {Investigation of the \ensuremath{\Lambda}(1405) line shape observed in pp collisions}, Phys. Rev. C 88 (2013) 055201.
\newblock \href {http://arxiv.org/abs/1306.5183} {\path{arXiv:1306.5183}}, \href {https://doi.org/10.1103/PhysRevC.88.055201} {\path{doi:10.1103/PhysRevC.88.055201}}.

\bibitem{Ren:2015bsa}
X.-L. Ren, E.~Oset, L.~Alvarez-Ruso, M.~J. Vicente~Vacas, {Antineutrino induced \ensuremath{\Lambda}(1405) production off the proton}, Phys. Rev. C 91~(4) (2015) 045201.
\newblock \href {http://arxiv.org/abs/1501.04073} {\path{arXiv:1501.04073}}, \href {https://doi.org/10.1103/PhysRevC.91.045201} {\path{doi:10.1103/PhysRevC.91.045201}}.

\bibitem{J-PARCE31:2022plu}
S.~Aikawa, et~al., {Pole position of \ensuremath{\Lambda}(1405) measured in d(K\ensuremath{-},n)\ensuremath{\pi}\ensuremath{\Sigma} reactions}, Phys. Lett. B 837 (2023) 137637.
\newblock \href {http://arxiv.org/abs/2209.08254} {\path{arXiv:2209.08254}}, \href {https://doi.org/10.1016/j.physletb.2022.137637} {\path{doi:10.1016/j.physletb.2022.137637}}.

\bibitem{Wickramaarachchi:2022mhi}
N.~Wickramaarachchi, R.~A. Schumacher, G.~Kalicy, {Decay of the \ensuremath{\Lambda}(1405) hyperon to \ensuremath{\Sigma}0\ensuremath{\pi}0 measured at GlueX}, EPJ Web Conf. 271 (2022) 07005.
\newblock \href {http://arxiv.org/abs/2209.06230} {\path{arXiv:2209.06230}}, \href {https://doi.org/10.1051/epjconf/202227107005} {\path{doi:10.1051/epjconf/202227107005}}.

\bibitem{Dalitz:1960du}
R.~H. Dalitz, S.~F. Tuan, {The phenomenological description of $K^-$-nucleon reaction processes}, Annals Phys. 10 (1960) 307--351.
\newblock \href {https://doi.org/10.1016/0003-4916(60)90001-4} {\path{doi:10.1016/0003-4916(60)90001-4}}.

\bibitem{Dalitz:1967fp}
R.~H. Dalitz, T.~C. Wong, G.~Rajasekaran, {Model calculation for $Y^{*0}(1405)$ resonance state}, Phys. Rev. 153 (1967) 1617--1623.
\newblock \href {https://doi.org/10.1103/PhysRev.153.1617} {\path{doi:10.1103/PhysRev.153.1617}}.

\bibitem{Dalitz:1959dn}
R.~H. Dalitz, S.~F. Tuan, {A possible resonant state in pion-hyperon scattering}, Phys. Rev. Lett. 2 (1959) 425--428.
\newblock \href {https://doi.org/10.1103/PhysRevLett.2.425} {\path{doi:10.1103/PhysRevLett.2.425}}.

\bibitem{Kaiser:1995eg}
N.~Kaiser, P.~B. Siegel, W.~Weise, {Chiral dynamics and the low-energy kaon - nucleon interaction}, Nucl. Phys. A 594 (1995) 325--345.
\newblock \href {http://arxiv.org/abs/nucl-th/9505043} {\path{arXiv:nucl-th/9505043}}, \href {https://doi.org/10.1016/0375-9474(95)00362-5} {\path{doi:10.1016/0375-9474(95)00362-5}}.

\bibitem{Kaiser:1996js}
N.~Kaiser, T.~Waas, W.~Weise, {SU(3) chiral dynamics with coupled channels eta and kaon photoproduction}, Nucl. Phys. A 612 (1997) 297--320.
\newblock \href {http://arxiv.org/abs/hep-ph/9607459} {\path{arXiv:hep-ph/9607459}}, \href {https://doi.org/10.1016/S0375-9474(96)00321-1} {\path{doi:10.1016/S0375-9474(96)00321-1}}.

\bibitem{Bhargava:1969jt}
S.~C. Bhargava, S.~H. Patil, {The nucleon and the Roper resonance}, Phys. Rev. 182 (1969) 1711--1713.
\newblock \href {https://doi.org/10.1103/PhysRev.182.1711} {\path{doi:10.1103/PhysRev.182.1711}}.

\bibitem{Oller:2000fj}
J.~A. Oller, U.~G. Mei{\ss}ner, {Chiral dynamics in the presence of bound states: Kaon nucleon interactions revisited}, Phys. Lett. B 500 (2001) 263--272.
\newblock \href {http://arxiv.org/abs/hep-ph/0011146} {\path{arXiv:hep-ph/0011146}}, \href {https://doi.org/10.1016/S0370-2693(01)00078-8} {\path{doi:10.1016/S0370-2693(01)00078-8}}.

\bibitem{Oset:2001cn}
E.~Oset, A.~Ramos, C.~Bennhold, {Low lying S = -1 excited baryons and chiral symmetry}, Phys. Lett. B 527 (2002) 99--105, [Erratum: Phys.Lett.B 530, 260--260 (2002)].
\newblock \href {http://arxiv.org/abs/nucl-th/0109006} {\path{arXiv:nucl-th/0109006}}, \href {https://doi.org/10.1016/S0370-2693(01)01523-4} {\path{doi:10.1016/S0370-2693(01)01523-4}}.

\bibitem{Jido:2002zk}
D.~Jido, E.~Oset, A.~Ramos, {Chiral dynamics of p wave in $K^- p$ and coupled states}, Phys. Rev. C 66 (2002) 055203.
\newblock \href {http://arxiv.org/abs/nucl-th/0208010} {\path{arXiv:nucl-th/0208010}}, \href {https://doi.org/10.1103/PhysRevC.66.055203} {\path{doi:10.1103/PhysRevC.66.055203}}.

\bibitem{Garcia-Recio:2002yxy}
C.~Garcia-Recio, J.~Nieves, E.~Ruiz~Arriola, M.~J. Vicente~Vacas, {S = -1 meson baryon unitarized coupled channel chiral perturbation theory and the S(01) $\Lambda(1405)$ and $\Lambda(1670)$ resonances}, Phys. Rev. D 67 (2003) 076009.
\newblock \href {http://arxiv.org/abs/hep-ph/0210311} {\path{arXiv:hep-ph/0210311}}, \href {https://doi.org/10.1103/PhysRevD.67.076009} {\path{doi:10.1103/PhysRevD.67.076009}}.

\bibitem{Jido:2003cb}
D.~Jido, et~al., {Chiral dynamics of the two $\Lambda(1405)$ states}, Nucl. Phys. A 725 (2003) 181--200.
\newblock \href {http://arxiv.org/abs/nucl-th/0303062} {\path{arXiv:nucl-th/0303062}}, \href {https://doi.org/10.1016/S0375-9474(03)01598-7} {\path{doi:10.1016/S0375-9474(03)01598-7}}.

\bibitem{Oller:2006jw}
J.~A. Oller, {On the strangeness -1 S-wave meson-baryon scattering}, Eur. Phys. J. A 28 (2006) 63--82.
\newblock \href {http://arxiv.org/abs/hep-ph/0603134} {\path{arXiv:hep-ph/0603134}}, \href {https://doi.org/10.1140/epja/i2006-10011-3} {\path{doi:10.1140/epja/i2006-10011-3}}.

\bibitem{Doring:2010rd}
M.~Döring, D.~Jido, E.~Oset, {Helicity Amplitudes of the $\Lambda(1670)$ and two $\Lambda(1405)$ as dynamically generated resonances}, Eur. Phys. J. A 45 (2010) 319--333.
\newblock \href {http://arxiv.org/abs/1002.3688} {\path{arXiv:1002.3688}}, \href {https://doi.org/10.1140/epja/i2010-11015-0} {\path{doi:10.1140/epja/i2010-11015-0}}.

\bibitem{Haidenbauer:2010ch}
J.~Haidenbauer, G.~Krein, U.-G. Mei{\ss}ner, L.~Tolos, {DN interaction from meson exchange}, Eur. Phys. J. A 47 (2011) 18.
\newblock \href {http://arxiv.org/abs/1008.3794} {\path{arXiv:1008.3794}}, \href {https://doi.org/10.1140/epja/i2011-11018-3} {\path{doi:10.1140/epja/i2011-11018-3}}.

\bibitem{Mai:2014xna}
M.~Mai, U.-G. Mei\ss{}ner, {Constraints on the chiral unitary $\bar KN$ amplitude from $\pi\Sigma K^+$ photoproduction data}, Eur. Phys. J. A 51~(3) (2015) 30.
\newblock \href {http://arxiv.org/abs/1411.7884} {\path{arXiv:1411.7884}}, \href {https://doi.org/10.1140/epja/i2015-15030-3} {\path{doi:10.1140/epja/i2015-15030-3}}.

\bibitem{Feijoo:2015yja}
A.~Feijoo, V.~K. Magas, A.~Ramos, {The $\bar{K} N \rightarrow K \Xi$ reaction in coupled channel chiral models up to next-to-leading order}, Phys. Rev. C 92~(1) (2015) 015206.
\newblock \href {http://arxiv.org/abs/1502.07956} {\path{arXiv:1502.07956}}, \href {https://doi.org/10.1103/PhysRevC.92.015206} {\path{doi:10.1103/PhysRevC.92.015206}}.

\bibitem{Cieply:2015pwa}
A.~Ciepl\'y, V.~Krej\v{c}i\v{r}\'\i{}k, {Effective model for in-medium $\bar{K}N$ interactions including the $L=1$ partial wave}, Nucl. Phys. A 940 (2015) 311--330.
\newblock \href {http://arxiv.org/abs/1501.06415} {\path{arXiv:1501.06415}}, \href {https://doi.org/10.1016/j.nuclphysa.2015.05.004} {\path{doi:10.1016/j.nuclphysa.2015.05.004}}.

\bibitem{Cieply:2016jby}
A.~Ciepl\'y, M.~Mai, U.-G. Mei\ss{}ner, J.~Smejkal, {On the pole content of coupled channels chiral approaches used for the $\bar{K}N$ system}, Nucl. Phys. A 954 (2016) 17--40.
\newblock \href {http://arxiv.org/abs/1603.02531} {\path{arXiv:1603.02531}}, \href {https://doi.org/10.1016/j.nuclphysa.2016.04.031} {\path{doi:10.1016/j.nuclphysa.2016.04.031}}.

\bibitem{Liu:2016wxq}
Z.-W. Liu, et~al., {Structure of the ${\Lambda(1405)}$ from Hamiltonian effective field theory}, Phys. Rev. D 95~(1) (2017) 014506.
\newblock \href {http://arxiv.org/abs/1607.05856} {\path{arXiv:1607.05856}}, \href {https://doi.org/10.1103/PhysRevD.95.014506} {\path{doi:10.1103/PhysRevD.95.014506}}.

\bibitem{Ramos:2016odk}
A.~Ramos, A.~Feijoo, V.~K. Magas, {The chiral $S = -1$ meson-baryon interaction with new constraints on the NLO contributions}, Nucl. Phys. A 954 (2016) 58--74.
\newblock \href {http://arxiv.org/abs/1605.03767} {\path{arXiv:1605.03767}}, \href {https://doi.org/10.1016/j.nuclphysa.2016.05.006} {\path{doi:10.1016/j.nuclphysa.2016.05.006}}.

\bibitem{Kamiya:2016jqc}
Y.~Kamiya, et~al., {Antikaon-nucleon interaction and $\Lambda$(1405) in chiral SU(3) dynamics}, Nucl. Phys. A 954 (2016) 41--57.
\newblock \href {http://arxiv.org/abs/1602.08852} {\path{arXiv:1602.08852}}, \href {https://doi.org/10.1016/j.nuclphysa.2016.04.013} {\path{doi:10.1016/j.nuclphysa.2016.04.013}}.

\bibitem{Miyahara:2018onh}
K.~Miyahara, T.~Hyodo, W.~Weise, {Construction of a local $\bar K N-\pi \Sigma-\pi \Lambda$ potential and composition of the $\Lambda(1405)$}, Phys. Rev. C 98~(2) (2018) 025201.
\newblock \href {http://arxiv.org/abs/1804.08269} {\path{arXiv:1804.08269}}, \href {https://doi.org/10.1103/PhysRevC.98.025201} {\path{doi:10.1103/PhysRevC.98.025201}}.

\bibitem{Bruns:2021krp}
P.~C. Bruns, A.~Ciepl\'y, {SU(3) flavor symmetry considerations for the $K^-N$ coupled channels system}, Nucl. Phys. A 1019 (2022) 122378.
\newblock \href {http://arxiv.org/abs/2109.03109} {\path{arXiv:2109.03109}}, \href {https://doi.org/10.1016/j.nuclphysa.2021.122378} {\path{doi:10.1016/j.nuclphysa.2021.122378}}.

\bibitem{Lu:2022hwm}
J.-X. Lu, L.-S. Geng, M.~Doering, M.~Mai, {Cross-Channel Constraints on Resonant Antikaon-Nucleon Scattering}, Phys. Rev. Lett. 130~(7) (2023) 071902.
\newblock \href {http://arxiv.org/abs/2209.02471} {\path{arXiv:2209.02471}}, \href {https://doi.org/10.1103/PhysRevLett.130.071902} {\path{doi:10.1103/PhysRevLett.130.071902}}.

\bibitem{Sadasivan:2022srs}
D.~Sadasivan, et~al., {New insights into the pole parameters of the $\Lambda(1380)$, the $\Lambda(1405)$ and the $\Sigma(1385)$}, Front. Phys. 11 (2023) 1139236.
\newblock \href {http://arxiv.org/abs/2212.10415} {\path{arXiv:2212.10415}}, \href {https://doi.org/10.3389/fphy.2023.1139236} {\path{doi:10.3389/fphy.2023.1139236}}.

\bibitem{Hyodo:2011ur}
T.~Hyodo, D.~Jido, {The nature of the $\Lambda(1405)$ resonance in chiral dynamics}, Prog. Part. Nucl. Phys. 67 (2012) 55--98.
\newblock \href {http://arxiv.org/abs/1104.4474} {\path{arXiv:1104.4474}}, \href {https://doi.org/10.1016/j.ppnp.2011.07.002} {\path{doi:10.1016/j.ppnp.2011.07.002}}.

\bibitem{Hyodo:2020czb}
T.~Hyodo, M.~Niiyama, {QCD and the strange baryon spectrum}, Prog. Part. Nucl. Phys. 120 (2021) 103868.
\newblock \href {http://arxiv.org/abs/2010.07592} {\path{arXiv:2010.07592}}, \href {https://doi.org/10.1016/j.ppnp.2021.103868} {\path{doi:10.1016/j.ppnp.2021.103868}}.

\bibitem{Nieves:2024dcz}
J.~Nieves, A.~Feijoo, M.~Albaladejo, M.-L. Du, {Lowest-lying ${\frac{1}{2}}^-$ and ${\frac{3}{2}}^-$$\Lambda_{Q}$ resonances: from the strange to the bottom sectors}, Prog. Part. Nucl. Phys. 137 (2024) 104118.
\newblock \href {http://arxiv.org/abs/2402.12726} {\path{arXiv:2402.12726}}, \href {https://doi.org/10.1016/j.ppnp.2024.104118} {\path{doi:10.1016/j.ppnp.2024.104118}}.

\bibitem{Zhuang:2024udv}
Z.~Zhuang, R.~Molina, J.-X. Lu, L.-S. Geng, {Pole trajectories of the {\ensuremath{\Lambda}}(1405) help establish its dynamical nature}, Sci. Bull. 70 (2025) 1953--1961.
\newblock \href {http://arxiv.org/abs/2405.07686} {\path{arXiv:2405.07686}}, \href {https://doi.org/10.1016/j.scib.2025.04.029} {\path{doi:10.1016/j.scib.2025.04.029}}.

\bibitem{Melnitchouk:2002eg}
W.~Melnitchouk, et~al., {Excited baryons in lattice QCD}, Phys. Rev. D 67 (2003) 114506.
\newblock \href {http://arxiv.org/abs/hep-lat/0202022} {\path{arXiv:hep-lat/0202022}}, \href {https://doi.org/10.1103/PhysRevD.67.114506} {\path{doi:10.1103/PhysRevD.67.114506}}.

\bibitem{Nemoto:2003ft}
Y.~Nemoto, N.~Nakajima, H.~Matsufuru, H.~Suganuma, {Negative parity baryons in quenched anisotropic lattice QCD}, Phys. Rev. D 68 (2003) 094505.
\newblock \href {http://arxiv.org/abs/hep-lat/0302013} {\path{arXiv:hep-lat/0302013}}, \href {https://doi.org/10.1103/PhysRevD.68.094505} {\path{doi:10.1103/PhysRevD.68.094505}}.

\bibitem{Burch:2006cc}
T.~Burch, et~al., {Excited hadrons on the lattice: Baryons}, Phys. Rev. D 74 (2006) 014504.
\newblock \href {http://arxiv.org/abs/hep-lat/0604019} {\path{arXiv:hep-lat/0604019}}, \href {https://doi.org/10.1103/PhysRevD.74.014504} {\path{doi:10.1103/PhysRevD.74.014504}}.

\bibitem{Takahashi:2009ik}
T.~T. Takahashi, M.~Oka, {Two-flavor lattice QCD study of $\Lambda(1405)$}, PoS LAT2009 (2009) 108.
\newblock \href {http://arxiv.org/abs/0911.2542} {\path{arXiv:0911.2542}}, \href {https://doi.org/10.22323/1.091.0108} {\path{doi:10.22323/1.091.0108}}.

\bibitem{Hall:2016kou}
J.~M.~M. Hall, et~al., {Light-quark contributions to the magnetic form factor of the $\Lambda(1405)$}, Phys. Rev. D 95~(5) (2017) 054510.
\newblock \href {http://arxiv.org/abs/1612.07477} {\path{arXiv:1612.07477}}, \href {https://doi.org/10.1103/PhysRevD.95.054510} {\path{doi:10.1103/PhysRevD.95.054510}}.

\bibitem{MartinezTorres:2012yi}
A.~Martinez~Torres, M.~Bayar, D.~Jido, E.~Oset, {Strategy to find the two $\Lambda(1405)$ states from lattice QCD simulations}, Phys. Rev. C 86 (2012) 055201.
\newblock \href {http://arxiv.org/abs/1202.4297} {\path{arXiv:1202.4297}}, \href {https://doi.org/10.1103/PhysRevC.86.055201} {\path{doi:10.1103/PhysRevC.86.055201}}.

\bibitem{Molina:2015uqp}
R.~Molina, M.~D\"oring, {Pole structure of the $\Lambda$(1405) in a recent QCD simulation}, Phys. Rev. D 94~(5) (2016) 056010, [Addendum: Phys.Rev.D 94, 079901 (2016)].
\newblock \href {http://arxiv.org/abs/1512.05831} {\path{arXiv:1512.05831}}, \href {https://doi.org/10.1103/PhysRevD.94.079901} {\path{doi:10.1103/PhysRevD.94.079901}}.

\bibitem{Pavao:2020zle}
R.~Pavao, et~al., {The negative-parity spin-1/2 \ensuremath{\Lambda} baryon spectrum from lattice QCD and effective theory}, Phys. Lett. B 820 (2021) 136473.
\newblock \href {http://arxiv.org/abs/2010.01270} {\path{arXiv:2010.01270}}, \href {https://doi.org/10.1016/j.physletb.2021.136473} {\path{doi:10.1016/j.physletb.2021.136473}}.

\bibitem{Anisovich:2020lec}
A.~V. Anisovich, et~al., {Hyperon III: $K^{-}p - \pi \Sigma $ coupled-channel dynamics in the $\Lambda (1405)$ mass region}, Eur. Phys. J. A 56~(5) (2020) 139.
\newblock \href {https://doi.org/10.1140/epja/s10050-020-00142-8} {\path{doi:10.1140/epja/s10050-020-00142-8}}.

\bibitem{Belle:2022ywa}
Y.~Ma, et~al., {First Observation of $\Lambda\pi^+$ and $\Lambda\pi^-$ Signals near the $K\,N(I=1)$ Mass Threshold in $\Lambda_c ^+\to\Lambda\pi^+\pi^+\pi^-$ Decay}, Phys. Rev. Lett. 130~(15) (2023) 151903.
\newblock \href {http://arxiv.org/abs/2211.11151} {\path{arXiv:2211.11151}}, \href {https://doi.org/10.1103/PhysRevLett.130.151903} {\path{doi:10.1103/PhysRevLett.130.151903}}.

\bibitem{Wang:2024jyk}
E.~Wang, et~al., {Review of the Low-Lying Excited Baryons $\Sigma^*(1/2^-)$}, Chin. Phys. Lett. 41~(10) (2024) 101401.
\newblock \href {http://arxiv.org/abs/2406.07839} {\path{arXiv:2406.07839}}, \href {https://doi.org/10.1088/0256-307X/41/10/101401} {\path{doi:10.1088/0256-307X/41/10/101401}}.

\bibitem{Lin:2025pyk}
J.-X. Lin, J.~Song, M.~Albaladejo, A.~Feijoo, E.~Oset, {Testing the nature of the $\Sigma^*(1430)$} (5 2025).
\newblock \href {http://arxiv.org/abs/2505.15650} {\path{arXiv:2505.15650}}.

\bibitem{BESIII:2024jgy}
M.~Ablikim, et~al., {Partial wave analysis of $\psi(3686)\to\Lambda\bar\Sigma^0\pi^0+c.c.$}, JHEP 02 (2025) 212.
\newblock \href {http://arxiv.org/abs/2408.00495} {\path{arXiv:2408.00495}}, \href {https://doi.org/10.1007/JHEP02(2025)212} {\path{doi:10.1007/JHEP02(2025)212}}.

\bibitem{Wang:2011kti}
Y.~F. Wang, M.~Ablikim, {Observation of $\chi_{cJ}$ decaying into the $p\bar{p}K^{+}K^{-}$ final state}, Phys. Rev. D 83 (2011) 112009.
\newblock \href {http://arxiv.org/abs/1103.2661} {\path{arXiv:1103.2661}}, \href {https://doi.org/10.1103/PhysRevD.83.112009} {\path{doi:10.1103/PhysRevD.83.112009}}.

\bibitem{Barnes:1977hg}
T.~Barnes, {Colored Quark and Gluon Constituents in the MIT Bag Model}, Nucl. Phys. B 158 (1979) 171--188.
\newblock \href {https://doi.org/10.1016/0550-3213(79)90194-9} {\path{doi:10.1016/0550-3213(79)90194-9}}.

\bibitem{Barnes:1982fj}
T.~Barnes, F.~E. Close, {Where are the Hermaphrodite Baryons?}, Phys. Lett. B 123 (1983) 89--92.
\newblock \href {https://doi.org/10.1016/0370-2693(83)90965-6} {\path{doi:10.1016/0370-2693(83)90965-6}}.

\bibitem{Golowich:1982kx}
E.~Golowich, E.~Haqq, G.~Karl, {Are there baryons which contain constituent gluons?}, Phys. Rev. D 28 (1983) 160, [Erratum: Phys.Rev.D 33, 859 (1986)].
\newblock \href {https://doi.org/10.1103/PhysRevD.28.160} {\path{doi:10.1103/PhysRevD.28.160}}.

\bibitem{Abdullah:1971aq}
T.~Abdullah, F.~E. Close, {Weak and electroproduction of nucleon resonances in the harmonic oscillator quark model}, Phys. Rev. D 5 (1972) 2332.
\newblock \href {https://doi.org/10.1103/PhysRevD.5.2332} {\path{doi:10.1103/PhysRevD.5.2332}}.

\bibitem{LeYaouanc:1972ju}
A.~Le~Yaouanc, L.~Oliver, O.~Pene, J.~C. Raynal, {Electroproduction of nucleon isobars in the quark model}, Nucl. Phys. B 37 (1972) 552--576.
\newblock \href {https://doi.org/10.1016/0550-3213(72)90519-6} {\path{doi:10.1016/0550-3213(72)90519-6}}.

\bibitem{Ping:2004wz}
R.~G. Ping, H.~C. Chiang, B.~S. Zou, {A Study of the Roper resonance as a hybrid state from $J /\psi$ decays}, Nucl. Phys. A 743 (2004) 149--169.
\newblock \href {http://arxiv.org/abs/nucl-th/0408007} {\path{arXiv:nucl-th/0408007}}, \href {https://doi.org/10.1016/j.nuclphysa.2004.07.004} {\path{doi:10.1016/j.nuclphysa.2004.07.004}}.

\bibitem{PMahlberg}
P.~Mahlberg, et~al., Decays of scalar excitations of the nucleon from new data on $\gamma p\to p\pi^0\pi^0$, in preparation (2025).

\bibitem{CBELSATAPS:2014wvh}
E.~Gutz, et~al., {High statistics study of the reaction $\gamma p\to p\pi^0\eta$}, Eur. Phys. J. A 50 (2014) 74.
\newblock \href {http://arxiv.org/abs/1402.4125} {\path{arXiv:1402.4125}}, \href {https://doi.org/10.1140/epja/i2014-14074-1} {\path{doi:10.1140/epja/i2014-14074-1}}.

\bibitem{Isgur:1985vy}
N.~Isgur, R.~Kokoski, J.~Paton, {Gluonic Excitations of Mesons: Why They Are Missing and Where to Find Them}, Phys. Rev. Lett. 54 (1985) 869.
\newblock \href {https://doi.org/10.1103/PhysRevLett.54.869} {\path{doi:10.1103/PhysRevLett.54.869}}.

\bibitem{Kokoski:1985is}
R.~Kokoski, N.~Isgur, {Meson Decays by Flux Tube Breaking}, Phys. Rev. D 35 (1987) 907.
\newblock \href {https://doi.org/10.1103/PhysRevD.35.907} {\path{doi:10.1103/PhysRevD.35.907}}.

\bibitem{E852:2004gpn}
J.~Kuhn, et~al., {Exotic meson production in the $f_1(1285)\pi^-$ system observed in the reaction $\pi^- p \to \eta \pi^+ \pi^- \pi^- p$ at 18-GeV/c}, Phys. Lett. B 595 (2004) 109--117.
\newblock \href {http://arxiv.org/abs/hep-ex/0401004} {\path{arXiv:hep-ex/0401004}}, \href {https://doi.org/10.1016/j.physletb.2004.05.032} {\path{doi:10.1016/j.physletb.2004.05.032}}.

\bibitem{E852:2004rfa}
M.~Lu, et~al., {Exotic meson decay to $\omega \pi^0 \pi^-$}, Phys. Rev. Lett. 94 (2005) 032002.
\newblock \href {http://arxiv.org/abs/hep-ex/0405044} {\path{arXiv:hep-ex/0405044}}, \href {https://doi.org/10.1103/PhysRevLett.94.032002} {\path{doi:10.1103/PhysRevLett.94.032002}}.

\bibitem{COMPASS:2018uzl}
M.~Aghasyan, et~al., {Light isovector resonances in $\pi^- p \to \pi^-\pi^-\pi^+ p$ at 190 GeV/${\it c}$}, Phys. Rev. D 98~(9) (2018) 092003.
\newblock \href {http://arxiv.org/abs/1802.05913} {\path{arXiv:1802.05913}}, \href {https://doi.org/10.1103/PhysRevD.98.092003} {\path{doi:10.1103/PhysRevD.98.092003}}.

\bibitem{Glozman:1999tk}
L.~Y. Glozman, {Parity doublets and chiral symmetry restoration in baryon spectrum}, Phys. Lett. B 475 (2000) 329--334.
\newblock \href {http://arxiv.org/abs/hep-ph/9908207} {\path{arXiv:hep-ph/9908207}}, \href {https://doi.org/10.1016/S0370-2693(00)00096-4} {\path{doi:10.1016/S0370-2693(00)00096-4}}.

\bibitem{Glozman:2002kq}
L.~Y. Glozman, {Chiral symmetry restoration and the string picture of hadrons}, Phys. Lett. B 541 (2002) 115--120.
\newblock \href {http://arxiv.org/abs/hep-ph/0204006} {\path{arXiv:hep-ph/0204006}}, \href {https://doi.org/10.1016/S0370-2693(02)02227-X} {\path{doi:10.1016/S0370-2693(02)02227-X}}.

\bibitem{Glozman:2002jf}
L.~Y. Glozman, {Chiral symmetry restoration in hadron spectra}, Prog. Part. Nucl. Phys. 50 (2003) 247--257.
\newblock \href {http://arxiv.org/abs/hep-ph/0210216} {\path{arXiv:hep-ph/0210216}}, \href {https://doi.org/10.1016/S0146-6410(03)00017-6} {\path{doi:10.1016/S0146-6410(03)00017-6}}.

\bibitem{Glozman:2007ek}
L.~Y. Glozman, {Restoration of chiral and U(1)A symmetries in excited hadrons}, Phys. Rept. 444 (2007) 1--49.
\newblock \href {http://arxiv.org/abs/hep-ph/0701081} {\path{arXiv:hep-ph/0701081}}, \href {https://doi.org/10.1016/j.physrep.2007.04.001} {\path{doi:10.1016/j.physrep.2007.04.001}}.

\bibitem{Klempt:2012fy}
E.~Klempt, B.~C. Metsch, {Multiplet classification of light-quark baryons}, Eur. Phys. J. A 48 (2012) 127.
\newblock \href {https://doi.org/10.1140/epja/i2012-12127-1} {\path{doi:10.1140/epja/i2012-12127-1}}.

\bibitem{Klempt:2002tt}
E.~Klempt, {Do parity doublets in the baryon spectrum reflect restoration of chiral symmetry?}, Phys. Lett. B 559 (2003) 144--152.
\newblock \href {http://arxiv.org/abs/hep-ph/0212241} {\path{arXiv:hep-ph/0212241}}, \href {https://doi.org/10.1016/S0370-2693(03)00338-1} {\path{doi:10.1016/S0370-2693(03)00338-1}}.

\bibitem{Klempt:2002vp}
E.~Klempt, {A Mass formula for baryon resonances}, Phys. Rev. C 66 (2002) 058201.
\newblock \href {http://arxiv.org/abs/hep-ex/0206012} {\path{arXiv:hep-ex/0206012}}, \href {https://doi.org/10.1103/PhysRevC.66.058201} {\path{doi:10.1103/PhysRevC.66.058201}}.

\bibitem{Hendry:1978cd}
A.~W. Hendry, {Analysis of Pion - Nucleon Scattering Up to 10\,{GeV}/$c$}, Phys. Rev. Lett. 41 (1978) 222.
\newblock \href {https://doi.org/10.1103/PhysRevLett.41.222} {\path{doi:10.1103/PhysRevLett.41.222}}.

\bibitem{GlueX:2020idb}
S.~Adhikari, et~al., {The GLUEX beamline and detector}, Nucl. Instrum. Meth. A 987 (2021) 164807.
\newblock \href {http://arxiv.org/abs/2005.14272} {\path{arXiv:2005.14272}}, \href {https://doi.org/10.1016/j.nima.2020.164807} {\path{doi:10.1016/j.nima.2020.164807}}.

\bibitem{Schlimme:2024eky}
S.~Schlimme, et~al., {The MESA physics program}, EPJ Web Conf. 303 (2024) 06002.
\newblock \href {http://arxiv.org/abs/2402.01027} {\path{arXiv:2402.01027}}, \href {https://doi.org/10.1051/epjconf/202430306002} {\path{doi:10.1051/epjconf/202430306002}}.

\bibitem{Accardi:2023chb}
A.~Accardi, et~al., {Strong interaction physics at the luminosity frontier with 22 GeV electrons at Jefferson Lab}, Eur. Phys. J. A 60~(9) (2024) 173.
\newblock \href {http://arxiv.org/abs/2306.09360} {\path{arXiv:2306.09360}}, \href {https://doi.org/10.1140/epja/s10050-024-01282-x} {\path{doi:10.1140/epja/s10050-024-01282-x}}.

\bibitem{Kobzarev:1962wt}
I.~Y. Kobzarev, L.~B. Okun, {Gravitational Interaction of Fermions}, Zh. Eksp. Teor. Fiz. 43 (1962) 1904--1909.

\bibitem{Pagels:1966zza}
H.~Pagels, {Energy-Momentum Structure Form Factors of Particles}, Phys. Rev. 144 (1966) 1250--1260.
\newblock \href {https://doi.org/10.1103/PhysRev.144.1250} {\path{doi:10.1103/PhysRev.144.1250}}.

\bibitem{Polyakov:2002yz}
M.~V. Polyakov, {Generalized parton distributions and strong forces inside nucleons and nuclei}, Phys. Lett. B 555 (2003) 57--62.
\newblock \href {http://arxiv.org/abs/hep-ph/0210165} {\path{arXiv:hep-ph/0210165}}, \href {https://doi.org/10.1016/S0370-2693(03)00036-4} {\path{doi:10.1016/S0370-2693(03)00036-4}}.

\bibitem{Kim:2024hhd}
J.-Y. Kim, K.~M. Semenov-Tian-Shansky, H.-Y. Won, S.~Son, C.~Weiss, {Complete definition of $N \rightarrow \Delta$ transition generalized parton distributions} (12 2024).
\newblock \href {http://arxiv.org/abs/2501.00185} {\path{arXiv:2501.00185}}.

\bibitem{Azizi:2020jog}
K.~Azizi, U.~\"Ozdem, {Gravitational form factors of N(1535) in QCD}, Nucl. Phys. A 1015 (2021) 122296.
\newblock \href {http://arxiv.org/abs/2012.06895} {\path{arXiv:2012.06895}}, \href {https://doi.org/10.1016/j.nuclphysa.2021.122296} {\path{doi:10.1016/j.nuclphysa.2021.122296}}.

\bibitem{Goharipour:2024atx}
M.~Goharipour, H.~Hashamipour, F.~Irani, K.~Azizi, {Impact of JLab data on the determination of GPDs at zero skewness and new insights from transition form factors $N\rightarrow \Delta$}, Phys. Rev. D 109~(7) (2024) 074042.
\newblock \href {http://arxiv.org/abs/2403.19384} {\path{arXiv:2403.19384}}, \href {https://doi.org/10.1103/PhysRevD.109.074042} {\path{doi:10.1103/PhysRevD.109.074042}}.

\end{thebibliography}

\end{document}